\documentclass[12pt,a4paper,twoside,openright]{book}

\usepackage{verbatim}
\usepackage{cancel}
\usepackage{amsmath}
\usepackage{mathrsfs}
\usepackage{amssymb}
\usepackage{amsfonts}
\usepackage{geometry}
\usepackage[pdftex]{graphicx}

\usepackage[figuresright]{rotating}
\usepackage[linktocpage=true,breaklinks=true]{hyperref}
\usepackage[style=numeric-comp,sorting=none]{biblatex}

\usepackage{lscape}
\usepackage{caption}
\usepackage{booktabs}
\usepackage{frontevr}

\def \met {\mbox{${\not}{E_T}$} }
\def \metr {\mbox{${\not}{E_T^{raw}}$}}
\def\lbl{\label}
\def\fb1{~fb$^{-1}$}
\def\gv{~GeV}
\def\gc{~GeV$/c$}
\def\gc2{~GeV$/c^{2}$}
\newcommand{\degree}{$^\circ$}

\renewcommand{\arraystretch}{1.15}

\bibliography{references} 
\begin{document}



\begin{frontespizio}
\Titolo{Evidence for Diboson Production in the Lepton plus Heavy Flavor Jets Final State at CDF}
\IlCandidato{Fedederico Sforza}
\Relatore{Prof. Giorgio Chiarelli}
\Date{XXIV Entrance, 2009--2011}
\end{frontespizio}

\thispagestyle{empty} 

\frontmatter 
\setcounter{page}{3}
\thispagestyle{plain}

\tableofcontents

\setcounter{secnumdepth}{3}
\setcounter{tocdepth}{3}
\sloppy
{\raggedright
\listoffigures
\listoftables
}
\fussy


\clearpage
\chapter{Introduction}

The topic of this thesis is the measurement of $W$ and $Z$ bosons associate production in the lepton plus neutrino plus Heavy Flavor ($HF$) quarks final state:
\begin{equation}
  p\bar{p}\to WW/WZ \to \ell\nu + HF,
\end{equation}
identified by the CDF II experiment at the Tevatron collider at $\sqrt{s}=1.96$~TeV.

The associate production of the massive vector bosons $W$ and $Z$, as well as the different final states, are predicted by the Standard Model of the elementary particles (SM). The SM, briefly described in Chapter~\ref{chap:physics}, is an extremely successful theory in which a minimal set of equations explains most of the known interactions. However the mechanism responsible for the mass of the particles still needs to be fully proved by the discovery of the (predicted) Higgs boson.

The CDF II experiment, described in Chapter~\ref{chap:detector}, set tight constraints on the existence of the Higgs boson. Most of the sensitivity for the low mass Higgs boson production comes from:
\begin{equation}
  p\bar{p}\to WH\to \ell\nu+b\bar{b}.
\end{equation}
This happens to be extremely similar to the process studied in this thesis\footnote{I personally contributed to the $p\bar{p}\to WH\to \ell\nu+b\bar{b}$ CDF result as a developer of the common analysis framework described in Appendix~\ref{chap:AppWHAM}.}, thus it allows to test, on a well known physics process, the correctness of the analysis procedure used in the Higgs search. 

However, the observation of diboson production in the  $\ell\nu + HF$ final state is not a simple task and, before this work (in particular its preliminary version described in Appendix~\ref{App:7.5} and presented in 2011), no evidence was observed at a hadron collider experiment.

The actual analysis procedure is divided in four steps: object identification, event selection, background estimate and statistical analysis. 

First, a set of advanced identification algorithms, described in Chapter~\ref{chap:objects}, is exploited for the recognition of the final state objects: one charged lepton ($\ell$), a neutrino ($\nu$), and two high energy jets, of which at least one tagged by the identification of the secondary decay vertex produced by a $HF$ hadron. Key element of the selection, the $HF$-tagging allows an efficient identification of \mbox{$Z\to c\bar{c}/b\bar{b}$} signal candidates together with $W\to c\bar{s}$ candidates.

Further specific event selection criteria, described in Chapter~\ref{chap:sel}, are imposed to maximise the accepted signal events while keeping the background under control. An original technique, based on the support vector machine algorithm and described in Appendix~\ref{chap:AppSvm}, was developed to suppress the multi-jet events, a background, difficult to model, due to events in which no real $W\to \ell\nu$ decay is present.

Successively, as described in Chapter~\ref{chap:bkg}, the total background is estimated. A variety of methods are used for the different background sources. We exploit Monte Carlo information for several backgrounds (e.g. for top quark production), a completely data-driven approach for the multi-jet contamination and a combination of data and Monte Carlo information for the $W+$ jets background estimate.

After the full selection a large irreducible background fraction (i.e. with the same final state signature) remains. In particular, the total $W+HF$ non-resonant production is estimated to be more than a factor twenty larger than the expected signal. The shape analysis of the di-jet invariant mass distribution, $M_{Inv}(jet1, jet2)$, allows the extraction of the combined diboson signal, but the separation between $WW$ and $WZ$ contributions is still not feasible due to the close mass of the $W$ and the $Z$ bosons. To overcome this last issue, we exploited the discriminative power of a flavor separation neural network (KIT-NN described in Section~\ref{sec:kitnn}) to classify the single-tagged events according to their $c$ quark or $b$ quark origin. This was not necessary for the double-tagged selection as it is dominated only by events with two $b$ quarks in the final state.

Chapter~\ref{chap:StatRes} describes the statistical analysis of the bi-dimensional distribution $M_{Inv}(jet1, jet2)$ $vs$ KIT-NN, of the single-tagged events, together with the simple $M_{Inv}(jet1, jet2)$ distribution, of the double-tagged events. This allowed both the measurement of the cross sections for the total diboson associate production processes ($WW+ WZ/ZZ$) and to separate the $WW$ and $WZ/ZZ$ contributions.

A summary of the results and the conclusions are reported in Chapter~\ref{chap:conclusions}.

\mainmatter 
\clearpage
\chapter{Theoretical and Experimental Overview}\label{chap:physics}

The goal of particle physics is the understanding of the principles of Nature.

The quest is pursued through the scientific method: the observation of a phenomenon is explained by a hypothesis that must be, successively, verified or rejected by experimental evidences.

In this prospect, the observed phenomenon is the existence itself of the atomic and sub-atomic structure of matter, the hypothesis is the Standard Model theory of Elementary Particles and Fundamental Interaction (SM) while the experimental tools are the high energy physics colliders and detectors, available nowadays.  

The main infrastructure of the SM~\cite{pdg2010, glasgow, weinberg_lep} was developed in the 70's and, since then, it showed to be a very successful theory. One of its main success was the prediction of new elementary particles, later observed at hadron collider experiments. The discovery of the $W$ and $Z$ force carrier vector bosons~\cite{ua1_w_obs, ua2_w_obs, ua1_z_obs,ua2_z_obs} and of the $top$ quark~\cite{CDF_top_obs,d0_top_obs} shed light on the fundamental structure of the matter. 

Strengthened by these results, SM describes the {\em electromagnetic}, {\em weak} and {\em strong} interactions, three of the four fundamental forces that compose the physics description of Nature. The fourth force, gravitational interaction, is left out but it is negligible at atomic and subatomic scale. 

Despite the great success of the SM, one predicted particle has not yet been observed: the \emph{Higgs boson}, an essential element for the inclusion of the mass of the particles in the equation of motion~\cite{higgs1964,englert1964}. Because of this, a considerable effort is ongoing to prove or disprove the existence of the Higgs boson\footnote{Some hints of its existence are confirmed by the present experiments~\cite{tev_cmb, atlas_cmb, cms_cmb}.} and the analysis of the data collected by the CDF II experiment, situated at the Tevatron $p\bar{p}$ collider, plays a relevant role in it~\cite{cdf_cmb}.

In the scenario~\cite{Tev4LHC} of a low-mass Higgs ($m_H\lesssim 135$\gc2), one of the most relevant CDF search channels is the $WH$ associate production with a $\ell\nu+b\bar{b}$ final state. In light of this, the  diboson decay channel considered for the presented analysis becomes a perfect benchmark for the Higgs boson search. The accurate SM prediction for the diboson production and decay can be used as a standard comparison for an unknown process. 

This Chapter introduces the relevant aspects of the SM theory (Section\ref{sec:sm_teo}), some of the latest results of the Higgs boson search (Section\ref{sec:higgs_search}) and several, diboson related, experimental confirmation of the SM validity (Section\ref{sec:dib_res}).

\section{The Standard Model of Elementary Particles}\label{sec:sm_teo}

The SM is defined by the language of mathematics and theoretical physics so that it can be used to produce accurate predictions that have to be verified by the experiments. 
In this language a particle is defined by a local quantum field. If no interaction is present, the free field is described by only two quantum numbers, the spin and the mass; if interactions are presents, the {\em Gauge symmetries} can elegantly describe them: new quantum numbers classify the type and the strength of force while new particles, force-mediator vector bosons, are used to propagate the interaction.

The fundamental building blocks of matter, observed up to now, are the spin-$1/2$ fields (\emph{fermionic}), named \emph{quarks} and \emph{leptons}, and the spin-$1$ ({\em vector}) fields, named \emph{gauge bosons}. The leptons are divided into three \emph{generations}, or \emph{families}, and are
grouped in a left weak isospin doublet\footnote{See Section~\ref{sec:SM} for 
the explanation of the weak isospin quantum number.} and a right weak isospin singlet. Also quarks are divided into three \emph{flavor families} but weak 
isospin classification mixes quark doublets of different families. Quarks are also subject to the strong interaction, described by the \emph{color} quantum number. Finally the {\em charge} quantum number is used, for both quarks and leptons, to describe the electromagnetic interaction. 

\begin{figure}[!ht]
\begin{center}
\includegraphics[width=0.7\textwidth]{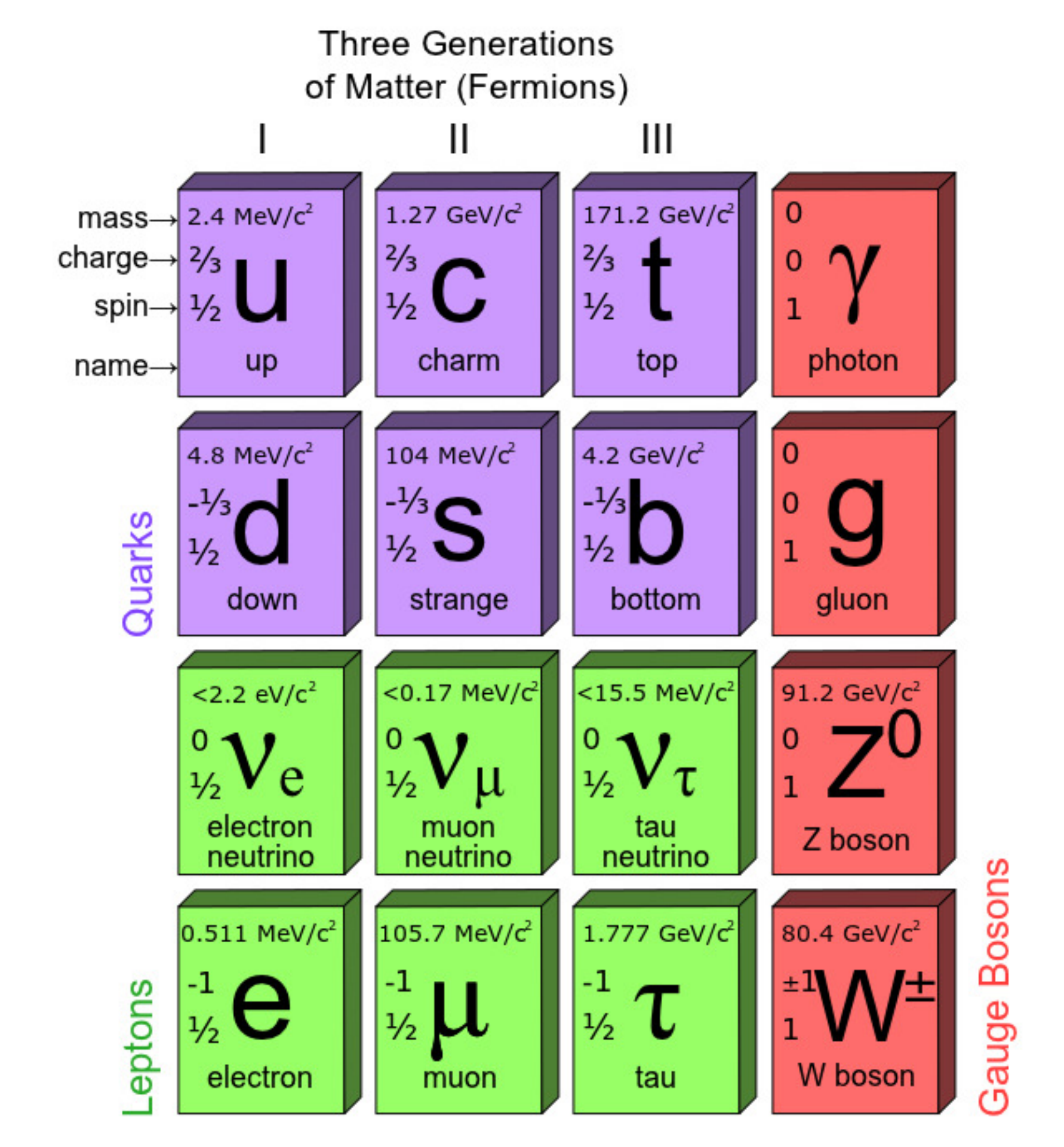}
\caption[SM Particles and Their Properties]{Quarks, leptons and gauge bosons in Standard Model and some of their 
characteristics~\cite{pdg2010}, for each particle the corresponding antiparticle exists.}\label{fig:particles}
\end{center}
\end{figure}

The force mediators are $W^\pm$, $Z^0$, $\gamma$, that carry electroweak 
force, and $g$ (\emph{gluons}), which mediate strong interaction. A short 
summary of the SM fundamental particles is reported in Figure~\ref{fig:particles}.

\subsection{Gauge Theory Example: QED}

The Quantum Electro-Dynamic (QED) is a perfect example to explain the importance of gauge invariance.

The equation that describes free fermionic fields is the {\em Dirac} Lagrangian:
\begin{equation}\label{eq:dirac}
\mathscr{L}(x)=\bar{\psi}(x)(i\gamma^{\mu}\partial_{\mu}- m)\psi(x),
\end{equation}
where $\psi$ is the Dirac field of mass $m$ and $\gamma^{\mu}$ are the Dirac's 
matrices. Equation~\ref{eq:dirac} satisfies the \emph{global} $U(1)$ symmetry
transformation: 
\begin{equation}
\psi(x)\to e^{iQ\alpha}\psi(x),
\end{equation}
with the electric charge $Q$ and the space independent parameter $\alpha$ ($x$ is a space-time 4-vector).
The \emph{Noether theorem}~\cite{qftbook_weinberg} states that when a symmetry 
appears in a Lagrangian there is a corresponding conserved current. In the case of the Dirac field:
\begin{equation}
\partial_{\mu}j^{\mu} = 0\quad\mathrm{with}\quad j^{\mu} = -Q\bar{\psi}\gamma^{\mu}\psi,
\end{equation}
describes the conservation of charge, i.e the time component of the current 4-vector $j^{\mu}$, integrated over the space, is a constant.

An elegant way to introduce interaction in the free Lagrangian is to shift from
the global, i.e. space independent, $U(1)$ transformation to a \emph{local} 
$U(1)$ transformation, i.e. with a space dependent parameter $\alpha(x)$:
\begin{equation}
\psi(x)\to e^{iQ\alpha(x)}\psi(x).
\end{equation}
To maintain the gauge invariance condition in the Lagrangian~\ref{eq:dirac}, a 
covariant derivative $D_{\mu}$ is introduced:
\begin{equation}\label{eq:em_derivative}
\partial_{\mu}\to D_{\mu}=\partial_{\mu}+iQA_{\mu}, \qquad
D_{\mu}\psi(x)\to e^{iQ\alpha(x)}D_{\mu}\psi(x),
\end{equation}
where the new vector field $A_{\mu}$ is defined to transform in the following manner:
\begin{equation}
A_{\mu}\to A_{\mu}-\frac{1}{Q}\partial_{\mu}\alpha(x)\mathrm{.}
\end{equation}
Equations~\ref{eq:em_derivative} and~\ref{eq:dirac} can be composed to give the final QED Lagrangian:
\begin{equation}\label{eq:L_qed}
\mathscr{L}_{QED}=\bar{\psi}(x)(i\gamma^{\mu}D_{\mu}- m)\psi(x) - \frac{1}{4}F_{\mu\nu}F^{\mu\nu},
\end{equation}
where $F_{\mu\nu}\equiv\partial_{\mu}A_{\nu}-\partial_{\nu}A_{\mu}$ is the covariant kinetic term of $A_{\mu}$.
The Dirac equation of motion for a field $\psi$ undergoing electromagnetic interaction is obtained by applying the Euler-Lagrange equation~\cite{qftbook_weinberg} to the QED Lagrangian:
\begin{equation}
(i\gamma^{\mu}\partial_{\mu}- m)\psi(x) = Q\gamma^{\mu}A_{\mu}\psi(x)\mathrm{,}
\end{equation}
the force is mediated by the massless vector field $A_{\mu}$. A mass term in 
the form $\frac{1}{2}m^{2}A_{\mu}A^{\mu}$ would break apart gauge invariance of
Equation~\ref{eq:L_qed}, indeed this is consistent with zero mass of the photon.

\subsection{Standard Model Theory}\label{sec:SM}

The leptonic sector of the SM\footnote{Only electroweak interaction on the leptons is considered here to simplify the discussion.} is based on the gauge group:
\begin{equation}
  SU(2)\otimes U(1)\textrm{,}  
\end{equation}
where $SU(2)$ is the non-Abelian group used in the spin algebra, and $U(1)$ is the Abelian group equivalent to the one used in QED. 
The quantum number arising from $SU(2)$ is the \emph{weak isospin}, $\vec{T}$, and the one arising from $U(1)$ is {\em hypercharge}, $Y$. They are related to the observed charge of real particles, $Q$, by the 
the Gell-Mann-Nishijima equation:
\begin{equation}
Q = T_{3}+\frac{Y}{2}\mathrm{.}
\end{equation}
where $T_{3}$ is the third component of weak isospin.

Electroweak interaction can be explained with a simplified model containing only two spin $1/2$, elementary, massless, fermions, $f$ and $f'$, such that 
$Q_{f}=Q_{f'}+1$ ($Q$ is the electric charge). Weak interaction is built from V-A currents, i.e. left and right components are 
defined and collected into a left doublet field and into two right singlet fields:
\begin{equation}\label{eq:smL}
\psi_{1} \equiv \left( \begin{array}{c} f_{L}(x) \\ f'_{L}(x) \end{array}\right), \qquad \psi_{2} \equiv f_{R}(x) 
\qquad \psi_{3} \equiv f'_{R}(x),
\end{equation}
with:
\begin{equation}
f_{L,R}(x)=\frac{1}{2}(1\pm\gamma_{5})f(x), \qquad 
\bar{f}_{L,R}(x)=\frac{1}{2}\bar{f}(x)(1\pm\gamma_{5}),
\end{equation}
\begin{equation}
f'_{L,R}(x)=\frac{1}{2}(1 \pm \gamma_{5})f'(x), \qquad 
\bar{f'}_{L,R}(x)=\frac{1}{2}\bar{f'}(x)(1 \pm \gamma_{5}).
\end{equation}
All the leptonic sector of the SM is explained by such pattern: the left doublet with $T_{3}=\pm 1/2$, $Y = 1$ is the charged lepton $f$
plus the corresponding neutrino $f'$, while the right singlet with $T_{3}=0$, $Y = -2$ is only the charged lepton.

The electroweak interaction is introduced through $SU(2)\otimes U(1)$ gauge transformation:
\begin{equation}
\psi_{j}(x)\to\psi'_{j}(x)= e^{i\frac{\tau}{2}\cdot\vec{\alpha}(x) + iY_{j}\beta(x)}\psi_{j}(x),
\end{equation}
of the free field Lagrangian:
\begin{equation}\label{eq:freeL}
\mathscr{L_{0}}(x)=\sum_{j=1}^{3}i\bar{\psi}_{j}(x)\gamma^{\mu}\partial_{\mu}\psi_{j}(x)\textrm{,}
\end{equation}
where a covariant derivative is also introduced to maintain gauge invariance. The result is:
\begin{equation}\label{eq:SML}
\mathscr{L}_{I}(x)=\sum_{j=1}^{3}i\bar{\psi}_{j}(x)\gamma^{\mu}D_{\mu}^{j}\psi(x)_{j} - \frac{1}{4}\vec{W}_{\mu,\nu} \vec{W}^{\mu,\nu} - \frac{1}{4} B_{\mu,\nu}B^{\mu,\nu},
\end{equation}
\begin{equation}\label{eq:bosons}
\mathrm{with}\qquad D_{\mu}^{j}= \partial_{\mu}-ig\frac{\tau}{2}\cdot\vec{W}_{\mu}(x)-ig'Y_{j}B_{\mu}(x)\mathrm{,}
\end{equation}
\begin{equation}\label{eq:tensor}
\mathrm{and}\quad \vec{W}_{\mu,\nu}=\partial_{\mu}\vec{W}_{\nu} - \partial_{\nu} \vec{W}_{\mu} + g \vec{W}_{\mu}\times\vec{W}_{\nu}\textrm{,}\quad B_{\mu,\nu}= \partial_{\mu}B_{\nu} - \partial_{\nu} B_{\mu}.
\end{equation}
Equation~\ref{eq:bosons} contains three vector bosons ($\vec{W}_{\mu}$) from the $SU(2)$ 
generators, one vector boson ($B_{\mu}$) from the $U(1)$ generator and four coupling constants: 
\begin{equation}
  g,\quad g'Y_{j} \quad \mathrm{with}\quad j=1,2,3\mathrm{.}
\end{equation}
After some algebra the Lagrangian~\ref{eq:SML} can be written in the form:
\begin{equation}
\mathscr{L}_{I}(x)= \mathscr{L}_{CC}(x) + \mathscr{L}_{NC}(x)\mathrm{,}
\end{equation}
with a {\em charged current} contribution ($\mathscr{L}_{CC}$) and a {\em neutral
current contribution} ($\mathscr{L}_{NC}$). The charged current contribution is seen only by left doublet fields:
\begin{equation}\label{eq:W}
\mathscr{L}_{CC}(x)=\frac{g}{2\sqrt{2}}\Big\{\bar{f}(x)\gamma^{\mu}(1-\gamma_{5})f'(x)\frac{1}{\sqrt{2}}W_{\mu}^{+}(x) + h.c.\Big\},
\end{equation}
with $W_{\mu}^{+}(x)$ defined by a linear combination of $W_{\mu}^{1}(x)$ and $W_{\mu}^{2}(x)$. Equation~\ref{eq:W} defines the Lagrangian for charged current 
interactions mediated by the $W$ boson.

The fermion coupling to $Z^{0}$ field and photon ($A$) field is produced in a similar way, by an appropriate orthogonal linear combination of neutral vector fields $B_{\mu}(x)$
and $W_{\mu}^{0}(x)$:
\begin{equation}
\mathscr{L}_{NC}(x)= \mathscr{L}_{NC}^{A}(x) + \mathscr{L}_{NC}^{Z}(x)\textrm{,}
\end{equation}
where:
\begin{eqnarray}
\mathscr{L}_{NC}^{A}(x)=\sum^{3}_{j=1}\bar{\psi}_{j}(x)\gamma^{\mu}\big[g\frac{\tau_{3}}{2}\sin \theta_{W}
+ g'Y_{j}\cos \theta_{W}\big]\psi_{j}(x)A_{\mu}(x),\\
\mathscr{L}_{NC}^{Z}(x)=\sum^{3}_{j=1}\bar{\psi}_{j}(x)\gamma^{\mu}\big[g\frac{\tau_{3}}{2}\cos \theta_{W}
+ g'Y_{j}\sin \theta_{W}\big]\psi_{j}(x)Z_{\mu}(x),
\end{eqnarray}
the parameter $\theta_{W}$ is named Weinberg angle and the generic four coupling constants, arising from the SM group structure, have now a physical meaning:
\begin{equation}
g\sin \theta_{W} = e,
\end{equation}
\begin{equation}
g'\cos \theta_{W}Y_{1}=e(Q_{f}-1/2), \qquad g'\cos \theta_{W}Y_{2}=eQ_{f},
\end{equation}
\begin{equation}
g'\cos \theta_{W}Y_{3}=eQ_{f'}\textrm{.}
\end{equation}
Previous equations are the core of the Standard Model. However one problem remains as no mass term appears for any of the fields: the spontaneous symmetry breaking and Higgs mechanism can generate the mass term without breaking the gauge invariance.

\subsection{Spontaneous Symmetry Breaking}

Spontaneous symmetry breaking can be applied to Equation~\ref{eq:smL} to give mass to $W^{\pm}$ and $Z^{0}$ bosons. The actual application procedure is named {\em Higgs mechanism}: two complex scalar fields are introduced such that they form an iso-doublet with respect to $SU(2)$:
\begin{equation}\label{eq:phi_doublet}
\phi(x)\equiv\left(\begin{array}{c} \phi^{+}(x) \\\phi^{0}(x) \end{array}\right),
\end{equation}
the field $\phi^{+}(x)$ is the charged component of the doublet and $\phi^{0}(x)$ is neutral component. The Higgs potential, $ V_{H}(x)$, is then defined as:
\begin{equation}\label{eq:h_pot}
 V_{H}(x)\equiv -\mu^{2}\phi^{\dagger}(x)\phi(x)-h\big[\phi^{\dagger}(x)\phi(x)\big] ^{2},
\end{equation}
with $h>0$ and $\mu^{2}<0$. The neutral scalar field $\phi^{0}(x)$ has an unconstrained (i.e. to be obtained from measurements) vacuum expectation value of $\frac{\lambda}{\sqrt{2}}$, so that (at first order)
the field  $\phi(x)$ is:
\begin{equation}\label{eq:doublet}
\phi(x)=e^{\frac{i}{\lambda}\vec{\tau}\cdot\vec{\theta(x)}}\left(\begin{array}{c} 0 \\\frac{1}{\sqrt{2}}\big(\lambda + \chi(x)\big) \end{array}\right),
\end{equation}
where the $SU(2)$ gauge freedom is explicit. This permits to \emph{gauge away} three of the four components of field $\phi(x)$ leaving only one real scalar field: 
\begin{equation}\label{eq:}
  \phi^{0}(x)= \frac{1}{\sqrt{2}}\big(\lambda + \chi(x)\big).
\end{equation}
The explicit evaluation of Equation~\ref{eq:h_pot} and the coupling of $\phi^{0}(x)$ with the electroweak force carriers ($W^{\pm}$, $Z^0$) gives the last piece of the SM Lagrangian:
\begin{eqnarray}\label{eq:finalL}
\mathscr{L}(x)&=& \frac{1}{4}g^{2}\lambda^{2}W_{\mu}^{\dagger}(x)W^{\mu}(x) + \frac{1}{1}(g^{2}+g'^{2})\lambda^{2}Z_{\mu}(x)Z^{\mu}\\
&&+\frac{1}{2}g^{2}\lambda W^{\dagger}_{\mu}(x)W^{\mu}(x)\chi(x)+\frac{1}{4}g^{2}W^{\dagger}_{\mu}W^{\mu}\chi^{2}(x)\nonumber\\
&&+\frac{1}{4}(g^{2}+g'^{2})\lambda Z_{\mu}(x)Z^{\mu}(x)\chi(x)+\frac{1}{8}g^{2}Z_{\mu}(x)Z^{\mu}(x)\chi^{2}(x)\nonumber\\
&&+\frac{1}{2}\big[\partial^{\mu}\chi(x)\partial_{\mu}\chi(x) + 2\mu^{2}\chi^{2}(x)\big]\nonumber\\
&&+\frac{\mu^{2}}{\lambda}\chi^{3}(x)+\frac{\mu^{2}}{4\lambda^{2}}\chi^{4}(x)-\frac{1}{4}\lambda^{2}\mu^{2}\nonumber\textrm{.}
\end{eqnarray}
We conclude that the $Z^{0}$ and $W^{\pm}$ bosons have acquired mass:
\begin{equation}
  M_{W}=\frac{1}{2}\lambda g,  
\end{equation}
\begin{equation}
  M_{Z}=\frac{1}{2}\lambda \sqrt{g^{x}+g'^{x}}=\frac{1}{2}\frac{\lambda g}{\cos \theta_{w} },  
\end{equation}
some parameters are now constrained, for example:
\begin{equation}
  M_{Z}=\frac{M_{W}}{cos\theta{w}}\geqslant M_{W},
\end{equation}
\begin{equation}
  \frac{G_{F}}{\sqrt{2}}=\frac{g^{2}}{8M^{2}_{W}},
\end{equation}
while the Higgs mass, $M_{\chi}=\sqrt{-2\mu^{2}}$ ($m_{H}$ is also used), remains a free parameter to be measured by the experiments. 
The Higgs mechanism can generate also fermion masses if a Yukawa coupling is added:
\begin{eqnarray}\label{eq:h_yukawa}
\mathscr{L_{f}}(x)&=&c_{f'}\Bigg[(\bar{f}(x),\bar{f}'(x))_{L} \left(\begin{array}{c} \phi^{+}(x) \\\phi^{0}(x) \end{array}\right)\Bigg]f_{R}'(x)\\
&&+ c_{f}\Bigg[(\bar{f}(x),\bar{f}'(x))_{L} \left(\begin{array}{c} -\bar{\phi}^{0}(x) \\\phi^{-}(x) \end{array}\right)\Bigg]f_{R}(x) + h.c.\mathrm{,}\nonumber
\end{eqnarray}
therefore, after symmetry breaking, fermion masses have the form:
\begin{equation}
m_{f}=-c_{f}\frac{\lambda}{\sqrt{2}}\mathrm{,}\qquad m_{f'}=-c_{f'}\frac{\lambda}{\sqrt{2}}\mathrm{,}
\end{equation}
where the constants $c_{f}$ and $c_{f'}$ can be derived by the measurements of the fermion masses.

\section{Higgs Boson Search and Results}\label{sec:higgs_search}

The mechanism that generates the mass of all the SM particles is a key element for the understanding of Nature, therefore it is not a surprise that the Higgs boson search is considered, by the High Energy Physics community, one of the most interesting research topics. 

\begin{figure}[!ht]
\begin{center}
\includegraphics[angle=-90,width=0.495\textwidth]{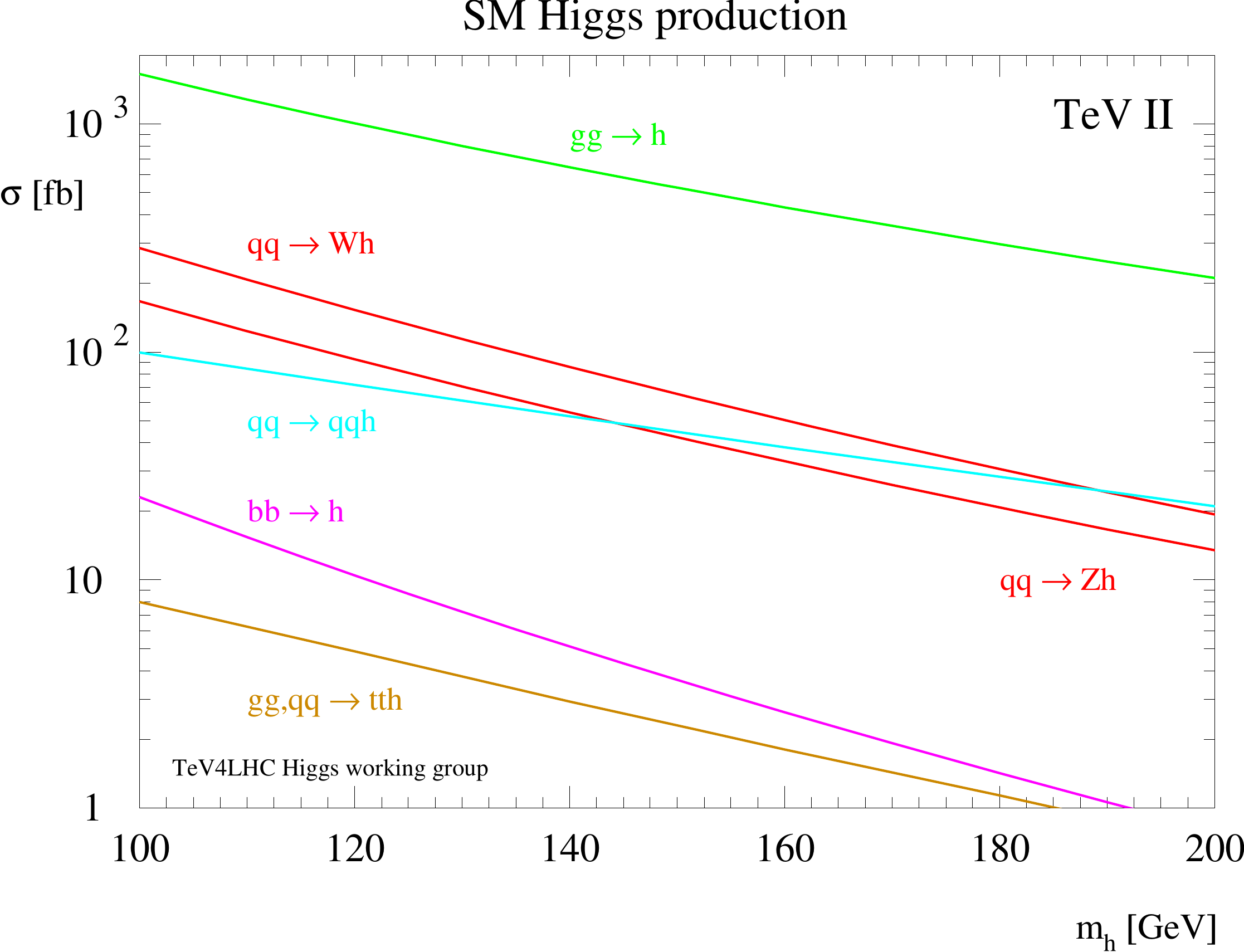}
\includegraphics[angle=-90,width=0.495\textwidth]{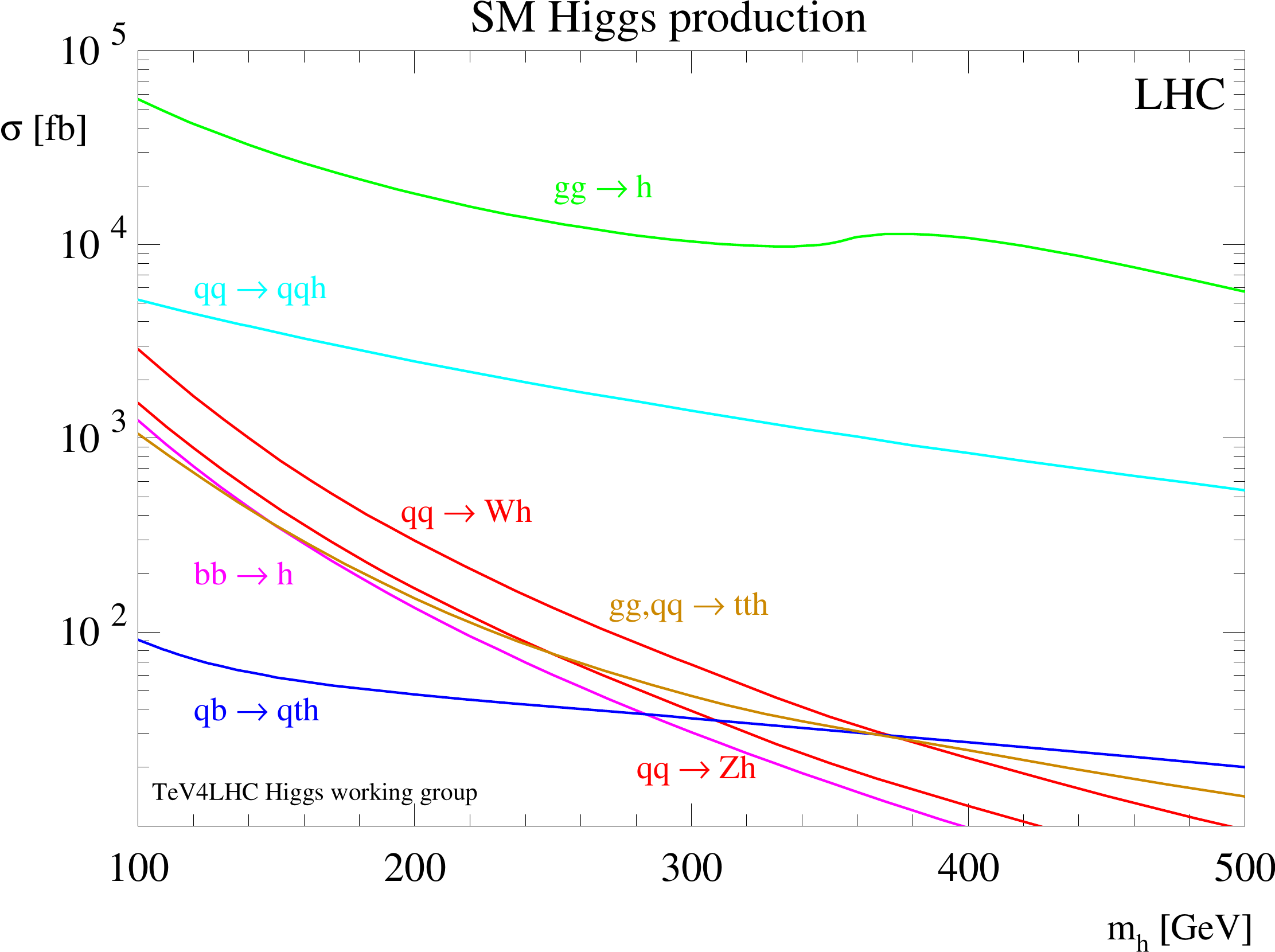}
\caption[Higgs Boson Production Cross Sections]{Higgs boson production cross sections in different modes~\cite{Tev4LHC}. At the Tevatron $p\bar{p}$, $\sqrt{s}=1.96$~TeV (left) and at the LHC $pp$, $\sqrt{s}=7$~TeV (right). In 2012 the LHC raised the collision energy to  $\sqrt{s}=8$~TeV increasing still more the Higgs production cross section.}\label{fig:higgs_cx}
\end{center}
\end{figure}

\begin{figure}[!ht]
\begin{center}
\includegraphics[width=0.75\textwidth]{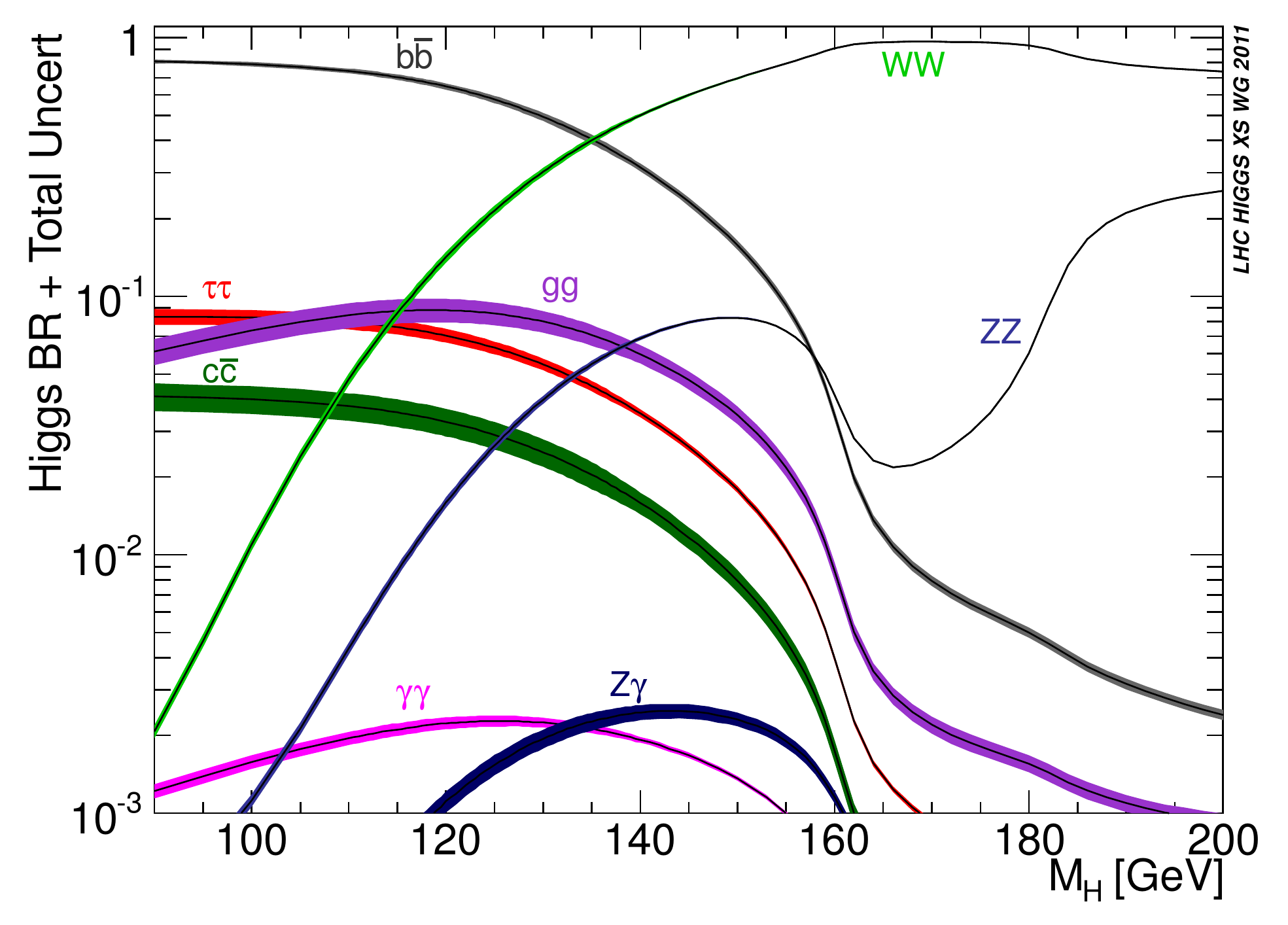}
\caption[Higgs Boson Decay Branching Ratios]{Higgs boson decay branching ratios~\cite{LHCHiggs_cx_wg:2011ti}. The Higgs boson couples to the mass of the particles therefore the decay to the $b\bar{b}$ quark pair is favored for $m_H\lesssim 135$\gc2 while $H\to W^+W^-$ decay dominates for larger masses.}\label{fig:higgs_br}
\end{center}
\end{figure}

Although the existence of the Higgs particle is unknown, its hypothetical couplings and decay properties are important for the interpretation of the experimental results: Figure~\ref{fig:higgs_cx} shows the Higgs production cross section~\cite{Tev4LHC}, at $\sqrt{s}=1.96$~TeV and $\sqrt{s}=7$~TeV, and Figure~\ref{fig:higgs_br} shows the Higgs decay Branching Ratios~\cite{LHCHiggs_cx_wg:2011ti} (BR) for a mass range $100< m_H<200$\gc2. 

\begin{figure}[!ht]
\begin{center}
\includegraphics[width=0.9\textwidth]{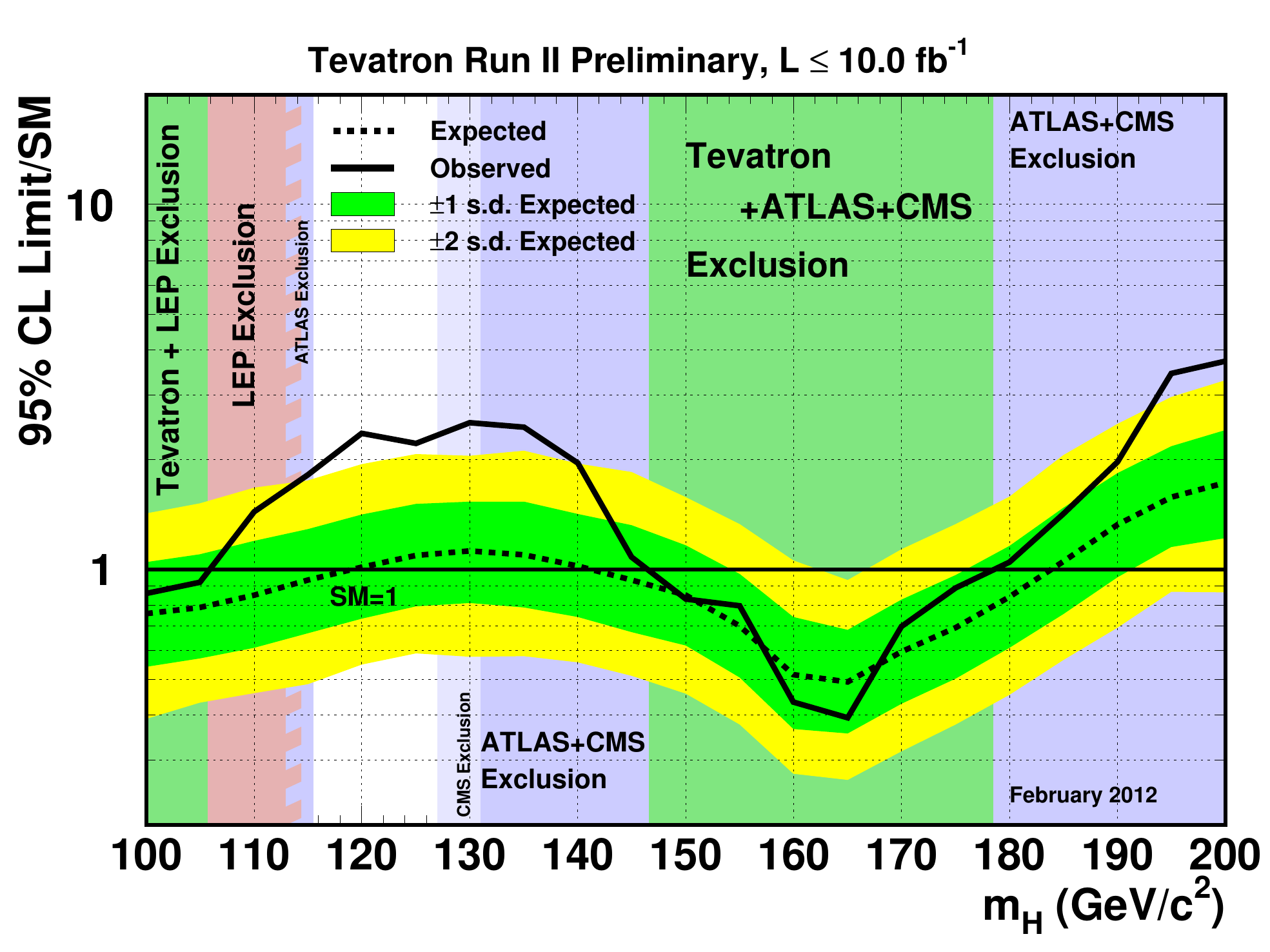}
\caption[Tevatron Combined 95\% Exclusion CL]{Mass exclusion limits (95\% CL) obtained from the combination of all the Tevatron searches for a SM Higgs boson~\cite{tev_cmb}. Exclusion limits obtained  from the CMS~\cite{cms_cmb} and Atlas~\cite{atlas_cmb} Collaboration are overlaid.}\label{fig:tev_cmb}
\end{center}
\end{figure}

The LEP experiments were the first to test the existence of  Higgs boson for masses larger that $100$\gc2, but, as no signal evidence was found~\cite{lep}, all the searches were combined to provide a lower mass limit of $m_H>114.4$\gc2, at 95\% Confidence Level (CL). In the latest years also the experiments situated at the Tevatron and LHC colliders provided several mass exclusion limits~\cite{tev_cmb, cms_cmb, atlas_cmb}. Figure~\ref{fig:tev_cmb} gives a summary of the 95\% CLs of all the three colliders, overlaid to the Tevatron result in the mass range $100< m_H<200$\gc2, only a tiny fraction of the phase space is still available to the Higgs presence and, interestingly, a broad excess appears in mass range $110\lesssim m_H \lesssim 140$\gc2. 

In a short time, as the LHC continues the data taking, a conclusive statement about the Higgs existence will be possible.

However, the Tevatron and LHC results are also complementary because they investigate different couplings of the Higgs boson.
The LHC experiments base most of the low-mass ($m_H\lesssim 135$\gc2) sensitivity on the $H\to\gamma\gamma$ final state. This channel offers an excellent mass resolution and background rejection although at the price of a very low BR (see Figure~\ref{fig:higgs_br}). This is optimal for the higher background rate and Higgs production cross sections availables at a the LHC $pp$ collisions of energy $\sqrt{s}=7$~TeV and $\sqrt{s}=8$~TeV.

\begin{figure}[!ht]
\begin{center}
\includegraphics[width=0.9\textwidth]{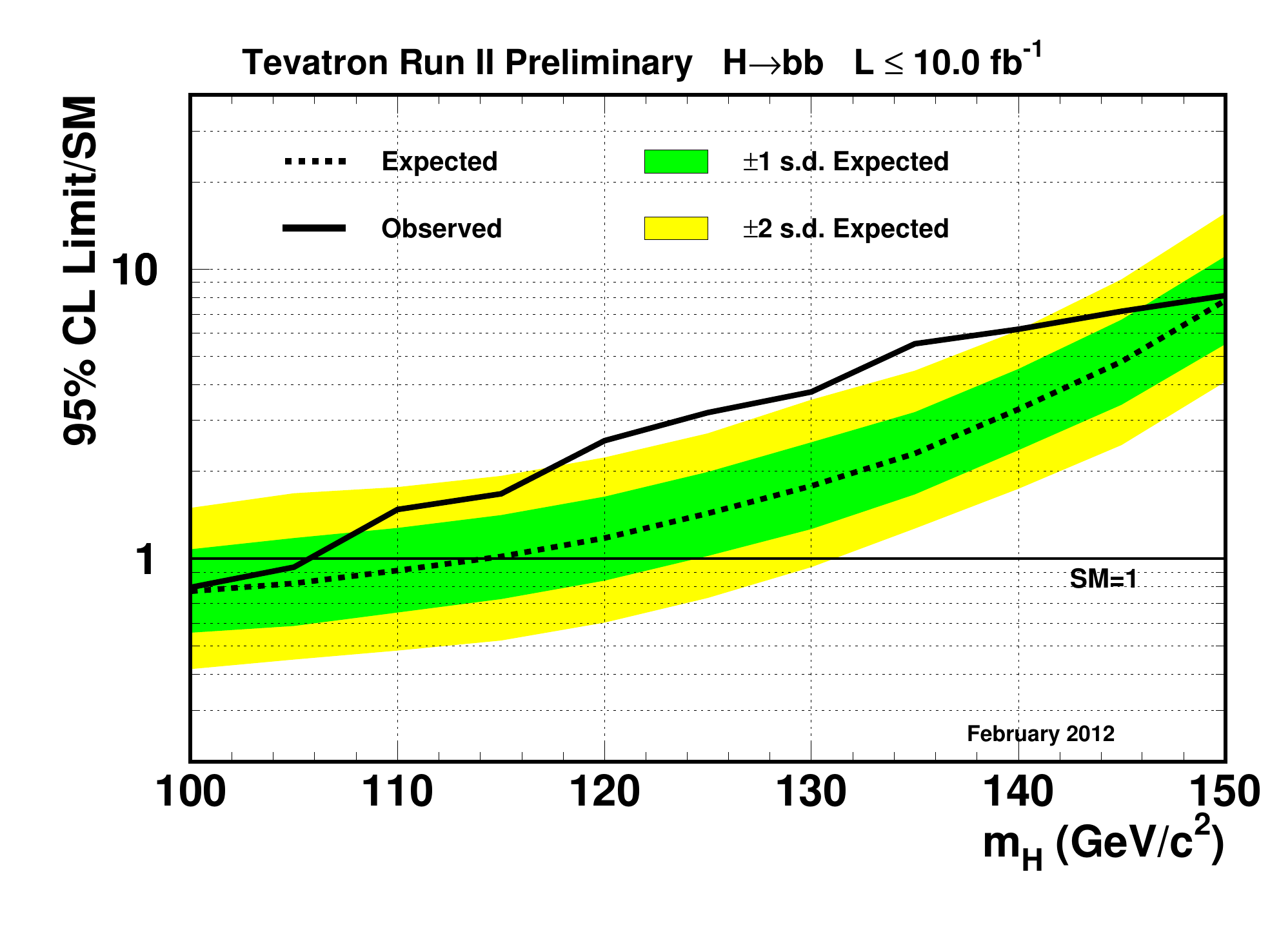}
\caption[Tevatron $H\to b\bar{b}$ Combined 95\% Exclusion CLs]{Mass exclusion limits (95\% CL) obtained from the combination of the Tevatron searches~\cite{tev_cmb} exploiting the $H\to b\bar{b}$ final state.}\label{fig:tev_bb}
\end{center}
\end{figure}

The Tevatron experiments rely more on the $H\to b\bar{b}$ final state, where the Higgs is produced in association with a vector boson ($WH$ and $ZH$ production). The lower production cross section is compensated by the larger BR (see Figures~\ref{fig:higgs_cx} and~\ref{fig:higgs_br}) while the presence a leptonic decay of the $W$ or $Z$ boson allows to keep the background under control. Figure~\ref{fig:tev_bb} shows the  $H\to b\bar{b}$ only Tevatron combined search result. Furthermore the investigation of the $H\to b\bar{b}$ BR is important to understand the coupling of the Higgs with the fermion masses and for the confirmation of the SM assumption coming from Equation~\ref{eq:h_yukawa}.

\section{Status of the Diboson Measurements}\label{sec:dib_res}

In the context of the Higgs searches at the Tevatron, the diboson observation in  $\ell\nu + b\bar{b}$  final state is particularly relevant as it is a direct check of the  $p\bar{p}\to WH\to \ell\nu + b\bar{b}$ analyses.

\begin{figure}[!ht]
\begin{center}
\includegraphics[width=0.75\textwidth]{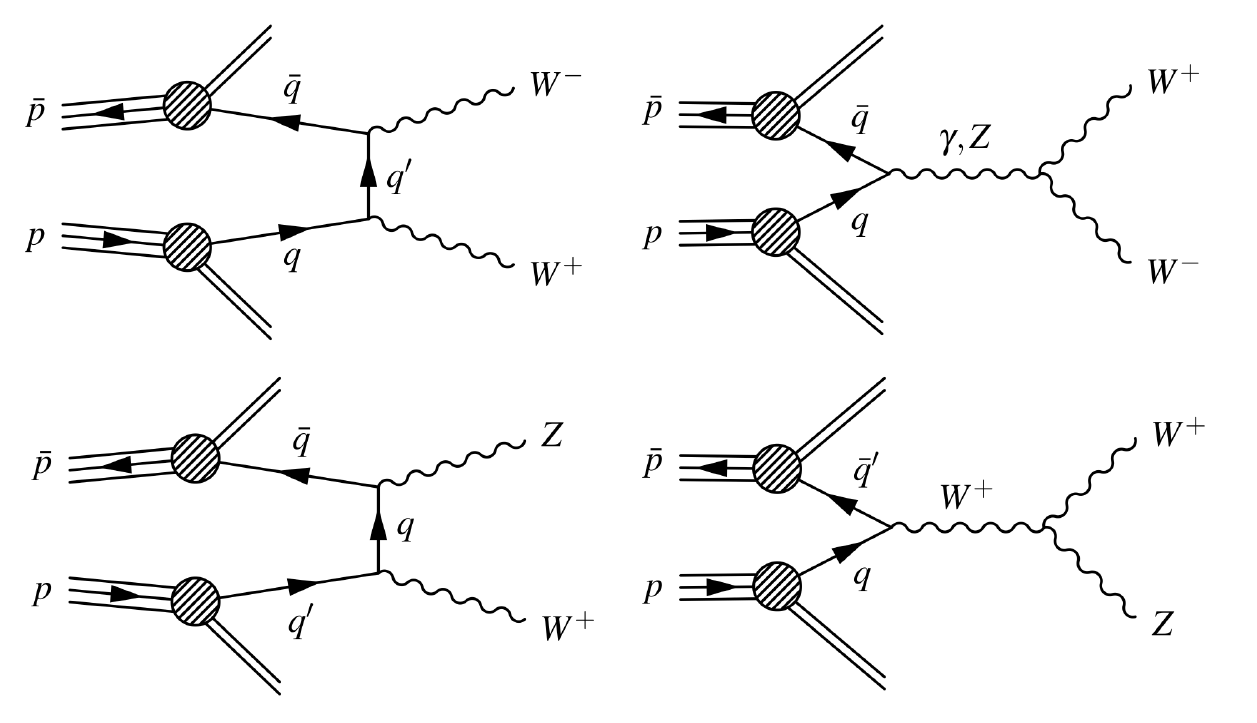}
\caption[$WW$ and $WZ$ Tree-Level Diagrams]{$WW$ and $WZ$ production Feynman diagrams at tree-level, $t$-channel
(left) and $s$-channel (right).}\label{fig:wz_prod}
\end{center}
\end{figure}

The relevant tree-level diagrams involved in dibosons production are shown in Figure~\ref{fig:wz_prod}. 
The simultaneous emission of a $W$ and the $Z$ vector bosons can happen in the $t$-channel (left of Figure~\ref{fig:wz_prod}), with the exchange of a virtual quark, or in the $s$-channel (right of Figure~\ref{fig:wz_prod}) with the exchange of a virtual force carrier. The second case is due to the {\em non-Abelian} characterisctics of the $SU(2)$ group that origins a {\em Triple Gauge Coupling} (TGC) in the kinetic term (Equation~\ref{eq:tensor}) of the SM Lagragnian. The cross sections for the $WW$ and $WZ$ production\footnote{In this analysis we consider also $ZZ\to \ell\ell + HF$ as a signal when a lepton is misidentified, however this contributes to less than 3\% of the total diboson signal yield.} calculated at NLO~\cite{mcfm_prd,mcfm_url}, for $p\bar{p}$ collision at $\sqrt{s}=1.96$~TeV, are:

\begin{equation}
  \sigma_{p\bar{p}\to WW} = 11.34 \pm 0.66\mathrm{~pb};\quad\sigma_{p\bar{p}\to WZ} = 3.47 \pm 0.21\mathrm{~pb};  
\end{equation}
An increase in the TGC, $s$-channel, production cross section would point to a possible contribution from New Physics (NP) processes. However, the precision that we can obtain in the $\ell\nu+HF$ final state is not comparable to the one achievable in other channels with higher leptonic multiplicity (see Table~\ref{tab:dib_measurements}).

\begin{figure}[!ht]
\begin{center}
\includegraphics[width=0.75\textwidth]{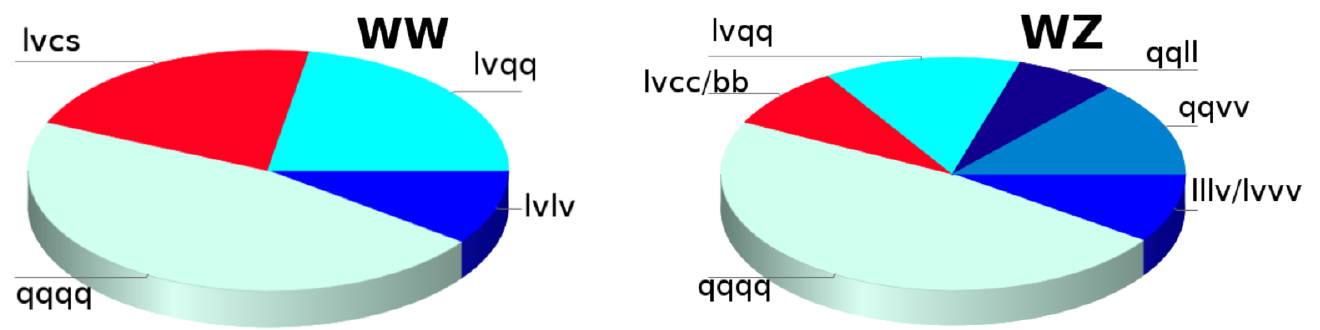}\vspace{0.2cm}
\caption[Diboson Final States BR]{Branching ratios into the different final states allowed to $WW$ and $WZ$ production. $\ell\nu+ HF$ final state is highlighted in red.}\label{fig:torte}
\end{center}
\end{figure}

Figure~\ref{fig:torte} shows the small dibosons BR in $b$ or $c$ quarks\footnote{The experimental identification of $HF$ quarks, described in Section~\ref{sec:secvtx} has also a low efficiency.}, furthermore 
the hadronic final state is background rich and has a low invariant mass resolution.  The search is challenging but it is a perfect standard candle to confirm the understanding of the $\ell\nu+HF$ dataset on a well known SM process.

Diboson related experimental results are widely present in literature. LEP~\cite{lep_diboson} performed the first measurements exploiting all the decay channels of the $WW$ and $ZZ$ processes: their cross sections were measured with good precision. The use of an $e^+e^-$ machine allowed also the observation of hadronically decaying $W$'s, in addition to the semi-leptonic $WW$ decays. However $WZ$ production was not allowed at LEP, since it is forbidden by charge conservation.

Hadron colliders, both Tevatron and LHC, observed $WW$, $WZ$ and $ZZ$ production in their fully leptonic decay modes, obtaining excellent agreement with the SM prediction~\cite{cdfWZlllv_2012, d0WZZZlep_2012, cmsDib_2011, cdfZZllll_2012, atlasZZ_2012, cdfWWllvv_2012, d0WWllvv_2009, cmsWW_2012, atlasWW_2012, cdfDibMetJet_2009, cdfDiblvJet_Mjj2010, cdfDiblvJet_ME2010, d0DiblvJet_2011, cdfDiblvHF_Mjj2011, cdfWZZZCombHF_2012, d0WZZZCombHF_2011,  tevWZZZCombHF_2012}.

The semi-leptonic final states, more difficoult to isolate due to the background rich hadronic environment, were observed at the Tevatron, both at CDF and D0 experiments. In this thesis we present the measurement in the channel $p\bar{p}\to WW/WZ\to \ell\nu +HF$ with an update and improvement of the analysis described in Appendix~\ref{App:7.5}~\cite{cdfDiblvHF_Mjj2011}, performed on a smaller dataset in 2011 ($7.5$~fb$^{-1}$). 

Recently~\cite{cdfWZZZCombHF_2012, d0WZZZCombHF_2011, tevWZZZCombHF_2012}, CDF and D0 produced the evidence for $WZ/ZZ$ production with $HF$ jets in the final state and $WW$ production considered as background. Both experiments produced single results in the three semileptonic diboson decay modes:
\begin{equation}
  WZ/ZZ \to \ell\ell+ HF\textrm{;}
\end{equation}
\begin{equation}
  WZ/ZZ \to \ell\nu + HF\textrm{;}
\end{equation}
\begin{equation}
  WZ/ZZ \to \nu\nu+ HF\textrm{;}
\end{equation}
where the $\nu$ may indicate a lepton failing the identification. The analyses were performed as an exact replica of corresponding Higgs searches in those channels, with the final signal discriminants re-optimized for $WZ/ZZ$ extraction.  Heavy usage of multivariate tecniques, as for example in the $HF$ selection strategy~\cite{HOBIT, BDT_D0} and in the final signal--background discrimination, is the key of the impressive sensitivity of these analyses. The final combined cross section measurement, with a significance of $4.6\sigma$, is: 
\begin{equation}
  \sigma_{WZ+ZZ}  =  4.47\pm 0.67^{+0.73}_{-0.72}\textrm{~pb,}
\end{equation}
where the SM ratio between $WZ$ and $ZZ$ is imposed. This confirms the SM production prediction\footnote{Both $\gamma$ and $Z$ components are assumed in the neutral current exchange and corresponding production of dilepton final states for $75\le m_{\ell^+\ell^{-}} \le 105$~\gc2.} of  $\sigma_{WZ+ZZ}=4.4\pm0.3$~pb~\cite{mcfm_prd,mcfm_url}. The evidence of the $WZ/ZZ$ signal, obtained independently by each experiment and their combination, strongly supports the Tevatron $H\to b\bar{b}$ search results.

A summary of all the present diboson measuremens is reported in Table~\ref{tab:dib_measurements}.

\renewcommand{\tabcolsep}{3pt}
\begin{table}[h] 
\begin{center}
  \begin{small}

\begin{tabular}{ccccc}\toprule
Channel & Experiment &  $\mathscr{L}$ (fb$^{-1}$) & Measured $\sigma$(pb) & Theory $\sigma$(pb) \\\midrule

$WZ\to \ell\ell\ell\nu$ & CDF II\cite{cdfWZlllv_2012} & $7.1$ & $3.9\pm 0.8$    &  $3.47 \pm 0.21$ \\
                                    & D0\cite{d0WZZZlep_2012} & $8.6$ & $4.5^{+0.6}_{-0.7} $  &  $3.47 \pm 0.21$ \\
                                    & CMS\cite{cmsDib_2011} & $1.1$ & $17.0 \pm 2.4\pm 1.5$    &  $ 17.3^{+1.3}_{-0.8}$ \\
                                    & Atlas\cite{cdfWZlllv_2012} & $$ & $20.5 \pm ^{+3.2+1.7}_{-2.8-1.5}$ & $ 17.3^{+1.3}_{-0.8}$ \\
\midrule

$ ZZ \to \ell\ell
+\ell\ell/\nu\nu$ 
 & CDF II\cite{cdfZZllll_2012} & $6.1$ & $ 1.64^{+0.44}_ {-0.38}$ & $1.4\pm 0.1$\\
&  D0\cite{d0WZZZlep_2012} & $8.6$ & $ 1.44^{+0.35}_{-0.34}$ & $1.4\pm 0.1$\\
\midrule

  $ZZ \to \ell\ell\ell\ell$ & Atlas\cite{atlasZZ_2012} & $1.02$ & $ 8.5^{+2.7+0.5}_ {-2.3-0.4}$ & $6.5^{+0.3}_{-0.2}$\\
                            & CMS\cite{cmsDib_2011} & $1.1$ & $ 3.8^{+1.5}_{-1.2}\pm 0.3 $ & $6.5^{+0.3}_{-0.2}$\\

\midrule

$WW\to \ell\ell\nu\nu$ & CDF II\cite{cdfWWllvv_2012} & $3.6$ & $12.1\pm 0.9 ^{+1.6}_{-1.4}$  &  $11.3 \pm 0.7$ \\
                                    & D0\cite{d0WWllvv_2009} & $1.0$ & $11.5\pm 2.2$   &  $11.3 \pm 0.7$ \\
                                    & CMS\cite{cmsWW_2012} & $4.92$ & $52.4\pm 2.0\pm 4.7$    &  $ 47.0\pm 2.0 $ \\
                                    & Atlas\cite{atlasWW_2012} & $1.02$ & $54.4\pm 4.0 \pm 4.4$  &  $ 47.0\pm 2.0$ \\

\midrule
$ Diboson\to \nu\nu + jets$ & CDF II\cite{cdfDibMetJet_2009} & $3.5$ & $18.0\pm 2.8 \pm 2.6$  &  $16.8 \pm 0.5$ \\

\midrule
$ WW/WZ\to \ell\nu + jets$ & CDF II\cite{cdfDiblvJet_Mjj2010} & $4.3$ & $18.1\pm 3.3\pm 2.5$  &  $16.8 \pm 0.5$ \\
                                      & CDF II\cite{cdfDiblvJet_ME2010} & $4.6$ & $16.5^{+3.3}_{-3.0}$  & \\

  & D0\cite{d0DiblvJet_2011} & $4.3$ & $19.6^{+3.2}_{-3.0}$  &  $16.8 \pm 0.5$ \\
 
\midrule

$ WW/WZ\to \ell\nu + HF$ & CDF II\cite{cdfDiblvHF_Mjj2011} & $7.5$ & $18.1^{+3.3}_{-6.7}$  &  $16.8 \pm 0.5$ \\

\midrule

 & CDF II\cite{cdfWZZZCombHF_2012} & $9.45$ & $4.1^{+1.8}_{-1.3}$  &  $4.4 \pm 0.3$ \\
$  WZ/ZZ \to \ell\ell/\ell\nu/\nu\nu+ HF$ & D0\cite{d0WZZZCombHF_2011} & $8.4$ & $5.0 \pm 1.0 ^{+1.3}_{-1.2}$  &  $4.4 \pm 0.3$ \\
 & D0,CDF II\cite{ tevWZZZCombHF_2012} & $7.5 - 9.5$ & $4.47\pm 0.67^{+0.73}_{-0.72}$  &  $4.4 \pm 0.3$ \\

\bottomrule
\end{tabular}
\caption[Recent Diboson Measurements]{Summary of the recent measurements of the diboson production cross section in leptonic and semi-leptonic final states. The reference to the individual measurements are reported in the table as well as the used integrated luminosity, the statistical uncertainty appears before the systematic uncertainties (when both are available), the theoretical predictions are calculated with NLO precision~\cite{mcfm_prd,mcfm_url}.}\label{tab:dib_measurements}
    
  \end{small}
\end{center}
\end{table}
\renewcommand{\tabcolsep}{6pt}

\clearpage
\chapter{The CDF II Experiment}\label{chap:detector}
In this Chapter the accelerator facility and the detector apparatus are described in their main features. 

The analyzed dataset presented in this thesis corresponds to $9.4$\fb1 of data collected by the CDF II (Collider Detector at Fermilab for Run II) experiment along its entire operation time\footnote{Most of the presented work was performed during the data taking period therefore some experimental features are presented as if the operations are not yet concluded.}, from February 2002 to September $30^{th}$, 2011.
The CDF multi-purpose detector was located at one of the two instrumented interaction points along the Tevatron accelerator ring where $p\bar{p}$ beams collided at an energy of $\sqrt{s}=1.96$~TeV. 

\section{The Tevatron}

The Tevatron collider was a proton-antiproton storage ring and circular accelerator located at the Fermi National Accelerator Laboratory (FNAL or Fermilab), $50$~Km west from Chicago (Illinois, U.S.A.). With a center-of-mass energy of $\sqrt{s}=1.96$~TeV, it was the world highest energy accelerator~\cite{tevatronweb} before the beginning of the Large Hadron Collider (LHC) era\footnote{The Tevatron collision energy record was exceed by the LHC on March $30^{th}$, 2010, when the first $pp$ collisions at $\sqrt{s}=7$~TeV took place.} and the largest anti-matter source in the world. The decommissioning of the accelerator started at the end of 2011 with the last collision and the stop of the operations on September $30^{th}$, 2011.

The history of the Tevatron is marked by impressive technology achievements and physics results. For example, starting the operations in 1983, it was the fist super-conducting magnet accelerator ring, in 1995, the {\em top} quark was discovery here~\cite{CDF_top_obs, d0_top_obs} or in 2006, $B_s$ mixing was observed~\cite{cdf_bsmix_obs}.

\begin{figure}[!ht]
\begin{center}
\includegraphics[width=0.99\textwidth]{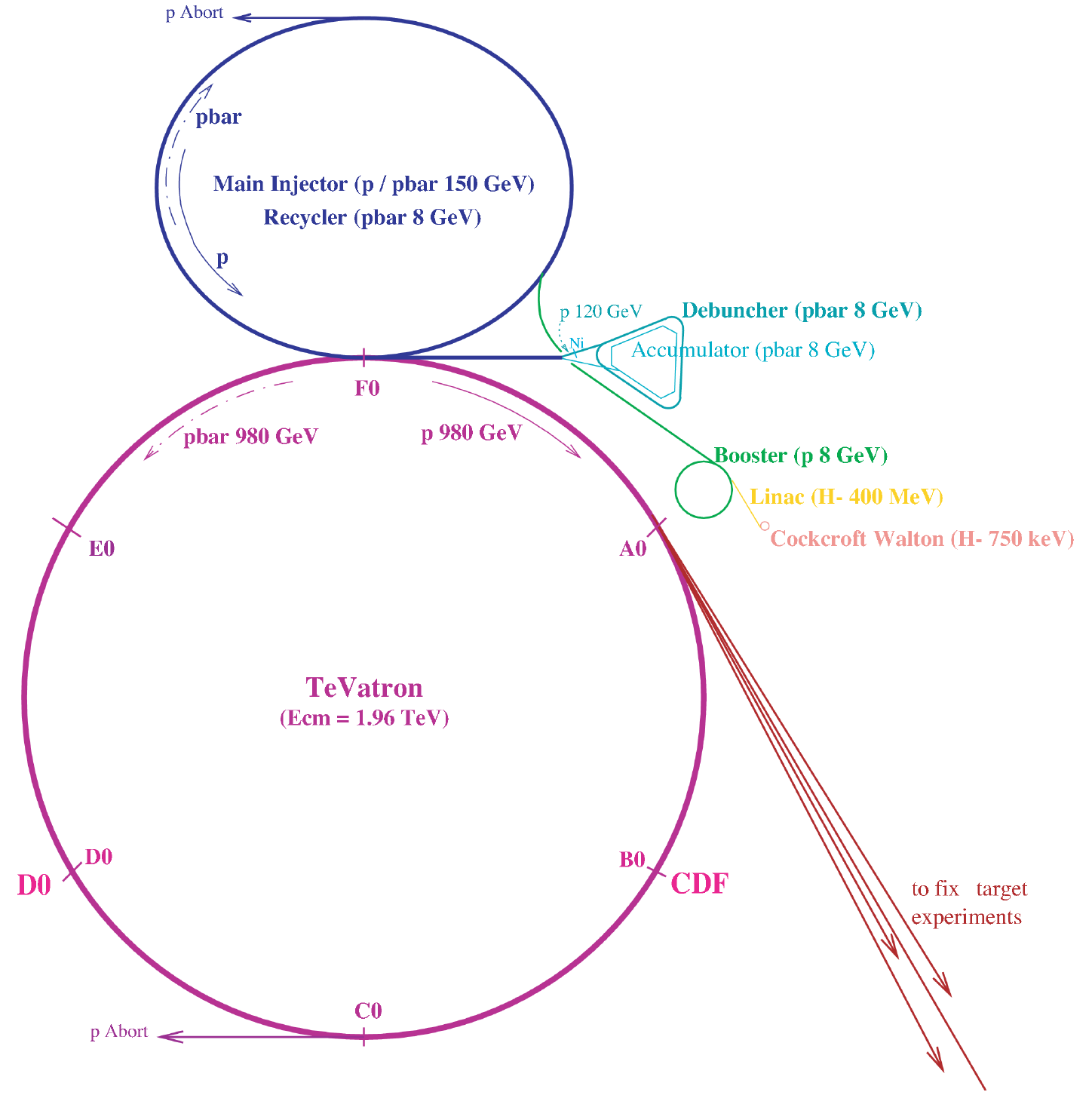}
\caption[CDF Detector Schema]{Schematic view of Tevatron accelerator complex at Fermilab, different colors mark different accelerator sections.}\label{accel_sec}
\end{center}
\end{figure}

One of most striking achievement of the Tevatron was the whole process of the proton-antiproton production and acceleration that involved the simultaneous operation of a chain of accelerator machines. Figure~\ref{accel_sec} shows a view of Tevatron complex and of its sections~\cite{tevatronweb,rookie_book}, next paragraphs summarize their operation and performances.

\subsection{Proton and Antiproton Production}

The first stage, proton extraction and initial acceleration, takes place in the Pre-Accelerator (\emph{PreAc}). Hot hydrogen gas molecules ($H_{2}$) are 
split by an intense local electrostatic field and charged with two electrons; $H^{-}$ ions are accelerated up to $750$~KeV by a Cockcroft-Walton accelerator every $66$~ms.

PreAc ion source\footnote{The ion sources are actually two, named H- and 
I-, and working alternatively.} constantly produces beams at $15$~Hz rate and 
send them to the \emph{Linac}: a linear accelerator that increases the ions 
energy $750$~KeV to $400$~MeV. It is made of two sections: a low energy 
drift tube and a high energy coupled cavity at the end.

Next acceleration stage is performed by a circular accelerator (synchrotron)
of $75$~m radius called \emph{Booster}. The insertion of a thin carbon foil 
strips off electrons from the $400$~MeV ions and a sweep, from $\sim38$ to 
$\sim53$~MHz in radio-frequency (RF), carries resulting protons to an energy 
of $8$~GeV. The use of negative ions permits injection of more particles from 
the Linac, otherwise the magnetic field needed to catch the protons would also
kick away protons already inside the Booster. Bunches are extracted when about
$8\cdot 10^{12}$ protons are collected.

Resulting bunches have the correct energy to be sent to the \emph{Main Injector}, a larger synchrotron with radius of $\approx 0.5$~Km): here conventional magnets and 18 accelerating cavities are employed to accelerate protons up to $120$~GeV or $150$~GeV, depending upon their use. The $120$~GeV protons are used to produce the antiprotons, while $150$~GeV protons are further accelerated into the Tevatron main ring. 

Antiproton production takes place in the \emph{Antiproton Source}. This machine is composed by several parts (see Figure~\ref{accel_sec}): first 
there is a target station where the $120$~GeV protons, extracted from Main Injector, collide with a Nickel target and $8$~GeV $\bar{p}$ are selected from all the resulting particles. Typically, $10\div 20$ $\bar{p}$ are collected for each $10^{6}$ protons on target. After production, antiprotons have a large spatial and momentum spread while acceleration into Main Injector requires narrow $8$~GeV packets. Therefore they are sent to the \emph{Debuncher}: a triangular shape synchrotron, with a mean radius of $90$~m. The bunch signal is picked up and analyzed at one side of the ring and then it is corrected on the other side in a process named \emph{stochastic cooling} and \emph{bunch rotation}~\cite{Marriner:2003mn_cooling}.
The final step of the $\bar{p}$ production is the accumulation: the beam is sent to a smaller synchrotron (with a mean radius of $75$~m) inside Debuncher ring called \emph{Accumulator}. Other cooling methods are applied here.

From the Accumulator, $8$~GeV antiprotons can be transferred either to the Main
Injector or to the Recycler ring. The latter is a $3.3$~Km long ring of permanent magnets, located in the Main Injector enclosure, which is used to gather antiprotons as a final storage
before the injection into the Tevatron, thus allowing the Accumulator to operate at its optimal efficiency.

During the last year of running, 2011, a store could start with up to $3.5\times 10^{12}$ antiprotons, collected in $10\div 20$ hours of production.

\subsection{Collision and Performance}

Last acceleration stage takes place into the \emph{Tevatron} Main Ring: with a radius of one kilometer this is the largest of the Fermilab accelerators, and, thanks to superconducting magnets, it can store and 
accelerate beams from an energy of $150$~GeV (Main Injector result) to $980$~GeV.
Table~\ref{tab:FermilabAcceleratorComplex} summarizes the acceleration characteristics of the different stages of the Fermilab $p\bar{p}$ Accelerator Complex.

\begin{table}[h] 
\begin{center}
\begin{small}
\begin{tabular}{cccccccc}\toprule
Acc. & H & $H^{-}$ & C-W & L & B & M & T\\
\midrule
E &0.04 eV & 25 KeV &750 KeV & 400 MeV&8 GeV& 150 GeV&0.98 TeV\\
$\beta$ & $9.1\cdot10^{-8}$ & $0.01$& $0.04$ & $0.71$ & $0.99$ & $1$ &$1$\\
$\gamma$ & $1$ & $1$ & $1$ & $1.43$ & $9.53$ & $161$ & $1067$\\
\bottomrule
\end{tabular}
\end{small}
\caption[Performance of Fermilab Accelerator Complex]
{Performances of the Fermilab Accelerator Complex. The steps along the accelerator chain (Acc.), with the corresponding labelling, are: Cockwroft-Walton (C-W), Linac (L), Booster (B), Debuncher and Recycler, Main Injector (M), Tevatron (T). The energy reached at the end of the step is E, $\beta=\frac{v}{c}$ expresses
the speed of the particle as a fraction of the speed of light in vacuum and \mbox{$\gamma=\frac{E}{pc}=\frac{1}{\sqrt{1-(\frac{v}{c})^{2}}}$} is the relativistic factor.}\label{tab:FermilabAcceleratorComplex}
\end{center}
\end{table}

When beams production and acceleration is complete, a Tevatron \emph{store} is started: 36 protons and 36 antiprotons bunches, containing respectively 
$\sim 10^{13}$ and $\sim 10^{12}$ particles, are injected into the Main Ring at location\footnote{The Tevatron is divided into six sections (see 
Figure~\ref{accel_sec}) and each junction zone, named form A to F, has a different function: most important areas are B0, D0 and F0, the first two are experimental areas where CDF and DO~detectors are placed, while F0 contains RF cavity for beam acceleration and switch areas to connect Main Injector and the Tevatron.} F0 to be collided. The $0.5$~mm thin proton and antiproton bunches share the same beam pipe, magnets and vacuum system and they follows two non intersecting orbits kept $5$~mm away from each other. Beam control is obtained through nearly 1000 superconducting magnets $6$~m long, cooled to $4.3$~K and capable of $4.2$~T fields. 
\\

\begin{figure}[!ht]
\begin{center}
\includegraphics[width=0.75\textwidth]{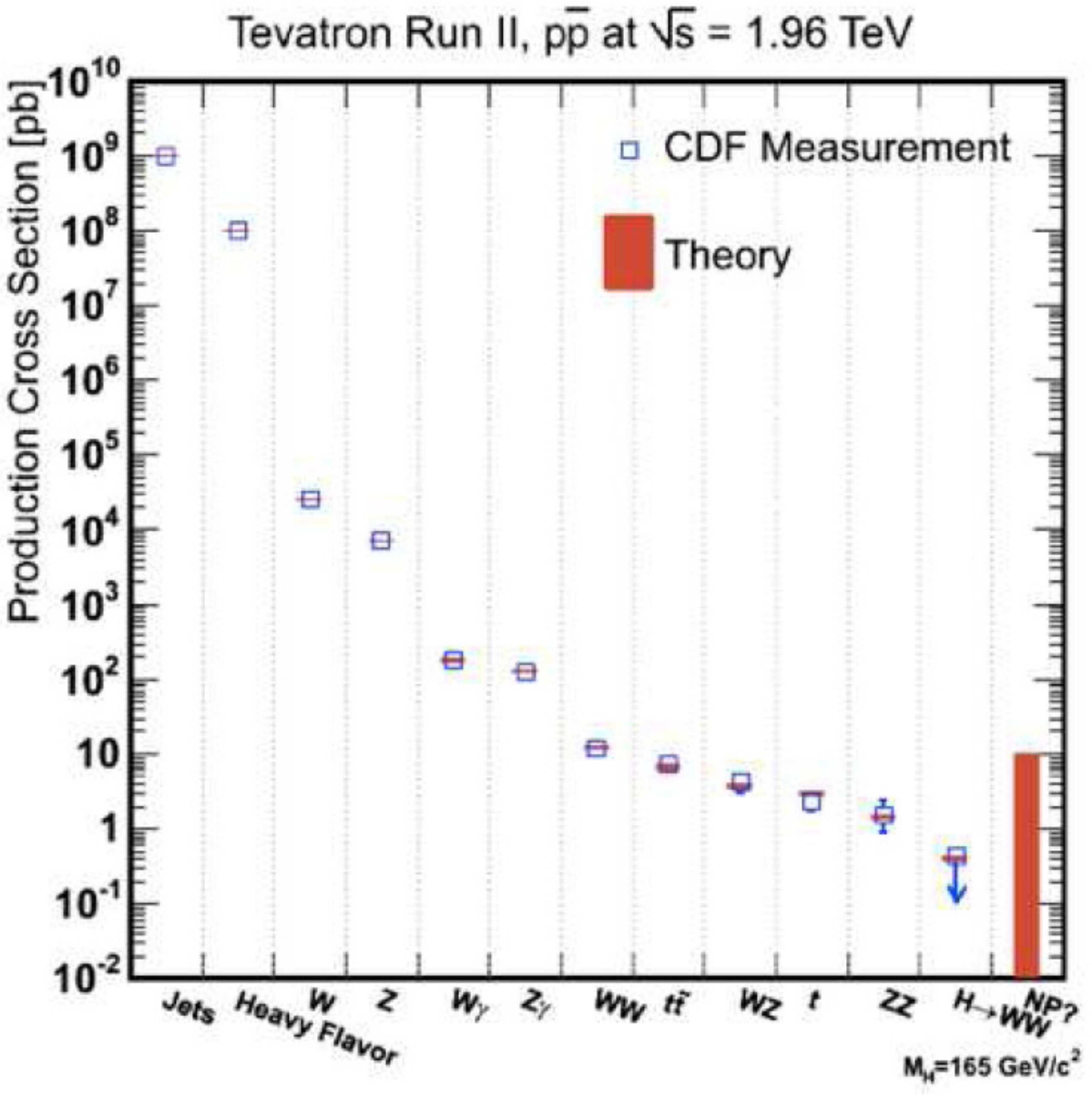} 
\caption[Production Cross Section of Physical Processes at CDF]{Predicted production cross section of physical processes at CDF along with their measured values. Higgs cross section varies with the Higgs mass.}\label{processes}
\end{center}
\end{figure}

Beside energy, the other fundamental parameter of an accelerator is the 
instantaneous luminosity ($\mathscr{L}$), as the rate of a physical process
with cross section $sigma$ is:
\begin{equation}\label{ist_ev}
\frac{dN}{dt}[\textrm{events s}^{-1}]=\mathscr{L}[\textrm{cm}^{-2}\textrm{s}^{-1}]\sigma[\textrm{cm}^{2}]\mathrm{.}
\end{equation}
High energy permits an insight to incredibly small scale physics but only very 
high instantaneous luminosity and very large integrated (in time) luminosity
allow to see rare events. Figure~\ref{processes} shows the production
cross section of different physical processes\footnote{Due to the tiny cross sections we deal with, through most of this work we will be using {\it picobarns} (pb) where $1$~pb$= 10^{-36}\mathrm{~cm}^{2}$.}. 

The instantaneous luminosity of an accelerator, usually measured in cm$^{-2}$s$^{-1}$, is given by:
\begin{equation}\label{ist_lum}
\mathscr{L}=\frac{N_{p}N_{\bar{p}}B f}{2\pi(\sigma^2_{p}+\sigma^2_{\bar{p}})} F\Big(\frac{\sigma_l}{\beta^\ast}\Big),
\end{equation}
where $N_{p}$ ($N_{\bar{p}}$) are the number of protons (antiprotons) per bunch, $B$ is the number of bunches inside accelerator, $f$ is the bunch 
crossing frequency, $\sigma_{p(\bar{p})}$ is the r.m.s. of the proton (antiproton) beam at the interaction point and $F$ is a beam shape form factor depending on the ratio between the the longitudinal r.m.s. of the bunch, $\sigma_l$,  and the beta function, $\beta^\ast$, a measure of the beam extension in the $x,y$ phase space.
Several of these parameters are related to the accelerator structure so they are (almost) fixed inside the Tevatron:
36 $p\bar{p}$ bunches crossed with frequency of $396$~ns and, at interaction points, quadrupole magnets focus beams in $\approx 30$~$\mu$m spots with $\beta^\ast\simeq 35$~cm and $\sigma_l \simeq 60$~cm.

\begin{figure}[!ht]
\begin{center}
\includegraphics[width=0.99\textwidth]{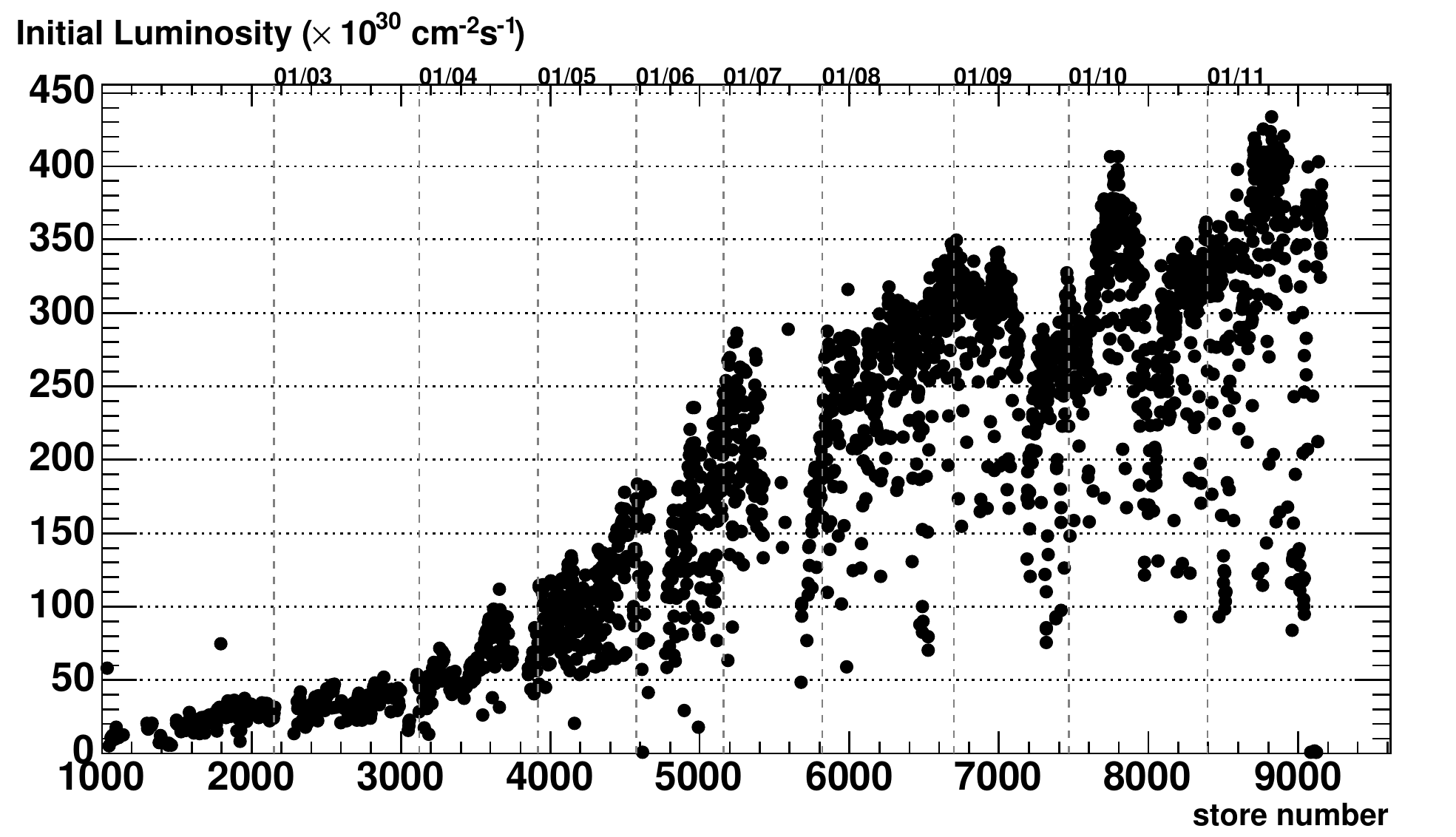}
\caption[Initial Instantaneous Luminosity Delivered by the Tevatron to CDF]{Initial instantaneous luminosity delivered by the Tevatron accelerator to the CDF II detector. Performances grown by two order of magnitude from the beginning of the operation (2002) to the final collisions (2011)~\cite{lum}.}\label{initLum}
\end{center}
\end{figure}

\begin{figure}[!ht]
\begin{center}
\includegraphics[width=0.99\textwidth]{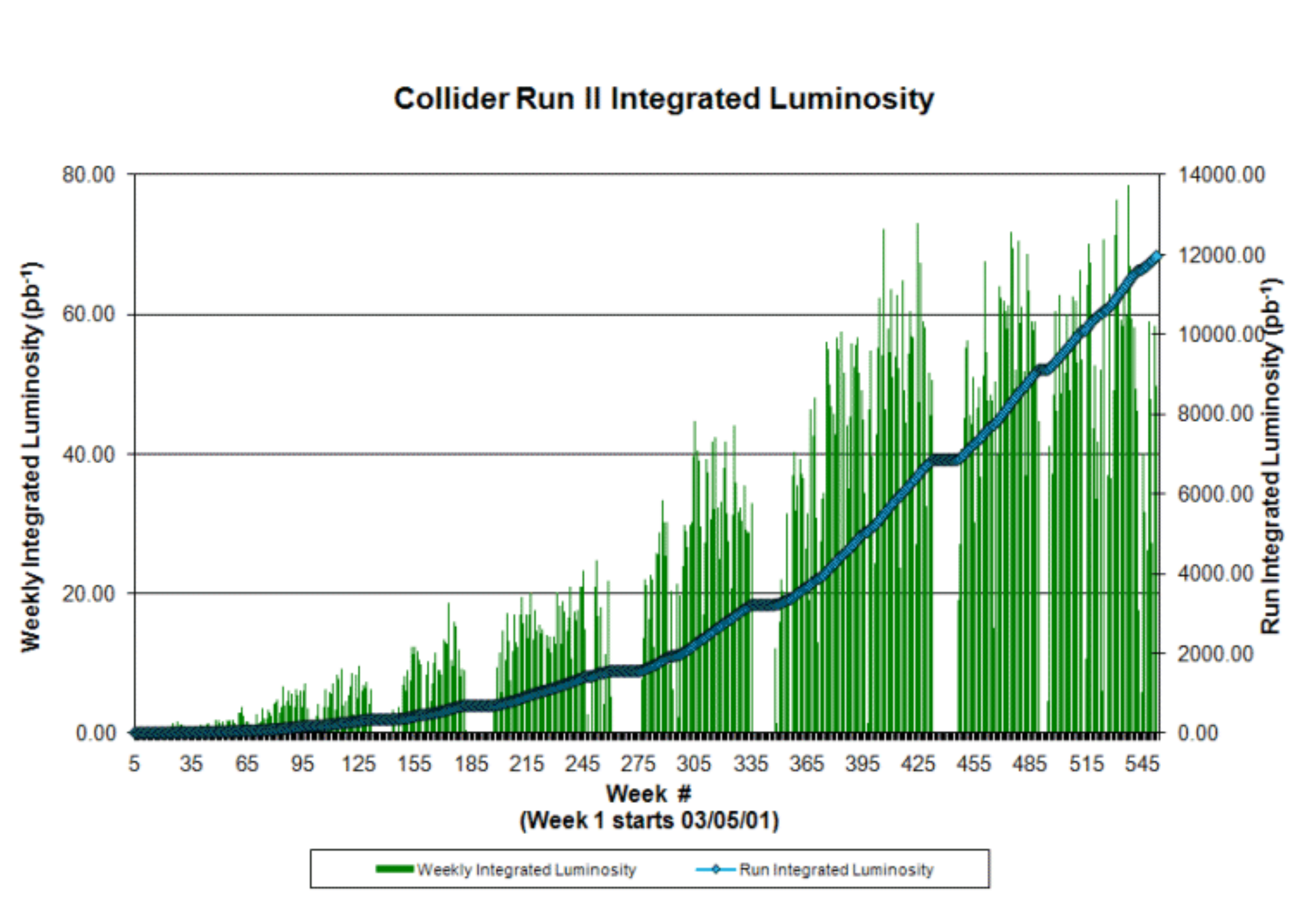}
\caption[Integrated Luminosity Delivered by the Tevatron for each Run]{Integrated luminosity delivered by the Tevatron for each physics run (blue) and averaged on a week by week basis (green bars)~\cite{lum}.}\label{lumi}
\end{center}
\end{figure}

Higher luminosities were achieved in Run II thanks to the increased antiproton stack rate. Figures~\ref{initLum} and~\ref{lumi} show the delivered luminosity and the constant progress in the performances of the machine with an instantaneous luminosity record of $4.3107\times 10^{32}$~cm$^{-2}$s$^{-1}$ on May, 3$^{rd}$ 2011~\cite{acc_records}. 

The total integrated luminosity produced by the Tevatron is more than $12$~fb$^{-1}$ and CDF wrote on tape, on average, about $85\%$ of it, with small inefficiencies due
to detector calibration, stores not used to collect data for physics or dead time during the start-up of the data taking. This analysis uses all the data collected by CDF in Run II, corresponding, after data quality requirements, to a integrated luminosity of about $9.4$~fb$^{-1}$.

\section{The CDF Detector}

CDF II is a multi-purpose solenoidal detector situated at the B0 interaction point along the Tevatron main accelerator ring. Thanks to accurate charged 
particle tracking, fast projective calorimetry and fine grained muon detection, the CDF II detector can measure energy, momentum and charge of most particles resulting from $\sqrt{s}=1.96$~TeV $p\bar{p}$ collisions. 

The first original design goes back to 1981 but CDF underwent many upgrades 
during the past twenty years. The last and most extensive one began in 
1996 and ended in 2001 when Tevatron \emph{Run II} started. At present the 
CDF II experiment is operated by an international collaboration
that embraces more than 60 institutions from 13 different countries, for a total
of about 600 researchers.

\subsection{Overview and Coordinate system}

CDF is composed by many parts (sub-detectors) for a total of about 
5000 tons of metal and electronics, a length of $\sim 16$~m and a diameter of 
$\sim 12$~m. It is approximately cylindrical in shape with axial and 
forward-backward symmetry about the B0 interaction point. Before going further
we describe the coordinate system used at CDF and through this thesis.

B0 is taken as the origin of CDF right-handed coordinate system: $x$-axis
is horizontal pointing North\footnote{Outward with respect to the center of 
Tevatron.}, $y$-axis is vertical pointing upward and $z$-axis is along beam 
line pointing along proton direction, it identifies {\em forward} and 
{\em backward} regions, respectively at $z>0$, East, and $z<0$, West. Sometimes 
it is convenient to work in cylindrical ($r$, $z$, $\phi$) coordinates where 
the azimuthal angle $\phi$ is on the $xy$-plane and is measured from the 
$x$-axis. 
The $xy$-plane is called \emph{transverse}, quantities projected on it are 
noted with a T subscript. Two useful variables are the transverse momentum, $p_{T}$, and energy, $E_{T}$, of a particle:
\begin{equation}
\vec{p}_{T}\equiv p\sin(\theta)\mathrm{,}\qquad E_{T}\equiv E \sin(\theta)\mathrm{,}
\end{equation}
in collider physics another widely used variable, used in place of $\theta$, is the \emph{pseudorapidity}: 
\begin{equation}
\eta\equiv-\ln\big(\tan (\theta/2)\big).
\end{equation}
If $(E,\vec{p})$ is the 4-momentum of a particle, the  pseudorapidity is the 
high energy approximation ($p\gg m$) of the \emph{rapidity}:
\begin{equation}
y=\frac{1}{2}\ln\frac{E+p\cos(\theta)}{E-p\cos(\theta)}\stackrel{p \gg m}{\rightarrow} \frac{1}{2}\ln\frac{p+p\cos(\theta)}{p-p\cos(\theta)}= -\ln\big( \tan (\theta/2)\big) \equiv \eta.
\end{equation}
A Lorentz boost along the $\hat{z}$ direction adds a constant $\ln(\gamma + \gamma\beta)$ to $y$, therefore rapidity differences are invariant.
The statistical distribution of final state particles is roughly flat in $y$ because, in hadronic colliders, the interactions between the (anti)proton 
constituents, which carry only a fraction of the nucleon energy, may have large momentum imbalances along $\hat{z}$.

\begin{figure}[!ht]
\begin{center}
\includegraphics[width=0.99\textwidth]{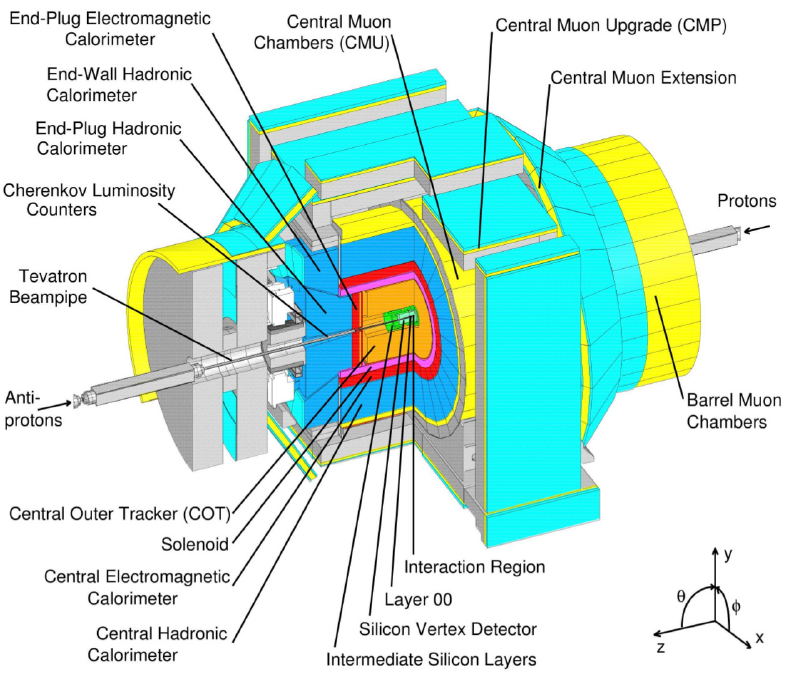}
\caption[Isometric View of the CDF II Detector]{Isometric view of the CDF II detector, the various sub-detectors are highlighted in different colors and listed.}\label{CDF_schema}
\end{center}
\end{figure}

Figure~\ref{CDF_schema} shows an isometric view of the CDF detector and of 
its various sub-detectors. The part inside the $1.4$~T superconducting 
solenoid contains the integrated tracking system: three silicon sub-detectors 
(the \emph{Layer00}, the \emph{Silicon Vertex detector II} and the 
\emph{Intermediate Silicon Layers}) are the inner core of CDF II. The high 
resolution capability of silicon microstrips is necessary to have good track 
resolution near the interaction point, where particle density is higher. Afterward 
an open cell drift chamber (the \emph{Central Outer Tracker}) covers until 
$r\simeq 130$~cm, in the region $|\eta|<1.0$, the extended lever arm provides 
very good momentum measurement (\mbox{$\Delta p_T/p^2_T \simeq 10^{-3}~$GeV$/c^{-1}$}).

Calorimeter systems are located outside the superconducting solenoid. They are
based on \emph{shower sampling calorimeters} made of sequential layers of 
high-Z passive absorbers and active signal generator plastic scintillators. 
The system is composed by towers with $\eta - \phi$ segmentation, each one 
divided in electromagnetic and hadronic part, they cover the region up to 
$|\eta|\simeq 3.6$ ($\theta\simeq3^{\circ}$) and are organized in two main
sections: the \emph{Central Calorimeter} covering the region 
$|\eta|\lesssim 1.1$ and the \emph{Plug Calorimeter} 
extending the coverage up to $|\eta|\simeq 3.6$. While the central calorimeter
is unchanged since 1985, the plug calorimeter active part was completely 
rebuilt  for Run II, replacing gas chambers with plastic scintillator tiles to
better cope with the higher luminosity.

The outermost part of CDF detector, outside calorimeters, is occupied by the
muon detectors. They are multiple layers of drift chambers arranged in various 
subsections which cover the region $|\eta| \lesssim 1.5$. Only high penetrating
charged particles, such as muons, can go across the entire detector.

Other detectors are used for a better particle identification, calibration or 
monitoring. However a detailed description of the entire CDF detector is far 
from the scope of this work. The next paragraphs will focus on tracking and 
calorimeter systems which play a significant role in the analysis. A complete 
description of CDF II detector can be found in~\cite{TDR}.

\subsection{Integrated Tracking System}\label{track_par}

\begin{figure}[!ht]
\begin{center}
\includegraphics[width=0.75\textwidth]{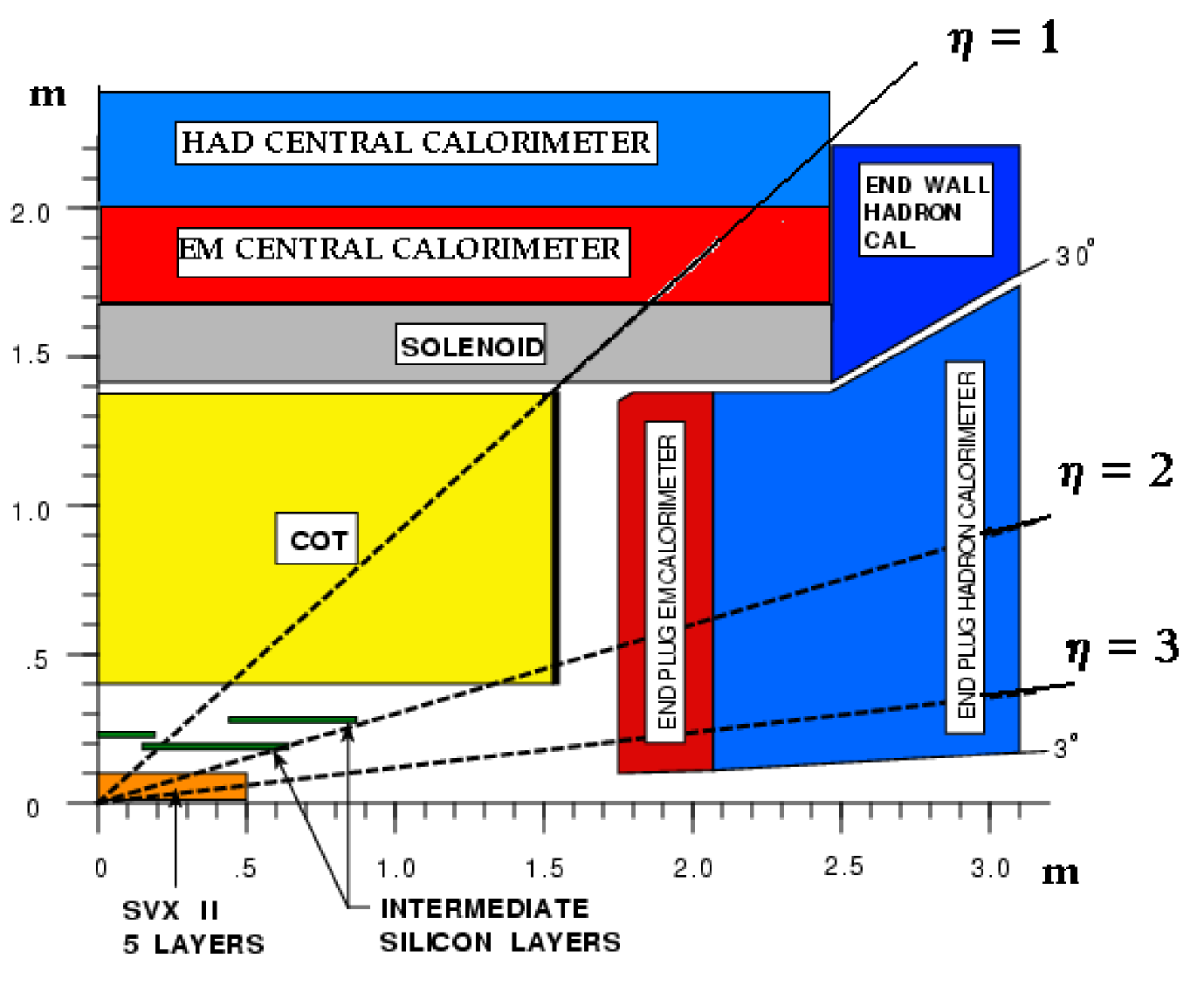}
\caption[CDF II Tracking Volume and Calorimeter]{View of CDF II tracking volume and calorimeter location.}\label{CDF_track}
\end{center}
\end{figure}

The trajectory of a charged particle in a uniform 
magnetic field in vacuum is a helix. A tracking detector identifies some points along particle path so 
that it is possible to obtain momentum measurements by reconstructing the 
helix parameters\footnote{See Section~\ref{sec:track} for track reconstruction 
details.}. A schematic view of CDF tracking volume can be seen in 
Figure~\ref{CDF_track}: the three main components are the superconducting 
magnet, the silicon sub-detectors and the central drift chamber.

The solenoidal magnet, made  by NbTi/Cu superconducting coils, maintains a 
bending magnetic field with a central value of $1.4116$~Tesla, oriented along 
the positive $\hat{z}$ direction and nearly uniform in all the tracking volume 
($r\lesssim 150$~cm and $|z|\lesssim 250$~cm). 
The momentum threshold for a particle to radially escape the magnetic field is 
$p_{T}\gtrsim 0.3$~GeV/$c$ and the radial thickness of the coil is $0.85$ 
radiation lengths ($X_{0}$).

\subsection*{Silicon System}

The silicon system is the first tracking sub-detector encountered by particles
exiting from the primary interaction vertex. Semiconductor detectors offer
excellent spatial resolution and fast response time.
Therefore it permits the reconstruction of secondary vertices displaced from the
primary, produced in the decay of long lived $b$-hadrons\footnote{Correct 
identification of $b$-hadrons is fundamental in many analyses e.g. $b$-hadrons 
are one of the decay products of $top$ quark and also Higgs boson has a high 
branching ratio to $b$ quarks for $m_H\lesssim 140$\gc2.}.

\begin{figure}[!ht]
\begin{center}
\includegraphics[width=0.49\textwidth]{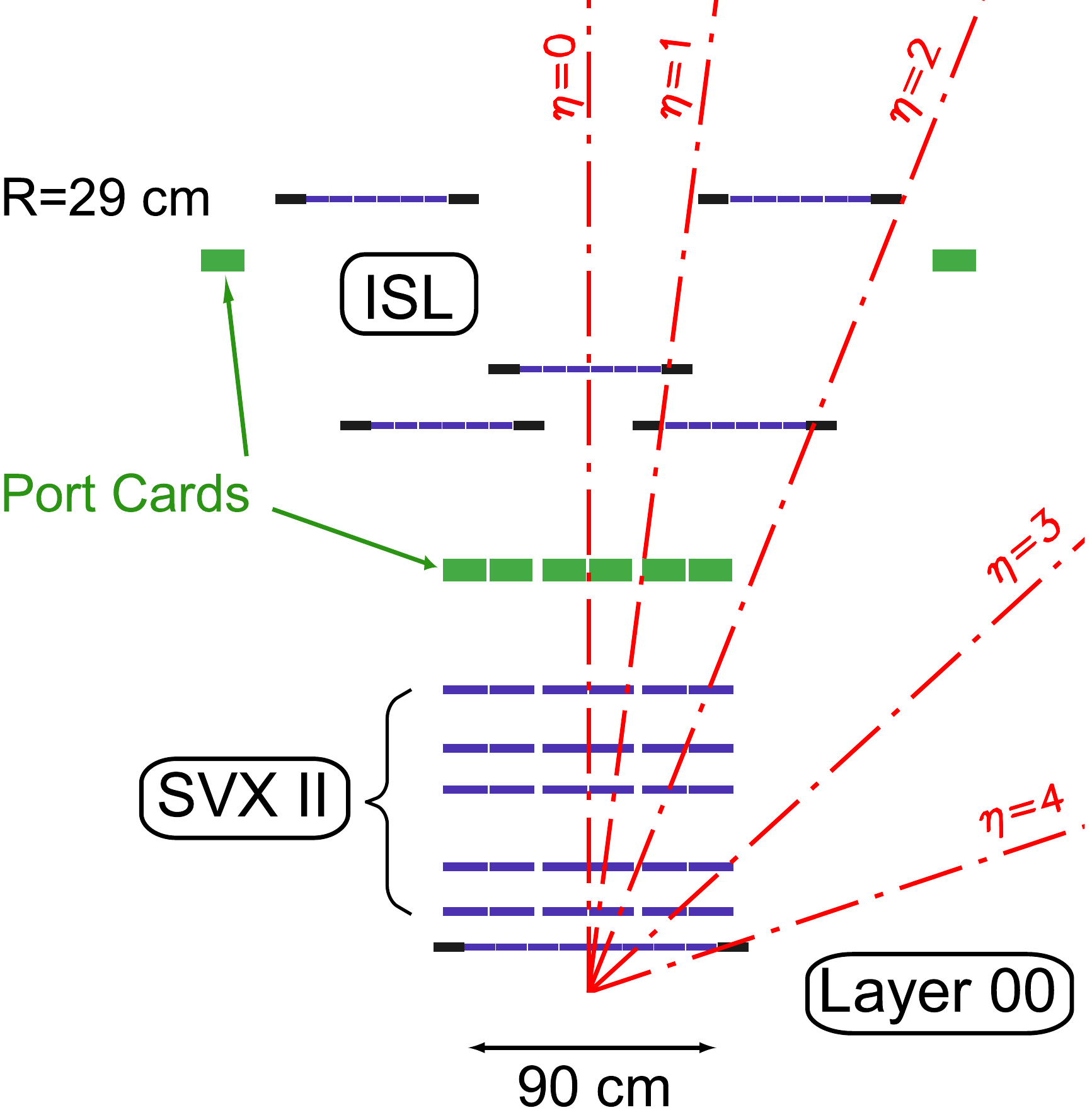}
\includegraphics[width=0.49\textwidth]{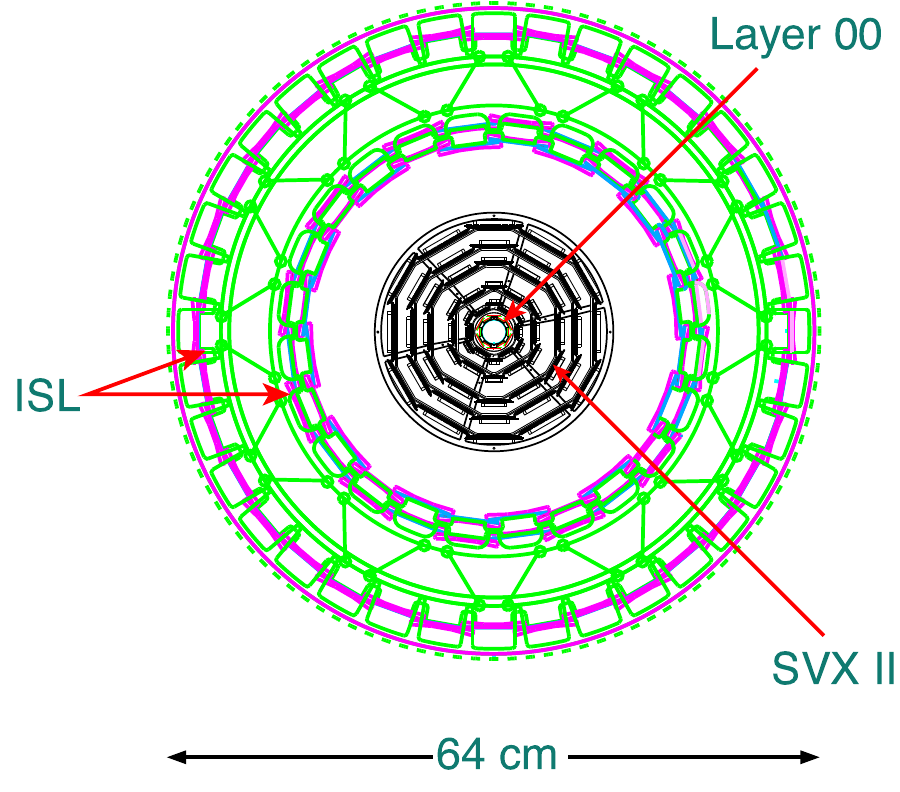}
\caption[Side and Front View of Silicon System]{Side and front view of silicon tracking system at CDF.}\label{silicons}
\end{center}
\end{figure}

CDF employs $\sim7$~m$^2$ silicon active-surface for a total 
of 722,432 different channels read by about 5500 integrated custom chips. The
complete silicon tracking detector is displayed in Figure~\ref{silicons}.
Of the three subsystems composing the core of CDF, the 
\emph{Layer00 }~\cite{l00} (L00 ) is the innermost.
It consists of a single layer of single-sided silicon sensors directly mounted 
on the beam pipe at radii, alternating in $\phi$, of $1.35$~cm or $1.62$~cm, 
covering the region $|z|\lesssim 47$~cm. During the construction of the SVX II 
microvertex (see below) CDF realized that the multiple scattering due to the 
presence of read-out electronics and cooling systems installed 
inside tracking volume was going to degrade the impact parameter resolution.
L\O\O~ was designed to recover it thanks to its proximity to the beam. 
Furthermore, being made of state-of-the-art radiation-tolerant sensors, it will
ensure a longer operating lifetime to the entire system.

The main component of the silicon system is SVX II~\cite{svx}, the 
\emph{Silicon VerteX detector}
is made of three cylindrical barrels for a total length of about $96$~cm along
$z$, covering the luminosity region until $\simeq 2.5~\sigma_{l}$, and with a 
pseudo-rapidity
range $|\eta|\lesssim 2$. Each barrel is divided in twelve identical wedges in
$\phi$, arranged in five concentric layers between radii 
$2.4$~cm and $10.7$~cm. Each layer is divided into independent longitudinal 
read-out units, called {\em ladders}. Each ladder consists of a low-mass support
for a double-sided silicon microstrip detector. Three out of five layers 
combine an $r-\phi$ measurement on one side with $90^{\circ}$ stereo measurement
on the other, the remaining two layers combine an $r-\phi$ measure with a small
angle $r-z$ stereo measurement (with tilt angle of $1.2^{\circ}$). 
The highly parallel fiber based data acquisition 
system reads out the entire sub-detector in approximately $10~\mu$s.

The \emph{Intermediate Silicon Layers} detector~\cite{isl} (ISL) is the outermost of the three silicon sub-detectors, radially
located between SVX II and the drift chamber covering the region $|\eta|\lesssim 2$. It is divided in three barrels segmented into $\phi$ 
wedges. The central barrel ($|\eta|\lesssim 1$) is made of one layer of silicon sensors at radius of $22$~cm, instead the two outer barrels 
($1\lesssim|\eta|\lesssim 2$) are made of two layers at radii of $20$~cm and 
$28$~cm. Its purpose is to strengthen the CDF tracking in the central region and to add precision hits in a region not fully covered by the 
drift chamber. Track reconstruction can be extended to the whole region $|\eta|<2$ using the silicon detector.

The complete silicon sub-detector (L00, SVX II and ISL) has an asymptotic 
resolution of $40~\mu$m in impact parameter and of $70~\mu$m along $z$ 
direction. The total amount of material varies roughly as:
\begin{equation}
\frac{0.1X_{0}}{\sin(\theta)}
\end{equation}
in the central region and doubles in the forward region because of the presence of read-out electronics, cooling system and support frames~\cite{sil_performance}.

\subsection{Central Outer Tracker}

The \emph{Central Outer Tracker}~\cite{cot} (COT) is an open-cell drift 
chamber used for particles tracking at large radii. It has an 
hollow-cylindrical geometry and covers $43.3<r<132.3$~cm, $|z|\lesssim 155$~cm.
Figure~\ref{CDF_track} shows that COT fully covers the central region 
($|\eta|\lesssim 1$) with some residual capability up to $|\eta| \approx 1.8$

\begin{figure}[!ht]
\begin{center}
\includegraphics[width=0.98\textwidth]{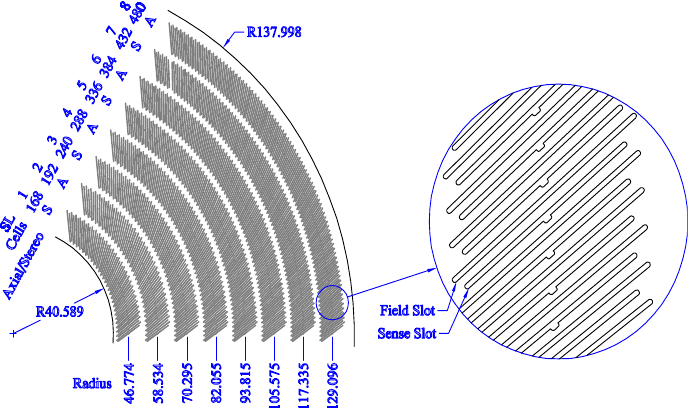}
\caption[A $1/6$ Section of the COT End-plate]{A $1/6$ section of the COT end-plate with the eight super-layers structure and the location of cell slots.}\label{cot1}
\end{center}
\end{figure}

The COT (see Figure~\ref{cot1}) is structured into eight {\em super-layers} each 
divided into $\phi$ cells; each cell contains twelve sampling wires, spaced 
$0.583$~cm, to collect the ions produced by passing charged particles. The 
arrangement of the cells has a $\chi=35^{\circ}$ tilt with respect to the 
chamber radius to partially compensate the Lorentz angle of the electrons 
drifting in the magnetic field and obtain the best 
resolution\footnote{Electrons drifting in a gas within an electromagnetic 
field $(\vec{E},\vec{B})$ move with an angle 
$\chi\simeq\arctan\big(\frac{v(E,B=0)B}{kE}\big)$, where $k$ is empirical 
parameter of gas and electric field and $v(E,B=0)$ is the velocity without the
magnetic field. The angle $\chi$ is also known as Lorentz angle.}.

\begin{figure}[!ht]
\begin{center}
\includegraphics[width=0.98\textwidth]{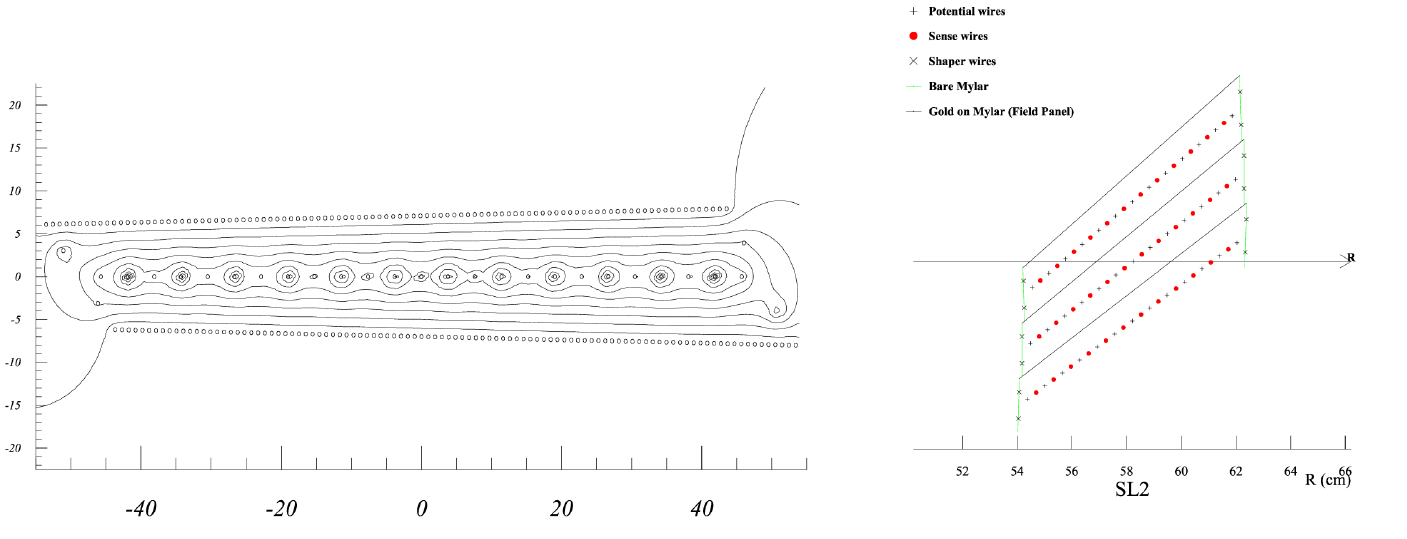}
\caption[Equipotential Line and Wire Layout Inside one of the COT Cells]{Left: equipotential line inside one of the COT super-layer cell. 
Right: layout of sense-wires, field-wires and shaper-wires inside one COT cell.}\label{cot2}
\end{center}
\end{figure}

The final structure has $8 \times 12$ sampling planes alternated with 
planes of potential wires (see Figure~\ref{cot2}), $96$ hits are measured for a 
particle crossing the entire COT ($|\eta|<1$). Four super-layers employ 
sense-wires parallel
to the beam axis for measurements in $r-\phi$ plane, the other four interspacing
super-layers are named {\em stereo} super-layers because their wires are 
alternately canted at angles of $+2^{\circ}$ and $-2^{\circ}$ with respect to 
the beam line and are used to measure $r-z$ coordinates. The electric 
drift field (see Figure~\ref{cot2}) is $1.9$~kV/cm. A $50:50$ gas admixture of 
argon and ethane bubbled through isopropyl alcohol ($1.7$\%) constantly flows 
in the chamber volume. The drift velocity is about $100~\mu$m/cm for a
maximum drift space of $0.88$~cm. The material of the COT is about 
$0.017X_{0}$, mostly concentrated in the inner and outer shell.

\subsection{Calorimeter System}

Located immediately outside the solenoid, the calorimeter system covers
a solid angle of nearly $4\pi$ around $p\bar{p}$ interaction point and has the
fundamental role to measure energies of electrons, photons, particle clusters 
(\emph{jets}) and the imbalance in transverse energy flow (signature of 
\emph{neutrinos}). 
The location of calorimeter sections is visible in Figure~\ref{CDF_track}.
Both plug and central are sampling calorimeters divided into an 
electromagnetic section (lead/scintillator), optimized to collect all the 
energy of electrons and photons, and a subsequent hadronic section 
(iron/scintillator), thicker and optimized for hadron energy measurement. 
Calorimeters have an in-depth segmentation, finer near the collision point and
coarser outward. The $\eta-\phi$ plane is mapped in rectangular cells, each one
corresponding to the independent read-out of a projective electromagnetic or 
hadronic tower. Thanks to the fast response of scintillators, signals from
calorimeters are quickly processed and used at various trigger levels. 
Following paragraphs explains in more detail the composition 
of the different subsections and Table~\ref{cal_tab} summarizes their main 
characteristics.
\begin{table}
  \begin{center}
    \begin{small}

      \begin{tabular}{ccccc}
        \toprule
        & En. Resolution & $\eta$ Coverage & Absorber & Longitudinal Depth\\
        \midrule
        CEM & $13.5\%/\sqrt{E} \oplus 2\% $ &  $|\eta|<1.1$  & lead  & $19X_{0}$,  $1\lambda$\\
        CHA & $50\%/\sqrt{E}\oplus 3\%$ &  $|\eta|<0.9$      & iron  & $4.5\lambda$\\
        WHA & $75\%/\sqrt{E}\oplus 4\%$ &  $0.7<|\eta|<1.3$  & iron  & $4.5\lambda$\\
        PEM & $16\%/\sqrt{E}\oplus 1\%$ &  $1.1<|\eta|<3.6$  & lead  & $21X_{0}$, $1\lambda$ \\
        PHA & $74\%/\sqrt{E}\oplus 4\%$ & $1.3<|\eta|<3.6$   & iron  & $7\lambda$ \\
        \bottomrule
      \end{tabular}      
    \end{small}
    \caption[CDF II Calorimeter System]{Main characteristics of CDF II calorimeter system.}\label{cal_tab}
  \end{center}
  \end{table}

\subsection*{Central Calorimeter}

The central region of the detector is covered by the \emph{Central Electromagnetic} (CEM) and the \emph{Central HAdronic} (CHA) calorimeters~\cite{cen_cal}, corresponding to the pseudo-rapidity region $|\eta|<1.1$ and $|\eta|<0.9$ respectively. 

The CEM is a hollow cylinder located at $173<r<208$~cm, divided in four $180^{\circ}$ arches each composed by 12 azimuthal sections ($\Delta\phi=15^{\circ}$) and 10 pseudo-rapidity sections ($\Delta\eta\simeq0.11$) for a total of 478 instrumented towers\footnote{Two towers are missing to permit access to the solenoid, the
so-called {\em chimney}.}. 
The CHA covers region $|\eta|<0.9$ and it is divided into 9x12 $\eta-\phi$ 
towers corresponding to CEM segmentation for a total of 384 towers. Central 
hadronic calorimeter covering is extended up to $|\eta|\simeq 1.3$ thanks to 
the \emph{Wall HAdron Calorimeter}~\cite{wha} (WHA). It has same $\phi$ 
segmentation and six additional $\eta$ towers: the first three overlap
CHA and the last three extend $\eta$ coverage.

\begin{figure}[!ht]
\begin{center}
\includegraphics[width=0.75\textwidth]{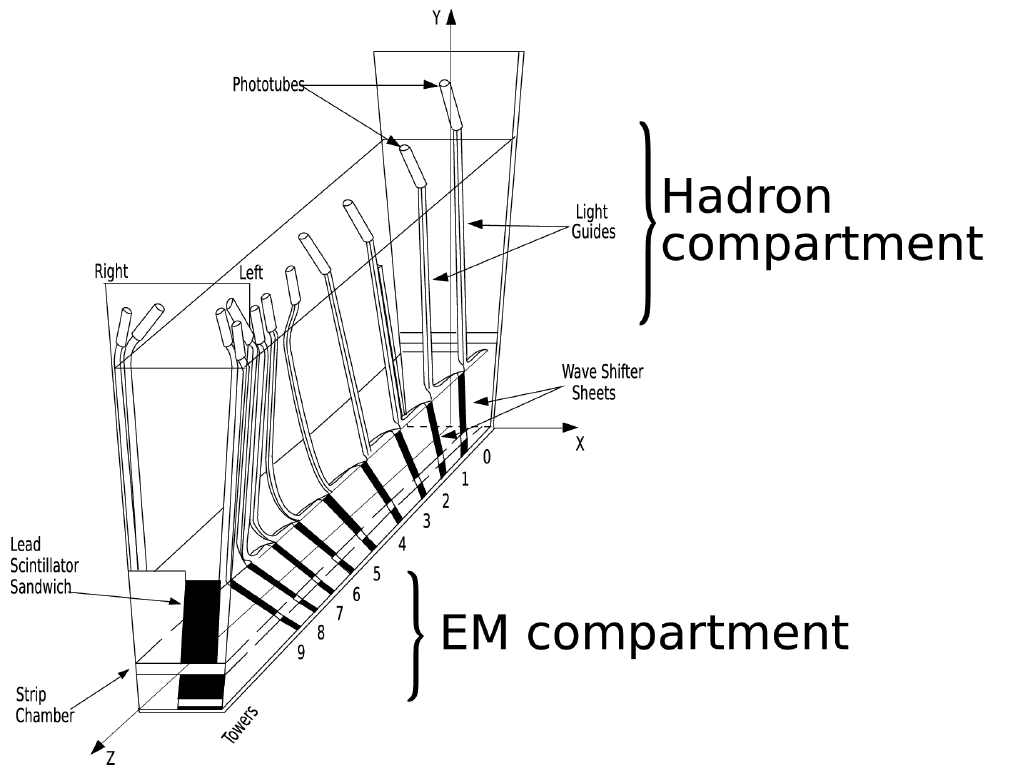}
\caption{Structure of a wedge of CDF central calorimeter.}\label{cen_sector}
\end{center}
\end{figure}

Figure~\ref{cen_sector} shows a wedge of the central calorimeter system. Each
CEM sector is a sampling device made of 31 layers of polystyrene scintillator 
($5$~mm thick) radially alternated with layers of aluminum-clad lead 
($3.18$~mm thick). Some of the 30 lead layers are replaced by acrylic 
(Plexiglas) as a function of $\theta$ to maintain a uniform thickness in 
$X_{0}$.
As particles loose energy into the absorber medium, the blue light emitted by 
active scintillator medium is collected by thin bars of blue-to-green 
wave-length shifter acrylic material placed on the sides of each tower that 
bring the light to two photomultiplier tubes (PMT) outside CHA.
CEM contains also the \emph{Central Electron Strip chambers} (CES) and the 
\emph{Central PReshower detector} (CPR). CES is a multi-wire proportional 
chamber placed at a radial depth of $\sim6X_{0}$ and is used to determine shower 
position and transverse shower development with an accuracy of $\sim 0.2$~cm. 
CPR is a layer of finely segmented scintillators located immediately outside 
the solenoid and is used to monitor photon conversion started in the tracking 
region.

The structure of hadronic calorimeters is similar to electromagnetic ones but
absorber materials are 32 steel, $2.5$~cm-thick, layers in CHA and 15 steel, 
$5.1$~cm thick, layers in WHA both alternated with acrylic scintillator, 
$1.0$~cm thick.

The total thickness of electromagnetic section is nearly uniform and 
corresponds to $19$ radiation lengths ($X_{0}$) or $1$ interaction length 
($\lambda_{int}$). Based on test beam data, the CEM energy resolution 
for an electron going through the center of a tower is found to be:
\begin{equation}\label{cem_res}
\frac{\sigma_{E}}{E}=\frac{13.5\%}{\sqrt{E(\mathrm{GeV})}}\oplus 2\%.
\end{equation}
The total thickness of hadronic section is $\sim4.5\lambda_{int}$ and the 
energy resolution is:
\begin{equation}\label{cha_res}
\frac{\sigma_{E}}{E}=\frac{50\%}{\sqrt{E(\mathrm{GeV})}}\oplus 3\%, \qquad \frac{\sigma_{E}}{E}=\frac{75\%}{\sqrt{E(\mathrm{GeV})}}\oplus 4\%.
\end{equation}
respectively for CHA and WHA.

\subsection*{Forward Calorimeter}\label{forward_cal_sec}

\begin{figure}[!ht]
\begin{center}
\includegraphics[width=0.75\textwidth]{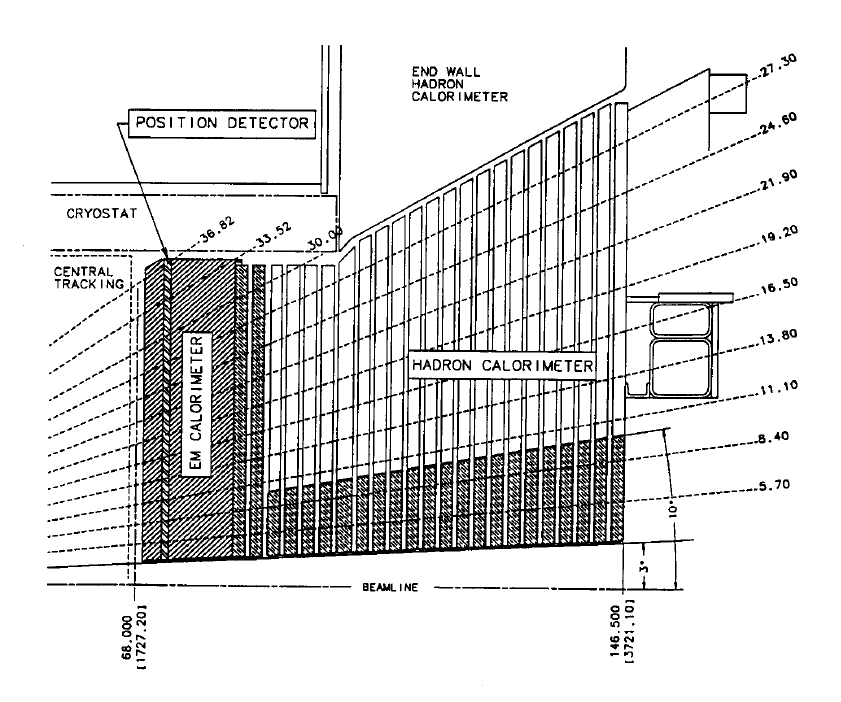}
\caption{Elevation view of one quarter of the CDF plug calorimeter.}\label{plug1}
\end{center}
\end{figure}

Plug calorimeters~\cite{plug} are two identical structures, East and 
West, covering region $1.1\lesssim|\eta|\lesssim 3.6$. Figure~\ref{plug1} 
shows the structure
of plug calorimeters, in a way similar to the central device: there is a 
\emph{Plug ElectroMagnetic calorimeter} section (PEM), a 
\emph{Plug PReshower} (PPR) detector before the calorimeter, a 
\emph{Plug Electromagnetic Shower-maximum} detector (PES) embedded (at 
$6X_{0}$)  and a subsequent \emph{Plug HAdronic calorimeter} section (PHA).

Electromagnetic section is $21X_{0}$ thick and is composed by 23 annular 
plates, of $2.77$~m outer diameter and an inner hole for the beam pipe made of
$4.5$~mm thick lead absorber. Towers have 
a segmentation with varying $\Delta\eta$ and $\Delta\phi$ as 
Table~\ref{cal_segment} shows, with an azimuthal-angle covering of 
$7.5^{\circ}$ down to $\eta=2.11$ and of $15^{\circ}$ further. Active elements
are $4$~mm thick scintillator tiles read-out by embedded wavelength shifters 
connected to PMT. All is assembled in triangular shape \emph{pizza-pans} that
enclose a slice of a $\Delta\phi=15^{\circ}$ sector. Two layers are different:
the first scintillator layer is $10$~mm thick and is used as a preshower detector,
and another layer, at about $5$~cm from the surface, is used as the Plug Electron Shower-max (PES) detector, it is made by two strips of scintillators that provide shower maximum 
position measurement with $\sim1$~mm accuracy. 

\begin{table}\begin{center}
\begin{tabular}{ccc}
\toprule
$|\eta|$ Range & $\Delta\phi$ & $\Delta\eta$ \\
\midrule
$0.-1.1 (1.3 H)$ & $15^{\circ}$ & $\sim 0.1$ \\
$1.1(1.3 H)-1.8$ & $7.5^{\circ}$ & $\sim 0.1$ \\
$1.8-2.1$ & $7.5^{\circ}$ & $\sim 0.16$ \\
$2.1-3.64$ & $15^{\circ}$ & $0.2 - 0.6$ \\
\bottomrule
\end{tabular}
\caption[CDF II Calorimeter Segmentation]{CDF II calorimeter segmentation, $H$ stands for the hadronic section.}\label{cal_segment}
\end{center}
\end{table}

\begin{figure}[!ht]
\begin{center}
\includegraphics[width=0.75\textwidth]{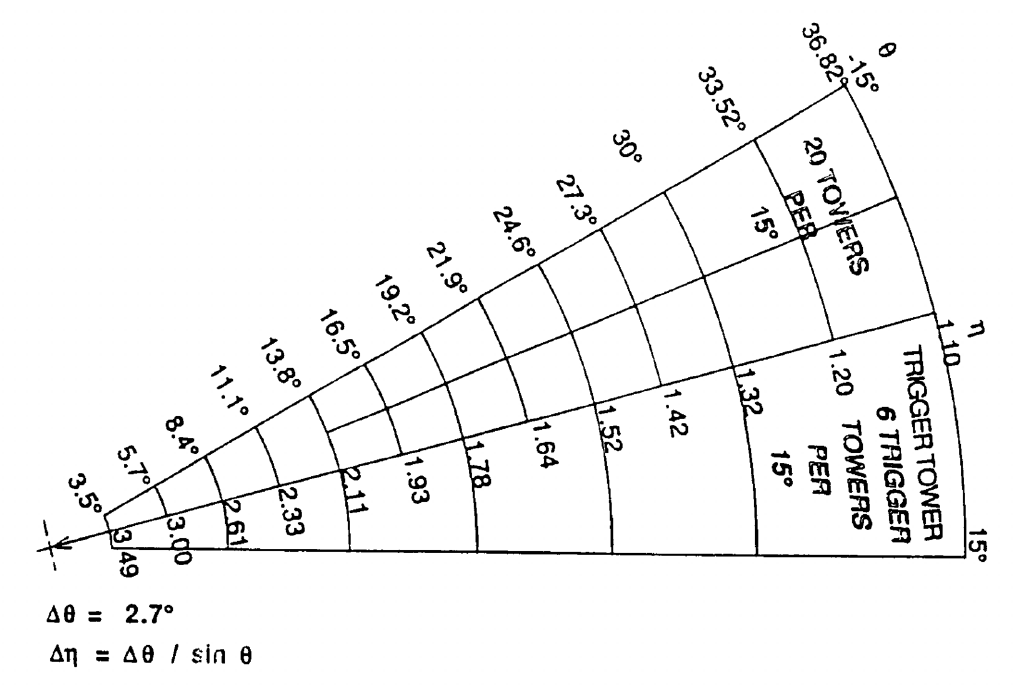}
\caption[Plug Calorimeter Segmentation]{Segmentation of the plug calorimeter and tower location inside one wedge.}\label{plug2}
\end{center}
\end{figure}

Hadronic section is about $7\lambda_{int}$ thick and segmented in 
$\Delta\phi=30^{\circ}$ for a total of 12 sections of 23 iron $5$~cm-thick 
layers alternated with $6$~mm scintillator active material layers. The 
characteristic {\em plug} shape is due to the growing radii of the layers far 
from interaction point to match WHA coverage. Energy resolution is:
\begin{equation}\label{pem_res}
\frac{\sigma_{E}}{E}=\frac{16\%}{\sqrt{E(\mathrm{GeV})}}\oplus 1\%, \qquad \frac{\sigma_{E}}{E}=\frac{74\%}{\sqrt{E(\mathrm{GeV})}}\oplus 4\%.
\end{equation}
respectively for PEM and PHA. Figure~\ref{plug2} shows the segmentation of a
$\Delta\phi=15^{\circ}$ sector and describes the distribution of trigger 
towers.

\subsection{Muon Detectors}\label{sec:mu_cham}
Although nearly all particles are absorbed by the calorimeter system, muons pass through the calorimeters as minimum ionizing particles and can exit the calorimeter system\footnote{Muons from $Z^0$ decays, for instance, deposit on average about $0.4$~GeV in the electromagnetic portion of the calorimeter and $4$~GeV in the hadronic one.}, therefore the outermost sub-detector of CDF is the muon detection system~\cite{muon}. It is made out of single wire drift chambers and scintillator counters for fast timing, located radially just outside the calorimeter system.

\begin{figure}[h]
\begin{center}
\includegraphics[angle=0,width=0.75\textwidth,clip=]{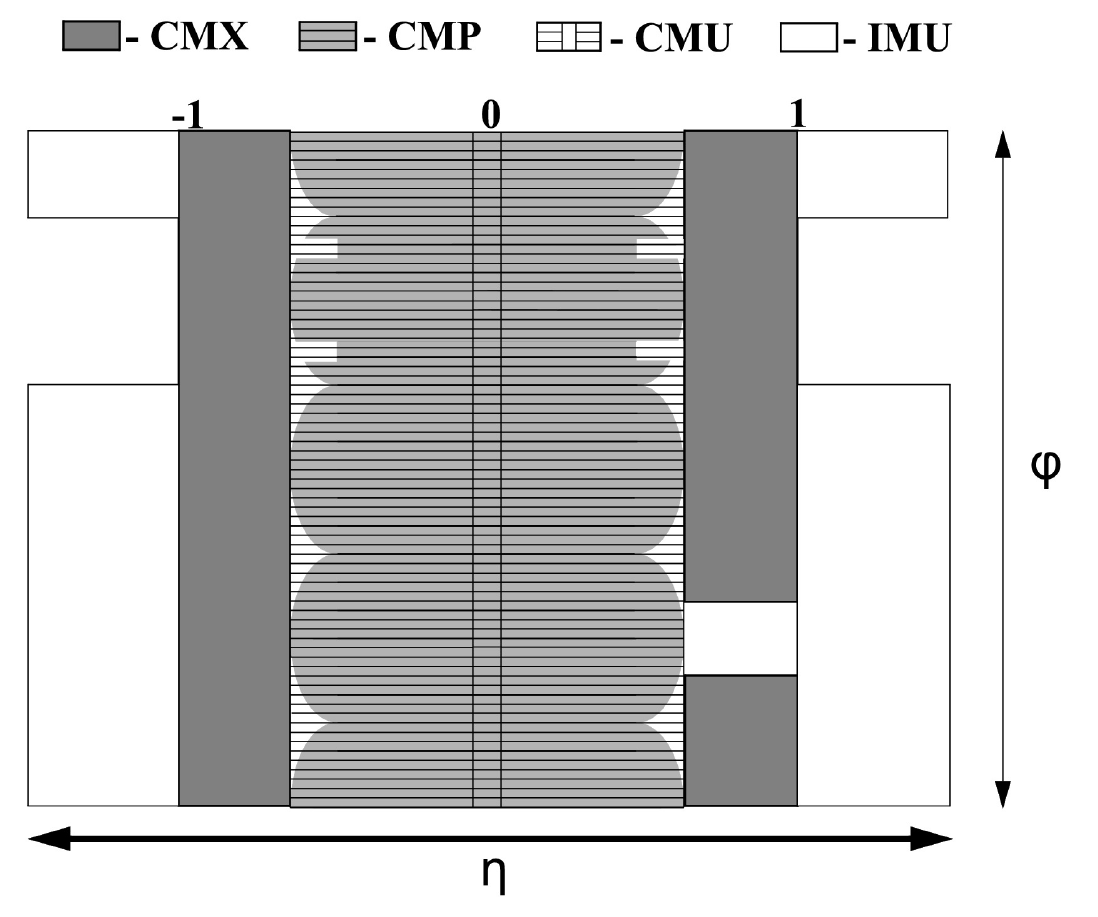}
\caption[CDF II Muon System $\eta$-$\phi$ Diagram]
{Diagram in the $\eta$-$\phi$ plane of the muon systems at CDF: CMU, CMP, CMX and BMU muon detectors. The BMU detector is referred in this diagram as IMU.}\label{figure:MuonCoverage}
\end{center}
\end{figure}

There are various muon subsystems with slightly different characteristics and named according to their locations: the Central Muon Detector (CMU), the Central Muon uPgrade Detector (CMP), the Central Scintillator uPgrade (CSP), the Central Muon eXtension Detector (CMX), the Central Scintillator eXtension (CSX), the Toroid Scintillator Upgrade (TSU), the Barrel Muon Upgrade (BMU) and the Barrel Scintillator Upgrade (BSU). 
The CMU, CMP and CSP systems cover an $\eta$ range of $|\eta|<0.6$, the CMX and CSX systems cover an $\eta$ range of $0.6<|\eta|<1.0$ and the TSU, BMU and BSU subsystems cover an $\eta$ range of $1.0<|\eta|<2.0$. A diagram of the muon subsystems coverage can be seen in Figure~\ref{figure:MuonCoverage} .

The innermost muon system is CMU, it was built for CDF I and is located just outside the CHA calorimeter, at a radius of $350$~cm and arranged in $12.6$\degree wedges in $\phi$. Each wedge is made of three layers ({\em stacks}) composed by four rectangular drift tubes. Each drift tube operates in proportional mode, with an Argon-Ethane mixture gas and a single 50~$\mu$m sense wire in the middle of the cell, parallel to the $z$ axis: absolute differences of drift arrival time between two tubes provide a measurement of the azimuthal coordinate, while the charge division at each end of a wire can be used to determine the $z$ coordinate. The CMU is followed by another muon system of similar structure, the CMP, installed beyond a $60$~cm thick layer of steel. The minimal $p_T$ threshold for the CMU (CMP) is $1.4$ ($2.2$)~GeV$/c$. 

Outside the CMP we find the CSP: a fast response detector used for triggering and formed by a single scintillator layer connected to a light guide and a PMT. 

The CMX muon system is located at each edge between the CDF barrel and forward regions. It has a conical geometry with drift chambers similar to the CMP. Also, it has a scintillating system called the CSX, similar to the CSP. The CMX system covers $360$\degree with 15 wedges in $\phi$. Each wedge is formed of eight layers of drift chambers in the radial direction. 
Various properties of the CMU/CMP/CMX subsystems are summarized in Table~\ref{table:MuonSubsystems}.

\begin{table}[h] 
\begin{center}
\begin{tabular}{cccc}\toprule
General Parameters & CMU & CMP & CMX\\
\midrule
$\eta$ coverage & 0-0.6 & 0-0.6   & 0.6-1.0\\
$p_T$ Threshold [GeV$/c$] & 1.4 & 2.2 & 1.4\\
\midrule 
Drift Tubes & CMU &  CMP & CMX\\
\midrule 
Thickness [cm] & 2.68 & 2.5 & 2.5 \\
Width [cm] & 6.35 & 15 & 15 \\
Length [cm] & 226 & 640 & 180 \\
Max. drift time [$\mu$s] & 0.8 & 1.4 & 1.4 \\
\midrule 
Scintillators & N/A & CSP & CSX\\
\midrule 
Thickness [cm] & N/A & 2.5 & 1.5 \\
Width [cm] & N/A & 30 & 30-40 \\
Length [cm] & N/A & 320 & 180 \\
\bottomrule
\end{tabular}
\caption[CDF II Summary of Muon Subsystem Properties]{Summary of the properties of the muon subsystems at CDF. \label{table:MuonSubsystems}}
\end{center}
\end{table}

Muon identification proceed on the base of short ionization tracks left in the drift chambers (called {\em stubs}) and reconstructed thanks to the timing information provided by the individual drift chambers. Then a COT track is matched to the stubs to confirm the muon candidate providing an accurate measurement of the the muon momentum.

\subsection{Cherenkov Luminosity Counters}\label{clc_par}

The \emph{Cherenkov Luminosity Counters}~\cite{clc_performances} (CLC) are two symmetrical detector modules designed to measure the instantaneous luminosity through the rate of $p\bar{p}$ interactions in the forward region.

\begin{figure}[!ht]
\begin{center}
\includegraphics[width=0.75\textwidth]{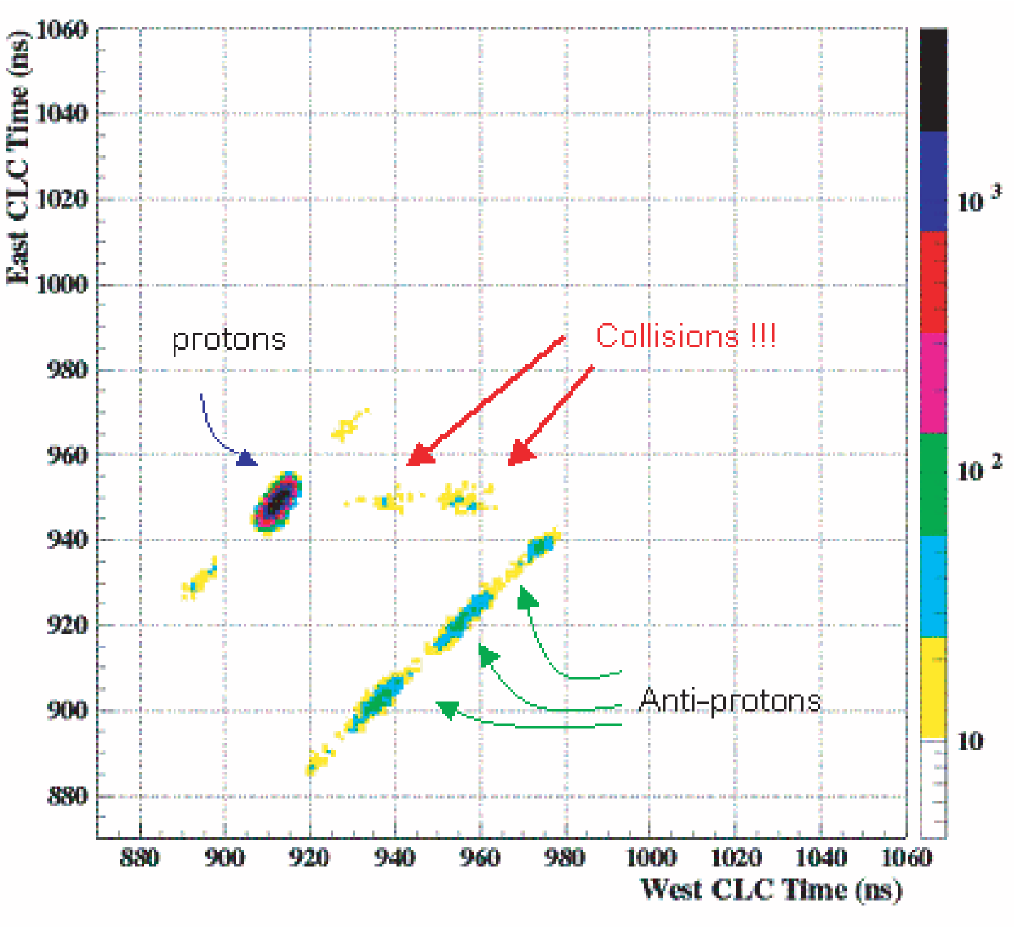}
\caption[Time Distribution of CLC Modules Signal]{Time distribution of East and West CLC modules signals. A $p\bar{p}$ collision deposits a coincidence signal in the two modules.}\label{clc_coincidence}
\end{center}
\end{figure}

Each counter is made of 48 conical, isobuthane gas filled, Cherenkov counters pointing to the nominal interaction region and located inside 
each plug calorimeter in a forward pseudo-rapidity region (\mbox{$3.7<|\eta|<4.7$}). Cones are disposed in a concentric way, with smaller counters at the center (length $110$~cm, initial diameter $2$~cm) and larger ones outward (length $180$~cm, initial diameter $6$~cm). The narrow shape and orientation is optimal to collect particles outgoing from the 
interaction point that produce an important Cherenkov light yield. On the other hand particles from beam halo or from secondary interactions have larger crossing angle, hence they produce a much smaller signal. The excellent time resolution (less than $100$~ps) allows the analysis of the coincidence between the two modules (East and West) and it is an additional tool to remove 
background interactions. Figure~\ref{clc_coincidence} shows the time distribution of the hits on the two modules. 

The CLC signal shape is used to measure the average number of inelastic interactions per bunch crossing ($\bar{N}$), the instantaneous luminosity $\mathscr{L}$ is inferred from the relation:
\begin{equation}\label{lum1}
\bar{N} f_{b.c.}=\mathscr{L} \sigma_{in} \epsilon ,
\end{equation}
where the bunch crossing frequency ($f_{b.c.}$) is precisely known from the Tevatron RF, $\epsilon$ is the CLC acceptance for inelastic scattering and $\sigma_{in}$ is the inelastic $p\bar{p}$ cross section. The last parameter comes from the averaged CDF and E811 luminosity independent measurements at $\sqrt{s}=1.80$~TeV~\cite{cx_ppbar_cdf, cx_ppbar_e811}, extrapolated to $\sqrt{s}=1.96$~TeV:
\begin{equation}\label{lum2}
\sigma_{in}(1.80\mathrm{~TeV})=60.4\pm 2.3\mathrm{~mb}\rightarrow\sigma_{in}(1.96\mathrm{~TeV})=61.7\pm 2.4\mathrm{~mb}.
\end{equation}
The combined systematic uncertainty on the luminosity measurement~\cite{clc_performance} is $6$\%: a
4\% due to the extrapolation applied in Equation~\ref{lum2} and about 4\% due to the uncertainty of the CLC acceptance.

\section{Trigger and Data Handling}\label{sec:trigger}

The purpose  of the trigger system is the on-line selection of useful physics events from the background of uninteresting processes produced at much higher rate. The online selection step is rather important because only a fraction of data can be stored for offline physics analysis.

At the B0 interaction point, with a bunch crossing frequency of $2.5$~MHz, an inelastic $p\bar{p}$ cross section of $\sigma_{in}\simeq 60$~mb and an instantaneous luminosity of 
\mbox{$\mathscr{L}\simeq 10^{32}$~cm$^{-2}$s$^{-1}$}, there are about $1\div 2$ inelastic collision in each bunch crossing. It is clearly impossible to store the entire detector information for each collision, as the maximum recording rate is $50\div 100$~Hz, and it would also be useless, because interesting processes have much smaller cross section than generic inelastic interactions (diboson production cross sections are $\mathscr{O}(10^{-9})$ w.r.t. generic jet production, see Figure~\ref{processes}). The CDF trigger system is designed for the efficient selection of the interesting events.

\begin{figure}[!ht]
\begin{center}
\includegraphics[width=0.525\textwidth]{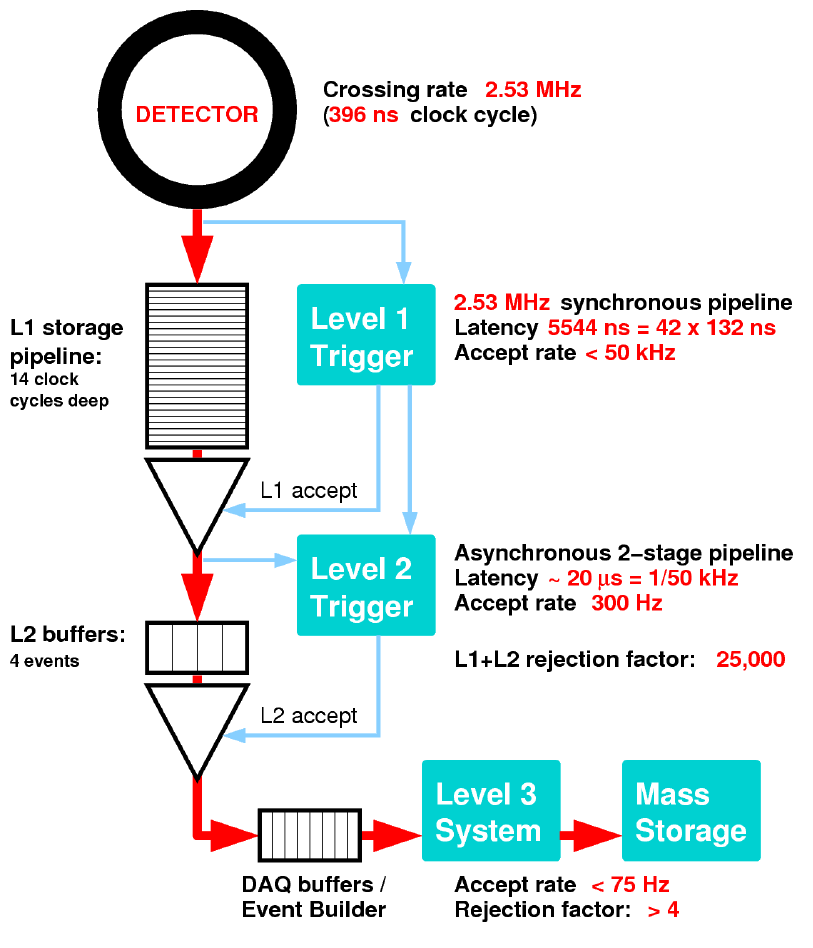}
\includegraphics[width=0.465\textwidth]{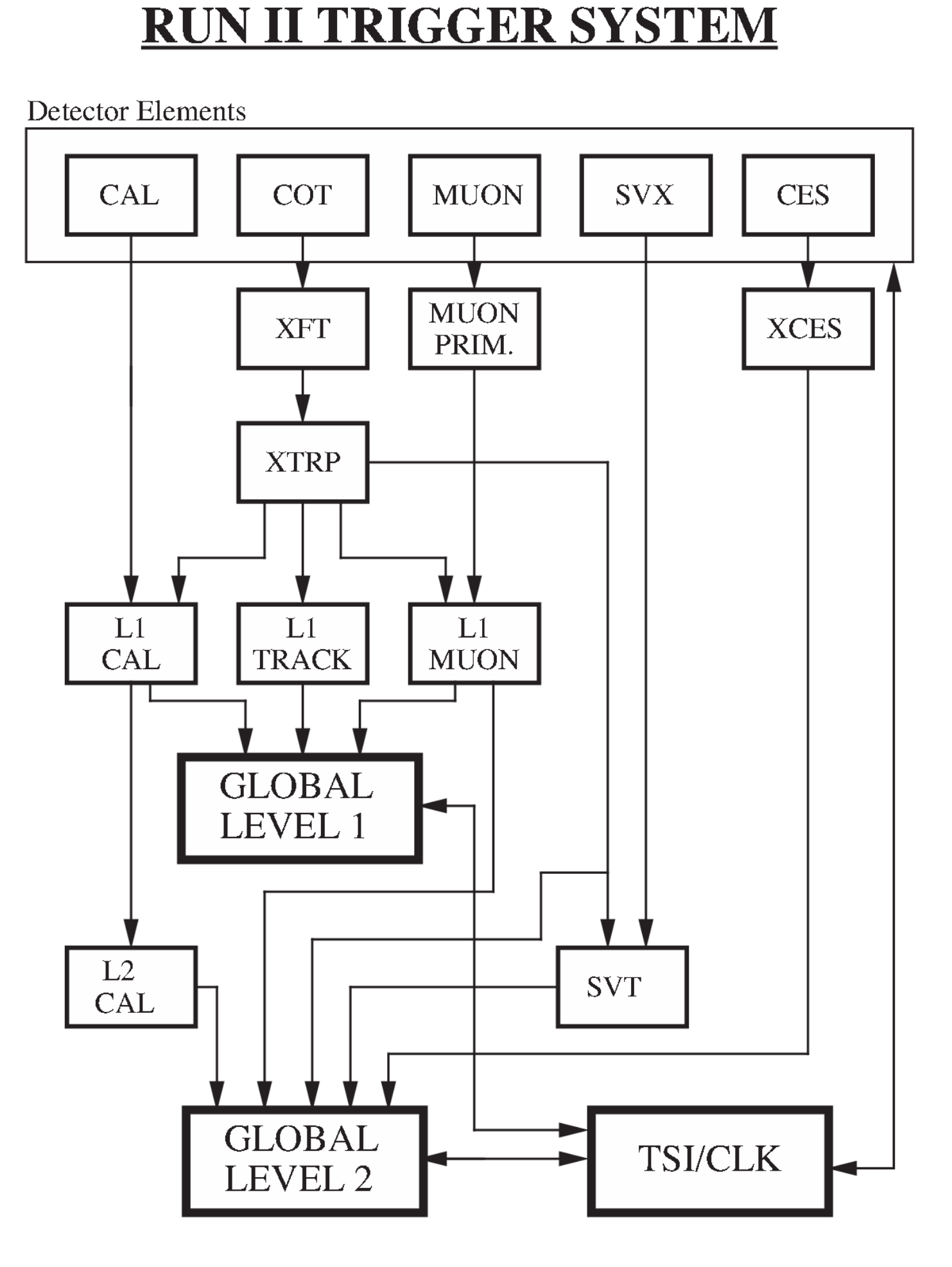}
\caption[CDF II Trigger System]{The CDF II trigger system. Left: block diagram of three-levels trigger
 and DAQ system. Right: L1 and L2 trigger streams.}\label{trigger_lev}
\end{center}
\end{figure}

The system is composed by three levels, L1, L2 and L3 (see 
Figure~\ref{trigger_lev}); each one provides a sufficient rate reduction to allow
the feeding and processing by the next level with, virtually, no dead-time\footnote{\emph{Dead-time} occurs when events must be rejected because trigger system is occupied processing a preceding event.}.

Each level filters the events using a set of programmable conditions, step by 
step more complex as the detector read-out completes and more elaboration time becomes available.
A, so-called, \emph{trigger path} is the logic combination of criteria from different levels. 

A peculiar requirement that needs to be described is the {\em PreScale} (PS) condition: a known fraction of the events selected by L1 or L2 are immediately discarded before the elaboration of the following trigger level. PS can be fixed or {\em Dynamic} (DPS): the first is applied to auxiliary trigger paths, used for efficiency estimates or data quality control, the second is used for physics trigger paths that would require excessive computing time, thus their scaling factor is optimized during the data taking according to the available bandwidth at each trigger level.

The several trigger paths used this analysis are described in Section~\ref{sec:onlineSel}, however the CDF experiment collects about 150 trigger paths. They are arranged in a \emph{trigger table} aimed to maximise the acceptance of interesting events allowing a maximum acquisition dead time of $5\%$.

\subsection{Level-1}

The first trigger level is a custom designed hardware system composed by three
parallel streams which feed inputs to Global Level-1 decision unit (see Figure~\ref{trigger_lev}). One stream, \verb'L1_CAL', collects prompt colorimetric response: it is divided into {\em object triggers}, i.e. single hadronic or electromagnetic deposits above threshold, and {\em global triggers}, i.e. total event transverse energy ($\sum{E_{T}}$) or the presence of raw missing transverse energy\footnote{Missing transverse energy is defined as $\vec{\mathrm{\cancel{E}}}_{T}\equiv - \sum_i{\vec{E}_T^i}$, with $i$ index of the calorimeter towers. See Section~\ref{sec:metObj} for more details.} (MET or \cancel{E}$_T$ ).
A second stream, \verb'L1_MUON', collects information from muon chambers thus identifying muon stubs. The last stream, \verb'L1_TRAK', comes from the eXtreme Fast Tracker (XFT), a powerful
parallel pattern recognition algorithm used to feed COT raw tracking information to L1 trigger. 
The collected information already allows a coarse but efficient reconstruction of candidate physics objects: a track plus matched EM deposit is an electron candidate, a track plus a matched stub is a muon candidate, MET can identify a neutrino, a hadronic cluster in the calorimeter can be a jet and so on. 

While the L1 trigger takes a decision, the events are stored in forty-two pipelined buffers synchronized with Tevatron clock cycles ($132$~ns). In a latency time of $132\cdot42\simeq 5.5~\mu$s, L1 drastically reduces the rate of accepted events from $2.5$~MHz to about $50$~KHz, the feed rate of L2.

\subsection{Level-2}

At the second trigger level there is enough time to readout the information of slower sub-detectors and perform more complex physic object identification algorithms. The main components of L2 are:
the readout of the shower-max trigger boards, the 3D reconstruction of the XFT tracks~\cite{xft_3d}, the \verb'L2CAL' hardware~\cite{l2cal_up} clustering of calorimeter towers and the Silicon Vertex Trigger~\cite{svt_up} (SVT) for the reconstruction of displaced secondary vertices. 

The shower-maximum detector information ensures a better electron or photon recognition with azimuthal information and better spatial resolution. Electron identification is also improved by the XFT 3D track matching. The requirement of a track matched to a muon chamber stub improves also the muon fake rejection.

\verb'L2CAL' is a custom hardware setup able to perform raw energy clustering. Adjacent towers above a predefined threshold (usually few GeV), are merged iteratively to build refined trigger objects like jets or EM clusters. The use of physics objects allows to define advanced selection criteria on the base of  detector $\eta-\phi$ position, multiplicity or $E_T$ threshold for the full objects.

The last fundamental piece of the L2 is SVT: the SVX II detector $r-\phi$ side is readout and the silicon hits, together with XFT information, are compared on on the fly with a large batch of Associative Memories (AM) where all the possible track configurations (of a certain resolution) are simulate and stored. The comparison with pre-processed simulation is the key for an extremely fast reconstruction of the track parameters and it allows the identification of displaced secondary vertices, a possible signature of beauty hadrons present in the event.

Starting from the L1 input rate of $\sim 50$~KHz, L2 must bring the accept rate to $300$~Hz exploiting four memory buffers and a short latency time of \mbox{$5.5\times 4 = 22~\mu$s}. The high luminosity delivered by the Tevatron after 2005 produced events with lager detector occupancy that required more time for L2 elaboration. To cope with this, an extensive upgrade of the L2 hardware  took place between 2005 and 2006. My first experience with CDF happened in this context and, in particular, I was part of the SVT~\cite{svt_up} upgrade team.

\subsection{Level-3} 

The last stage of the trigger system, L3, is composed by a farm 
of several hundreds processors exploiting LINUX OS and C++ based software for the full event reconstruction. The complete detector information is assembled by the \emph{EVent Builder}~\cite{evb}
(EVB) and processed by a simplified version of the offline reconstruction code.

If the L3 trigger requirements are satisfied, the {\em event record} corresponding to the given bunch crossing is transferred to the Consumer Server/Data Logger (CSL) that streams the data to disk, while a fraction of the output is also used for real time monitoring. The L3 accept rate suitable for disk storage is $100$~Hz.

\subsection{Data Structure}\label{sec:data-struc}

When an event record is saved on disk, it is labelled with a progressive number, grouped and classified.

All the events belonging to a continuous data taking period  are expected to have  very similar detector conditions (i.e. active sub-detecors, calibration parameters, trigger table, etc) so they define a {\em run}. A large set of runs is grouped into a {\em data period}, that usually corresponds to an integrated luminosity of a few hundreds of pb$^{-−1}$. The thirty-eight data periods that compose the complete CDF dataset are summarized in Table~\ref{tab:datasets} together with the corresponding integrated luminosities and run ranges.

Events are also classified into {\em data-streams}, a four-character label, describing similar trigger properties or common physical interest. This analysis uses four of the several data-streams available at CDF: 
\begin{description}
  \item[\texttt{bhel}:] high $E_T$ central electron stream;
  \item[\texttt{bpel}:] high $E_T$ forward electron stream;
  \item[\texttt{bhmu}:] high $p_T$ muon stream;
  \item[\texttt{emet}:] high \met stream.
\end{description}

\renewcommand{\arraystretch}{0.95}
\begin{table}
  \begin{center}
    \begin{small}
    \begin{tabular}{cccc}\toprule
      Data Period & Run range & Period $\int\mathscr{L}dt$ (pb$^{-1}$) & Total $\int\mathscr{L}dt$ (pb$^{-1}$)\\

\midrule
p0	&138425-186598	&550	&550   \\
p1	&190697-195408	&130	&680   \\
p2	&195409-198379	&130	&810   \\
p3	&198380-201349	&100	&910   \\
p4	&201350-203799	& 95	&1005  \\
p5	&203819-206989	&135 	&1140 \\
p6	&206990-210011	& 110 	&1250 \\
p7	&210012-212133	& 50 	&1300 \\
p8	&217990-222426	& 210 	&1510 \\ 
p9	&222529-228596	&180 	&1690 \\
p10	&228664-233111	&280 	&1970  \\
p11	&233133-237795	&264 	&2234  \\
p12	&237845-241664	&185 	&2419  \\
p13	&241665-246231	&317 	&2736  \\
p14	&252836-254683	&44.5 	&2780  \\
p15	&254800-256824	&159 	&2939  \\
p16	&256840-258787	&142 	&3081  \\
p17	&258880-261005	&188 	&3269  \\
p18	&261119-264071	&407 	&3676  \\
p19	&264101-266513	&287 	&3963  \\
p20	&266528-267718	&256 	&4219  \\
p21	&268155-271047	&520 	&4739  \\
p22	&271072-272214	&292 	&5031  \\
p23	&272470-274055	&232 	&5263  \\
p24	&274123-275848	&283 	&5546  \\
p25	&275873-277511	&236 	&5782  \\
p26	&282976-284843	&189 	&5971  \\
p27	&284858-287261	&422 	&6393  \\
p28	&287294-289197	&333  	&6726  \\
p29	&289273-291025	&360  	&7086  \\
p30	&291294-293800	&460 	&7546  \\
p31	&293826-294777	&172 	&7718  \\
p32	&294778-299367	&435 	&8153  \\
p33	&299368-301303	&357 	&8510  \\
p34	&301952-303854	&359 	&8869  \\
p35	&304266-306762	&364 	&9233  \\
p36	&306791-308554	&462 	&9695 \\
p37	&308570-310441	&174 	&9869 \\
p38	&310472-312510	&252 	&10121 \\
\bottomrule
    \end{tabular}
    \end{small}
    \caption[Data Period and Luminosity]{Summary of data taking periods for the complete CDF dataset. Corresponding run range and collected integrated luminosity are reported. \label{tab:datasets}}
  \end{center}
\end{table}  
\renewcommand{\arraystretch}{1.}

After that the event information is saved on disk and properly classified, it is possible to start the offline processing, also named {\em production}. At this stage, the low-level detector data is extracted, corrected with the calibration constants and appropriate algorithms are used to reconstruct high-level physics objects (tracks, electrons, muons and jets described in Chapter~\ref{chap:objects}).

The software is an object oriented framework where all the algorithms are defined by a self consistent \texttt{C++} module~\cite{data_proc}; this allows an independent testing and development of the separate algorithms and improves flexibility. For example the same {\em track module} can be used both to reconstruct a single track information or as an element of a more complex algorithm, like the electron identification module. The final format of the physics data is a large array ({\em n-tuple}) that can be analyzed with commercial software\footnote{The open source \texttt{ROOT}~\cite{rootr} analysis software is widely used in the high energy physics community.}.  

Although each of the \texttt{AC++} module can be improved or modified, the analyzed data should be as much stable and uniform as possible therefore there is always a {\em recommended} analysis prescription for the production code version to use in each data period: Table~\ref{tab:prod_prescription} reports the one used in this thesis and approved by the Higgs Discovery  Group (HDG).

\begin{table}
  \begin{center}
    \begin{tabular}{cccc}\toprule
      Data Period & p0 & p1-p17 & p18-p38 \\\midrule
      Production Version & 5.3.1 & 6.1.1 & 6.1.6p+ \\      
      \bottomrule 
    \end{tabular}
    \caption[Data Code Production Version]{Production code versions used the different data periods.}\label{tab:prod_prescription}
  \end{center}
\end{table}

\section{Monte Carlo Simulation}\label{sec:mcGen}

An accurate Monte Carlo (MC) simulation of the physics processes of interest and of the detector response are a fundamental tool for most of the high energy physics experiments. A wide variety of MC samples is available to the CDF users and, as explained in Chapters~\ref{chap:sel} and~\ref{chap:bkg}, this analysis relies on MC both for signal evaluation and for part of the background estimate. 

The simulation of an high energy hadron collision event proceeds through four independent phases:
\begin{description}
\item[Parton Density Function Application:] quarks and gluons are the initial states of any interaction. Unluckily they are {\em confined} by the strong interaction within the $p$ and the $\bar{p}$ making impossible a perfect knowledge of the initial parameters. The Particle Density Functions (PDFs) overcome this problem by giving a parametrization of the interaction probability as a function of the momentum transfer ($Q^2$) during the $p\bar{p}$ collision. When a MC event generator starts the evaluation of a process, the first step is the extraction of two partons of given $Q^2$ from the PDFs. Several parametrization of the PDFs exist but all the MCs used in this analysis employ the CTEQ5L~\cite{cteq5l} PDFs. Systematic variations are obtained by using different  PDFs sets and varying the prediction within the theoretical uncertainties.

\item[Event Generation:] once that two initial partons are extracted from the PDFs functions, a hard interaction between them is calculated to obtain the simulation of the desired final state.
Within the perturbative approximation~\cite{qftbook_weinberg}, the Matrix Element (ME) equation is derived with numerical integration of Leading-Order (LO) or Next-Leading Order (NLO) equations. The softwares performing such calculations are named {\em generators}. Here we exploit a variety of them depending of the different final states: \verb'ALPGEN'~\cite{alpgen}, v2.1, for $W+$jets and $Z+$Jets prediction, \verb'PYHTIA'~\cite{pythia} v6.216 for $WW$, $WZ$, $ZZ$ and $t\bar{t}$ prediction and, finally, \verb'POWHEG'~\cite{powheg1,powheg2, powheg3} v6.510, for single-top $s$ and $t$ channel NLO prediction.

\item[Parton Shower:] away from the hard interaction, higher order QCD processes are needed. The perturbative approach breaks and simulation is based on analytical parametrization and QCD models. Parton Shower (PS) programs like \verb'PYHTIA' and \verb'HERWIG'~\cite{herwig} are used to simulate quark hadronization, soft gluon emission or underlying-event processes (i.e. secondary soft interactions and spectator quark interactions).
Partons are evolved until they form real final state particles that can be undergo physical interaction in the detector. 

\item[Detector Simulation:] the MC receives, as input, the positions, the four-momenta, and the identities of all particles and it reproduces the response of the different sub-detectors, including resolution effects, passage through passive material (such as cables or support structures) and secondary decays. CDF uses the \verb'GEANT3'~\cite{geant}, V.3.15, program to model the tracking volume of the detector. A mathematical model is used with full  simulation of charged particles passage, showering and secondary or tertiary particle production. The calorimeter section is not completely simulated because it would be too much time and CPU consuming. A much faster parametric response program, called \verb'GFLASH'~\cite{geant4}, tuned on test beam data, is used.
\end{description}
When all the simulation is completed, the MC events are saved into a data structure identical to the one used for collision data, thus allowing reconstruction algorithms to work in the same way on data and MC events.

\clearpage

\chapter{Physics Objects Identification}\label{chap:objects}

The final state topology of this analysis presents four different high-$p_T$ physics objects reconstructed combining the data of several sub-detectors. The identification and reconstruction methodologies are introduced here and discussed more in depth in the rest of the Chapter.

Events are selected online by the three-level trigger system described in Section~\ref{sec:trigger} with two trigger strategies (described more in detail in next Chapter): {\em single high-$p_T$ lepton} triggers and {\em multiple-objects} triggers. After this, the digitalized electrical pulses recorded by the CDF sub-detectors are analyzed to reconstruct the physics objects of interest: the primary vertex of the interaction, one charged lepton, missing transverse energy (signaling the escape of an undetected neutrino) and two high-$E_T$ jets containing a reconstructed secondary vertex that {\em tags} the presence of a Heavy Flavor ($HF$) hadron decay.

Given the small expected signal yield, the efficient identification of charged lepton candidates is a key feature of the analysis. We identify three main lepton categories: electrons, muons and isolated tracks. Electrons are defined as electromagnetic energy clusters matched to a charged track, muons are defined by ionization deposits in the muon chambers ({\em stubs}) matched to a charged track and, finally, isolated tracks are just high-quality charged tracks isolated from other detector activity. The variety of identification criteria, with a total of {\em eleven} different selection algorithms, allows to identify a large fraction of the $W\to\ell\nu$ decays: factorizing the trigger efficiency and the jet and neutrino selection, we estimate an approximate lepton acceptance of $35$\% for electrons, $45$\% for muons and $4$\% for taus\footnote{$\tau$ identification is not enforced but leptonic and one prong $\tau$ decays contributes to the selected lepton categories.}.

$W$ selection is completed by the neutrino identification (Section~\ref{sec:metObj}). A neutrinos is the only particle that leaves the experimental apparatus completely undetected, however its presence is revealed by a large imbalance in the total transverse energy of the event (MET or \met), since the total transverse energy of the $p\bar{p}$ interaction is expected to be zero.

Another key feature of the analysis is the identification of $HF$ jets (Section~\ref{sec:jetObj}). Jets are the experimental signature of high momentum quark and gluon production that hadronizes in a narrow shower of particles. They appear as a energy deposit (clustered) in both electromagnetic and hadronic calorimeters.
Heavy flavor quarks hadronize to meta-stable particles that can travel a distance away from the primary interaction vertex before they decay in other particles. The identification of a secondary decay vertex, displaced from the primary, is used to {\em tag} the jet as coming from a $HF$, $b$ or $c$ quark, hadronization. The selection efficiency for an event containing a $b\bar{b}$ quark pair is approximately $50$\%, going down to about $12$\% for $c\bar{c}$ pairs; a residual contamination of $1\div 2$\% Light-Flavor ($LF$) quarks selection is also present. A Neural-Network flavor separator (named KIT-NN and described in Section~\ref{sec:kitnn}) is also applied on each tagged jet to separate $b$ quarks from $c$ and $LF$ quark components, thus giving further discrimination between $W\to cs$ and \mbox{$Z\to b\bar{b}$} signals.

\section{Charged Tracks Reconstruction}\label{sec:track}

The ability to detect and reconstruct charged particle trajectories is essential for particle identification and momentum reconstruction. Precise, high efficiency tracking is the first step in the lepton identification, moreover track reconstruction allows the measurement of the track impact parameter, thus the identification of secondary vertices.

\begin{figure}[!ht]
\begin{center}
\includegraphics[width=0.7\textwidth]{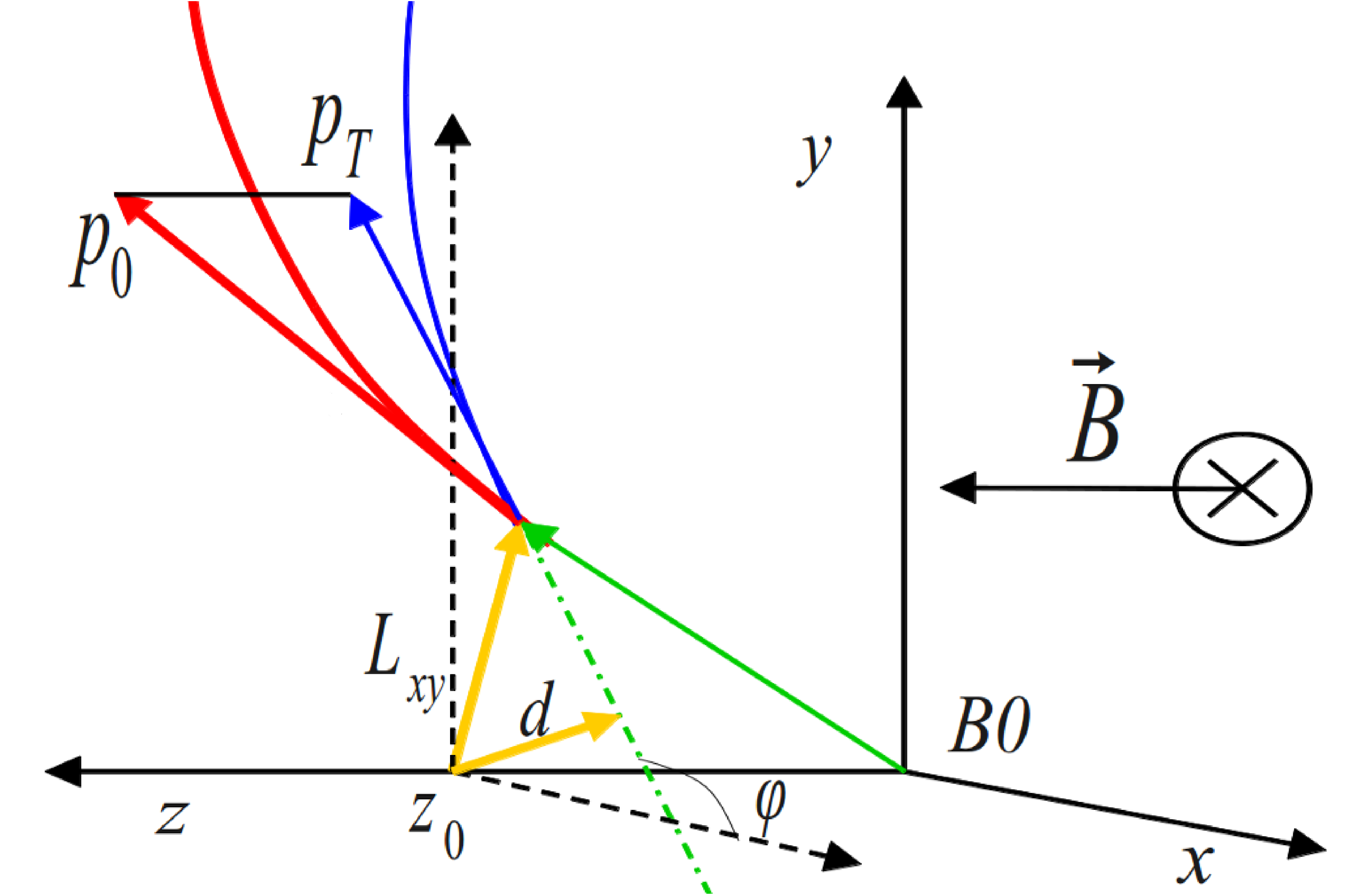}
\caption[Track Reconstruction Parameters]{CDF track parameters and coordinate system.}\label{coordinate}
\end{center}
\end{figure}

A charged particle moving in a uniform magnetic field ($\vec{B}$ with $|\vec{B}|=1.4$~T for CDF) produces an helicoidal trajectory that can be uniquely described by five parameters. At CDF we use (see Figure~\ref{coordinate}):
\begin{itemize}
\item $C$: the half-curvature of the trajectory, $C\equiv 1/2qr$ with $r$ equal to the helix radius and $q$ the measured charge of the particle. It has
  the same sign of the particle charge and it is related to the transverse momentum of the track:
  \begin{equation}
    p_T=\frac{cB}{2 |C|}
  \end{equation}
\item $d_0$: the impact parameter, i.e. the distance of closest approach in the transverse plane between the helix and the origin. It is defined as:
  \begin{equation}
    d_0=q\Big(\sqrt{x_0^2+y_0^2}-r\Big)\mathrm{,}
  \end{equation}
  where $x_0$ and $y_0$ are the coordinates of the center obtained by the projection of the helix on the transverse plane and $r=1/2C$. The quality of $d_0$ measurement is often parametrized by the {\em impact parameter significance} defined as $|d_0/\sigma_{d_0}|$.
\item $\lambda$: the helix pitch, i.e. the cotangent of the polar angle between the track and the $z$-axis ($\cot \theta_0$). The longitudinal component of the momentum is given by:
 \begin{equation}
   p_z=p_T\cot \theta_0\mathrm{.}
 \end{equation}
\item $z_0$: the $z$ position of the track vertex.
\item $\phi_0$: the azimuthal angle of the track at its vertex. 
\end{itemize}
The helix is completely described by these five parameters. Indeed every point along the trajectory satisfies the following equations~\cite{track_eq}:
\begin{eqnarray}
x&=& r\sin \phi -(r-d_0)\sin \phi_0\mathrm{,}\\
y&=& -r\cos \phi +(r-d_0)\cos \phi_0\mathrm{,}\\
z&=& z_0 +s\lambda\mathrm{,}
\end{eqnarray}
where $s$ is the length projected along the track, and $\phi= 2Cs+\phi_0$.

CDF exploits several tracking algorithms~\cite{cdf_trk}, optimized for different detector regions, to derive the previously defined parameters.
The main ones, described in the following paragraphs, are: the Outside-In algorithm (OI), the Silicon-Stand-Alone (SiSA) algorithm and the Inside-Out (IO) algorithm.

\paragraph{Outside-In Algorithm}

The Outside-In is the most reliable of CDF tracking algorithms as it is based on COT coverage, efficient up to $|\eta|\simeq 1$.

Track pattern recognition starts in the COT outer layers (lower hit density) and proceeds through four steps: first
each super-layer is searched for groups of three aligned hits that are fit to a straight line with the least squares method. Then the 
tracks are reconstructed from the information of the {\em axial} super-layers that are linked by two different algorithms ({\em segment linking} and 
{\em histogram linking} algorithms~\cite{hist_alg}). During the third step, the information of the stereo layers is added and the algorithm searches for the vertex of the track. As final step, a global refit of the track is performed taking into account corrections for the non-uniformity of the magnetic field and for the modeling of the electrons drift.

At second stage, the track found in the COT is propagated into the silicon system. A road around a track is defined using the uncertainties on the COT track parameters and silicon hits are added if they lie inside this predefined road. When a hit is added, the track parameters are recalculated and the search is performed again. The impact parameter resolution of COT + SVX tracks is found to be $\sigma_{d_0}\simeq 20~\mu$m.

\paragraph{Silicon-Stand-Alone Algorithm}

The hits in silicon sub-detectors not used by OI tracking are available to the Silicon-Stand-Alone algorithm~\cite{sisa}, it covers the region $|\eta|<2$ with a small residual capability up to $|\eta|\simeq 2.8$. 

The SiSA algorithm starts from a collection of at least four hits in the SVX II detector in the $r-\phi$ plane (SVX has five axial layers, three $90^{\circ}$ layers and two small angle layers) and fits the $C$, $d_0$ and $\lambda$ parameters to obtain a projection of the helix on the transverse plane. Then the algorithm creates a 3-D seed track adding small angle hits and the primary vertex information. At this point the $90^{\circ}$ stereo hits are added and a global refit is performed. 

SiSA tracks reconstructed using only SVX II have a poor resolution for high $p_T$ tracks so hits are searched in LOO and ISL with the SVX II track as seed. The track is refit if other layers can be added. However, the performances on momentum and impact parameter resolution are limited and indeed SiSA tracks are not used for secondary vertexing.

\paragraph{Inside-Out Algorithm}

The third tracking algorithm, the Inside-Out~\cite{io}, tries to recover efficiency and $p_T$ resolution in the region $1.2<|\eta|<1.8$ where the COT coverage is limited. SiSA tracks are used as seeds which are extrapolated to the COT inner cylinder. Matching hits in the COT are added, the track is refitted and all duplicates are removed.

\section{Primary Vertex Identification}\label{sec:primVtx}

Precise identification of the primary interaction vertex (PV) is the very first step in the event reconstruction process. Due to the relatively long $\sigma_Z$ of the beam ($\sigma_z\simeq 28$~cm), important correction to $E_T$ and \cancel{E}$_T$ may be needed. Furthermore, PV position allows the individuation of displaced secondary vertices in an event, the signature of long living $HF$ hadrons.

The algorithm used to reconstruct primary vertices is \verb'PrimVtx'~\cite{primVtx}: a seed vertex is calculated as the average $z$ position measured during 
collisions and is provided as input, then all tracks with $|z_{0}-z_{PV}|< 1$~cm, $|d_0|<1$~cm and $|d_0/\sigma_{d_0}|<3$ are collected and ordered in decreasing
$p_T$. They are fitted to a new $3D$ vertex and the tracks with $\chi^2 > 10$ are removed. The procedure is iterated until all accepted tracks have $\chi^2< 10$.
A quality index (see Table~\ref{tab:pVtx}) is assigned to the primary vertex depending on track multiplicity and type: a quality $\ge 12$ is required for primary interaction vertex reconstruction. 

The PV position is defined by $\left(x_{PV},y_{PV},z_{PV}\right)$. Typical $x_{PV}$ and $y_{PV}$ are of the order of tens of microns while a cut of $|z_{PV}|\leq 60$~cm (luminosity region fiducial selection) is applied to constrain the collisions in the geometrical region where the detector provides optimal coverage.

  \begin{table}[h] 
\begin{center}
    \begin{tabular}{cc}\toprule
       Criterion & Quality Value\\\midrule
       Number Si -tracks$\ge$3 & 1\\
       Number Si -tracks$\ge$6 & 3\\
       Number COT-tracks$\ge$1 & 4\\
       Number COT-tracks$\ge$2 & 12\\
       Number COT-tracks$\ge$4 & 28\\
       Number COT-tracks$\ge$6 & 60\\
      \bottomrule
    \end{tabular}
    \caption[Primary Vertex Quality Criteria]{Primary Vertex quality criteria: a quality $\ge 12$ is required for primary interaction vertex reconstruction.}\label{tab:pVtx}
\end{center}
  \end{table}

\section{Lepton Identification Algorithms}\label{sec:lep_id}

\begin{figure}[h!]
\begin{center}
\includegraphics[width=0.495\textwidth]{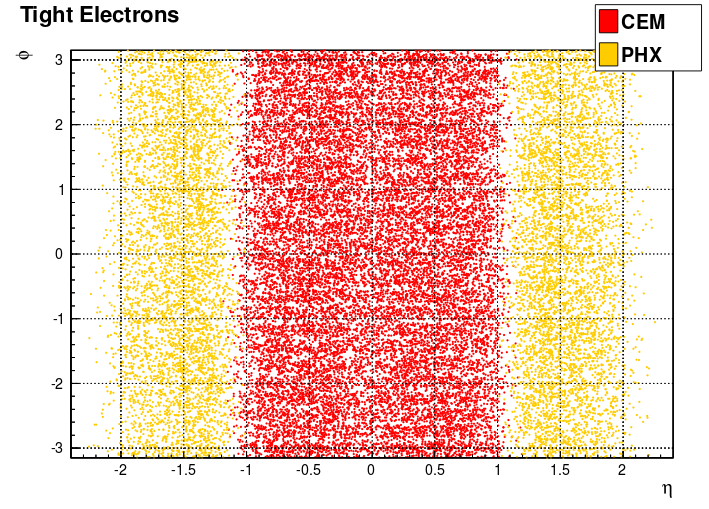}
\includegraphics[width=0.495\textwidth]{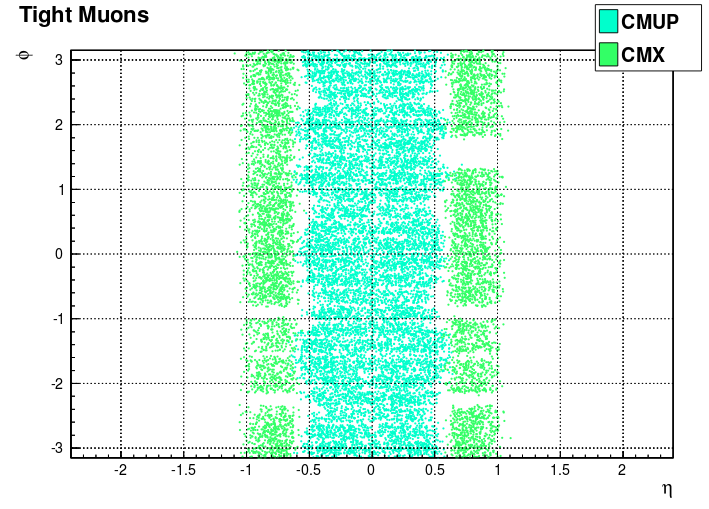}\\
\includegraphics[width=0.495\textwidth]{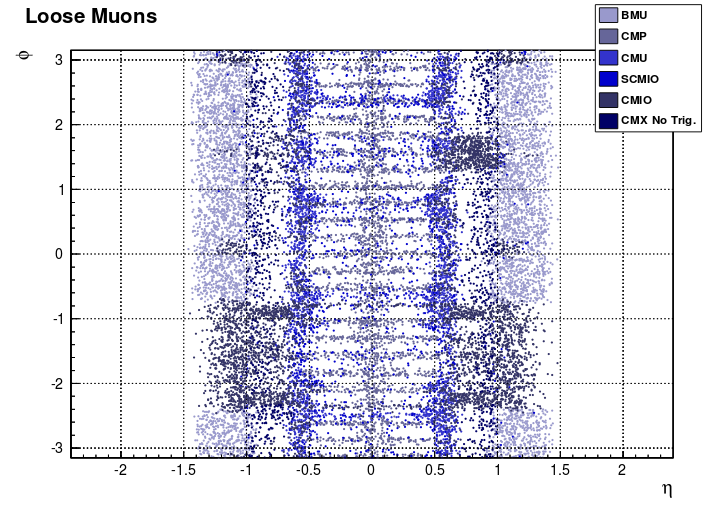}
\includegraphics[width=0.495\textwidth]{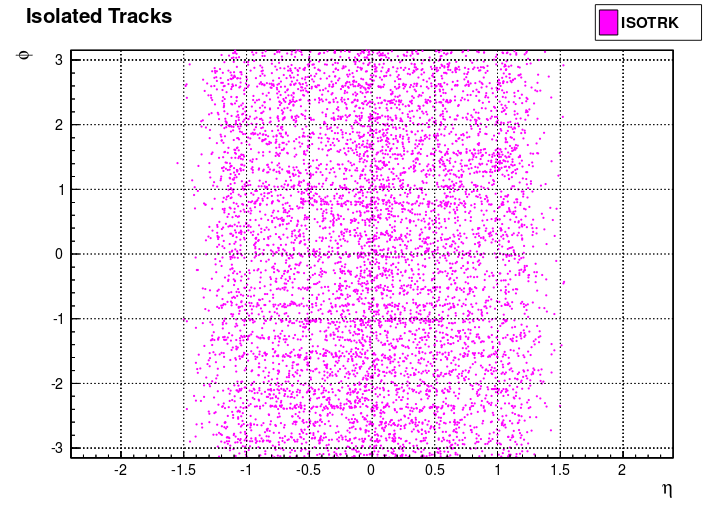}
\caption[Coverage, in the $\eta - \phi$ Plane, for the Lepton Identification Algorithms]
        {Coverage of the detector $\eta - \phi$ plane for the different lepton identification algorithms evaluated on a $WZ$ Monte Carlo: CEM and PHX tight electrons (top left); CMUP and CMX tight muons (top right); BMU, CMU, CMP, SCMIO, CMIO, CMX-NotTrig loose muons (bottom left); ISOTRK isolated tracks (bottom right).}\label{fig:split_alg}
\end{center}
\end{figure}

\begin{figure}[h!]
\begin{center}
\includegraphics[width=0.99\textwidth]{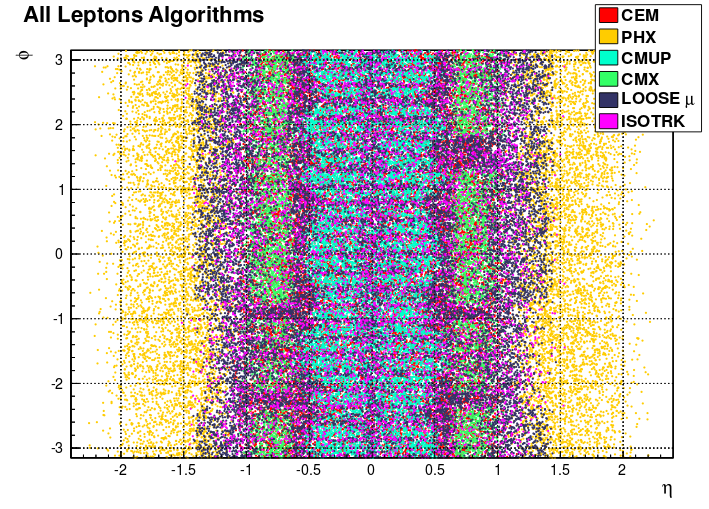}
\caption[Combined coverage, in the $\eta - \phi$ Plane, for the Lepton Identification Algorithms Combined]
{Coverage of the detector $\eta - \phi$ plane for all the combined lepton identification algorithms evaluated on a $WZ$ Monte Carlo.}\label{fig:all_alg}
\end{center}
\end{figure}

In this section the specific lepton identification algorithms\footnote{A lepton is, by the experimental point of view, an electron or a $\mu$ with no distinction between a particle and its anti-particle. Also $\tau$ leptonic decays can enter in the lepton sample but the algorithms are not optimized for them.} are discussed. We distinguish two {\em tight electron} identification algorithms (CEM, PHX), two {\em tight muon} identification algorithms (CMUP, CMX), six {\em loose muon} identification algorithms (BMU, CMU, CMP, SCMIO, CMIO, CMX-NotTrig) and one isolated track identification algorithm (ISOTRK). 
Figure~\ref{fig:split_alg} shows, for a $WZ$ MC, the detector $\eta-\phi$ coverage separately for each lepton identification category. Figure~\ref{fig:all_alg} shows them for all the categories together. The eleven lepton identification algorithms are described in the following sections.

\subsection{Electron Identification}\label{sec:eleSel}

A candidate electron is ideally defined as an energy deposit ({\em cluster}) in the EM section of the calorimeters and a charged track matched to it. The track requirements removes most of the ambiguity due to photon EM showers. 

The two electron identification algorithms used in this analysis are divided according to the CDF calorimeter segmentation in CEM (Central ElectroMagnetic object) and PHX (Plug electromagnetic object with pHoeniX tracking), for $|\eta|<1.1$ and $1.2<|\eta|<3.6$ respectively.

The CDF EM clustering algorithm~\cite{ele_cal} works in a simple but efficient way. The physical space corresponding to the calorimeter towers is mapped in the $\eta-\phi$ plane and the algorithm creates two lists of the calorimeter towers ordered by decreasing energy measurement: the {\em usable list} (working towers with energy $>100$~MeV) and the {\em seed list} (towers with energy $>2$~GeV). Then, it takes the first seed tower and creates an $\eta-\phi$ cluster by adding the neighboring towers to form a $2\times 2$ or $3\times 3$ $\eta-\phi$ area. An EM cluster is found if:
\begin{equation}
  E^{Had}/E^{EM}<0.125\textrm{,}
\end{equation}
where $E^{Had}$ is the energy deposited in the hadronic calorimeter section and $E^{EM}$ is the corresponding quantity for the EM section. As final step, the $\eta-\phi$ centroid of the cluster is calculated and the used towers are removed from the list. The algorithm selects the next seed tower and iterates the process until there are no more seed towers available. 

The $3\times 3$ clustering is used for the CEM algorithm while the $2\times 2$ clusters are used in the Plug region. A cluster is not allowed to cross the boundary between different sub-detectors. 

Several corrections are applied to reconstruct the final energy: lateral leakage, location inside the physical tower, on-line calibration and response curve drawn by the test beam data. Also the energy measured in the shower-max (CES, PES) and preshower (CPR, PPR) detectors is added to the final reconstructed energy. The shower-max profile is also compared to the calibration profiles of electrons or photons and, last but not least, it is used to measure the position of the EM shower centroid. 

\begin{figure}[!h]
\begin{center}
\includegraphics[width=0.92\textwidth]{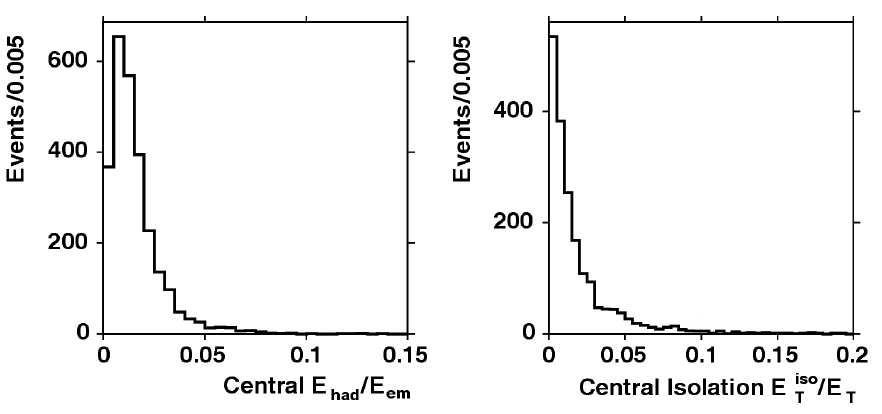}\\
\includegraphics[width=0.445\textwidth]{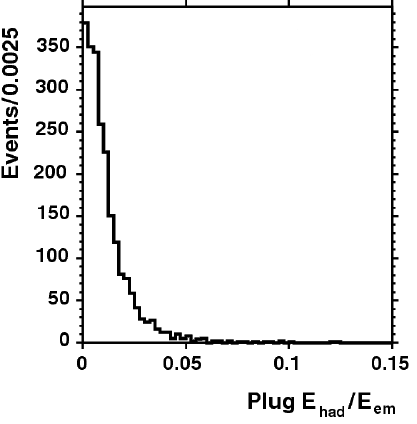}
\includegraphics[width=0.445\textwidth]{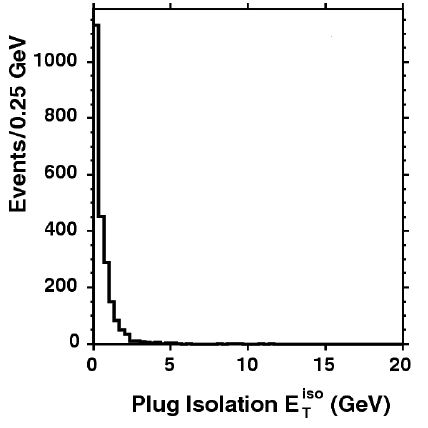}
\caption[Electron Identification Criteria for $Z^0\to e^+e^-$ Candidate Events]{$E^{Had}/E^{EM}$ (left) and isolation (right) distribution of central (top) and plug (bottom) calorimeter electron selection from unbiased, second legs of $Z^0\to e^+e^-$ candidate events in data~\cite{w_z_prl}.}\label{ele_deposit}
\end{center}
\end{figure}

Beyond the EM energy measurement, the calorimeter information is further exploited for a better particle identification. The following variables are used:
\begin{itemize}
\item $E^{Had}/E^{EM}$ ratio: studies performed with candidate $Z^0\to e^+e^-$ events~\cite{w_z_prl} show that electrons detected in the central or in the plug region have a little deposit in the hadronic part of the calorimeter (Figure~\ref{ele_deposit}). 
\item {\em Lateral shower sharing} variable, $L_{shr}$, compares the sharing of energy deposition between the towers in the CEM to the expected in true electromagnetic showers taken with test beam data:
\begin{equation}
  L_{shr}=0.14\sum_i \frac{E_i^{adj}- E_i^{expect}}{\sqrt{(0.14\sqrt{E_i^{adj}})^2+(\Delta E_i^{expect})^2}}\mathrm{,}
\end{equation}
where the sum is over the towers adjacent (\emph{adj}) to the seed tower of the cluster, $0.14\sqrt{E_i^{adj}}$ is the error on the energy measure and $\Delta E_i^{expect}$ is the error on the energy estimate.
\item The $\chi^2$ of the fit between the energy deposit and the one obtained from test beam data ($\chi^2_{strip}$ for CEM and $\chi^2_{towers}$ for PHX).
\end{itemize}

Finally, a reconstructed track matched to the EM cluster is used to suppress photons in the central region. Due to poor tracking in the forward region, PHX electrons are defined by a different strategy that relies on the {\em Phoenix} matching scheme~\cite{phx_alg}. 

Tracks candidate electrons in the central region must satisfy the following requirements:
\begin{itemize}
\item track quality: $\ge 3$ ($\ge 2$) COT Axial (Stereo) segments with at least five hits each associated with the track.
\item $E^{cluster}/p^{trk}<2$: bremsstrahlung photons are emitted colinearly to the electron and energy is radiated in the same EM cluster, therefore the measured track momentum is lower than the original $p_T$ of the electron.
\item Track-shower matching: 
  \begin{equation}
   3.0<q \Delta x<1.5\textrm{~cm}\quad\textrm{and}\quad\Delta z<3\textrm{~cm,}
  \end{equation}
  where $\Delta x=|x^{CES}-x^{trk}|$, $\Delta z=|z^{CES}-z^{trk}|$ and $q$ is the charge. 
\end{itemize}

Electron candidates in the PEM region do not have any track matching requirement due to the limited coverage of the tracking system for $|\eta|>1.1$. However to provide some fake-electron rejection a road is built starting from the $x, y$ position in the PES detector, the PV position and the curvature provided by the $E_T$ cluster. The Phoenix matching succeeds if at at last three hits in the silicon detectors are found, thus allowing the selection of a PHX electron candidate. This is not considered a {\em real} track matching because only four points are used in the fit and it is impossible to reconstruct a five-parameter helix, moreover some events have a high density of silicon hits, increasing the ambiguity of the matching. 

The other important identification requirement is the {\em Isolation} or $IsoRel$, a variable describing how much calorimeter activity surrounds the lepton. It is defined as: 
\begin{equation}
  IsoRel\equiv E_T^{iso}/E_T^{cluster}<0.1,\quad E_T^{iso}=E_T^{0.4}-E_T^{cluster},
\end{equation}
$E_T^{0.4}$ is the energy collected by the calorimeters within a radius $\Delta R=0.4$ from the centroid of the EM cluster. Isolation is used in analyses involving a $W^{\pm}$ or $Z^0$ boson because the kinematic region allowed to leptons coming from the bosons decay is usually far from jets or other particles (see Figure~\ref{ele_deposit}).

Table~\ref{tab:ele_alg} summarizes all the CEM and PHX identification criteria while the top left part of Figure~\ref{fig:split_alg} shows the detector coverage of the two algorithms in the $\eta - \phi$ plane for a $WZ$ Monte Carlo.
\begin{table}\begin{center}
\begin{tabular}{cc}
\toprule
Electrons & Identification Cuts\\
\midrule
& EM fiduciality \\
& $E_T^{EM}>20$~GeV \\
& $E^{Had}/E^{EM}<0.055+0.0045 E^{EM}$ \\
& $L_{shr}<0.2$ \\
& $\chi^2_{\mathrm{{CES strip}}}<10$ \\
CEM &$\ge 3$ axial $\ge 2$ stereo COT Layers \\
& $\Delta z(CES,trk)<3$~cm \\
& $-3.0<q \Delta x(CES,trk)<1.5$~cm \\
&$E/p^{trk}<2.0$ or $E_T^{\mathrm{cluster}}>100$~GeV \\
&$|z_0^{trk}-z_0^{PV}|<5$~cm \\
& $|z_0^{trk}|<60$~cm \\
&$IsoRel<0.1$ \\
\midrule
& $1.2<|\eta^{PES}|<2.8$ \\
& $E_T^{EM}>20$~GeV \\
& $E^{Had}/E^{EM}<0.05$ \\
& PEM$3\times 3$Fit $\ne 0$ \\
& $\chi^2_{PEM3\times 3} <10$ \\
PHX & $PES5\mathrm{by}9U>0.65$ \\
& $PES5\mathrm{by}9V>0.65$ \\
& $\Delta R(PES,\mathrm{cluster})<3$~cm \\
& N Si Hits $\ge 3$ \\ 
& $\mathrm{Phoenix~Track}$\\
&$|z_0^{PHX}-z_0^{PV}|<5$~cm \\
& $|z_0^{PHX}|<60$~cm \\
&$IsoRel<0.1$ \\
\bottomrule
\end{tabular}
\caption[CEM and PHX Electron Selection Requirements]{Summary of CEM and PHX \emph{tight electron} selection requirements.}\label{tab:ele_alg}
\end{center}\end{table}

\subsection{Muon Identification}

A muon behaves like a minimum ionizing particle due to its rest mass, which is about 200 times larger than the electron one. Therefore muons deposit very little energy in the calorimeter systems and can leave a signal in the outer layer of the CDF detector which is instrumented with arrays of gas detector and scintillators (muon chambers). Ionization deposits from a muon candidate in a given muon detector constitute a {\em stub} and the candidate naming depends by the coverage of the muon detector that records them (see Section~\ref{sec:mu_cham}). 

The basic selection for a muon candidate~\cite{muon_id} is a high quality COT track pointing to a Minimum Ionizing Particle (MIP) energy deposit in the EM and HAD calorimeters and matched to a stub in the muon chambers. The precise requirements are the following:
\begin{itemize}
\item track quality: the reconstructed COT track must have a minimal amount of axial ($\ge 3$) and stereo ($\ge 2$) COT super-layers;
\item the $\chi^2_{trk}$ returned by the track fitting algorithm should be less than $2.3$;
\item track/stub matching in the $\phi$-plane is required by an appropriate $\Delta x(trk,\mathrm{stub}) = | x^{trk} - x^{\mathrm{stub}}|$ with cut values different for each specific muon sub-detector
\end{itemize}

If we are interested to a muon coming from $Z$ or $W$ decay, we expect the muon to be isolated from other detector activity (as in the electron case). The isolation is defined exploiting the muon candidate momentum and MIP energy deposit in the calorimeter:
\begin{equation}
  IsoRel\equiv E_T^{iso}/p_T<0.1;\quad E_T^{iso}=E_T^{0.4}-E_T^{MIP},
\end{equation}
where the $p_T$ is the COT track momentum and $E_T^{MIP}$ is the transverse energy deposited in the towers crossed by the track.

The two {\em tight} muon reconstruction algorithms are the CMUP and CMX. The first covers the region $|\eta|<0.6$ where a track is required to match stubs in both CMU and CMP muon detectors. The second covers the region $0.65<|\eta|<1.0$: it requires a stub in the CMX muon detector and a minimum curvature of the COT track, $\rho_{COT}$, of $140$~cm\footnote{This last requirement  ensures appropriate efficiency of the CMX trigger.}. Table~\ref{tab:central_mu_alg} summarizes all the CMUP and CMX identification criteria while the top right section of Figure~\ref{fig:split_alg} shows the detector coverage of the two algorithms in the $\eta - \phi$ plane for a $WZ$ Monte Carlo.
\begin{table}\begin{center}
\begin{tabular}{cc}
\toprule
Muons & Identification Cuts\\
\midrule
&$p_T^{trk}>20$~GeV$/c$ \\
&$E_T^{EM}<2+\mathrm{max}\big(0, ((p^{trk}-100)0.0115)\big)$~GeV \\
&$E_T^{Had}<6+\mathrm{max}\big(0, ((p^{trk}-100)0.028)\big)$~GeV \\
& $\ge 3$ axial $\ge2$ stereo COT Layers \\
CMUP & $|d_0^{trk}|<0.2$~cm (track w/o silicon) \\
CMX & $|d_0^{trk}|<0.02$~cm (track with silicon) \\
& $|z_0^{trk}-z_0^{PV}|<5$~cm \\
& $|z_0^{trk}|<60$~cm \\
& $\chi^2_{trk}<2.3$\\
&$IsoRel<0.1$ \\ \midrule
& CMU Fiduciality \\ 
CMUP & CMP Fiduciality \\
& $\Delta X_{CMU}(trk, \mathrm{stub})<7$~cm \\
& $\Delta X_{CMP}(trk, \mathrm{stub})<5$~cm \\\hline
& CMX Fiduciality \\ 
CMX & $\rho_{COT}>140$~cm \\
& $\Delta X_{CMX}(trk, \mathrm{stub})<6$~cm \\
\bottomrule
\end{tabular}
\caption[CMUP and CMX Muon Selection Requirements]{Summary of CMUP and CMX \emph{tight muons} selection requirements.}\label{tab:central_mu_alg}
\end{center}\end{table}

\subsection{Loose Muons Identification}\label{sec:LooseMuId}

Muons can be faked by cosmic rays or hadrons showering deep inside the calorimeters or not showering at all, however the muon candidates usually have a very clean signature because of the many detector layers used to identify them. On the other hand the signal acceptance for the tight muon categories is geometrically limited, therefore a set of lower quality muon identification criteria, {\em loose muons}, was developed. Loose muons were used in other analysis~\cite{single_top} to increase the signal yield. These are the main requirements of the loose muons algorithms:
\begin{itemize}
\item BMU: forward isolated muons ($1.0<|\eta|<1.5$) with hits in {\em Barrel} muon chambers.
\item CMU: central isolated muons with hits {\em only} in the CMU chambers and not in CMP.
\item CMP: central isolated muons with hits {\em only} in the CMP chambers and not in CMU.
\item SCMIO: a good quality track matched to an isolated MIP deposit and a {\em non-fiducial} stub in the muon detector (Stubbed Central Minimum Ionizing particle).
\item CMIO: a good quality track {\em only} matched to an isolated MIP deposit and failing any other muon identification criteria (Central Minimum Ionizing particle).
\item CMX-NotTrigger: also named CMXNT, are isolated muon detected in the CMX chamber but which are {\em not} triggered by the CMX specific triggers because of the geometrical limits of the COT ($\rho_{COT}<140$~cm).
\end{itemize}
Table~\ref{tab:loose_alg} summarizes all the Loose muons identification criteria while the bottom left section of Figure~\ref{fig:split_alg} shows the detector coverage of the six algorithms in the $\eta - \phi$ plane for a $WZ$ Monte Carlo.

\begin{table}\begin{center}
\begin{tabular}{cc}
\toprule
Muons & Identification Cuts \\
\midrule
&$p_T^{trk}>20$~GeV$/c$ \\
&$E_T^{EM}<2+\mathrm{max}\big(0, ((p^{trk}-100)0.0115)\big)$~GeV \\
BMU &$E_T^{Had}<6+\mathrm{max}\big(0, ((p^{trk}-100)0.028)\big)$~GeV \\
CMU & $|d_0^{trk}|<0.2$~cm (track w/o silicon) \\
CMP & $|d_0^{trk}|<0.02$~cm (track with silicon) \\
SCMIO& $|z_0^{trk}-z_0^{PV}|<5$~cm \\
CMIO& $|z_0^{trk}|<60$~cm \\
CMXNT& $\chi^2_{trk}<2.3$\\
&$IsoRel<0.1$ \\ \midrule

& BMU Fiduciality \\ 
BMU & $\Delta X_{BMU}(trk, \mathrm{stub})<9$~cm \\
& COT Hits Frac. $>0.6$ \\\midrule

& CMU Fiduciality \\ 
CMU & $\Delta X_{CMU}(trk, \mathrm{stub})<7$~cm \\
& $\ge 3$ axial $\ge 2$ stereo COT Layers \\\midrule

& CMP Fiduciality \\ 
CMP & $\Delta X_{CMP}(trk, \mathrm{stub})<7$~cm \\
& $\ge 3$ axial $\ge 2$ stereo COT Layers \\\midrule

& Stub Not Fiducial  \\ 
SCMIO & $\ge 3$ axial $\ge 2$ stereo COT Layers \\
&$E_T^{EM} + E_T^{Had} > 0.1$  \\\midrule

& No Stub  \\ 
CMIO & $\ge 3$ axial $\ge 2$ stereo COT Layers \\
&$E_T^{EM} + E_T^{Had} > 0.1$  \\\midrule

& CMX Fiduciality \\ 
CMXNT & $\Delta X_{CMX}(trk, \mathrm{stub})<6$~cm \\
& $\ge 3$ axial $\ge 2$ stereo COT Layers \\
& $\rho_{COT}<140$~cm, Not trigger CMX \\
\bottomrule
\end{tabular}
\caption[Loose Muons Selection Requirements]{Summary of the BMU, CMU, CMP, SCMIO, CMIO, CMX-Not-Trigger (CMXNT) {\em Loose muons} selection requirements.}\label{tab:loose_alg}
\end{center}\end{table}

\subsection{Isolated Track Identification}\label{sec:IsoTrk}

The isolated tracks are last lepton category used in this analysis. They are defined to be high-$p_T$ good quality tracks isolated from energy deposits in the tracking systems. The track isolation is defined as:
\begin{equation}\label{TrackIsolation}
\rm{TrkIso}=\frac{p_T({\rm{trk\ candidate}})}{p_T({\rm{trk\ candidate}})+\sum p_T({\rm{other\ trk)}}}\,\rm{,}
\end{equation}
where $p_T({\rm{trk\ candidate}}$ is the transverse momentum of the specific track we analyze (candidate) and $\sum p_T({\rm{other\ trk)}}$ is the sum of the transverse momenta of all good quality tracks within a cone radius of 0.4 of the candidate track. The isolation requirement is necessary in order to ensure that the track corresponds to a charged lepton produced in a decay of a $W$ boson and it is not part of an hadronic jet. A track is fully isolated if $TrkIso = 1.0$, thus a cut value of $0.9$ is used in the analysis.
A selection requirement of $\Delta R>0.4$ between the track and any tight jet\footnote{Central, high $E_T$ jets are classified as {\em tight}: $|\eta|<2.0$, $E_T^{cor}>20$ (see Section~\ref{sec:jetObj}).} is also applied to remove jets with low track multiplicity.

The fact that the isolated track is not required to match a calorimeter cluster or a muon stub allows to recover real charged leptons that arrive in non-instrumented regions of the calorimeter or muon detectors, as seen in Figure~\ref{fig:split_alg}. 

Table~\ref{iso_tab} summarizes the criteria used to select good quality isolated tracks while the bottom right section of Figure~\ref{fig:split_alg} shows the detector coverage of the algorithms in the $\eta - \phi$ plane for a $WZ$ Monte Carlo.

\begin{table}\begin{center}
\begin{tabular}{cc}
\toprule
Track Leptons & Identification Cuts \\
\midrule
&$p_T^{trk}>20$~GeV$/c$ \\
& $\ge 24$ axial $\ge 20$ stereo COT Hits \\
& $|d_0^{trk}|<0.2$~cm (track w/o silicon) \\
& $|d_0^{trk}|<0.02$~cm (track with silicon)\\
Isolated & $|z_0^{trk}-z_0^{PV}|<5$~cm \\
Tracks & $|z_0^{trk}|<60$~cm \\
& $\chi^2_{trk}$ probability $<10^{-8}$\\
& N Si Hits $\ge 3$ \\ 
&$TrkIso>0.9$ \\
& $\Delta R(trk, \mathrm{tight~jets})>0.4$\\
& $\Delta \phi(trk, \mathrm{jet 1})>0.4$\\
\bottomrule
\end{tabular}
\caption[Isolated Tracks Selection Requirements]{Summary of ISOTRK \emph{isolated track} leptons selection requirements.}\label{iso_tab}
\end{center}\end{table}

\section{Jet Identification}\label{sec:jetObj}

\begin{figure}[t]
\begin{center}
\includegraphics[width=0.6\textwidth]{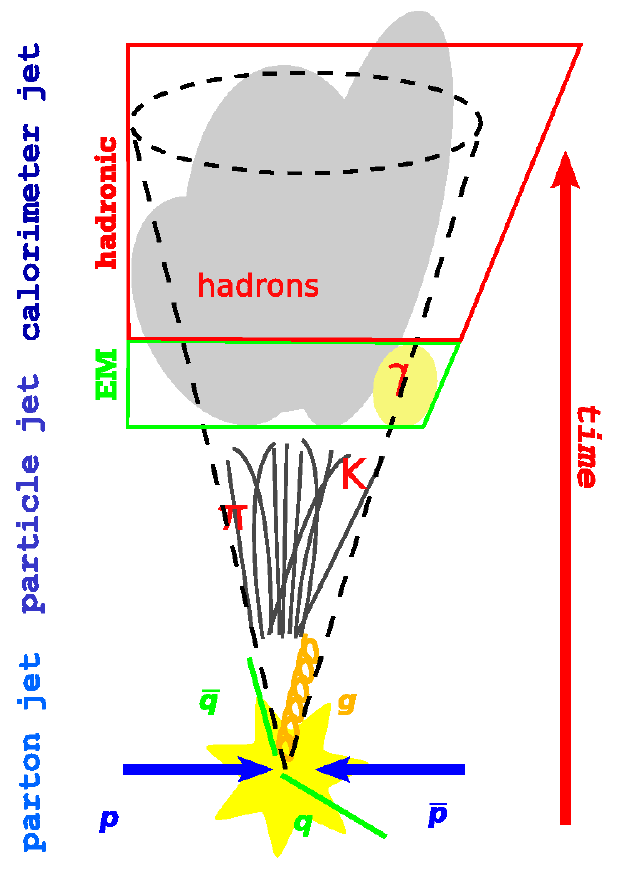}
\caption[Schematic Representation of a Hadronic Jet]{A parton originating from a hard scattering hadronizes and generates a
a narrow spray of particles identified as a {\em jet}.}\label{jet1}
\end{center}
\end{figure}

The QCD theory tells us that the partons composing the (anti)proton can be treated perturbatively as free particles if they are stuck by an external 
probe\footnote{I.e. a lepton, a photon or a parton from a different hadron.} with sufficient high energy (so called {\em hard scattering}). However partons resulting from the 
interaction can not exist as free particles because at longer distances (i.e. lower energies) the \emph{strong potential} can not be treated perturbatively and 
partons must form colorless hadrons. This process, called {\em hadronization} or {\em showering}, produces a {\em jet}, i.e. a narrow spray of stable particles that retains the information of the initial parton (for a pictorial representation see Figure~\ref{jet1}). 

\begin{figure}[t]
\begin{center}
\includegraphics[width=0.85\textwidth]{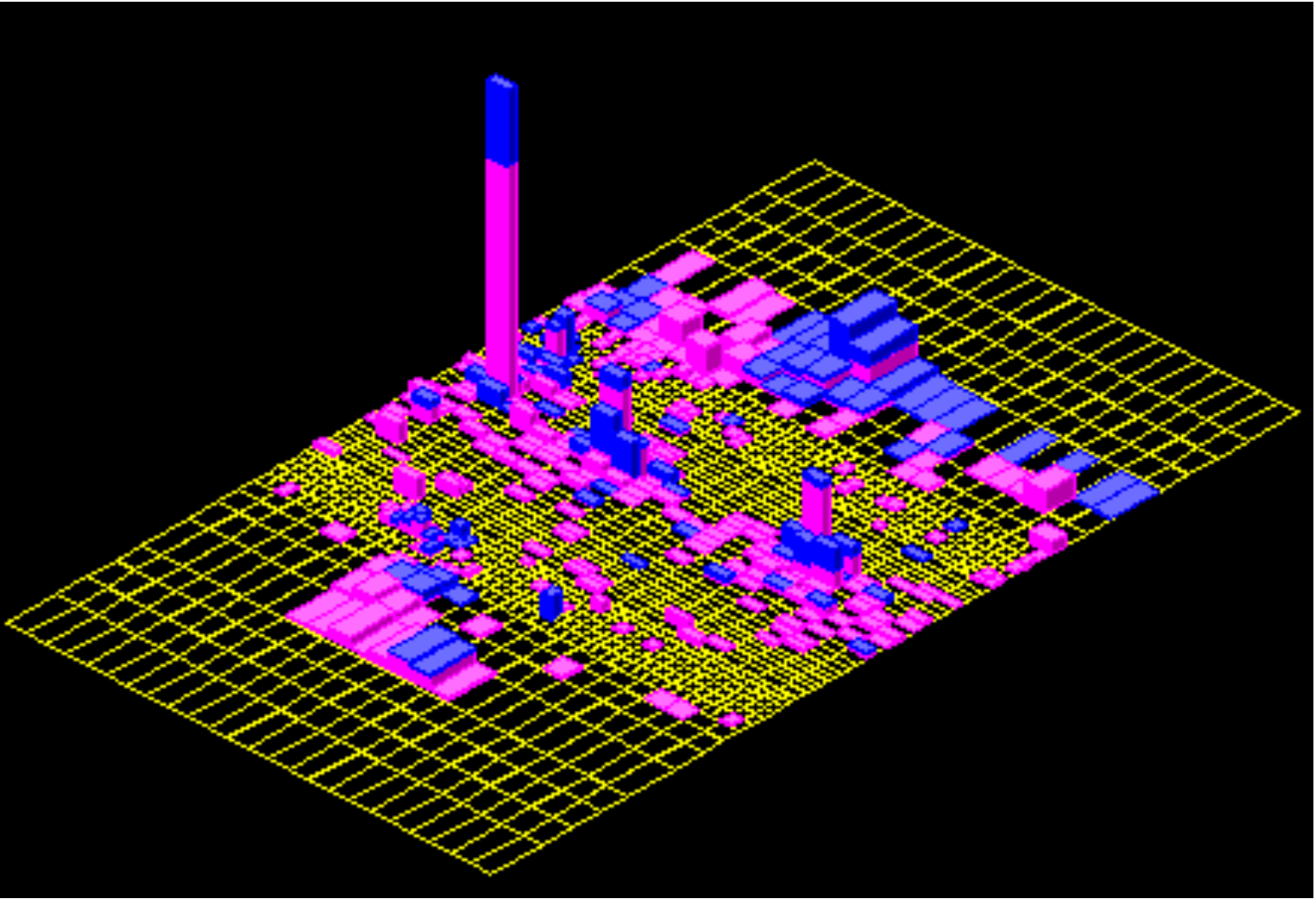}
\caption[Event Display with Jet Deposits]{Calorimetric deposit in the $\eta-\phi$ plane as represented in the CDF event display. EM deposits are red while HAD deposits are blue.}\label{lego_jet}
\end{center}
\end{figure}

From an experimenter's point of view a jet is defined as a large energy deposit in a localized area of the detector (see Figure~\ref{lego_jet}). The challenge of a physics analysis
is to recover from detector information the initial energy, momentum and, possibly, the kind of the parton produced in the original interaction. 
A {\em jet identification algorithm} is a tool to reconstruct such information and it should satisfy at best the following requirements~\cite{jet_req}:
\begin{itemize}
\item\emph{Infrared safety}: the presence of soft radiation between two jets may cause a merging of the two jets. This should not occur to avoid an incorrect parton attribution.
\item\emph{Collinear safety}: the jet reconstruction should be independent from any collinear radiation in the event, i.e. different energy distribution of particles inside calorimetric towers.
\item\emph{Invariance under boost}: the same jets should be found independently from boosts in longitudinal direction.
\item\emph{Boundary stability}: kinematic variables should be independent from the details of the final state.
\item\emph{Order independence}: the same reconstructed quantities should appear looking at parton, particle and detector levels.
\item\emph{Straightforward implementation}: algorithm should be easy to implement in perturbative calculations.
\end{itemize}
Beyond this theoretical aspects a jet algorithm should be experimentally efficient with a high reconstruction efficiency, good resolution and robust at high instantaneous luminosity.

\subsection{CDF Cone Algorithm}\label{sec:jetCone}

CDF uses several algorithms, none of them completely satisfying all the above requirements. The most common one, that is also the one used in this ana\-lysis,
is \verb'JETCLU'~\cite{jetclu}, an iterative fixed cone jet reconstruction algorithm based only on calorimetric information.

The algorithm starts by creating a list of the seed towers from all the calorimeter towers with transverse energy above the threshold of $1$~GeV. 
Starting with the highest-$E_T$ seed tower, a precluster is formed by combining
together all adjacent seed towers within a cone of given radius $R$ in the $\eta-\phi$ space\footnote{In this analysis we use $R=0.4$.}. This procedure is repeated, starting with the next unused seed tower, until the list is exhausted. The $E_T$-weighted centroid is then formed from the towers in the precluster and a new cone of radius $R$ is formed around this centroid. All towers with energy above the lower threshold of $100$~MeV within this new cone are added to the cluster. Then, a new centroid is calculated from the set of towers within the cluster and a new cone drawn. This process is iterated until the centroid of the energy deposition within the cone is aligned with the geometric axis of the cone (stable solution).

The initial clusters found can overlap so the next step is to merge or separate overlapping clusters, since each tower may belong only to one jet, each particle should not be assigned to more than one jet. Two clusters are merged if the total energy of the overlapping towers is greater than $75\%$ of the energy of the smallest cluster. If the shared energy is below this cut, the shared towers are assigned to the cluster that is closer in $\eta-\phi$ space. The process is iterated again until the list of clusters remains fixed. 

The final step of the jet identification happens on a second stage, after electron candidate identification and it is called {\em reclustering}. If an EM calorimeter cluster is found to be compatible with a tight electron identification (CEM, PHX both isolated or not), the EM calorimeter towers are removed and the jet clustering algorithm is iterated.

Massless four-vector momenta are assigned to the towers in the clusters for EM and HAD components with a magnitude equal to the energy deposited in the tower and the direction defined by a unit vector pointing from the event vertex to the center of the calorimeter tower at the depth that corresponds to the shower maximum. A cluster four-vector is then defined summing over the towers in the cluster:
\begin{eqnarray}
&& E=\sum_{i=1}^N(E_i^{EM}+E_i^{HAD})\mathrm{,}\\
&& p_x=\sum_{i=1}^N(E_i^{EM}\sin\theta_i^{EM}\cos\phi_i^{EM}+E_i^{HAD}\sin\theta_i^{HAD}\cos\phi_i^{HAD})\mathrm{,}\\
&& p_y=\sum_{i=1}^N(E_i^{EM}\sin\theta_i^{EM}\sin\phi_i^{EM}+E_i^{HAD}\sin\theta_i^{HAD}\sin\phi_i^{HAD})\mathrm{,}\\
&& p_z=\sum_{i=1}^N(E_i^{EM}\cos\theta_i^{EM}+E_i^{HAD}\cos\theta_i^{HAD})\mathrm{.}
\end{eqnarray}
where the index $i$ runs over the towers in the cluster.

In order to study jet characteristics, other variables (number of tracks, energy deposited in the HAD and EM calorimeters, etc.) are reconstructed and associated to the final jet analysis-object.

\subsection{Jet Energy Corrections}

The ultimate goal of the jet reconstruction algorithm is the determination of the exact energy of the outgoing partons coming from the hard interaction, i.e. the Jet Energy Scale (JES). Clearly many factors produce a mismatch between the raw energy measured by the algorithm and the one of the parton before the hadronization. 

\begin{figure}[t]
\begin{center}
\includegraphics[width=0.99\textwidth]{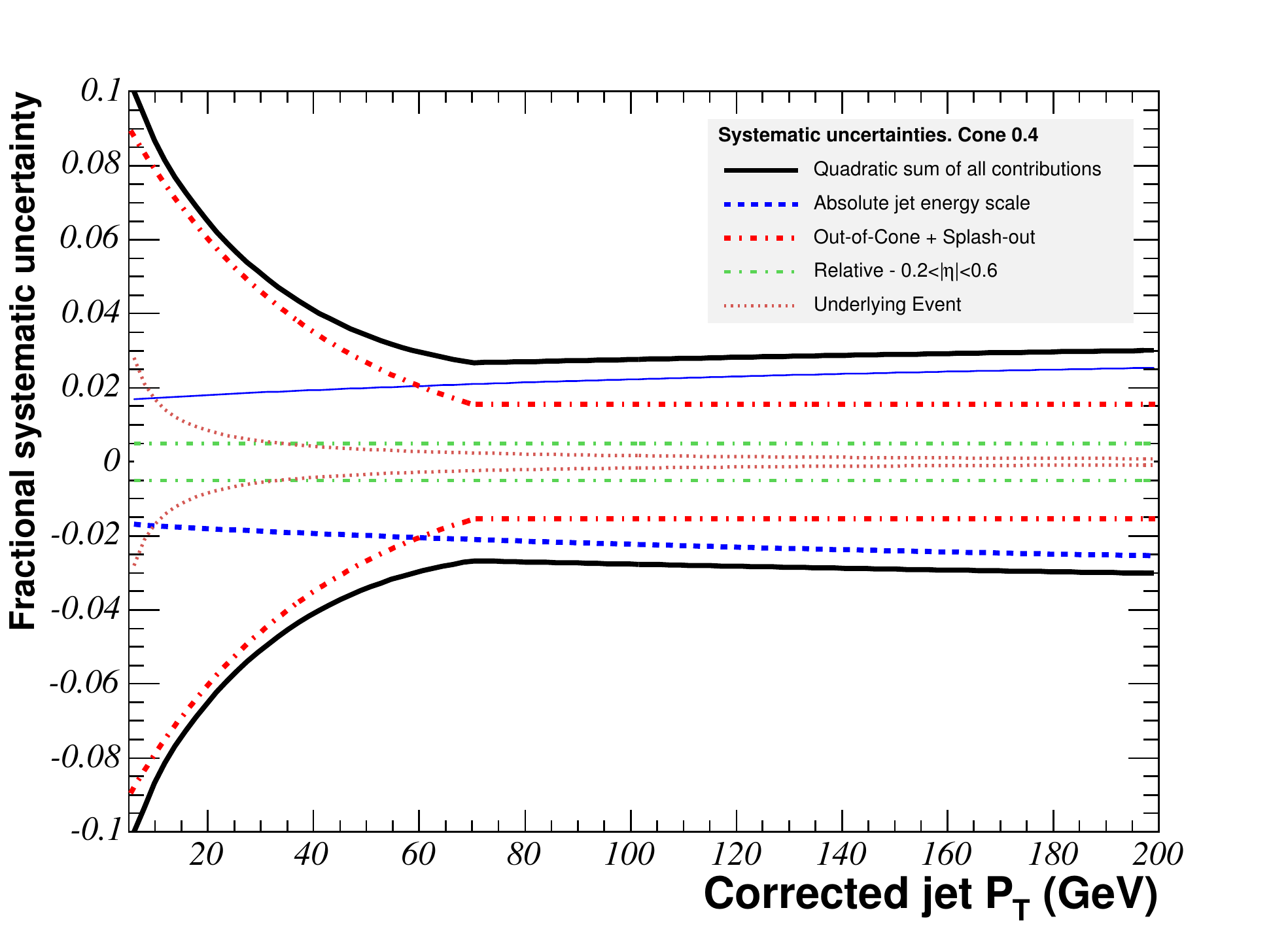}
\caption[Jet Energy Scale Systematic Uncertainty]{Jet $p_T$ dependence of the systematic uncertainties ($\sigma_{JES}$) relative to the different terms of Equation~\ref{eq:fullCor} for the \texttt{JETCLU} algorithm with $R=0.4$ cone size.}\label{totsys}
\end{center}
\end{figure}

CDF developed a set of jet energy corrections depending of $\eta$, $E_T^{raw}$ and $R$ of the jet reconstructed by \verb'JETCLU' algorithm. The corrections are divided into five levels\footnote{The actual naming skips $L2$, because it is absorbed in $L1$, and $L3$, as it was introduced as a temporary MC calibration in Run II.} ($L$-levels) that can be applied in a standard way to different analyses~\cite{jet_corr1, jet_corr2}: $\eta$-dependent response($L1$), effect of multiple interactions ($L4$), absolute energy scale ($L5$), underlying event ($L6$) and out-of-cone ($L7$) corrections. The correction $L1$ and $L5$ are multiplicative factors ($f_{L1}$ and $f_{L5}$) on the raw $E_T$ of the jet, the others are additive constants ($A_{L4}$, $A_{L6}$ and $A_{L7}$). The equation for the complete correction is:
\begin{equation}\label{eq:fullCor}
E_T^{FullCor}(\eta, E_T^{raw}, R)=(E_T^{raw} f_{L1}-A_{L4})f_{L5}-A_{L6}+A_{L7}.
\end{equation}
A description of each term is given in the following paragraphs while Figure~\ref{totsys} shows the separate systematic uncertainties ($\sigma_{JES}$) associated to each term of Equation~\ref{eq:fullCor}.

Recent studies~\cite{gluonJES_10829} have shown that the simulated detector response to jets originating from high-$p_T$ gluons is improved by lowering the reconstructed $E_T$ by two times the uncertainty used in Equation~\ref{eq:fullCor} ($-2\sigma_{JES}$). More accurate studies are ongoing within the CDF collaboration to fully understand the effect of a {\em parton dependent} jet energy correction. The $-2\sigma_{JES}$ gluon-jets prescription  has been applied also to this analysis, however the result is  negligible due to the small fraction of gluon-jets present in the $HF$-enriched signal region.

In this analysis, like in many others~\cite{single_top, wh94_note10796}, we choose to use jet corrections only up to L5. Therefore we define the Level-5 only correction:
\begin{equation}
E_T^{cor}(\eta, E_T^{raw}, R)=(E_T^{raw} f_{L1}-A_{L4})f_{L5}.
\end{equation}

L6 and L7 corrections are fundamental for the measurement of the unknown mass of a particle in the hadronic final state (for example top quark mass measurements) but they are much less relevant in a cross section measurement where only the relative data/MC energy scale matters  and not its absolute value. Furthermore, Figure~\ref{totsys} shows that a large uncertainty is associated to the out-of-cone correction for $R=0.4$ jets, which would decrease the di-jet invariant mass resolution, lowering the sensitivity of the present measurement.

Depending on, L5 corrected, $E_T^{cor}$ and the jet centroid position in the detector, jets are classified as {\em tight} or {\em loose}:
\begin{itemize}
\item{\bf Tight:} $|\eta^{det}|<2.0$, $E_T^{cor}>20$~GeV.
\item{\bf Loose:} $|\eta^{det}|<2.4$, $E_T^{cor}>12$~GeV and the jet is not tight.
\end{itemize}
Event selection (see Section~\ref{sec:jetSel}) is based on the number of tight jets while both tight and loose jets are used in the \met correction. Jet clusters of lower energy are not analyzed, it is assumed that they produce a negligible noise on the top of the unclustered energy of the event.

\paragraph{Level-1: $\eta$ Dependent Corrections}

\begin{figure}[t]
\begin{center}
\includegraphics[width=0.85\textwidth]{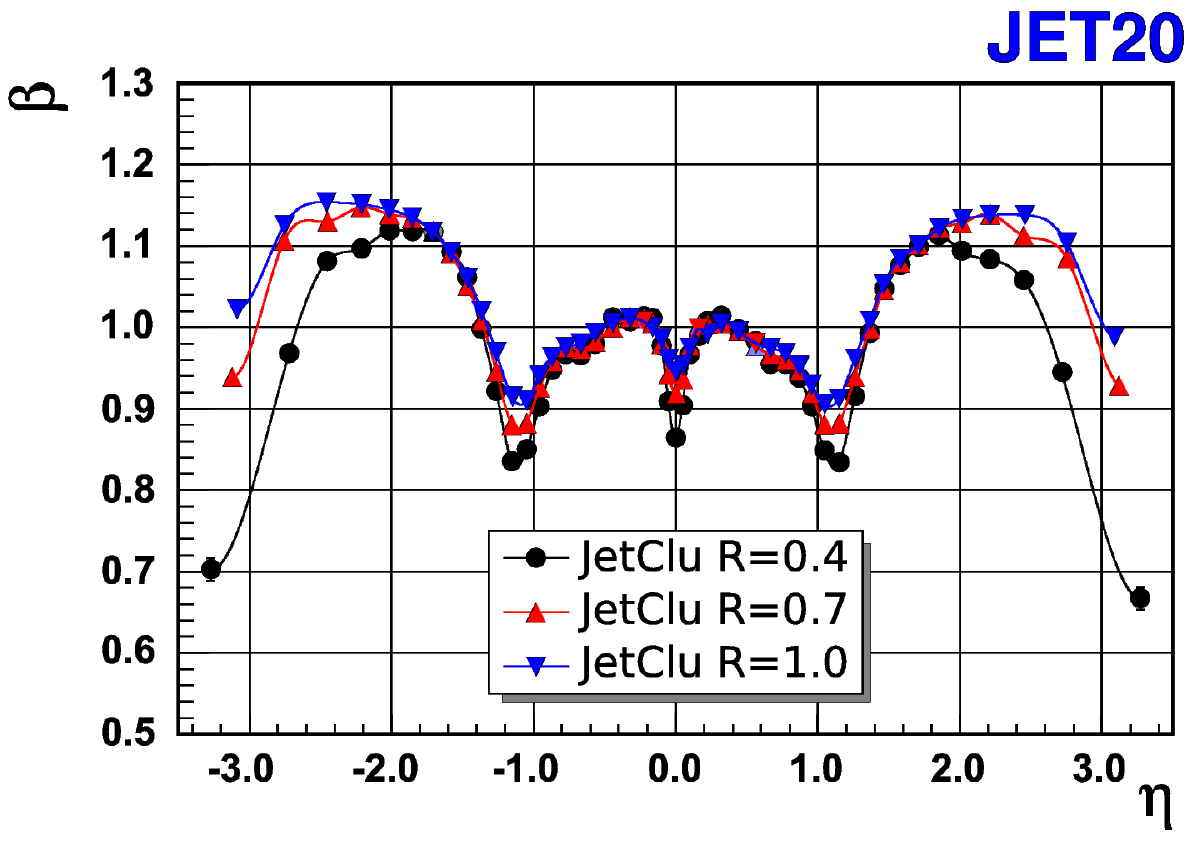}
\caption[Jet Energy Correction as a Function of $\eta$]{$\eta$-dependence of $\beta$ factor for cone radii $R=0.4$, $0.7$ and 
$1.0$, measured in the di-jet component of $\mathtt{jet20}$
sample.}\label{eta_correction}
\end{center}
\end{figure}

$L1$ correction is applied to raw jet energy measured in the calorimeters to 
make the detector response uniform in $\eta$, it takes into account aging of
the sub-detectors\footnote{This was the $L2$ correction during Run I} and other 
hardware non-uniformities (for example the presence of cracks).
This correction is obtained using a large di-jet sample:
events with one jet (\emph{trigger jet}) in the central region of the 
calorimeter ($0.2<|\eta|<0.6$), where the detector response is well known and 
flat in $\eta$, and a second jet (\emph{probe jet}), allowed to range anywhere
in the calorimeter ($|\eta|<3.6$). In a perfect detector the jets should be
balanced in $p_T$, a balancing fraction is formed:
\begin{equation}
f_b\equiv \frac{\Delta p_T}{p_T^{ave}}=\frac{p_T^{probe}-p_T^{trigger}}{(p_T^{probe}+p_T^{trigger})/2}\mathrm{,}
\end{equation}
the average of $f_b$ in the analyzed $\eta$ bin is used to define the $\beta$
factor\footnote{The definition of Equation~\ref{beta_eq} has a average value equal
to $p_T^{probe}/p_T^{trig}$ but is less sensitive to presence of non-Gaussian 
tails in the usual $p_T^{probe}/p_T^{trig}$ ratio.} (Figure~\ref{eta_correction} 
shows the $\beta$ distribution for different cone radii):
\begin{equation}\label{beta_eq}
\beta\equiv\frac{2+<f_b>}{2-<f_b>}\mathrm{.}
\end{equation}
The final $L1$ correction is defined as $f_{L1}(\eta, E_T^{raw}, R)=1/\beta$ 
and reproduces an approximately flat response in $\eta$ with an error varying 
from $0.5\%$ to $7.5\%$.

\paragraph{Level-4: Multiple Interactions Corrections}

\begin{figure}[t]
\begin{center}
\includegraphics[width=0.99\textwidth]{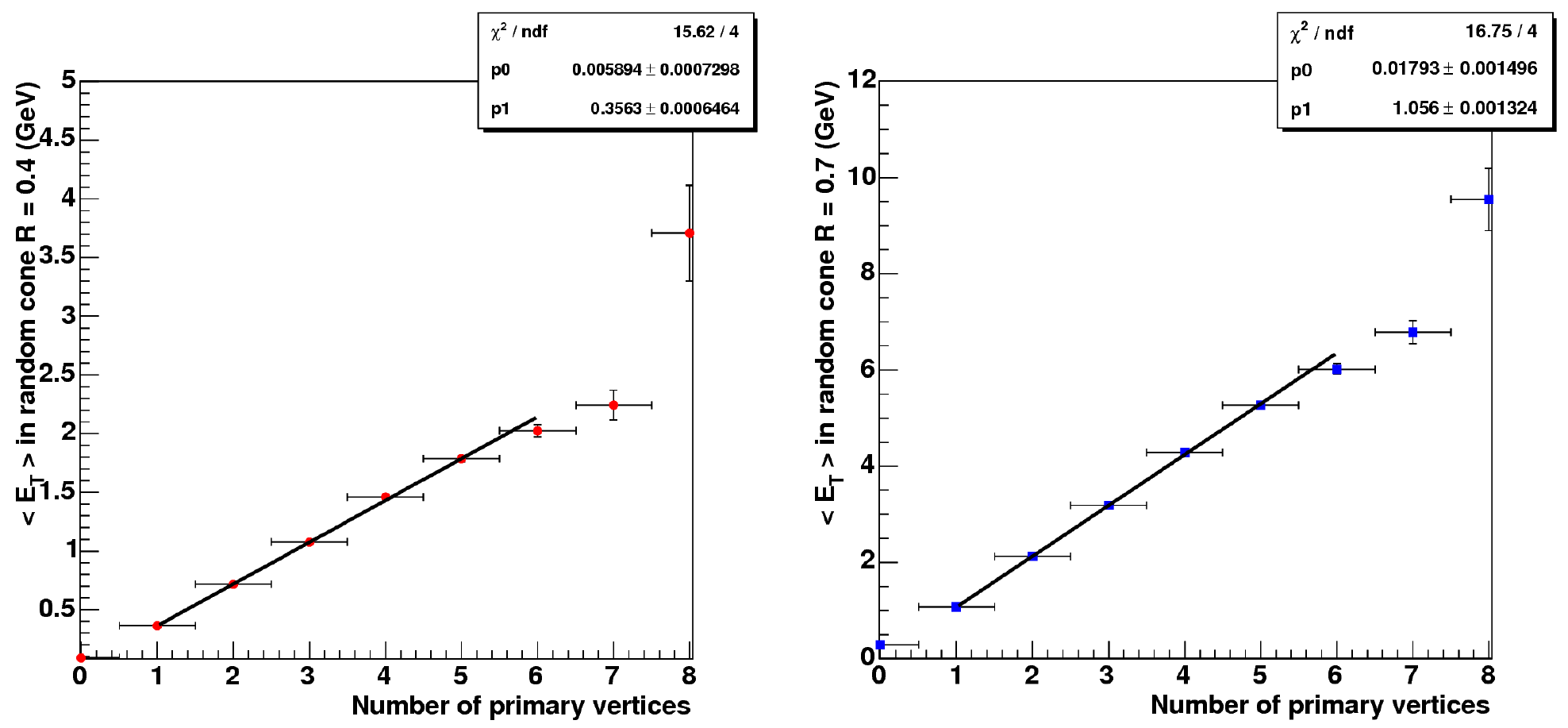}
\caption[$E_T$ Correction due to Multiple Interactions]{$E_T$ correction deriving from multiple interactions as a function of primary vertex number for cones with $R=0.4$ (right) and $R=0.7$ (left).}\label{mp_correction}
\end{center}
\end{figure}

Jet energy measurement is also degraded by the presence of minimum-bias events
that come from multiple $p\bar{p}$ interactions. This correction becomes more 
relevant at high luminosity, indeed the number of $p\bar{p}$ interactions is 
Poisson distributed with mean value approximately linear with instantaneous luminosity:
\begin{equation} 
\langle N(\mathscr{L}\simeq 10^{32}~\mathrm{cm}^{-2}\mathrm{s}^{-1})\rangle\simeq3\mathrm{,}\quad \langle N(\mathscr{L}\simeq 3\cdot 10^{32}~\mathrm{cm}^{-2}\mathrm{s}^{-1})\rangle\simeq 8\mathrm{.}
\end{equation}
The energy of particles coming from those processes is estimated from 
minimum-bias events drawing a cone in a random
position in the region $0.1<\eta<0.7$. Figure~\ref{mp_correction} shows that
the measured minimum-bias $E_T$ grows linearly with the number of primary 
vertices\footnote{Good quality primary vertices are reconstructed through at 
least 2 COT tracks.}, such quantity, $A_{L4}$, must be subtracted by jet raw 
energy. The total uncertainty is about $15\%$, it mostly depends on luminosity 
and event topology.

\paragraph{Level-5: Absolute Energy Scale Corrections}

While $L1$ and $L4$ make jet reconstruction uniform over the whole detector and
over the global behavior of $p\bar{p}$ beam interaction, $L5$ correction 
($f_{L5}$) aims to derive, from the detector jet energy measurement, the $E_T$ of particles originating the jet.

The study is MC driven: first jet events are generated with full CDF detector 
simulation, then jets are reconstructed both at calorimeter and hadron
generation levels (HEPG) with the use of same clustering algorithm. A 
calorimeter jet (C) is associated to the corresponding hadron jet (H) if 
$\Delta R <0.1$. For both HEPG and detector jets the transverse momentum, 
$p_T^C$ and $p_T^H$, is calculated. 
The absolute jet energy is defined as $\mathcal{P}(p_T^C|p_T^H)$, the probability\footnote{Different 
$p_T^H$ can give the same $p_T^C$, in this case the maximum is taken.} to measure $p_T^C$ with a given $p_T^{H}$.

\begin{figure}[t]
\begin{center}
\includegraphics[width=0.85\textwidth]{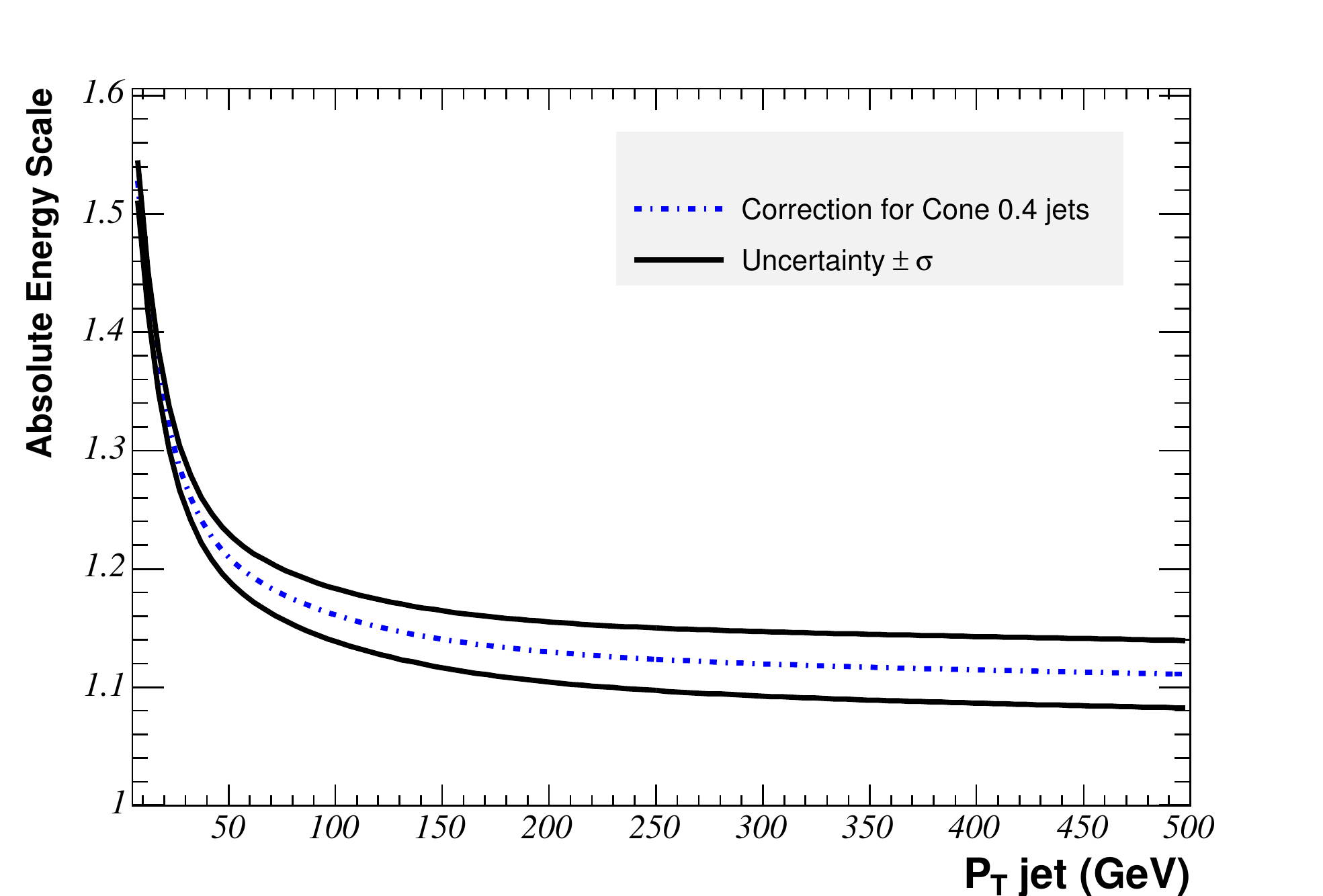}
\caption[Absolute Jet Energy Correction as a Function of $p_T$]{Absolute jet energy scale correction and systematic uncertainty ($f_{L5}$) for \texttt{JETCLU} algorithm with $R=0.4$ cone size.}\label{abs_correction}
\end{center}
\end{figure}

Figure~\ref{abs_correction} shows the correction factor $f_{L5}$ for different 
cone sizes as function of the different jet transverse energies. The total 
uncertainty is about $3\%$ and it mainly arises from the determination of
calorimetric response to single particles and MC fragmentation modeling.

\paragraph{Level-6 \& Level-7: Underlying Event and Out-of-cone Corrections}

Although we do not use L6 and L7 corrections, their description is reported for completeness. They are the last two corrections needed to infer the initial energy of the parton originating the jet. 

The underlying event correction ($L6$) takes into account the interaction 
processes which can occur between spectator partons or that originates from initial 
state radiation (usually soft gluon radiation) while the out-of-cone correction
($L7$) considers the fraction of particles coming from the original parton that fall outside the jet cone. 

The underlying event energy ($A_{L6}$) must be subtracted to the total jet 
energy. It was measured studying minimum-bias events during Run I and is 
parametrized with a constant value that scale with the cone radius. Out of cone
energy ($A_{L7}$) must be added to the total jet energy, studies are carried 
out with the same jet-parton matching method of $L5$.

\section{Neutrino Reconstruction}\label{sec:metObj}

Neutrinos are the only subatomic particles that leaves the detector completely undetected, therefore their signature is {\em missing energy}. 

Since the longitudinal energies of the colliding partons are unknown and not necessarily equal, we can only say that the total transverse energy of the $p \bar{p}$ collision is zero. Therefore the total amount of missing transverse energy \cancel{E}$_T$ (or MET) gives a measurement of the neutrino transverse momentum\footnote{For a massless neutrino $p_{T}=E_{T}$.} and it is defined as:
\begin{equation}\label{rawmet}
\vec{\mathrm{\cancel{E}}}_T\equiv - \sum_i{\vec{E}_T^i}
\end{equation}
where $\vec{E}_T^i$ is a vector with magnitude equal to the transverse energy 
collected by the i-th calorimeter tower and pointing from the interaction
vertex to the center of the tower. The sum involves all the towers with total 
energy above $0.1$~GeV in the region $|\eta|<3.6$.

The \met obtained from Equation~\ref{rawmet}, close to the online reconstructed missing energy, is often referred to as {\em raw} \cancel{E}$_T$ (or \metr). The fully reconstructed \met is corrected for the true $z$ vertex position, for the difference between the raw and corrected $E_T$ of the tight and loose jets, for the presence of muons in the event, by subtracting the momenta of minimum ionizing high-$p_T$ muons and adding back the transverse energy of the MIP deposit in the calorimeter towers. These corrections can be summarized in the following equation:
\begin{equation}
\met = \metr - \sum_{\text{muon}} p_T + \sum_{\text{muon}} (E_T^{EM} + E_T^{HAD}) - \sum_{jet}(E_T - E_T^{cor}).
\end{equation}

\section{Secondary Vertex Tagging}\label{sec:secvtx}

The algorithms able to select a jet coming from a Heavy-Flavor ($HF$) quark hadronization process are called {\em $b$-taggers} or {\em heavy-flavor taggers} and they are of fundamental importance in this and in many other analyses. For example both the top quark and the SM Higgs boson (for $m_H\lesssim 135$\gc2) have large branching fraction in $b$-quark, therefore an efficient $b$-tagging can dramatically reduce the background of uninteresting physical processes which contain only light-flavor ($LF$) hadrons in their final state.

We employ the Secondary Vertex Tagger\footnote{Historically it was the most important 
component in top discovery in 1995.} algorithm (\verb'SecVtx') to select a $HF$ enriched sample by requiring a $b$-tag on one or both the selected jets. In this way, it is possible to discriminate the $W\to cs$ and $Z\to c\bar{c}/b\bar{b}$ decays against the generic $W/Z\to u,d, s$ decays. 

In a successive phase of the analysis a flavor separator Neural Network~\cite{Richter_stop2007,kit_note7816}, developed by the Karsrhue Institute of Technology (also named KIT Flavor Separator), is used to separate jets originating from $c$ and $b$ quarks, allowing a separate measurement of $W\to cs$ and $Z\to c\bar{c}/b\bar{b}$.

\subsection{The SecVtx Algorithm}\label{sec:secvtx_alg}

\begin{figure}[t]
\begin{center}
\includegraphics[width=0.7\textwidth]{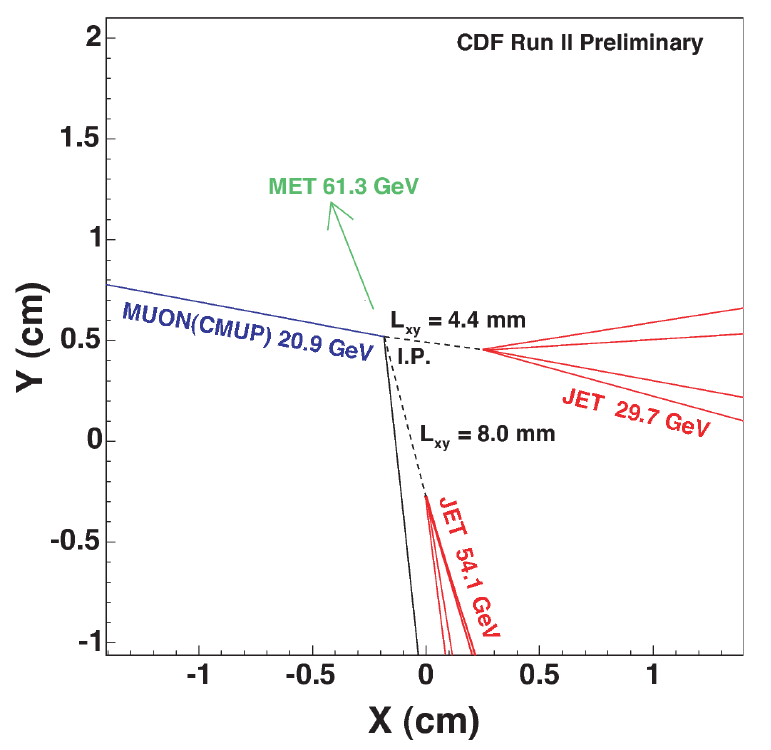}
\caption[$W +$ Jets Candidate Event with Two Reconstructed Secondary Vertices]{$W+$ jets candidate event with two secondary vertices tagged by 
$\mathtt{SecVtx}$ (run 166063, event 279746). The \cancel{E}$_T$ direction, a 
muon track, a prompt track and tracks from the secondary vertices are shown.}\label{svx_example}
\end{center}
\end{figure}

The \verb'SecVtx' algorithm takes advantage of the long life time of $b$-hadrons: a $c\tau$ value of about $450~\mu$m together with a 
relativistic boost due to a momentum of several GeV$/c$ permits to a $b$-hadron to fly several millimeters\footnote{The average transverse momentum of a $b$-hadron coming from a $WH$ events is about $40$~GeV$/c$ for a Higgs boson mass of $120$~GeV$/c^2$; in that condition a neutral $B^0$ meson of mass $5.28$~GeV$/c^2$ undergoes a boost $\beta\gamma=7.6$ and the average decay length is $3.5$~mm.} away from the primary interaction 
vertex. The relevant quantity is the $c\tau$ which is approximately the average
impact parameter of the outgoing debris of $b$-hadron decays. 
The decay produces small sub-jets composed by tracks with large impact 
parameter ($d_0$). The silicon detectors (see 
section~\ref{track_par}) are able to reconstruct $d_0$ with adequate 
precision to separate displaced tracks from the prompt tracks coming from the 
primary interaction. Figure~\ref{svx_example} shows as a $W+$ jets candidate
event with two displaced secondary vertices is identified by \verb'SecVtx' and 
reconstructed by the CDF event display.

Tagging is performed for all the jets with $|\eta|<2.4$ in an event. The algorithm searches for secondary vertices using the tracks within the jet cone of radius \mbox{$\Delta R=0.4$} . The usable tracks must satisfy the following requirements:
\begin{itemize}
\item $p_T>0.5$~GeV$/c$;
\item $|d_0|<0.15$~cm and $|d_0/\sigma_0|>2.0$;
\item $|z_0-z_{PV}|<2.0$~cm;
\item have a minimum number (depending on track reconstruction quality and position) of hits in the silicon detector;
\item be seeded or confirmed in the COT;
\end{itemize}
a {\em taggable} jet is defined as a jet containing at least two usable tracks.

The algorithm works on a two step basis and has two main operation modes\footnote{An {\em ultra-tight} operation mode exists but it is rarely used.}, {\em tight} (the standard one) and {\em loose}. The operating modes are defined by track and vertex quality criteria~\cite{secvtx_opt7578} but the two-step selection algorithm remains identical.

In the \emph{Pass 1} at least three tracks are required to pass loose selection
criteria. At least one of the tracks used is required to have $p_T > 1.0$~GeV$/c$. The selected tracks are 
combined two by two until a seed secondary vertex is reconstructed, then all 
the others are added one by one and a quality $\chi^2$ is computed. Tracks are 
added or removed depending of their contribute to the $\chi^2$.

The \emph{Pass 2} begins if \emph{Pass 1} does not find a secondary vertex. Now only 
two tracks are required to form a secondary vertex but they must pass tighter requirements: $p_t>1.0$~GeV$/c$, $|d_0/\sigma_0|>3.5$ and one of the tracks must have $p_T > 1.5$~GeV$/c$.

\begin{figure}
\begin{center}
\includegraphics[width=0.99\textwidth]{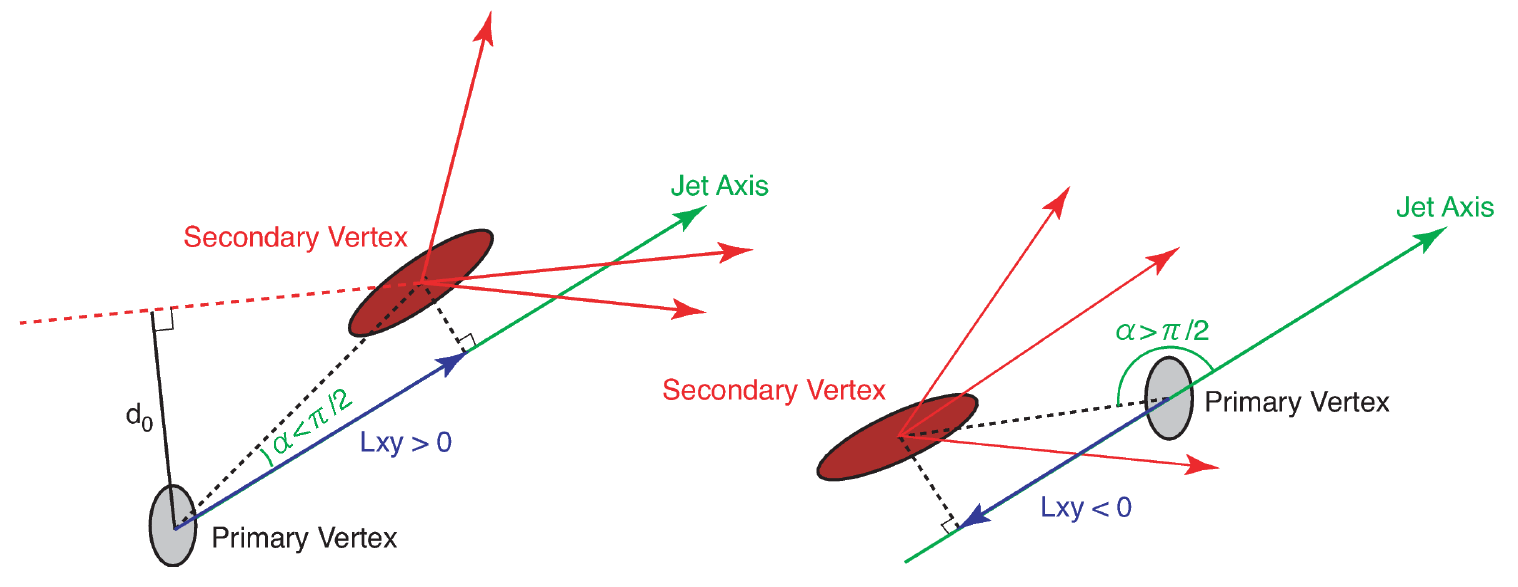}
\caption[Reconstructed Secondary Vertex: True and Fake $\mathtt{SecVtx}$]{Schematic representation of $\mathtt{SecVtx}$ variables and geometry. Left: true reconstructed secondary vertex. Right: negative $\mathtt{SecVtx}$ tag, falsely reconstructed secondary vertex.}\label{fig:svx}
\end{center}
\end{figure}

If a secondary vertex is identified in a jet, the jet is {\em tagged}. The bi-dimensional decay length $L_{xy}$ is calculated as the projection on the jet axis, in the $r-\phi$ 
plane, of the \verb'SecVtx' vector, i.e. the one pointing from the primary vertex 
to the secondary. The sign of $L_{xy}$ is defined by the angle $\alpha$ between
the jet axis and the \verb'SecVtx' vector. Figure~\ref{fig:svx} explains the 
geometry.

A secondary vertex coming from a $HF$ hadron is expected to have 
large $L_{xy}$. To reduce background due to mismeasured tracks 
\mbox{$|L_{xy}/\sigma_{L_{xy}}|>7.5$} is required\footnote{Negative $L_{xy}$ has
no physical meaning but it is important to estimate the mistag probability due 
to resolution effects.}. Other requirements are applied on the 
invariant mass of the pair of tracks, to avoid $K$ and $\lambda$ decays, and on
vertex multiplicity and impact parameter to reject secondary vertices due to 
interaction with material inside the tracking volume.

\subsection{Tagging Performances and Scale Factors}\label{sec:btag_sf}

\begin{figure}
\begin{center}
\includegraphics[width=0.99\textwidth]{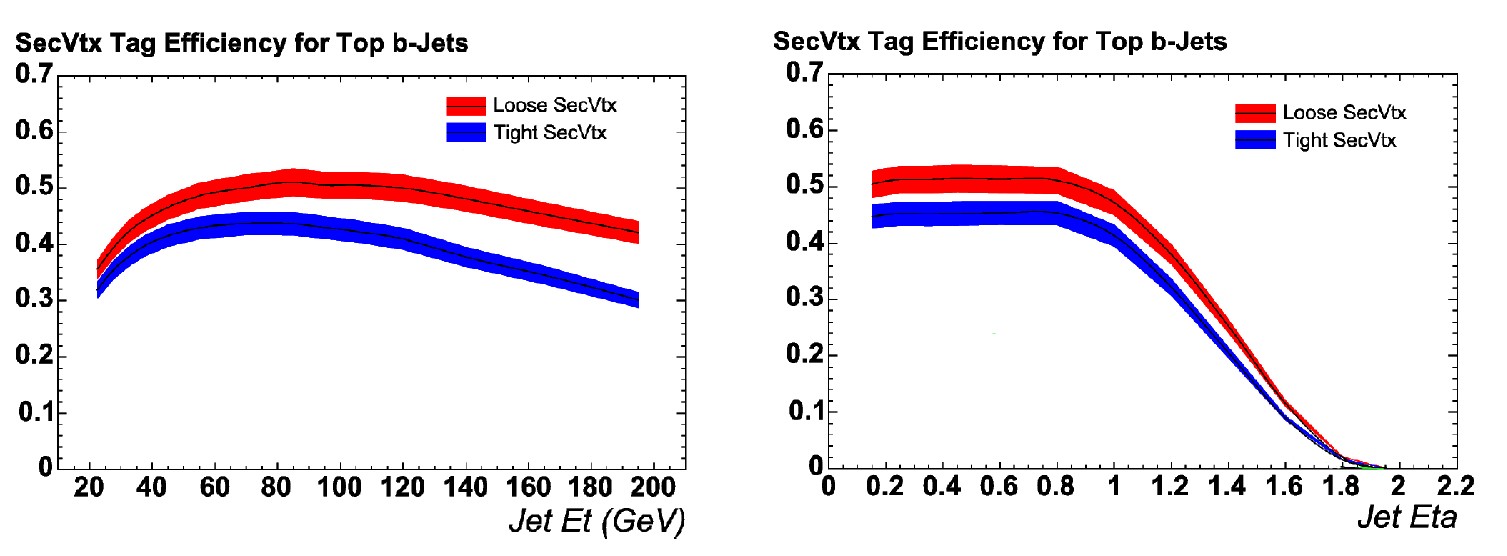}
\caption[\texttt{SecVtx} Efficiency]{Efficiency of the  \texttt{SecVtx} $b$-tagging algorithm for the tight and loose operating modes.}\label{fig:tag_eff}
\end{center}
\end{figure}

The performances of a $b$-tagger are evaluated on its efficiency, i.e the rate 
of correctly identified $b$-hadrons over all the produced $b$-hadrons, and on 
its purity, i.e the rate of falsely identified $b$-hadrons in a sample with no 
true $b$-hadrons. CDF uses $b\bar{b}$ QCD MC to evaluate \verb'SecVtx' efficiency 
relying on detector and physical processes simulation. Figure~\ref{fig:tag_eff} 
shows the $b$-tagging efficiency as a function of jet $\eta$ and $E_T$ for the 
tight and loose \verb'SecVtx' operating modes. Tagging efficiency drops at large $|\eta|$ because of tracking acceptance.

As MC does not reproduce the exact $b$-tagging efficiency of \verb'SecVtx' a {\em Scale Factor} ($SF_{Tag}$) is introduced to account for data/MC difference in the form:
\begin{equation}
SF_{Tag}\equiv\frac{\epsilon_{data}}{\epsilon_{MC}}.
\end{equation}

CDF uses two methods to calculate $SF_{Tag}$: both use di-jet samples where one of the two jets contains a low-$p_T$ lepton, electron or muon, which increases the presence of a $HF$ hadron decay, while the other jet can be $b$-tagged by \verb'SecVtx' or not. 

The method exploiting the electrons~\cite{eleSF_10178} finds algebraically the relative number of  $b$-tagged $HF$-jets with respect to the number of $b$-jets with a semileptonic electron decay. To extract the fraction, it also needs the number of jets with a low-$p_T$ electron but not containing $HF$ hadrons: this is obtained from a jet sample where the $\gamma\to e^+e^-$ conversion process is identified, and therefore, with depleted  $HF$ content. 

The second method, used as a validation of the first one, exploits the muon semileptonic decay of the $HF$ hadrons~\cite{muSF_8640}. The algorithm is similar to the electron based one, but the fraction of  low-$p_T$ muons not coming from $HF$ hadrons is extracted from a fit of the muon $p_T$ relative to the jet axis (named $p_T^{rel}$) that has a peculiar distribution for $HF$ hadron decays. 

\begin{figure}[!h]
\begin{center}
\includegraphics[width=0.99\textwidth]{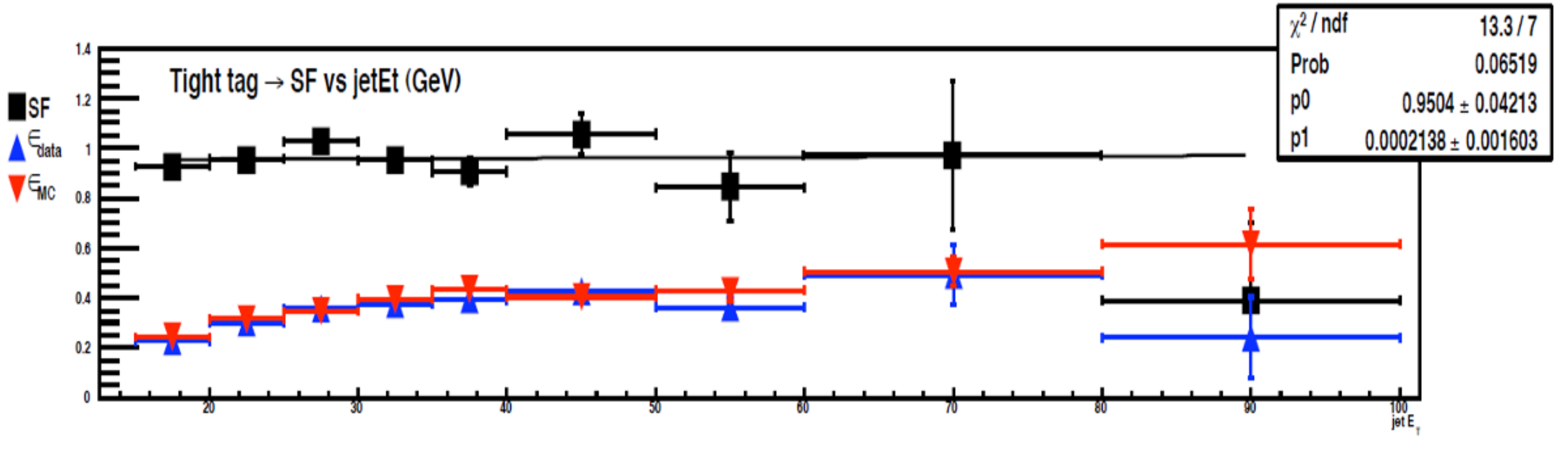}
\caption[\texttt{SecVtx} Scale Factor]{Evaluation of the \texttt{SecVtx} $SF_{Tag}$ with the electron method~\cite{eleSF_10178} and its dependence over the $E_T$ of the jet. The linear fit shows almost no dependency from the $E_T$ of the jet. Table~\ref{tab:tag_sf} reports the evaluation of $SF_{Tag}$ integrated over the variables of interest for the $HF-$tagger parametrization.}\label{fig:tag_sf}
\end{center}
\end{figure}

The efficiency on data are therefore compared to the MC results and the $SF_{Tag}$ is extracted.
Figure~\ref{fig:tag_sf} shows the $SF_{Tag}$ dependency with respect to the jet $E_T$ as determined by the electron method.
Table~\ref{tab:tag_sf} reports the $SF_{Tag}$ for the loose and tight \verb'SecVtx' operation modes integrated over the variables of interest for the $SF_{Tag}$ parametrization (no strong dependency is seen in any of them).
 The total per-jet tagging efficiency, deconvoluted from tracking effects is about $40$\% for $b$-jets and $6$\% for $c$-jets.
\begin{table}\begin{center}
\begin{tabular}{cccc}
\toprule
 mode & $SF_{Tag}$ & stat. err. & sys. err.\\
\midrule
Tight &  $0.96$ & $0.01$ & $0.05$ \\
Loose & $0.98$ & $0.01$ & $0.05$ \\
\bottomrule
\end{tabular}
\caption[\texttt{SecVtx} Efficiency Scale Factors]{$\mathtt{SecVtx}$ Scale Factors ($SF_{Tag}$) for the tight and loose $\mathtt{SecVtx}$ operation modes.}\label{tab:tag_sf}
\end{center}\end{table}

The number of falsely \verb'SecVtx' tagged jets is dubbed \emph{mistags}.
Mistags can be due to track resolution, long living $LF$ hadrons or 
secondary interactions with detector material.

\begin{figure}
\begin{center}
\includegraphics[width=0.99\textwidth]{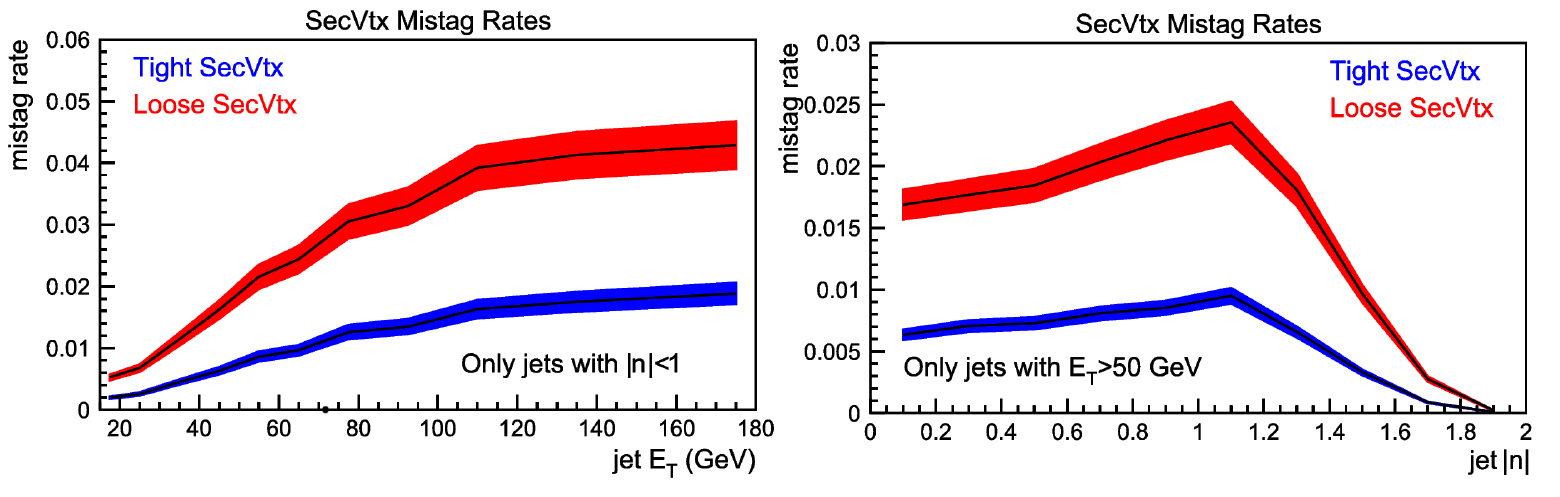}
\caption[$E_T$ and $\eta$ Mistag Rate]{Rate of wrongly $\mathtt{SecVtx}$ tagged jets (\emph{mistags}) as a function of $E_T$ and $\eta$ of the jets for the tight and loose $\mathtt{SecVtx}$ operation modes. The rate is derived from an inclusive jet sample.}\label{fig:tag_mistag}
\end{center}
\end{figure}

The rate of $W$ plus mistag jets is derived from a sample of events collected with 
an inclusive jet-based trigger with no $HF$ requirement\footnote{The presence of a 
small $HF$ contamination in the sample is a source of systematic uncertainty.}. 
The mistag parametrization~\cite{mistag} is obtained from an inclusive jet sample using {\em negative tags} (see Figure~\ref{fig:svx}), i.e. $b$-jets which appear to travel back toward the 
primary vertex. Resolution and material effects are expected to produce false 
tags in a symmetric pattern around the primary interaction vertex. The mistag 
rate is then corrected for the effects of long-lived $LF$ hadrons to take
into account the LF contamination giving real secondary vertices. 
A {\em per-jet} mistag probability, $p^j_{Mistag}$ is parametrized in bins of:
\begin{itemize}
\item total number of jets in the event, 
\item jet $E_T$, 
\item jet $\eta$, 
\item track multiplicity within the jet, 
\item total $E_T$ of the event, 
\item number of interaction vertices,
\item the $z$ vertex position.
\end{itemize}
This defines the, so called, {\em Mistag Matrix}. Figure~\ref{fig:tag_mistag} shows the mistag rate as function of $E_T$ and $\eta$ of the tagged jets for the tight and loose \verb'SecVtx'
operation modes. The approximate per-jet fake rate is about $1$\% for the tight operation mode and $2$\% for the loose one.

\section{Neural Network Flavor Separator (KIT-NN)}\label{sec:kitnn}

\begin{figure}[!h]
\begin{center}
\includegraphics[width=0.99\textwidth]{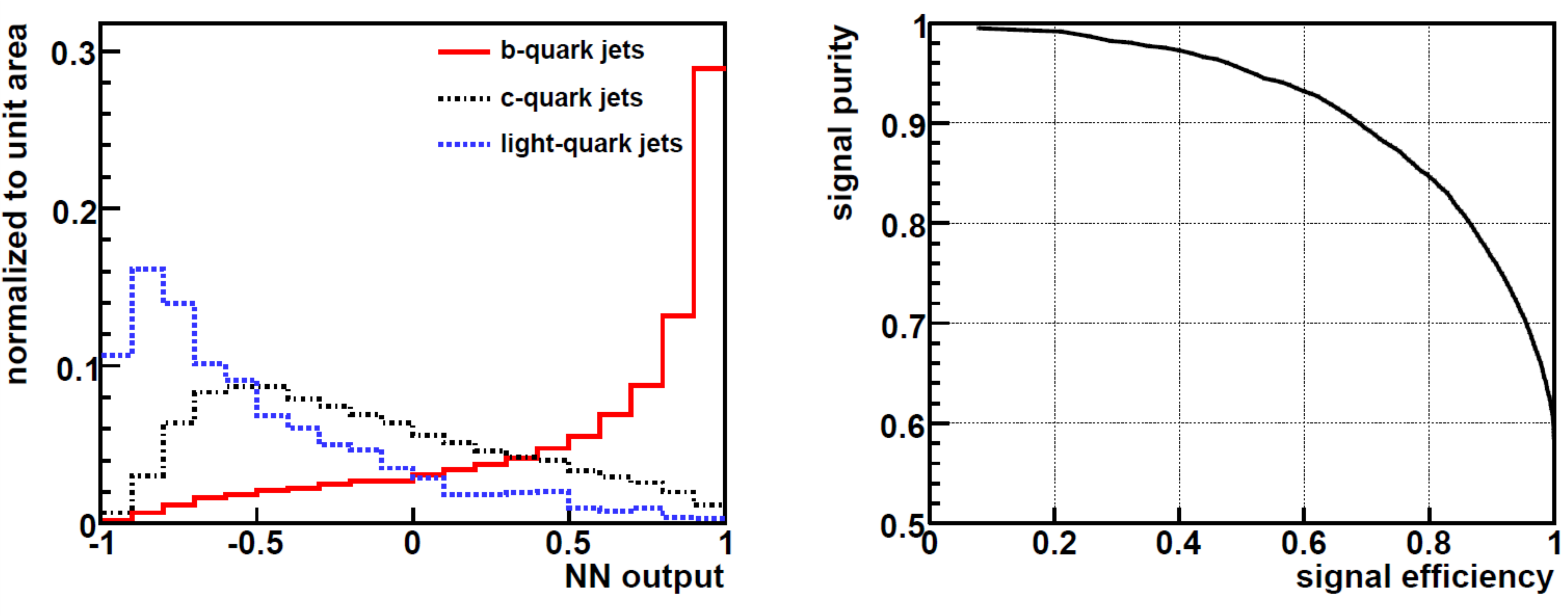}
\caption[KIT-NN Output Distribution and $b$-jet Efficiency/Purity Curve]{Left: Output value of the KIT-NN function for $b$, $c$ and $LF$ jets. Right: curve of the $b$-jet efficiency {\em vs} $b$-jet purity for different selection thresholds of the KIT-NN output.}\label{fig::kit_nn_out}
\end{center}
\end{figure}

The $b$-tag requirement drastically reduces the contamination of $LF$ jets, however also a per-jet fake rate of $1$\% can produce enough mistag background to decrease the sensitivity to rare processes. On the other hand a more stringent $b$-tagging requirement would reduce the signal yield. The KIT Flavor Separator Neural Network~\cite{Richter_stop2007,kit_note7816} (KIT-NN) offers a possible solution to these problems: after a \verb'SevVtx' tag, the Neural-Network (NN) exploits a broad range of $b$ quark discriminative variables to obtain a continuous distribution with a good separation power between real $b$-jets and $LF$ jets (see Figure~\ref{fig::kit_nn_out}). 

Beyond the $LF$ separation power, in this analysis we exploit the KIT-NN distribution in an original way. Thanks to the substantial shape difference between $c$-jets and $b$-jets KIT-NN values, we managed to measure separately $W\to cs$ and $Z\to c\bar{c}/b\bar{b}$ contributions (as shown in Chapeter~\ref{chap:StatRes}).

A short description of the KIT-NN implementation, developed for the single-top search at CDF~\cite{single_top}, can be useful to understand the physical meaning of such variable. 

A Bayesian NN is a supervised learning algorithm~\cite{BShop_machine_learning} that associates a score to an {\em event}, according to its likeliness to be {\em signal} or {\em background}. An event is defined by an array of $K$ input variables, $\alpha_i$, and the ouput score, $o \in [-1,1]$, is obtained from the following function:
\begin{equation}
  o =S  \Big( \sum^H_{j=0} \omega_j S \Big( \sum_{i=0}^{K}\omega_{i,j} \alpha_i +\mu_{0,j} \Big) \Big),
\end{equation}
where $\omega_i$, $\omega_{i,j}$ and $\mu_{0,j}$ are tunable parameters and $S$ is a sigmoid, or activation, function:
\begin{equation}
  S(x) =\frac{2}{1+e^{-x}} -1,
\end{equation}
The parameters are derived from a target function optimized over the training set of signal and background labeled events. 

The final result is a per-event signal-or-background {\em posterior probability} that takes into account non-linear correlations between the input variables. 

In our case, after that a jet has been tagged by \verb'SevVtx', twenty-five input variables, relative to the secondary vertex and to the jet, are fed into a NeuroBayes\textsuperscript{\textregistered} NN~\cite{neurobayes}. The training signal sample is composed by $t\bar{t}$, single-top and $W+b\bar{b}$ MC samples, while the background is built with $W+c\bar{c}$ and $W+LF$ MC samples. The variables exploit lifetime, mass, and decay multiplicity of the $b$ hadrons using several characteristics of the identified secondary vertex, the properties of the tracks inside the jet and information from the $b$-tag algorithm, for example if {\em Pass 1} or {\em Pass 2} reconstruction is used (see Section~\ref{sec:secvtx_alg}).
Figure~\ref{fig::kit_nn_out} shows the obtained classification values for  $b$, $c$ and $LF$ jets.

\begin{figure}[!h]
\begin{center}
\includegraphics[width=0.99\textwidth]{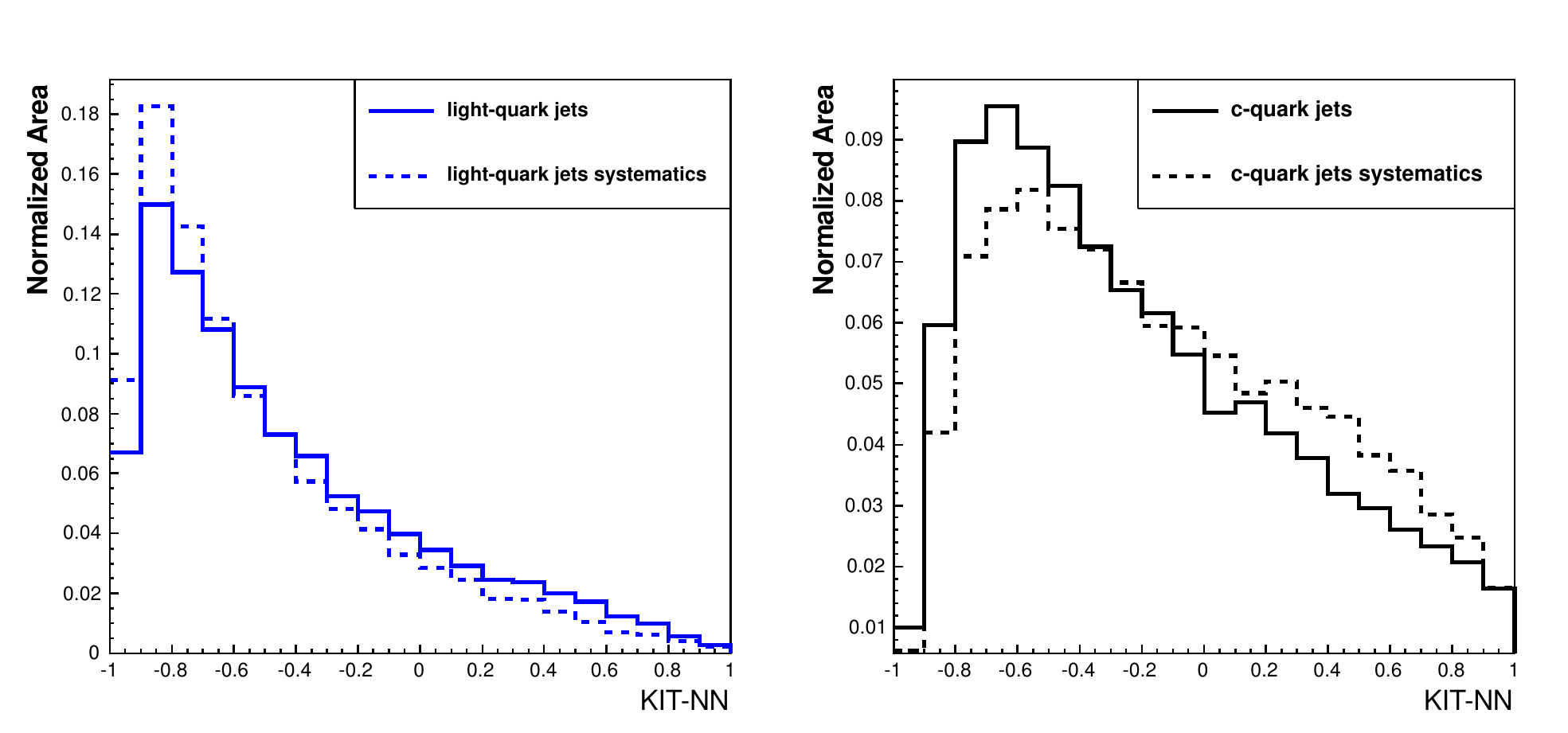}
\caption[KIT-NN Distribution and Systematic Variation for $LF$ and $c$ Jets]{KIT-NN default distribution (continuous lines) and systematic variations (dashed lines) for $LF$-jets (left) and $c$-jets (right).}\label{fig:kit_sys}
\end{center}
\end{figure}

Although a careful validation of the input variables was performed, the use of a MC-driven $LF$ training sample introduced a relevant discrepancy in the KIT-NN evaluation of {\em real} mistagged jets. Therefore a  $LF$ correction function was derived from a fake-enriched data sample selected by negative $\mathtt{SecVtx}$ tags. The final $LF$ KIT-NN distribution is obtained by assigning, to each $LF$-jet, a random value extracted from the corrected $LF$ distribution. The uncorrected template distribution is used as an {\em optimistic} systematic variation. A similar effect has been hypothized also for $c$-jets but it appears in a less relevan way. For $c$-jets the KIT-NN output is evaluated per-jet each jet and the $LF$-like variation is used as a {\em pessimistic} systematic variation. Figure~\ref{fig:kit_sys}
shows the default (central) KIT-NN distribution and systematic variations for $LF$ and $c$ jets.

\clearpage
\chapter{Events Selection}\label{chap:sel}

In the following we describe the selection requirements applied to produce samples of data enriched in $diboson\to \ell\nu+ HF$ candidates.

Every physics analysis starts with some kind of candidate signal selection. In the case of collider experiments, it is possible to see three ingredients in it: online (trigger) selection (Section~\ref{sec:onlineSel}), offline selection (Sections from~\ref{sec:w_sel} to~\ref{sec:svm_sel}) and the efficiency evaluation of each of the steps.

The variety of different physics object that compose the diboson decay channel under investigation is a characteristic of this analysis: charged leptons (electrons, muons and isolated tracks), missing energy, jets and $HF$ jets. These represent a good fraction of all the objects that can be reconstructed at a hadron collider experiment. Furthermore the need to relax the selection requirements in order to increase the statistics and the background-rich $W+2$ jets sample increase the challenges. It worths noticing that the same issues arise also in other primary analyses at CDF: for example single-top or $WH$ searchs.

In order to deal with these difficulties, one of the major task of this thesis work was the development of a robust analysis framework that could exploit the $\ell\nu+HF$ channel in an efficient and reliable way. The package, named $WH$ Analysis Modules (WHAM), is described in Appendix~\ref{chap:AppWHAM}. Its strengths are modularity, a wide set of configuration options and the use, with improvements, of CDF most advanced tools. The modular and object-oriented approach of the software allowed the handling of the complex selection formed by eleven different lepton types and four data-streams. At the same time it was also possible to develop a set of tools useful also in other contexts: for example a new multivariate multi-jet background rejection tool (see Appendix~\ref{chap:AppSvm}) or a versatile $b$-tag efficiency evaluation algorithm.

The actual selection process is divided into two main steps. First we select candidates that have the $\ell\nu+2$ jets signature, this defines the {\em pretag control} sample. Successively we require one or both the jets to be $b$-tagged, obtaining two signal samples ({\em single} and {\em double} tagged) enriched in $HF$. Section~\ref{sec:data_sig} shows the final result of the selection on the CDF dataset and on the diboson signal.

\section{Online Event Selection}\label{sec:onlineSel}
The trigger system, described in Section~\ref{sec:trigger}, is in charge of the online selection.

In order to maximize the signal acceptance we use a complex trigger strategy based on several trigger paths collected in four data-streams:
\begin{itemize}
\item high energy central electrons or \verb'bhel';
\item high energy forward electrons or \verb'bpel';
\item high momentum central muons or \verb'bhmu';
\item high missing transverse energy or \verb'emet'.
\end{itemize}
After online trigger selection but before offline analysis, the quality of the recorded data is crosschecked and the final luminosity is calculated (see Section~\ref{sec:data-struc}).

A detailed description of the trigger selection, the trigger efficiency ($\epsilon_{trig}$) evaluation and of the data quality requirements follows in the Section.

\subsection{High Energy Central Electron Trigger}

The trigger path used for the selection of tight central electron candidate events (CEM) is named \verb'ELECTRON_CENTRAL_18' and it is stored in the \verb'bhel' data-stream together with other auxiliary triggers. Single high energy electron identification is based on a calorimeter electromagnetic cluster matched with a reconstructed track.

The selection proceeds through the three trigger levels, first with minimal requirements and then with more and more sophisticated object reconstruction. L1 requirements are a track with $p_T > 8$~GeV/c, a central
($|\eta| < 1.2$) calorimeter tower with $E_T > 8$~GeV and the ratio between the
energy deposited in the hadronic calorimeter to that in the electromagnetic
calorimeter ($E^{HAD}/E^{EM}$) less than $0.125$. At L2, it is required a calorimeter cluster with
$E_T > 16$~GeV matched to a track of $p_T > 8$~GeV$/c$. At L3, it requires an
electron candidate with $E_T > 18$~GeV matched to a track of $p_T > 9$~GeV$/c$.

\begin{figure}[!ht]
\begin{center}
\includegraphics[width=0.495\textwidth, height=0.36\textwidth]{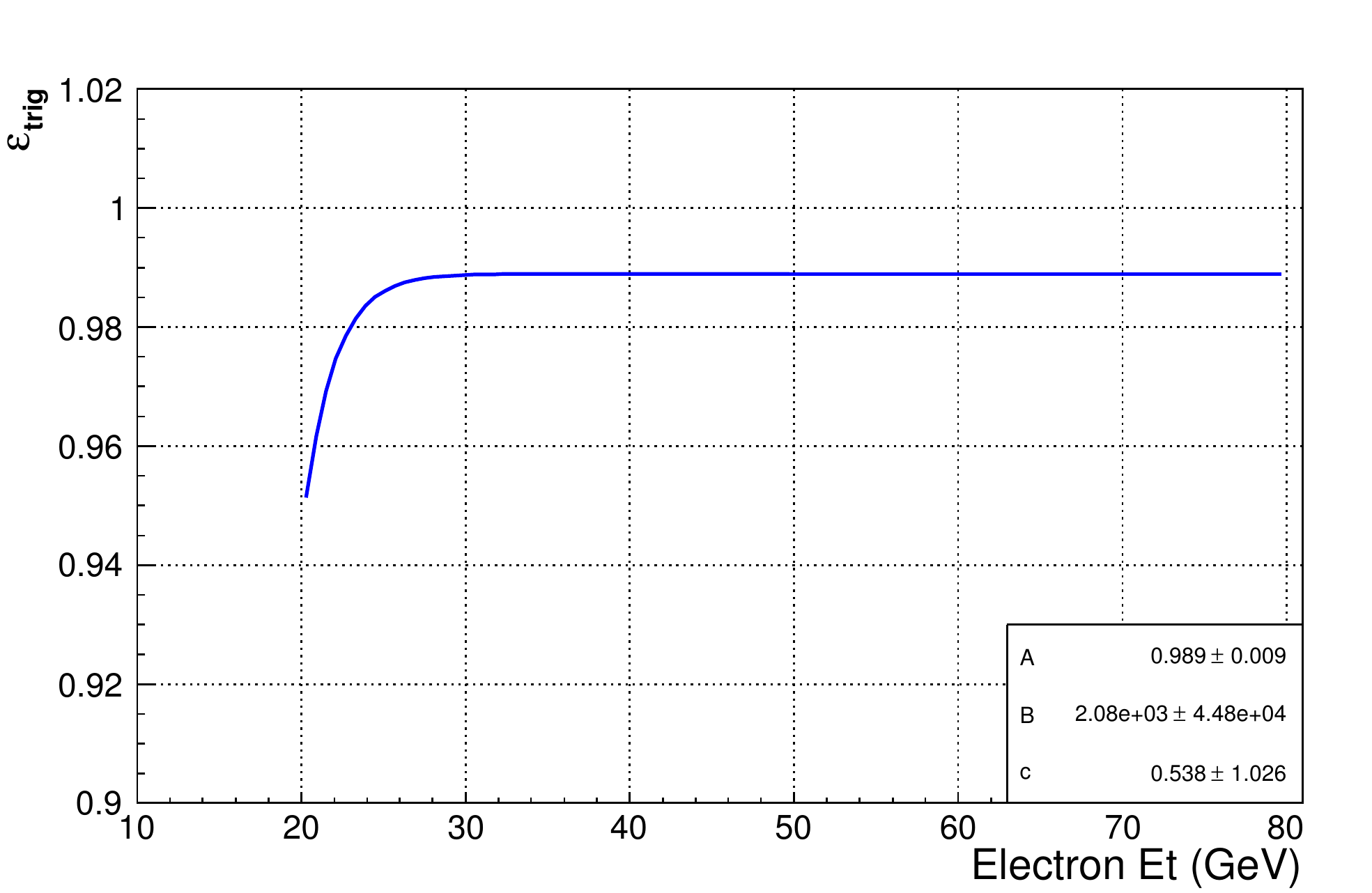}
\includegraphics[width=0.495\textwidth, height=0.36\textwidth]{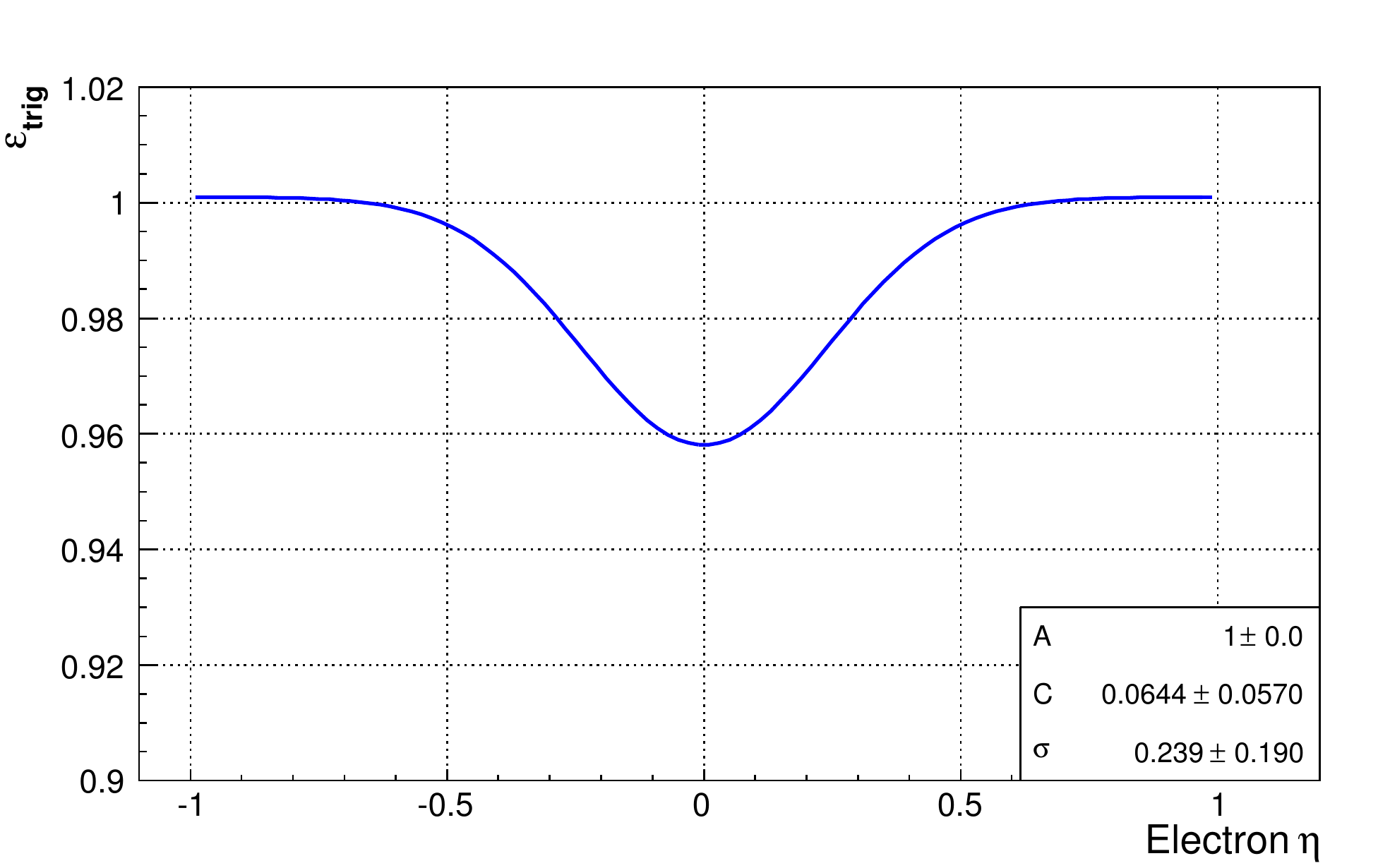}
\caption[\texttt{ELECTRON\_CENTRAL\_18} Trigger Efficiency Parametrization]{\texttt{ELECTRON\_CENTRAL\_18} trigger efficiency parametrization as a function of $E_T$ (left) and $\eta$ (right) of the candidate electrons. $E_T$ dependence is parametrized by Equation~\ref{eq:cem_et}, $\eta$ dependence by Equation~\ref{eq:cem_eta}. The fit parameters are reported in the legends.}\label{fig:cem_trig}
\end{center}
\end{figure}

The \verb'ELECTRON_CENTRAL_18' trigger path was extensively studied in precision CDF measurements involving $W$ and $Z$ bosons~\cite{w_z_prl} and now, after each data taking period, the efficiency is evaluated with a standard set of tools~\cite{perfidia}. A data sample of $W$ boson collected with a \verb'NO_TRACK' trigger is used to measure the tracking efficiency while a backup trigger is used to measure the efficiency of the calorimeter clustering. The efficiency is \mbox{$\epsilon^{CEM}_{trig} =0.982\pm 0.003$} across the complete dataset with a small $E_T$ and $\eta$ dependency (see Figure~\ref{fig:cem_trig}). The trigger turn-on $E_T$ dependency is parametrized with the following function:
\begin{equation}\label{eq:cem_et}
  \epsilon_{trig}(x) = A - Be^{-c x};
\end{equation}
where $x \equiv E_T$ and $A$, $B$, $c$ are the parameters of the fit. The following equation is used to model the small effect across $\eta$:
\begin{equation}\label{eq:cem_eta}
  \epsilon_{trig}(x) = A - \frac{C}{2\pi\sigma} e^{-\frac{x^2}{2\sigma^2}};
\end{equation}
where $x \equiv  \eta$ and $A, C, \sigma$ are the parameters of the fit. Figure~\ref{fig:cem_trig} shows the parametrization, derived from the full CDF dataset, and applied to MC events.

\subsection{High Energy Forward Electron Trigger}

The candidate high energy forward electrons (in $1.2 < |\eta_{Det}| < 2.0$ or Plug region) are stored in the \verb'bpel' data-stream. 

Electron triggering in the plug region is particularly challenging because it lacks of the COT coverage and because a large number of soft interactions are boosted in the region closer to the
beam-line. A partial solution to these problems is the use of a multiple-object trigger, named \verb'MET_PEM', dedicated to $W\to e\nu$ selection instead of an inclusive electron trigger: there is no track selection but large \met is required together with a high energy cluster in the forward electromagnetic calorimeter. The L1 requirements are a EM tower of $E_T > 8$~GeV in the forward region and \met$ > 15$~GeV. L2 confirms \met requirements but a reconstructed EM cluster of $E_T > 20$~GeV is required and, finally, L3 confirms again the same quantities (\mbox{\met$> 15$}~GeV and EM Cluster \mbox{$E_T > 20$}~GeV) with L3 reconstruction algorithms. 

\begin{figure}[!ht]
\begin{center}
\includegraphics[width=0.495\textwidth]{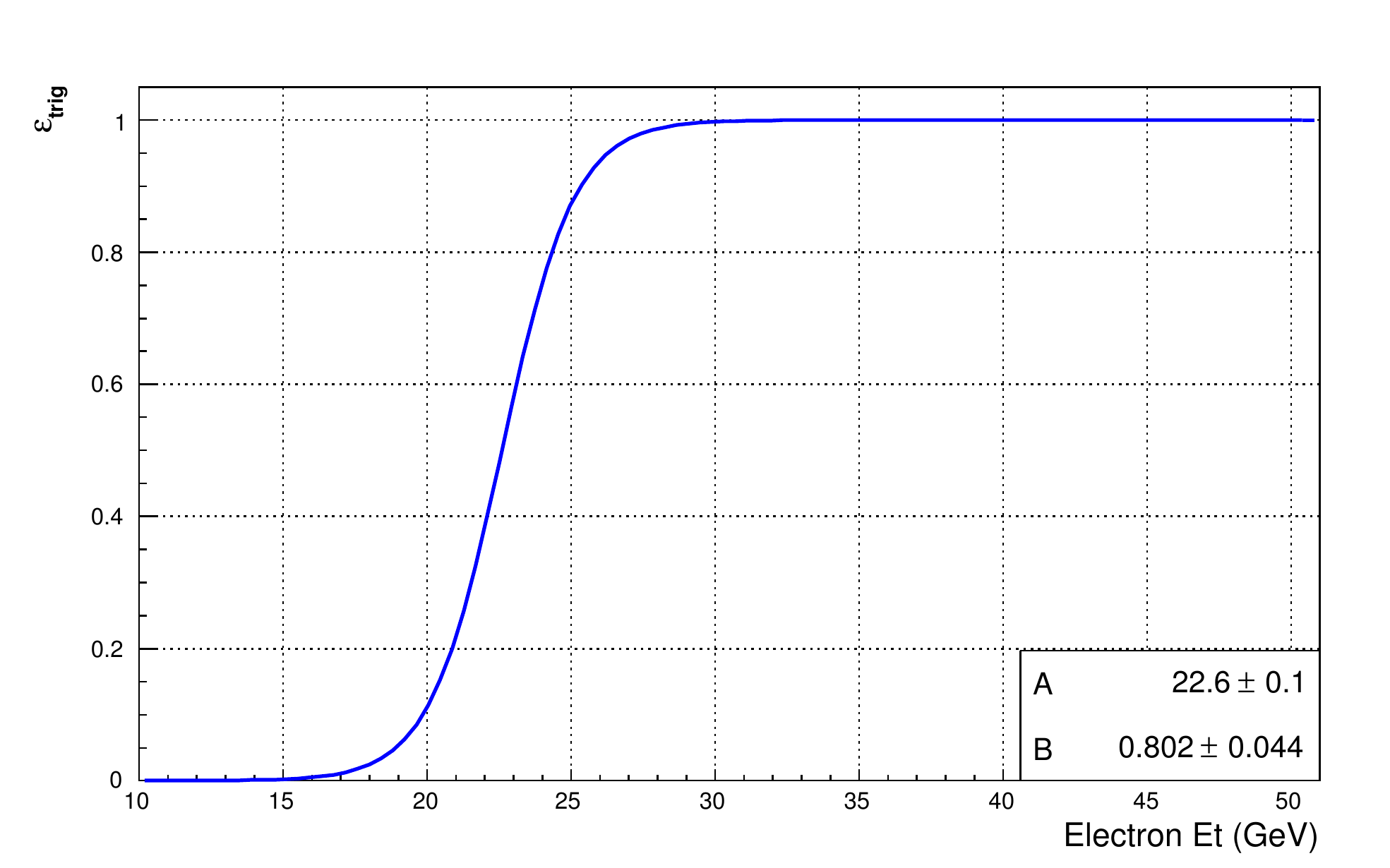}
\includegraphics[width=0.495\textwidth]{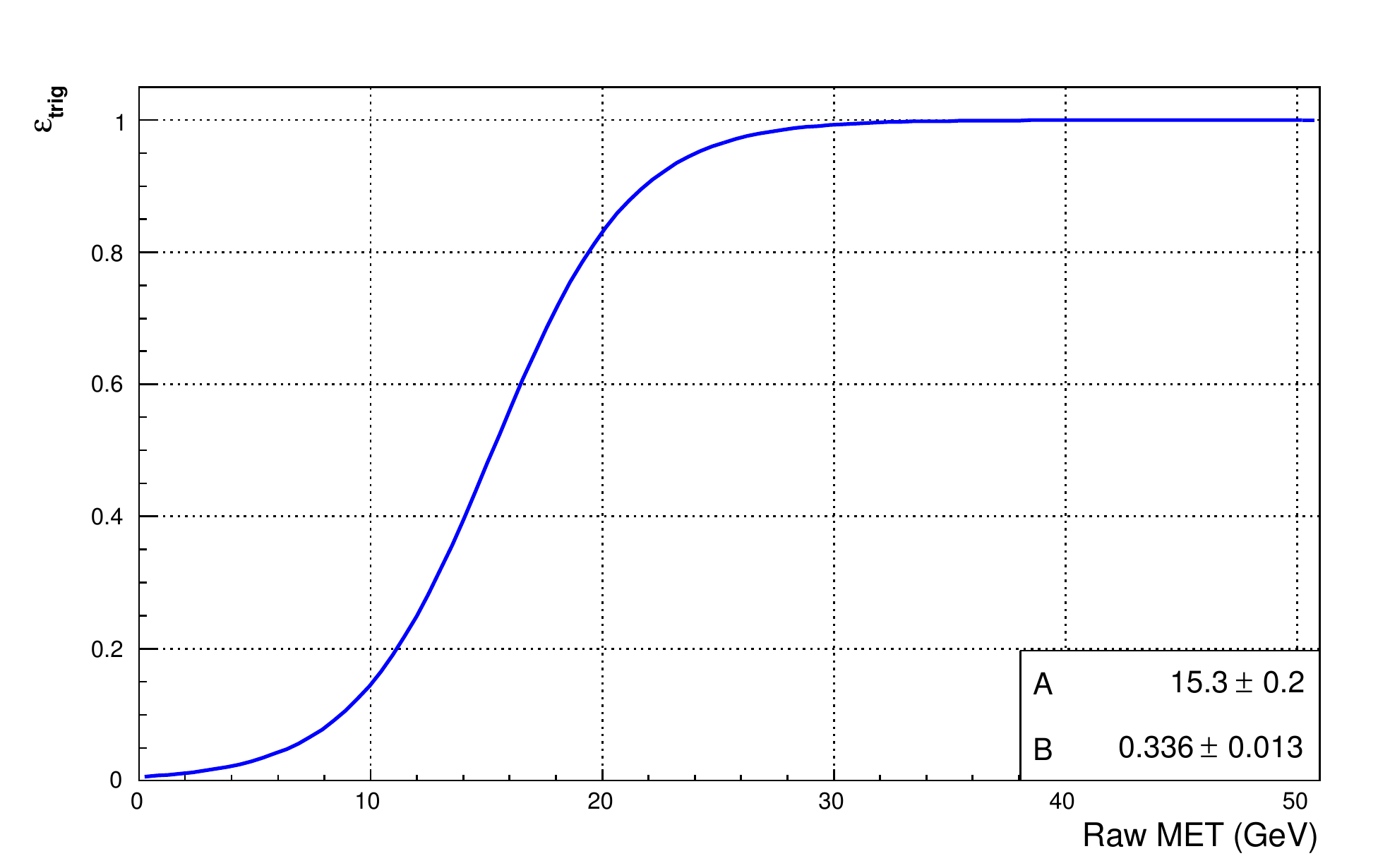}
\caption[\texttt{MET\_PEM} Trigger Efficiency Parametrization]{\texttt{MET\_PEM} trigger efficiency parametrization as a function of $E_T$ (left) of the electron candidate and \metr (right) of the event. The \metr is corrected only for the position of the primary vertex and therefore it is closer to the trigger level \met. The $E_T$ and \metr dependence is parametrized by Equation~\ref{eq:phx_trig} and the fit parameters are reported in the legends.}\label{fig:phx_trig}
\end{center}
\end{figure}

The trigger shows a turn-on for both the $E_T$ of the cluster and the \metr, i.e. corrected only for the position of the primary vertex. The following equation is used for the parametrization of the turn-on:
\begin{equation}\label{eq:phx_trig}
  \epsilon_{trig}(x) = \frac{1}{1+ e^{-A(x-B)}};
\end{equation}
where $x\equiv E_T$ or $x\equiv$\metr and $A$ and $B$ are the parameters of the fit, reported in Figure~\ref{fig:phx_trig}.

The \verb'MET_PEM' trigger parametrization was part of my initial work in
the CDF collaboration~\cite{met_pem_trig9493} and is now included in the CDF analysis
tools~\cite{perfidia}.

\subsection{High Momentum Central Muon Trigger}

Central tight muon candidate events (CMUP and CMX) are collected by the \verb'MUON_CMUP18' and the \verb'MUON_CMX18' triggers and saved into the \verb'bhmu' data-stream. 

For \verb'MUON_CMUP18' the main trigger requirements are: at L1, a track (\mbox{$p_T > 4$}~GeV$/c$) matched to a stub in both the CMU and CMP detectors. At L2, a confirmed track ($p_T > 15$~GeV$/c$) and a calorimeter deposit consistent with a MIP in the direction of the track and muon stubs. Finally at L3, a fully reconstructed COT track ($p_T > 18$~GeV$/c$) whose extrapolation matches the CMU (CMP) hits within $\Delta x_{CMU} < 10$~cm ($\Delta x_{CMP} < 20$~cm). 
Similar requirements are made also for \verb'MUON_CMX18' but the sub-detector of interest is the CMX (covering $0.6 < |\eta| < 1.1$) and the L1 requirement is a track with $p_T > 8$~GeV$/c$ matched to a stub in the CMX chamber.  

As for tight electron candidates, also the central muon triggers were extensively studied~\cite{w_z_prl} and the efficiency is
measured with a standard procedure~\cite{perfidia}. Using a data sample of $Z\to \mu^+\mu^-$, a muon passes the trigger requirements and the second muon is checked to pass the trigger selection or not. 

As the CMUP sample represents a relevant fraction of the good quality muon data, an additional effort was done, in the context of the $WH$ analysis~\cite{hao_trig10747}, to improve the CMUP trigger collection efficiency. Basically we exploited all the CMUP trigger candidates present in the \verb'bhmu' stream and collected by secondary and auxiliary triggers. The purity of those events was found to be the same of the \verb'MUON_CMUP18' selected, and we evaluated, with the bootstrapping method, a recover of about $12$\% efficiency on CMUP online selection. 

The trigger efficiencies are \mbox{$\epsilon^{CMUP}_{trig} = 0.976\pm 0.001$} and \mbox{$\epsilon^{CMX}_{trig}=0.873\pm0.001$}, flat in the kinematic variables.

\subsection{High Missing Transverse Energy Triggers}

In order to recover lepton acceptance lost because of the limited geometrical coverage of the single-lepton triggers, we rely on data collected by the \texttt{emet} data-stream. 

The \texttt{emet} stream contains samples with large online \met or \met plus reconstructed jets. No charged lepton information is used, therefore the acceptance of $\ell\nu + 2$~jets events can be recovered. In particular we use the \verb'emet' stream to collect Loose Muons (Section~\ref{sec:LooseMuId}) and Isolated Tracks (Section~\ref{sec:IsoTrk}) candidates, altogether dubbed Extended Muon Categories (EMC).

The trigger optimization~\cite{higgs_ttf8875} was performed with the aim to increase the sensitivity to Higgs boson production, but, as diboson and $WH$ production share the same $HF$ final state, an identical strategy was employed for this analysis. The three main triggers used are: \verb'MET45', \verb'MET2J' and \verb'METDI'. The exact naming and requirements of the trigger paths changed many times along the data taking period due to different instantaneous luminosity conditions but the main classification and characteristics are the following:
\begin{itemize}
\item \verb'MET45' requires large \met in the event, first at L1 and, with increased quality and cut levels, at L2 and L3. The trigger was present along all the data taking with slightly different specifications, however in the last 2/3 of the data taking the L3 \met cut was $40$~GeV.
\item \verb'MET2J' requires high \met in the event at L1 (\met$ > 28$~GeV), two jets at
L2 with one of them reconstructed in the central region of the detector ($|\eta| < 0.9$) and, finally, L3 object confirmation and \met$ > 35$~GeV. The trigger was Dynamically Pre-Scaled (DPS) to cope with the increasing luminosity in the second half of the data taking.
\item \verb'METDI' requires two jets and \met in the event, very similarly to \verb'MET2J', but the L1 requirement, with \met$ > 28$~GeV and at least one jet, is different.
This trigger was introduced after the first $2.4$~fb$^{-−1}$ of integrated luminosity.
\end{itemize}

The overlap and composite structure of the \verb'emet' triggers makes their use less straightforward than the inclusive lepton ones. Furthermore the use of multiple-objects requires to model, on signal and background MC, the possible selection efficiency correlations (e.g. \met and jets correlations).
Efficiency studies were performed in~\cite{Nagai:2010zza, Buzatu:2011zz}. The main results are the definition of appropriate kinematic regions where $\epsilon_{trig}$ is flat (plateau region) with respect to the jet variables and can be parametrized as a function of the \met only. \met is corrected for primary vertex and jet energy but not for the presence of muons. In this way, tight muon candidates, collected from
other triggers (e.g. \verb'MUON_CMUP18'), can be used to evaluate the turn-on curves as their contribution to the \met is not accounted at trigger level. 

An optimal trigger combination strategy for \texttt{emet} triggers was evaluated in~\cite{Buzatu:2011zz}. The trigger selection maximizes the $WH$ acceptance after that the \mbox{$\epsilon_{trig}(\met)$} function is evaluated for each data period and kinematic configuration of the three triggers. 

The final combined trigger efficiency reaches $\epsilon_{trig}\simeq 0.5$ for $WZ$ simulated events.

\subsection{Data Quality Requirements and Luminosity Estimate}\label{sec:lumi}

The last step before offline event selection is the application of strict data quality requirements. These are ensured by the Good Run List (GRL) selection: i.e. it is possible to use only the run periods where the performances of the sub-detectors are optimal and well parametrized.

A conservative approach, used in many of the CDF analyses, is to consider a run as good only if all the sub-detectors used to define the physics objects of the analysis are fully functional. However here, as in the latest $WH$ search results~\cite{wh94_note10796}, an ad-hoc approach is used~\cite{ext_grl}. We require a full functionality of the silicon detector (used for jet $b$-tagging) and calorimeters, while, as we are interested in the one-lepton final state, we de-correlate inefficiencies due to shower-max detectors functionality (CES and PES used for CEM and PHX identification) from muon chambers (CMU, CMP, CMX problematic runs). In this way it was possible to recover 8\% of the collected luminosity.

Once that the trigger paths and the appropriate GRLs are defined, it is possible to calculate the final integrated luminosity used for the various lepton types:
\begin{itemize}
\item CEM and PHX electrons use an integrated luminosity of $\int \mathcal{L}\mathrm{dt} = 9.446$~fb$^{−-1}$.
\item CMUP muons use an integrated luminosity of $\int \mathcal{L}\mathrm{dt} = 9.494$~fb$^{-−1}$. 
\item CMX muons use an integrated luminosity of $\int \mathcal{L} \mathrm{dt}= 9.396$~fb$^{-1}$.
\item EMC charged leptons use an integrated luminosity of $\int \mathcal{L}\mathrm{dt} = 9.288$~fb$^{-−1}$.
\end{itemize}
The calculated luminosity has systematic uncertainty of $6$\% (see Section~\ref{clc_par}).

\section{Offline Event Selection}

Now, on good data, we apply the object identification algorithms described in Chapter~\ref{chap:objects} to select the desired final state of $\ell\nu + HF$ jets. 

Table~\ref{tab:cuts} summarizes the complete selection {\em cut flow} and the following Sections (from \ref{sec:w_sel} to~\ref{sec:svm_sel}) describe in detail the selection criteria. Part of the cut flow is common to many other CDF analyses while the applied multi-jet rejection algorithm, as well as some other specific choices, are completely original.

Because we rely on simulation to evaluate the signal acceptance, the data/MC agreement is cross checked for each one of the employed algorithms with data driven methods. Well known physics processes are selected in control samples and the efficiency on data ($\epsilon^{Data}_{alg}$) and MC ($\epsilon^{MC}_{alg}$) is estimated. The ratio between them, with appropriate correlations and systematic errors evaluation, is a correction Scale Factor ($SF_{alg}$) to be applied on the simulation:
\begin{equation}
  SF_{alg} =\frac{ \epsilon^{MC}_{alg}}{\epsilon^{Data}_{alg}}
\end{equation}
The $SF_{alg}$ associated to the lepton identification ($SF_{l}$) and to the $b$-tagging ($SF_{Tag}$) are particularly important in this analysis because of the large number of different lepton reconstruction algorithms and because of the $HF$ final state of the signal. The $SF$s are discussed along with the selection criteria in the following Sections.

\begin{table}[h] 
  \begin{center}
   \begin{tabular}{c}
        \toprule
  {\bf Selection Cuts}\\\midrule

   Trigger Fired \\
   Good Run Requirement\\
  $|z_0^{PV}| \le 60$~cm \\
  Charged Lepton Selection \\
  $|z_0^{lep} −- z_0^{PV}| < 5$~cm\\
  Di-Lepton Veto\\
  $Z\to \ell\ell$ Veto\\
  Cosmic Ray Veto \\
  $\gamma\to e^+e^-$ Conversion Veto\\
  \met$ > 15$~GeV\\
  2 Tight Jets Selection \\
  $M_{Inv}(jet1, jet2)>20$~GeV$/c^2$\\
  1 or 2 \texttt{SecVtx} Tags\\
  {\bf Multi-jet Rejection:} \\
 {\em SVM}$_{CEM, EMC}>0$, {\em SVM}$_{CMUP, CMX}>-0.5$, {\em SVM}$_{PHX}>1$\\
      \bottomrule
    \end{tabular}
    \caption[Selection Summary]{Summary of all the selection requirements applied in the analysis.}\label{tab:cuts}
  \end{center}
\end{table}

\subsection{$W\to\ell\nu$ Offline Selection}\label{sec:w_sel}

The leptonic decay of the $W$ boson is selected by requiring exactly one high energy charged lepton and missing transverse energy signaling the presence of the neutrino.

The correct application of these two basic requirements exploits several of the identification algorithms described in the previous Chapter and other selection criteria in order to purify the lepton sample.
The summary of all the selection steps, applied on data and MC, is the following:
\begin{itemize}
\item identification of the primary interaction vertex (Section~\ref{sec:primVtx}) within the interaction fiducial region of 60 cm from the detector center ($|z_0^{PV}| \le 60$~cm). Minimum bias events are used to evaluate on data the efficiency of this cut. The result, averaged over all the data taking periods, is\footnote{The initial number of MC events, used to normalize all the simulated acceptances, is derived after this cut so that $\epsilon_{z_0}$ should be applied to obtain the final yields.}:
  \begin{equation}
    \epsilon_{z_0} = 0.9712\pm 0.0006.
  \end{equation}
\item Identification of one charged and isolated lepton candidate originating
from the primary vertex ($|z_0^{lep} −- z_0^{PV}| < 5$~cm). A charged lepton candidate is one of the eleven lepton identification algorithms described in Sections from~\ref{sec:eleSel} to~\ref{sec:IsoTrk}: CEM, PHX (tight electrons), CMUP, CMX (tight muons), BMU, CMU, CMP, CMIO, SCMIO, CMXNT (loose muons) and ISOTRK (isolated tracks). This requirement does not produce any appreciable signal loss.
\item Veto of di-leptonic candidate events: if an event contains two tight or loose electrons or muons both isolated or not, it is rejected as a Drell-Yan event or $t\bar{t}$ dileptonic candidate. Events with an ISOTRK candidate are vetoed if reconstructed together with another ISOTRK or a tight lepton. An efficiency of approximately 90\% is estimated from signal MC.
\item Further rejection of $Z\to\ell\ell$ candidates is obtained for CEM, PHX, CMUP and CMX leptons by vetoing events in which the tight lepton can be paired with a track or a EM calorimeter cluster that form an invariant mass within the range $[76, 116]$~GeV$/c^2$.  An efficiency of approximately 90\% is estimated from signal MC.
\item A cosmic tagger was implemented within the CDF offline code~\cite{cosmic_tag}
to reject high-$p_T$ cosmic muons which interact with the detector simultaneously with a bunch crossing. The algorithm exploits muon chambers, COT and TOF, calorimeter energy and timing information, providing
almost 100\% rejection of cosmic ray events, with a negligible loss of signal efficiency.
\item Rejection of electrons originating from photon conversions, $\gamma\to e^+e^{-−}$,
is obtained by applying the CDF conversion tagger~\cite{conversion_tag}. The algorithm looks for two opposite-sign tracks (one of them belonging to the identified electron) and requires $\Delta \cot(\theta) \le 0.02$ and the distance at the closest approach between them $D_{xy} \le 0.1$~ cm. The rejection efficiency of the conversion veto is about $65$\% with a $1\div 2$\% signal loss.
\item Finally, \met$ > 15$~GeV is required to signal the presence of a neutrino. The \met is fully corrected for position of the primary interaction vertex, presence of jets and muons in the event (see Section~\ref{sec:metObj}). Approximatly 7\% of the signal is lost with this requirement.
\end{itemize}

For each run period and lepton identification algorithm, the agreement
between data and MC is measured comparing the $Z\to\ell\ell$ data sample against the simulation. Table~\ref{tab:lepId} reports, averaged on all the run range, the yield correction, $SF_{l}$ that covers the
data/MC differences. The main differences arise in the simulation of the isolation and of the muon chamber response.

   \begin{table}[h] 
 \begin{center}
     \begin{tabular}{cccc}
       \toprule

       CEM & PHX & CMUP & CMX \\
       $0.973 \pm 0.005$ &   $0.908 \pm 0.009$ & $0.868\pm 0.008$ & $ 0.940\pm 0.009$\\
     \end{tabular}
  
      \begin{tabular}{cccccc}
        \midrule
        BMU & CMU & CMP & SCMIO & CMIO & ISOTRK \\

        $ 1.06\pm 0.02$ & $ 0.88\pm 0.02$ & $ 0.86\pm0.01$ & $ 1.02\pm 0.01$ & $ 0.97 \pm 0.02$  &  $ 0.94 \pm 0.04$ \\
        \bottomrule 
      \end{tabular}

     \caption[Lepton Algorithm Data/MC Scale Factors ($SF_l$)]{Data/MC Scale Factors ($SF_l$) for each of the single lepton identification algorithms used in the analysis, tight charged leptons in the first row and EMC categories in the lower row.}\label{tab:lepId}
 \end{center}
   \end{table}

If several lepton identification algorithms cover the same detector region they can not be considered independent and a further correction\footnote{This correction was introduced for the first time in this analysis and in the $WH\to \ell\nu+ b\bar{b}$ search~\cite{wh94_note10796}.} to the simulation is needed. As shown in Figure~\ref{fig:split_alg}, loose muon and ISOTRK largely overlap with the tight lepton selection and, often, a MC lepton candidate is reconstructed by more than one lepton selection algorithm. Because of this and to avoid ambiguity, we apply a prioritized selection, i.e. we label a lepton candidate with the highest priority algorithm according to:
  \begin{equation}
\begin{gathered}
 CEM > PHX > CMUP > CMX > BMU > CMU >\\
 CMP > SCMIO > CMIO > CMXNT >  ISOTRK
\end{gathered}
  \end{equation}

In most of the cases, selection cuts are orthogonal and only one lepton reconstruction is possible, however, if this is not true {\em and} $SF_{l}\ne 1$, the same simulated lepton has a probability greater than zero of not being identified by the highest priority algorithm\footnote{Quality criteria and historical reasons are at the base of the prioritization of the algorithms. Prioritization is indirecly present also in the $SF_{l}$ measurements: for example in the ISOTRK $SF_{l}$ measurement the lepton candidates are checked not to be reconstructed as tight leptons, while the check is not performed in the opposite case, i.e. no ISOTRK identification is performed during the tight $SF_{l}$ measurement.}. In this case the MC event under consideration is classified under more lepton categories with different weights, according to:
\begin{equation}\label{eq:composite_sf}
  \omega_{l} = SF_{l} + SF_{l}\prod_{l'>l} \big(1- SF_{l'} \big)
\end{equation}
where $\omega_{l}$ is a composite $SF_{l}$. Equation~\ref{eq:composite_sf} reduces to $\omega_{l} = SF_{l}$ when a lepton candidate is reconstructed only by one identification algorithm, otherwise it takes into account the probability of {\em not} identification from algorithms with higher priority ($l'>l$).

\begin{figure}[!ht]
\begin{center}
\includegraphics[width=0.8\textwidth]{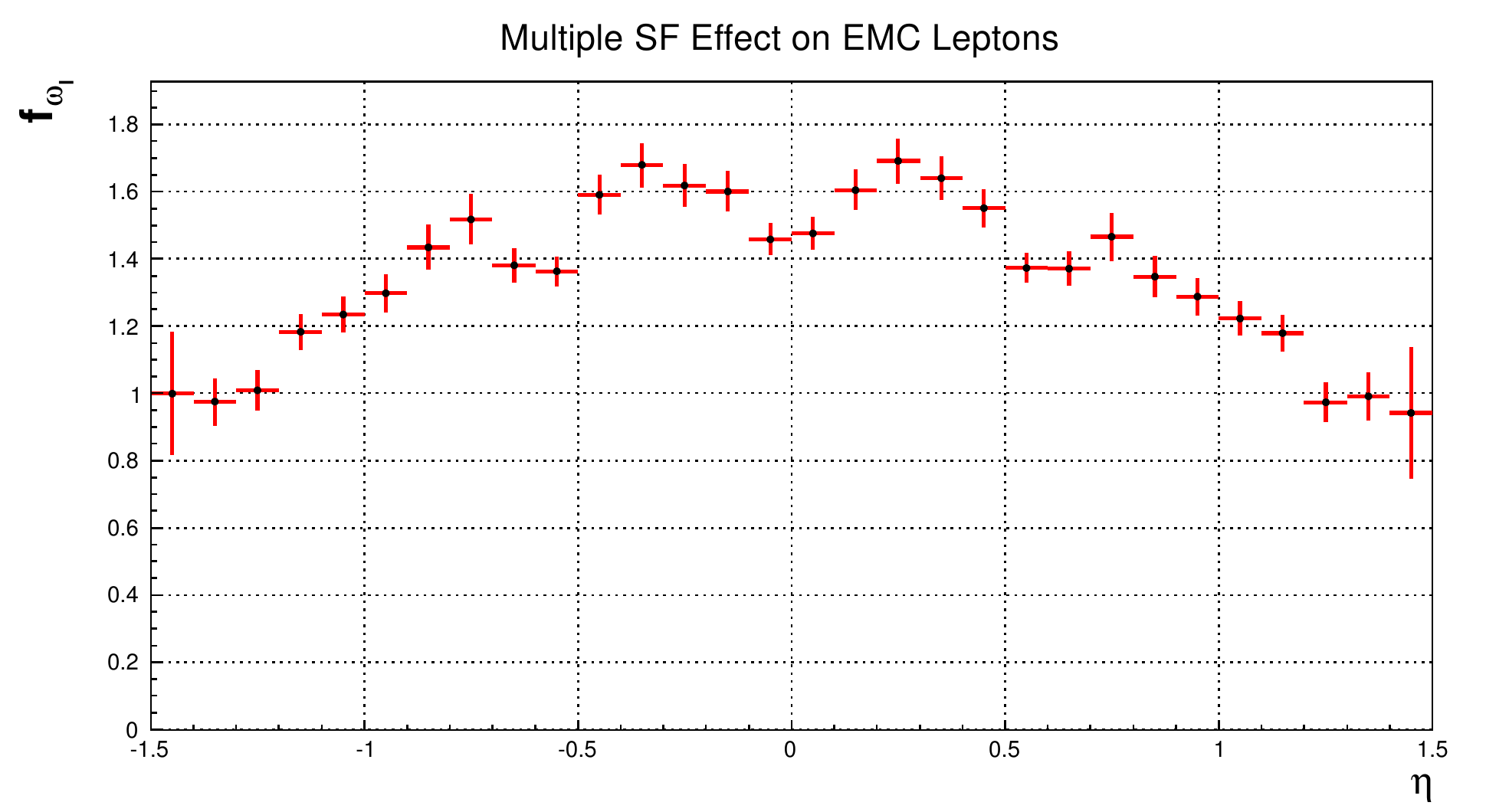}
\caption[$\eta$ Distribution of the $\omega_{l}$ Correction on EMC]{Effect of the $\omega_{l}$ correction described in Equation~\ref{eq:composite_sf} across the $\eta$ distribution of the EMC lepton category selected on $WZ$ MC. The correction appears as a factor $f_{\omega_l}\ne 1$ for each event where the reconstructed lepton is identified by more than one algorithm.}\label{fig:eta_multisf}
\end{center}
\end{figure}

\begin{figure}[!ht]
\begin{center}
\includegraphics[width=0.495\textwidth]{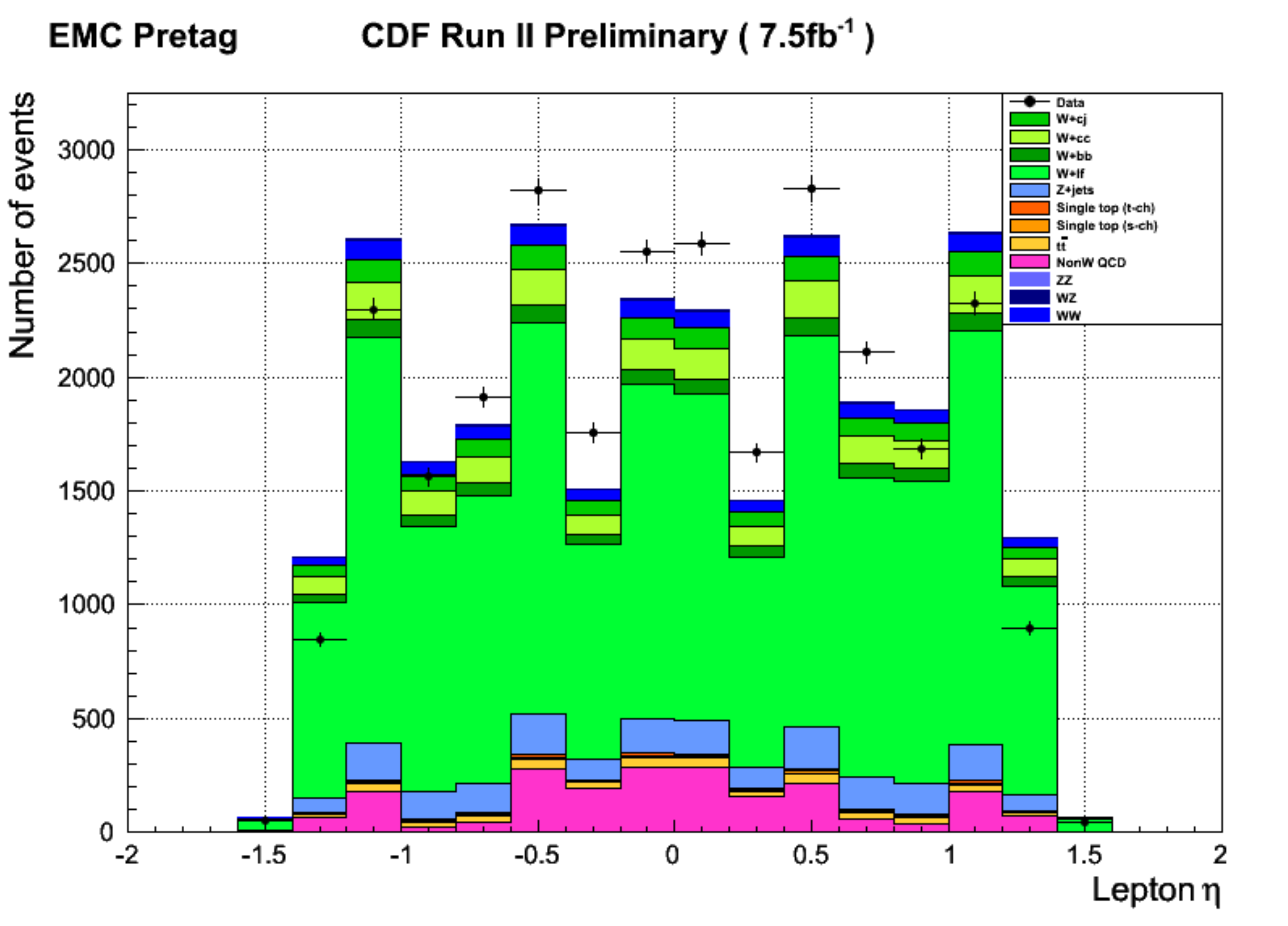}
\includegraphics[width=0.495\textwidth]{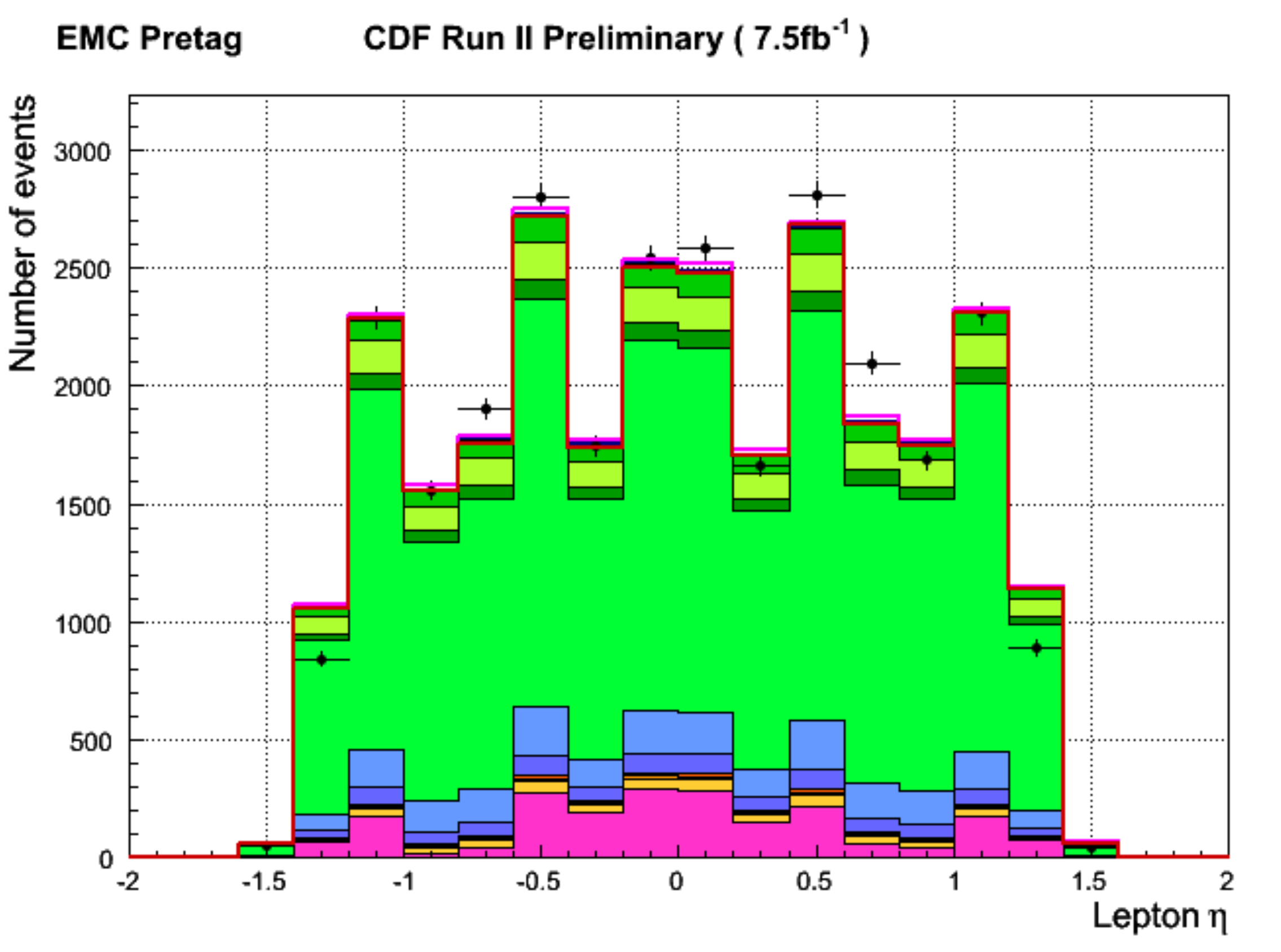}
\caption[Effect of $\omega_{l}$ on EMC $\eta$ Distribution after Background Estimate]{$\eta$ distribution of the EMC lepton category with full background evaluation before (left) and after (right) the application of the multiple $SF_{l}$ correction defined by Equation~\ref{eq:composite_sf}.}\label{fig:eta_pre_post}
\end{center}
\end{figure}

Figure~\ref{fig:eta_multisf} shows the corrected  $\eta$  distribution for the EMC lepton category (combination of loose muon and ISOTRK algorithms) while Figure~\ref{fig:eta_pre_post} shows the complete background estimate distribution for the combined $\eta$ of the EMC category before and after the correction of Equation~\ref{eq:composite_sf}.

\subsection{Offline Jets Selection}\label{sec:jetSel}

The final aim of this analysis is the study of the invariant mass spectrum of a two jet, $HF$ enriched, final state. 

Jet selection is based on the \verb'JETCLU' $R = 0.4$ algorithm, described in Section~\ref{sec:jetObj}, and the \texttt{SecVtx} $HF$-tagger algorithm, described in Section~\ref{sec:secvtx}.
One of the goals of this analysis is to demonstrate the ability of the CDF experiment to identify a resonance in the background rich sample of $W+HF$ jets. Therefore the jet selection was kept simple with the aim to understand the sample itself. The specific requirements are the following:
\begin{itemize}
\item  selection of exactly two tight jets. Jets central in the detector and with large $E_T$ (after L5 energy correction) are named tight jets if:

\begin{equation}
  E^{jet,cor}_T > 20\mathrm{~GeV}\quad\textrm{and}\quad |\eta^{jet}_{Det}| < 2.0;
\end{equation}

\item a minimal lower bound on the invariant mass distribution of the jet pair: $M_{Inv}(jet1, jet2) > 20$~GeV$/c^2$;
\item one or both of the selected jets $HF$-tagged by the \texttt{SecVtx} algorithm;
\item jets are ordered according to their $E_T^{jet,cor}$ and classified as: $jet1$ and $jet2$. 
\end{itemize}
No other requirements are applied on the jets and, in particular, we do not apply any criteria to the loose jets present in the event. 

The $HF$-tag requirement divides the $\ell\nu+ 2$ jets sample in three analysis regions:
\begin{description}
\item[Pretag:] no $HF$-tag requirement is applied and, due to the small $HF$ contamination, the sample is used as a control region (see Section~\ref{sec:wjets_pretag}).
\item[Single-tag:] events with exactly one jet tagged as $HF$ compose this signal region. The statistics of the selected sample is quite large therefore it is important for the sensitivity of this analysis to  $WW\to\ell\nu + c\bar{s}$ decay.
\item[Double-tag:] events with both jets tagged as $HF$. The statistics of this sample is small because of the strict double-tag requirement, however a large fraction of the sensitivity for the $WZ\to\ell\nu + b\bar{b}$ signal comes from this region as the background contamination is also low.
\end{description}

According to Section~\ref{sec:btag_sf}, the $HF$ selection efficiency of the \texttt{SecVtx} algorithm must be corrected with an appropriate data/MC $SF_{Tag}$. Furthermore a contamination due to $LF$ quarks producing secondary vertices will contribute to the signal sample. The $HF$ and $LF$ components for $1$−-Tag and $2$--Tag selection are accounted in the signal MC with a per-event tagging probability, $\omega_{k-−Tag}$ obtained by the combination of per-jet tagging probabilities:
\begin{equation}\label{eq:1tag}
  \omega_{1-Tag}=\sum_{i=1}^{n}p^i_{Tag}\cdot\Big(\prod^n_{j=1,j\ne i}\big( 1 - p^j_{Tag}\big)\Big),
\end{equation}
\begin{equation}\label{eq:2tag}
  \omega_{2-Tag}=\sum_{i=1}^{n-1}p^i_{Tag}\cdot\Big( \sum_{j>i}^{n}p^i_{Tag} \cdot \Big( \prod^n_{h=1,h\ne i,h\ne j}\big( 1 - p^k_{Tag}\big)\Big)\Big),
\end{equation}
where the per-jet tagging probability is defined by $p^j_{Tag} \equiv SF_{Tag}$, if the jet is matched\footnote{The matching is satisfied if $\Delta R(jet, had) < 0.4$ where we test the fully reconstructed jet 4-vector against all the final state hadrons 4-vectors.}  to a $HF$ hadron, or $p^j_{Tag} \equiv p^j_{Mistag}$, if the event is matched to a $LF$ hadron.

Equations~\ref{eq:1tag} and~\ref{eq:2tag} are applied to obtain \texttt{SecVtx} 1-Tag or 2-Tags probabilities.  A further generalization of the previous equations was implemented in the WHAM analysis framework (see Appendix~\ref{chap:AppWHAM}) so that efficiencies can be calculated for any $b$-tagging algorithm (and combinations of them), any tag and jet multiplicity.

\subsection{Multi-Jet Background Rejection}\label{sec:svm_sel}

Multi-jet events can fake the $W$ boson signature when one jet passes the high $p_T$ lepton selection criteria and fake \met is generated through jets mis-measurement. Although the probability for such an event is small, the high rate of multi-jet events, combined with the small cross section of the processes of interest, make this an important background. Additional difficulties arise in the simulation of the  mixture of physics and detector processes contributing to this background. Therefore it is desirable to reduce the multi-jet events as much as possible.

For this purpose, an original and efficient multi-variate method based on the Support Vector Machine (SVM) algorithm was developed for this analysis.

The basic concept behind a SVM classifier is quite simple: given two sets of $n$-dimensional vectors (the training sets) the algorithm finds the {\em best separating hyper-plane} between the two. The plane is then used again to classify a newly presented vector in one of the two sets. Non linear separation can be achieved with the use of appropriate transformations called {\em Kernels}. SVM classification performs better than other multivariate algorithms, like Neural Networks, in the case of statistically limited training sets and shows more stability with respect to over-training~\cite{svmbook}.

The work described here improves a previously developed software package\footnote{The previously created SVM based multi-jet rejection~\cite{CHEP_svm} was already tested and employed in previous analyses~\cite{wh75_note10596, wh94_note10796, cdfDiblvHF_Mjj2011}.}, based on the \texttt{LIBSVM}~\cite{libsvm} library, able to perform algorithm training, variable ranking, signal discrimination and robustness test. Here we report only a summary of the results while a detailed description of optimization, training procedure and variable definition is given in Appendix~\ref{chap:AppSvm}. 

We trained two specific SVMs, one for the {\em central} and one for the {\em forward} region of the detector. 
The SVMs aimed to improve the purity of the $W\to e\nu+2$ jets sample, however we avoided the use of input variables related to the electron identification (i.e. cluster or track related) while we focused on the kinematic variables. This gave an additional result as the multi-jet rejection proved to be optimal also for other lepton identification algorithms. For example the central SVM was used on all the lepton identification algorithms in the region $|\eta|<1.2$. 

Appendix A describes in details all the input variables used in the trainings.
The SVM for the {\em central} region exploits the following eight input variables:
\begin{itemize}
\item $W$ related variables: $M^W_T$, \metr, ${\not}{p_T}$, $\Delta \phi(lep,$ \met$)$, $\Delta R(\nu^{min},$ $lep)$;
\item global variables:  $MetSig$, $\Delta \phi(jet1,$ \met), $\Delta \phi({\not}{p_T},$ \met).
\end{itemize} 
The specific SVM for the {\em forward} region of the detector (used only for the PHX electron selection) exploits six input variables:
\begin{itemize}
\item $W$ related variables: $M^W_T$, \metr, ${\not}{p_T}$; 
\item global variables:  $MetSig$, $\Delta \phi({\not}{p_T},$ \metr), $\Delta \phi({\not}{p_T},$ \met).
\end{itemize}

Table~\ref{tab:cuts} shows different SVM selection values for the different lepton types:
\begin{equation}
\textrm{{\em SVM}}_{CEM, EMC}>0\textrm{;}
\end{equation}
\begin{equation}
  \textrm{{\em SVM}}_{CMUP, CMX}>-0.5\textrm{;}
\end{equation}
\begin{equation}
  \textrm{{\em SVM}}_{PHX}>1\textrm{.}
\end{equation}
We choose different thresholds due to the different probabilities to fake a $W$ signature: lower for the tight muon selection (CMUP, CMX), higher for CEM electrons and EMC lepton category, and way higher for PHX electrons, identified in a detector region with low track reconstruction efficiency and large calorimeter occupancies. 

\begin{figure}[!ht]
\begin{center}
\includegraphics[width=0.495\textwidth]{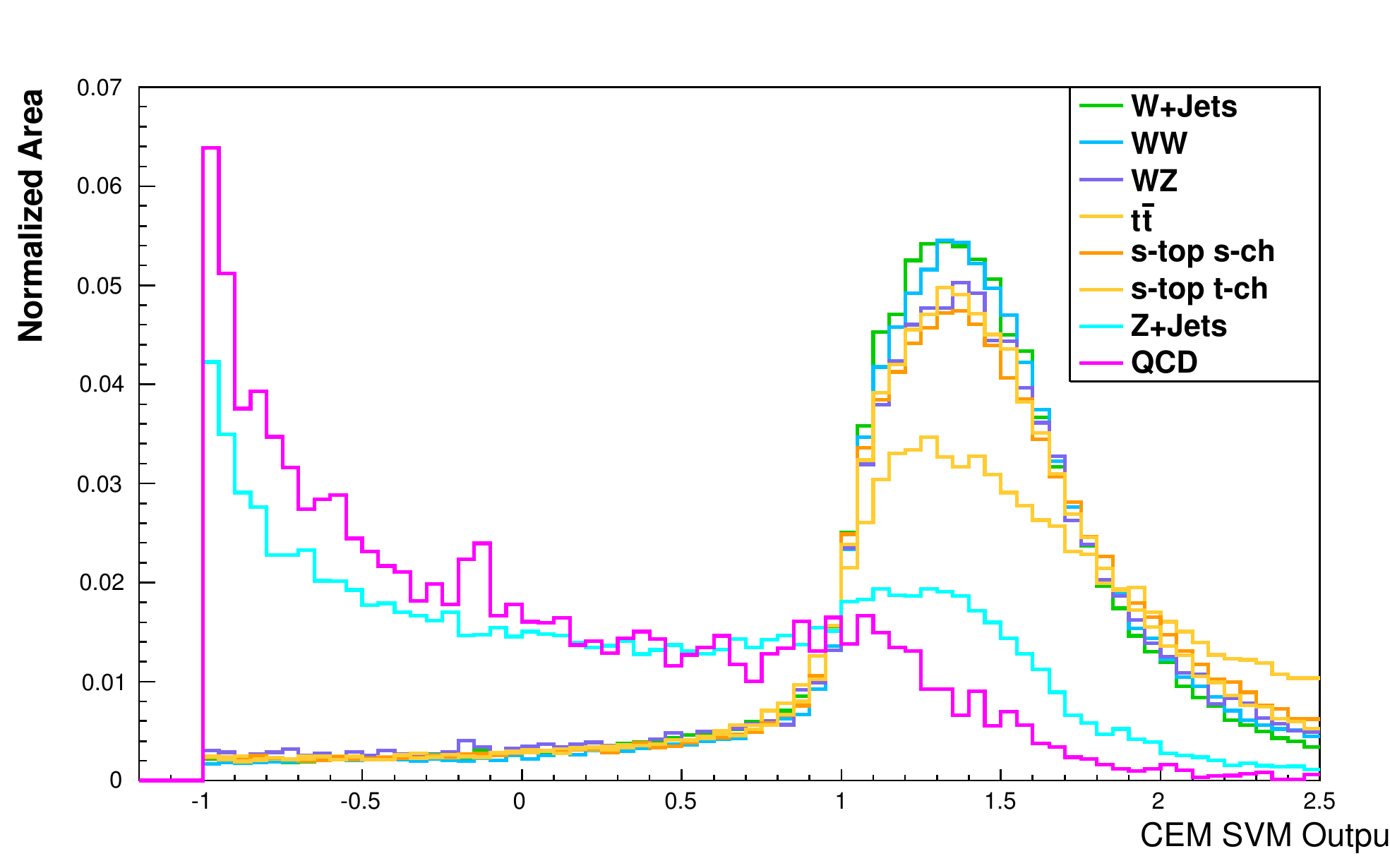} \includegraphics[width=0.495\textwidth]{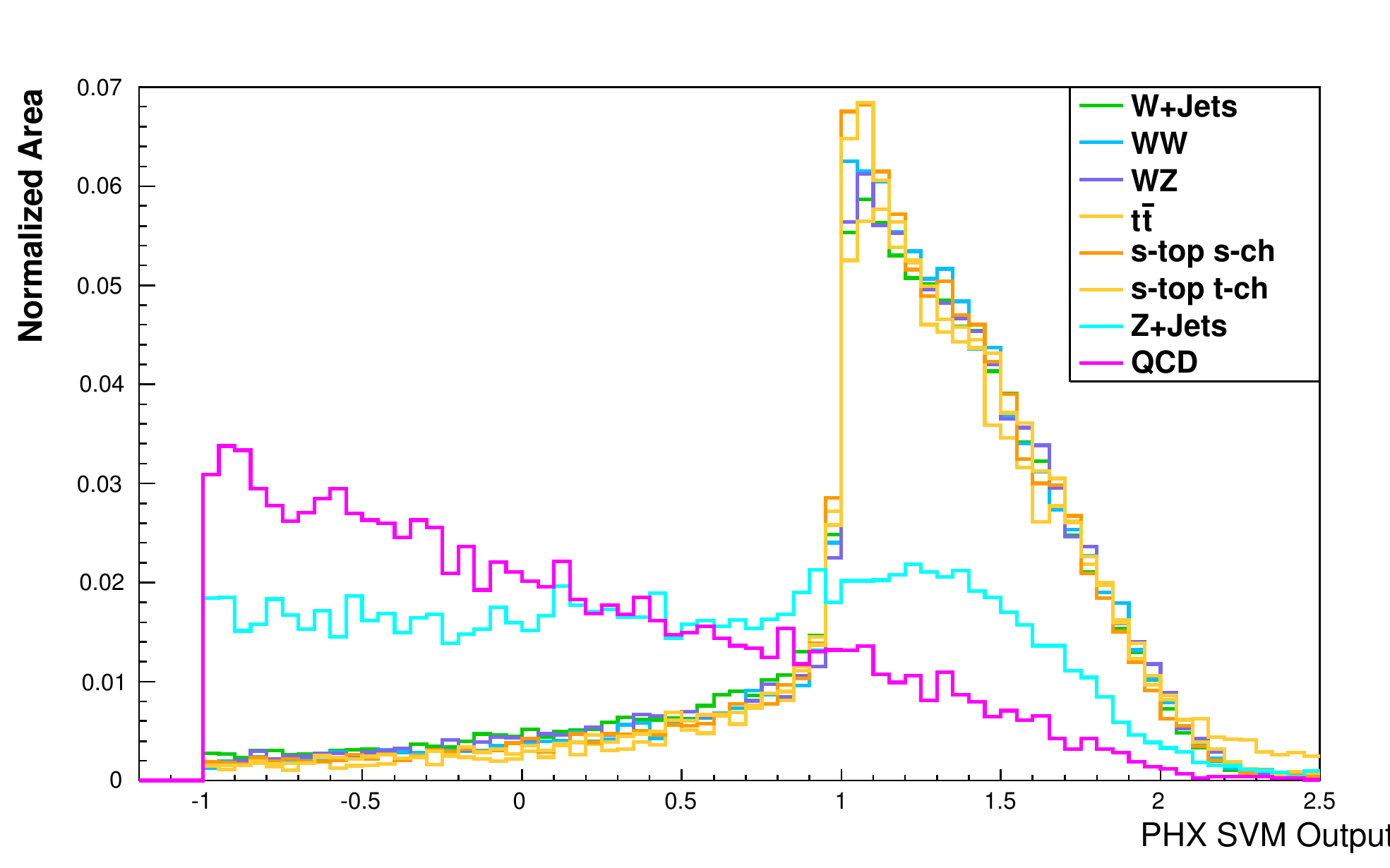}
\caption[SVM Output Distribution for CEM and PHX Leptons]{SVM output distribution for CEM (left) and PHX (right) leptons for the signal and background MC samples as well as for the multi-jet background model (see Section~\ref{sec:qcd} for the description).}\label{fig:svm_templates}
\end{center}
\end{figure}

Figure~\ref{fig:svm_templates} shows the SVM output distribution for CEM and PHX leptons for the signal and background MC samples as well as for the multi-jet background models (see Section~\ref{sec:qcd} for their description).

Table~\ref{tab:svm_eff} summarizes the signal selection efficiencies for the SVM cuts in the different lepton categories; all the analysis selection cuts are applied except for the tagging requirement. The evaluation of the final multi-jet background contamination is described in Section~\ref{sec:qcd} and a summary of the results is reported in Tables~\ref{tab:wjet_pretag_qcd_fit} and~\ref{tab:wjet_tag_qcd_fit}.

The training procedure requires the SVM to pass several quality criteria about data/MC agreement. We also checked the stability of the selection efficiency. We found it stable against a variation of up to 5\% of the SVM input variables. Therefore we did not find necessary to add a specific SF for the SVM selection.

  \begin{table}
\begin{center}
    \begin{tabular}{cccccc}
      \toprule
      Lepton & CEM & PHX & CMUP & CMX &  EMC \\\midrule
      $\epsilon_{SVM}^{WW}$ & $0.95 \pm 0.02$ & $0.82 \pm 0.03$ & $0.97 \pm 0.03$ & $0.97 \pm 0.04$ &  $0.94 \pm 0.03$ \\
      $\epsilon_{SVM}^{WZ}$ & $0.96 \pm 0.03$ & $0.83 \pm 0.04$ & $0.98 \pm 0.04$  &  $0.99 \pm 0.05$ & $0.95 \pm 0.03$ \\
      \bottomrule
    \end{tabular}
    \caption[$WW$ and $WZ$ SVM Selection Efficiency]{$WW$ and $WZ$ signal selection efficiencies for the SVM cuts used in the lepton selection. Reported errors are statistical only.}\label{tab:svm_eff}
\end{center}
  \end{table}

\section{Final Signal Estimate and Data Selection}\label{sec:data_sig}

The complete event selection can now be  applied to both the full data sample and the signal MC. 

The final signal yield is derived from the following equation, for each signal process ($Sig=WW$, $WZ$, $ZZ$), each lepton identification algorithm ($l$) and for single and double tagged ($k=1$, $2$) selections:
\begin{equation}\label{eq:mc_estimate}
  N^{Sig} = \epsilon_{l, k}^{Sig}\cdot\sigma_{p\bar{p}\to Sig}\cdot\int\mathcal{L}_{l}\mathrm{dt};
\end{equation}
the factors are:
\begin{itemize}
\item the cross sections, $\sigma_{p\bar{p}\to Sig}$, for inclusive diboson production, reported in Table~\ref{tab:mc_dib}, are evaluated at NLO~\cite{mcfm_prd,mcfm_url};
  \begin{table}
    \begin{center}
    \begin{tabular}{ccc}
      \toprule
      Signal & $\sigma$ (pb), NLO &  Initial Events \\
      \midrule
      $WW$    & 11.34 $\pm$ 0.66  & 4.6 M\\ 
      $WZ$    & 3.47  $\pm$ 0.21  & 4.6 M\\  
      $ZZ$    & 3.62  $\pm$ 0.22  & 4.8 M\\ 
      \bottomrule
    \end{tabular}
    \caption[Diboson MC Information and NLO Cross Section]{MC information about the diboson signal. The reported cross sections are evaluated at NLO precision~\cite{mcfm_prd,mcfm_url}.  \texttt{PYTHIA}~\cite{pythia} v6.216 generator and parton-showering is used for the MC simulation.}\label{tab:mc_dib}
    \end{center}
  \end{table}

\item the integrated luminosity, $\int\mathcal{L}_{l}dt$, is calculated in Section~\ref{sec:lumi} for the different leptons/data-streams;
\item $\epsilon_{l, k}^{Sig}$ is the simulated selection acceptance accounting for trigger and $|z_0|$ cut efficiencies ($\epsilon_{trig}$, $\epsilon_{z_0}$), lepton and $b$-tag efficiency corrections ($\omega_l$, $\omega_{k-Tag}$):
  \begin{equation}\label{eq:all_w_tag}
    \epsilon_{l, k}^{Sig} = \epsilon_{trig}\cdot\epsilon_{z_0}\cdot\omega_{l}\cdot\omega_{k-Tag}\cdot\frac{N^{Sig}_{l,k}}{N^{Sig}_{z_0}}
  \end{equation}
where $N^{Sig}_{l,k}$ is the number of MC events passing the lepton and tag selection and $N^{Sig}_{z_0}$ is the number of events generated in the fiduciality region of the detector. Equation~\ref{eq:all_w_tag} simplifies slightly in case we are evaluating the pretag region yield:
\begin{equation}\label{eq:all_w_pretag}
    \epsilon_{l, pretag}^{Sig} = \epsilon_{trig}\cdot\epsilon_{z_0}\cdot\omega_{l}\cdot\frac{N^{Sig}_{l,pretag}}{N^{Sig}_{z_0}}
\end{equation}
where $N^{Sig}_{l,pretag}$ is the number of MC events passing the lepton selection at pretag level. 
\end{itemize}
The final result of the event selection is reported in Table~\ref{tab:data_sig} for the CDF dataset and the signal expectation; also the pretag control region is shown for comparison. 
The selection suppresses the $Z\to\ell\ell$ events making the total $ZZ$ contribution almost negligible ($\approx 3$\% of the total diboson yield), however this component is included in the signal because no distinction is possible between the $Z\to HF$ decay originating from $WZ$ or $ZZ$ production.

\begin{table}[h] 
  \begin{small}
    \begin{center}
    \begin{tabular}{cccccc}
      \toprule
 \multicolumn{6}{c}{\bf Pretag Selection}\\
 Lepton & CEM & PHX & CMUP & CMX &  EMC \\\midrule
    {\bf Data} & {\bf 80263}  &{\bf 27759}  &    {\bf 39045}  & {\bf 22465}      & {\bf 35810} \\                      
  $WW$            & 2012.4 $\pm$ 168.7    &  705.5 $\pm$ 59.3     & 1047.7 $\pm$ 88.5   & 537.0 $\pm$ 45.5  &1157.7 $\pm$ 115.3 \\
  $ZZ$            & 22.4 $\pm$ 1.9        &  6.1 $\pm$ 0.5        & 25.2 $\pm$ 2.2      & 13.2 $\pm$ 1.1    &37.5 $\pm$ 3.8\\
  $WZ$            &  307.0 $\pm$ 26.2     & 135.7 $\pm$ 11.6      & 168.6 $\pm$ 14.5    & 95.3 $\pm$ 8.2    &225.7 $\pm$ 22.8\\
    \midrule
   \multicolumn{6}{c}{\bf Single-Tag Selection}\\
 Lepton & CEM & PHX & CMUP & CMX &  EMC \\ \midrule
    {\bf Data}        & {\bf 3115}           &{\bf 1073 }     & {\bf 1577}          & {\bf 830}          & {\bf 1705}          \\                           

   $WW$            & 84.35 $\pm$ 11.8 & 25.05 $\pm$ 3.54          & 43.7 $\pm$ 6.17        & 23.68 $\pm$ 3.35      & 53.11 $\pm$ 8           \\
   $ZZ$            & 1.85 $\pm$ 0.19  & 0.21 $\pm$ 0.02           & 2.45 $\pm$ 0.25        & 1.37 $\pm$ 0.14       & 3.45 $\pm$ 0.39        \\ 
   $WZ$            & 29.2 $\pm$ 2.95  & 12.32 $\pm$ 1.22          & 16 $\pm$ 1.66          & 9.21 $\pm$ 0.94       & 20.54 $\pm$ 2.39        \\

\midrule                                                                                        
  \multicolumn{6}{c}{\bf Double-Tag Selection}\\
 Lepton & CEM & PHX & CMUP & CMX &  EMC \\\midrule
  {\bf Data}               &{\bf 175 }       & {\bf 62 }     & {\bf 92 }            & {\bf 49 }             & {\bf 126}              \\     
       $WW$           & 0.72 $\pm$ 0.19   & 0.18 $\pm$ 0.05        & 0.35 $\pm$ 0.09        & 0.2 $\pm$ 0.05           & 0.49 $\pm$ 0.13      \\
       $ZZ$           & 0.26 $\pm$ 0.04   & 0.03 $\pm$ 0.01           & 0.46 $\pm$ 0.06        & 0.29 $\pm$ 0.04          & 0.63 $\pm$ 0.10     \\  
       $WZ$           & 5.28 $\pm$ 0.75   & 2.6 $\pm$ 0.37         & 2.52 $\pm$ 0.36        & 1.67 $\pm$ 0.24          & 3.52 $\pm$ 0.54      \\
       \bottomrule

    \end{tabular}
    \caption[Final Signal Estimate and Data Selection]{Final result of the event selection for the $W+2$ jets dataset and signal expectation in the different lepton categories. The pretag control region is shown together with single and double-tagged signal regions. The uncertainties from MC statistics, productionn cross section, luminosity, lepton identification, trigger and $HF$-tagging are considered in the table.}\label{tab:data_sig}
    \end{center}
  \end{small}
  \end{table}

\clearpage
\chapter{Background Estimate}\label{chap:bkg}

In this analysis an accurate background evaluation is crucial since a low signal over background ratio is expected and several components contribute to the final $M_{Inv}$ line shape used for the signal extraction. 

Background estimate should be as independent as possible from the signal regions, the single and double-tagged $\ell\nu+2 $~jets samples. One possible approach is the simulation of background events with Monte Carlo generators, another approach is the extrapolation of background components from data control regions. Both methods present advantages and disadvantages, for example a MC simulation is completely under control but can be limited by the knowledge of the underlying physics phenomena parametrization. Data-driven techniques, on the other hand, are based on the assumption that a background in the control region can be simply extrapolated to the signal region. This brings some approximations, eventually covered by systematic uncertainties. 

In our case we use a mixture of MC and data-driven estimates depending on the different background component being addressed.

The complete machinery used for the $\ell\nu+HF$ jets background estimate is named {\em Method II}\footnote{The name derives from a methodology developed for the lepton plus jets $t\bar{t}$ decay channel. Method II appeared after the {\em Method I}, the data-driven background estimate used in the top discovery~\cite{cdf_top_evidence}.} and it was extensively studied at CDF, first on $t\bar{t}$~\cite{top_lj2005} analyses, and later in single-top~\cite{single_top} and $WH$~\cite{wh75_note10596, wh94_note10796} analyses. Together with the MC, two data control samples play a major role in this kind of background estimate: the {\em pretag} control sample, without any $b$-tag requirement, and the fake-$W$ enriched sample, without any multi-jet rejection requirement.

The background evaluation can be divided in four categories: 
\begin{description}
\item[Electroweak and Top:] sometimes dubbed $EWK$, these are backgrounds due to contributions of well known physical processes:  $t\bar{t}$, single-top and $Z+ $~jets  production. For these processes, described in detail in Section~\ref{sec:mc_bkg}, we completely rely on MC simulation for the line-shape and theoretical prediction or previous measurements for the normalization. 
\item[$W+$ Heavy Flavors:] also named $W+HF$, is the part of the $W+$~jets sample where the $W$ is produced in association with $HF$ quarks. It is classified as {\em irreducible} because it has the same final state ($W+c/b$ quarks) of the $diboson\to \ell\nu+HF$ signal. The line shape is derived by a LO $W+n$~partons MC while the normalization is obtained from the {\em pretag} data control region and from MC derived $W+HF$ fractions. Section~\ref{sec:whf} describes in detail the procedure.
\item[$W+$ Light Flavors:] also named {\em mistags} or $W+LF$. These are events where the $W$ is produced in association with one or more Light Flavor ($LF$) jets. This component is drastically reduced by the requirement of one or more $\mathtt{SecVtx}$ tagged jets. Remaining events are due to mis-measured jet tracks and long living $LF$ hadrons. The background evaluation, explained in Section~\ref{sec:mistag}, is obtained using LO $W+ n$~partons MC where appropriate reweighting and normalization are derived from a data control sample of multi-jet events.
\item[Fake-$W$:] sometimes also named multi-jet or non-$W$ background, is composed by events without a real $W$ passing the selection. Those are due to multi-jet QCD production where the jets fake the charged lepton and the \met of the neutrino. Section~\ref{sec:qcd} describes the normalization and line-shape evaluation procedure, both completely data-driven.
\end{description}
The normalization of the four categories proceeds through sequential steps: the $EWK$ contribution is evaluated first, then $W+$~jets normalization is extracted from the pretag control region so that the $W+HF$ fraction can be scaled to the single and double tag signal regions. 

Next step is the estimate, from data, of $W+LF$ contribution and, finally, the {\em tagged} non-$W$ contribution is evaluated before the application of the SVM requirement.

\section{Electroweak and Top Processes}\label{sec:mc_bkg}

\begin{figure}[!ht]
\begin{center}
\includegraphics[width=0.25\textwidth]{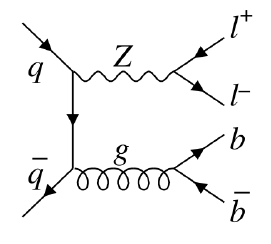}
\includegraphics[width=0.7\textwidth]{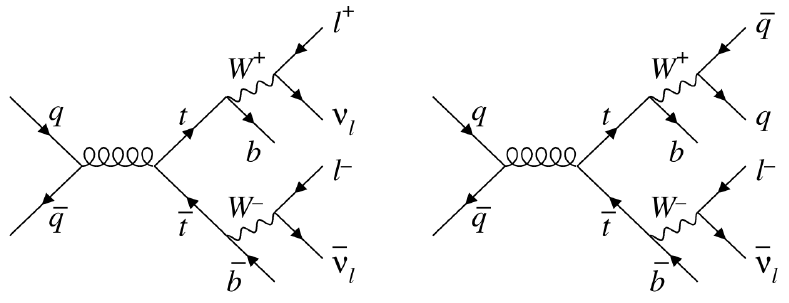}\\
\includegraphics[width=0.7\textwidth]{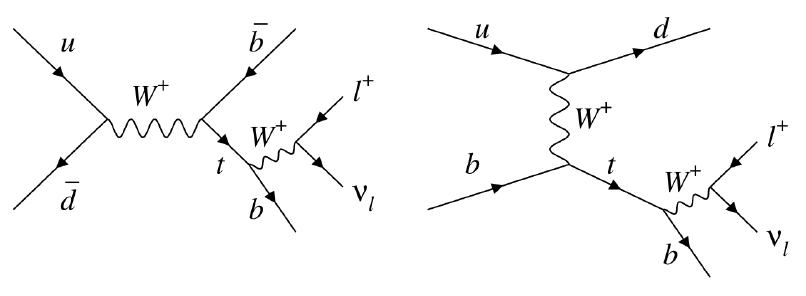}
\caption[Some Tree-Level Diagrams Contributing to the $EWK$ Backgrounds]{Examples of tree-level Feynman diagrams contributing to the $EWK$ background set. They are $Z+$jets production (top left in association with a $b\bar{b}$ quark pair), $t\bar{t}$ production and decay in semi-leptonic (top center) and di-leptonic channels (top right), single-top production in the $s$ (bottom left) and $t$ (bottom right) channels and semi-leptonic decay.}\label{fig:fey_ewk}
\end{center}
\end{figure}

This category contains all the processes whose shape is evaluated directly from MC samples and whose normalization is obtained from theory prediction or experimental measurements. Figure~\ref{fig:fey_ewk} shows some of the tree-level Feynman diagrams contributing to this background set. 

The components of $EWK$ backgrounds are: 
\begin{description}
\item[$Z/\gamma$ + jets:] Drell-Yan and $Z\to \ell\ell$ events can be produced in association with jets emitted by QCD radiation. If one lepton is misidentified, they enter as a background in our selection. The simulation of  $Z/\gamma+n$ partons is available only at LO level (with \texttt{ALPGEN}~\cite{alpgen} Matrix Element generator) therefore the process is evaluated with the same procedure of $W+$~jets background (see Section~\ref{sec:whf}). Luckily, the small acceptance and the possibility to use a cross section measured at CDF~\cite{z_jets_cx} allows us to evaluate the $Z/\gamma+$jets background directly from MC. The $Z+$jets cross section measurement refers only to the on-shell $Z$ contribution, therefore a further scaling factor is derived by comparing the \texttt{ALPGEN} LO predictions for the on-shell and the inclusive $Z/\gamma+$ jets MC sample. 

\item[Top pair:] $t\bar{t}$ production and decay in the lepton plus jets ($\ell\nu+b\bar{b}+2$ jets) and di-lepton ($2\ell+2\nu+b\bar{b}$) channels can enter in our selection if one or more of the final state objects is misidentified. Although the probability is small, the background becomes sizable in the $HF$-tagged sample because of the presence of two $b$-quarks.  
\item[Single-top:] $t$ and $s$ channel production have cross sections comparable to the diboson one and the semi-leptonic decay channel produces the same final state we are interested in. On the other hand, the signal to background discrimination is performed by the $M_{Inv}(jet1,jet2)$ where the single-top has no peak.
\end{description}

The normalization, $N_{l,k}^{EKW}$, of each $EWK$ process is obtained for each lepton category, $l$, in the pretag and $1,2$-Tag regions, $k$, by using Equation~\ref{eq:mc_estimate}.  A summary of the production cross sections and the MC generator information are reported in Table~\ref{tab:mc_ewk}.

  \begin{table}[h] 
    \begin{center}
    \begin{tabular}{cccc}
      \toprule
      Sample & $\sigma$ (pb)& Source & Initial Events \\
      \midrule
      $Z/\gamma+$jets    & 787.4 $\pm$ 85    & Measured~\cite{z_jets_cx} & Several MCs \\ 
      $t\bar{t}$         & 7.04  $\pm$ 0.49  & NNLO~\cite{ttbar_cx} & 6.7 M\\  
       single-top, $s$   & 1.017 $\pm$ 0.065 & NNLO~\cite{stops_kidonakis} & 3.5 M\\ 
       single-top, $t$   & 2.04  $\pm$ 0.18  & NNLO~\cite{stopt_kidonakis} & 3.5 M\\ 
      \bottomrule
    \end{tabular}
    \caption[MC Information for the $EWK$ Backgrounds]{MC information for backgrounds of the $EWK$ category, i.e. the ones estimated from simulation. $Z+$ jets cross section~\cite{z_jets_cx} refers to the on-shell $Z$ contribution and \texttt{ALPGEN} LO MC prediction is used to derive a correction factor. \texttt{HERWIG} plus \texttt{PYTHIA} PS are used for $t\bar{t}$, \texttt{POWHEG} plus \texttt{PYTHIA} PS are used for single-top $s$ and $t$ channels. The generated top mass is $m_t=172.5$\gc2.}\label{tab:mc_ewk}
    \end{center}
  \end{table}

As the three MC derived background processes and the diboson signal are estimated in the same way, the label $EWK$ will be used to indicate generically all of them:
\begin{equation}
  \sum\limits_{EWK} N_{l,k}^{EWK} = N_{l,k}^{Zjets} + N_{l,k}^{t\bar{t}} + N_{l,k}^{s-top,s} + N_{l,k}^{s-top,t} +  \sum\limits_{diboson} N_{l,k}^{diboson}\mathrm{.}
\end{equation}

\section{$W$ plus Heavy Flavor Background}\label{sec:whf}

\begin{figure}[!ht]
\begin{center}
\includegraphics[width=0.9\textwidth]{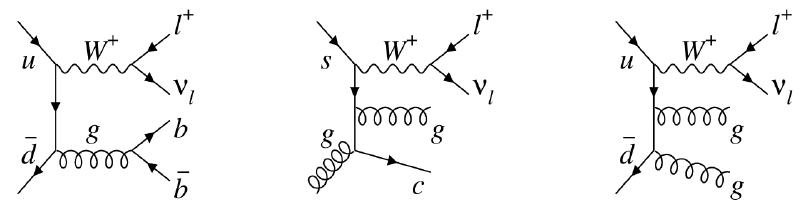}
\caption[Some Tree Level Diagrams Contributing to $W+HF$ and $W+LF$ Backgrounds]{Examples of tree-level Feynman diagrams contributing to the $W+HF$ and the $W+LF$ backgrounds. NLO diagrams also play a relevant role in the $W+$~QCD radiation emission.}\label{fig:fey_wjets}
\end{center}
\end{figure}

The associated $W+HF$ production (Figure~\ref{fig:fey_wjets} shows few examples of tree-level diagram contributions) is the most significant contribution to the total background in the single and double tag signal samples. For its estimate we rely partially on MC and partially on data. 

Normalization and kinematic distributions of the $W+$~jets sample, and of the $W+HF$ sub-sample, present several theoretical difficulties.  NLO MC simulations became available lately for $W+b\bar{b}$ processes~\cite{wbb_powheg} but large higher order contributions enhance the theoretical uncertainty. Furthermore the prediction of the complete $HF$ spectrum ($W+c\bar{c}$, $W+b\bar{b}$, $W+c$) is beyond the current capabilities. 

The two aspects of the problem can be solved by factorizing the evaluation of the kinematic properties from the estimate of the normalization:
\begin{itemize}
\item a solution to estimate the kinematic properties and the composition of $W+$~jets QCD emission was proposed in~\cite{first_lo_match} in 2001: the appropriate matching scheme of a LO calculation with a properly tuned Parton Shower (PS) simulation can reproduce the experimental results. Section~\ref{sec:alpgen_comp} describes in detail our use of the \texttt{ALPGEN} LO generator, interfaced with \texttt{PHYTIA} PS. $W+HF$ relative components or $HF$ fractions, $f^{HF}$, can be extracted with this method.
\item the only free parameter remaining is, now, the $W+$~jets normalization. We derive it completely from data using a maximum likelihood fit on the SVM output distribution: this variable, described in Appendix~\ref{chap:AppSvm}, has a high discriminating power between non-$W$ and real $W$ events. Therefore, as all the other $EWK$ components are known, both $N^{nonW}_{pretag}$ and $N^{Wjets}_{pretag}$  can be extracted from the fit. Section~\ref{sec:wjets_pretag} describes the procedure.
\end{itemize}

\subsection{$W +$ Jets Spectrum Composition}\label{sec:alpgen_comp}

$W+$~jets composition and kinematic properties are evaluated from MC. We employ a large set of $W+n$~partons \texttt{ALPGEN}~\cite{alpgen} LO MC that includes $HF$ production in the ME calculation and is interfaced with \verb'PYTHIA'~\cite{pythia} PS:
\begin{itemize}
\item[-]$W$ plus $0$, $1$, $2$, $3$, $\ge 4$ partons (for $W+ n$~generic jets description);
\item[-]$W+b\bar{b}$ plus $0$, $1$, $2$ partons;
\item[-]$W+c\bar{c}$ plus $0$, $1$, $2$ partons;
\item[-]$W+c$ (or $W+\bar{c}$) plus $0$, $1$, $2$, $3$ partons.
\end{itemize}
In total we used ninety samples: fifteen to be multiplied by the three leptonic $W$ decay modes ($W\to e,\mu,\tau +\nu_e,\nu_{\mu}\nu_{\tau}$) and by another factor two for {\em low luminosity} and {\em high luminosity} running periods. Table~\ref{tab:samples} summarizes the total of ninety samples used along with the respective LO cross sections and the approximate number of events initially generated.
\begin{table}\begin{center}
\begin{tabular}{ccc}
\toprule
Sample& $\sigma$ (pb)& Initial Events\\
\midrule
$W+0p$& 1810& 7 M\\
$W+1p$& 225 & 7 M\\
$W+2p$& 35.3& 1.4 M\\
$W+3p$& 5.59& 1.4 M\\
$W+4p$& 1.03& 0.5 M\\
$W+b\bar{b}+0p$& 2.98& 2.1 M\\
$W+b\bar{b}+1p$& 0.888& 2.1 M\\
$W+b\bar{b}+2p$& 0.287& 2.1 M\\
$W+c\bar{c}+0p$& 5& 2.8 M\\
$W+c\bar{c}+1p$& 1.79& 2.8 M\\
$W+c\bar{c}+2p$& 0.628& 2.8 M\\
$W+c+0p$& 17.1& 2.8 M\\
$W+c+1p$& 3.39& 2.8 M\\
$W+c+2p$& 0.507& 2.8 M\\
$W+c+3p$& 0.083& 2.8 M\\
\bottomrule
\end{tabular}
\caption[ALPGEN $W + $ Partons MC Samples]{\texttt{ALPGEN} LO plus \texttt{PYTHIA} (PS) MC samples used to estimate the various contributions to the $W+HF$ and $W+LF$ backgrounds. The second column lists the LO cross section of the samples, the third column shows the approximate number of generated events. Different MC samples are used for the three lepton flavors ($e, \mu, \tau$) of the $W$ decay.}\label{tab:samples}
\end{center}\end{table}

The basic composition algorithm is simple: the acceptance of each sample, for a given jet multiplicity and for each lepton category, is weighted by its own LO production cross section, then a normalization factor equal to the sum of all the samples is applied. The key idea is that, if the PS matching is well tuned, the higher order corrections to the LO $W+n$~partons simulation, should simplify when the normalization factor is applied. 

The procedure is more complex in our case because we want an accurate estimate of the $W+HF$ spectrum: we need to derive the $HF$ fractions, $f^{HF}$, from the $W+n$~partons samples and relate them to the physical observables of $W+b\bar{b}/c\bar{c}/c$ jets with one or two \texttt{SecVtx} $HF$-tags.

The $W+HF$ production is evaluated explicitly at ME level in some of the $W+n$ partons samples, however a certain amount of $HF$ quarks is also produced by the PS algorithm in all the samples. Some care must be used when combining all the samples because the double counting of a process will degrades the relative weights.

\begin{figure}[!ht]
\begin{center}
\includegraphics[width=0.495\textwidth]{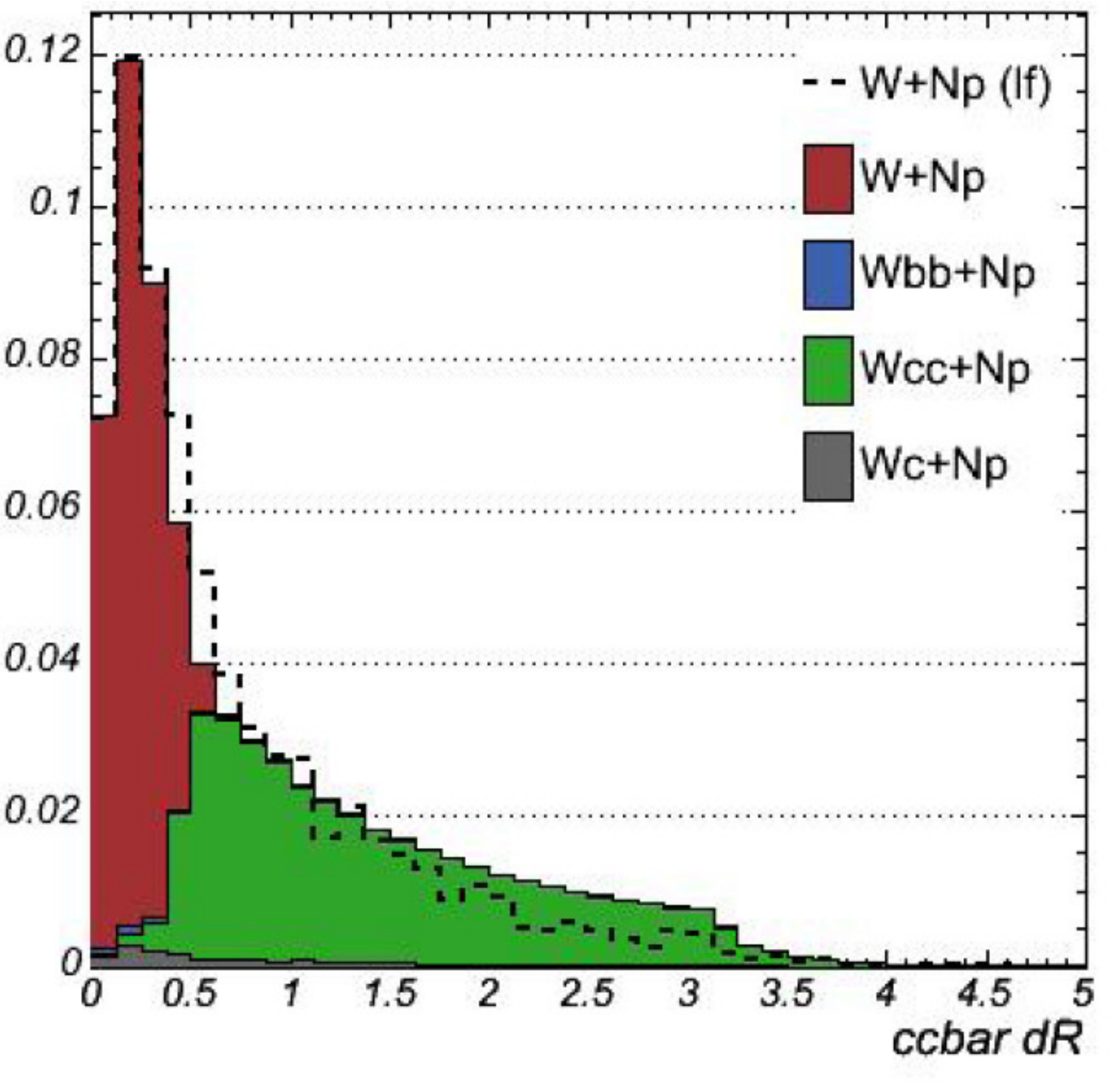}
\includegraphics[width=0.495\textwidth]{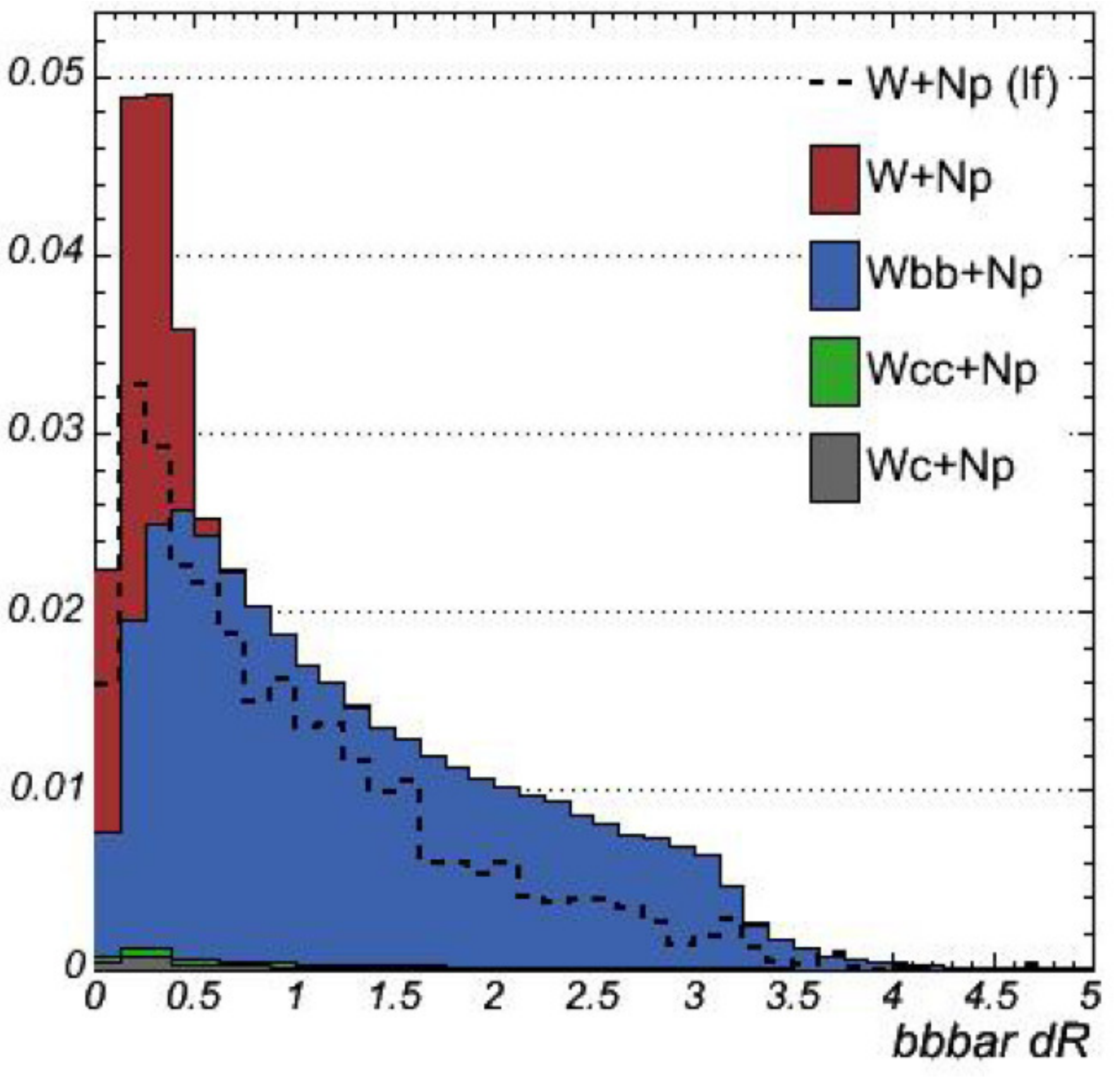}
\caption[Generator Level $\Delta R$ Distributions of $b\bar{b}$ and $c\bar{c}$ ]{Generator level $\Delta R$ distributions of $b\bar{b}$ (left) and $c\bar{c}$ (right) pair production after the composition of the $W+n$~partons samples and the overlap removal procedure~\cite{hf_8765}. The $\Delta R $ from PS-only is also reported showing disagreement at large $\Delta R $ w.r.t. the $HF$ generated with ME.}\label{fig:overlap_rem_spectrum}
\end{center}
\end{figure}

The {\em overlap removal} technique used in this work was previously developed by the CDF collaboration and it is documented in~\cite{hf_8765}. It consists in the selection of $HF$ production from the PS {\em or} from the ME on the base of detector level reconstructed jets. We explicitly veto events where the ME heavy flavor quarks wind up in the same jet, and we also remove events where the $HF$ quark pair from the shower is divided in two jets. In both cases, the distinction is made with simple kinematic cuts that define a parton to be {\em in a jet} if:
\begin{equation}
  \Delta R(parton, jet)<0.4\quad\textrm{and}\quad E_T^{jet,cor}> 12~\textrm{GeV}
\end{equation}
The reason for this choice is that PS and ME predict different $\Delta R$ (separation) between heavy quarks: PS model is tuned on the more collinear gluon splitting quark pairs but fails in the limit of large opening angles, while ME generation works better in latter regime. Figure~\ref{fig:overlap_rem_spectrum} shows the generator level $\Delta R$ distribution of $b\bar{b}$ and $c\bar{c}$ pair production after the composition of the $W+n$~partons samples and the overlap removal procedure. The $\Delta R $ from PS-only is also reported in Figure~\ref{fig:overlap_rem_spectrum} showing disagreement at large $\Delta R $ with respect to the $HF$ generated with ME. The smooth transition between the different  $W+n$~partons contributions indicates that the overlap removal scheme is working properly.

The complete procedure for the determination of the $f^{HF}$'s is the following:
\begin{itemize}
\item each MC sample is assigned a weight:
  \begin{equation}\label{eq:weight}
    w^{m}=\frac{\sigma^m}{N_{z_0}^m} \mathrm{,}
  \end{equation}
  where $m$ varies on all the different samples of Table~\ref{tab:samples}, $\sigma^m$ is the cross section and $N_{z_0}^m$ is the total number of generated events with $|z_0^{vtx}|<60$~cm.
\item After pretag selection, the denominators of the $HF$ fractions are calculated as:
  \begin{equation}
    Den=\sum_{m=1}^{samples} w^{m}N^m\mathrm{,}
  \end{equation}
  where $N^m$ are the {\em unique} events (i.e. not vetoed by the overlap removal) falling in each jet-bin for sample $m$.
\item Each selected jet is classified as $HF$ jet or not: $HF$ jets have a $b$ or $c$ parton (at generation level) {\em match}, i.e. laying inside the cone (\mbox{$\Delta R(jet,parton)<0.4$}). We distinguish four $HF$ categories depending on the number of matched jets and on the kind of $HF$ parton ($b$ or $c$). The categories are: $1B$, $2B$, $1C$ and $2C$ depending if we have 1 or $\ge 2$, $b$ or $c$ quarks matched; no $b$ quarks should be matched in the case of $1,2C$ categories.
\item The numerators are defined by the sum of the events in each $HF$ category (with weights given by Equation~\ref{eq:weight}) over all the samples, in each jet-bin:
  \begin{equation}
    Num^{HF}=\sum_{m=1}^{samples}w^m N^{m,HF}\mathrm{.}
  \end{equation}
\end{itemize}
The computation of the $f^{HF}$'s is now straightforward. Table~\ref{tab:hf_frac} summarizes the final $HF$ fractions averaged on all the lepton categories.
\begin{table}\begin{center}
\begin{tabular}{cccc}
\toprule
\multicolumn{4}{c}{Heavy Flavor Fractions}\\
\midrule
$f^{1B}$ & $f^{2B}$ & $f^{1C}$ & $f^{2C}$ \\
$0.024$ & $ 0.014$  & $ 0.114$  & $ 0.022$ \\ 
\bottomrule
\end{tabular}
\caption[$HF$ fractions, $f^{HF}$, for Two Jets Selection]{$HF$ fractions, $f^{HF}$, for the two jets selection derived from \texttt{ALPGEN} MC composition. They are classified in the different categories. $1B$: events with $1$ $b$-jet. $2B$: events with $\ge 2$ $b$-jet. $1C$: events with $1$ $c$-jet and no $b$-jet. $2C$: events with $\ge 2$ $c$-jet and no $b$-jet.}\label{tab:hf_frac}
\end{center}\end{table}

However $f^{HF}$'s  are not immediately usable to predict $W+HF$ background. They need an additional correction, $K$, to account for different $HF$ composition in data and in the LO ME simulation:
\begin{equation}\label{eq:k}
K^{HF}=\frac{f^{HF}_{data}}{f^{HF}_{MC}}\mathrm{,}
\end{equation}
The $HF$ calibration is carried out on a data sample not used in this measurement: we base our correction on the studies performed in~\cite{single_top}. There, the $W+1$~jet data sample is used to derive a correction of \mbox{$K^{c\bar{c}/b\bar{b}}=1.4$} to the $W+c\bar{c}/b\bar{b}$ processes, while no correction ($K^{c}=1.0$) is quoted for the $W+c$ process\footnote{$W+c$ production is an electroweak production process measured at CDF in~\cite{cdf_wc}.}. A large uncertainty of $30$\% is associated to all these $K$ factors as the correction is extrapolated from the $W+1$~jet data sample to the $W\ge 2$ jets data sample. A previous study~\cite{k_8768}, on a multi-jet control sample, obtained a $K$ closer to $1$ and it is consistent within the uncertainty. Table~\ref{tab:k-fac} shows a summary of the applied $K$ corrections.
\begin{table}\begin{center}
\begin{tabular}{ccc}
\toprule
\multicolumn{3}{c}{Heavy Flavor $K-$Factor Corrections}\\
\midrule
$K^{W+b\bar{b}}$ & $K^{W+c\bar{c}}$ & $K^{W+c}$ \\
$1.4\pm 0.42$ & $1.4\pm 0.42$ & $1.0\pm 0.33$ \\ 
\bottomrule
\end{tabular}
\caption[$HF$ $K-$Factor Correction]{$K-$Factor corrections needed for the calibration of the $HF$ production in data. Studies performed in~\cite{single_top}.}\label{tab:k-fac}
\end{center}\end{table}

The final estimate of $f^{HF}$ is:
\begin{equation}
f^{HF}=K\frac{Num^{HF}}{Den}\mathrm{.}
\end{equation}

The $HF$ fractions can be used in the construction of the pretag sample shapes but are of marginal importance because of the overwhelming $LF$ contribution.
On the other hand $HF$ are fundamental in the tagged sample composition. To evaluate it we need the tagging efficiency of each $HF$ category ($\epsilon^{HF}$): they are derived from the MC samples requiring a \verb'SecVtx' tag, then we apply the $\omega_{k-Tag}$ correction as described in Section~\ref{sec:jetSel}. Tagging efficiencies for each $HF$ category are summarized in Table~\ref{tab:hf_tag_ef}.
\begin{table}\begin{center}
\begin{tabular}{ccccc}
\toprule
\multicolumn{5}{c}{Heavy Flavor Tagging Efficiency}\\
\midrule
& $\epsilon^{1B}$ & $\epsilon^{2B}$ & $\epsilon^{1C}$ & $\epsilon^{2C}$ \\
1-Tag & $ 22$\%  & $ 30$\% & $6.7$\%  & $ 9.1$\% \\ 
2-Tag & $ 0.4$\%  & $8.6$ & $ 0.06$\%  & $ 0.4$\% \\ 
\bottomrule
\end{tabular}
\caption[Heavy Flavor Tagging Efficiency $\epsilon^{HF}$ ]{Heavy Flavor tagging efficiency $\epsilon^{HF}$ in each category. $1B$: events with $1$ $b$-jet. $2B$: events with $\ge 2$ $b$-jet. $1C$: events with $1$ $c$-jet. $2C$: events with $\ge 2$ $c$-jet.}\label{tab:hf_tag_ef}
  \end{center}
\end{table}

Finally we can combine all the pieces to obtain the {\em final} $HF$ tagged contribution. For each lepton $l$ and tag category $k$, it is:
\begin{equation}\label{eq:tag_hf}
N^{HF}_{l,k}=N^{Wjets}_{l,k}\cdot f^{HF} K^{HF}\epsilon^{HF}_{k}\mathrm{,}
\end{equation}
Where the $HF$ label distinguishes the number of jet-matched heavy flavor hadrons: $1,2B$, $1,2C$. The $1,2B$ fractions represent $W+b\bar{b}$, $2C$ represents $W+c\bar{c}$ prediction and $1C$ is related to $W+c$ production. The extraction of the $W+$~jets pretag normalization ($N^{Wjets}$) is discussed in the next Section.

\subsection{Pretag $W +$ Jets Normalization Estimate}\label{sec:wjets_pretag}

According to Equation~\ref{eq:tag_hf}, the $W+$~jet normalization, $N^{Wjets}$,  is the only missing piece in the complete $W+HF$ estimate.  $N^{Wjets}$ is obtained solving:
\begin{equation}\label{eq:pretag_comp}
N^{Data}_{Pretag} = F^{nonW}_{Pretag} \cdot N^{Data}_{Pretag} + F^{Wjets}_{Pretag} \cdot N^{Data}_{Pretag} + \sum_{EWK}N^{EWK}_{Pretag} \mathrm{,}
\end{equation}
where:
\begin{itemize}
\item $N^{EWK}$ is derived from Section~\ref{sec:mc_bkg};
\item the fractions $F^{nonW}_{Pretag}$ and $F^{Wjets}_{Pretag}$ are unknown parameters obtained by a maximum likelihood fit in the pretag region. This provides a $W+$~jets estimate unbiased by the $b$-tag requirement;
\item the SVM output distribution is used in the fit before the application of the multi-jet rejection cut (Section~\ref{sec:svm_sel}) so that the fake-$W$ fraction of the sample can be extrapolated from the sideband. 
\end{itemize}
We build a binned likelihood function on the SVM distribution on the base of Equation~\ref{eq:pretag_comp}. Events are Poisson distributed and a Gaussian constraint is imposed to the  $N^{EWK}$ normalizations while $F^{nonW}_{Pretag}$ and $F^{Wjets}_{Pretag}$ are left free to vary.

Templates used in the fit are derived in different ways according to the background source: $EWK$ and $diboson$ SVM distributions are derived from MC, $W+$~jets template is derived from the $W+n$~partons MC composition described in previous Section. Non-$W$ template is obtained in data by reversing appropriate lepton identification cuts, a detailed description is given in Section~\ref{sec:qcd}.
Templates are built for each lepton category but we decided to combine all the EMC algorithms in one template to reduce statistical fluctuations exploiting the homogeneous behaviour of the different components of EMC.

The validation of the SVM output distribution is another important step of the fit procedure. Most of the previous analyses~\cite{single_top, wh94_note10796} used a fit to the \met distribution to derive the $W+$ jets fraction. Here a slightly different strategy is used as the SVM was trained to discriminate the $W+$~jets sample against multi-jet events. This physics interpretation was checked during the training process (see Appendix~\ref{chap:AppSvm}). A simplified version of the fit (slightly different selection and composed only by non-$W$ and $W+$~jets samples) was performed at each training step to reject unreliable configurations. We based the rejection on two figures of merit: 
\begin{itemize}
\item the $\chi^2$ evaluation of post fit template/data shapes; 
\item the comparison of the non-$W$ and $W+$~jets expectation and fit estimate.
\end{itemize}

\begin{figure}[!ht]
\begin{center}
\includegraphics[width=0.495\textwidth]{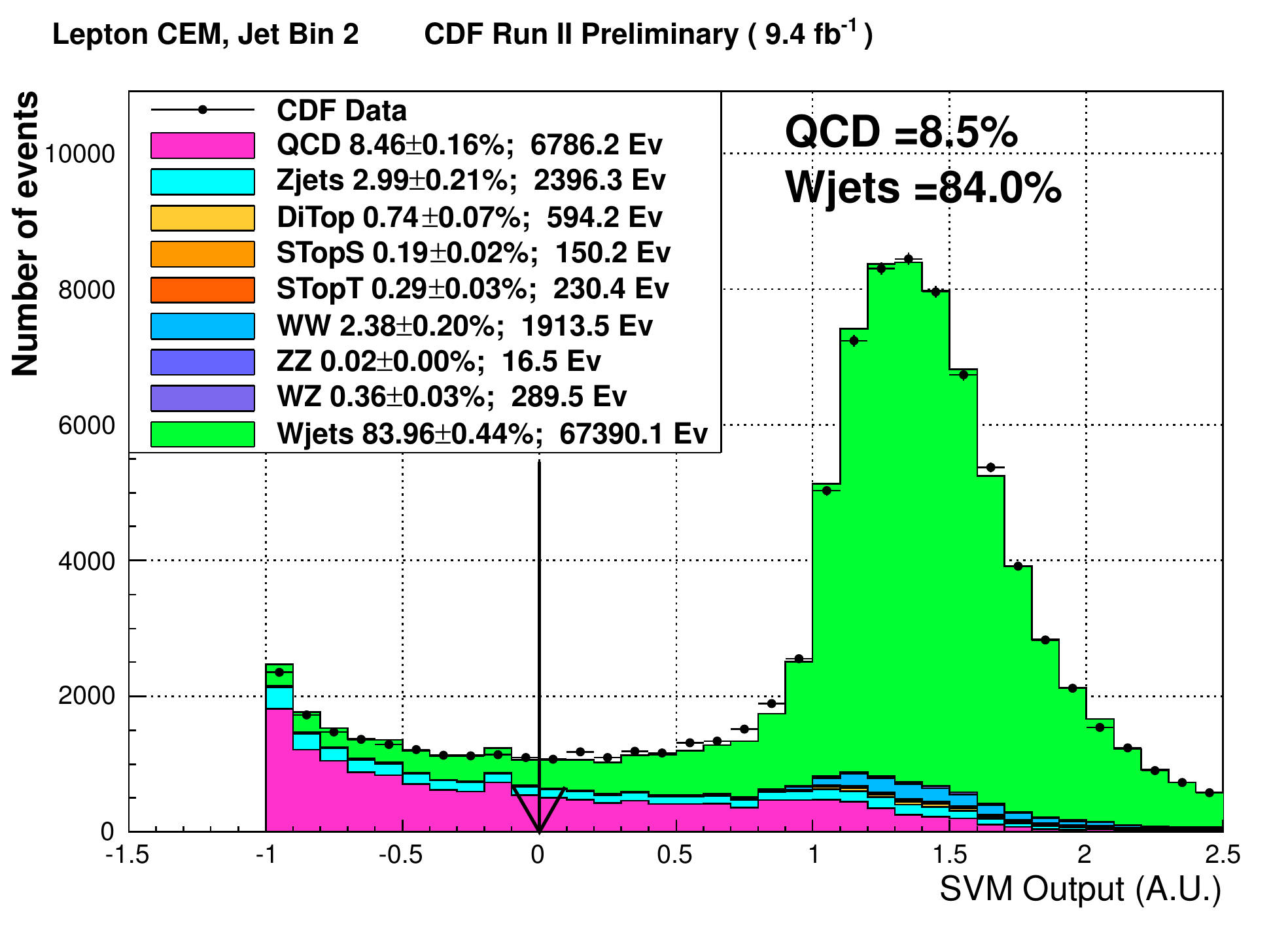}
\includegraphics[width=0.495\textwidth]{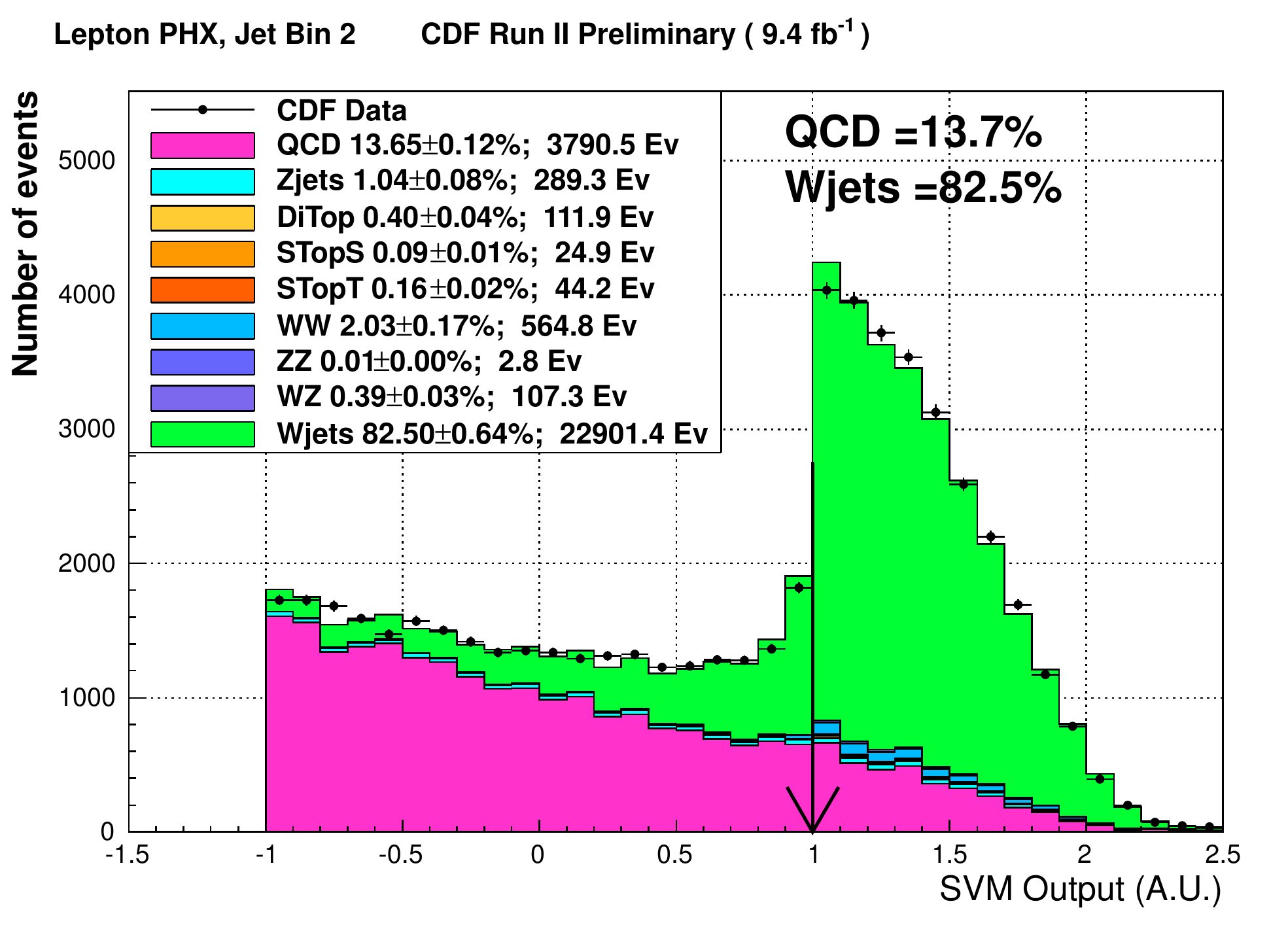}\\
\includegraphics[width=0.495\textwidth]{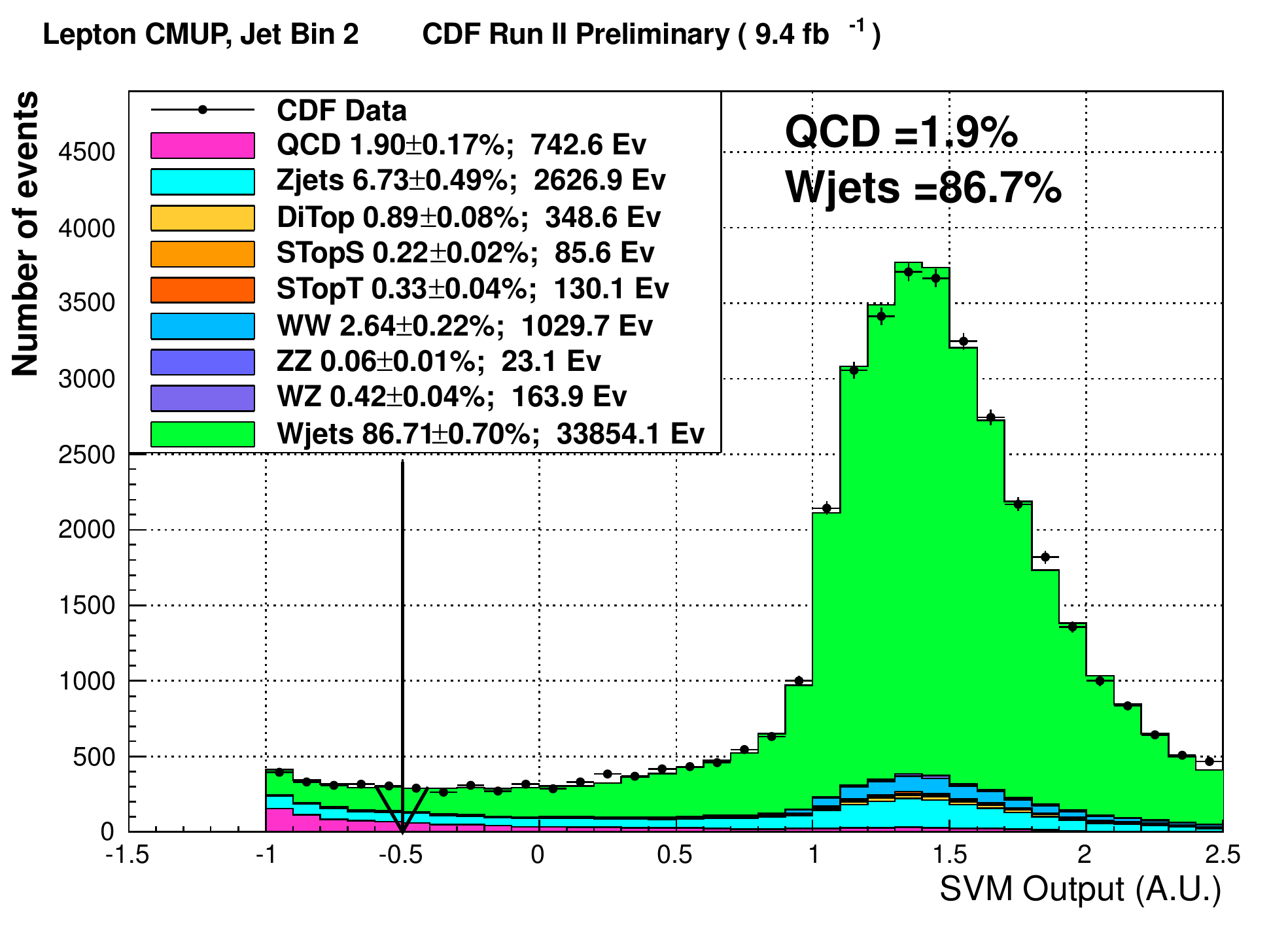}
\includegraphics[width=0.495\textwidth]{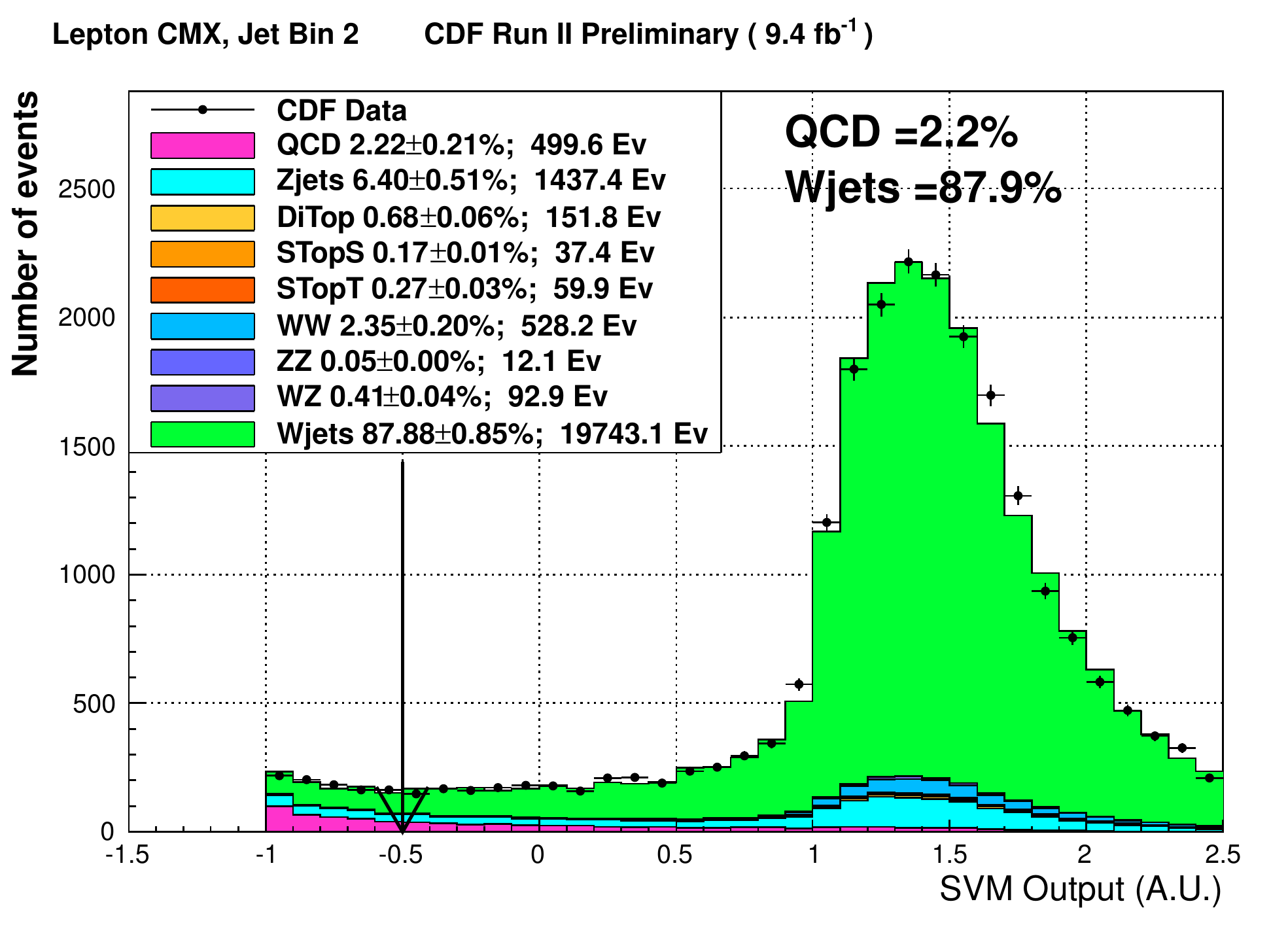}\\
\includegraphics[width=0.495\textwidth]{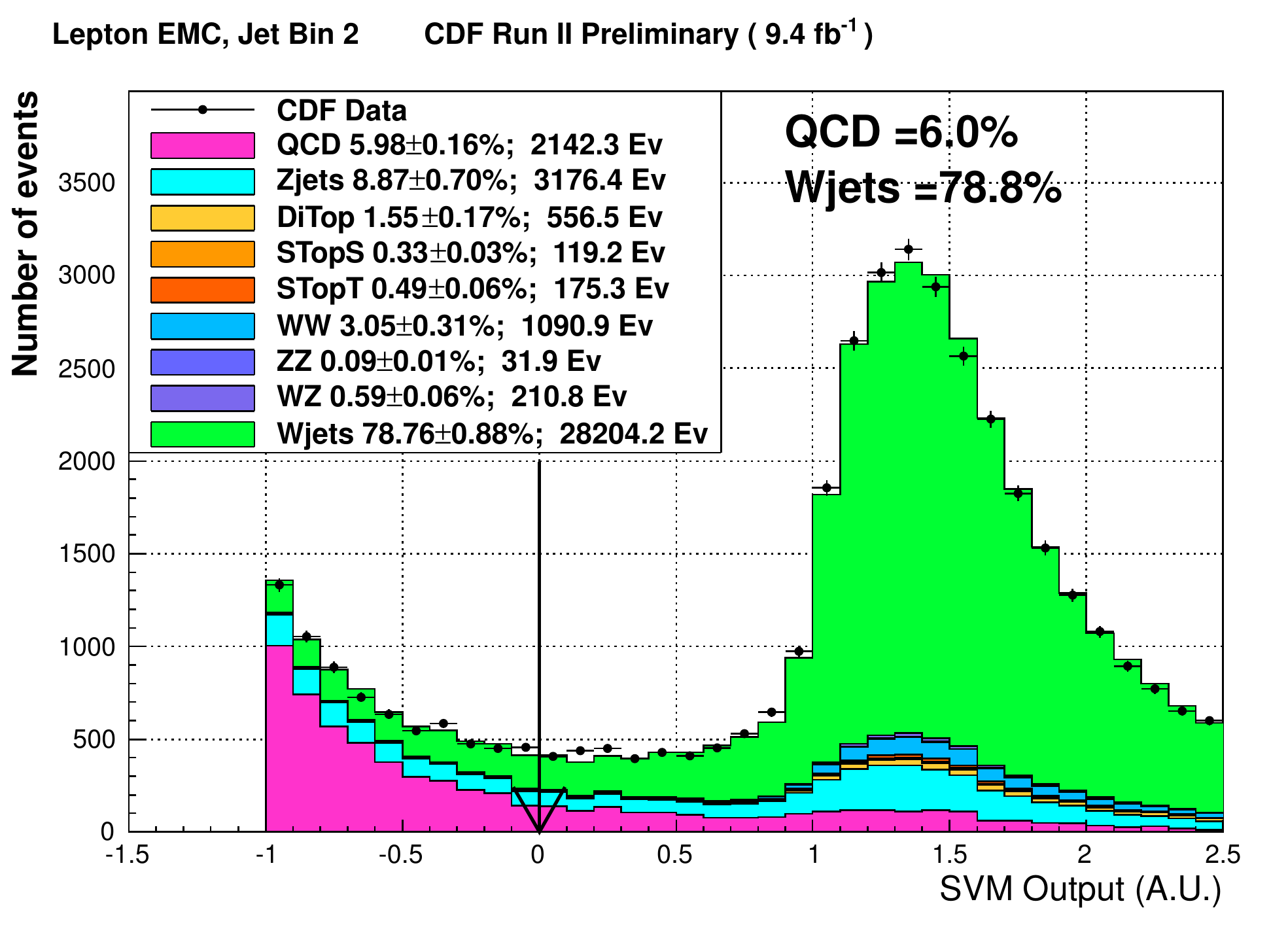}
\caption[$W +$ Jets and Non-$W$ Pretag Sample Estimate]{Background templates composition with proportions returned by the maximization likelihood fit on the SVM pretag distribution for the different lepton categories: CEM (top left), PHX (top right), CMUP (center left), CMX (center right), EMC (bottom). An arrow indicates the selection cut value in each lepton category.}\label{fig:pretag_fit}
\end{center}
\end{figure}

The final step is the actual fit evaluation for the CEM, PHX, CMUP, CMX and EMC lepton categories. Figure~\ref{fig:pretag_fit} shows the result: the template composition with the fractions returned by the maximization likelihood fit. Table~\ref{tab:wjet_pretag_qcd_fit} summarizes $F^{Wjets}_{Pretag}$ and $F^{nonW}_{Pretag}$ extracted from the fit. 

The pretag non-$W$ fraction, $F^{nonW}_{Pretag}$, is not used any more in the background estimate but the total non-$W$ pretag normalization, $N^{nonW}_{Pretag}$, is needed to cross check the kinematic of the pretag control sample. We derive it with the following equation:
\begin{equation}
  N^{nonW}_{Pretag} = F^{nonW}_{Pretag} \cdot N^{Data}_{Pretag} + \sigma^{EWK}\textrm{;}
\end{equation}
where $\sigma^{EWK}$ is the change in the normalization of the $EWK$ backgrounds as returned by the Likelihood fit.

\begin{table}\begin{center}
\begin{tabular}{cccccc}
\toprule
 Lepton & CEM & PHX & CMUP & CMX &  EMC \\\midrule
$F^{nonW}_{Pretag}$ & $ 8.5 \pm 0.2$\% &  $ 13.6\pm 0.1 $\% &  $ 1.9\pm 0.2 $\% &  $2.2 \pm 0.2 $\% & $5.9\pm 0.2$\%\\
$F^{Wjets}_{Pretag}$ &  $ 84.0 \pm 0.4$\% &  $ 82.5 \pm 0.6$\% &  $86.7 \pm 0.7 $\% &  $ 87.9 \pm 0.9$\% &  $ 78.7\pm 0.9 $\%\\
\bottomrule
\end{tabular}
\caption[$W +$Jets and Non-$W$ Pretag Sample Compositions]{$W+$ jets and non-$W$ fractional compositions of the pretag selection sample estimated by a maximum likelihood fit on the SVM output distributions reported in Figure~\ref{fig:pretag_fit}. The statistical error of the fit is reported.}\label{tab:wjet_pretag_qcd_fit}
\end{center}\end{table}

\subsection{$W +$ Jets Scale Uncertainty}
The simulation of the $W+$~jets sample should take into account one more effect, the uncertainty on the factorization and renormalization scale, $Q^2$, of the ALPGEN event generator. 

The $Q^2$ of the generator can be seen as the momentum scale of the hard interaction. Technically this value is used in two contexts:
\begin{itemize}
\item in the evaluation of the PDFs of the hard interaction, the {\em factorization scale};
\item in the calculation of the ME perturbative expansion of a given QCD process, the {\em renormalization scale};
\end{itemize}
A reasonable choice for it, in the $W+n$~partons MC, is given by the following equation: 
\begin{equation}\label{eq:scale}
  Q^2 = M^2_W + \sum^{partons} p_T^2  
\end{equation}
where $M_W$ is the $W$ boson mass, the sum extends over all the partons and $p_T$ is their transverse energy.

However the $Q^2$ is not a physical observable, as it is an artifact of the perturbative approximation needed to solve QCD problems, therefore an appropriate uncertainty should be taken into account. 
Table~\ref{tab:q2samples} shows the characteristics of two new sets of $W+n$~partons MC samples used for systematic variation and generated with the $Q^2$ parameter doubled and halved.
\begin{table}\begin{center}
\begin{tabular}{lcccc}
\toprule
&\multicolumn{2}{c}{\bf $\mathbf{Q^2=2.0}$ (Up)}& \multicolumn{2}{c}{\bf $\mathbf{Q^2=0.5}$ Down}\\
Sample& $\sigma$ (pb)& Initial Events &  $\sigma$ (pb)& Initial Events\\
\midrule
$W+0p$        & 1767  & 3.5 M & 1912  & 3.5 M\\
$W+1p$        & 182.9 & 3.5 M & 303   & 3.5 M\\
$W+2p$        & 24.6  & 0.7 M & 57.6  & 0.7 M\\
$W+3p$        & 3.36  & 0.7 M & 10.8  & 0.7 M\\
$W+4p$        & 0.54  & 0.7 M & 2.24  & 0.7 M\\
$W+b\bar{b}+0p$& 2.30  & 0.7 M & 4.14  & 0.7 M\\
$W+b\bar{b}+1p$& 0.550 & 0.7 M & 1.64  & 0.7 M\\
$W+b\bar{b}+2p$& 0.152 & 0.7 M & 0.615 & 0.7 M\\
$W+c\bar{c}+0p$& 3.89  & 1.4 M & 6.88  & 1.4 M\\
$W+c\bar{c}+1p$& 1.08  & 1.4 M & 3.18  & 1.4 M\\
$W+c\bar{c}+2p$& 0.323 & 1.4 M & 1.30  & 1.4 M\\
$W+c+0p$       & 13.8  & 1.4 M & 23.3  & 1.4 M\\
$W+c+1p$       & 2.30  & 1.4 M & 5.72  & 1.4 M\\
$W+c+2p$       & 0.294 & 1.4 M & 1.03  & 1.4 M\\
$W+c+3p$       & 0.042 & 1.4 M & 0.189 & 1.4 M\\
\bottomrule
\end{tabular}
\caption[ALPGEN $W + $Partons MC Samples for $Q^2$ Systematics]{\texttt{ALPGEN} LO plus \texttt{PYTHIA} (PS) MC samples used to estimate the $Q^2$ systematic variation of the $W+HF$ and $W+LF$ backgrounds. The $Q^2$ is halved and doubled with respect to the default value given by Equation~\ref{eq:scale}. Different MC samples are used for the three lepton flavors ($e, \mu, \tau$) of the $W$ decay.}\label{tab:q2samples}
\end{center}\end{table}

\begin{figure}[!ht]
\begin{center}
\includegraphics[width=0.495\textwidth]{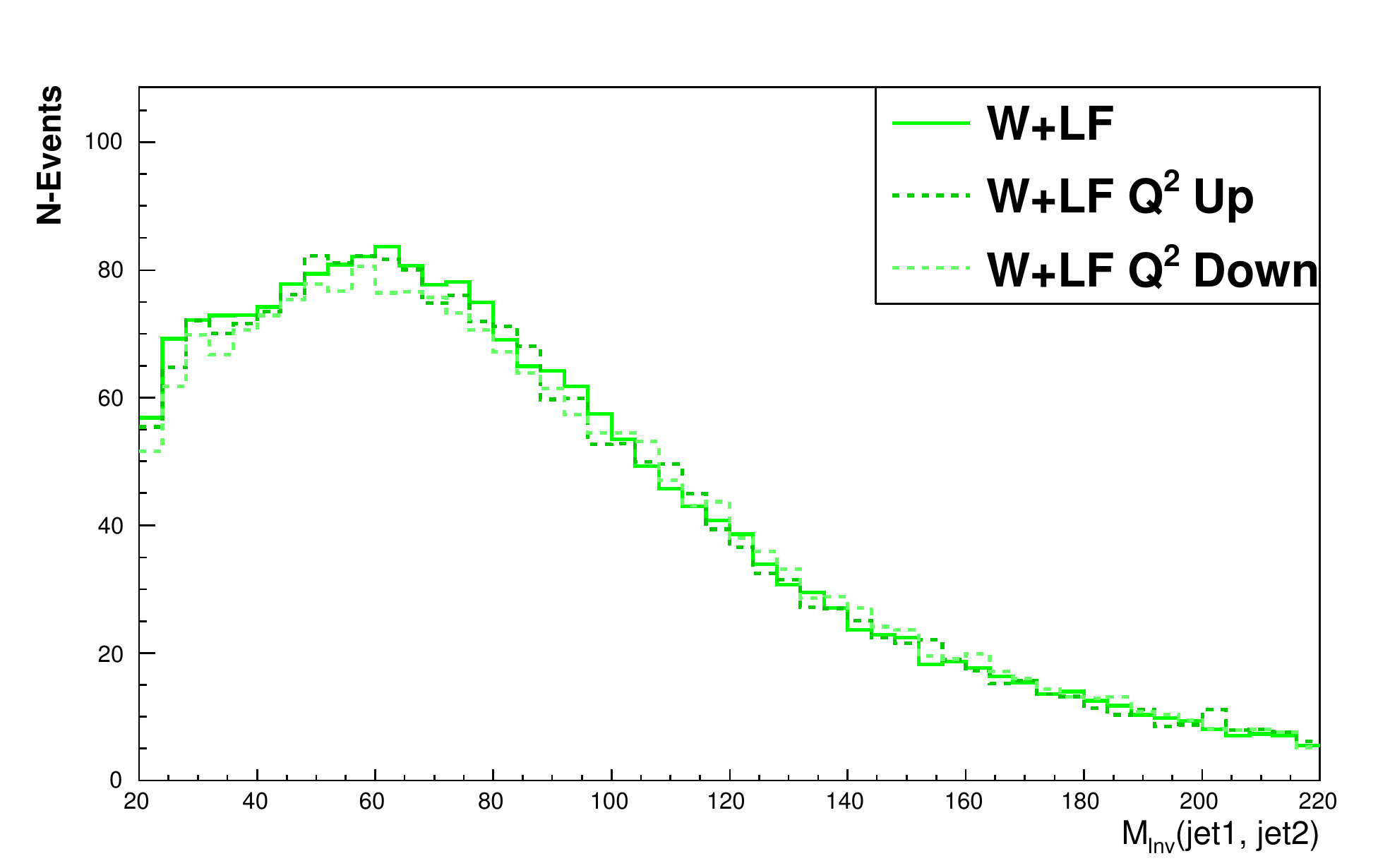}
\includegraphics[width=0.495\textwidth]{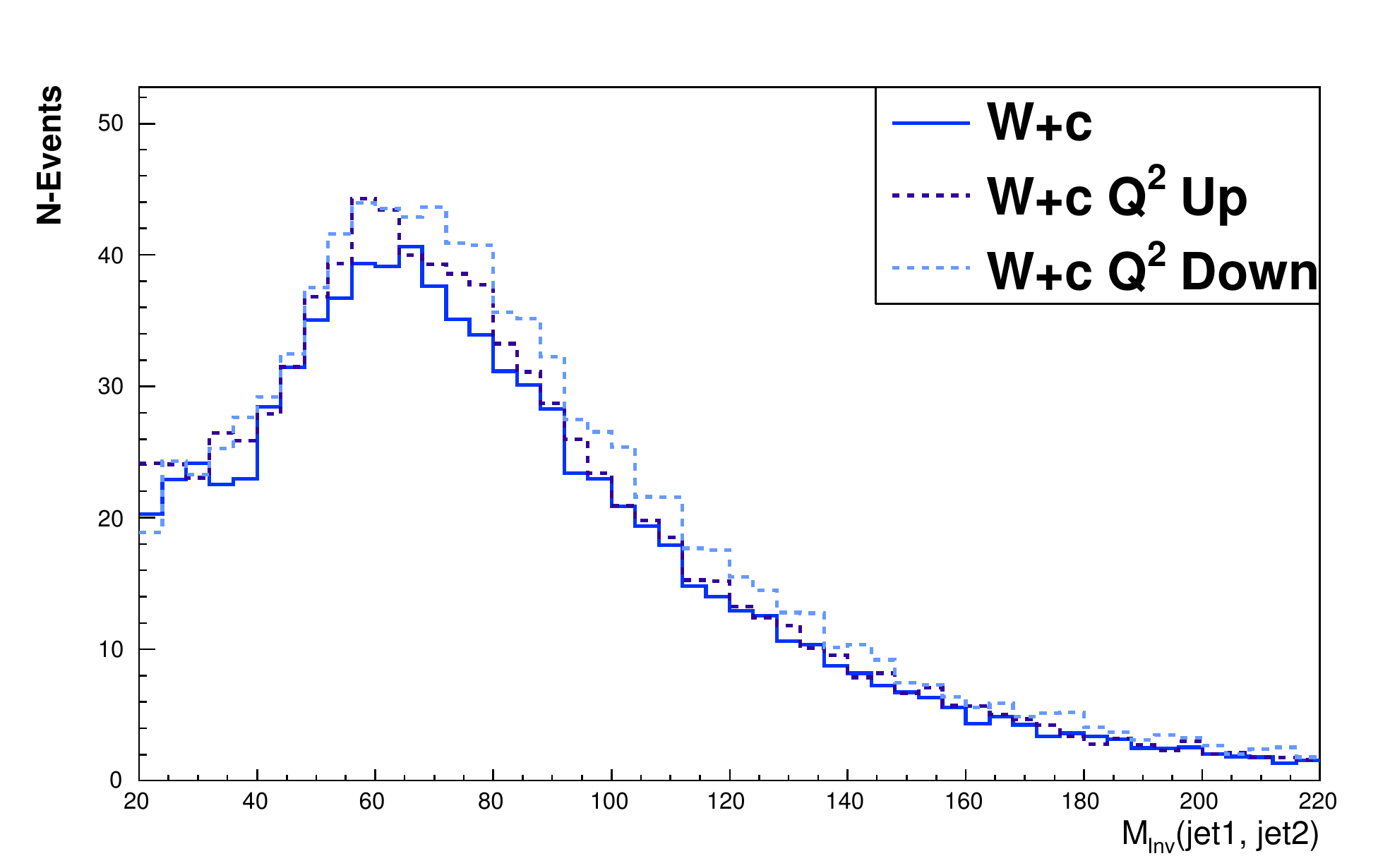}\\
\includegraphics[width=0.495\textwidth]{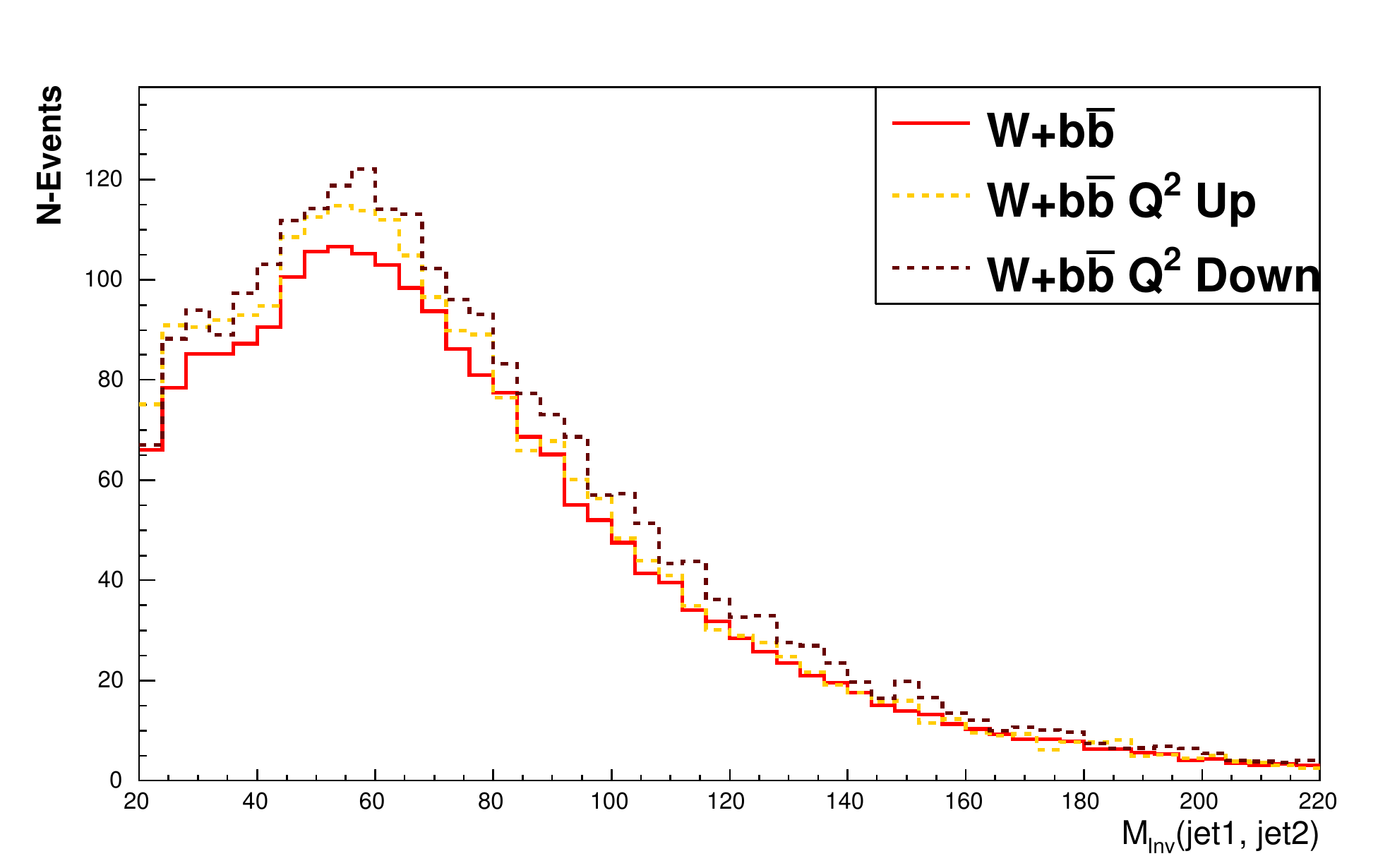}
\includegraphics[width=0.495\textwidth]{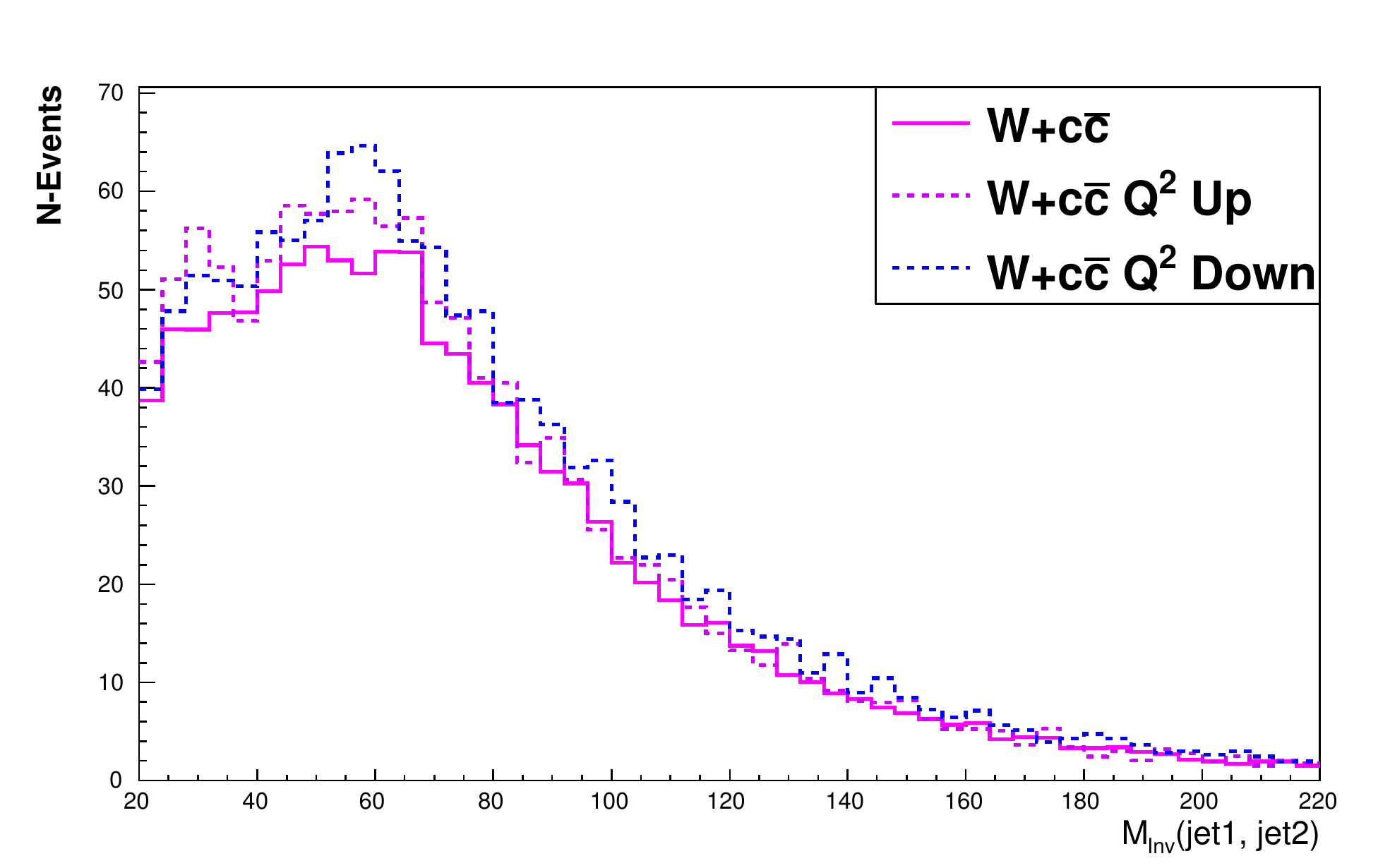}
\caption[$M_{Inv}(jet1,jet2)$ Distribution and $Q^2$ Variations for $W + ~LF/c/c\bar{c}/b\bar{b}$ Samples]{$M_{Inv}(jet1,jet2)$ distribution for single \texttt{SecVtx} tagged events, after combination of all the lepton categories: the default estimate of the four $W+$~jets samples is marked by the continuous line while the $Q^2$ systematic variations are marked by the dashed lines. Top left: $W+LF$. Top right: $W+c$. Bottom left: $W+c\bar{c}$. Bottom right: $W+b\bar{b}$.}\label{fig:wjets_mjj_q2}
\end{center}
\end{figure}

The $Q^2$ variation affects both the kinematic and the parton composition of the $W+$~jets sample, therefore, for each $Q^2$ systematic variation, the background estimate was redone as described Section~\ref{sec:wjets_pretag}. The results are different shapes and rates for the $W+$~jets components. Figure~\ref{fig:wjets_mjj_q2} shows the effect on the single-tag $M_{Inv}(jet1,jet2)$ distributions.

\section{$W$ plus Light Flavors}\label{sec:mistag}

The $W+$~jets events originating from a $LF$ quark can produce a secondary vertex for several reasons. Long living $LF$ hadrons produce a small amount of real \verb'SecVtx' tags while false \verb'SecVtx' tags are due to track reconstuction errors or interaction with the detector material and the beam pipe.

The $LF$ pretag fraction ($f_j^{LF}$) is derived from the $W+n$~partons MC composition, in the same way as the $HF$ fraction. However after requiring a $b$-tag, only a very small fraction of $LF$ jets remains in the sample. As the effects that generate mistags are not adequately simulated, appropriate parametrization is obtained studying the {\em mistag} behaviour of the \verb'SecVtx' algorithm. 

\begin{figure}[!ht]
\begin{center}
\includegraphics[width=0.495\textwidth]{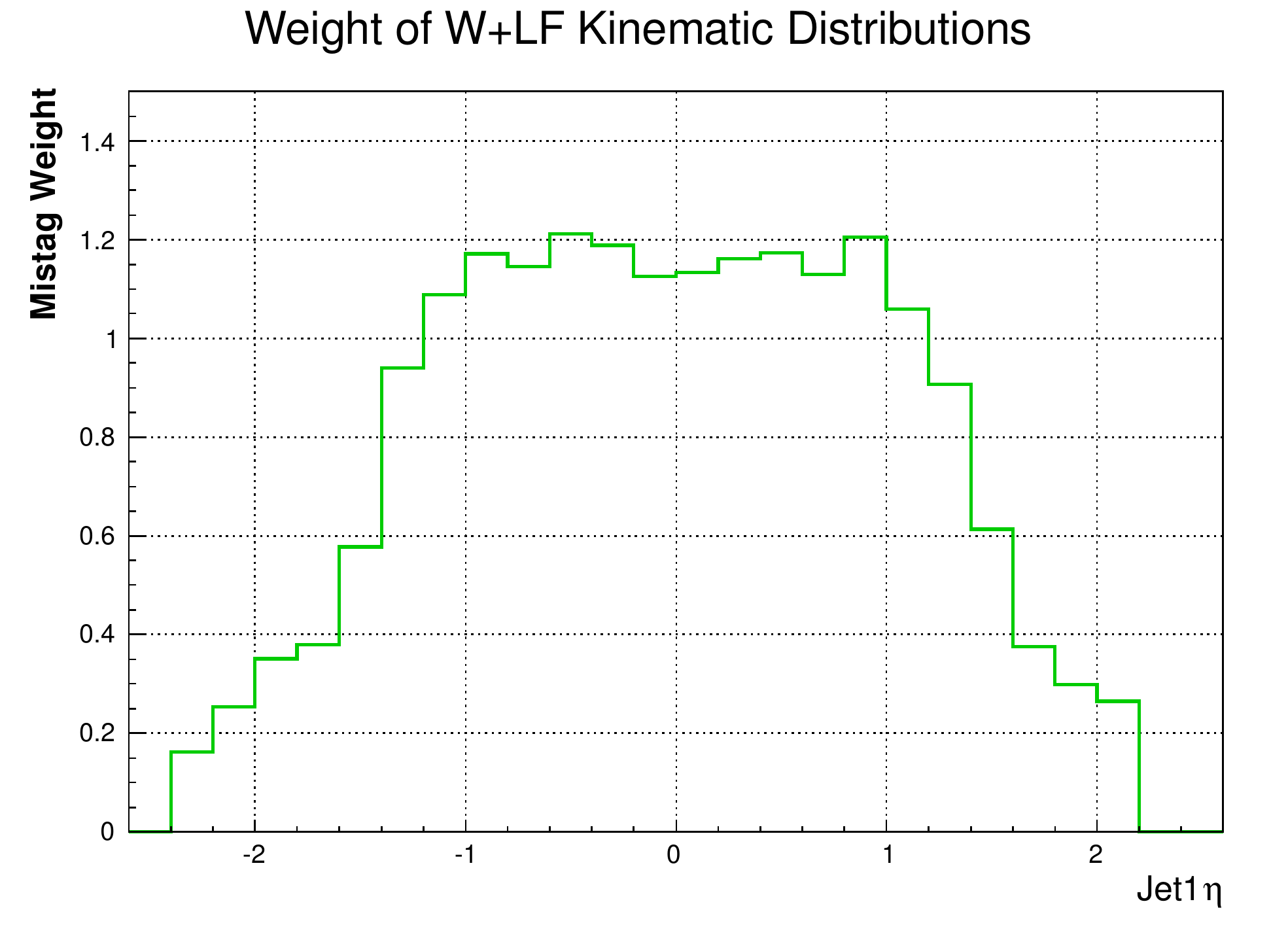}
\includegraphics[width=0.495\textwidth]{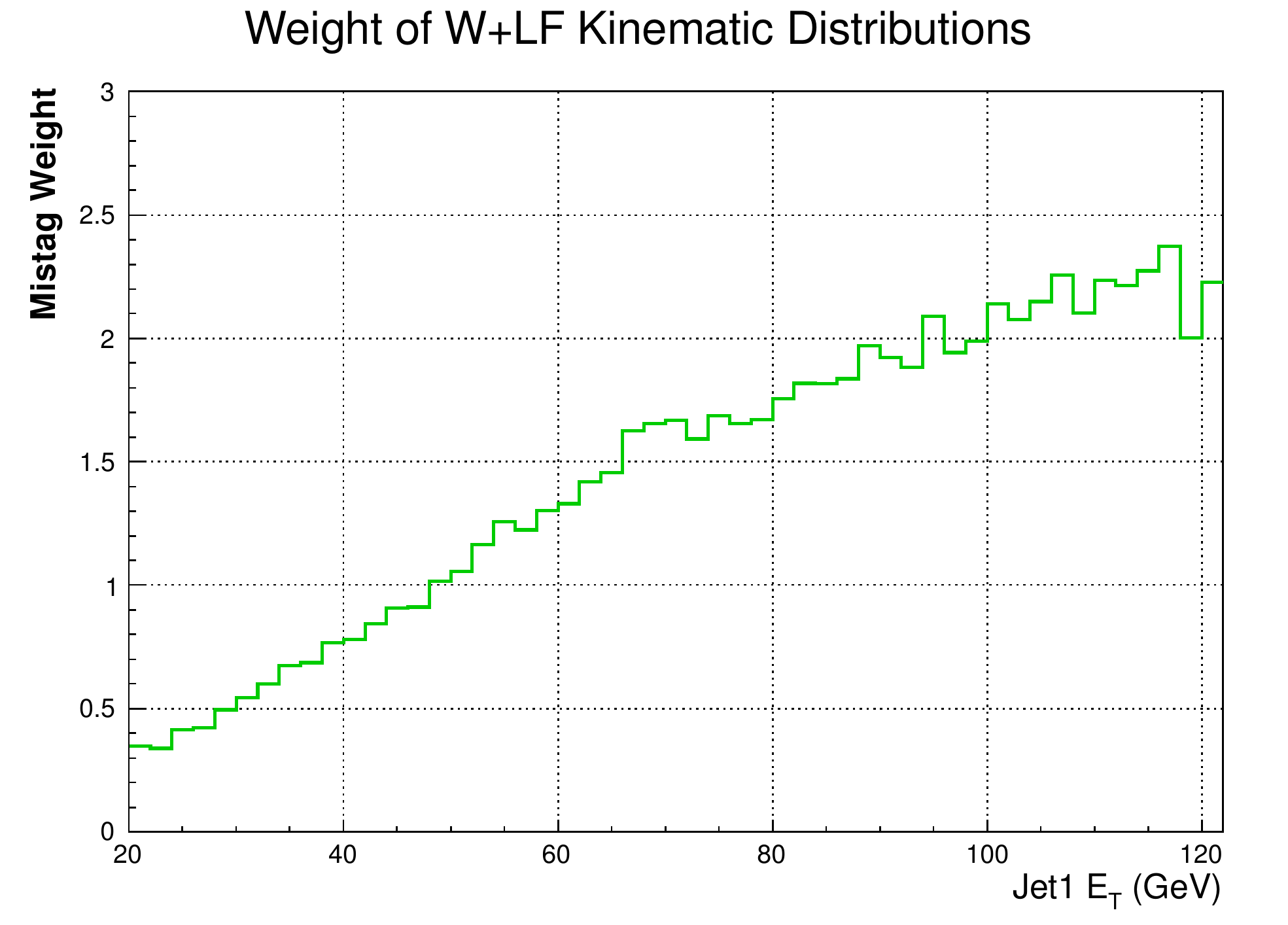}\\
\includegraphics[width=0.7\textwidth]{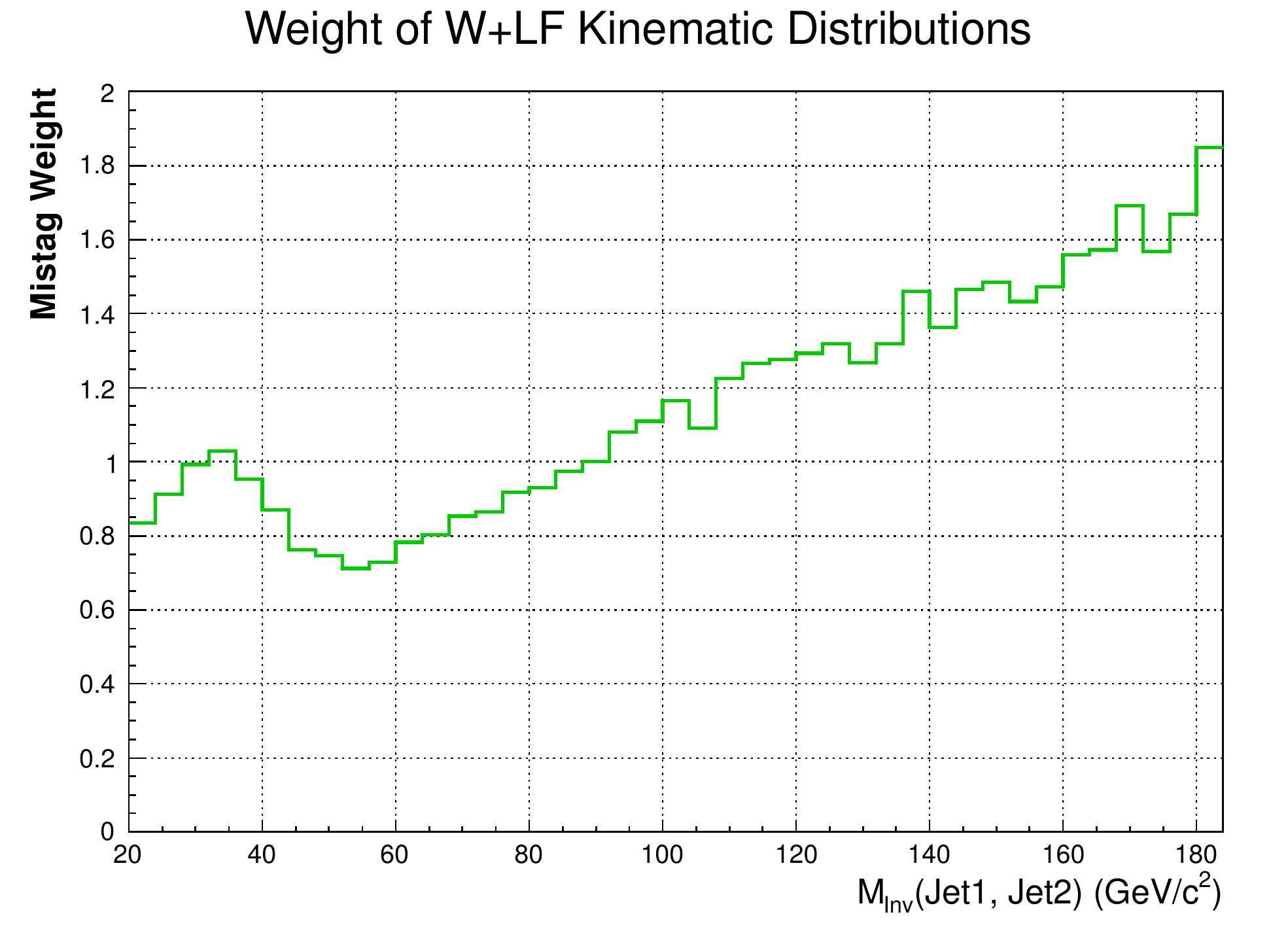}
\caption[Normalized Ratios of $W+LF$ Pretag and Tag Distributions]{Normalized ratios between the single-tagged $W+LF$ weighted MC and the original pretag $W+LF$ MC distribution: jet 1 $\eta$ distribution (top left), jet 1 $E_T$ distribution (top right) and $M_{Inv}(jet1, jet2)$ (bottom).}\label{fig:wlf_weight}
\end{center}
\end{figure}

As explained in Section~\ref{sec:btag_sf}, a multi-jet control sample is used to parametrize a per-jet mistag probability,  $p^j_{Mistag}$, function of six variables specific of the event and the jet. The $p^j_{Mistag}$ evaluation can be applied to any data sample or to MC. This gives the possibility to divide the normalization and the shape evaluation issues:
\begin{itemize}
\item the normalization is derived directly from the selected data sample. For each event we calculate an {\em event-mistag} estimate, $w_{Mistag}^{ev}$, by using  Equations~\ref{eq:1tag} and~\ref{eq:2tag} with the $p^j_{Mistag}$ of the jets as inputs. The sum of all the event-mistag gives a raw normalization of the total $LF$ contribution:
  \begin{equation}
    N^{rawLF}=\sum_{ev} w_{Mistag}^{ev}\textrm{,}
  \end{equation}
that needs to be corrected for the contribution of the other backgrounds. The $W+LF$ only part is:
\begin{equation}
  N^{W+LF} = N^{rawLF} \frac{N^{Wjets} - N^{W+HF}- \sum\limits_{EWK}{N^{EWK}}}{N^{Wjets}+ \sum\limits_{EWK}N^{EWK}}\textrm{.}
\end{equation}

\item Shape is derived from the $W+LF$ composition of ALPGEN MC (no $HF$ matched jets are used) where each simulated event is weighted by its own $w_{Mistag}$ to reproduce the kinematic beaviour of the mistagged jets. Figure~\ref{fig:wlf_weight} shows the normalized ratios between the single-tagged $W+LF$ weighted MC and the original pretag $W+LF$ distribution: $LF$ jets have a higher mistag probability in the central $\eta$ region of the detector (where also \texttt{SecVtx} $b$-tagging efficiency is higher) and for large $E_T$. The effect on the $W+LF$ $M_{Inv}(jet1, jet2)$ spectrum after the $b$-tagging requirement is also shown.

\end{itemize}

A rate uncertainty of $11$\% and $21$\% is included in the single and double-tagged mistag estimates respectively. It is derived from the mistag matrix parametrization and takes into account the statistical uncertainties and correlations between jets which fall in the same jet-bin of the mistag matrix. 

\section{Multi-jet Background}\label{sec:qcd}

Another source of background comes from fake-$W$ events, where a QCD multi-jet event fakes the charged lepton and produces enough \met to pass the event selection (\met$>15$~GeV). 

The probability to fake the lepton identification both at trigger and at offline level selection is very small but, at hadron colliders, multi-jet events are produced at such a high rate that this background becomes important. The selection of a fake lepton can happen for a mixture of physics and detector effects: for example a jet with large EM fraction and low track multiplicity can easily fake an electron or a high $p_T$ hadron can reach the muon chambers leaving little energy in not well instrumented sections of the calorimeter. A simulation of these effects would need an extremely high statistics, a perfect simulation of the detector and an accurate QCD prediction at high orders. This is not feasible therefore we rely on data to model the multi-jet sample (both shape and normalization). 

A fake-$W$ sample is obtained by reversing one or more $W\to \ell \nu$ identification criteria so that the final sample will be enriched in multi-jet events and orthogonal to the signal selection. However the correlation between the reversed identification variables and the related kinematic quantities are lost in the process. The fake-$W$ sample may present discrepancies with respect to the truly multi-jet events selected in the signal region. 

\begin{figure}[!ht]
\begin{center}
\includegraphics[width=0.7\textwidth]{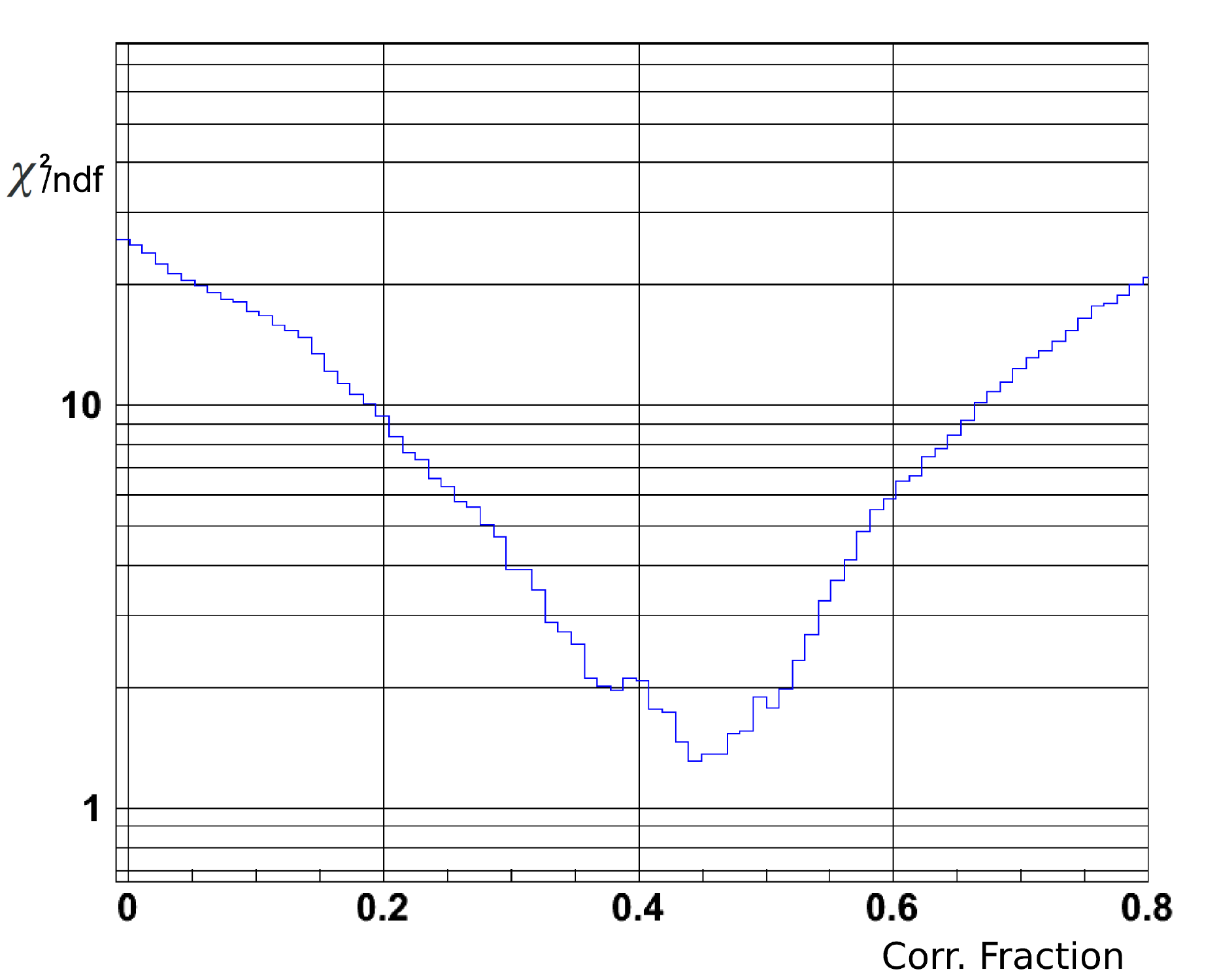}
\caption[Reduced $\chi^2$ Used to Tune \met Correction on Fake-$W$ Model]{Reduced $\chi^2$ obtained comparing the normalized $M^W_T$ shapes of data and anti-CEM fake-$W$ model in the multi-jet enriched region  \met$<15$~GeV. The test changes as we correct the \met for fractions of the energy associated with the jet corresponding to the anti-CEM.}\label{fig:chi2}
\end{center}
\end{figure}

The production mechanism is strictly related to the lepton identification algorithm, therefore we employ three different models:
\begin{description}
\item[Not isolated muons:] the $IsoRel <0.1$ cut is one of the most important $W$ identification requirements therefore we can employ real muons but in the sideband with $IsoRel>0.2$ to select fake-$W$ candidates for CMX, CMUP and EMC categories\footnote{EMC is composed by 7 lepton identification algorithms but the non-isolated EMC category is build only by four of them: CMU, CMP, CMXNT and BMU. The multi-jet sample properties are reproduced well enough also with this simpler model.}. 
 For a better simulation, the jet associated with the non-isolated muon is removed from the jet-multiplicity count and its energy (without the muon) accounted for the \met correction (Section~\ref{sec:metObj}).
\item[Fake-CEM:] CEM electron case is different because the isolation cut is correlated with the EM cluster energy, used both at trigger level selection and in the calorimeter \met calculation. A fake-$W$ model that maintains $IsoRel<0.1$ is obtained reversing at least two out of five shower identification requirements: Table~\ref{tab:antiEle} shows the reversed cuts. As in the muon case, jet multiplicity should be corrected for the jet associated with the lepton, however it is not straightforward to decide if any correction needs to be applied to the \met calculation. 

We studied the behaviour of the fake model on data, after full selection, but with no SVM cut and for \mbox{\met$<15$~GeV}. The region is supposed to be dominated by multi-jet events (except for a small $Z$ contamination) therefore we compared data to the fake-$W$ model.  Figure~\ref{fig:chi2} shows how the reduced $\chi^2$ test changes when comparing $M^W_T$ shapes for data and fake-$W$ model with different \met corrections. The best $\chi^2$ value is obtained when the \met is corrected as if the jet corresponding to the anti-CEM is present in the event but with an energy of $0.45\cdot E_T^{jet,raw}$. We apply this correction.

\item[Fake-PHX:] also the fake-$W$ PHX model is obtained with an anti-PHX selection where 2 out of 5 shower identification cuts are reversed (listed in Table~\ref{tab:antiEle}). In this case no special prescription to the \met correction was found to drastically improve the kinematic model. Although some properties of the multi-jet sample are not perfectly reproduced, the high rejection power of the SVM discriminant reduces this to a minor problem.
\end{description}
The accurate description of the multi-jet background and its rejection by using the SVM discriminant played a major role in this analysis and in the latest $WH$ search results~\cite{wh75_note10596,wh94_note10796}.

\begin{table}\begin{center}
\begin{tabular}{c|c}
\toprule
\multicolumn{2}{c}{\bf Anti-Electron: $\ge 2$ Failed Cuts }\\
CEM  & PHX \\\midrule
$E^{Had}/E^{EM}<0.055+0.0045 E^{EM}$& $E^{Had}/E^{EM}<0.05$ \\        
$L_{shr}<0.2$                      & PEM$3\times 3$Fit $\ne 0$ \\    
$\chi^2_{\mathrm{{CES strip}}}<10$ & $\chi^2_{PEM3\times 3} <10$ \\  
$\Delta z(CES,trk)<3$~cm           & $PES5\mathrm{by}9U>0.65$ \\     
$-3.0<q \Delta x(CES,trk)<1.5$~cm  & $PES5\mathrm{by}9V>0.65$ \\     
\bottomrule
\end{tabular}
\caption[CEM and PHX Multi-jet Model Selection Cuts]{Multi-jet models for CEM and PHX electrons should fail at least 2 of the shower identification cuts listed here. The models built in this way are named {\em anti-CEM} and {\em anti-PHX} leptons.}\label{tab:antiEle}
\end{center}\end{table}

The fake-$W$ models can now be used in the background evaluation. Section~\ref{sec:wjets_pretag} already gives the formula for the pretag $N^{nonW}$ estimate, however this result is not used in this analysis except for control purposes and to plot the background composition in the pretag region. On the other hand, the tagged region normalization of the fake-$W$ sample is the only missing piece in the full background evaluation. To estimate it we use again the multi-jet enriched sideband of the SVM output distribution.

\begin{figure}[!ht]
\begin{center}
\includegraphics[width=0.495\textwidth]{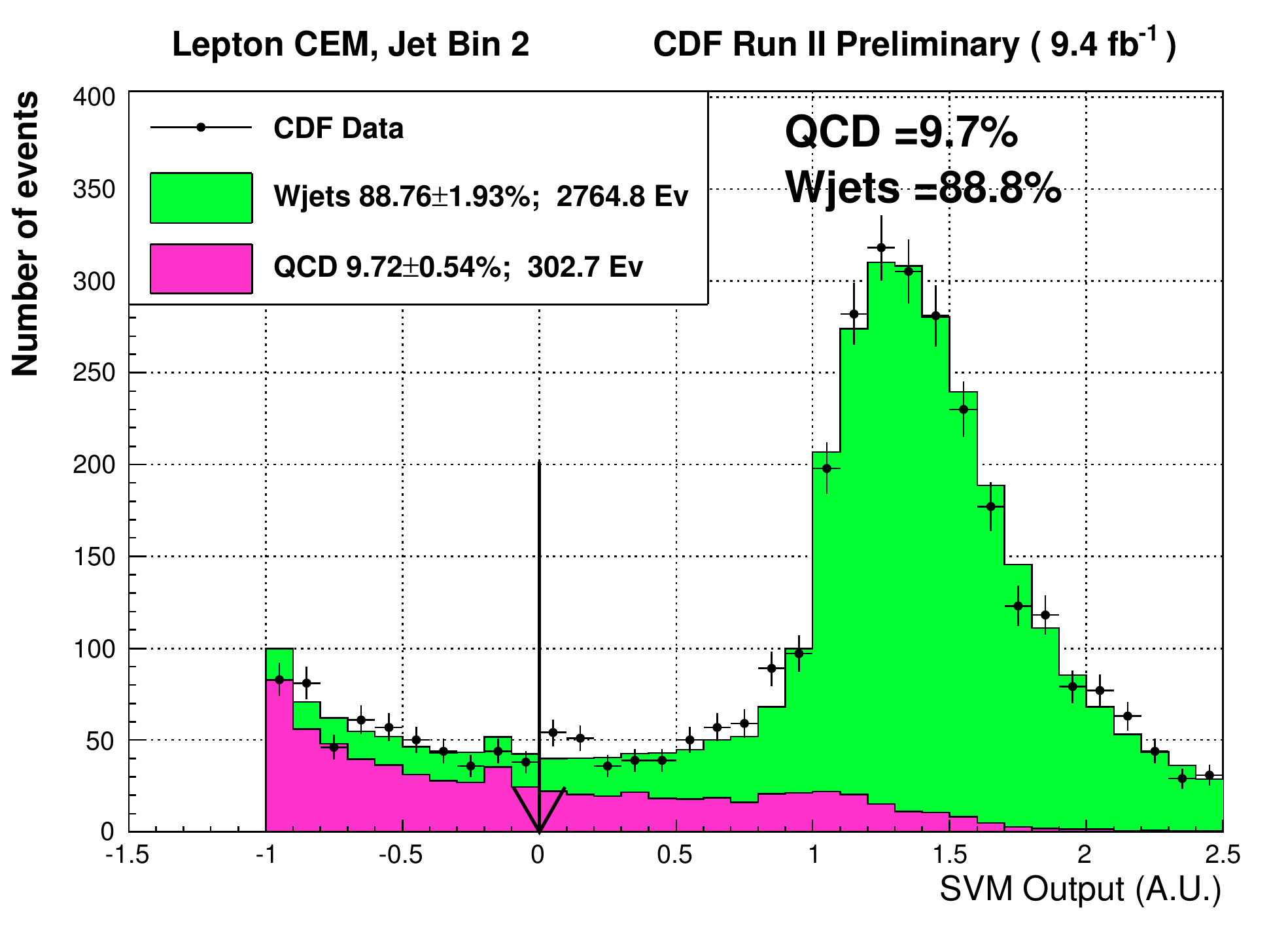}
\includegraphics[width=0.495\textwidth]{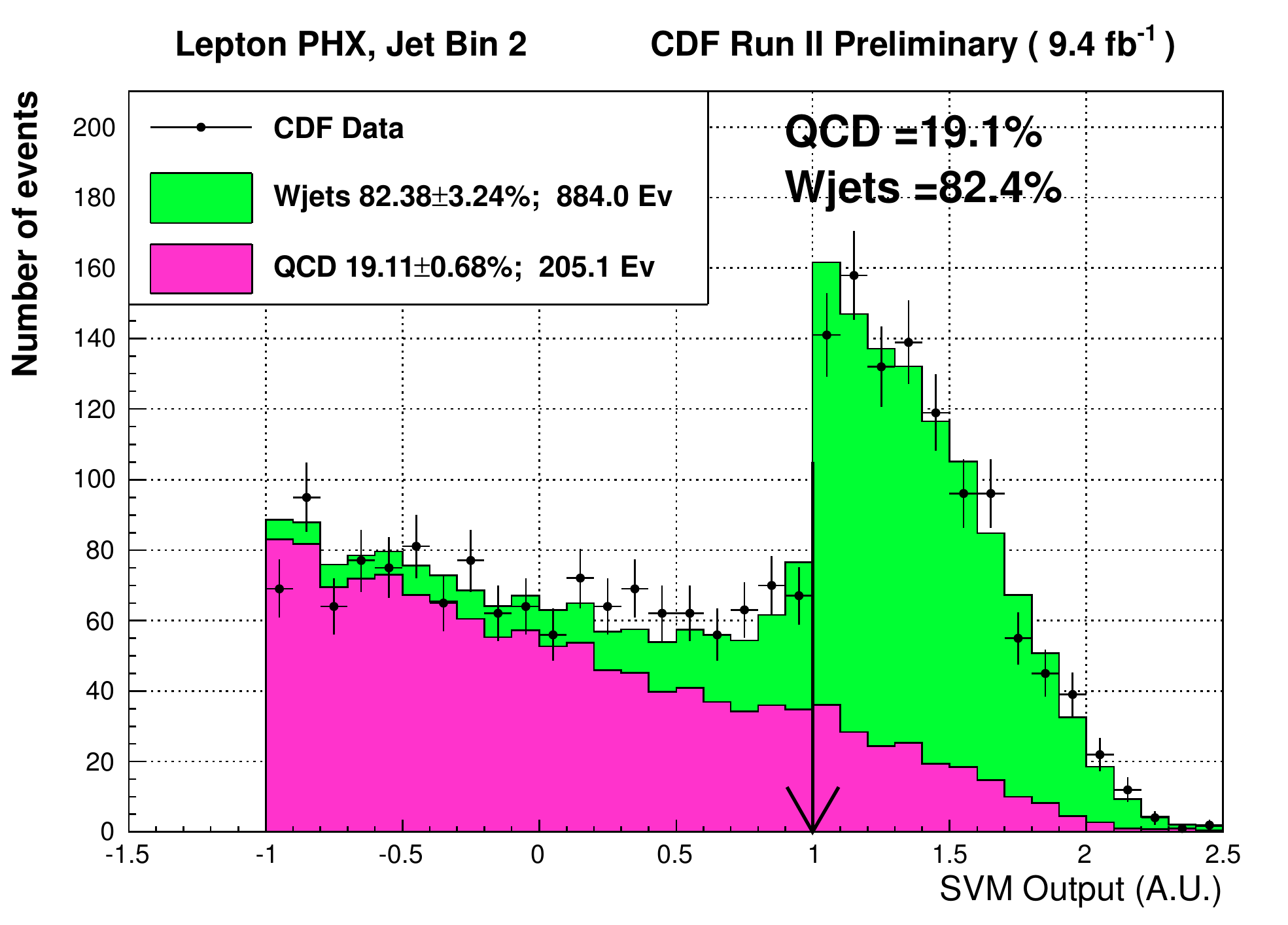}\\
\includegraphics[width=0.495\textwidth]{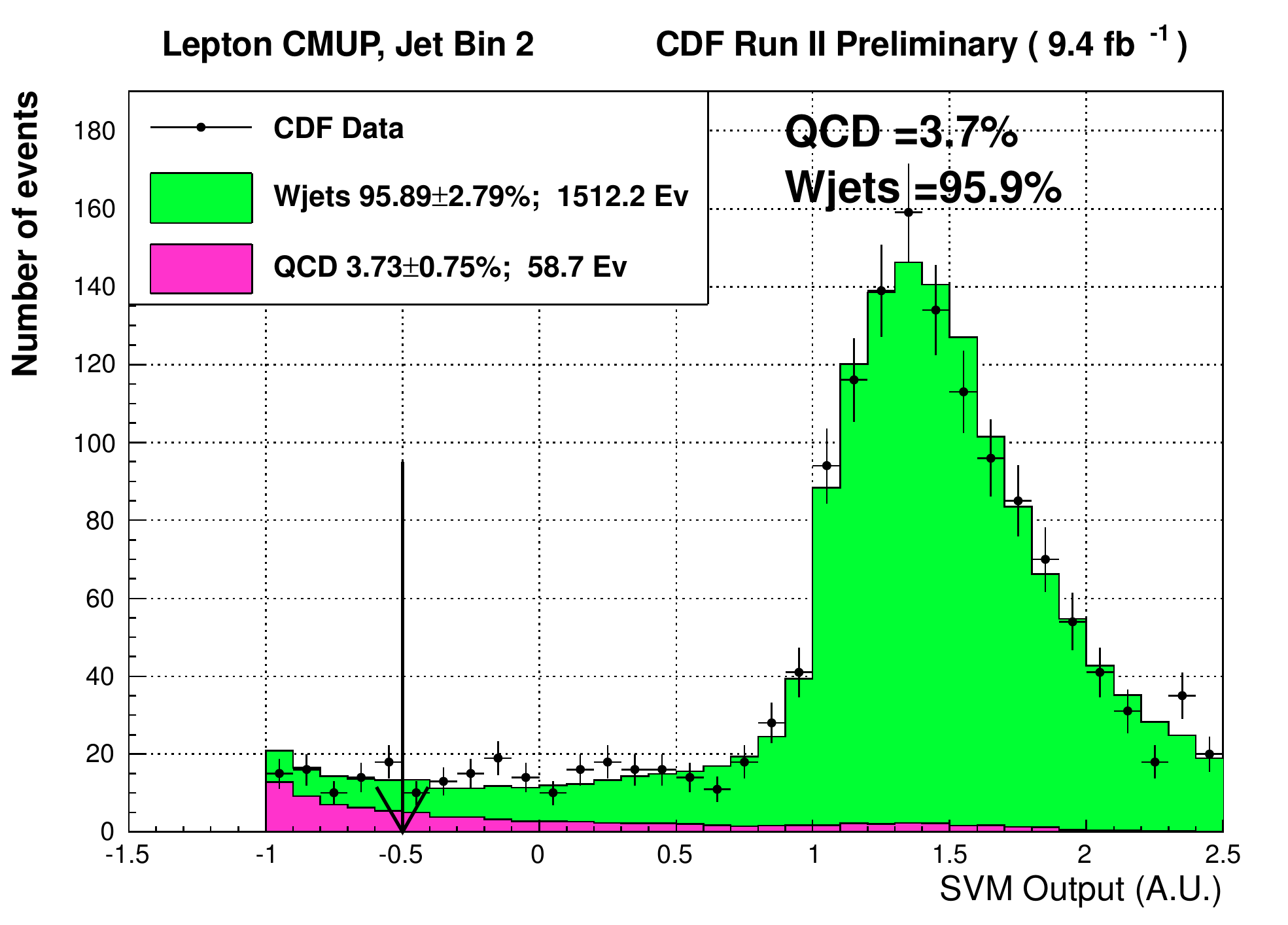}
\includegraphics[width=0.495\textwidth]{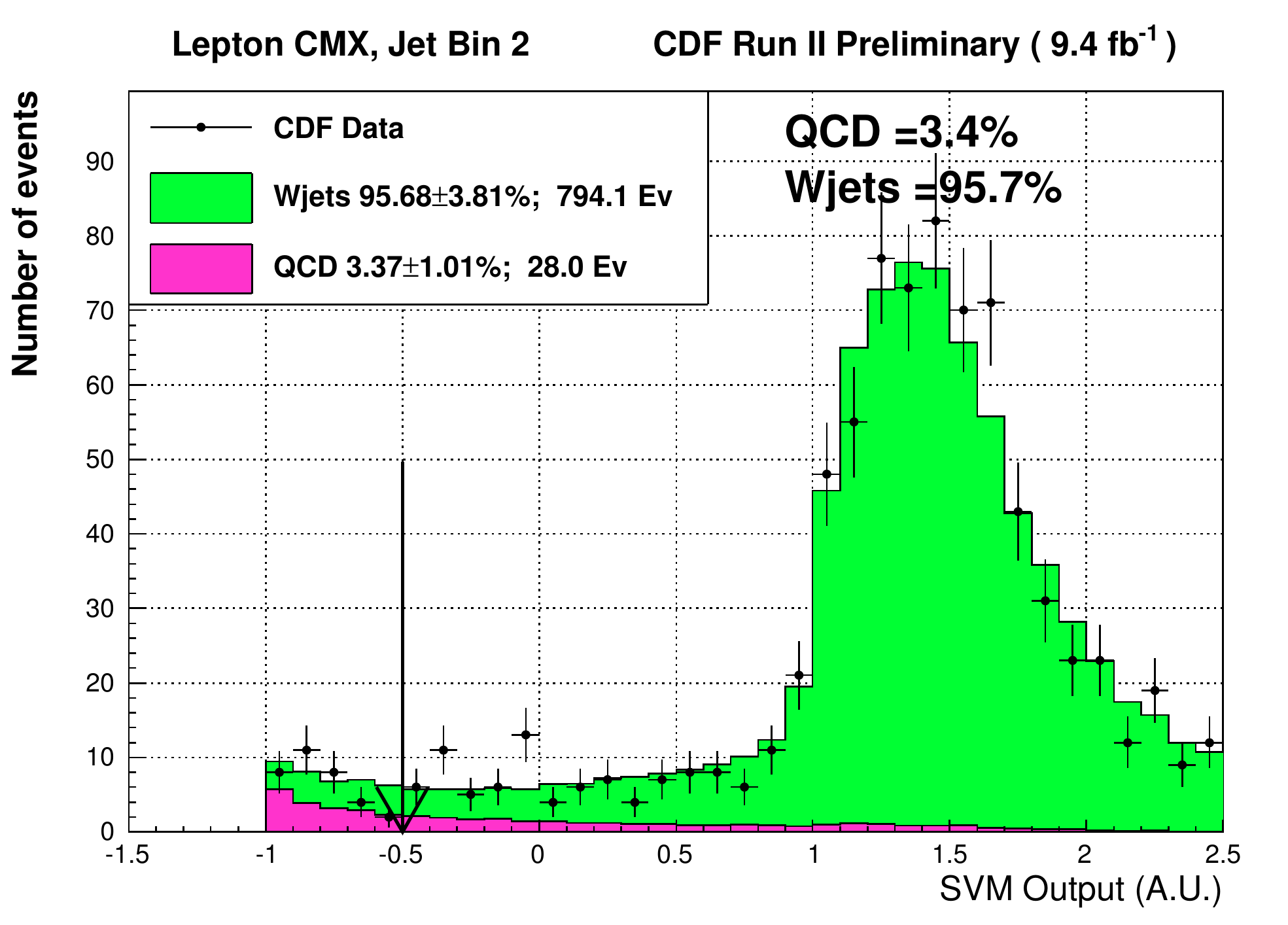}\\
\includegraphics[width=0.495\textwidth]{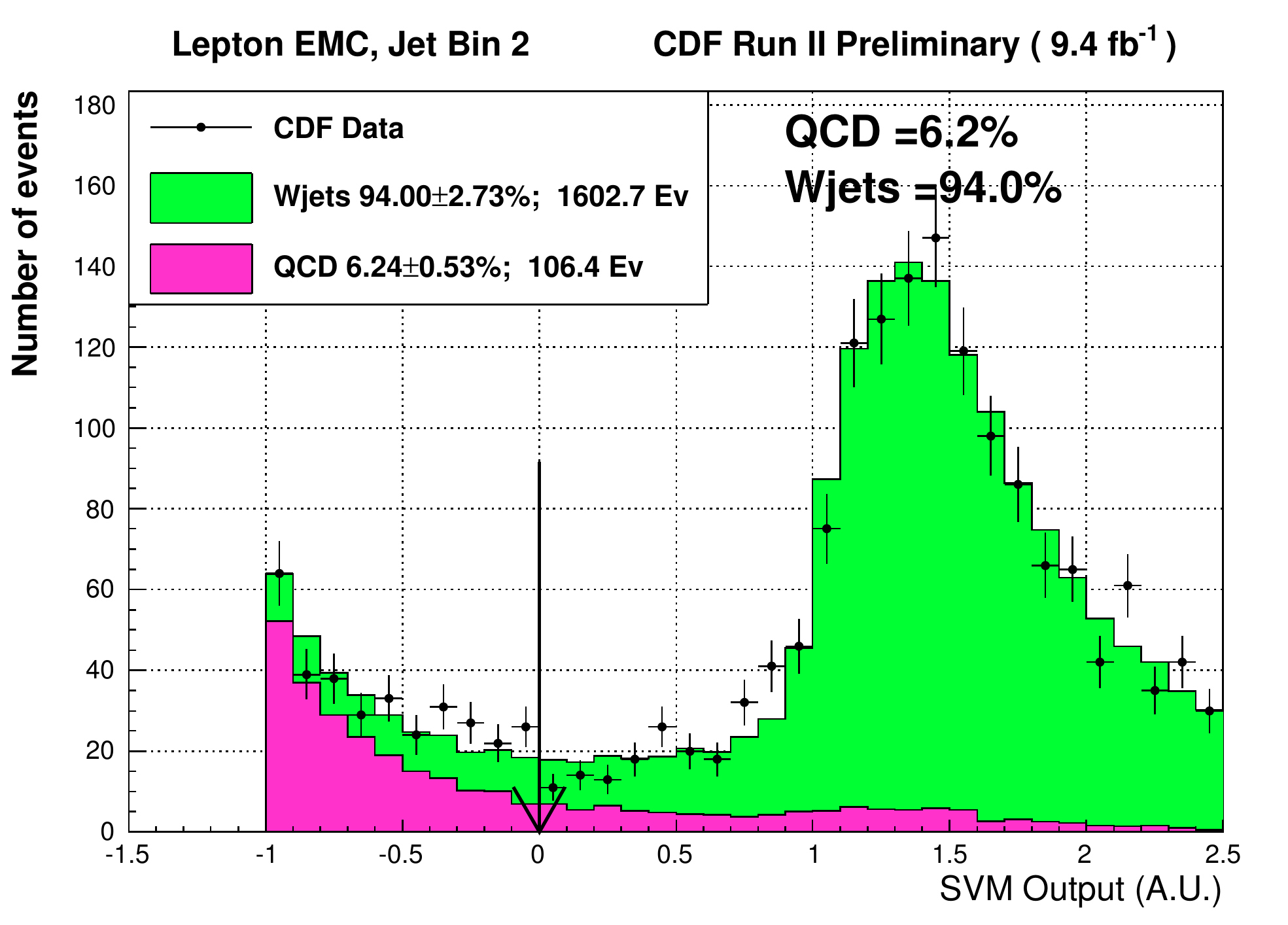}
\caption[Single Tag SVM Distribution for Non-$W$ Background Estimate]{Result of the maximum likelihood fit performed on the single-tagged SVM distribution to estimate the non-$W$ background normalization: $N^{nonW}_{1-Tag} = F^{nonW}_{1-Tag}\cdot N^{Data}_{k-Tag}$. Results are reported for the different lepton categories: CEM (top left), PHX (top right), CMUP (center left), CMX (center right), EMC (bottom). An arrow indicates the used selection cut value in each lepton category.}\label{fig:nonw_tag1}
\end{center}
\end{figure}

\begin{figure}[!ht]
\begin{center}
\includegraphics[width=0.495\textwidth]{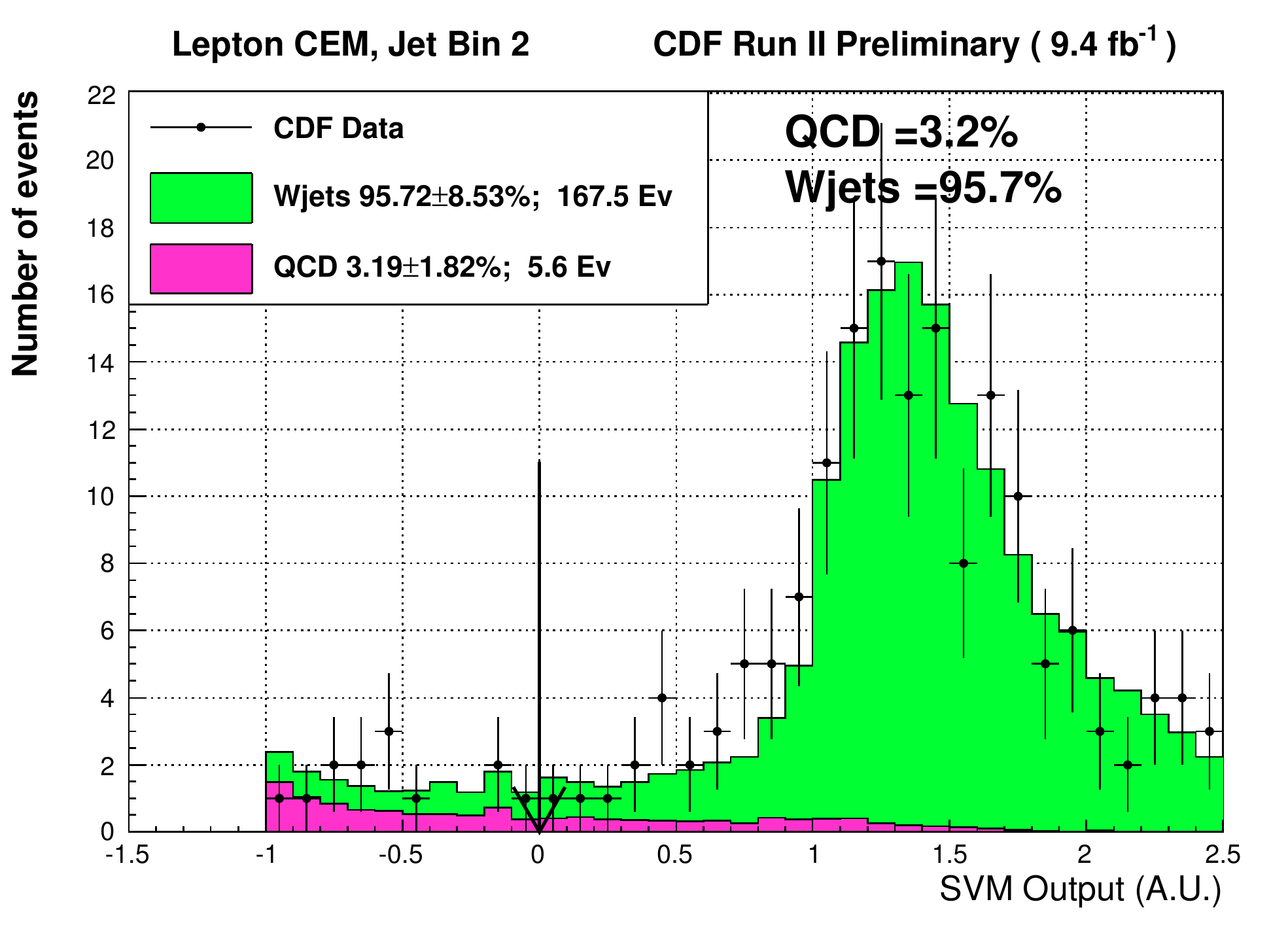}
\includegraphics[width=0.495\textwidth]{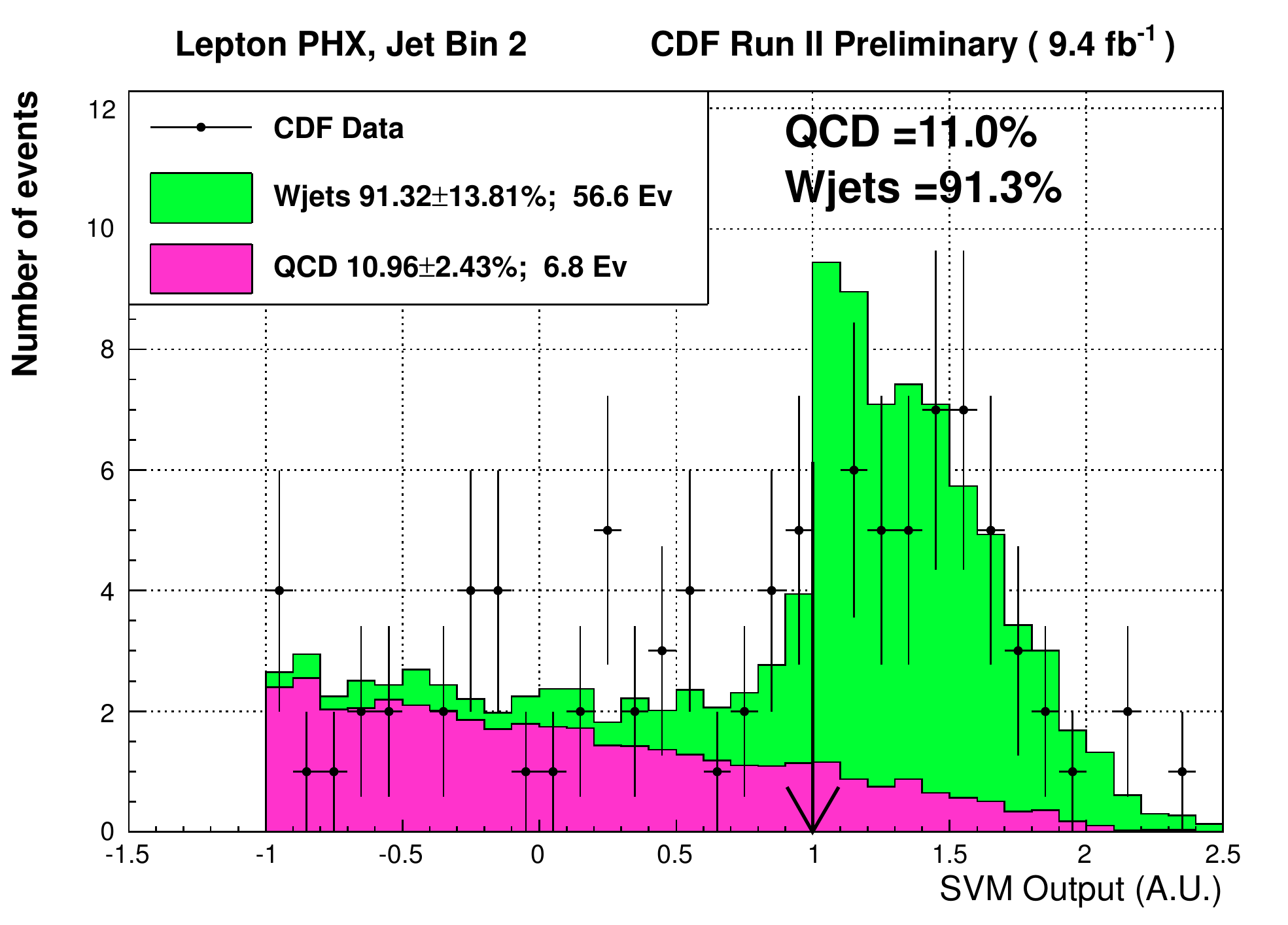}\\
\includegraphics[width=0.495\textwidth]{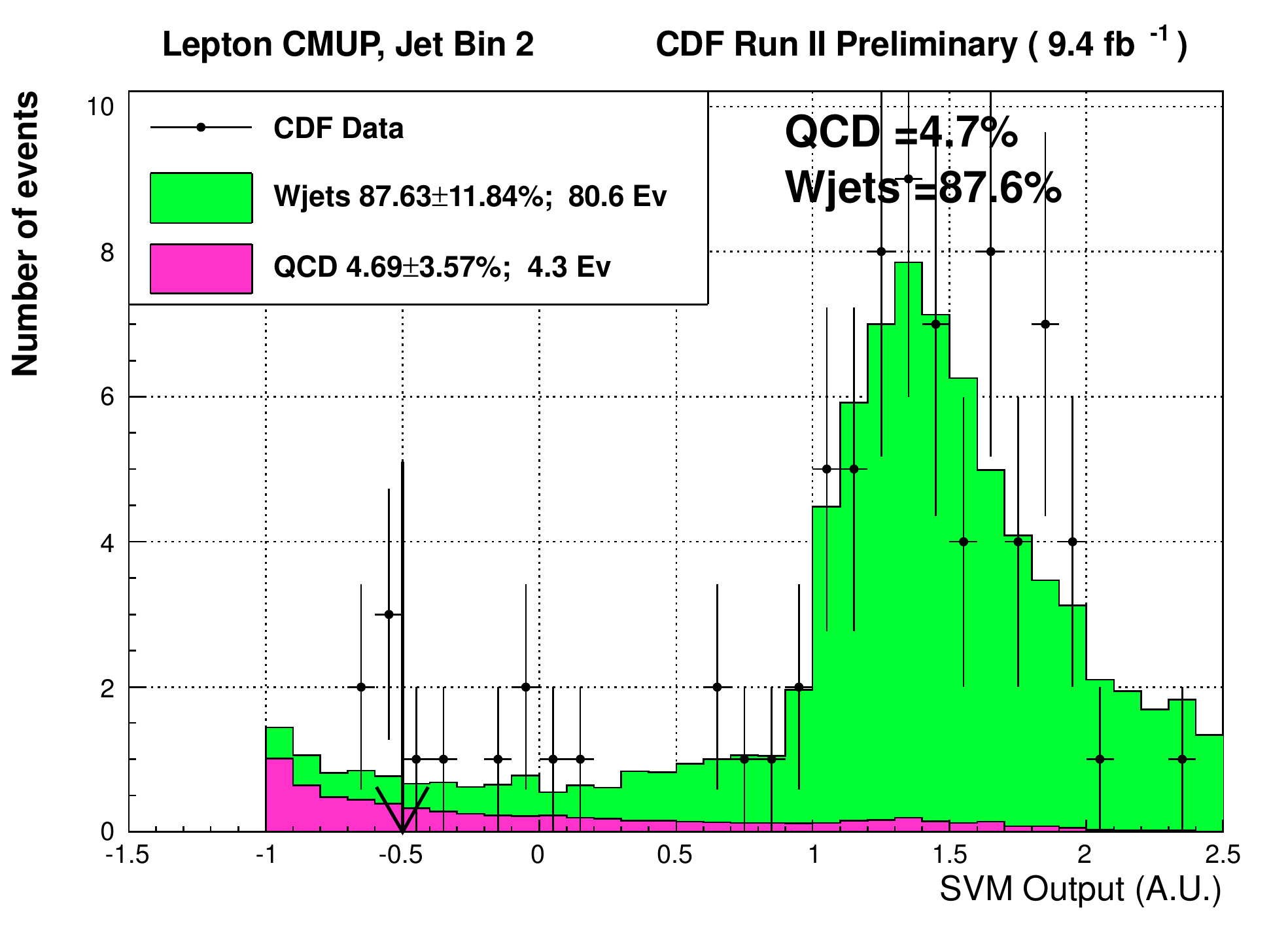}
\includegraphics[width=0.495\textwidth]{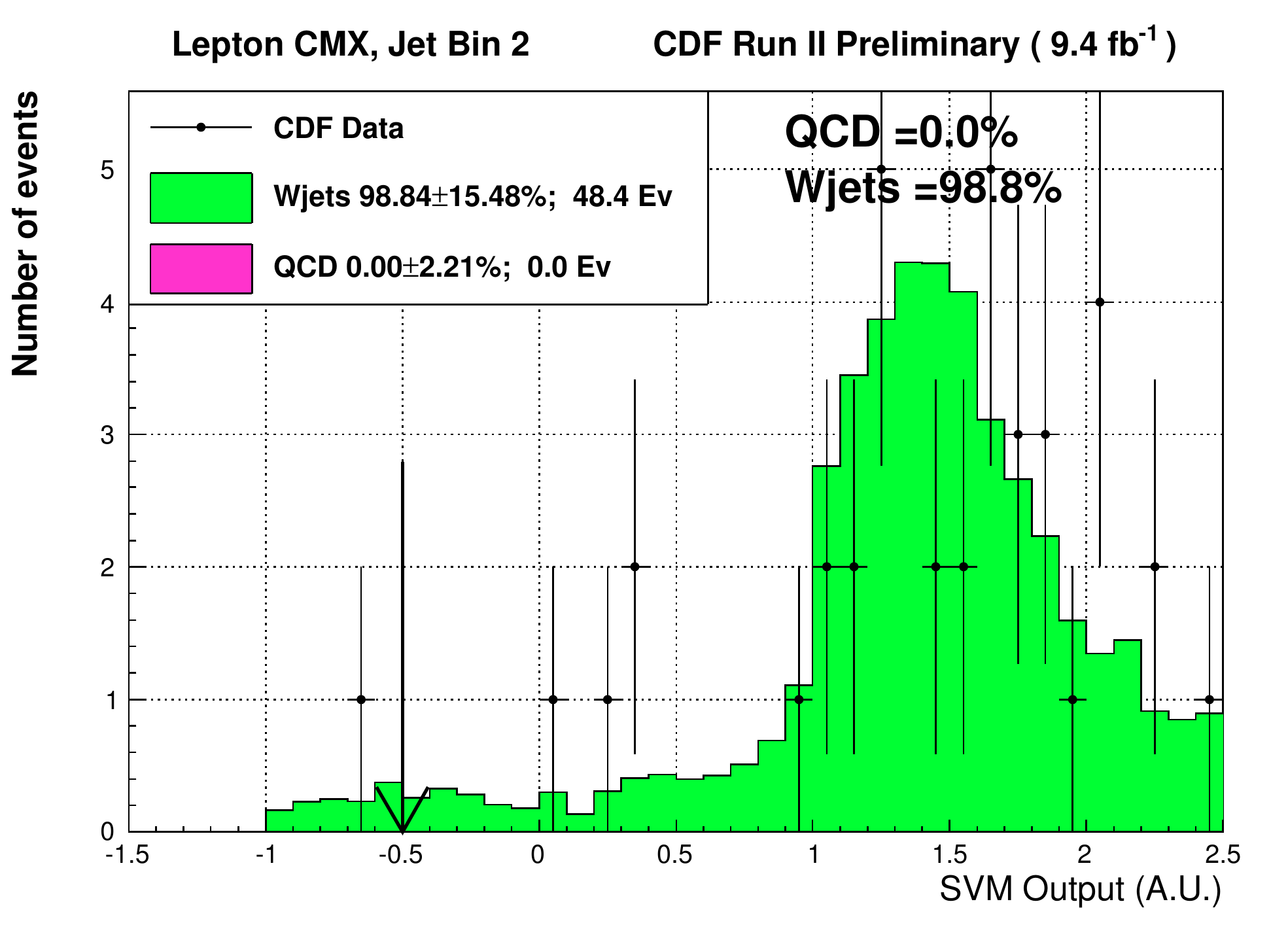}\\
\includegraphics[width=0.495\textwidth]{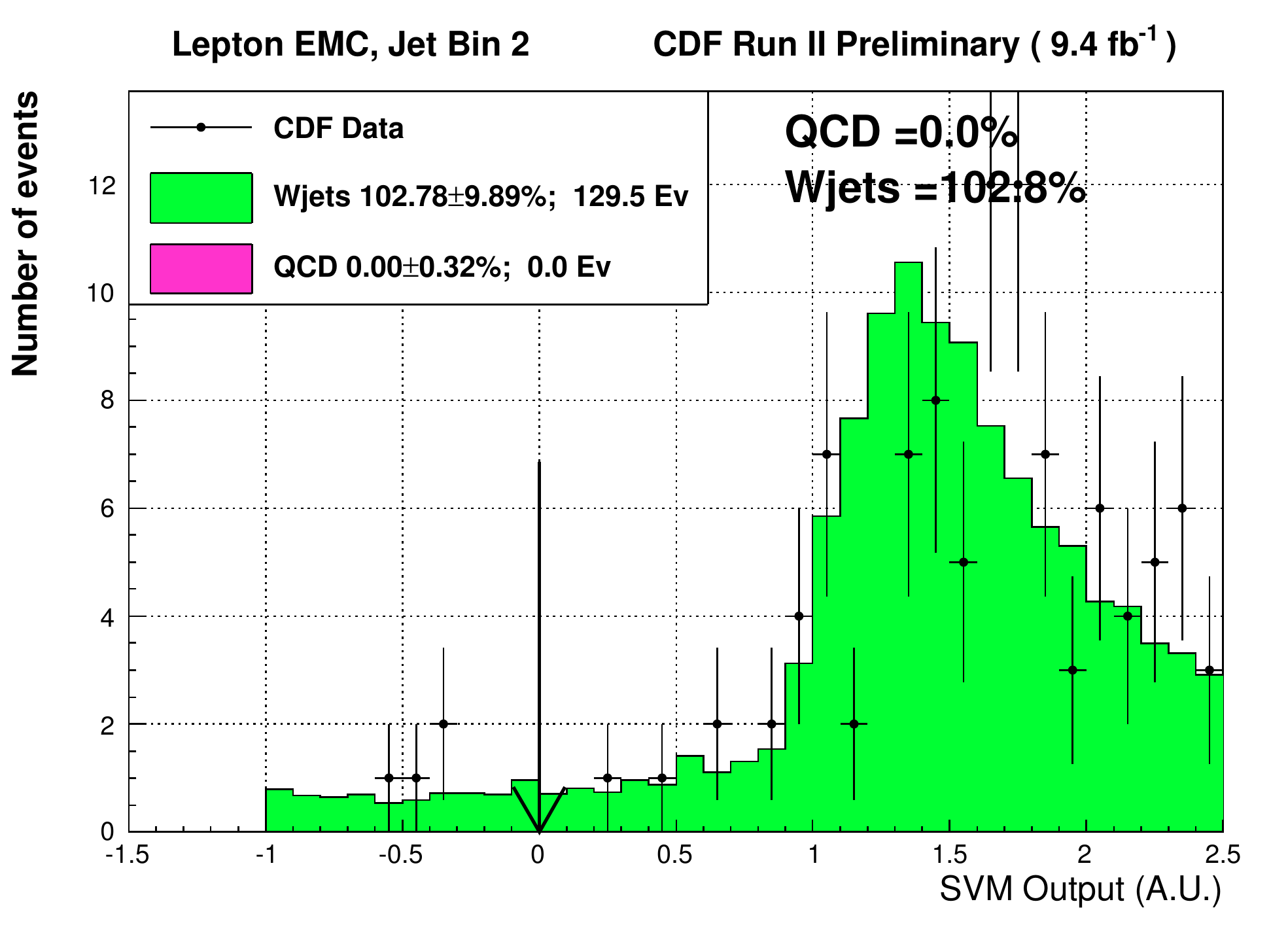}
\caption[Double Tag SVM Distribution for Non-$W$ Background Estimate]{Result of the maximum likelihood fit performed on the double-tagged SVM distribution to estimate the non-$W$ background normalization: $N^{nonW}_{1-Tag} = F^{nonW}_{1-Tag}\cdot N^{Data}_{k-Tag}$. Results are reported for the different lepton categories: CEM (top left), PHX (top right), CMUP (center left), CMX (center right), EMC (bottom). An arrow indicates the used selection cut value in each lepton category.}\label{fig:nonw_tag2}
\end{center}
\end{figure}

We perform another maximum likelihood fit similar to the one described in Section~\ref{sec:wjets_pretag} with two differences. The first is a modification of the template: we require events with at least $1$ ($2$) {\em taggable} jets when evaluating the single (double) tag fake-$W$ templates, this allows an approximate simulation of the $b$-tag requirement retaining most of the statistics of the samples. The second difference is in the fit strategy: all the real-$W$ samples are added into one template whose normalization is left free to float in the fit together with $N^{nonW}_{k-Tag}$. This procedure and the use of the sidebands make the fake-$W$ determination as much uncorrelated as possible from the other backgrounds and from a possible unknown signal. Figures~\ref{fig:nonw_tag1} and~\ref{fig:nonw_tag2} show the results of the $N^{nonW}_{k-Tag}$ fit on the single and double tagged SVM distributions for the CEM, PHX, CMUP, CMX and EMC lepton categories.
A conservative rate uncertainty of 40\% is applied uniformly on all the lepton categories to account for low statistics in some of the fits and for the extrapolation from the sideband region. Fake-$W$ contamination is summarized in Table~\ref{tab:wjet_tag_qcd_fit}, confirming the excellent rejection power of the SVM discriminant.
\begin{table}\begin{center}
\begin{tabular}{cccccc}
\toprule
 Lepton & CEM & PHX & CMUP & CMX &  EMC \\\midrule
$F^{nonW}_{1-Tag}$ & $ 9.7 \pm 0.5$\% &  $ 19.1 \pm 0.7  $\% &  $ 3.7 \pm 0.7  $\% &  $ 3.3\pm 1.0 $\% & $ 6.2 \pm 0.5$\%\\
 $F^{nonW}_{2-Tag}$ &  $ 3.2 \pm 1.8$\% &  $ 10.8 \pm 2.4$\% &  $ 4.7 \pm 3.6$\% &  $ 0.0 \pm 2.2$\% &  $ 0.0\pm 0.3$\%\\
\bottomrule
\end{tabular}
\caption[Non-$W$ Background Contamination in Single and Double Tag Samples]{Estimate of the non-$W$ contamination in the single and double-tagged selection samples. Only the statistical error of the fit is reported, a systematic uncertainty of $40$\% is used in the final estimate.}\label{tab:wjet_tag_qcd_fit}
\end{center}\end{table}


\section{Correction of Luminosity Effects}

The effect of instantaneous luminosity should be corrected in the available MC samples before completing of the background estimate. 

The instantaneous luminosity of the collisions influences several parameters like the occupancy of the detector, the number of vertices in the interactions and so on. Most of these effects are parametrized within the CDF simulation (see Section~\ref{sec:mcGen}) and they are simulated in a run-dependent way (i.e. each MC section corresponds to a period of data taking with appropriate tuning). Unluckily the available MC samples reproduce only a first section (up to 2008) of the CDF dataset while the highest luminosities (easily above $\mathscr{L} = 2\times 10^{32}$~cm$^{-2}$s$^{-1}$) were delivered by the Tevatron in the latest years. 

\begin{figure}[!ht]
\begin{center}
\includegraphics[width=0.495\textwidth]{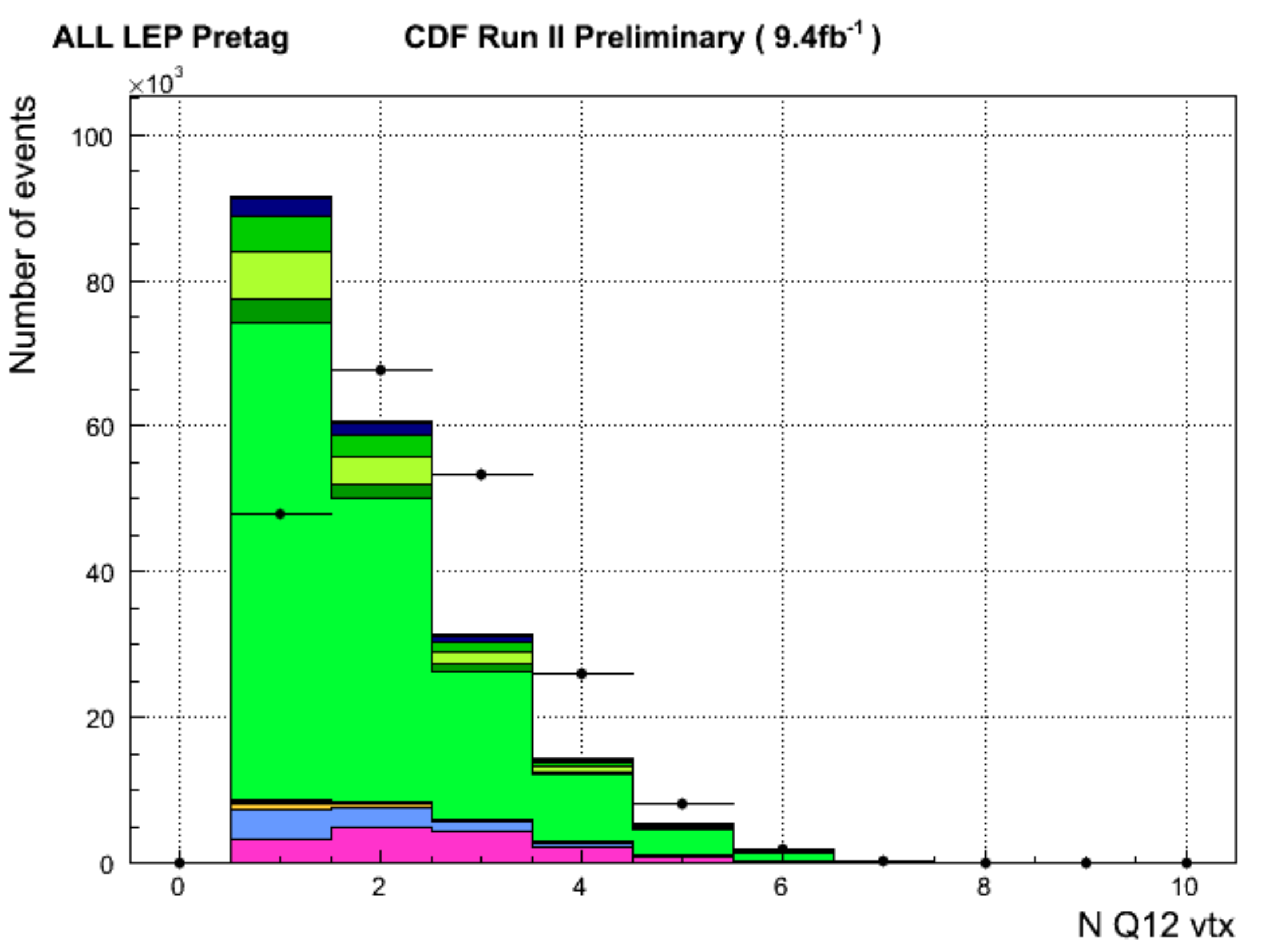}
\includegraphics[width=0.495\textwidth]{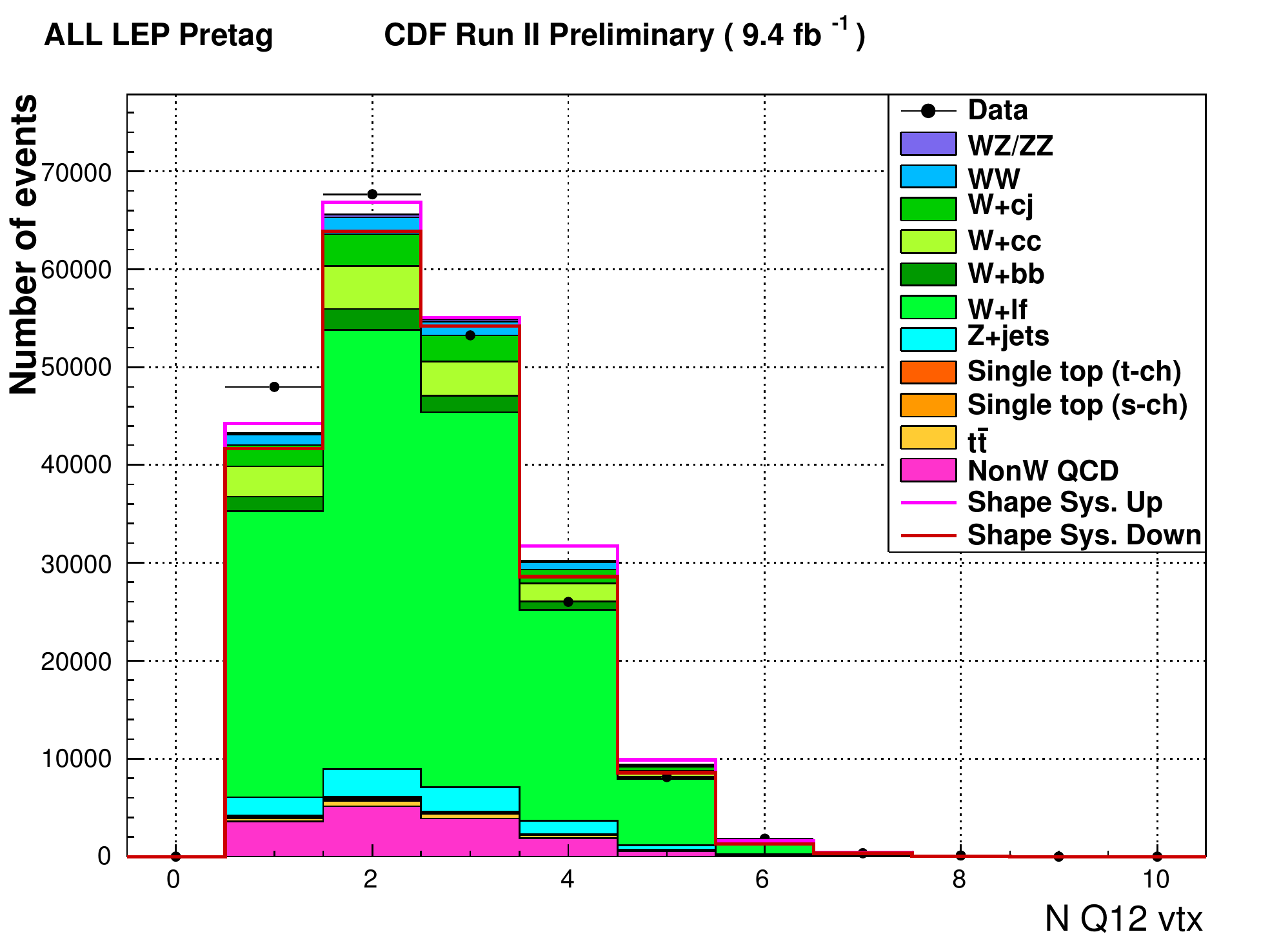}
\caption[Number of \mbox{{\em Quality} $\ge 12$} Before and After Reweighting]{Distribution (for all the lepton categories) of the number of \mbox{{\em Quality} $\ge 12$} primary vertices for the events selected in the pretag control region before (left) and after (right) the application of a reweighting function to the MC events.}\label{fig:nvtx_rew}
\end{center}
\end{figure}

An approximate solution to the problem is the {\em reweighting} of the MCs as a function of the observed number of good quality primary vertices\footnote{Quality$\ge 12$, see Table~\ref{tab:pVtx}.} (a variable highly correlated with the instantaneous luminosity). The reweighting function is evaluated comparing, in the pretag control region, the number of vertices estimated in MC against the one observed in data. Figure~\ref{fig:nvtx_rew} shows the distribution of the number of primary vertices withouth any correction and after correcting the background estimate with the reweighting function that we apply to the final evaluation.

\section{Final Background Evaluation}

As explained in the previous part of this Chapter, we derive the shape and normalization contributions of each background in an independent way. The composition of them should reproduce, within uncertainties, the observed number of selected events and the kinematic distributions of data in control and signal regions. If the background estimate is proven to be solid, then it is possible to investigate the presence of a signal with adequate tools. 

We present here the final background estimate. Statistical analysis and the measured properties of the $diboson\to \ell \nu + HF$ signal will be described in the next Chapter.

\begin{figure}
    \begin{center}    
  \includegraphics[width=0.495\textwidth]{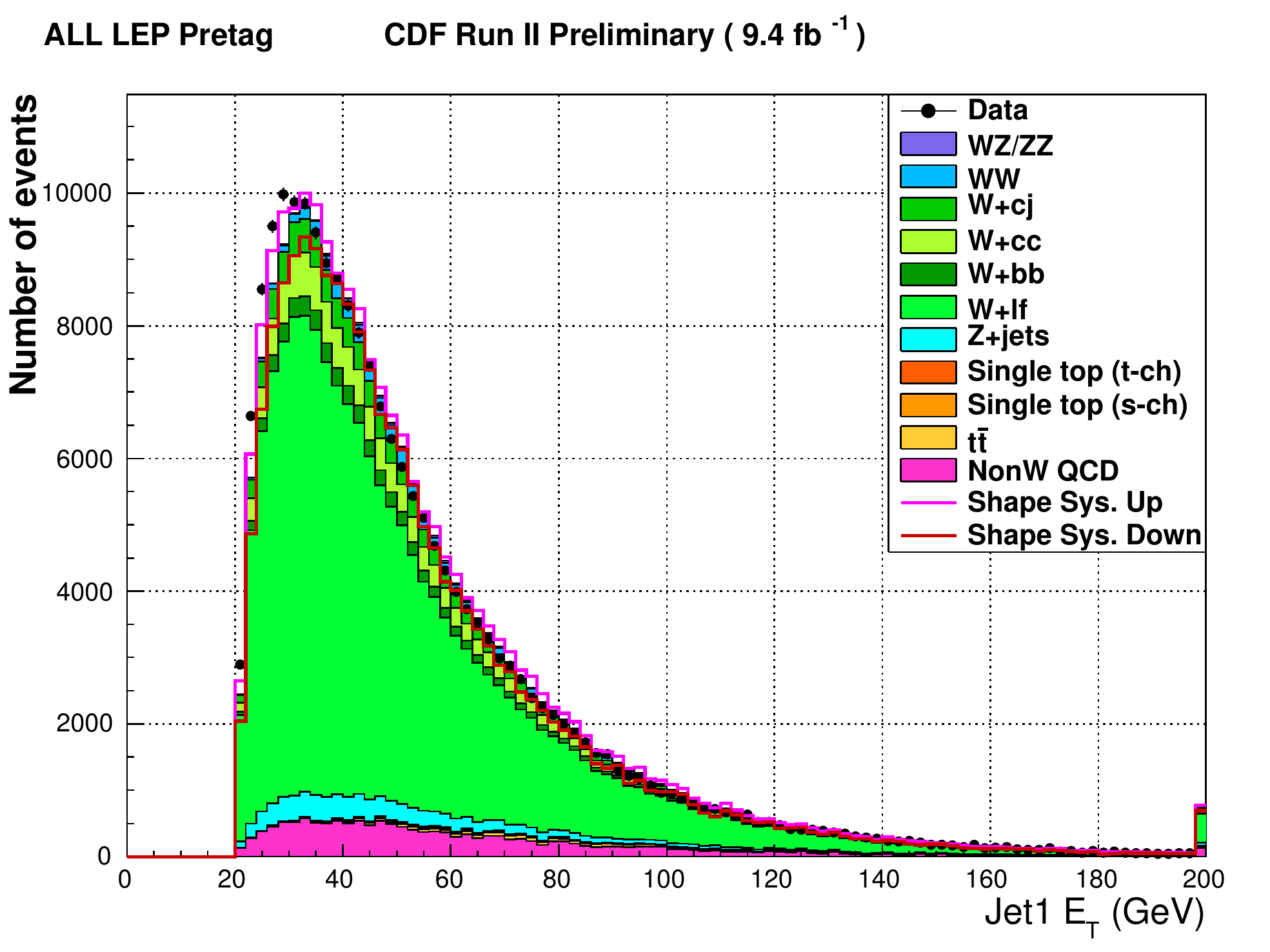}
  \includegraphics[width=0.495\textwidth]{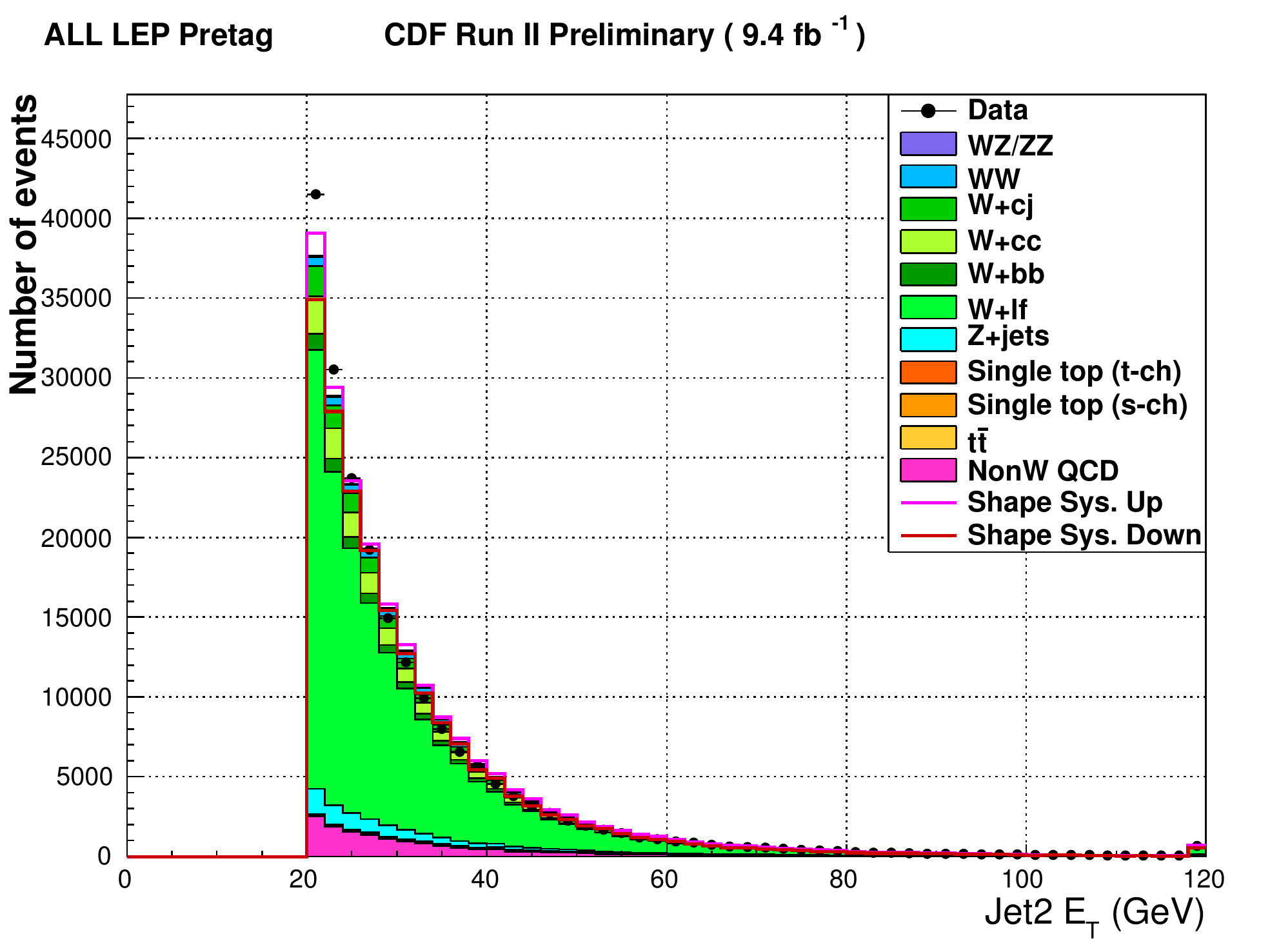}
    \end{center}
  \caption[Jets $E_T$ Distributions in Pretag Control Region]{Jets corrected $E_T$ distributions in the pretag control region, for all the lepton categories combined: jet 1 $E_T$ (left) and jet 2 $E_T$ (right) for all the lepton categories combined, in the pretag control region.}\label{fig:pret_jetEt}
  \end{figure}

\begin{figure}
    \begin{center}    
    \includegraphics[width=0.495\textwidth]{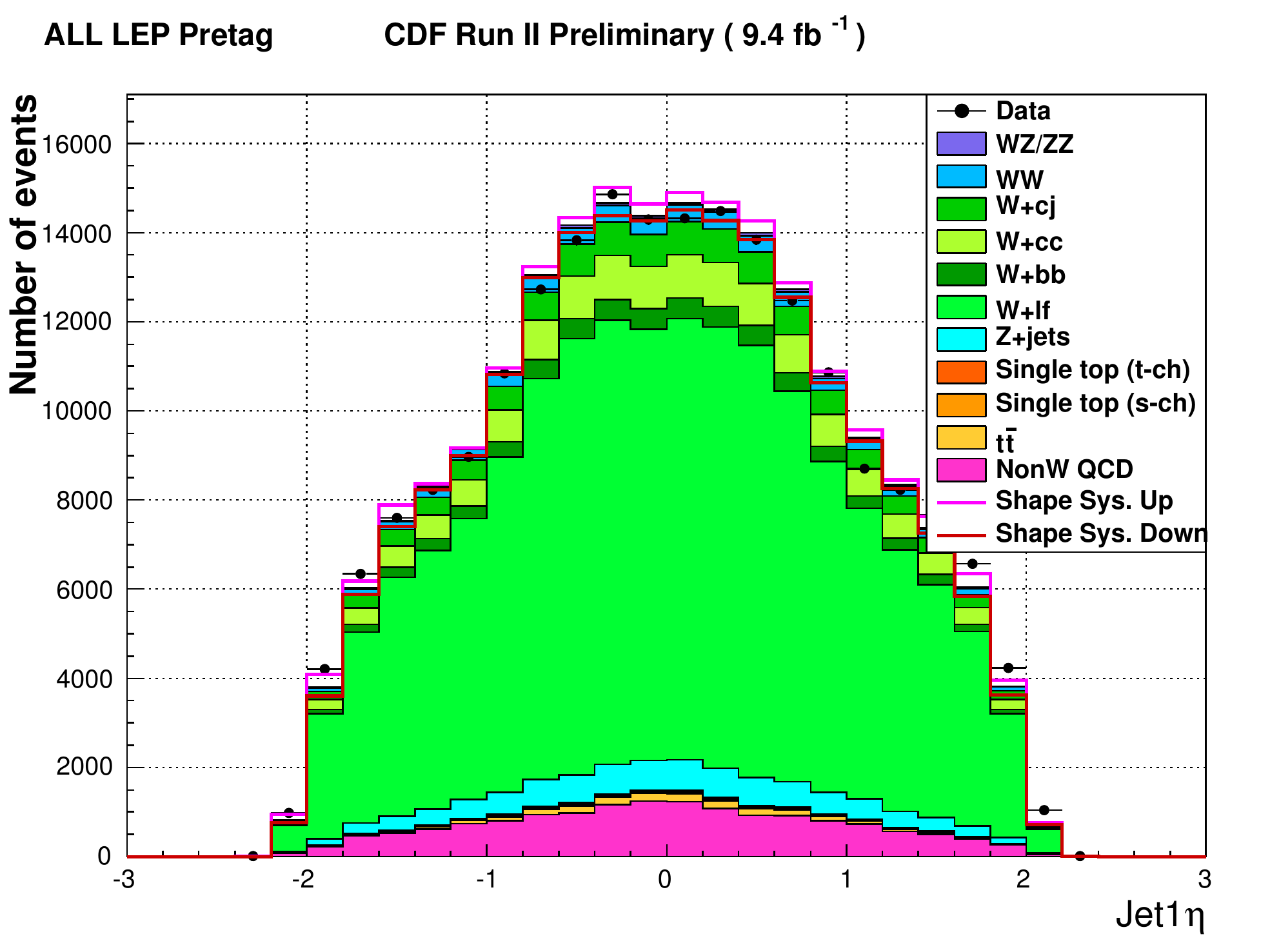}
    \includegraphics[width=0.495\textwidth]{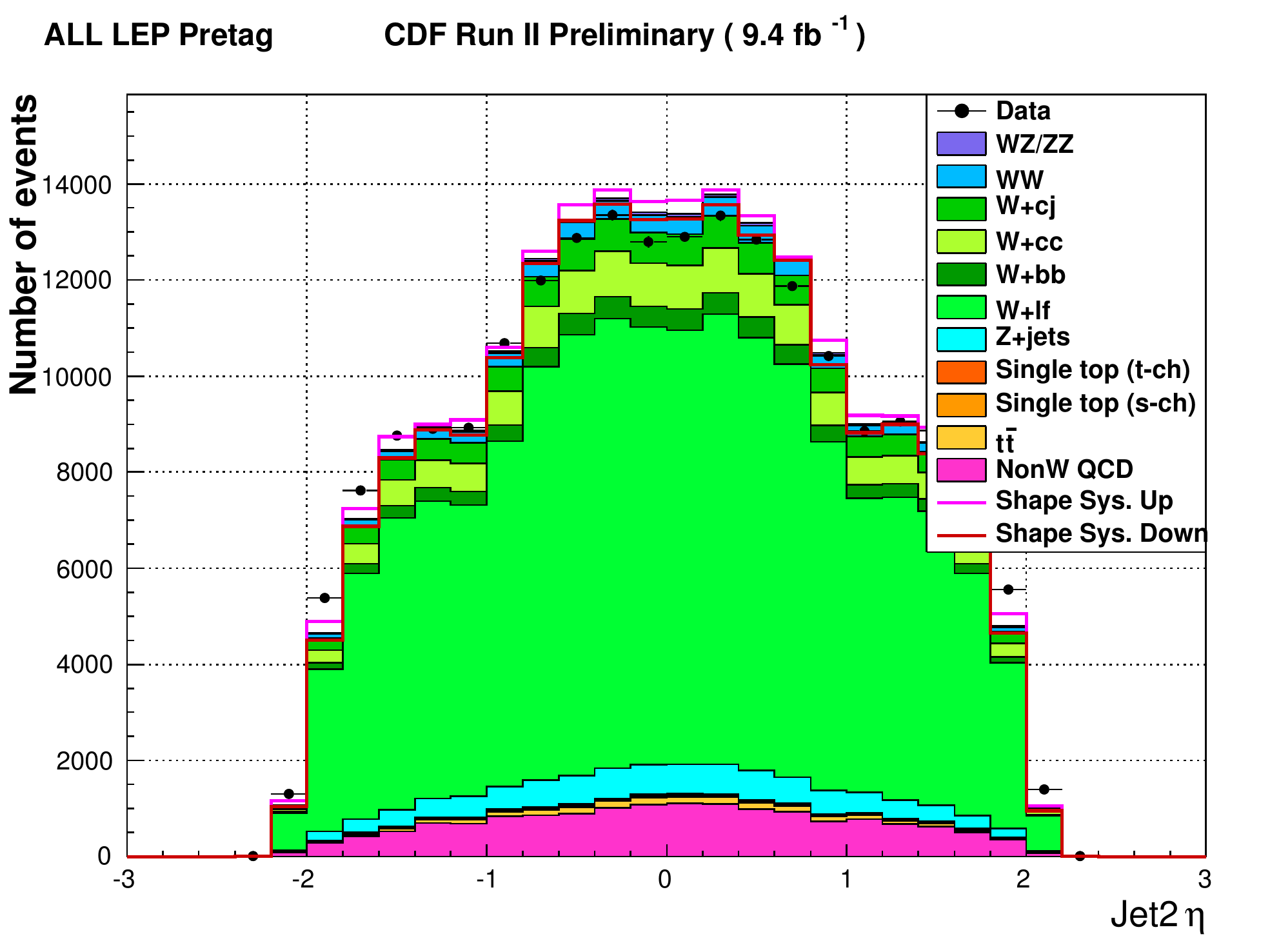}
    \end{center}
    \caption[Jets $\eta$ Distributions in Pretag Control Region]{Jets $\eta$ distributions in the pretag control region, for all the lepton categories combined: jet 1 $\eta$ (left) and jet 2 $\eta$ (right).}\label{fig:pret_jetEta}
  \end{figure}

\begin{figure}
    \begin{center}    
    \includegraphics[width=0.495\textwidth]{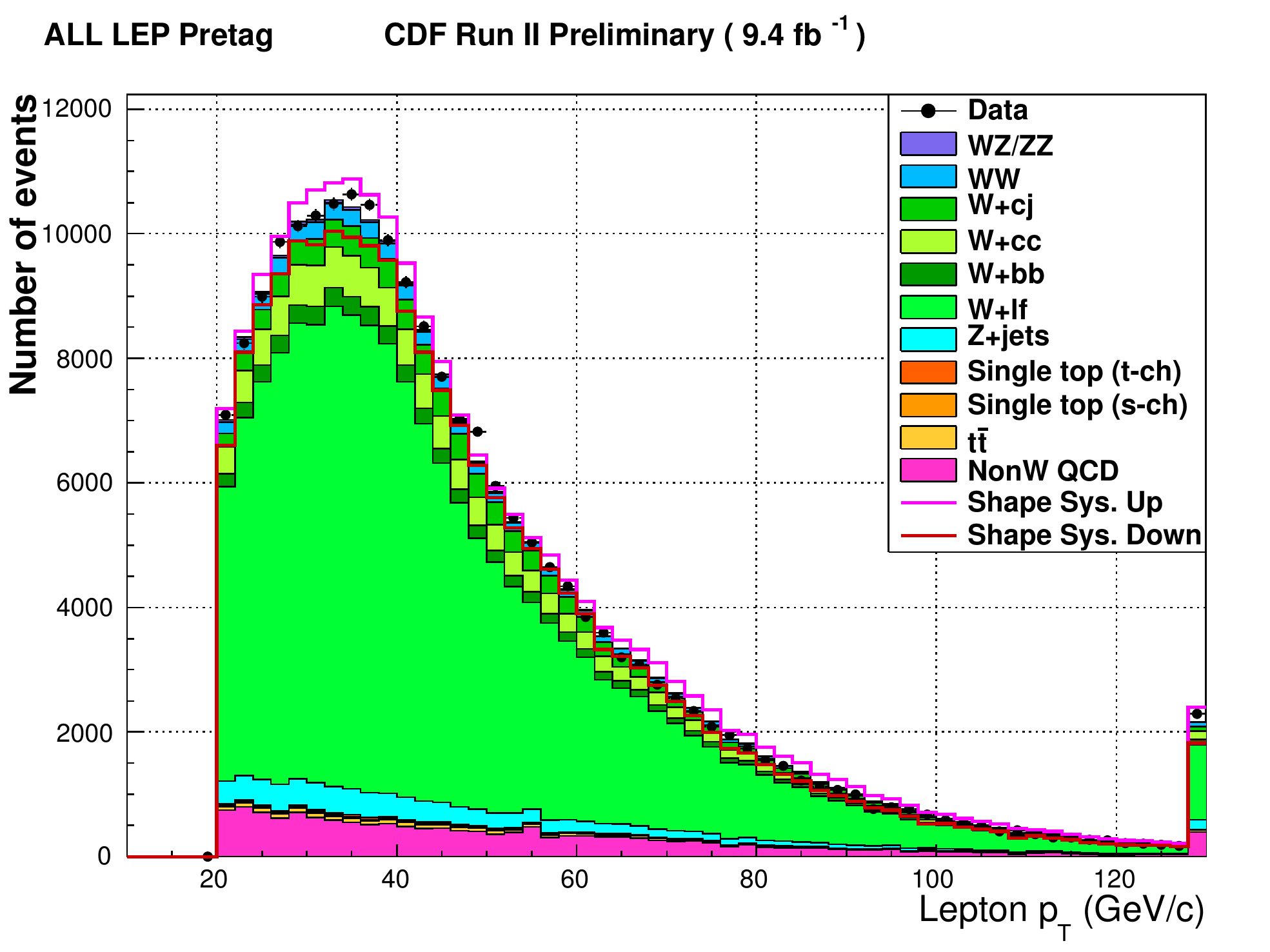}
    \includegraphics[width=0.495\textwidth]{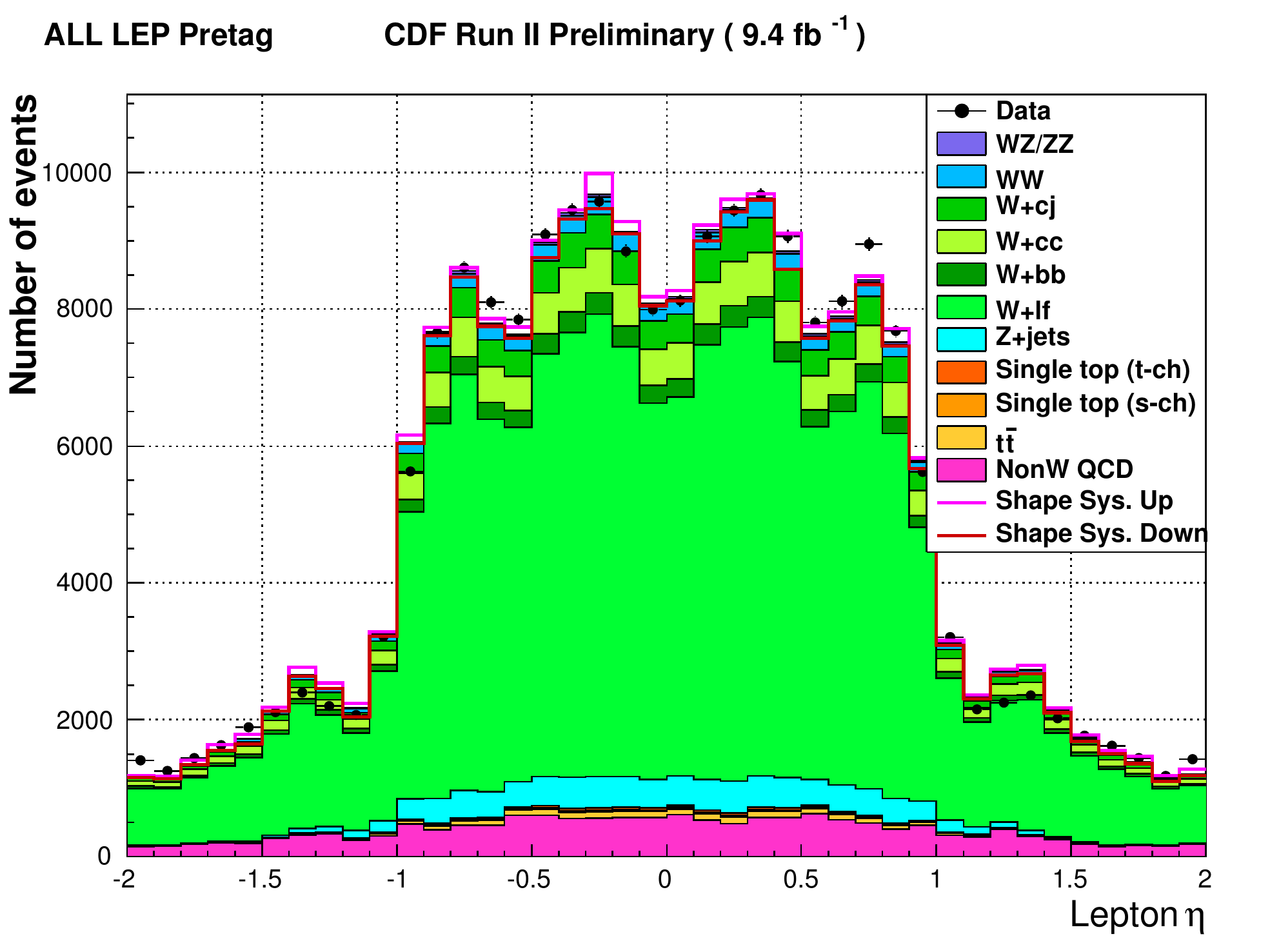}
    \end{center}
    \caption[Lepton $p_T$ and $\eta$ Distributions in Pretag Control Region]{Two lepton related kinematic distributions in the pretag control region, for all the lepton categories combined: $p_T$ (left) and $\eta$ (right).}\label{fig:pret_lepKin}
  \end{figure}

\begin{figure}
    \begin{center}
  \includegraphics[width=0.495\textwidth]{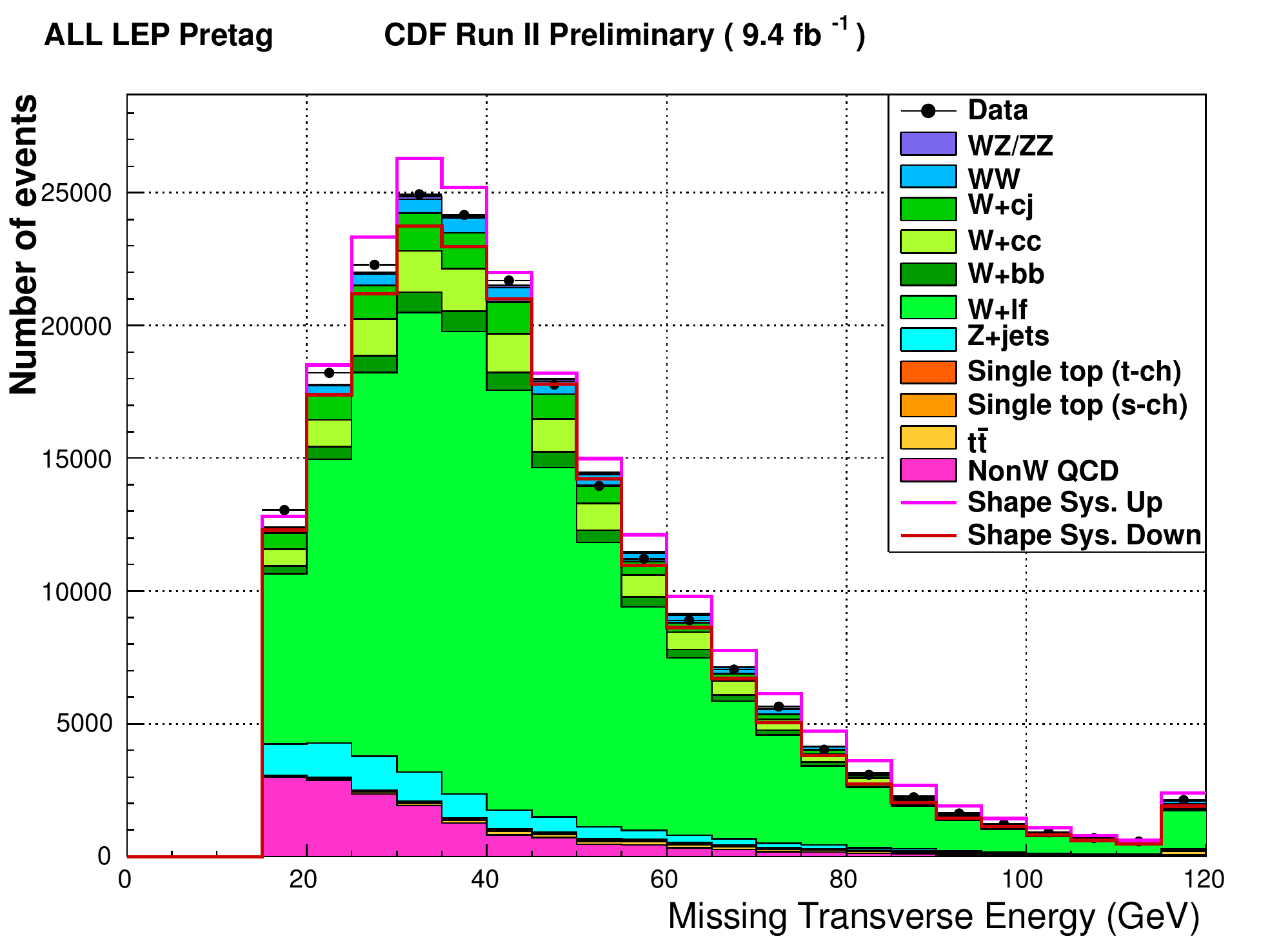}
  \includegraphics[width=0.495\textwidth]{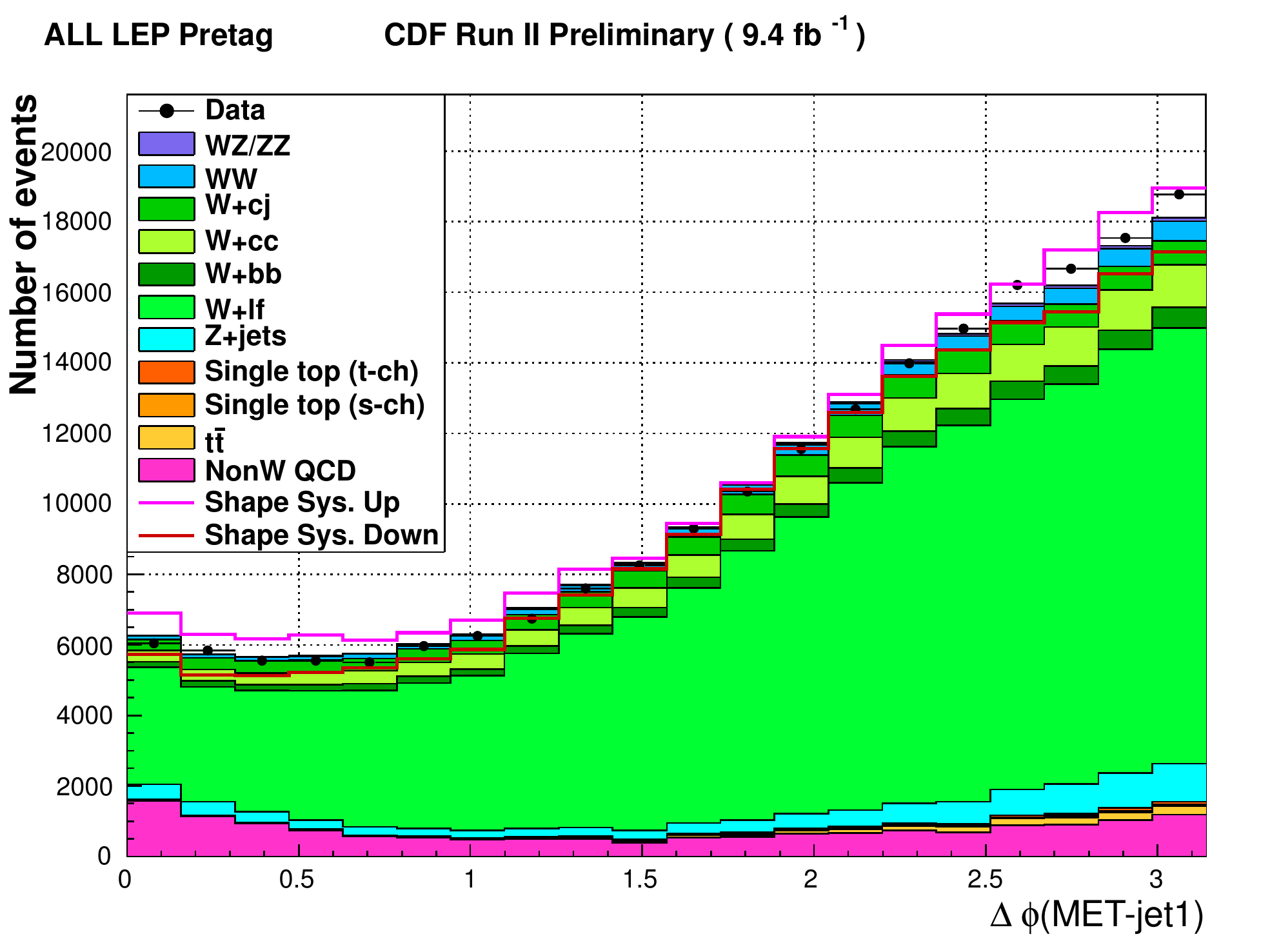}
    \end{center}
  \caption[$\nu$ Kinematic Distributions in Pretag Control Region]{Two $\nu$ related kinematic distributions in the pretag control region, for all the lepton categories combined: \met (left) and $\Delta\phi(jet1, $\met$)$ (right).}\label{fig:pret_metKin}
  \end{figure}

\begin{figure}
    \begin{center}      
      \includegraphics[width=0.49\textwidth]{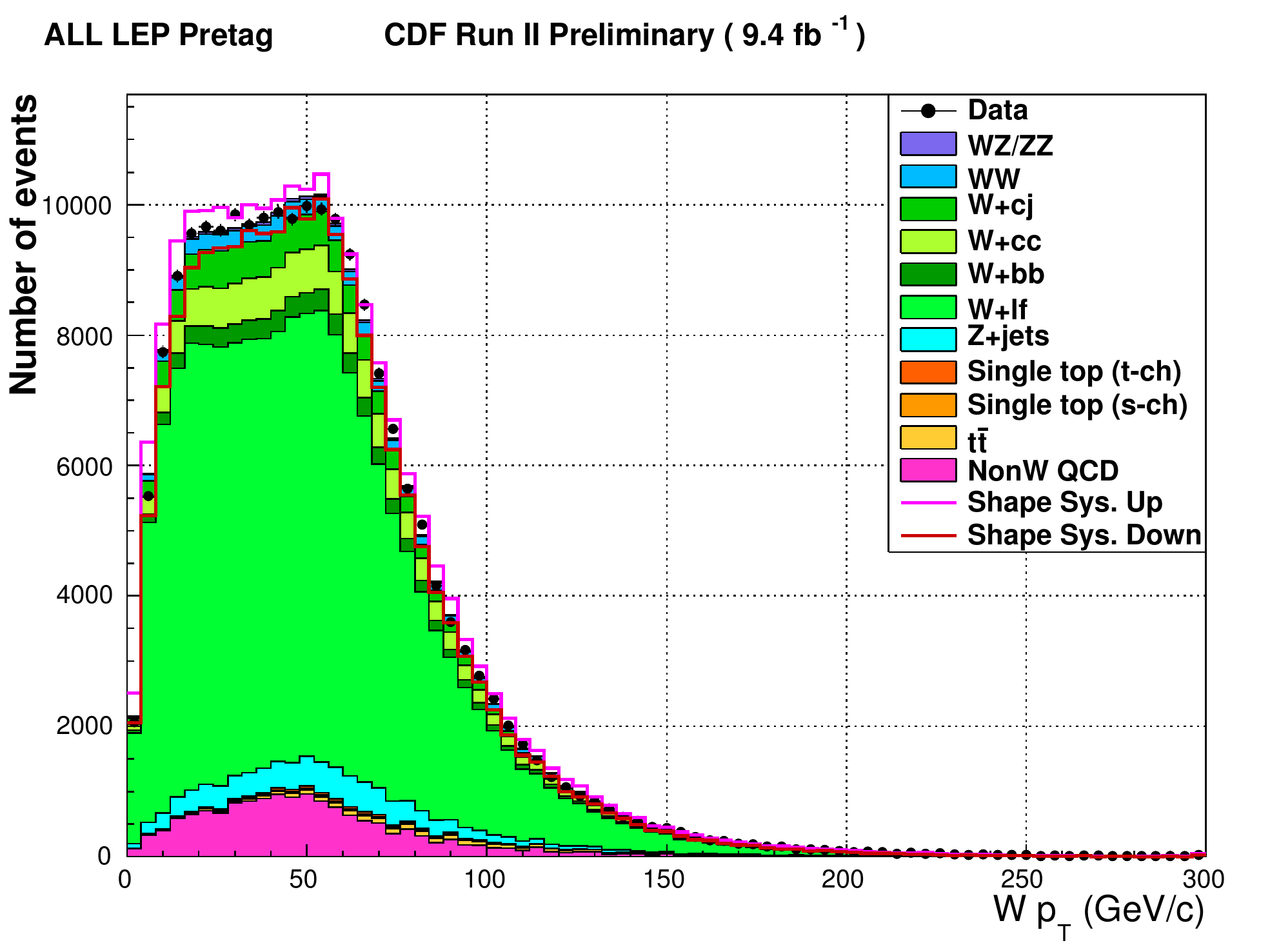}
      \includegraphics[width=0.49\textwidth]{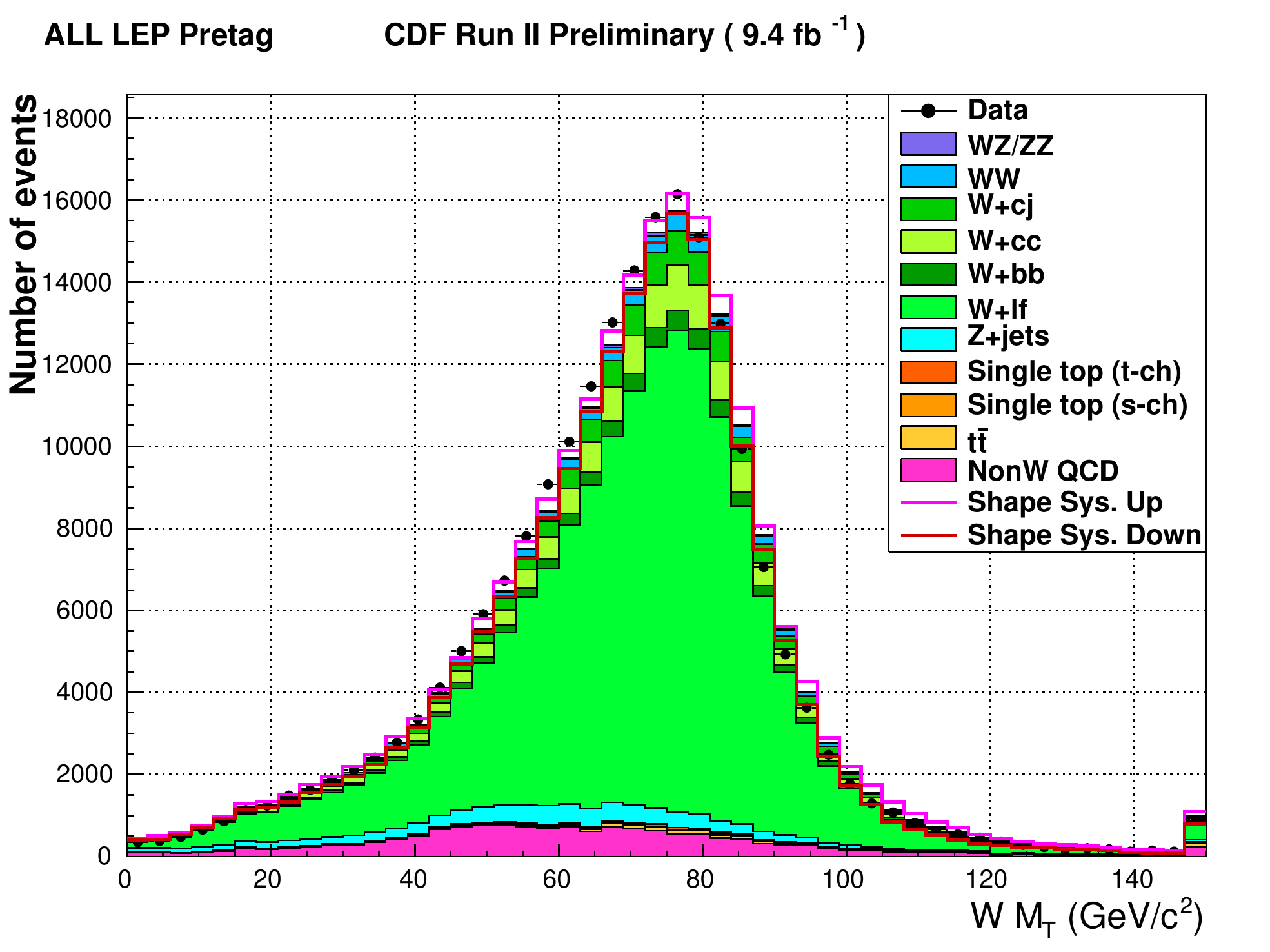}
      \caption[$W$ Kinematic Distributions in Pretag Control Region]{Two $W$ related kinematic distributions in the pretag control region, for all the lepton categories combined: $p_T^W$ (left) and $M_T^W$ (right).}\label{fig:pret_wkin}
    \end{center}
  \end{figure}

\begin{figure}
    \begin{center}
      \includegraphics[width=0.49\textwidth]{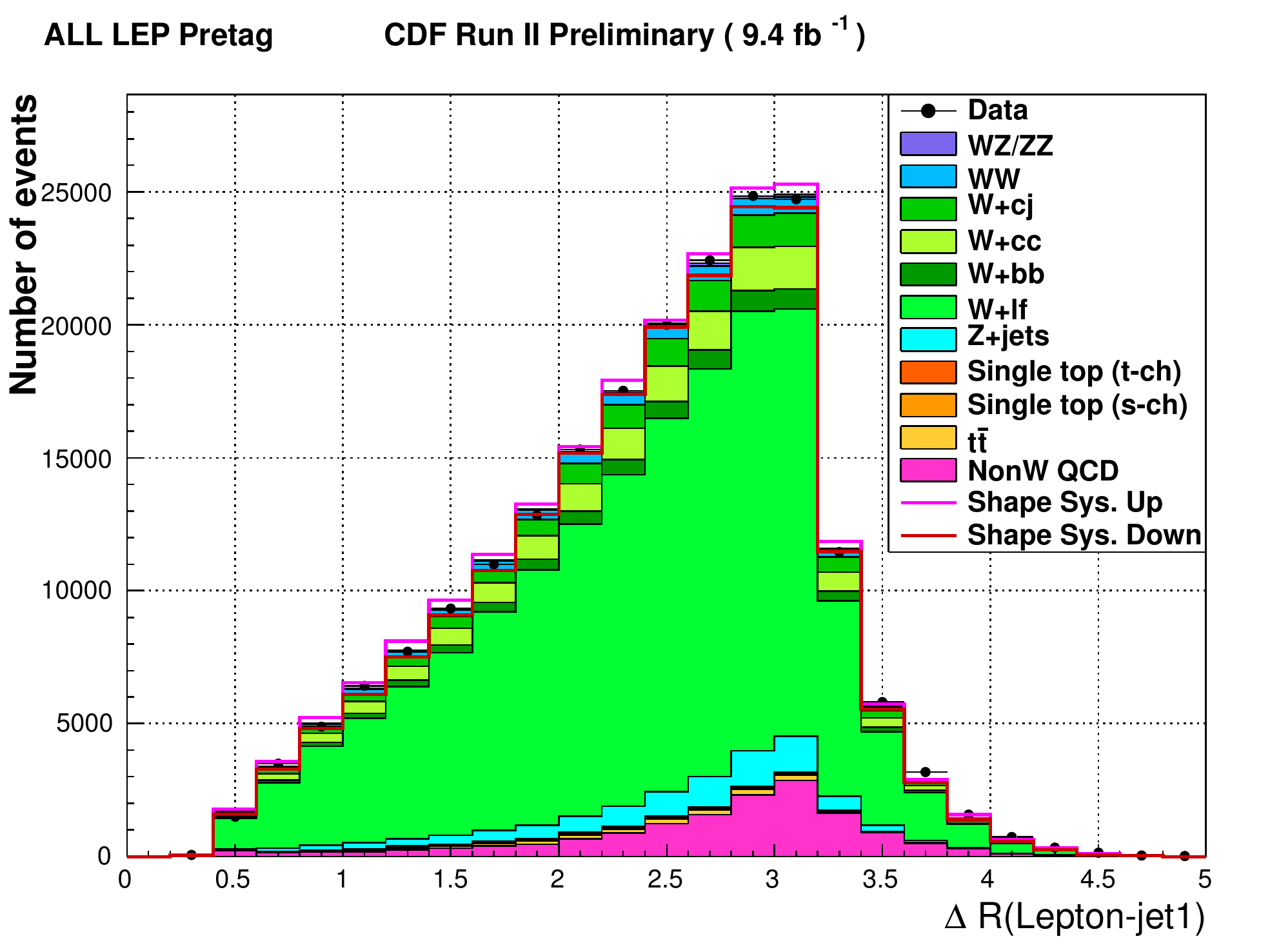}
      \includegraphics[width=0.49\textwidth]{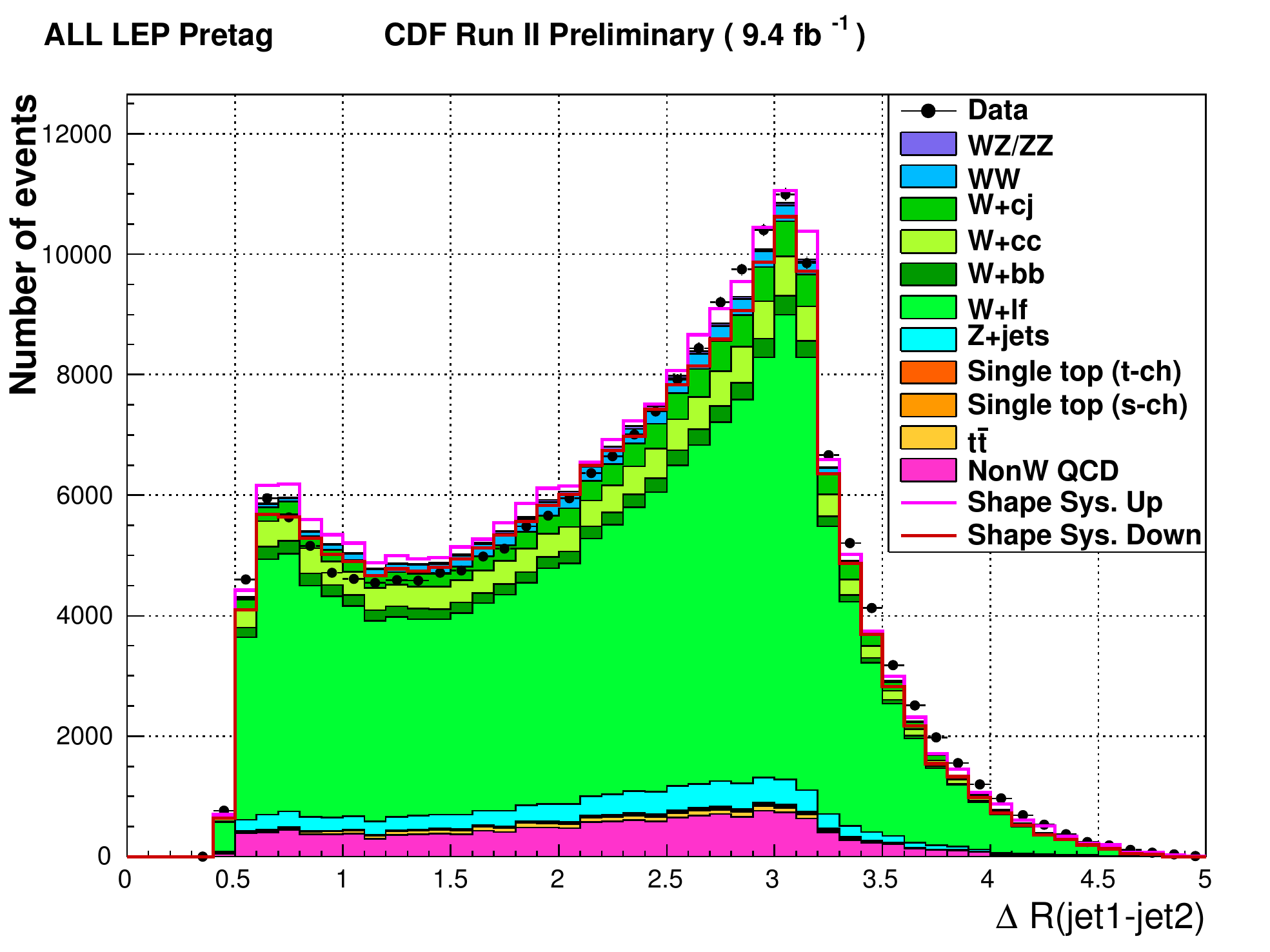}     
    \end{center}
    \caption[Angular Separation Distributions in Pretag Control Region]{Two angular separation distributions in the pretag control region, for all the lepton categories combined: $\Delta R(lep,jet1)$ (left), $\Delta R(jet1,jet2)$, (right).}\label{fig:pret_dphi}
  \end{figure}

\begin{figure}
\begin{center}
\includegraphics[width=0.98\textwidth]{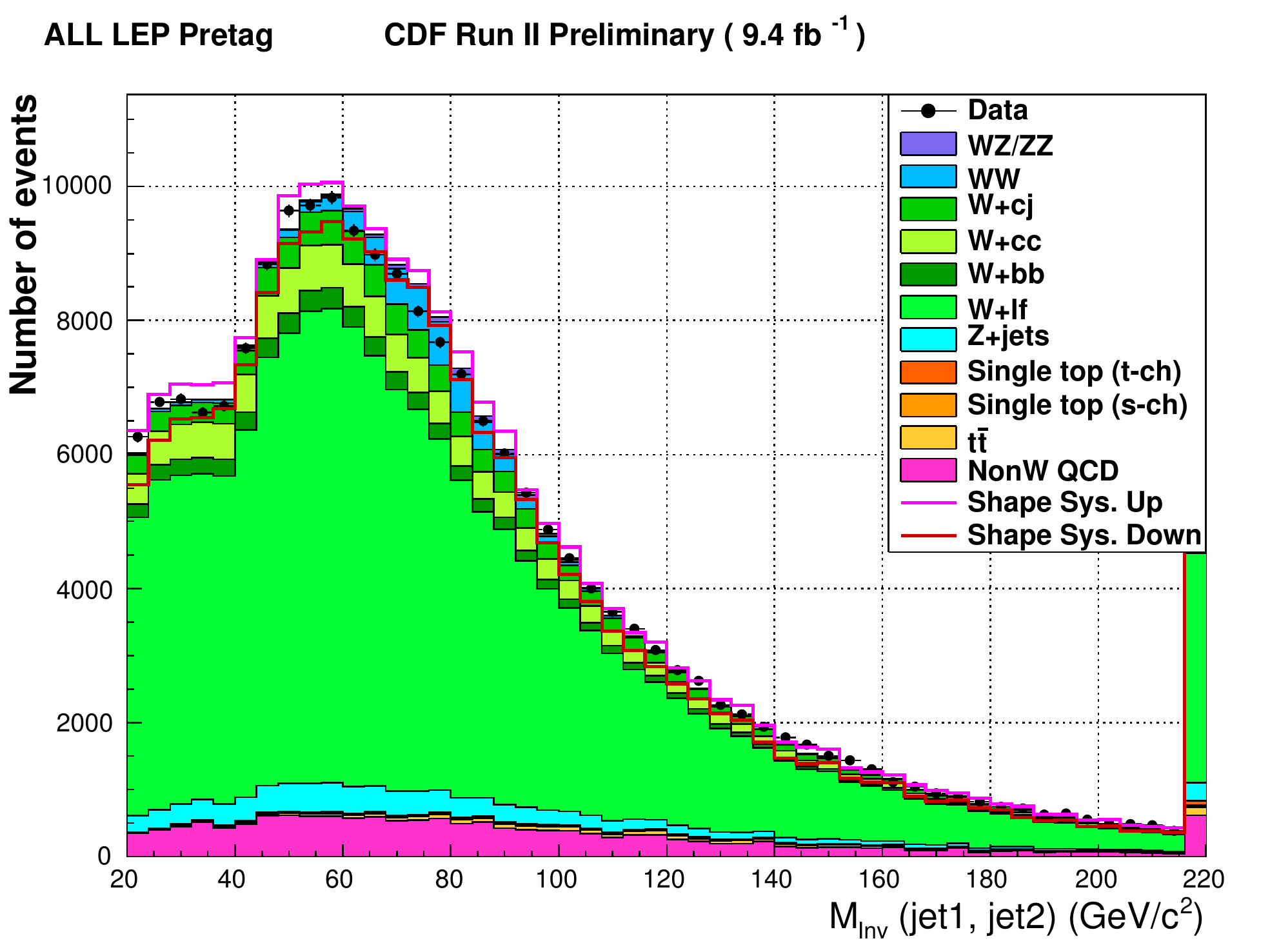}
\caption[$M_{Inv}(jet1, jet2)$ Distribution in Pretag Control Region]{Invariant mass distribution of the selected di-jet pair in the pretag control region, for all the lepton categories combined.}\label{fig:pret_mjj}
\end{center}
\end{figure}

The first important validation criteria is the agreement of the kinematic distributions in the pretag control region. As the normalization of the region is constrained by data the most important effect comes from the evaluation of the shapes. Figures from~\ref{fig:pret_jetEt} to~\ref{fig:pret_mjj} show several kinematic variables after the composition of all the background for all the lepton categories. For each variable, we estimated a central pretag background value and four systematic variations: 
\begin{itemize}
\item jet energy scale plus and minus one sigma (JES $+1\sigma$, JES $-1\sigma$);
\item $W+n$~partons renormalization scale variation: $Q^2$ is doubled ($Q^2_{Up}= 2.0 Q^2 $) and halved ($Q^2_{Down}= 0.5 Q^2$).
\end{itemize}
The positive (negative) systematic variations are added in quadrature and the result is shown together with the central background evaluation. The agreement is excellent within uncertainties. Few distributions that show mis-modeling, like the $\eta$ of the second jet shown in Figure~\ref{fig:pret_jetEta}, become irrelevant after the tagging requirement that enforces a more central selection of the jets.

%

The other validation method is the actual comparison of the expected backgrounds in the signal region against the observed selection. Tables~\ref{tab:1tag_final} and~\ref{tab:2tag_final} report the complete background composition in each of the separate lepton categories for the single and double \texttt{SecVtx}-tagged selections, neither JES nor $Q^2$ variations are reported in the tables. All the estimates are within the uncertainties, that are, however, rather large because of the $K^{HF}$ factor uncertainties listed in Table~\ref{tab:k-fac}. 

\begin{sidewaystable}
  \begin{center}
    \begin{tabular}{lccccc}
  \toprule
        \multicolumn{6}{c}{\bf Single-tag Event Selection}\\
        Lepton ID & CEM                  &  PHX         &             CMUP             & CMX    & EMC    \\  
\midrule                                                         
   $Z+$jets           & 55.53 $\pm$ 4.73     & 7.76 $\pm$ 0.67       & 65.3 $\pm$ 5.73        & 37.14 $\pm$ 3.28    & 104.18 $\pm$ 10.8  \\    
   $t\bar{t}$      & 237.55 $\pm$ 23.3    & 46.93 $\pm$ 4.59      & 139.68 $\pm$ 13.8      & 62.22 $\pm$ 6.14    & 228.58 $\pm$ 25.6   \\   
   Single-top $s$  & 64.23 $\pm$ 5.88     & 11.08 $\pm$ 1.02      & 36.42 $\pm$ 3.35       & 16.11 $\pm$ 1.49    & 51.68 $\pm$ 5.48    \\   
   Single-top $t$  & 84.95 $\pm$ 9.99     & 16.86 $\pm$ 1.98      & 47.75 $\pm$ 5.64       & 22.5 $\pm$ 2.66     & 66.22 $\pm$ 8.56    \\   
   $WW$              & 84.35 $\pm$ 11.8          & 25.05 $\pm$ 3.54       & 43.7 $\pm$ 6.17        & 23.68 $\pm$ 3.35      & 53.11 $\pm$ 8     \\
   $ZZ$              & 1.85 $\pm$ 0.19           & 0.21 $\pm$ 0.02        & 2.45 $\pm$ 0.25        & 1.37 $\pm$ 0.14       & 3.45 $\pm$ 0.39  \\ 
   $WZ$              & 29.2 $\pm$ 2.95           & 12.32 $\pm$ 1.22       & 16 $\pm$ 1.66          & 9.21 $\pm$ 0.94       & 20.54 $\pm$ 2.39  \\
   $W+b\bar{b}$    & 858.71 $\pm$ 258        & 263.08 $\pm$ 79.1       & 428.58 $\pm$ 129       & 238.99 $\pm$ 71.9    & 409.97 $\pm$ 123   \\
   $W+c\bar{c}$    & 441.77 $\pm$ 134        & 145.08 $\pm$ 44.1       & 212.98 $\pm$ 64.8      & 120.26 $\pm$ 36.6    & 214 $\pm$ 65       \\
   $W+cj$            & 342.86 $\pm$ 104        & 96.42 $\pm$ 29.3       & 171.77 $\pm$ 52.2      & 93.01 $\pm$ 28.3     & 143.47 $\pm$ 43.6  \\ 
   $W+LF$         & 809.01 $\pm$ 87.1       & 302.78 $\pm$ 32.2       & 408.74 $\pm$ 43.3      & 230.75 $\pm$ 24.7    & 463.65 $\pm$ 53.5  \\
   Non-$W$           & 302.69 $\pm$ 121        & 205.06 $\pm$ 82       & 58.75 $\pm$ 23.5       & 27.96 $\pm$ 11.2     & 106.35 $\pm$ 42.5    \\
\midrule                                                                                                                                    
   {\bf Prediction}      & {\bf 3312.68 $\pm$ 521}       & {\bf 1132.63 $\pm$ 176 }      & {\bf 1632.11 $\pm$ 253 }     & {\bf 883.21 $\pm$ 140}     & {\bf 1865.2 $\pm$ 248}   \\
   {\bf Observed }       & {\bf 3115}                    & {\bf 1073}                  & {\bf 1577 }                  & {\bf 830}                   & {\bf 1705}               \\
\midrule                                                                                                                                
   {\bf Dibosons}     & {\bf 115.39 $\pm$ 13}           & {\bf 37.57 $\pm$ 4.05}      & {\bf 62.15 $\pm$ 6.87}       & {\bf 34.26 $\pm$ 3.75}    & {\bf 77.09 $\pm$ 9.35}    \\
\bottomrule
    \end{tabular}  
  \end{center}
    \caption[Observed and Expected Events with One \texttt{SecVtx} Tags]{Summary of observed and expected events with one secondary vertex tag (\texttt{SecVtx}), in the $W+2$~jets sample, in 
      $9.4$~fb$^{-1}$ of data in the different lepton categories. Statistical and systematics rate uncertainties are included in the table except for the contribution of the JES and the $Q^2$  variations. Diboson expected yield is added in the prediction.}\label{tab:1tag_final}
\end{sidewaystable}

\begin{sidewaystable}
  \begin{center}

  \begin{tabular}{lccccc}
        \toprule 
        \multicolumn{6}{c}{\bf Double-tag Event Selection}\\
        Lepton ID         & CEM                     & PHX                    &      CMUP                   & CMX                     & EMC               \\ 
        \midrule                                                                                                                             
       $Z+$jets                  & 1.43 $\pm$ 0.13      & 0.27 $\pm$ 0.02      & 2.84 $\pm$ 0.27        & 1.41 $\pm$ 0.13          & 4.6 $\pm$ 0.5           \\
       $t\bar{t}$             & 48.21 $\pm$ 6.99     & 9.91 $\pm$ 1.44      & 27.31 $\pm$ 3.98       & 12.37 $\pm$ 1.8          & 44.89 $\pm$ 6.95        \\
       Single-top $s$         & 16.89 $\pm$ 2.36     & 2.87 $\pm$ 0.4       & 9.68 $\pm$ 1.36        & 4.17 $\pm$ 0.58          & 13.72 $\pm$ 2.05        \\
       Single-top $t$         & 5.07 $\pm$ 0.81      & 1.13 $\pm$ 0.18      & 2.85 $\pm$ 0.46        & 1.34 $\pm$ 0.22          & 4.16 $\pm$ 0.71         \\
       $WW$                     & 0.72 $\pm$ 0.19      & 0.18 $\pm$ 0.05      & 0.35 $\pm$ 0.09        & 0.2 $\pm$ 0.05           & 0.49 $\pm$ 0.13         \\
       $ZZ$                     & 0.26 $\pm$ 0.04      & 0.03 $\pm$ 0         & 0.46 $\pm$ 0.06        & 0.29 $\pm$ 0.04          & 0.63 $\pm$ 0.1          \\  
      $WZ$                     & 5.28 $\pm$ 0.75      & 2.6 $\pm$ 0.37       & 2.52 $\pm$ 0.36        & 1.67 $\pm$ 0.24          & 3.52 $\pm$ 0.54         \\
       $W+b\bar{b}$           & 114.7 $\pm$ 35.1     & 33.92 $\pm$ 10.4     & 59.06 $\pm$ 18.1       & 29.49 $\pm$ 9.04         & 60.71 $\pm$ 18.6        \\
       $W+c\bar{c}$           & 6.68 $\pm$ 2.1       & 2.16 $\pm$ 0.68      & 3.41 $\pm$ 1.08        & 1.63 $\pm$ 0.51          & 4 $\pm$ 1.25            \\     
       $W+cj$                   & 5.18 $\pm$ 1.63      & 1.43 $\pm$ 0.45      & 2.75 $\pm$ 0.87        & 1.26 $\pm$ 0.4           & 2.69 $\pm$ 0.84         \\   
       $W+LF$           & 4.53 $\pm$ 0.94      & 1.7 $\pm$ 0.36       & 2.35 $\pm$ 0.48        & 1.28 $\pm$ 0.27          & 2.98 $\pm$ 0.66         \\    
       Non-$W$                  & 5.58 $\pm$ 2.23      & 6.79 $\pm$ 2.72      & 4.31 $\pm$ 1.73        & 0 $\pm$ 0.5              & 0 $\pm$ 0.5             \\ 
       \midrule                                                          
      {\bf  Prediction}            & {\bf 214.53 $\pm$ 40.5}    & {\bf 62.99 $\pm$ 12.1}     & {\bf 117.92 $\pm$ 21.1}      & {\bf 55.11 $\pm$ 10.4}         & {\bf 142.38 $\pm$ 23.3}       \\
       {\bf Observed}                 & {\bf 175}                 & {\bf 62 }                   & {\bf 92}                        & {\bf 49 }                        & {\bf 126}              \\     
       \midrule                                                          
       {\bf Dibosons }          & {\bf 6.26 $\pm$ 0.79}           & {\bf 2.8 $\pm$ 0.38}       & {\bf 3.34 $\pm$ 0.4}         & {\bf 2.15 $\pm$ 0.26}          & {\bf 4.64 $\pm$ 0.61 }        \\
       \bottomrule
      \end{tabular}        
  \end{center}
  \caption[Observed and Expected Events with Two \texttt{SecVtx} Tags]{Summary of observed and expected events with two secondary vertex tags (\texttt{SecVtx}), in the $W+2$~jets sample, in 
      $9.4$~fb$^{-1}$ of data in the different lepton categories. Statistical and systematics rate uncertainties are included in the table except for the contribution of the JES and the $Q^2$  variations. Diboson expected yield is added in the prediction.}\label{tab:2tag_final}
\end{sidewaystable}

Background is still large although the diboson signal is sizable. An analysis of the $diboson\to \ell\nu+HF$ properties with a counting experiment is not feasible. Therefore, in the next Chapter, we exploit the separation power of two variables: 
\begin{itemize}
\item the di-jet invariant mass, $M_{Inv}(jet1,jet2)$, improves the separation between the non-resonant $W+HF$ production and the diboson.
\item KIT-NN improves the $b$-jets $vs$ $c$-jets separation.
\end{itemize}
The shape analysis and the fitting procedure also constraint the large normalization uncertainties, thereby increasing the significance of the measurement.

\clearpage
\chapter{Statistical Analysis and Results}\label{chap:StatRes}

The measurement of a potential signal (and its properties) over a predicted background requires a statistical analysis of the selected events.

The detection of diboson events in the $\ell\nu+HF$ final state is a challenging problem as it combines a small expected signal yield, a sizable irreducible background and large systematic uncertainties typical of the hadronic environment.

After the full event selection (Chapter~\ref{chap:sel}), the remaining background processes are mainly of irreducible nature ($W+c/c\bar{c}/b\bar{b}$) and still overwhelm the signal by more than a factor twenty. 

As the diboson production is a resonant process while $W+HF$ is not, a feasible and optimal signal extraction strategy is the shape analysis of the di-jet invariant mass distribution, $M_{Inv}(jet1,jet2)$.

The use of the di-jet mass has two effects. Firstly, we can not distinguish $WZ\to\ell\nu+b\bar{b}/c\bar{c}$ decay from $ZZ\to\ell\not{\ell}+b\bar{b}/c\bar{c}$ decay where one lepton is lost. Therefore, from now on, the $WZ$ diboson production is considered together with the small amount of selected $ZZ$ events\footnote{Table~\ref{tab:data_sig} shows that $ZZ$ is about $1/10$ of the $WZ$ expected events and 1/30 of the total diboson sample.}. A more important effect is that, due to the low mass resolution, typical of the hadronic final states, we can not separate contribution from $WW\to \ell\nu+c\bar{s}$ from the $WZ\to \ell\nu+b\bar{b}/c\bar{c}$, in the single-tagged event selection\footnote{In the double-tagged event selection the only signal contribution comes from $WZ\to \ell\nu+b\bar{b}$.}. We solve this problem with the use of a second variable, the NN flavor-separator (KIT-NN) described in Section~\ref{sec:kitnn}. We use it on single-tagged events to build a bi-dimensional distribution:
\begin{equation}
  M_{Inv}(jet1,jet2)\textrm{ {\em vs }KIT-NN}  \textrm{.}
\end{equation}
This, together with the simple $M_{Inv}(jet1,jet2)$ distribution of the double-tagged events, allows to measure $WW$ and $WZ/ZZ$ contributions separately.

The statistical analysis is performed with a methodology and a software tool, named {\em mclimit}~\cite{mclimit, single_top}, also used in the CDF and Tevatron Higgs searches~\cite{cdf_cmb, tev_cmb} and adapted to the present analysis.

The technique, described in Section~\ref{sec:likelihood}, is based on the integration, over data, of a binned likelihood function where the diboson production cross section is a free parameter. The systematic uncertainties, listed in Section~\ref{sec:sys_desc}, are included with a Bayesian approach and marginalized to increase the sensitivity. 

\begin{figure}
  \includegraphics[width=0.495\textwidth]{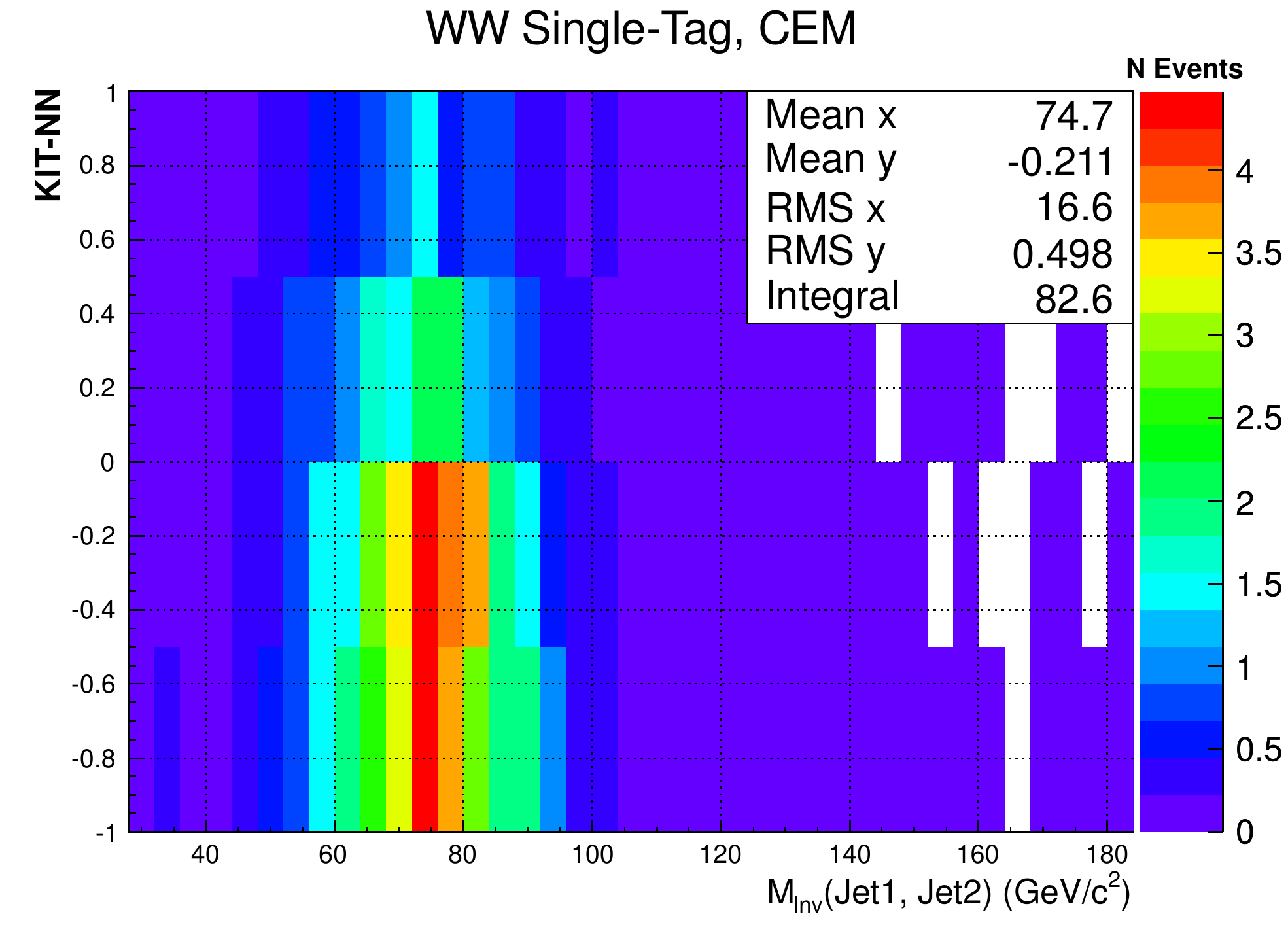}
  \includegraphics[width=0.495\textwidth]{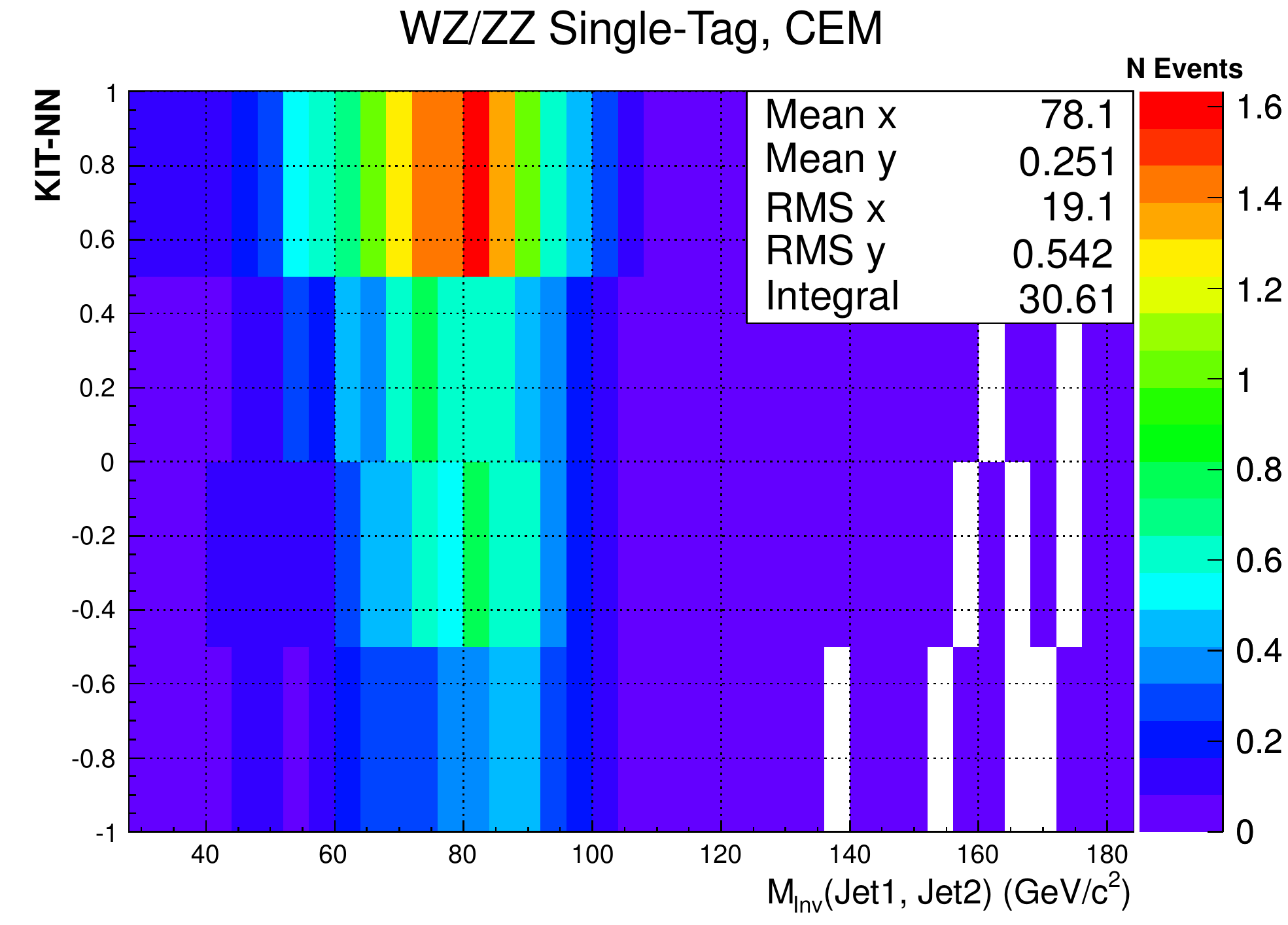}\\
  \includegraphics[width=0.495\textwidth]{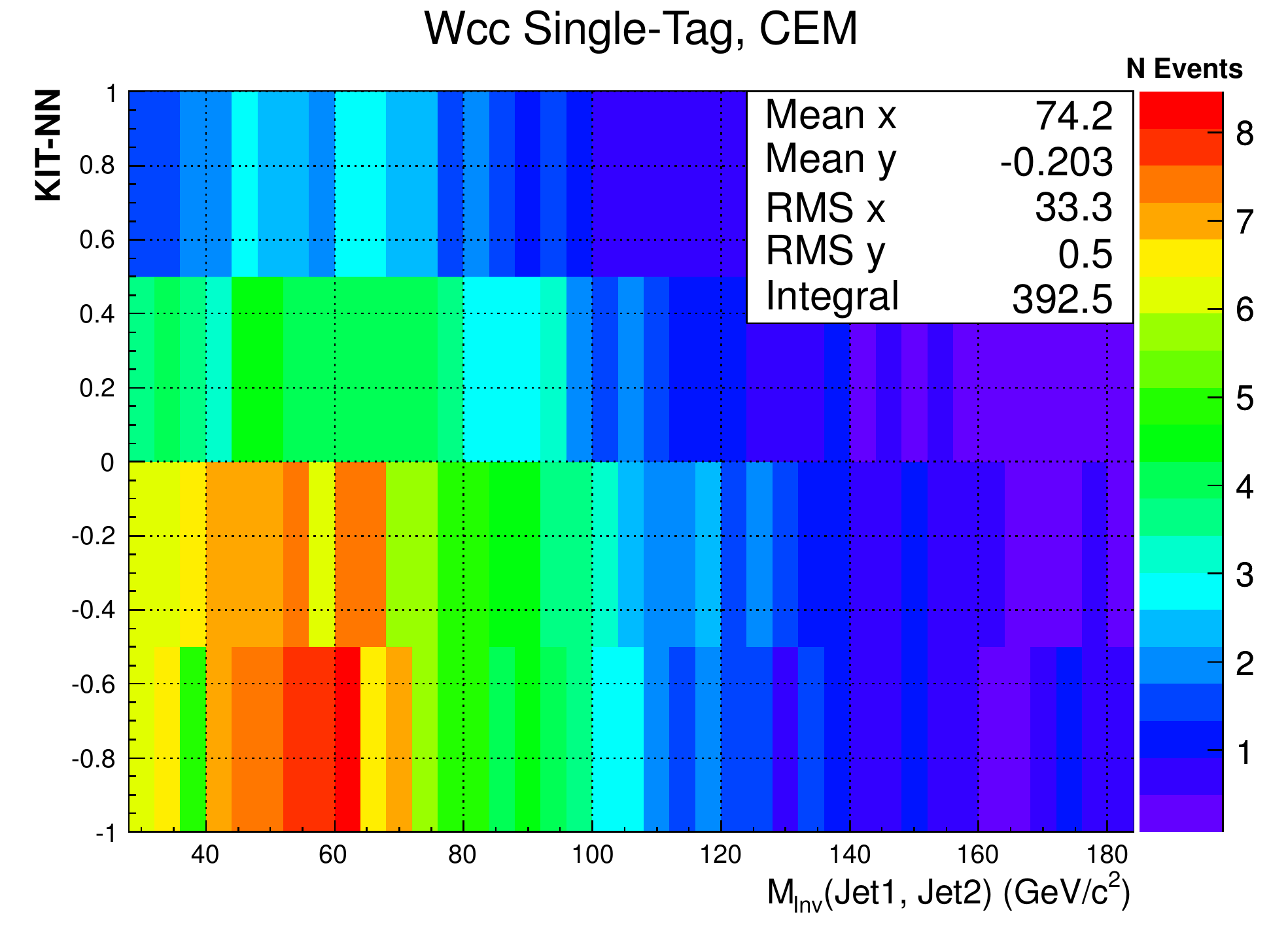}
  \includegraphics[width=0.495\textwidth]{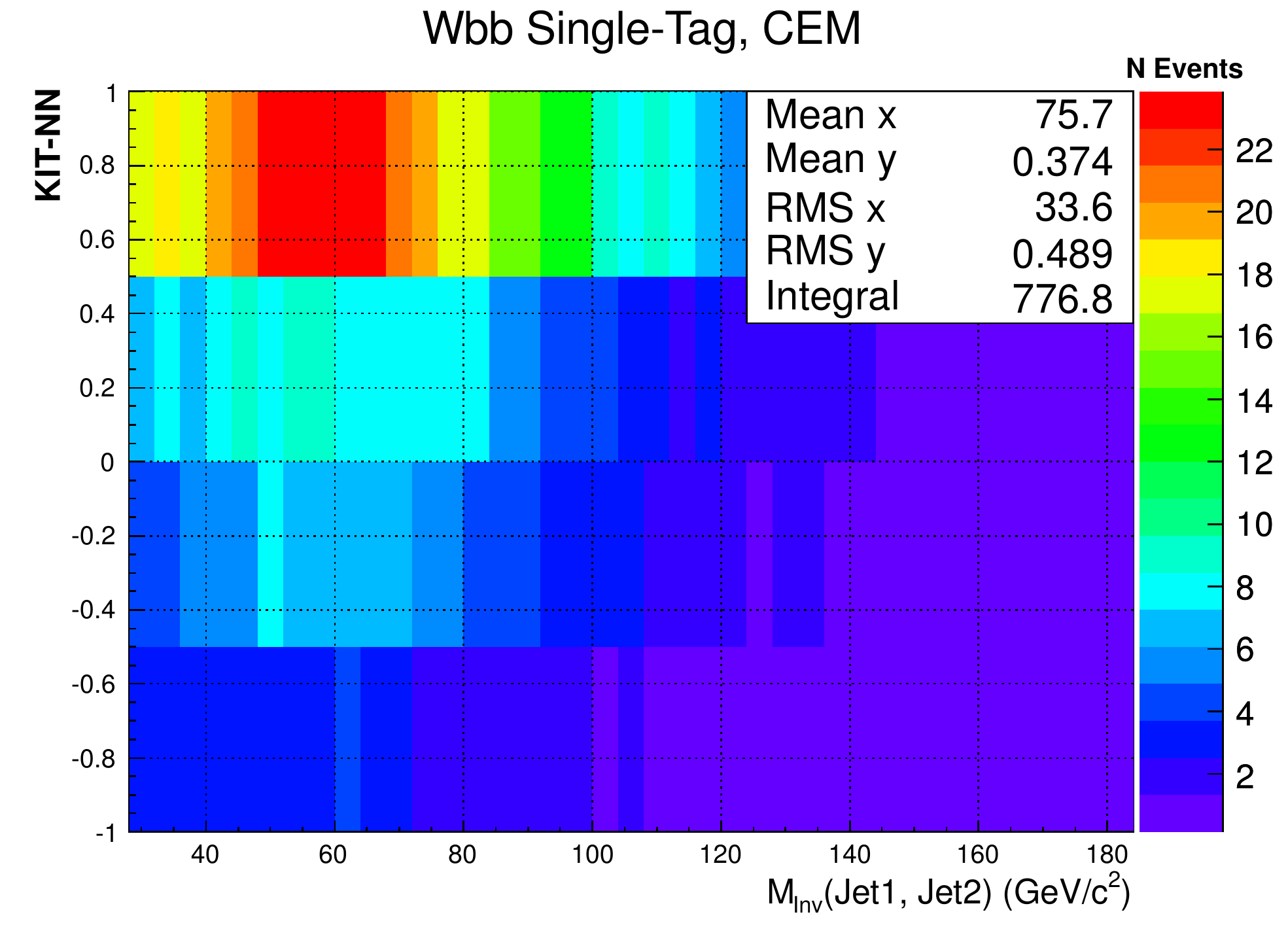}
  \caption[Example of $M_{Inv}(jet1,jet2)$ {\em vs } KIT-NN Templates]{Four examples of the bi-dimensional templates, $M_{Inv}(jet1,jet2)$ {\em vs } KIT-NN, used for the statistical analysis of the  single-tagged channels in the CEM lepton selection: $WW$ signal (top left), $WZ/ZZ$ signals (top right), $W+c\bar{c}$ background (bottom left), $W+b\bar{b}$ background (bottom right). The contribution of $c$ and $b$ quarks cluster at opposite sides of the KIT-NN distribution, the diboson production cluster around an invariant mass resonant peak while the generic $HF$ production spreads along the $M_{Inv}(jet1,jet2)$ direction.}\label{fig:2d_templates}
\end{figure}

The likelihood is built by dividing the selected data in eight orthogonal channels depending on the kinematic properties and on the background composition:
\begin{itemize}
\item {\bf 4 Lepton categories} are derived from the data-streams used to collect the events (see Section~\ref{sec:onlineSel}): central electrons (CEM), forward electrons (PHX), central muons (CMUP+CMX) and extended muons (EMC). The kinematic is homogeneous within each sample.
\item {\bf 2 Tag categories} are derived from the single or double \texttt{SecVtx} tagged event selection for each lepton category. The sample composition drastically changes in the two regions. Section~\ref{sec:mjj_kit_distrib} shows the $M_{Inv}(jet1,jet2)$ distribution in the double-tagged categories and slices of the bi-dimensional distribution, $M_{Inv}(jet1,jet2)$ {\em vs} KIT-NN, used for the single-tagged categories. Figure~\ref{fig:2d_templates} shows four examples (for the CEM lepton selection) of the bi-dimensional templates used for the single-tagged channels evaluation: $WW$, $WZ/ZZ$ signals and $W+c\bar{c}$, $W+b\bar{b}$ backgrounds.
\end{itemize}

The measured diboson cross sections, $\sigma_{WW}$ and  $\sigma_{WZ/ZZ}$, are obtained in Section~\ref{sec:cx_measure} from the evaluation of the Bayesian posterior distribution of the combined likelihood. 

As a first result, we measure the total diboson production cross section, fixing the $WW$ and the $WZ/ZZ$ relative contributions to the SM prediction. Then we remove the constraint on the $WW$ and the $WZ/ZZ$ relative fractions. We let them free to float independently so to obtain a measurement where the full correlation between $\sigma_{WW}$ and  $\sigma_{WZ/ZZ}$ is accounted. Finally we also obtain two individual measurements integrating out $\sigma_{WW}$ or  $\sigma_{WZ/ZZ}$ one at the time from the Bayesian posterior distribution.

In Section~\ref{sec:sigma_eval}, we discuss the significance of the different measurements and we show the evidence of the total diboson production. The previous version of this analysis~\cite{cdfDiblvHF_Mjj2011}, performed with a dataset of $7.5$~fb$^{-1}$ and reported in Appendix~\ref{App:7.5}, was the first evidence of this process at a hadron collider.

In general the result confirms the SM prediction for diboson production also when they are detected in the $HF$ final state. This provides additional confidence on the capability and the methods used by the CDF collaboration for the Higgs search in the $H\to b\bar{b}$ channel.

\section{Construction of the Likelihood Function}\label{sec:likelihood}

The comparison of observed data with the templates derived from signal and background is possible with the use of a likelihood function in which we use a Bayesian approach to include the prior probabilities of the background and systematic effects.

We assume that the outcome of the data observation $n_{i,j}$, in the $j^{th}$ bin of the $i^{th}$ input histogram follows the Poisson statistics and the expectation value, $\mu_{i,j}$, depends by the estimated backgrounds and signals. Therefore the complete likelihood function, $\mathscr{L}$, of the bin by bin outcome is:
\begin{equation}\label{eq:likelihood}
  \mathscr{L}(\vec{\alpha}, \vec{s}, \vec{b}| \vec{n}, \vec{\beta}) =\prod^{Hists}_{i} \prod^{Bins}_{j} \frac{e^{-\mu_{i,j}}\mu_{i,j}^{n_{i,j}}}{n_{i,j}!}
\end{equation}
where the first product runs over the total number of input channels, $Hist$ (i.e. the different histograms provided for them), and the second runs on the bins of each distribution. The bin expectation value contains the signal and background dependence: 
\begin{equation}\label{eq:mu_likelihood}
  \mu_{i,j}(\vec{\alpha}, \vec{s}, \vec{b}| \vec{\beta}) = \sum_p^{Sgn} \alpha_{p} s_{i,j,p}(\vec{\beta}) + \sum_q^{Bkg} b_{i,j,q}(\vec{\beta})\textrm{,}
\end{equation}
where $s_{i,j,p}$ represents the $p^{th}$ unknown signal, $b_{i,j,q}$ is the $q^{th}$ estimated background process and $\alpha_{p}$ is the {\em scaling parameter} used to measure the amount of the different signals. The unknown expectation of the signal is parametrized by flat, uniform, positive prior distribution in $\vec{\alpha}$. The last element of Equations~\ref{eq:likelihood} and~\ref{eq:mu_likelihood} is $\vec{\beta}$, the vector of the nuisance parameters: it incorporates in the likelihood the systematic uncertainties. They are parametrized as fractional variations on the $s_{i,j,p}$ and $b_{i,j,q}$ rates: 
\begin{equation}
   s_{i,j,p}(\vec{\beta})= s_{i,j,p}^{central}\prod_{k}^{Sys}(1+\frac{\sigma_{i,j,k}}{s_{i,j,p}^{central}}\beta_k)
\end{equation}
\begin{equation}
   b_{i,j,p}(\vec{\beta})= b_{i,j,p}^{central}\prod_{k}^{Sys}(1+\frac{\sigma_{i,j,k}}{b_{i,j,p}^{central}}\beta_k)  
\end{equation}
where $\frac{\sigma_{i,j,k}}{s_{i,j,p}^{central}}$, and $\frac{\sigma_{i,j,k}}{b_{i,j,p}^{central}}$ are the relative uncertainties related to the systematic effect $k$ and the $\beta_k$ variables. It is important to notice that we distinguish:
\begin{itemize}
\item {\em total rate uncertainties}, like the luminosity dependence of the signal, where there is no bin by bin variation of the systematic effect: $\sigma_{i,j,k}\equiv \sigma_{i,k}$.
\item {\em shape uncertainties} where the rate variation can change on a bin by bin basis, evaluated by the ratios between the central and the varied histograms.
\end{itemize}
To account for the nuisance parameters effect and correlate them across different channels we introduce in Equation~\ref{eq:likelihood}, for each systematic, a Gaussian probability constraint centered in zero and with unitary variance. The final result is:
\begin{equation}\label{eq:likelihood_full}
  \mathscr{L}(\vec{\alpha}, \vec{s}, \vec{b}| \vec{n}, \vec{\beta}) =\prod^{Hists}_{i} \prod^{Bins}_{j} \frac{e^{-\mu_{i,j}}\mu_{i,j}^{n_{i,j}}}{n_{i,j}!}\prod_{k}^{Sys}e^{-\beta_k^{2}/2}\textrm{,}
\end{equation}
the previous equation is the full likelihood used in the analysis of the results.

\subsection{Parameter Measurement and Likelihood Integration}\label{sec:likelihood_meas}

By maximizing the likelihood, we obtain the {\em measurement} of the unknown $\vec{\alpha}$.

The maximum of Equation~\ref{eq:likelihood_full} can be found in different ways: with a fit in the multidimensional space of $\vec{\alpha}$ and $\vec{\beta}$ or with the integration ({\em marginalization}) of the nuisance parameters over their prior probability distributions. 

The second technique, exploited here for the measurements, returns a {\em Bayesian posterior distribution} of $\vec{\alpha}$: the minimal extension that covers $68$\% of the distribution around the maximum gives the one-standard-deviation confidence band and the uncertainty on the measurement.

Section~\ref{sec:cx_measure} reports the results of the marginalization, where the integration of the nuisance parameters is performed numerically with a Markov-chain adaptive integration~\cite{mk-chain_book}.

However we perform also a fit of the likelihood function because it gives the {\em best} outcome for all the nuisance parameters. Those values, the result of the fit and the reduced $\chi^2$, should be consistent with the results obtained by the marginalization.

\section{Sources of Systematic Uncertainties}\label{sec:sys_desc}

Along the preceding Chapters, we discussed several sources of rate and shape systematics and we need to include them in the likelihood Equation~\ref{eq:likelihood_full}. 

The {\em rate-only} systematic effects that we include are summarized in the following points:
\begin{description}
\item{\em Initial and Final State Radiation} uncertainties are estimated by changing (halving and doubling) the parameters related to ISR and FSR emission on the signal MC. Half of difference between the two shifted samples is taken as the systematic uncertainty on the signal samples. The total effect on the signal acceptance is about $4\%$.
\item{\em Parton Distribution Functions} uncertainties are evaluated by reweighting each event of the signal MC according to several PDFs parametrisation and to the generator level information of the event. Then the acceptance is evaluated again giving a rate variation with respect to the original PDFs. The exact procedure is described in~\cite{pdf_note7051}. The effect on the signal acceptance was evaluated in previous similar analyses to be around $1\div 2$\%. It is added in quadrature with the ISR/FSR systematic as they both influence only the signal acceptance.
\item{\em $b$-tag Scale Factor} uncertainty  comes from the measured \texttt{SecVtx} $b$-tag efficiency variation. The $SF_{b-Tag}$ is $0.96\pm0.5$ for a single $b$-matched and \texttt{SecVtx} tagged jet. The uncertainty, propagated through the per-event tagging probability, is assumed to be double in the case of $c$-jets. The uncertainty is applied on the signal and on all the EWK backgrounds described in Section~\ref{sec:mc_bkg}.
\item{\em Luminosity} measurement uncertainty contributes for an overall 6\% rate uncertainty on the signal and on all the EWK backgrounds.
\item{ \em Lepton Acceptance} uncertainty derives from the quadrature sum of trigger efficiency measurements and lepton ID Scale Factors. They range from $1 \div 2$\% for tight lepton categories (CEM, CMUP+CMX, PHX) to $6$\% for the EMC leptons collected by the \met plus jets triggers. These uncertainties are applied on signal and EWK samples.
\item {\em Top Production} uncertainty  is a $10$\% rate uncertainty applied on all the top-related processes ($t\bar{t}$, single-top $s$ and $t$ channels). It covers the cross section theoretical prediction uncertainty and the acceptance differences due to systematic top quark mass variation. 
\item {\em QCD} normalization, independent for each lepton channel, is constrained to be within 40\% of the value extracted in Section~\ref{sec:qcd}.
\item{\em Mistag} uncertainty is derived from the mistag matrix $\pm 1\sigma$ variation and propagated to the final $W+LF$ sample. The rate variations are $11$\% and 21\% for single and double-tagged channels respectively.
\item{\em K-Factor} uncertainty is of 30\% on $W+HF$ rate estimate. It also includes the $b$-tag SF uncertainty used to derive this correction. We consider an uncorrelated uncertainty for $W+c\bar{c}/W+b\bar{b}$  and  $W+c$ backgrounds as the first process is produced by strong interaction while the latter is of electroweak nature.
\item {\em $Z+$Jets} normalization uncertainty is, conservatively, set to $45$\% as it includes the uncertainty on $Z+HF$ production.
\end{description}
Appendix~\ref{app:rate_sys} summarizes the systematic rate variation on the different templates.

We also account for {\em shape} systematic variations from three sources: JES, $Q^2$ and KIT modeling.
\begin{itemize}
\item {\em Jet Energy Scale} shape uncertainty (JES) is estimated  by shifting the JES of the input templates by $\pm 1\sigma$ from the nominal value. The acceptance of the process is allowed to change, therefore a new background estimate is performed with the JES varied templates. This produces a simultaneous rate and shape uncertainty. All the templates are affected except the fake-$W$ sample that is derived from data.
\item {\em $W+$ Jets $Q^2$ Scale} uncertainty is obtained by halving and doubling the nominal generation $Q^{2}$ (defined in Equation~\ref{eq:scale}) of all the $W+$ jets samples ($W+b\bar{b}$, $W+c\bar{c}$, $W+c$, $W+LF$). Again, we have a shape and rate uncertainty as a new background estimate is performed with the  varied templates.
\item {\em KIT-NN} shape uncertainty is the last systematic that we take into account. We apply two kind of independent variations: one on the fake-$W$ templates and the other on $c$-quarks and mistag templates. 
For the multi-jet template, a variation of the flavor composition is applied allowing more or less $b$-quark like events in the fake-$W$ KIT-NN template. For templates with relevant $c$-quark component  ($WW$, $WZ$, $ZZ$, $W+c$ and $W+c\bar{c}$) and $W+LF$ a correction to the KIT-NN output is derived from negative tags in data. This systematic variation influences only the shape and not the rate of the final discriminant because it leaves untouched the background composition.

\end{itemize}

\begin{figure} [!h]
\centering
\includegraphics[width=0.4\textwidth,height=0.35\textwidth]{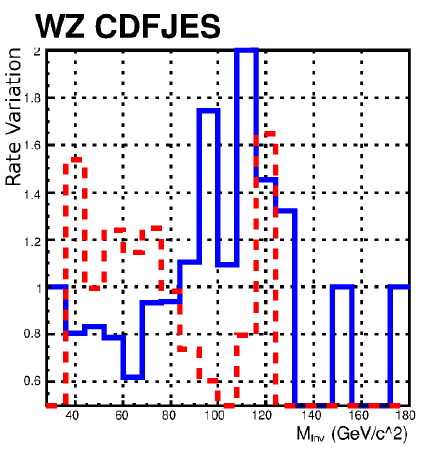}\hspace{0.7cm}
\includegraphics[width=0.4\textwidth,height=0.35\textwidth]{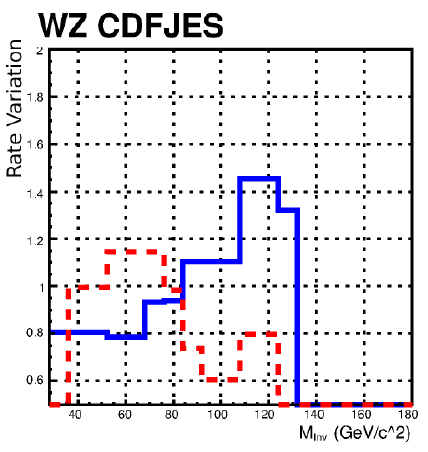}
\caption[Example of $M_{Inv}(jet1,jet2)$ Template Before and After Smoothing]{Example of the double-tagged CEM lepton $WZ$ $M_{Inv}(jet1,jet2)$ template rate variation due to JES systematic before (left) and after (right) a 3-bin median filter smoothing~\cite{median_filter}. The red (blue) line indicates the one sigma down (up) JES variation.}\label{fig:smoothing}
\end{figure}

In the case of templates with few events passing all requirements\footnote{The MC events can be spread across a large number of bins, especially in the bi-dimensional case where each template is composed by $50\times 4$ bins.} large statistical fluctuations can introduce a bias in the evaluation of the shape systematics. Therefore, shape variations are filtered to reduce the statistical noise. Filters are widely employed in modern image processing for different purposes. We choose a {\em median filter smoothing}~\cite{median_filter} because it maintains long range correlations among the histogram bins. We use a 5-bin filter along the $M_{Inv}(jet1,jet2)$ direction for the bi-dimensional, single-tagged, templates while a 3-bin filter is used for double-tagged templates\footnote{A double bin width is used for double-tagged $M_{Inv}(jet1,jet2)$ templates w.r.t. single-tagged ones as a 5-bin long range correlation would spoil the systematic variation.}. We also apply a low and high boundary to the possible template variations to reduce them in $[0.5,2.0]$ range.
Figure~\ref{fig:smoothing} shows an example of the double-tagged CEM lepton $WZ$ template before and after the smoothing filtering.

\section{$M_{Inv}(jet1,jet2)$ and KIT-NN Distributions}\label{sec:mjj_kit_distrib}

The final templates, including all the systematic effects, are then used to build the likelihood of Equation~\ref{eq:likelihood_full}. 

We combine a total of eight different channels: {\em four} lepton sub-samples (CEM, PHX, Tight Muons, EMC) times {\em two} $b$-tag prescriptions (single and double \texttt{SecVtx} tags). For the double \texttt{SecVtx} tagged events, the signal discrimination is based only on the di-jet invariant mass, $M_{Inv}(jet1,jet2)$, while for the single-tagged events we exploit the bi-dimensional distribution of $M_{Inv}(jet1,jet2)$ {\em vs} KIT-NN flavor separator. The KIT-NN output ranges from $-1$ to $1$ and it is divided in four equal size bins: the rightmost is highly enriched in $b$-like jets, while the others have variable composition of $b$-like, $c$-like and $LF$-like jets. 

\begin{figure}
\centering
\includegraphics[width=0.99\textwidth]{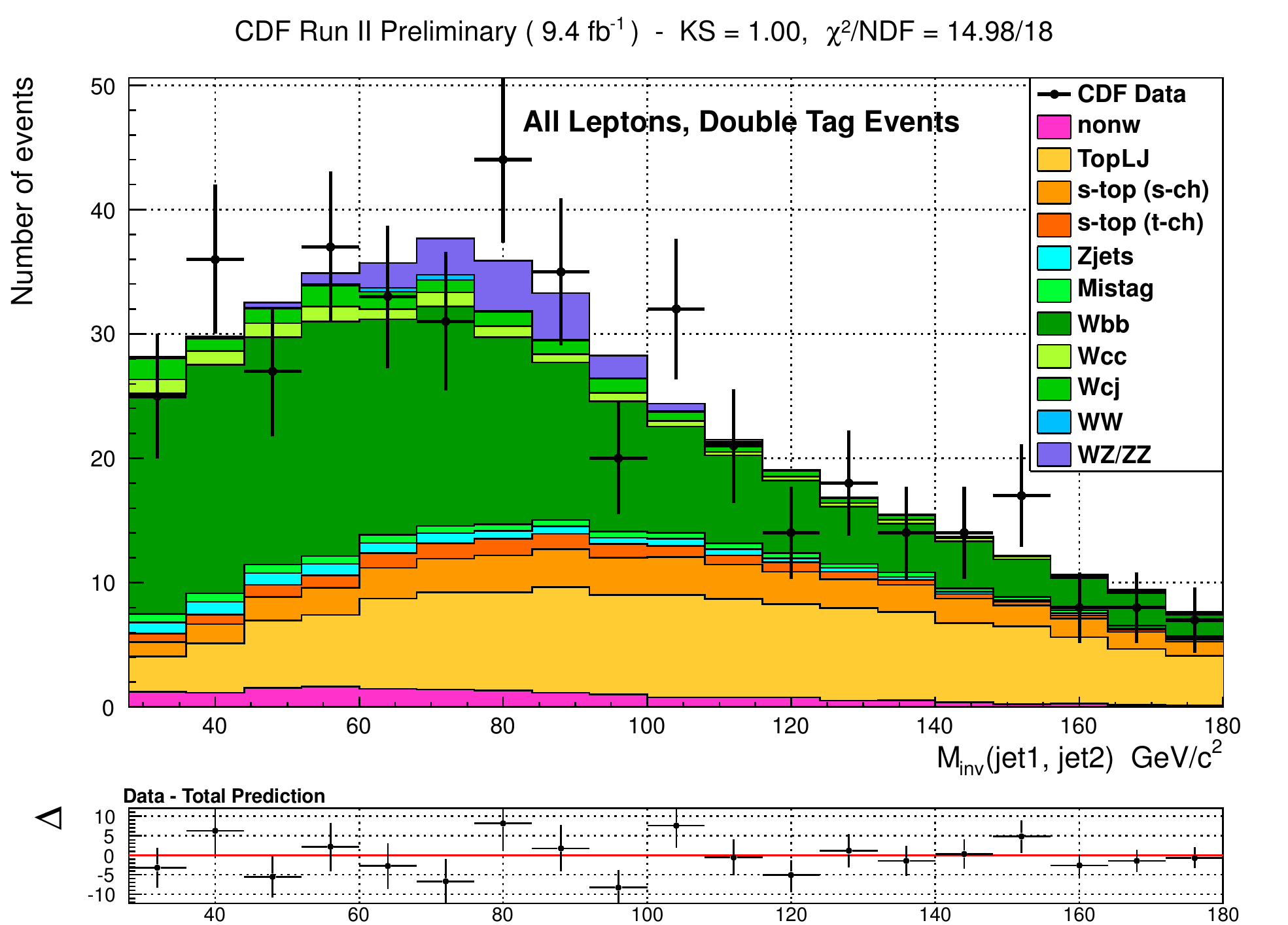}
\caption[$M_{Inv}(jet1,jet2)$ Distribution for 2 \texttt{SecVtx} Tag Events]{$M_{Inv}(jet1,jet2)$ distribution for candidate events with two \texttt{SecVtx} tags, all the lepton categories have been added together. The best fit values for the rate and shape of the backgrounds are used in the figure. }  
\label{fig:mjj_2tag_nokit}
\end{figure}

\begin{figure}
\centering
\includegraphics[width=0.99\textwidth]{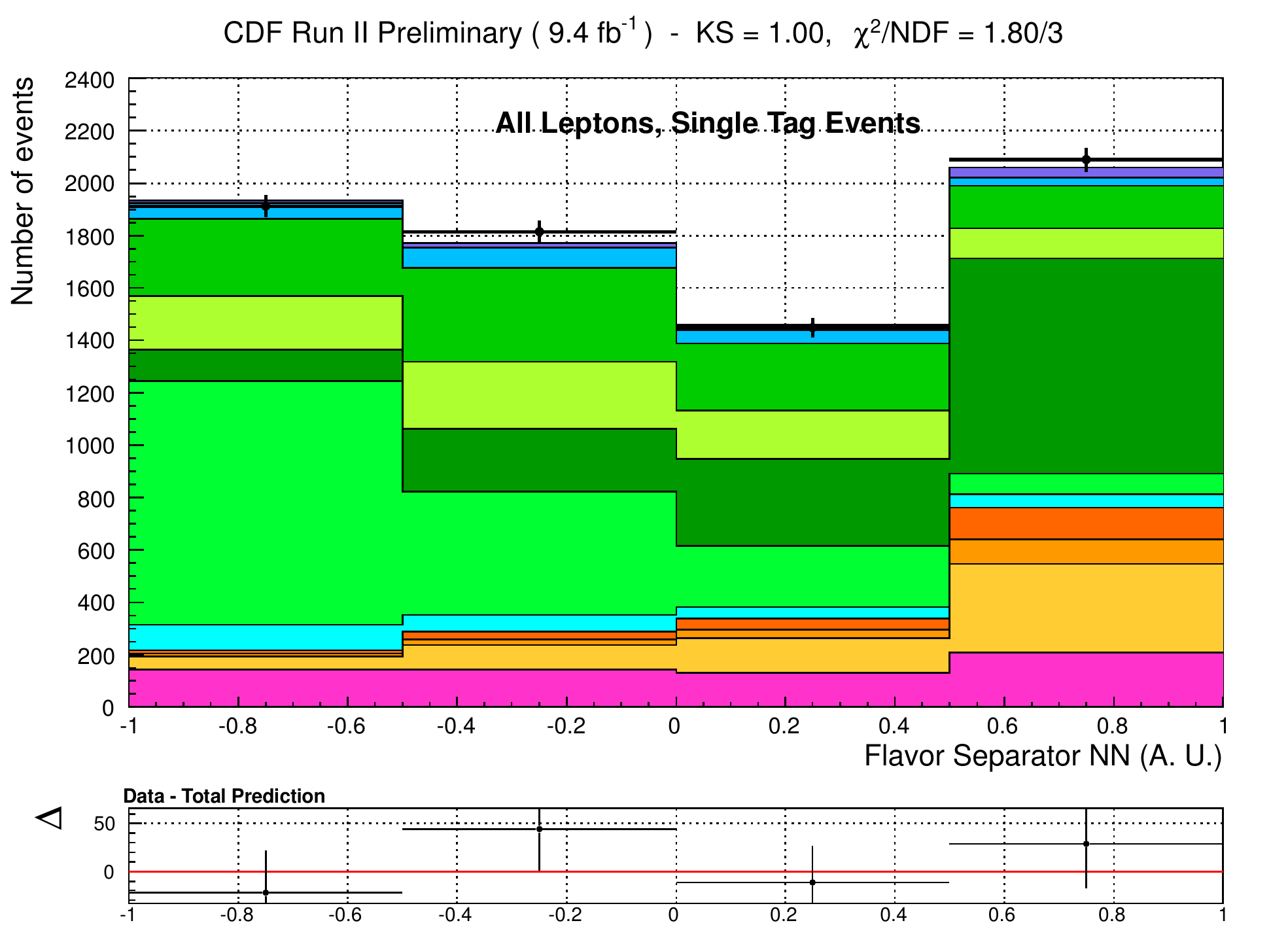}
\caption[KIT-NN Distribution for 1 \texttt{SecVtx} Tag Events]{KIT-NN distribution for candidate events with a single \texttt{SecVtx} tag, all the lepton categories have been added together. The best fit values for the rate and shape of the backgrounds are used in the figure.}  
\label{fig:kit_1tag}
\end{figure}

\begin{figure}
\centering
\includegraphics[width=0.99\textwidth]{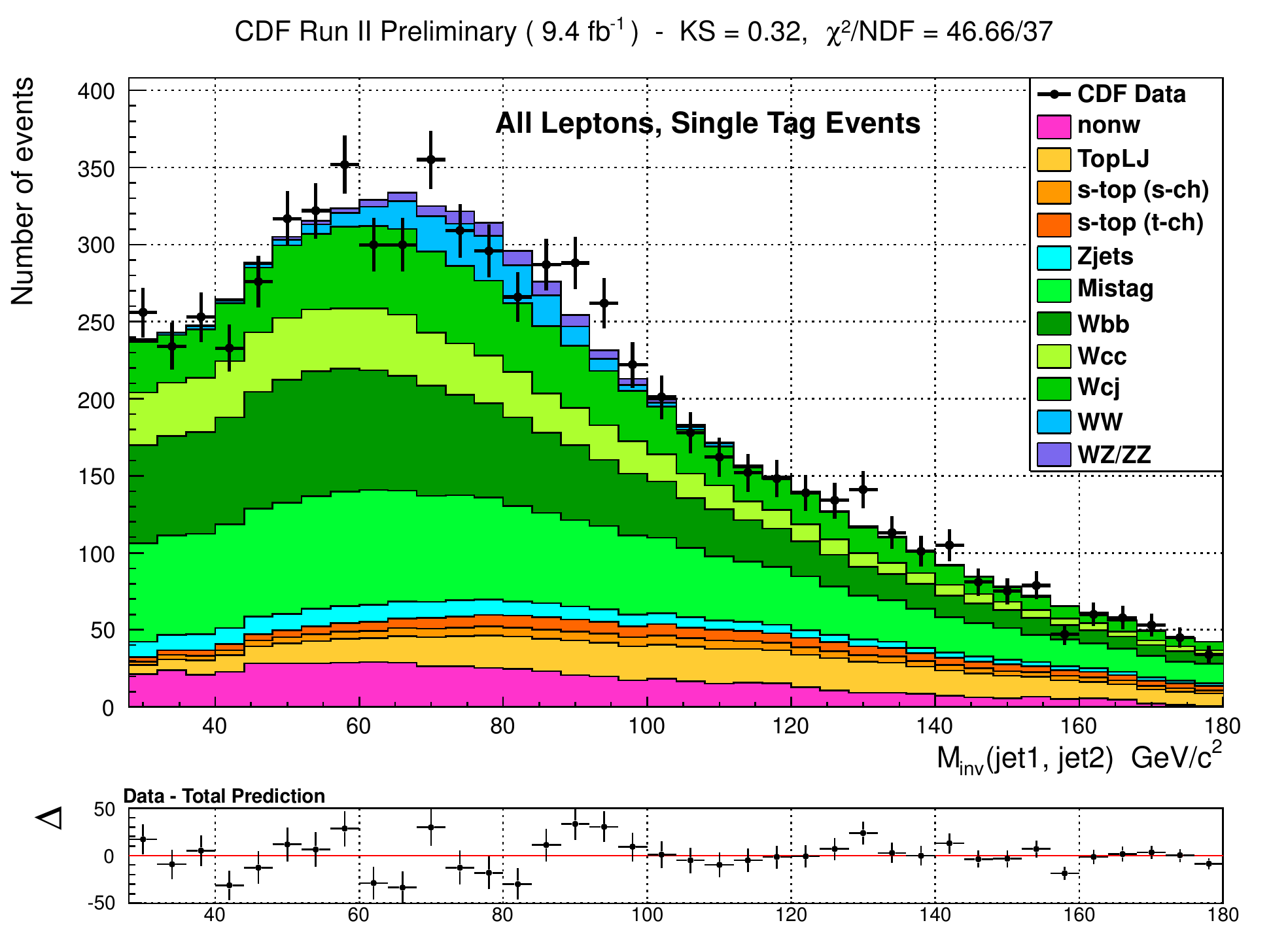}\\\vspace{0.5cm}
\includegraphics[width=0.495\textwidth]{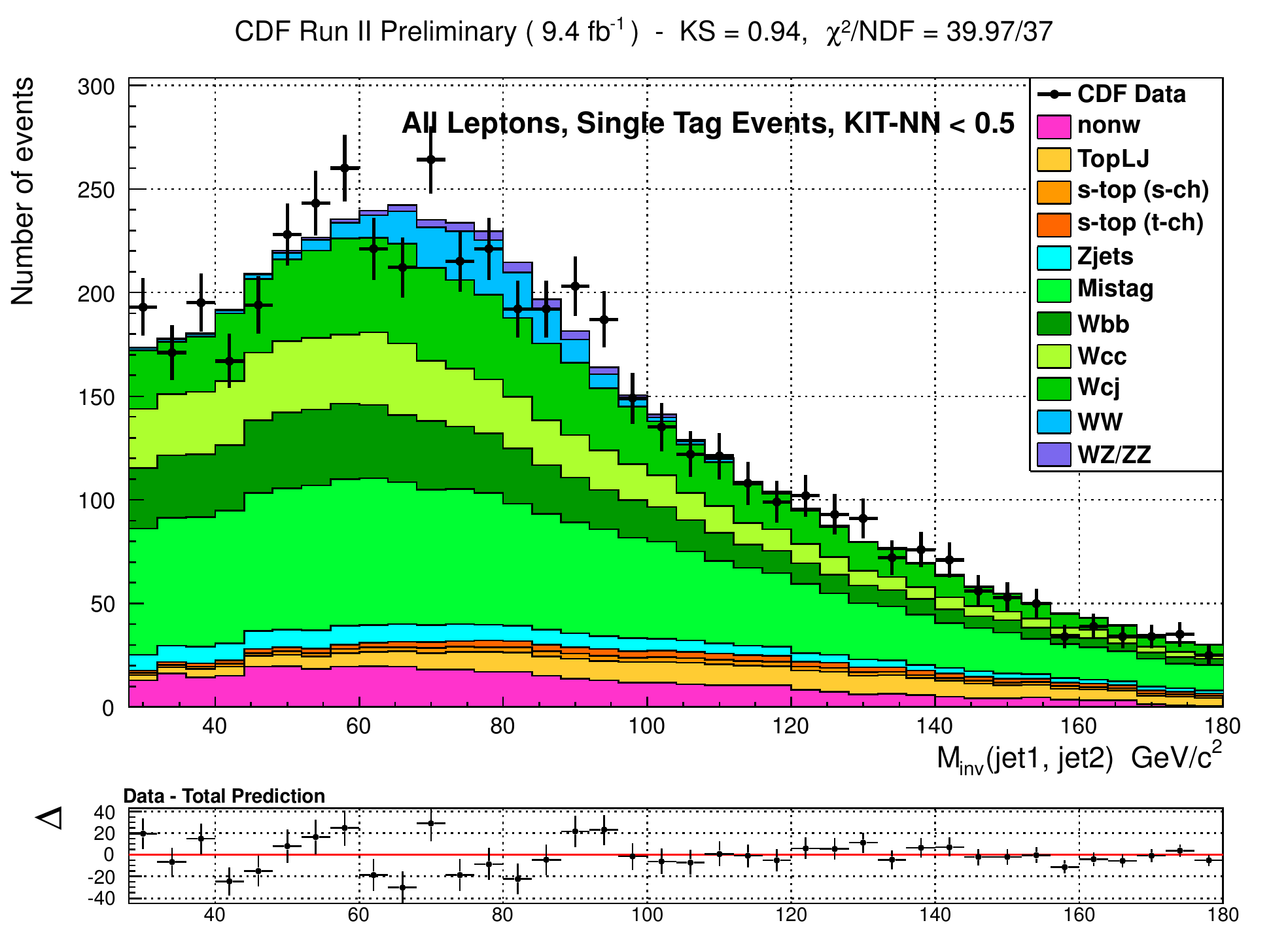}
\includegraphics[width=0.495\textwidth]{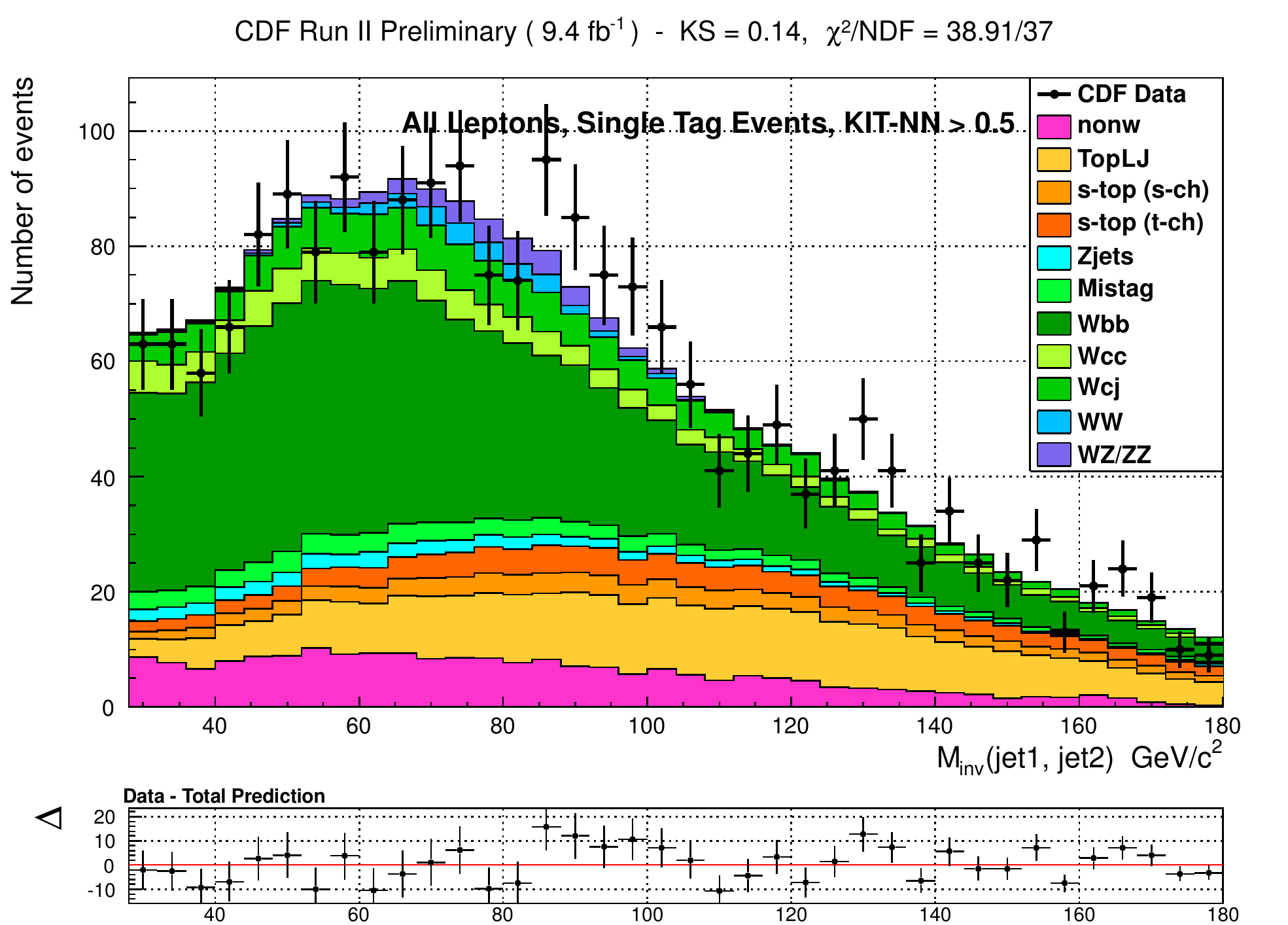}
\caption[$M_{Inv}(jet1,jet2)$ Distribution for 1 \texttt{SecVtx} Tag Events]{$M_{Inv}(jet1,jet2)$ distribution for candidate events with a single \texttt{SecVtx} tag, all the lepton category have been added together. Top distribution shows the results for all KIT-NN values while the bottom distributions are separated for KIT-NN$<0.5$ (bottom left) and KIT-NN$>0.5$ (bottom right). The best fit values for the rate and shape of the backgrounds are used in the figures.}
\label{fig:mjj_1tag_allkit}
\end{figure}

As noted in Section~\ref{sec:likelihood_meas}, we first obtain the maximum value of the likelihood with a fit. The  best fit result allows to check the agreement of the predicted distribution with data. The reduced $\chi^2$ of the fit is:
 \begin{equation}
   \frac{\chi^2}{ NDoF}=\frac{744.4}{664} = 1.12\textrm{,}
 \end{equation}
corresponding to a probability $P=0.984$. To further investigate the agreement of data and prediction, several post-fit distributions are shown from Figure~\ref{fig:mjj_2tag_nokit} to~\ref{fig:mjj_1tag_allkit}. We evaluate on them both the $\chi^2$ the Kolmogorov-Smirnov tests.
\begin{itemize}
\item Figure~\ref{fig:mjj_2tag_nokit} shows the  $M_{Inv}(jet1,jet2)$ distribution for double-tagged events;
\item Figure~\ref{fig:kit_1tag} shows the KIT-NN distribution for single-tagged events after the integration of all the $M_{Inv}(jet1,jet2)$ values used in the bi-dimensional distribution;
\item Figure~\ref{fig:mjj_1tag_allkit} shows the $M_{Inv}(jet1,jet2)$ distribution  for the single-tagged channel, added for all the lepton categories and integrated across all the KIT-NN values as well as the two most interesting KIT-NN regions: KIT-NN$>0.5$, $b$-enriched, and KIT-NN$<0.5$ with contribution from $c$ and $LF$ quarks.
\end{itemize}

\section{Cross Section Measurement}\label{sec:cx_measure}

The actual cross section measurement is performed by marginalizing the likelihood with respect to the nuisance parameters and studying the resulting Bayesian posterior as a function of the diboson signal cross section.

\begin{figure}
\centering
\includegraphics[width=0.9\textwidth]{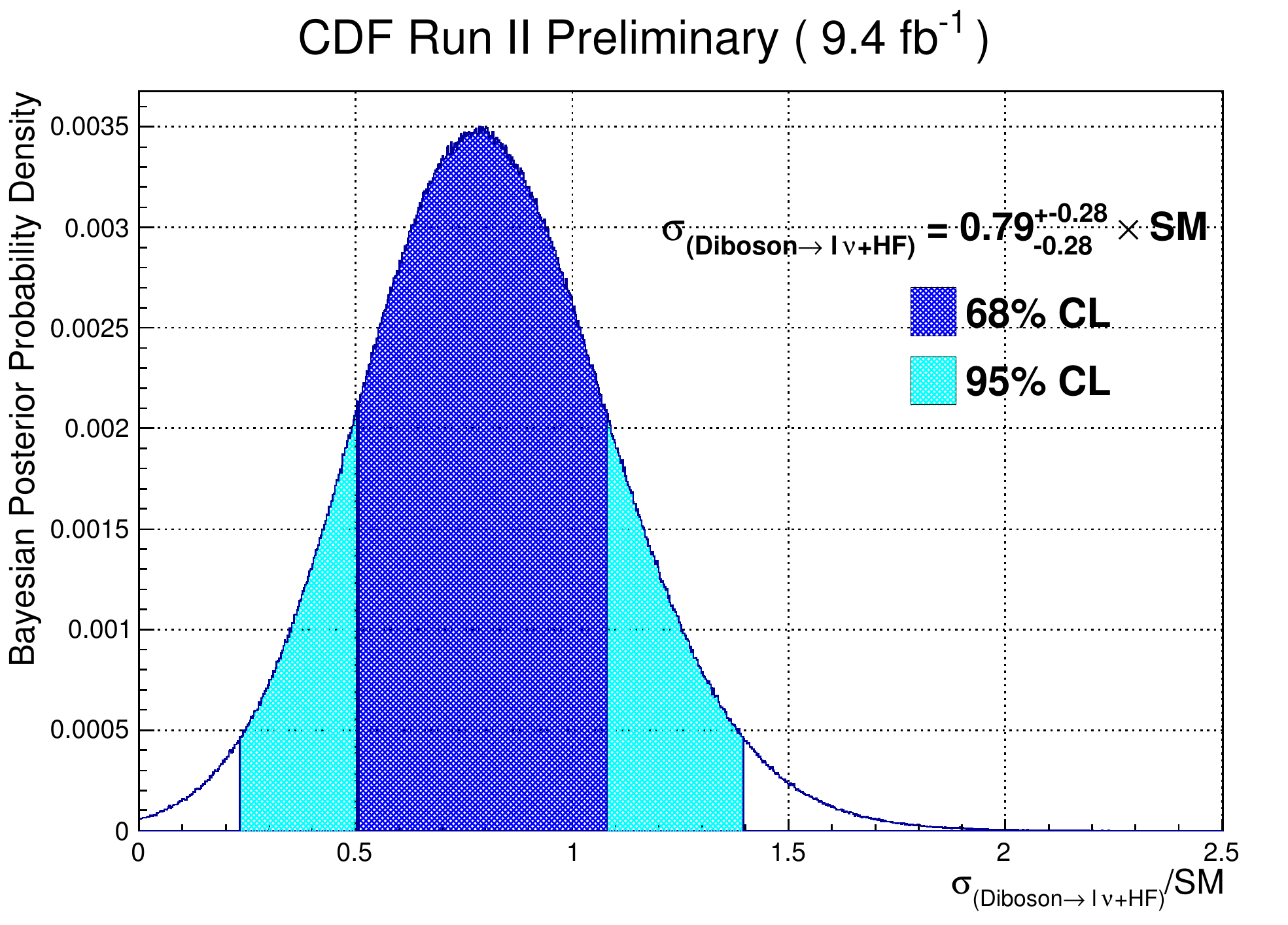}
\caption[Bayesian Posterior Distribution of $\sigma_{Diboson}$ Measurement]{The Bayesian posterior, marginalized over nuisance parameters, is shown. The maximum value is the central value of the cross-section, the blue and azure areas represent the smallest intervals enclosing 68\%, 95\% of the posterior integrals, respectively. The final cross section measurement is
\mbox{$\sigma_{Diboson}^{Obs}=\left(0.79\pm 0.28\right)\times \sigma_{Diboson}^{SM} = \left(14.6\pm 5.2\right)\textrm{~pb}$}.}
\label{fig:posterior_dib}
\end{figure}

First we measure the total diboson cross section, $\sigma_{Diboson}^{Obs}$, constraining the relative  $WW$ and $WZ/ZZ$ cross sections to the SM prediction. The resulting Bayesian posterior distribution is shown in Figure~\ref{fig:posterior_dib} together with the 68\% and 95\% confidence intervals. The measured cross section is:
\begin{equation}
 \sigma_{Diboson}^{Obs}=\left(0.79\pm 0.28\right)\times \sigma_{Diboson}^{SM} = \left(14.6\pm 5.2\right)\textrm{~pb,}
\end{equation}
where the errors include statistical and systematic uncertainties and $\sigma^{SM}_{Diboson}$ is the SM predicted cross section derived from Table~\ref{tab:mc_dib}:
\begin{equation}
  \sigma_{Diboson}^{SM}= \sigma^{SM}_{WW}+ \sigma^{SM}_{WZ/ZZ} = \left( 18.43\pm0.73\right)\textrm{~pb.}
\end{equation}

In order to separate the different components, we exploit the $c$ versus $b$ classification power of KIT-NN and the different sample composition of single and double-tagged events to obtain a separate measurement of $WW$ and $WZ/ZZ$. 

\begin{figure}[h!]
  \centering
\includegraphics[width=0.99\textwidth]{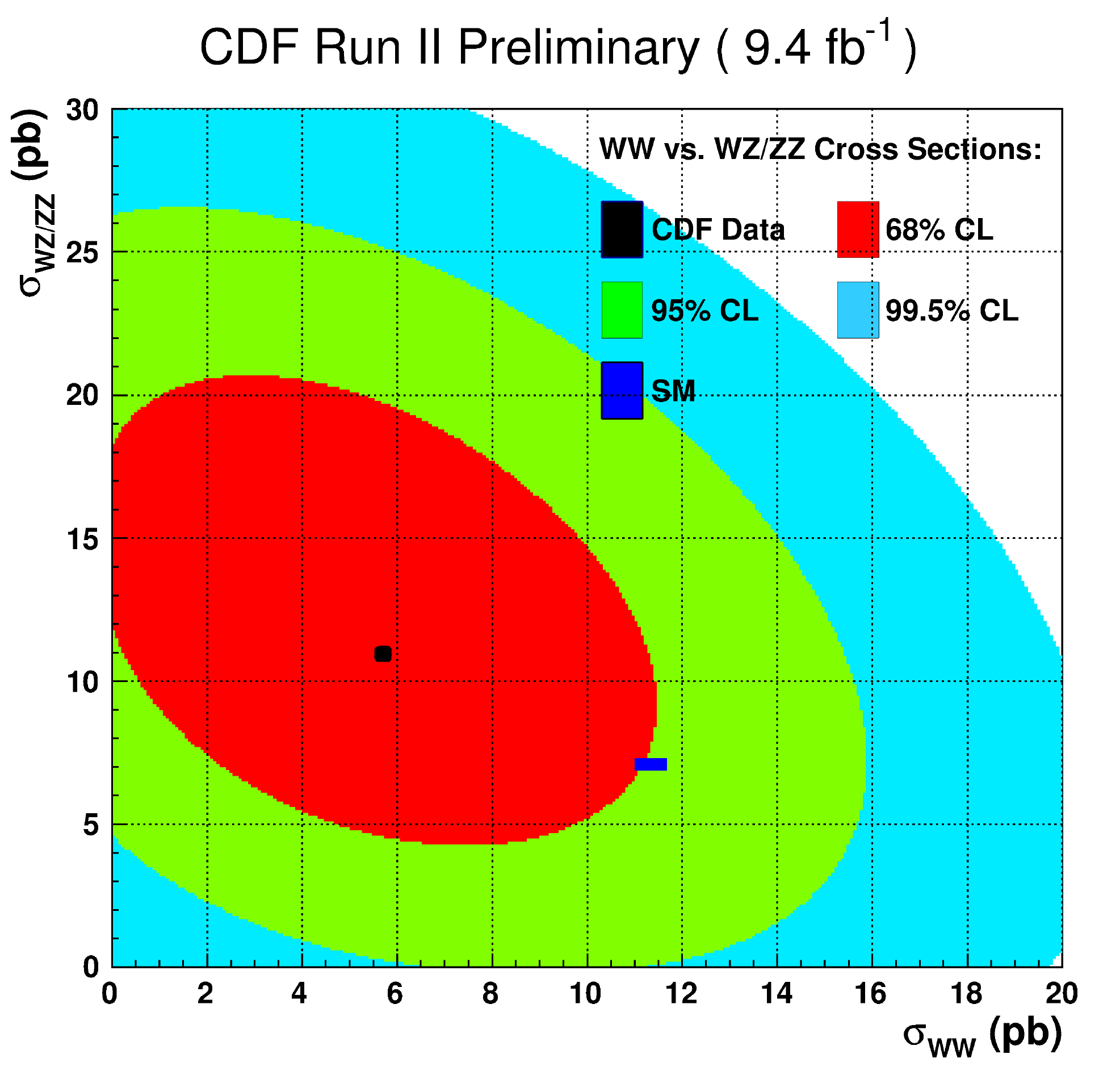}
\caption[2-Dim Bayesian Posterior Distribution of $\sigma_{WW}$ and $\sigma_{WZ/ZZ}$ Combined Measurement]{The Bayesian posterior, marginalized over nuisance parameters (scaled to SM expectation), is shown in the plane $\sigma_{WW}$ {\em vs} $\sigma_{WZ/ZZ}$. The measured cross sections correspond to the maximum value of $\sigma^{Obs,2D}_{WW}=0.50\times\sigma^{SM}_{WW} = 5.7\textrm{~pb}$ and $\sigma^{Obs,2D}_{WZ/ZZ}=1.56\times\sigma^{SM}_{WZ/ZZ}= 11.1\textrm{~pb}$. The red, blue and azure areas represent smallest areas enclosing 68\%, 95\% and 99\% of the posterior integrals, respectively.}
\label{fig:posterior_2d}
\end{figure}

We iterate the cross section measurement procedure but, this time, $\sigma_{WW}$ and $\sigma_{WZ/ZZ}$ are left free to float independently (i.e. not constrained to the SM ratio). Figure~\ref{fig:posterior_2d} shows the resulting Bayesian posterior distribution. The maximum of the posterior distribution gives the value of the measured cross sections:
\begin{equation}
\sigma^{Obs,2D}_{WW}=\left(0.50^{+0.51}_{-0.46}\right)\times\sigma^{SM}_{WW} = \left(5.7^{+5.8}_{-5.2}\right)\textrm{~pb}
\end{equation}
and
\begin{equation}
  \sigma^{Obs,2D}_{WZ/ZZ}=\left(1.56^{+1.22}_{-0.73}\right)\times\sigma^{SM}_{WZ/ZZ}= \left(11.1^{+8.7}_{-6.3}\right)\textrm{~pb,}
\end{equation}
where the SM predictions are $\sigma^{SM}_{WW} = 11.34\pm0.66$~pb, $\sigma^{SM}_{WZ/ZZ} =7.09\pm0.30$~pb and the errors are evaluated by the intersection of the $x$ and $y$ position of the maximum with the boundary of the smallest area enclosing the 68\% of the posterior distribution. The smallest areas enclosing 68\%, 95\% and 99\% of the posterior integrals give the contours of one, two and three standard deviations and are explicitly shown in Figure~\ref{fig:posterior_2d} with correlation between $\sigma^{Obs,2D}_{WW}$ and  $\sigma^{Obs,2D}_{WZ/ZZ}$.

\begin{figure}[h!]
\centering
\includegraphics[width=0.495\textwidth]{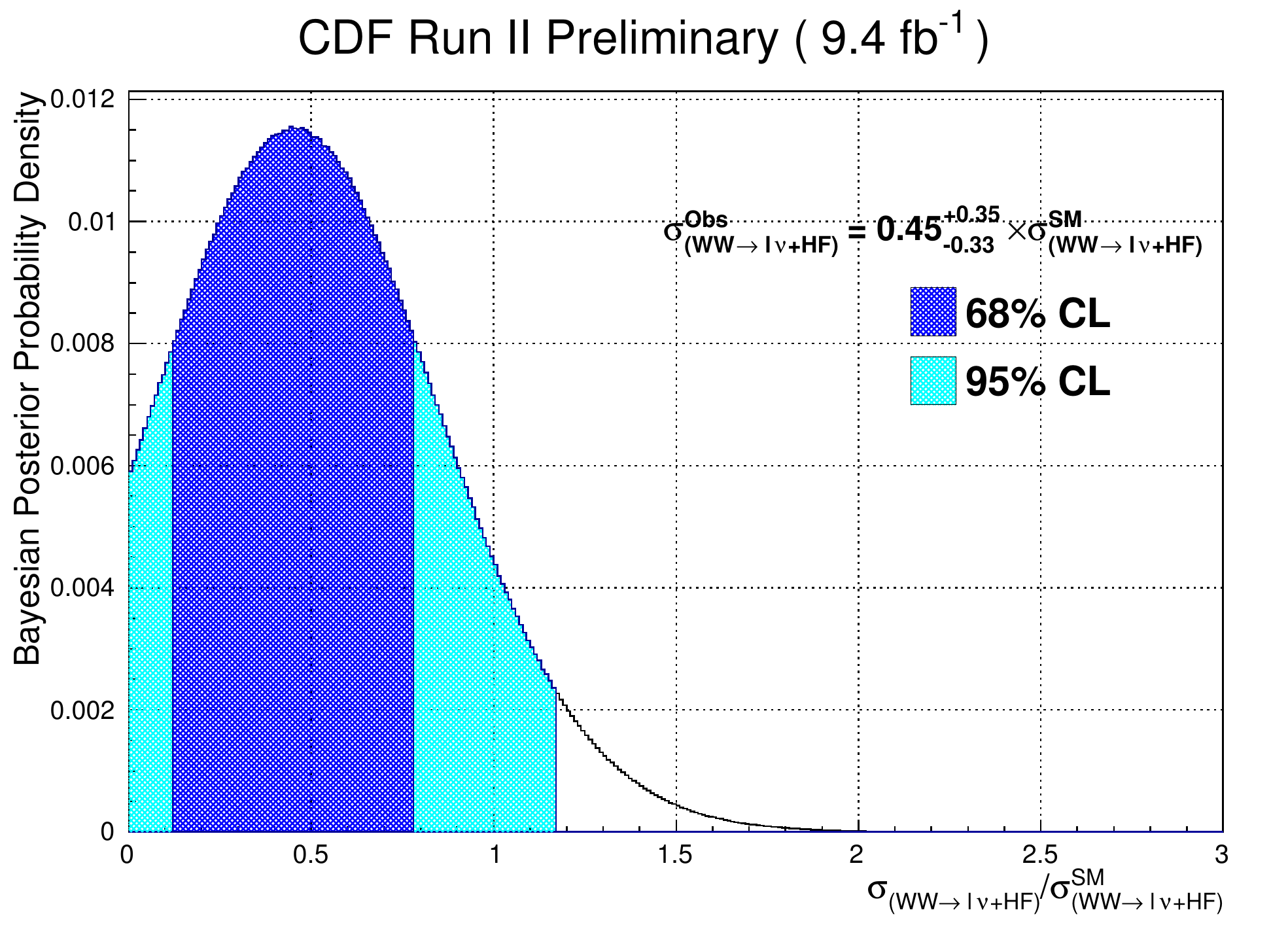}
\includegraphics[width=0.495\textwidth]{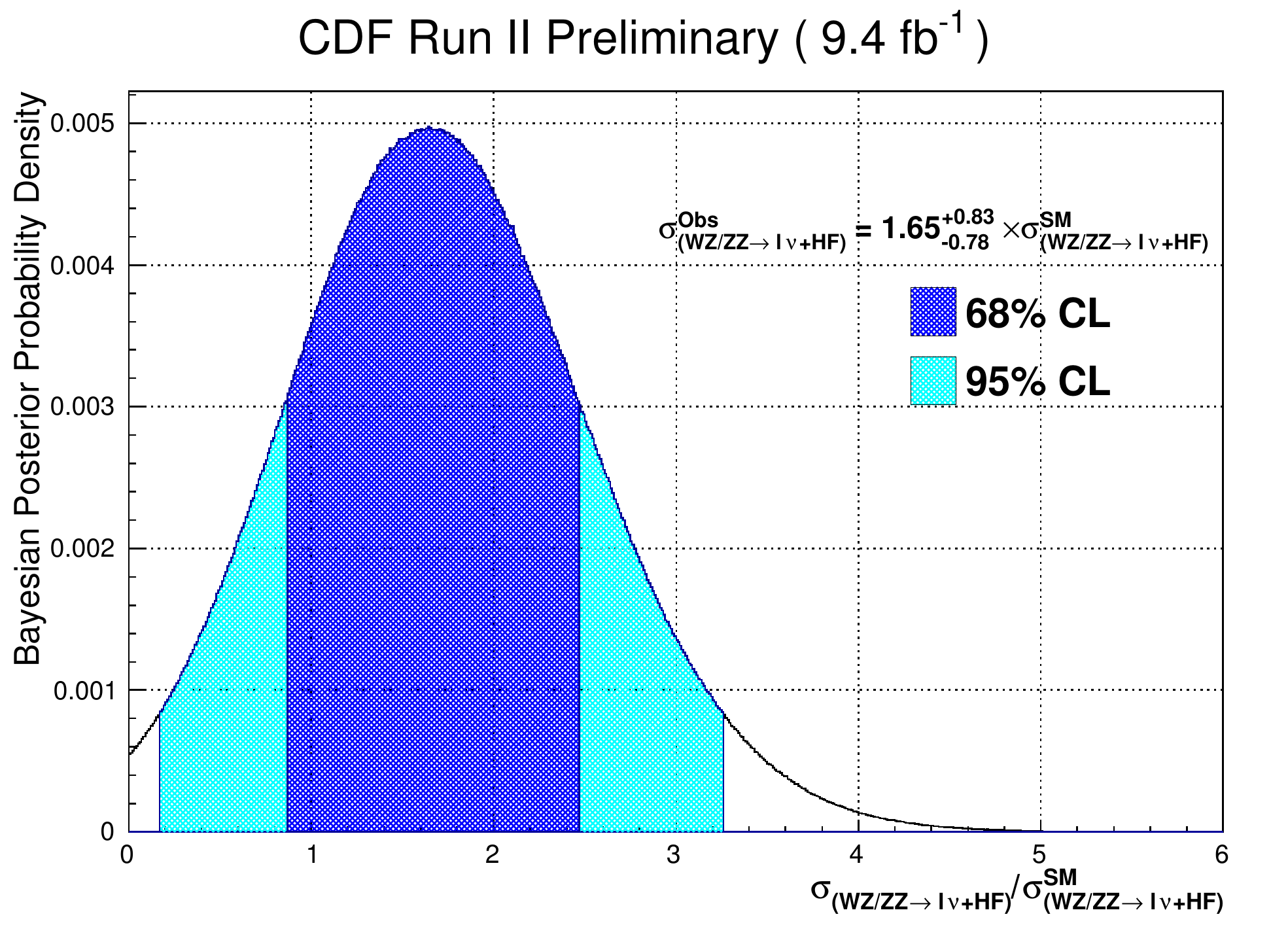}
\caption[Bayesian Posterior Distributions of $\sigma_{WW}$ and $\sigma_{WZ/ZZ}$ Separate Measurements]{The Bayesian posterior, function of $\sigma_{WW}$ and $\sigma_{WZ/ZZ}$ marginalized over nuisance parameters, is shown after projection on the $\sigma_{WW}$ (left) and $\sigma_{WZ/ZZ}$ (right) axes.
The maximum value is the central value of the cross-section, the blue and azure areas represent the smallest intervals enclosing 68\%, 95\% of the posterior integrals, respectively. The final cross section measurements are: \mbox{$\sigma^{Obs}_{WW}=\left(0.45^{ +0.35}_{ -0.32}\right)\times\sigma^{SM}_{WW} = \left(5.1^{+4.0}_{-3.6}\right)\textrm{~pb}$} and \mbox{$\quad \sigma^{Obs}_{WZ/ZZ}=\left(1.64^{+0.83}_{-0.78}\right)\times\sigma^{SM}_{WZ/ZZ} = \left(11.6^{+5.9}_{-5.5}\right)\mathrm{~pb}$}.}\label{fig:posterior_2d_proj}
\end{figure}

The $WW$ and $WZ/ZZ$ channels are also analyzed separately by projecting the two-dimensional Bayesian posterior on the $\sigma_{WW}$ and the $\sigma_{WZ/ZZ}$ axes. In this way, the two processes are considered as background one at the time. For both $WW$ and $WZ/ZZ$ we re-computed the maximum values and confidence intervals. Figure~\ref{fig:posterior_2d_proj} shows the results, the measured cross sections are:
\begin{equation}
\sigma^{Obs}_{WW}=\left(0.45^{ +0.35}_{ -0.32}\right)\times\sigma^{SM}_{WW} = \left(5.1^{+4.0}_{-3.6}\right)\textrm{~pb}
\end{equation}
and
\begin{equation}
  \quad \sigma^{Obs}_{WZ/ZZ}=\left(1.64^{+0.83}_{-0.78}\right)\times\sigma^{SM}_{WZ/ZZ} = \left(11.6^{+5.9}_{-5.5}\right)\mathrm{~pb,}    
\end{equation}
where the errors include statistical and systematic uncertainties.

\section{Evaluation of the Statistical Significance}\label{sec:sigma_eval}

To compute the significance of the measurements we perform a {\em hypothesis test} comparing data observation to the {\em null hypothesis} ($H_0$).

Random generated Pseudo Experiments (PEs) are extracted from the predicted background processes distribution in the $H_0$ hypothesis (i.e. excluding  the diboson production): this is straightforward once we know the probability distribution of the background and of the nuisance parameters. Then we repeat the cross section measurements with the complete marginalization of the likelihood. We expect a distribution peaking at $\sigma_{Diboson}^{PEs}/\sigma_{Diboson}^{SM} =0$ and we compare it to the measured cross section.

\begin{figure}
\centering
\includegraphics[width=0.9\textwidth]{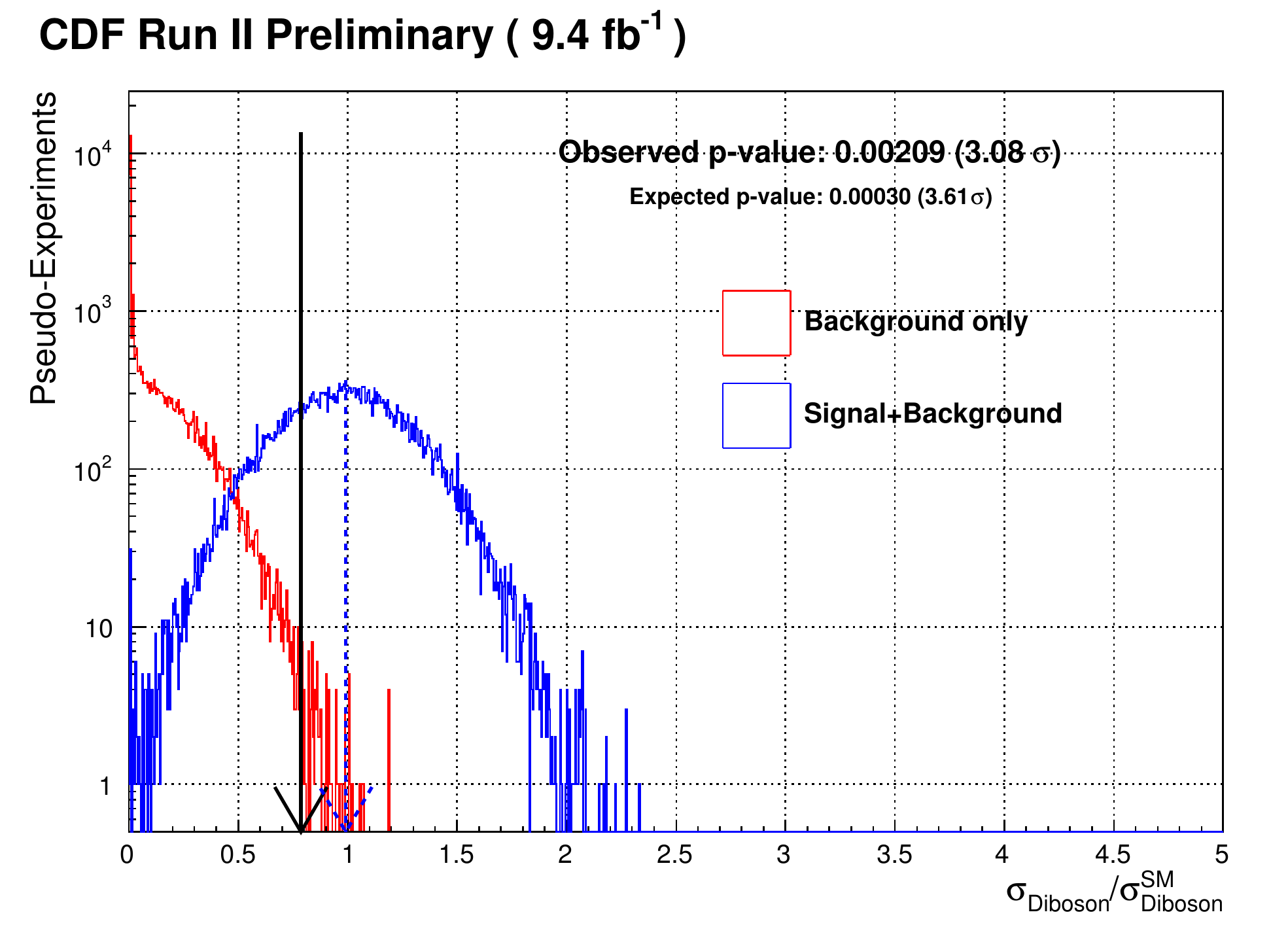}
\caption[Diboson Significance Evaluation]{Possible outcomes of many diboson cross section measurements from Pseudo Experiments (PEs) generated in a background-only and in a background-plus-signal hypothesis. The $p-$value for $\sigma^{Obs}_{Diboson}=0.79\times\sigma^{SM}_{Diboson}$ is $0.00209$, corresponding to a significance of $3.08\sigma$.}
\label{fig:significance_dib}
\end{figure}

Figure~\ref{fig:significance_dib} shows the possible outcomes of many cross section PEs in a back\-ground-only and in a background-plus-signal hypothesis. The number of times that a background fluctuation produces a cross section measurement greater than \mbox{$\sigma^{Obs}_{Diboson}=0.79\times\sigma^{SM}_{Diboson}$} has a $p-$value of $0.00209$. 

The result is an evidence for the diboson production in $\ell\nu +HF$ final state with a significance\footnote{A two sided significance estimate is used because both upper and lower fluctuation of the cross section measurement are considered in the integral of the {\em null} hypothesis cross section distribution above the measured value.} of $3.08\sigma$.

\begin{figure}
\centering
\includegraphics[width=0.495\textwidth]{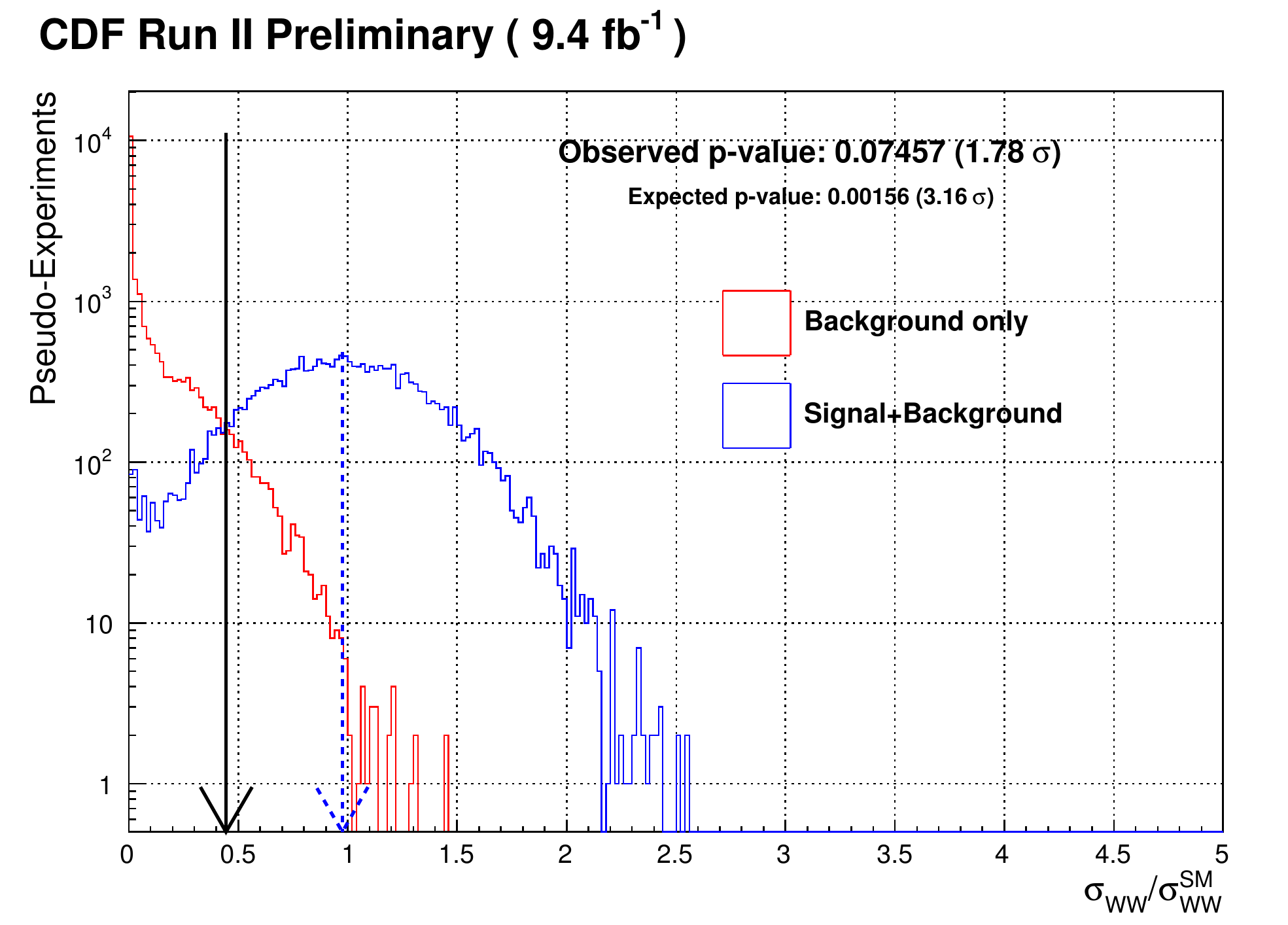}
\includegraphics[width=0.495\textwidth]{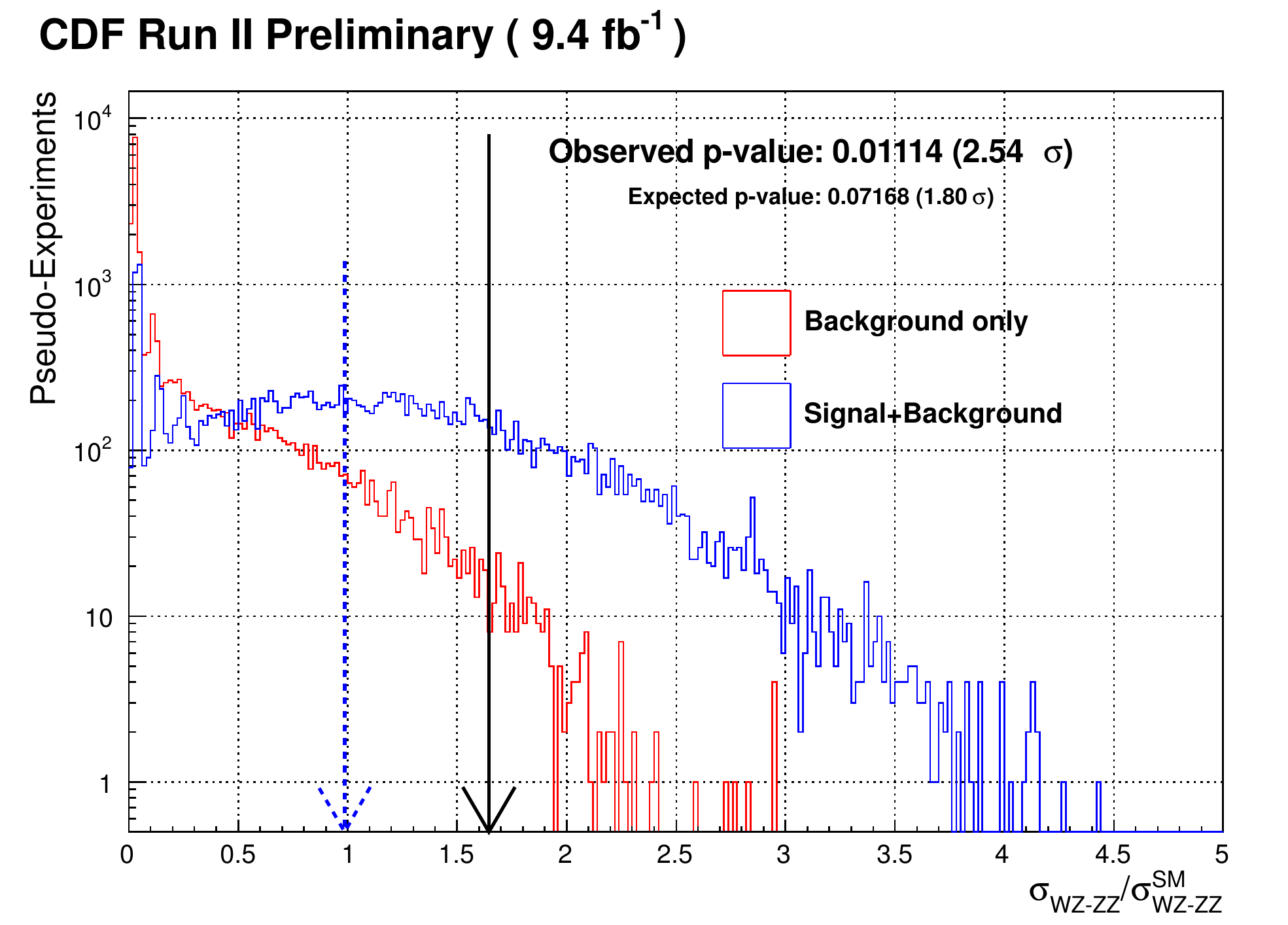}
\caption[$WW$ and $WZ$ Separate Significance Evaluation]{Possible outcomes of many diboson cross section measurements from Pseudo Experiments (PEs) generated in a background-only and in a background-plus-signal hypothesis in the $\sigma_{WW}$ vs $\sigma_{WZ/ZZ}$ plane and then projected on the  $\sigma_{WW}$ (left) and $\sigma_{WZ/ZZ}$ (axis). The $p-$values of $0.074565$ and $0.011145$ correspond to a significance of $1.78\sigma$ and $2.54\sigma$ for $WW$ and $WZ/ZZ$ respectively.}
\label{fig:significance_2d_proj}
\end{figure}

Then, we evaluate the single $WW$ and $WZ/ZZ$ significances in a similar way: PEs are generated with null hypothesis for both $WW$ and $WZ/ZZ$ signals. Then, the cross section PEs measurements are projected along the $\sigma_{WW}$ vs $\sigma_{WZ/ZZ}$ axes and compared with $\sigma^{Obs}_{WZ}$ and $\sigma^{Obs}_{WZ/ZZ}$. The result of the $p-$value estimates are shown in Figure~\ref{fig:significance_2d_proj}. 

We obtain: $p-$value$_{WW}=0.074565$ and $p-$value$_{WZ/ZZ}=0.011145$. They correspond to a significance of $1.78\sigma$ and $2.54\sigma$ for $WW$ and $WZ/ZZ$ respectively.

\clearpage
\chapter{Conclusions}\lbl{chap:conclusions}

The result of this thesis is the evidence, at $3.08\sigma$, for the associate production of massive vector bosons ($W$ and $Z$) detected at the CDF II experiment in a final state with one lepton, \met and $HF$-tagged jets. 

Such a result, obtained on the complete CDF II dataset ($9.4$\fb1 of data), was possible thanks to a simultaneous effort in several directions. The signal acceptance was extended both at online and offline selection level; the multi-jet background was strongly suppressed using a SVM-based multi-variate algorithm; second vertex $HF$-tagging was used in conjunction with a flavor-separator NN (KIT-NN). 

The signals, both inclusive diboson production and $WW$ {\em vs} $WZ/ZZ$ separately, were extracted from the invariant mass distribution, $M_{Inv}(jet1,jet2)$, of single and double $HF$-tagged jet pairs. For single-tagged  events the $b$ quark {\em vs} $c$ quark discrimination was obtained by using a bi-dimensional distribution $M_{Inv}(jet1,jet2)$ {\em vs} KIT-NN.

We measure the total diboson production cross section, fixing the $WW$ and $WZ/ZZ$ relative contribution to the SM prediction. We obtain:
\begin{equation}
 \sigma_{Diboson}^{Obs}=\left(0.79\pm 0.28\right)\times \sigma_{Diboson}^{SM} = \left(14.6\pm 5.2\right)\textrm{~pb,}
\end{equation}
where the errors include statistical and systematic uncertainty and $\sigma^{SM}_{Diboson}$ is the SM predicted cross section. 

Then, after removing the constraint on the $WW$ and $WZ/ZZ$ relative contribution, we leave them free to float independently. We perform a simultaneous measurement of $\sigma_{WW}$ and $\sigma_{WZ/ZZ}$ where all correlations are included. We can also perform a separate measurement of each contribution one at the time by considering the other as a background. In this case we obtain:
\begin{equation}
\sigma^{Obs}_{WW}=\left(0.45^{ +0.35}_{ -0.32}\right)\times\sigma^{SM}_{WW} = \left(5.1^{+4.0}_{-3.6}\right)\textrm{~pb}
\end{equation}
and
\begin{equation}
  \sigma^{Obs}_{WZ/ZZ}=\left(1.64^{+0.83}_{-0.78}\right)\times\sigma^{SM}_{WZ/ZZ} = \left(11.6^{+5.9}_{-5.5}\right)\mathrm{~pb,}    
\end{equation}
where the errors include statistical and systematic uncertainties.
The significance of the measurements is $1.78\sigma$ for $WW$ signal and $2.54\sigma$ for $WZ/ZZ$ signal.

All the results are consistent with the SM prediction and they confirm the CDF capability of identify a small signal in this challenging final state. In particular the previous version of this analysis, performed with a dataset of $7.5$~fb$^{-1}$ and reported in Appendix~\ref{App:7.5}, was the first evidence of diboson production in $\ell\nu+HF$ final state at a hadron collider.

Beyond the pure testing of SM predicted processes, several of the techniques developed for this thesis were also applied to the $WH$ search at CDF, with a relevant improvement of the sensitivity to this process.

\appendix

\clearpage
\chapter[SVM Multi-Jet Rejection]{Support Vector Machines Multi-Jet Rejection}\label{chap:AppSvm}

An innovative multivariate method, based on the Support Vector Machines algorithm (SVM), is used in this thesis to drastically reduce the multi-jet background. 

One of the crucial points in the search for diboson production in the $\ell\nu + HF$ final state is the maximization of the signal acceptance while keeping the background under control. This is a challenge because, in hadronic collider environment, jets are produced with a rate several order of magnitude larger than $W\to \ell \nu$ events, therefore, as a jet can fake the lepton identification with not negligible probability (especially for electrons identification algorithms), multi-jet events are introduced in the sample.
 
The multi-jet background, a mixture of detector and physics processes, is challenging to parametrize and, usually, approximate data-driven models are obtained by appropriate fake-enriched selections (see Section~\ref{sec:qcd}). These models are often statistically limited and the use of a different selection can produce unexpected biases in the simulated variables. It is obviously not trivial the use of multivariate techniques to tackle such a problem.

The SVM algorithm, described in Section~\ref{sec:SVM}, is considered to perform well in this case as it offers good non-linear separation and stable solutions also on low statistical training samples~\cite{svmbook, BShop_machine_learning}. It was never used before and we had to develop original solutions to address the major challenges: evaluate the robustness of the SVM against biases in the training set and establish the best, minimal set of input variables providing optimal performances. Section~\ref{sec:method} describes how we solved the first problem while Section~\ref{sec:var_sel} describes the input variable selection criteria.

The results, reported in Section~\ref{sec:svm_res}, in terms of signal efficiency and background rejection, are superior to any other cut based or multi-variate method previously applied at CDF.

\section{Support Vector Machines}\label{sec:SVM}

The SVM is a supervised learning binary classifier whose basic concept is the identification of the {\em best separating hyper-plane} between two classes of $n$-dimension vectors. 

In the case of linear separation the algorithm produces, given a training set of the vectors of the two classes, an {\em unique} solution where the plane is defined by the minimum amount of vectors, called {\em support vectors}, at the boundary of the two classes. In the case of non-linear separation, the plane is found in an abstract space, defined by a transformation of the input vectors. However it is not necessary to know the exact transformation, but just its effect on the scalar product between the vectors, named {\em Kernel}, thus allowing a feasible solution. Finally the cases of not perfect separability of the two samples are included by introducing a penalty parameter accounting for the contamination. 

The main advantages of the SVM with respect to other machine learning algorithms are the unique convergence of the problem, a small number of free tunable parameters (usually related to the Kernel choice) and good performances for low statistics training sets because only a small number of training vectors (the support vectors) are important  for the final solution.

It is possible to find more details in~\cite{svmbook, BShop_machine_learning}, but a short overview of the algorithm is also given in the following. For the actual, numerical, implementation of the SVM algorithm we relied on the \texttt{LIBSVM} open source library~\cite{libsvm}.

\subsection{The Linear Case}
\begin{figure}
  \begin{center}  
    \includegraphics[width=0.8\textwidth]{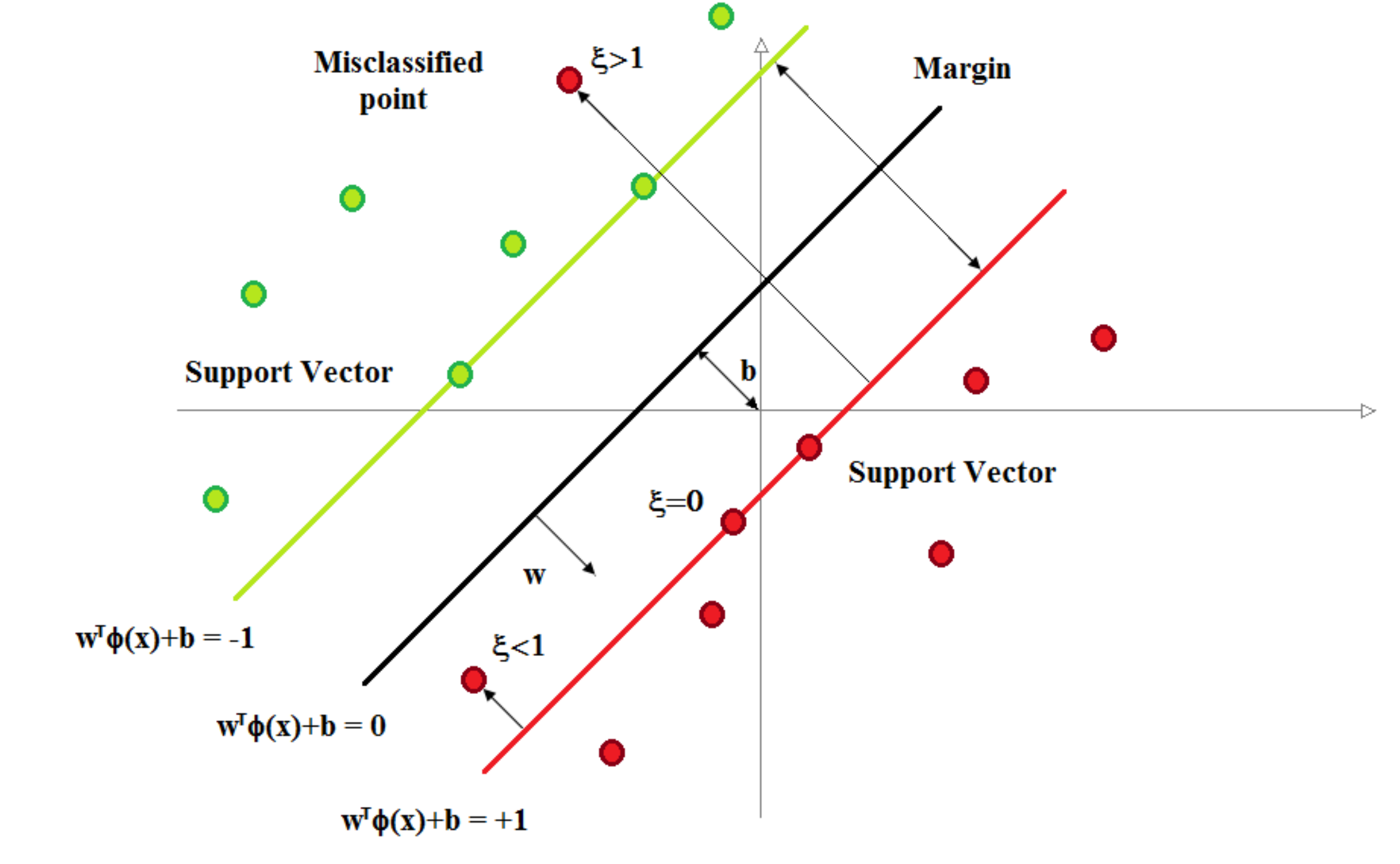}
    \caption[Example of SVM Linear Separation]{An example of SVM: two linearly separable classes of data are represented with red and blue dots. The hyperplane (in this case a simple line) leading to a maximum margin separation is defined by the weight vector $w$ and the bias vector $b$.}\label{fig:svm_linear}
  \end{center}
\end{figure}

Figure~\ref{fig:svm_linear} shows a basic example of the SVM linear classification separating two classes of bi-dimensional training vectors with a maximum margin hyperplane (a line for this simple case)

The problem can be formalized in a general way as the minimization of $|\vec{w}|^2$ (with $\vec{w}$ = vector normal to the plane) with the constraint:
\begin{equation}\label{eq:constraint}
  y_i(\vec{x_i}\cdot\vec{w} + b)-1\geq 0\qquad
  \left\{\begin{array}{ll}
  y_i=+1; & i \in \textrm{signal} \\
  y_i=-1; & i \in \textrm{background}\\
  \end{array} \right.
\end{equation}
The problem has an unique solution obtained by the maximization of:
\begin{equation}\label{eq:lagrange_1}
  L=\sum_i \alpha_i - \frac{1}{2}\sum_{i,j}\alpha_i\alpha_j y_i y_j\vec{x_i}\cdot\vec{x_j} ,
\end{equation}
obtained with the application of the Lagrange multipliers to Equation~\ref{eq:constraint}.
The solution identifies, for some $i$, $\alpha_i>0$. The associate vectors are the \textit{support vectors}, i.e. a subset of the training sample that define the best hyper-plane (see Figure~\ref{fig:svm_linear}).

To solve the case of not completely separable classes of vectors, a {\em penalty parameter}, $C$, is added into the target function to account for the contamination. So that we have a new minimization condition:
\begin{equation}
  |\vec{w}|^{2} + C \sum_{i}{\xi_{i}};
\end {equation}
and a new constraint (derived again from Equation~\ref{eq:constraint}):
\begin{equation}\label{eq:lagrange_2}
  y_i(\vec{x_i}\cdot\vec{w} + b) \geq 1-\xi_{i}\quad\textrm{with}\quad \xi \geq 0\textrm{.}
\end{equation}

The parameter $C$ defines, before the training, the SVM implementation therefore it represents as one of the {\em hyper parameter} of the SVM.

A newly seen vector, $\vec{X}$, is classified according the position with respect to the plane defined by the support vectors $\vec{x_i}$ and the parameters $\alpha_i$:
\begin{equation}\label{eq:distance}
  D(\vec{X})= \sum_i\alpha_i y_i \vec{x_i}\cdot \vec{X_i} - b,
\end{equation}
where $b$ is a bias term of the solution. The sign of $D(\vec{X})$ defines the classification but the value itself can be seen as the distance of a test vector $\vec{X}$ from the classification plane. However, as we will see in the next paragraph, a non-linear classification is possible only thanks to a not-explicit transformation in a different vector space, where $D$ looses it immediate geometrical meaning. 

Commonly SVMs are used as binary classifiers but, here, we add a large degree of flexibility by exploiting the full information of the variable $D$. We see it as a dimensionality reducer that summarizes all the information obtained during the training and classification process.

\subsection{Kernel Methods}

Non-linearly separable classes of vectors can be classified by transforming them into linearly separable
classes. An opportune function, $\Phi(\vec{x})$, can be used to map the elements into another
space, usually with higher dimension where the separation is possible. 

However the  identification of $\Phi(\vec{x})$ is {\em non trivial} and the, so called, {\em Kernel trick} is often used: a Kernel function, $\mathbf{K}(x_i,x_j)$, generalizes the scalar product appearing in Equation~\ref{eq:lagrange_1} (or Equation~\ref{eq:lagrange_2}) without the need of explicitly know $\Phi(\vec{x})$. Or in equations, we compose the mapping $\Phi(\vec{x})$ with the inner product:
\begin{equation}
  \mathbf{K}(x_i,x_j)= \Phi(x_i)\cdot\Phi(x_j)\quad\mathrm{with}\quad \Phi : \Re^n\mapsto \mathcal{H}.
\end{equation}

The function $\mathbf{K}$ should satisfy to a general set of roles to be a Kernel, but we want only to briefly describe the {\em Gaussian} Kernel we used in this work. It is expressed as:
\begin{equation}\label{eq:gauss_k}
K(x_{i},x_{j}) = e^{-\gamma |\vec{x}_{i}-\vec{x}_{j}|^2}
\end{equation}
The corresponding $\Phi(x)$ maps to an infinite dimension space and it is not known. The Kernel is defined only by one hyper-parameter, $\gamma$, that should be defined before the training.

\section{SVM Training in a Partially Biased Sample}\label{sec:method}

The assumption behind the supervised learning is that the labelled samples, used for the classifier training, are drawn from the same probability distribution of the unclassified events. 
However in our case of study, where only an approximate and statistically limited model of the background processes is available (see Sections~\ref{sec:cdf_set} and~\ref{sec:qcd} for the multi-jet background description), we do not expected the previous assumption to hold for every region of the phase space. To cope with this problem, we developed an original methodology to evaluate the SVM training performances.

Section~\ref{sec:SVM} shows that, for each choice of hyper-parameters and training vectors, only one optimal SVM solution exists and, for it, we need to evaluate the performances. 

As a performances estimator we use the {\em confusion matrix} of the classifier: the element $(i,j)$ of the matrix is the fraction of the class $i$ classified as member of class $j$. Figure~\ref{fig:confusion} shows a representation of it in the two classes case, where one class is considered the {\em background} and the other the {\em signal}. We obtain a reliable estimate of the classifier quality by filling the confusion matrix in two independent ways and combining all the available information.

\begin{figure}[h!]
  \begin{center}   
    \begin{tabular}{|c|c|}
      \toprule
      {\em Sgn} classified as {\em Sgn} & {\em Bkg} classified as {\em Sgn}\\
      \midrule
      {\em Sgn} classified as {\em Bkg} & {\em Bkg} classified as {\em Bkg}\\
      \bottomrule
    \end{tabular}   
    \caption[Definition of Confusion Matrix]{Definition of confusion matrix for a two classes ($Sgn$ and $Bkg$) classification problem. This reproduces the case of an algorithm used to discriminate signal {\em vs} background: the elements of the matrix are the signal and background classification performances and the cross contamination.}\label{fig:confusion}
  \end{center}
\end{figure}

The first performance evaluation method is the {\em $k$-fold cross-validation}: the training set is divided into $k$ sub-samples of which one is used as a validation set and the remaining $k - 1$ are used in the training; the confusion matrix is then evaluated applying the trained discriminant to the validation set. The cross-validation process is repeated $k$ times, the {\em folds}, and the final performance is given by the average on all the folds. This method is solid against over fitting but it has no protection against biases on the complete training sample.

\begin{figure}[!h]
  \begin{center}
  \includegraphics[width=0.9\textwidth]{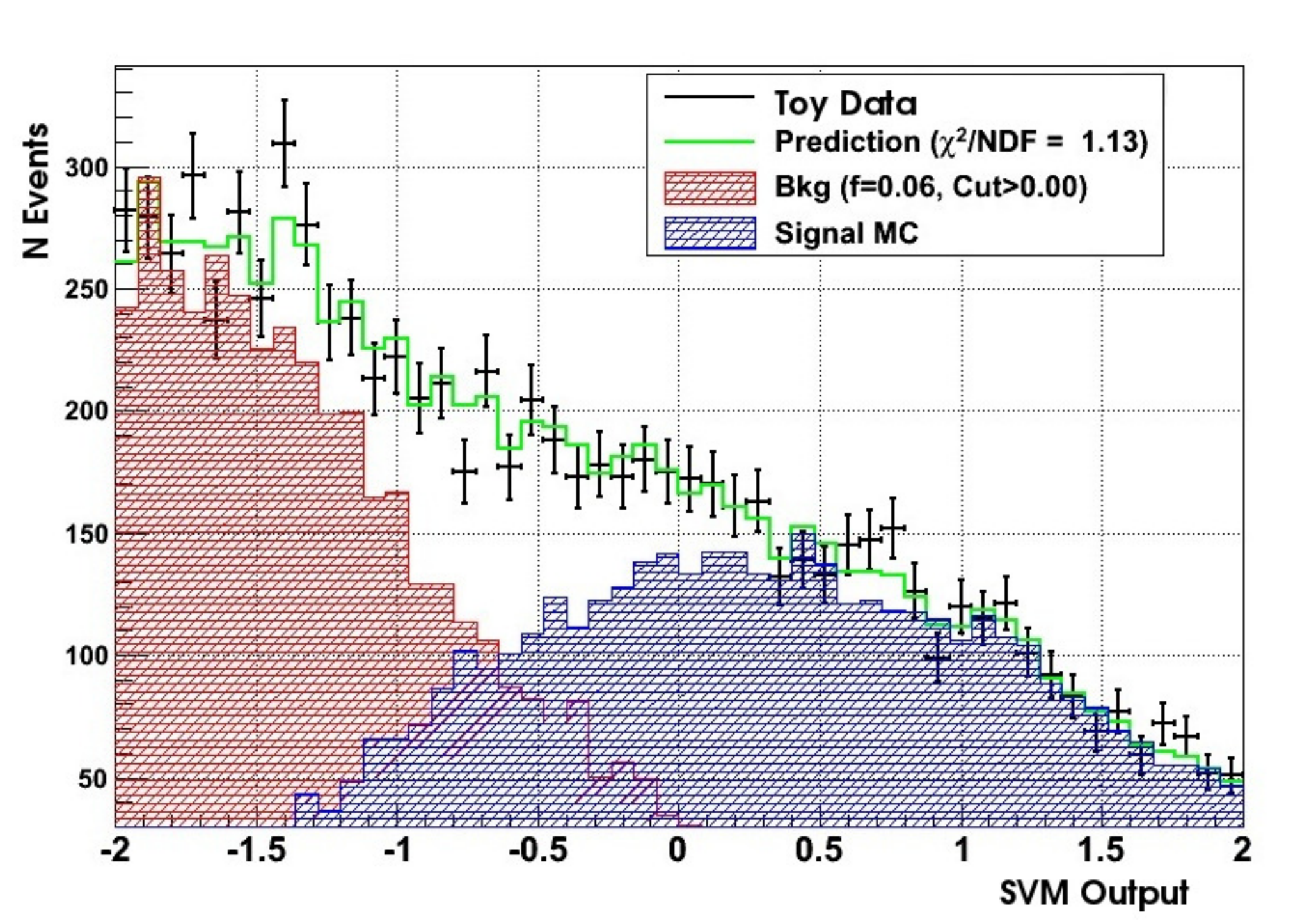}
  \caption[Bi-component Template Fit on Toy Data]{A bi-component fit is performed on the SVM distance, $D$ (Equation~\ref{eq:distance}), of a signal (blue) and background (red) templates toy data.}\label{fig:toysamplefit}    
  \end{center}
\end{figure}

The second method, a {\em key feature} of this work, is based on a bi-component fit that uses signal and background templates and it is performed on a significant distribution of the unclassified events, the {\em data}. While the signal and background templates are derived in the same way of the training samples, the unclassified data events are, by definition composed by an unknown mixture of the {\em true} signal and background events. The fit is performed by maximizing a binned likelihood function, $\lambda$, where the Poisson statistic of the templates is used and the fractions of the signal and of background templates, from which we can derive the elements of the confusion matrix, are free parameters. The fitting function is implemented in the ROOT~\cite{rootr} analysis package and it is derived from~\cite{Barlow_1993}. 
Figure~\ref{fig:toysamplefit} shows an example of the fit used on the toy model described in the next Section.

If the variable considered in the fit is not well reproduced in the simulation then we expect that the fitted fractions will differ greatly from the results obtained with the $k$-fold cross-validation. At the same time we can evaluate quantitatively the agreement between the data shape and the fitted templates because the quantity:
\begin{equation}
  \chi^2 = -2\ln(\lambda),
\end{equation}
follows a $\chi^2$ probability distribution (under general assumptions).

The last critical point is the identification of a {\em sensitive} variable to be used in the fit. In a previous work~\cite{CHEP_svm} we exploited the \met distribution as it is sensitive to the multi-jet contamination. A much more general approach, by the machine learning point of view, is the use of the SVM distance value, $D$, defined in Equation~\ref{eq:distance}. If the SVM training performances are optimal, also the variable $d$ offers an optimal discrimination, furthermore the cross check on the $\chi^2$ of the template fit ensure a good shape agreement between the data and signal and background templates. We verified the validity of the fit procedure with a toy example reported in the following.

\subsection{A Toy Example} 

We built a toy example in order to verify the robustness of the proposed method for an SVM performances evaluation when partially biased samples are available.

\begin{figure}[!h]
   \begin{center}
   \includegraphics[width=0.75\textwidth,height=0.5\textwidth]{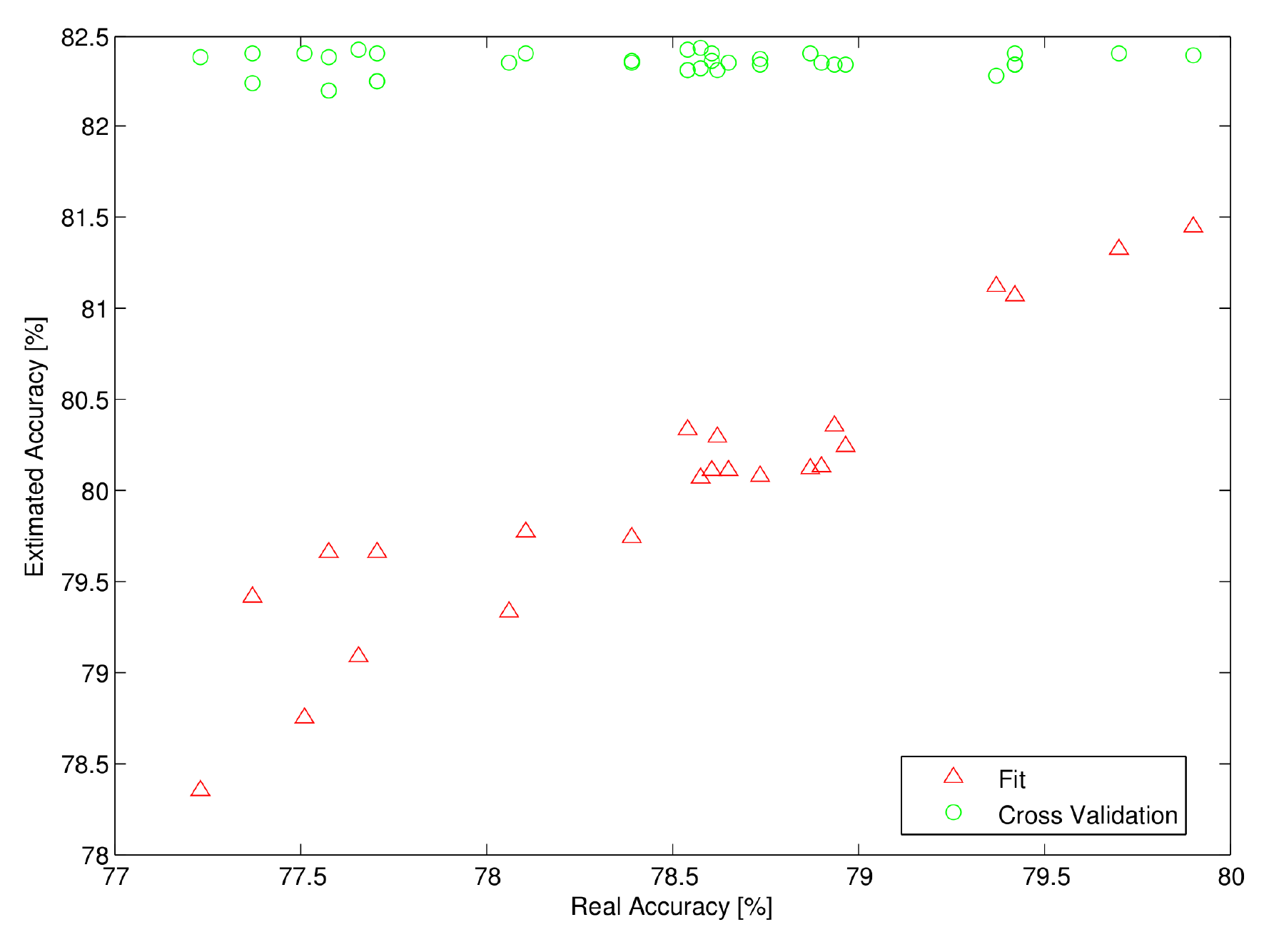}
   \caption[Toy Model SVM Performances Estimate with Fit and Cross-Validation]{SVM performances estimate with a $k$-fold cross-validation (green circles) and with a bi-component signal and background template fit on the SVM distance (red triangle) of toy data of known composition. The {\em true} performances of the SVM classifier are reported on the $x$ axis. The fit evaluation appears on the diagonal of the plane, signaling a more realistic estimate of the true performances of the classifier. }\label{fig:ToyResult}     
   \end{center}
 \end{figure}

The toy is composed by three data-sets generated with known probability distributions:
\begin{description}
\item[signal model:] $10^5$ vectors generated from a $2-$Dim Gaussian distribution with the following mean, $\vec{\mu}_{Sgn}$ and standard deviation, $\tilde{\sigma}_{Sgn}$:
  \begin{equation}\label{eq:sgn_gaus}
    \vec{\mu}_{Sgn} = \left[
      \begin{array}{c}
        - 3  \\
        0  \end{array} \right]
    ,\qquad 
     \tilde{\sigma}_{Sgn} = \left[
       \begin{array}{cc}
         8 & 0 \\
       0 & 8  \end{array} \right].
  \end{equation}
  
 \item[Background model:] $10^5$ vectors generated from a $2-$Dim Gaussian distribution with the following mean, $\vec{\mu}_{Bkg}$ and standard deviation, $\tilde{\sigma}_{Bkg}$:
   
   \begin{equation}\label{eq:bkg_gaus}
     \vec{\mu}_{Bkg} = \left[ 
     \begin{array}{c}
       3  \\
       0  \end{array} \right]
 ,\qquad 
     \tilde{\sigma}_{Bkg} = \left[
     \begin{array}{cc}
       8 & 0 \\
       0 & 8  \end{array} \right].
   \end{equation}
 \item[Data:] a mixture of $5 \cdot 10^4$ vectors generated from the same distribution of the signal model (Equation~\ref{eq:sgn_gaus}) and $5\cdot 10^4$ vectors generated from a {\em true} background distribution similar to the background model (Equation~\ref{eq:bkg_gaus}) but with $\tilde{\sigma}_{Bkg}$ increased by 20\% in one direction to simulate a mismatch between the real background and the model.
 \end{description}
 We tested several combinations of the hyper-parameters $C$ and $\gamma$ (over a grid) using the signal and background model in the training. For obtained SVM we evaluated the $k$-fold cross validation and we performed the template fit on the SVM distance, $D$, evaluated on the data sample. The result is reported in Figure~\ref{fig:ToyResult} with the real performances reported on the $x$ axis of the diagram (we know the true label of the data vectors). The evaluation of the performances obtained with the fit is on the diagonal of the plane, therefore it gives a much more realistic estimate of the true performances of the classifier.

\section{Performances on the CDF Dataset}\label{sec:cdf_set}

The final goal of the SVM discriminant we discussed is the realization of a tool able to reject the multi-jet background in a wide range of searches performed in the lepton plus jets channel in a hadron collider environment. 

We performed the SVM training on $W\to e\nu + $jets candidate, as the electron identification is more tamed by the multi-jet background. Furthermore  we performed the training process two times, one in the central ($|\eta|<1.1$) and one in the forward ($1.2<|\eta|<2.0$) region of the detector, as both the electron identification algorithm (see Section~\ref{sec:lep_id}) and the kinematic of the events are different.

\subsection{Training Sample Description}\label{sec:training}

We defined both a central and a forward training set using $7000$ $W\to e\nu +$jets signal events and $3500$ multi-jet background events:
\begin{description}
  \item[Signal:] $W + 2,3$ partons ALPGEN~\cite{alpgen} MC, where the $W$ is forced to decay into electron and neutrino. We have about $10^5$ generated events and we keep approximately  $9\times 10^4$ events as a control sample (i.e. not used for training). The CEM and PHX electron identification algorithms are used for the central and forward sample selection. 
\item[Background:] we obtain a suitable background sample with a data-driven approach. The anti-electrons selection described in Section~\ref{sec:qcd} is used for the central background training while for the forward training we had study a improved multi-jet model. 

In particular, we noticed that the anti-PHX sample produced large over estimates in the detector region of $1.2<|\eta|<1.4$ where two different calorimeter sub-systems are connected. These events are clear fake leptons and they are usually rejected by very loose kinematic requirements, nevertheless the quantity of them produced a too large bias in the SVM training efficiency estimate. Non-Isolated PHX electrons (with $IsoRel>0.1$) were found to give a reliable training set after a correction to the lepton $E_T$ equal to the amount of energy in the outer isolation cone ($E_T^{0.1<\Delta R<0.4}$).
\item[Data:] The data sample used for the bi-component fit validation corresponds to the data periods (see Table~\ref{tab:datasets}) from p18 to p24. They represent only a small fraction of the total CDF data with intermediate luminosity profile and run conditions.
\end{description}
A close to final lepton plus jet selection with loose kinematic cuts is applied to all the samples. We require two or more jets reconstructed (see Section~\ref{sec:jetCone}) in the central region of the detector ($\eta<2.0$), with energy corrected at L5 and $E_T> 18$~GeV, also a minimal amount of \met$>15$~GeV is used as signature of the escaping neutrino. We also removed events with not understood behaviour like electron $E_T>300$~GeV, $M_T^W>200$~\gc2 and, only for the forward sample, \metr$<20$~GeV (this last cut was needed to avoid trigger turn-on effects).

\subsection{Variable Selection}\label{sec:var_sel}

A multivariate algorithm relies on a given set of input variables. The {\em feature selection} problem is fundamental in machine-learning and, if possible, even more in the present case where the background sample does not guaranteed a perfectly model of all the variables. 

We started from a large set of twenty-four variables chosen according to two basic criteria: no correlation with respect to the lepton identification variables and exploit of the kinematic difference between real $W+$ jets events and multi-jet fakes. These requirements allowed the development of a flexible multi-jet rejection algorithm, applied with very good performances also to muon and isolated track lepton selections.

Table~\ref{tab:input_vars} shows all the input variables that we used in the optimization process. Many of them were introduced in Chapter~\ref{chap:objects} but the following are new:
\begin{itemize}
\item \mbox{${\not}{p_T}$} is the missing momentum defined as the momentum imbalance on the transverse plane. It is computed adding all the reconstructed charged tracks transverse momenta, $\vec{p_i}$:
  \begin{equation}
    \vec{{\not}{p_T}}\equiv - \sum_i{\vec{p^i}_T}\quad\textrm{with }|\vec{p^i}_T|>0.5\textrm{GeV}/c\textrm{;}
  \end{equation}
\item $M_T^W$ is the {\em transverse mass} of the reconstructed $W$ boson:
  \begin{equation}
    M_T^W=\sqrt{2(E_T^{lep}\cancel{E}_T - E_x^{lep}\cancel{E}_x - E_y^{lep}\cancel{E}_y)}\mathrm{.}
  \end{equation}
\item $MetSig$ is the \met {\em significance}, a variable that relates the reconstructed \met with the detector activity (jets and unclustered energy):
  \begin{equation}
    MetSig =\frac{\cancel{E}_T}{\sqrt{ \Delta E^{jets}+ \Delta E^{uncl}}}\mathrm{,}
  \end{equation}
  where:
  \begin{eqnarray}
    \Delta E^{jets} =\sum_{j}^{jets}(cor_{j}^2\cos^2\Big(\Delta\phi\big(\vec{p_j},\cancel{\vec{E}}_T\big)\Big)E_{T}^{raw, j}\mathrm{,}\\
    \Delta E^{uncl} = \cos^2\Big(\Delta\phi\big(\vec{E}_T^{uncl}\cancel{\vec{E}}_T\big)\Big)E_T^{uncl}\mathrm{,}
  \end{eqnarray}
  $uncl$ refers to the calorimeter energy not clustered into electrons or jets and $cor_j$ is the total correction applied to each jet.
\item $\nu^{Min},\quad \nu^{Max}$ are the two possible reconstruction of the neutrino momenta. As the $p_z^{\nu}$ component is not directly measurable we infer it from the $W$ boson mass and the lepton momentum. The constraints lead to a quadratic equation which may have two real solutions, one real solution, or two complex solutions\footnote{The real part is chosen in this case}. The reconstructed $\nu^{Min}$, $\nu^{Max}$ derive from the distinction of $p_z^{\nu,Max}$ and $p_z^{\nu,Min}$.
\end{itemize}

\begin{table}
  \begin{tabular}{cl| cl|  cl| cl }
    \toprule
    \multicolumn{8}{c}{Possible Input Variables }\\
    \midrule
    1 & $p_T^{lep}$           &  7 & $E_T^{raw, jet1} $   & 13 &  $\Delta\phi( {\not}{p_T},$ $lep)$ &   19 & $\Delta R( lep,$ $jet 2)$ \\
    2 & \met                 &  8 & $E_T^{raw, jet2} $   & 14 &  $\Delta\phi( {\not}{p_T},$ \met$)$ & 20 & $\Delta R( \nu^{min},$ $jet 1)$ \\ 
    3 & \metr                &  9 & $E_T^{cor,jet1} $       & 15 & $\Delta\phi( {\not}{p_T},$ \metr$)$ &  21 & $\Delta R( \nu^{min},$ $jet 2)$ \\ 
    4 & \mbox{${\not}{p_T}$} & 10 & $E_T^{cor,jet2} $       &  16 & $\Delta\phi( lep,$ \met$)$ &    22 & $\Delta R( \nu^{min},$ $lep)$ \\ 
    5 & $M_T^W$              & 11 & $\Delta\phi( jet 1,$ \met$)$ & 17 & $\Delta\phi( lep,$ \metr$)$ &  23  & $\Delta R( \nu^{max},$ $jet 1)$ \\ 
    6 & $MetSig$             & 12 & $\Delta\phi( jet 2,$ \met$)$ &  18 & $\Delta R( lep,$ $jet 1)$ &  24 & $\Delta R( \nu^{max},$ $jet 1)$ \\ 
\bottomrule
  \end{tabular}
  \caption[Possible SVM Input Variables]{All the possible input variables used for the SVM training and optimization. See Section~\ref{sec:var_sel} for a detailed description.}\label{tab:input_vars}  
\end{table}

\begin{figure}[!h]
  \begin{center}    
    \includegraphics[width=0.9\textwidth]{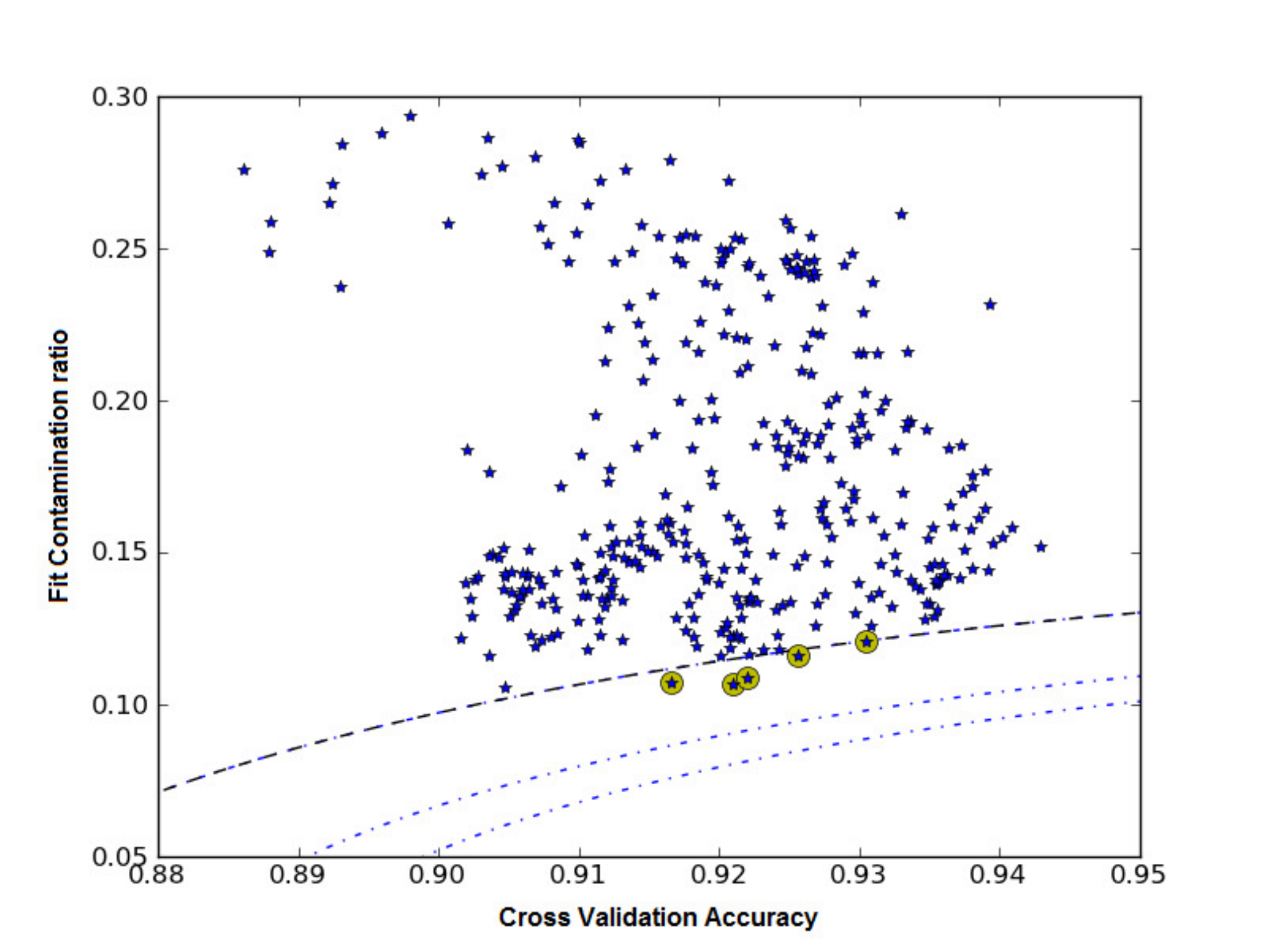}
  \caption[Signal Efficiency {\em vs} Background Contamination for Multiple SVM]{Different SVM configurations (i.e. with different input variables), obtained for the central region training, are displayed on a {\em signal-efficiency vs background-contamination} scatter plot. The signal efficiency is directly estimated from simulation while the background contamination is obtained from the bi-component template fit of the SVM distance, $D$, described in Section~\ref{sec:method}. The blue starts represent the performances obtained from a three (out of twenty-four) input variables training. The best five configurations w.r.t. the Euclidean distance from the optimal point ($\epsilon_{Sig}=1$, $f_{Bkg}=0$) are circled in yellow. The Euclidean distance of the fifth best SVM is represented by a dotted line. Three iterations, with {\em three, six and nine} input variables are represented by dotted lines, after that no more appreciable improvement occurs.}\label{fig:discriminants}
  \end{center}
\end{figure}

Unluckily the extensive research over all the possible combinations of variables across all the $C, \gamma$ phase space of a given SVM training, is computationally unfeasible. To scan the most relevant sectors of the phase space we applied factorized and incremental optimization:
\begin{itemize}
\item for all the configuration of {\em three} variables and the given training set, we evaluate a grid of  $C, \gamma$ values in the intervals\footnote{The use of a logarithmic scale allows to scan the parameters across different orders of magnitude.}:
  \begin{equation}
   \log_2 C\in [-3,8] \quad\textrm{and}\quad \log_2\gamma\in[-4,5].
  \end{equation}
We select only the best training configuration according to the confusion matrix evaluation.
\item For each {\em best} SVM of a given variable configuration we perform a bi-component fit on the SVM distance $D$. We evaluate the  $\chi^2$ of the fit, reduced by the Number of Degrees of Freedom ($NDoF$), and we compare the fitted background contamination, $f_{Bkg}^{Fit}$, against the one obtained from the $n$-fold cross-validation, $f_{Bkg}^{n-fold}$. The SVM under exam is rejected if:
  \begin{equation}
    \frac{\chi^2}{NDoF}>3\quad  or \quad\frac{f_{Bkg}^{Fit}}{f_{Bkg}^{n-fold}}>2
  \end{equation}

\item The remaining SVMs are displayed on a {\em signal-efficiency vs background-contamination} scatter plot like the one in Figure~\ref{fig:discriminants}. The 5 best variable combinations are 
selected for further processing.
\item We add other 2 or 3 variables to the best variables combinations obtained in the previous step and we iterate the chain.
\end{itemize}
After a couple of iterations the best variable combination and $C,\gamma$ hyper-parameters choice remains stable within $1\div 2\%$.

\section{Final SVM Results}\label{sec:svm_res}
The SVM configurations obtained by the process described in this appendix are finally used in the multi-jet rejection phase of the analysis (Section~\ref{sec:svm_sel}) and in the background normalization estimates\footnote{By construction a fit on the SVM distance, $D$, offers a reliable estimate of the multi-jet background normalization.} (see Sections~\ref{sec:wjets_pretag} and~\ref{sec:qcd}).

\begin{figure}[!h]
  \begin{center}    
    \includegraphics[width=0.495\textwidth]{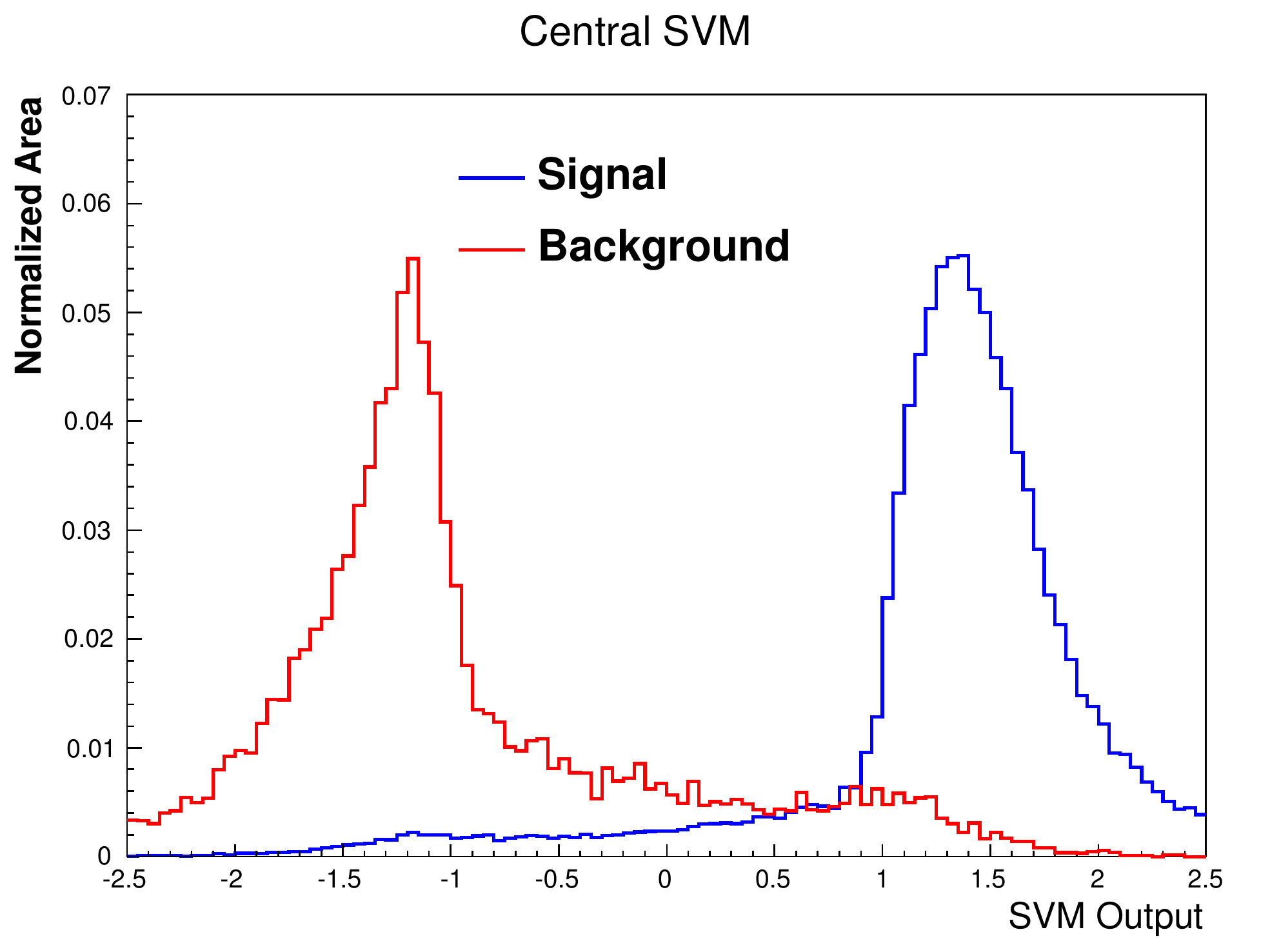}
    \includegraphics[width=0.495\textwidth]{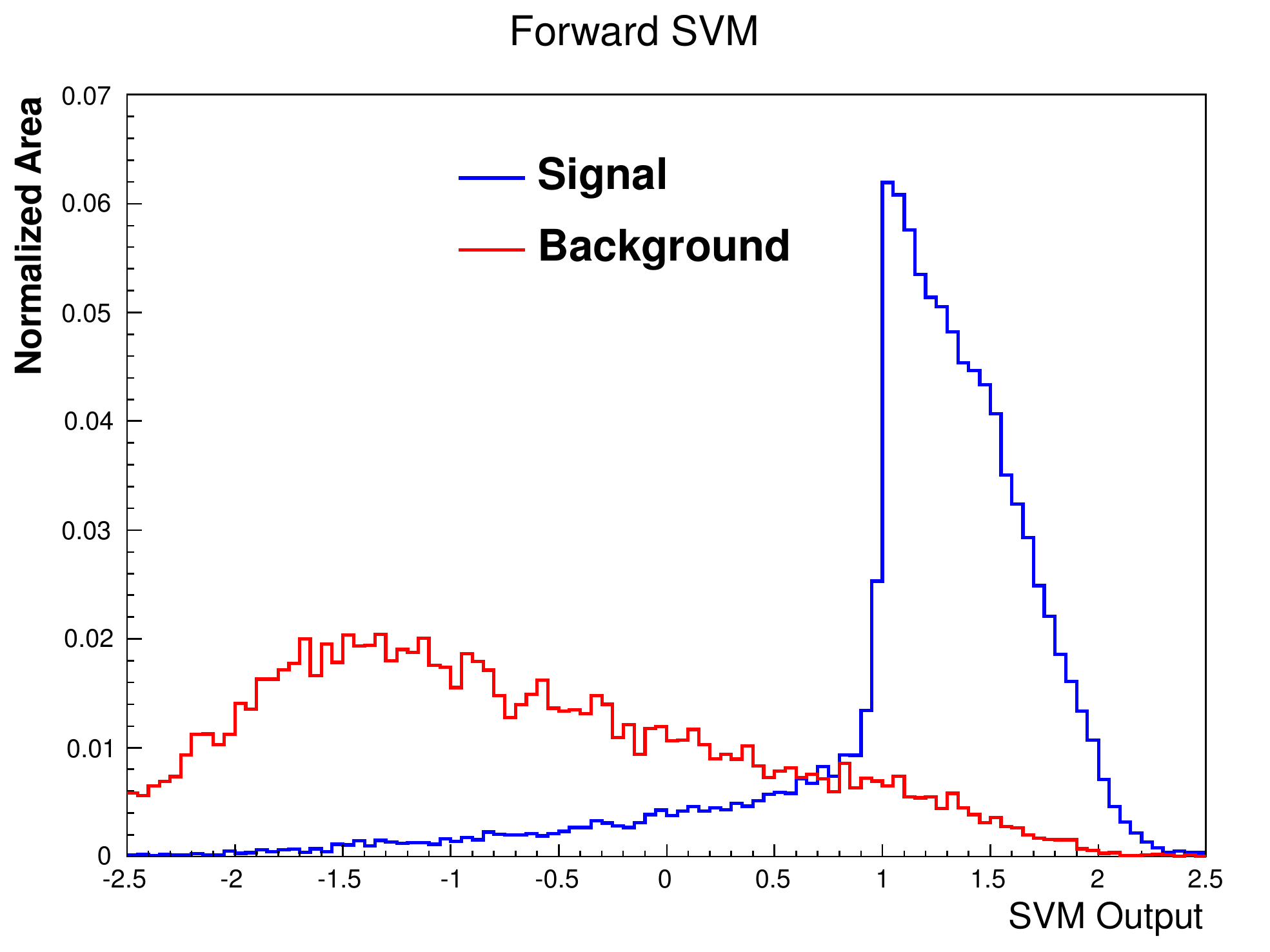}
  \caption[Final SVM Discriminants for Central and Forward Detector Regions]{Distribution of the SVM distance, $D$, described in Equation~\ref{eq:distance} for the central (left) and the forward (right) SVM discriminants obtained from the optimization process. Multi-jet background models are shown in red, $W+2$ partons MC signal is shown in blue.}\label{fig:final_svm}
  \end{center}
\end{figure}

The optimal hyper-parameter configurations that we obtain for the central (superscript $c$) and forward (superscript $f$) SVMs are:
\begin{equation}
C^c = 7,\quad  \gamma^c = −- 1;
\end{equation}
\begin{equation}
C^f = 8,\quad  \gamma^f = −- 1.
\end{equation}
Table~\ref{tab:svm_fin_vars} reports the final eight input variables used for the central SVM and the six ones used for the forward SVM.
\begin{table}
  \begin{center}   
  \begin{tabular}{lccc}
    \toprule
    \multicolumn{4}{c}{ Final SVM Input Variables}\\
    \midrule
    Central SVM: &  $M^W_T$ & \metr &  ${\not}{p_T}$\\ 
                 & $MetSig$ & $\Delta \phi({\not}{p_T},$ \met) & $\Delta \phi(lep,$ \met$)$ \\
                 & $\Delta R(\nu^{Min},$ $lep)$  & $\Delta \phi(Jet1,$ \met)  & \\
    \midrule
    Forward SVM: & $M^W_T$ & \metr & ${\not}{p_T}$ \\
                 &  $MetSig$ &  $\Delta \phi({\not}{p_T},$ \met) & $\Delta \phi({\not}{p_T},$ \metr)  \\
    \bottomrule
  \end{tabular}
  \caption[Final SVM Input Variables]{Input variables used for the configuration of the {\em central} and {\em forward} SVM multi-jet discriminants.}\label{tab:svm_fin_vars}
  \end{center}
\end{table}
Figure~\ref{fig:final_svm} shows the complete shape of the two discriminants for the multi-jet background models and the $W+2$ partons signal. A quantitative measurement of the performances can be seen in Table~\ref{tab:svm_eff}.

We can conclude that we successfully built a multi-jet rejection tool based of the SVM algorithm, nevertheless the challenges of a multi-variate approach to this problem. The CDF II dataset was a perfect test-bench for this problem with very good performances, however the procedure can be exported also to any of the LHC experiments.


\clearpage

\chapter{WHAM: WH Analysis Modules}\label{chap:AppWHAM}

High energy physics data analysis is done with the help of complex software frameworks which allow to manage the huge amount of information collected by the detector. 
The analysis framework which was used for this analysis is named $WH$ {\em Analysis Modules} or WHAM and I was one of the main developers of the package.  

The aim of the software is a reliable event selection and background estimate in the $\ell\nu+HF$ channel, with the possibility to easily implement new features and studies that can improve the Higgs search.
WHAM plays a relevant role in the CDF low-mass Higgs boson search, especially in the $WH\to \ell\nu+b\bar{b}$ channel but also in other contexts, like $t\bar{t}H$ and lately also $ZH$ searches. 

The analysis package tries to incorporate the CDF knowledge about the $\ell\nu+HF$ channel, most of it coming from the top~\cite{top_lj2005} and single-top~\cite{single_top} analyses. An effort was also made in the direction of code modularization and analysis customization with option loading at run-time.

A detailed explanation of the package is beyond the goal of this thesis, however a general overview of the package structure and functionality is given. More information is available in the CDF internal pages~\cite{wham_twiki} although a comprehensive documentation is not yet available.

\section{Package Structure}

The package is organized in a folder structure organized according to the purpose of each of the sub-elements. The {\em first level} directories are:
\begin{description}
\item[Setup:] contains the scripts needed to setup the analysis environment, both first installation and every day use, the references to all the external tools and any patche that needs to be applied.
\item[Documentation:] contains all the internal and public documentation. It is easily accessible and customizable by all the analysis group collaborators.
\item[Inputs:] contains the database files of the MC samples, the parametrization of the triggers and $SF$s, the re-weighting templates, the option configuration files. Basically every input to the analysis elaboration is here, except the data and the MC ntuples themselves.
\item[Commands:] this is more an utilities repository, it contains scripts to run the analysis on the CDF Grid for parallel computing (named CAF~\cite{caf_Sfiligoi2007}), plus a wide set of macros and scripts used for single-sample studies, text file processing or small data-handling tasks. 
\item[Results:] contains all the information elaborated by the rest of the analysis packages. This includes pre-processed data and MC samples as well as the final templates obtained after the complete elaboration. Several commands expect to find the input files here.
\item[Modules:] this is the core of the analysis package. It contains the C++ code used for the selection, the background estimate and the final production of the templates used for the statistical analysis and the validation of the kinematic distributions. Next section will describe it in more details.
\end{description}

\subsection{WHAM Modules}

The core of the WHAM analysis framework is the Modules directory. Here each functional step of the analysis is classified in a {\em module}, i.e. a self consistent C++ class built with standardized structure to allow straightforward compilation and testing. 

The modules are of two kinds: {\em functional modules} and {\em construction modules}. The formers are in charge of actually perform an operation, for example the event selection or the drawing of stacked histograms. The second kind of modules are the sub-components used by the first, for example the selection code needs to known the format of the input and output data as well as the definition of the lepton-object or jet-object.

The level of abstraction offered by the building modules proved to be extremely powerful. Two minimal examples (on which I contributed) that revealed to be extremely useful are: the handling of the configuration options and of the $b$-tag efficiency estimate. 
For the first, I implemented, using the \verb'libconfig' library~\cite{libconfig}, a text file reading utility that allows a single location definition and the run-time loading of all the options needed by the functional modules. For the second, it was necessary to identify the minimal amount of information needed to define a $b$-tagging algorithm. The only two values needed\footnote{Functional dependencies and correlations should be already taken into account.} are: $SF_{Tag}$ and $p_{Mistag}^j$ (see Section~\ref{sec:tag-sel}). With this information, it was possible to develop a single algorithm for the combination of any number of different $b$-tagging algorithms, for any required tag and jet multiplicity. The code works iteratively on the jets of the event requiring the definition of a $b$-tag in priority order, defined by the user.


The functional modules are four and, in the directory structure, are identified by the \verb'process' prefix:
\begin{description}
\item[Sample Selection:] access to the production ntuples (see Section~\ref{sec:data-struc}), as well as lepton and jet selections, are performed here. The result is a small size ntuple, the \texttt{EvTree}, containing the 4-vectors of the identified particles plus all the relevant information needed in the next steps. The \texttt{EvTree} is defined by a class with complex {\em methods} working on the simple stored variables: this allows a huge saving of disk space and computing time at selection stage. Furthermore the portability of the ntuple class ease the reproducibility of the same algorithms and allows faster checks.
\item[Sample Pre-processing:] here the complete selection is applied to the \texttt{EvTree}'s and the pre-processing with more analysis-specific algorithms is performed. For example, MC samples are scaled to the expected yield with the application of the latest available $SF$'s and trigger efficiencies. Also, the multivariate discriminants are evaluated here. The pre-processing has the possibility to be interfaced to other ntuples than the \texttt{EvTree}.
\item[Background Estimate:] this is the last step in the analysis of the $\ell\nu+HF$ channel. The background estimate described in Chapter~\ref{chap:bkg} is applied here and the templates of the different signals, backgrounds and data samples are stored in a ROOT file with the derived normalizations. In general any other background estimate method can be plugged at the end of the analysis chain but for the moment only the one described in Chapter~\ref{chap:bkg} is available.
\item[Result Display:] this step completes the analysis in the sense that it allows the comparison of the final estimate with the observed data distributions. Histograms are produced together with statistical indicators of the shape agreement: $\chi^2$, Kolmogorov-Smirnov tests, systematic overlay, background subtracted plots. Tables and histograms can appear in several formats, from html pages to simple eps files.
\end{description}
The very last step in each analysis, the statistical interpretation of the results, is implemented in a different software tool~\cite{mclimit} that is interfaced with the templates produced by the WHAM background estimate.

\section{Relevant Results}

Beyond the analysis presented here, a wide range of other analyses exploit the WHAM package. Between the most relevant: the new single-top cross section measuremens~\cite{single_top_benwu_10793}, the $t\bar{t}H$~\cite{tth_10801} search in the lepton plus jets channel and, lately, also top-properties~\cite{topR_10887} and SUSY searches are exploiting the package.

\begin{figure}[!ht]
\begin{center}
\includegraphics[width=0.85\textwidth,height=0.75\textwidth]{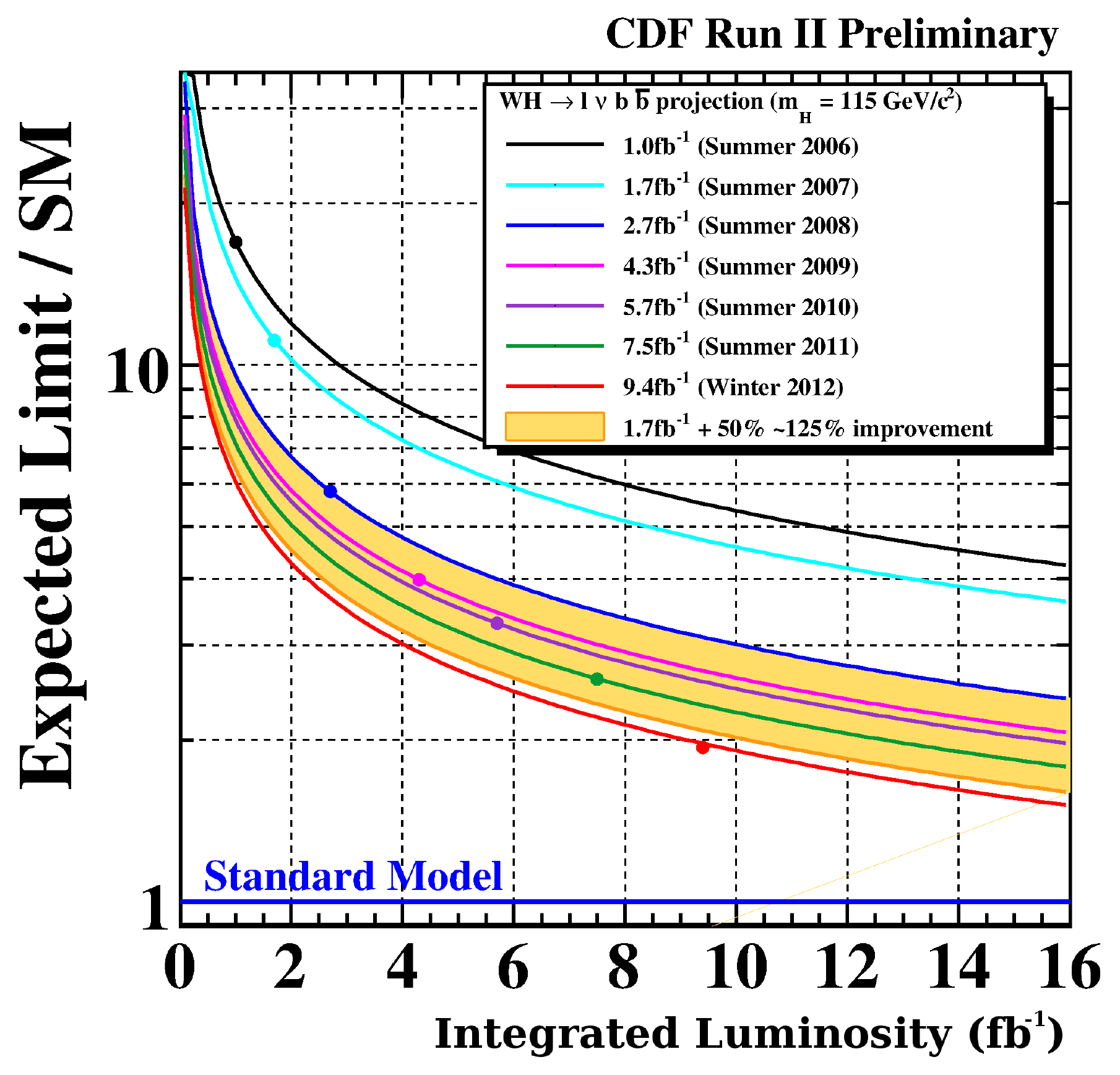}
\caption[Improvements to the CDF $WH$ Search Sensitivity]{Improvements to the CDF $WH\to \ell\nu+ b\bar{b}$ search sensitivity for a Higgs boson of $m_H=115$\gc2. Both the 2011~\cite{wh75_note10596} and 2012~\cite{wh94_note10796} results are mainly obtained within the WHAM analysis framework. The sensitivity reached improves more than 125\% over the first analysis of the same channel.}\label{fig:WH_improve}
\end{center}
\end{figure}

Figure~\ref{fig:WH_improve}, probably, shows the most striking result: the improvement of the $WH$ search sensitivity (7.5\fb1 and 9.4\fb1 versions~\cite{wh75_note10596, wh94_note10796}). The several improvements produced for the Higgs search were readily implemented and tested thanks to the backbone of a reilable framework. 

Nevertheless the shrinking of the CDF collaboration, the final sensitivity to the Higgs boson exceeded the best expectations.

\clearpage
\chapter[First Evidence of Diboson in $\ell\nu+ HF$ ($7.5~fb^{-1}$)]{First Evidence of Diboson in Lepton plus Heavy Flavor Channel ($7.5~\textrm{fb}^{-1}$)}\label{App:7.5}

The first evidence of diboson production in the  $\ell \nu + HF$ final state was already obtained in the Summer of 2011 with a preliminary version of this analysis performed on a smaller dataset, $7.5$\fb1 of CDF data. 

The differences with respect to the work described in the main parts of the thesis are the followings:
\begin{itemize}
\item a dataset of $7.5$\fb1 of CDF data.
\item The forward electrons category (PHX) was not used.
\item A previous version~\cite{CHEP_svm} of the SVM multi-jet rejection algorithm was employed. The algorithm was optimized only for binary classification and it was not possible to use the shape of the output distribution. Therefore the \met distribution was used to in the normalization of the $W + $ jets and the multi-jet backgrounds.
\item  No flavor separator ($KIT-NN$) information was used for single-tag events. Therefore only the di-jet invariant mass distribution was used as the final signal to background discriminator both for the single and double tagged signal regions.
\item The statistical analysis of the significance of the observation was performed in a different way. The likelihood ratio~\cite{mclimit} of the signal and {\em test} hypothesis, after the fit over the nuisance parameter, was used\footnote{This method can not be easily applied to the significance estimate of two signals (for example $WW$ {\em vs} $WZ/ZZ$) therefore, in the main part of the thesis, we moved to the method described in Section~\ref{sec:sigma_eval} for the statistical analysis}. 
\item We also estimated $95$\% Confidence Level (CL) limits, both in the case of diboson signal and no diboson signal. The first case can be used to constrain new physics models which produces an increase of the TGC couplings.
\end{itemize}
A summary of the event selection, the background estimate and the statistical analysis is reported in the following.

\section{Event Selection and Background Estimate}\label{sec:Data}

We select events consistent with the $\ell\nu+HF$ signature. 

The charged lepton candidate online and offline identification is described in Chapter~\ref{chap:sel} but we consider only the tight central lepton candidates (CEM, CMUP, CMX), the loose lepton candidates (BMU, CMU, CMP, CMIO, SCMIO, CMXNT) and the isolated track candidates (ISOTRK). Loose leptons and ISOTRK are classified together in the EMC category. The $W\to \ell \nu$ selection is completed by a cut on the \met variable, corrected for the presence of muons and jets. We require a \mbox{\met$>20$\gv} for CEM and EMC leptons while we relax the cut down to \mbox{\met$>10$\gv} for the tight muons categories. 

The $HF$ jets selection is also identical to the one reported in Chapter~\ref{chap:sel}: two central ($|\eta_{Det}| < 2.0$) jets with E$_T^{cor} > 20$~GeV (energy corrected for detector effects) 
on which we require the identification of one or two secondary vertices with the \texttt{SecVtx} $b$-tagging algorithm. The $b$-tagging requirement divides the selected sample in a {\em pretag} control region (no $b$-tag requirement) and two signal regions characterized by exactly one or two $b$-tagged jets.

A relevant difference in the event selection comes from the multi-jet background rejection strategy. We used a previous version of the SVM algorithm~\cite{CHEP_svm} based on the following six variables (described in Section~\ref{sec:var_sel}):
\begin{itemize}
\item Lepton $p_T$, \met, $MetSig$, $\Delta\phi(lep,\metr)$,  $E_T^{raw,jet2}$ and $E_T^{cor,jet2}$.
\end{itemize}
The discriminant was optimized to work in the central region of the detector and, as binary classifier, the performances were similar to the present algorithm described in Appendix~\ref{chap:AppSvm}.

\begin{figure}[!ht]
\begin{center}
\includegraphics[width=0.495\textwidth]{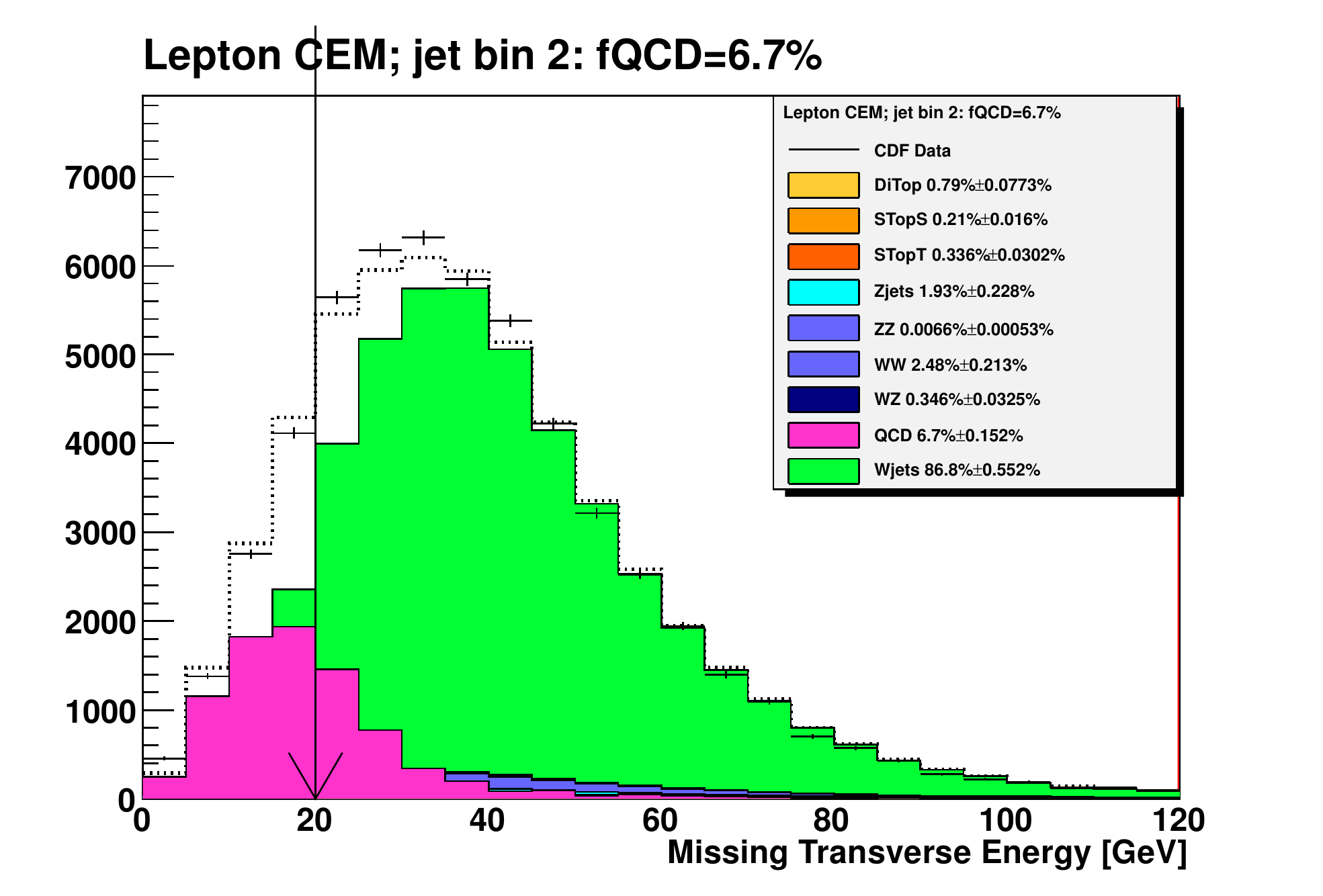}
\includegraphics[width=0.495\textwidth]{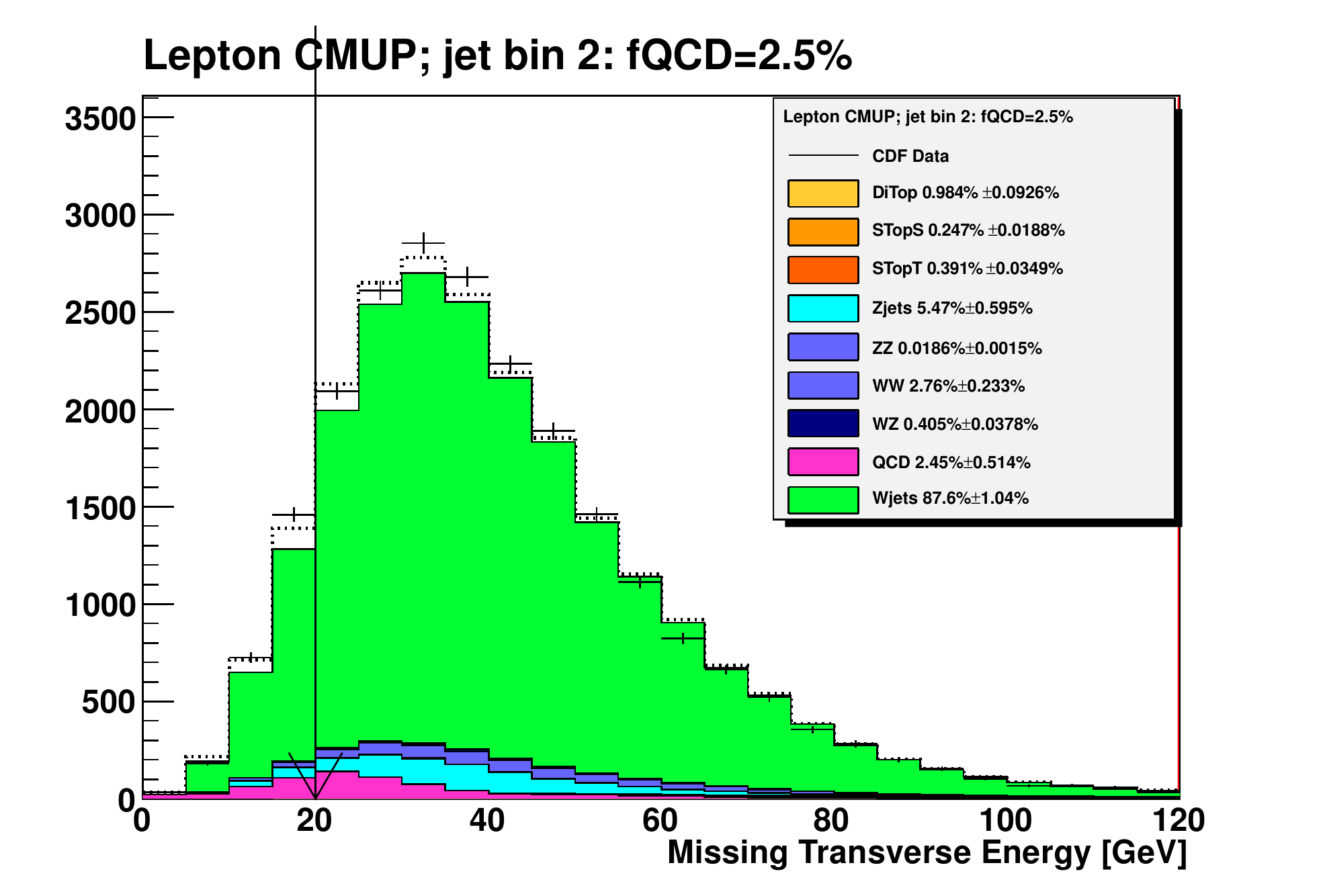}
\includegraphics[width=0.495\textwidth]{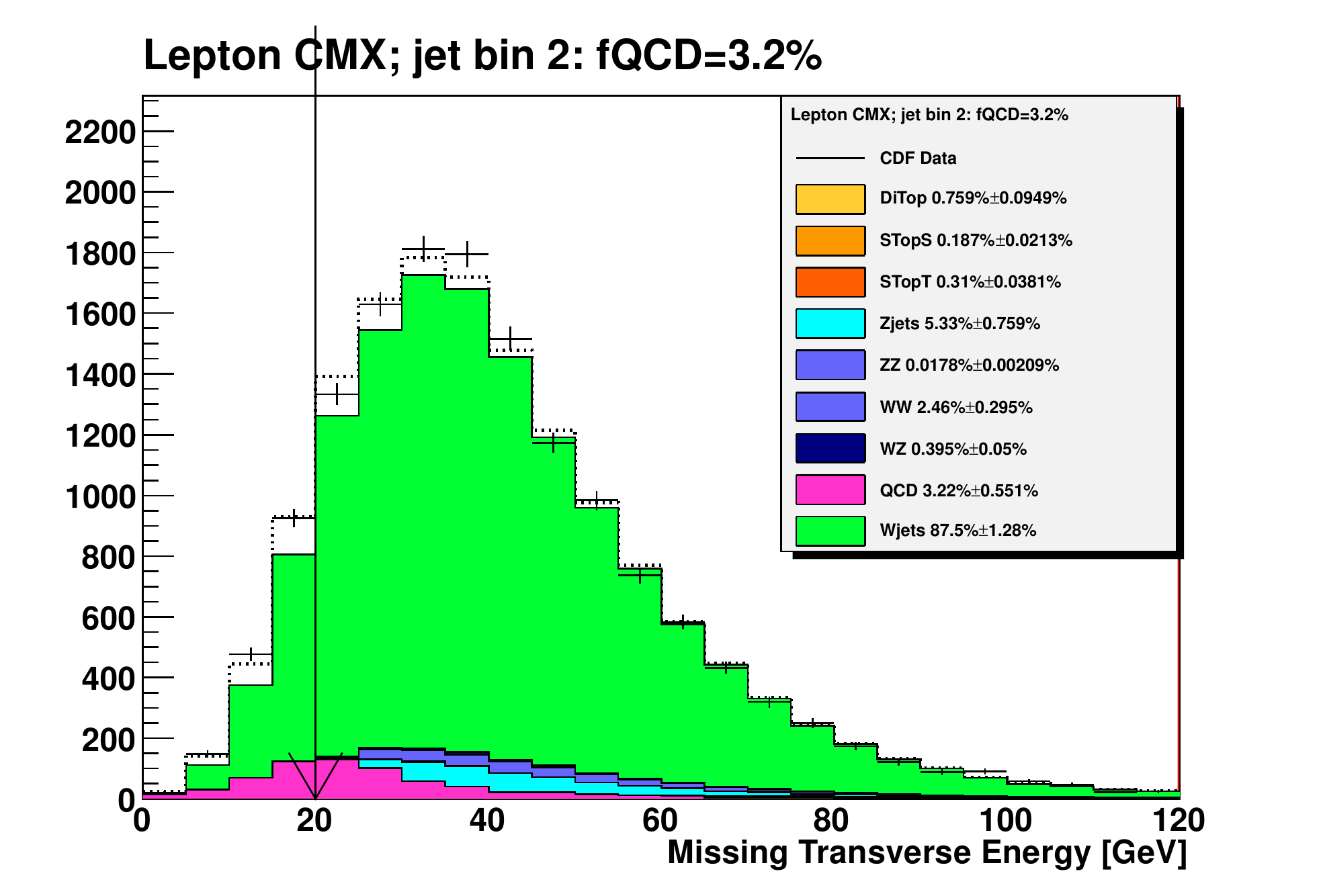}
\includegraphics[width=0.495\textwidth]{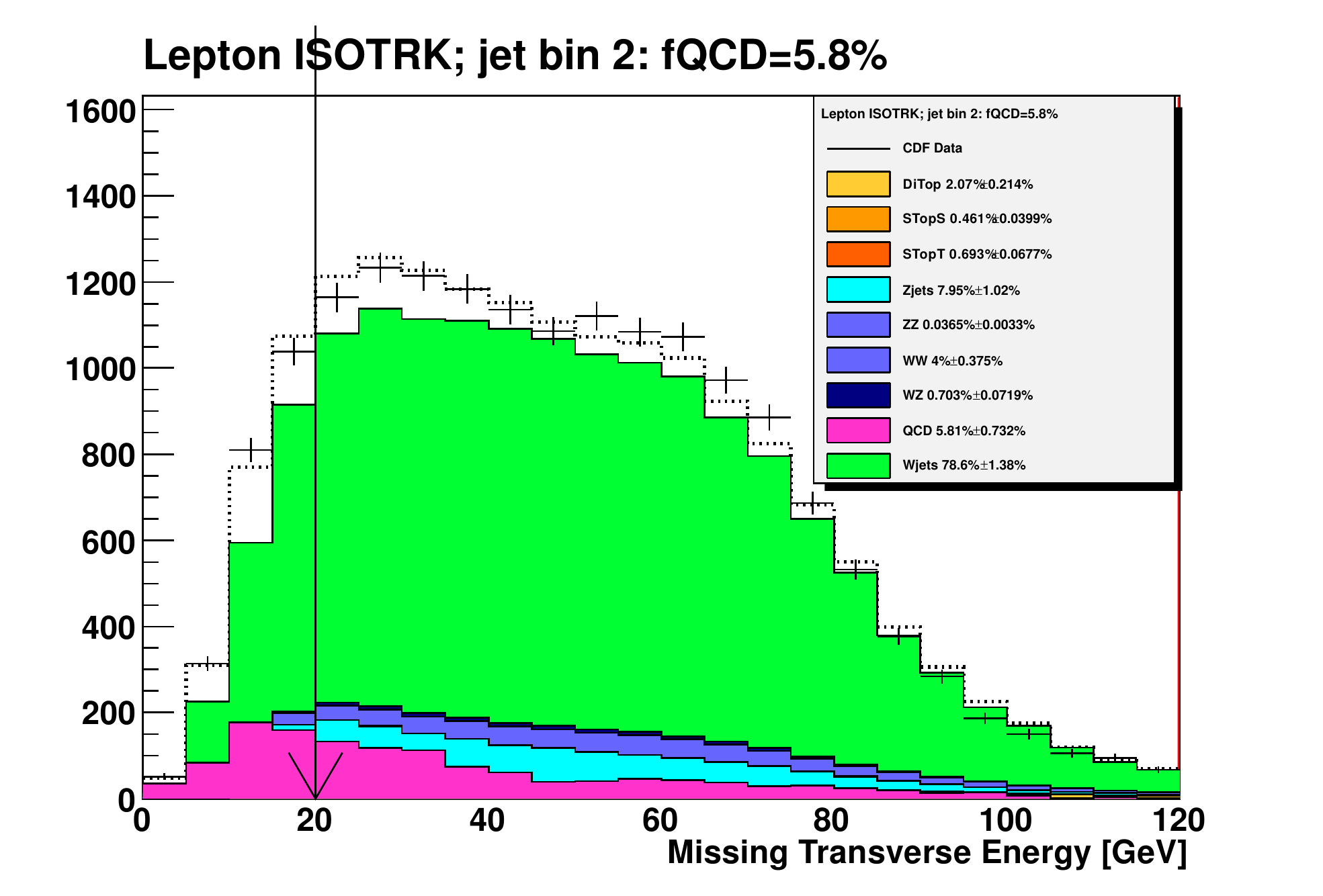}
\caption[\met Fit Used for $W +$ Jets Estimates in the Pretag Sample]{$W + $ jets  and non-$W$ fraction estimates, on the pretag sample, with a maximum likelihood fit on the \met distributions. The non-$W$ background is shown in pink while the $W + $ jets component is in green. The dashed line is the sum of all the backgrounds and the points are the data. The figures show (left to right and top to bottom) the CEM, CMUP, CMX and EMC 
  charged lepton categories.} \label{fig:qcd_fit_met}
\end{center}
\end{figure}

The background estimate was done with the same methodology described in Chapter~\ref{chap:bkg}. The only difference is that the fits, used to estimate $W + $ jets and multi-jet normalizations in the different lepton categories and tag regions, were performed on the \met distributions. Figure~\ref{fig:qcd_fit_met} shows the result of the pretag maximum likelihood fit on the \met, pretag sample, distribution of the different lepton categories: CEM, CMUP, CMX, EMC.

\begin{figure} [!ht]
\centering
\includegraphics[width=0.495\textwidth]{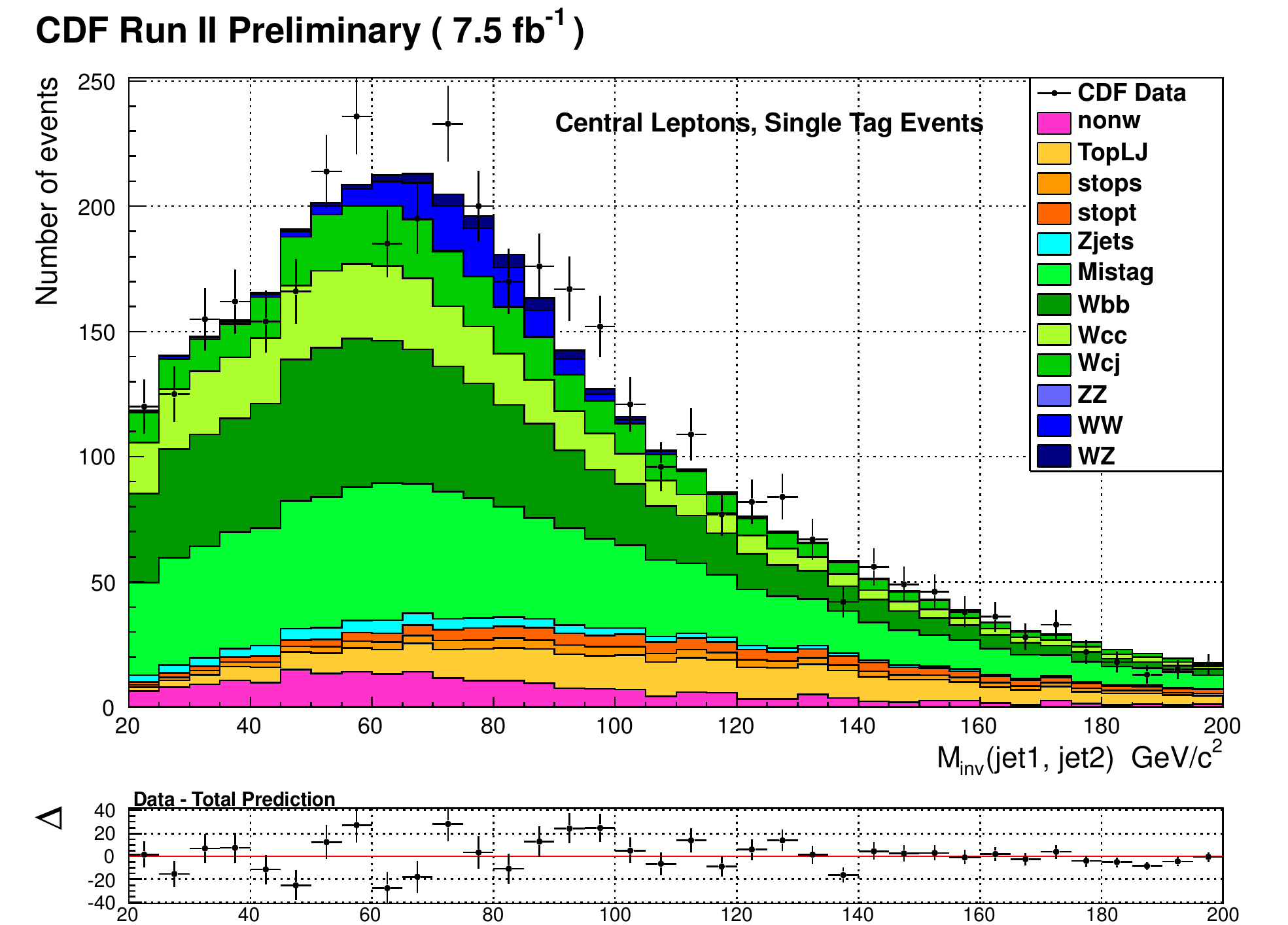}
\includegraphics[width=0.495\textwidth]{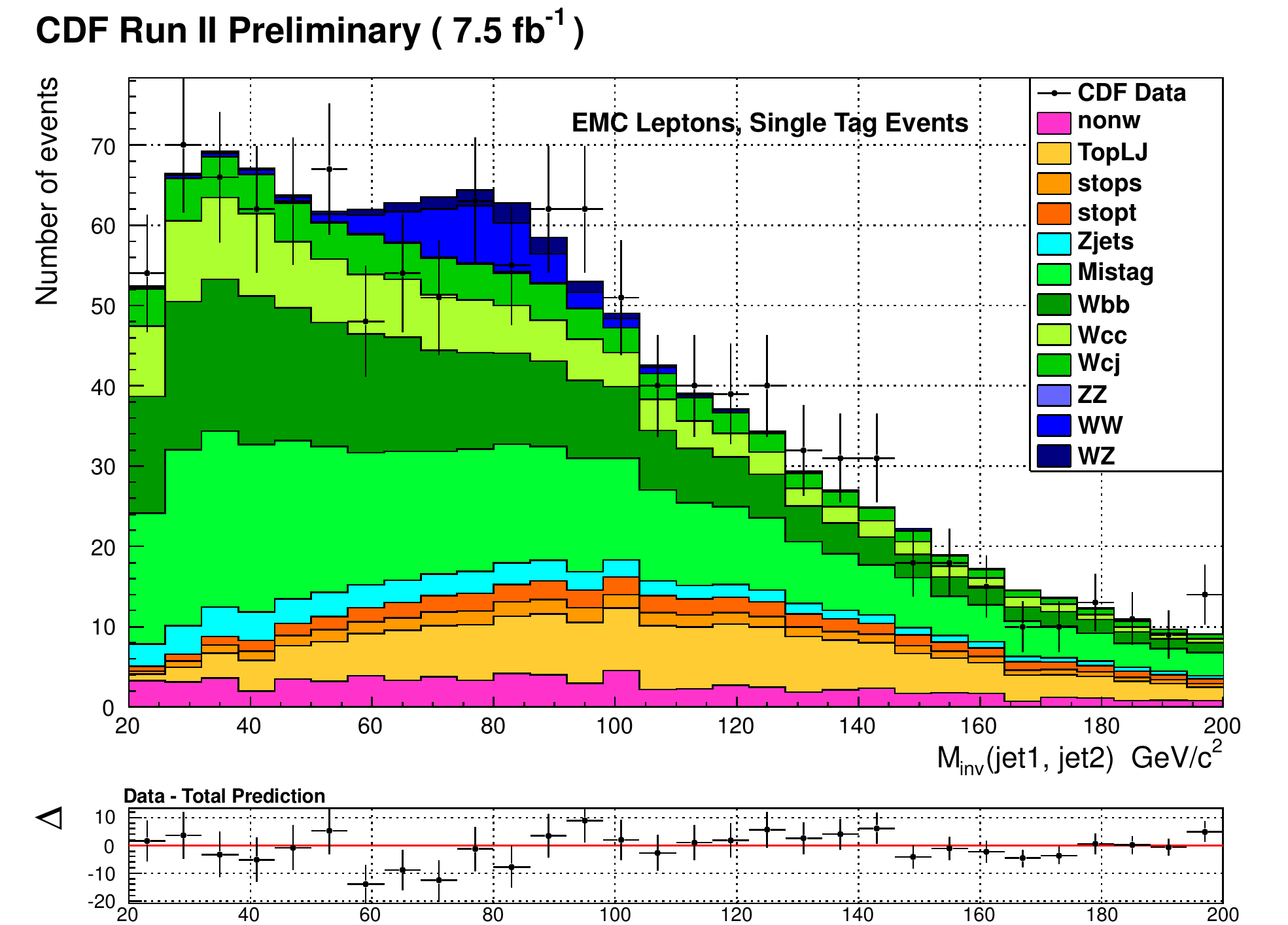} 
\caption[$M_{Inv}(jet1,jet2)$ Distribution for Single \texttt{SecVtx} Tag Events]{$M_{Inv}(jet1,jet2)$ distribution for the single \texttt{SecVtx} tagged events. Tight leptons (CEM+CMUP+CMX combined) on the left and EMC leptons on the right. The best fit values for the rate and shape of the backgrounds are used in the figures.}  
\label{fig:mjj_1tag}
\end{figure}

The total background estimate is reported in Tables~\ref{tbl:1tag} and~\ref{tbl:2tags}. The final statistical analysis of the selected events is performed on the $M_{Inv}(jet1, jet2)$ distribution of four channels: single and double tagged candidates for the central tight leptons (CEM+CMUP+CMX) and EMC leptons. Figure~\ref{fig:mjj_1tag} shows the high-statistics single-tag  $M_{Inv}(jet1, jet2)$ distribution of central tight leptons and of the EMC leptons.

\begin{table}[htb]
  \begin{center}
    \begin{tabular}{lcccc}
      \toprule
      \multicolumn{5}{c}{\bf Single-tag Event Selection}\\
      Lepton ID & CEM                       &             CMUP             & CMX    & EMC    \\  
      \midrule    

     Pretag Data       & 61596  & 29036 & 18878  &  27946 \\\midrule
      $Z+$jets            &27.9 $\pm$ 3.5     & 43.0 $\pm$ 5.5    & 27.3 $\pm$ 4.2     & 65.0 $\pm$ 8.4     \\  
      $t\bar{t}$        &201.3 $\pm$ 19.6   &109.8 $\pm$ 10.7   & 55.0 $\pm$ 7.1     & 171.9 $\pm$ 16.9    \\  
      Single Top $s$    &52.9 $\pm$ 4.8     &28.2 $\pm$ 2.6     & 14.0 $\pm$ 1.8     & 38.0 $\pm$ 3.5       \\  
      Single Top $t$    &71.4 $\pm$ 8.4     &37.4 $\pm$ 4.4     & 19.8 $\pm$ 2.9     & 49.5 $\pm$ 5.8       \\  
      $WW$                &68.0 $\pm$ 9.4     &33.3 $\pm$ 4.6     & 20.3 $\pm$ 3.3     & 38.4 $\pm$ 5.3       \\  
      $WZ$                &21.8 $\pm$ 2.3     & 11.5 $\pm$ 1.25   & 7.4 $\pm$ 1.0      & 14.1 $\pm$ 1.6     \\ 
      $ZZ$                &0.44 $\pm$ 0.04    & 0.65 $\pm$ 0.06   & 0.42 $\pm$ 0.05    & 0.86 $\pm$ 0.08    \\    
      $W + b\bar{b}$    &632.9 $\pm$ 254.2  & 309.8$\pm$ 124.1  & 192.3 $\pm$ 77.1   & 308.9 $\pm$ 124.3  \\
      $W + c\bar{c}$    &331.0 $\pm$ 133.7  & 155.1 $\pm$ 62.5  & 96.2 $\pm$ 38.8    & 164.2 $\pm$ 66.4   \\
      $W + cj$            &259.9 $\pm$ 105.0  & 127.8 $\pm$ 51.5  & 75.3 $\pm$ 30.4    & 106.4 $\pm$ 43.0   \\
      $W+LF$            &605.2 $\pm$ 71.3   & 283.8 $\pm$ 31.7   & 181.0 $\pm$ 20.6  & 346.2 $\pm$ 39.2   \\   
      Non-$W$             &173.9 $\pm$ 69.6   & 45.8 $\pm$ 18.3  & 2.8 $\pm$ 1.1       & 100.9 $\pm$ 40.4   \\   
      \midrule                                                                                               
       {\bf Prediction}        &{\bf 2446 $\pm$ 503} & {\bf 1186 $\pm$ 242 } &{\bf  692 $\pm$ 149}  & {\bf 1404 $\pm$ 243} \\
       {\bf Observed}          &{\bf 2332}               & {\bf 1137}              & {\bf 699}                & {\bf 1318}               \\
      \midrule                                                                                            
      {\bf Dibosons}             &{\bf 89.7 $\pm$ 10.2}    & {\bf 44.8 $\pm$ 5.05}   & {\bf  27.7 $\pm$ 3.9}   &  {\bf 52.5 $\pm$ 5.9}     \\
      \bottomrule                                                   
    \end{tabular}

    \caption[Observed and Expected Events with One \texttt{SecVtx} Tags]{Summary of  observed and expected events with one \texttt{SecVtx} tag, in the $W+2$ jets sample, in 
      7.5 fb$^{-1}$ of data.}\label{tbl:1tag}
  \end{center}
\end{table}

\begin{table}[htb]
\begin{center}
    \begin{tabular}{lcccc}
    \toprule
    \multicolumn{5}{c}{\bf Double-tag Event Selection}\\
    Lepton ID & CEM                       &             CMUP             & CMX    & EMC    \\  
    \midrule    
      
       Pretag Data       & 61596   &  29036  &  18878  &  27946 \\\midrule
       $Z+$jets            & 0.9 $\pm$ 0.1   &2.0 $\pm$ 0.3    & 1.2 $\pm$ 0.2   & 3.1 $\pm$ 0.4  \\    
       $t\bar{t}$        & 42.2 $\pm$ 6.1  &22.2 $\pm$ 3.2   & 11.1 $\pm$ 1.9  & 34.4 $\pm$ 5.0  \\
       Single Top $s$    & 14.1 $\pm$ 2.0  &7.6 $\pm$ 1.1    & 3.7 $\pm$ 0.6   & 10.2 $\pm$ 1.4  \\ 
       Single Top $t$    & 4.2 $\pm$ 0.7   &2.3 $\pm$ 0.4    & 1.2 $\pm$ 0.2   & 3.1 $\pm$ 0.5   \\ 
       $WW$                & 0.6 $\pm$ 0.1   & 0.26 $\pm$ 0.07 & 0.16 $\pm$ 0.04 & 0.33 $\pm$ 0.08\\  
       $WZ$                & 4.0 $\pm$ 0.6   & 1.9 $\pm$ 0.3   & 1.4 $\pm$ 0.2   & 2.4 $\pm$ 0.4  \\  
       $ZZ$                & 0.06 $\pm$ 0.01 & 0.12 $\pm$ 0.02 & 0.09 $\pm$ 0.01 & 0.16 $\pm$ 0.02  \\
       $W + b\bar{b}$    & 81.9 $\pm$ 33.2 &42.2 $\pm$ 17.1  & 23.4 $\pm$ 9.5  & 44.9 $\pm$ 18.2\\
       $W + c\bar{c}$    & 4.7 $\pm$ 1.9   &2.3 $\pm$ 1.0    & 1.3 $\pm$ 0.5   & 2.8 $\pm$ 1.1   \\ 
       $W + cj$            & 3.7 $\pm$ 1.5   &1.9 $\pm$ 0.8    & 1.0 $\pm$ 0.4   & 1.8 $\pm$ 0.7  \\    
       $W+LF$            & 3.2 $\pm$ 0.7   &1.6 $\pm$ 0.3    & 0.9 $\pm$ 0.2   & 2.2 $\pm$ 0.4  \\  
       Non-$W$             & 7.9 $\pm$ 3.2   &4.8 $\pm$ 1.9    & 0.1 $\pm$ 0.5   & 0.0 $\pm$ 0.5   \\   
       \midrule                                                                               
       {\bf Prediction}        & {\bf 167.3 $\pm$ 38.0} &{\bf 88.9 $\pm$ 19.6}  & {\bf 45.4 $\pm$ 10.9} & {\bf 105.3 $\pm$ 21.5}\\
       {\bf Observed}          & {\bf 147}             &{\bf 74}               &{\bf  39}              & {\bf 106}             \\  
       \midrule
       {\bf Dibosons}             &{\bf  4.6 $\pm$ 0.6}   &{\bf  2.1 $\pm$ 0.3}   & {\bf 1.5 $\pm$ 0.2}   &{\bf  2.7 $\pm$ 0.4}   \\
       \bottomrule
       \end{tabular}  

       \caption[Observed and Expected Events with Two \texttt{SecVtx} Tags]{Summary of observed and expected events with two \texttt{SecVtx} tags, in the $W+2$ jets sample, in $7.5$~fb$^{-1}$ of data.}\label{tbl:2tags}
\end{center}
\end{table}

\section{Statistical Analysis}

At the time of this analysis, the process $WZ/WW \rightarrow \ell \nu $ + Heavy Flavors was not yet observed at hadron colliders. Therefore we started by evaluating a 95\% exclusion CL on a potential signal and then we evaluated the cross section of the process and its significance. 

Most of the statistical analysis procedure has been described in Chapter~\ref{chap:StatRes}. We build a likelihood function (Equation~\ref{eq:likelihood_full}) with templates derived from the selected data and backgrounds, both shape and rate systematics are taken into account as described in Section~\ref{sec:sys_desc}. The only relevant change concerns the $b$-tag SF for $c$-marched jets where we applied the same prescriptions of the $WH$ CDF search, it does not double the uncertainty for $c$-matched quarks. 

The last relevant difference is the evaluation of the 95\% CLs for the diboson signal. The CLs are evaluated by integrating the likelihood distribution over the unknown parameter $\alpha$ (i.e. the diboson cross section) up to cover 95\% of the total possible outcomes. In formulas:
\begin{equation}
\int^{\bar{\alpha}}_{0} \mathscr{L}(\alpha)\mathrm{d}\alpha.  
\end{equation}
where $\mathscr{L}(\alpha)$ is derived from Equation~\ref{eq:likelihood_full} after the integration over the nuisance parameters, $\vec{\alpha}$ is reduced to just one dimension because we perform CLs on only one signal and the limit on the integration, $\bar{\alpha}$, is given by the condition of 95\% coverage.

A first set of expected CLs are obtained assuming no SM diboson production and generating Pseudo Experiments (PEs) on the base of the expected background yields varied within the assigned systematics. 
Combining single--tagged and double--tagged results for all lepton categories, we find an expected limit of:
\begin{equation}
\left( 0.575^{+0.33}_{-0.31} \right) \times \textrm{SM prediction.}
\end{equation}

We also calculated a second set of expected CLs assuming the predicted SM diboson yield in the PEs generation. In this case an excess in the observed CLs would indicate the presence of {\em new physics}.
Combining single--tagged and double--tagged results for all lepton categories, we find an expected limit of:
\begin{equation}
\left( 1.505 ^{+0.46}_{-0.61}\right) \times \textrm{SM prediction.}
\end{equation}

The observed limit of $1.46$ times the SM prediction is thereby consistent with the existence of a signal and no presence of new physics.

Table~\ref{tab:limits} summarizes the expected (and observed) 95\% production limits in units of the SM prediction. 

\begin{table}
\begin{center}
\begin{tabular}{lccc}
\toprule
 Category     & Expected 95\% CL & Expected 95\% CL & Observed \\ 
&  (No Diboson) &  (With Diboson)  & 95\% CL \\\midrule
Single Tag  &&&\\
CEM+CMUP+CMX          & $0.715 ^{+0.42}_{-0.38}$ & $1.625 ^{+0.54}_{-0.69}$ & 2.07 \\
EMC                   & $1.215 ^{+0.57}_{-0.46}$ & $2.125 ^{+0.76}_{-0.6}$ & 1.70 \\ \midrule
Double Tag &&&\\
CEM+CMUP+CMX         & $4.015 ^{+1.94}_{-1.50}$ & $4.875 ^{+2.15}_{-1.7}$ & 3.03 \\ 
EMC                  & $6.075 ^{+2.92}_{-2.20}$ & $7.015 ^{+3.15}_{-2.42}$ & 8.59 \\ \midrule
All combined                & $0.575 ^{+0.33}_{-0.31}$ & $1.505 ^{+0.46}_{-0.61}$ & 1.46 \\
\bottomrule                                                                                                       
\end{tabular}
\caption[Expected and Observed 95\% CL for Diboson Production]{Expected and observed 95\% exclusion confidence levels for each lepton category, single and double tagged events, in units of the SM cross section for diboson production. Expected CLs are produced also including the diboson signal generated PEs, thus probing the contribution of new physics processes.}\label{tab:limits}
\end{center}
\end{table}

\subsection{Significance and Cross Section Measurement}

To compute the significance of the signal, we performed a hypothesis test comparing the data to the likelihood ratio of the {\em null} and {\em test} hypotheses. 

The {\em null} hypothesis, $H_0$, assumes all the predicted processes except diboson production. The {\em test} hypothesis, $H_1$, assumes that the diboson production cross section and the branching ratio into $HF$ are the ones predicted by the SM. The likelihood ratio is defined as:
\begin{equation}\label{eq:l_ratio}
-2 \ln Q = -2 \ln \frac{ \mathscr{L}(data|H_1,\hat{\theta})}{ \mathscr{L}(data|H_0,\hat{\hat{\theta}})}
\end{equation}
where $\mathscr{L}$ is defined by Equation~\ref{eq:likelihood_full}, $\theta$ represents the nuisance parameters describing the uncertain values of the quantities studied for systematic error, $\hat{\theta}$ the best fit values of $\theta$  under $H_1$ and $\hat{\hat{\theta}}$ are the best fit values of the nuisance parameters under $H_0$.

\begin{figure}
\centering
\includegraphics[width=0.95\textwidth]{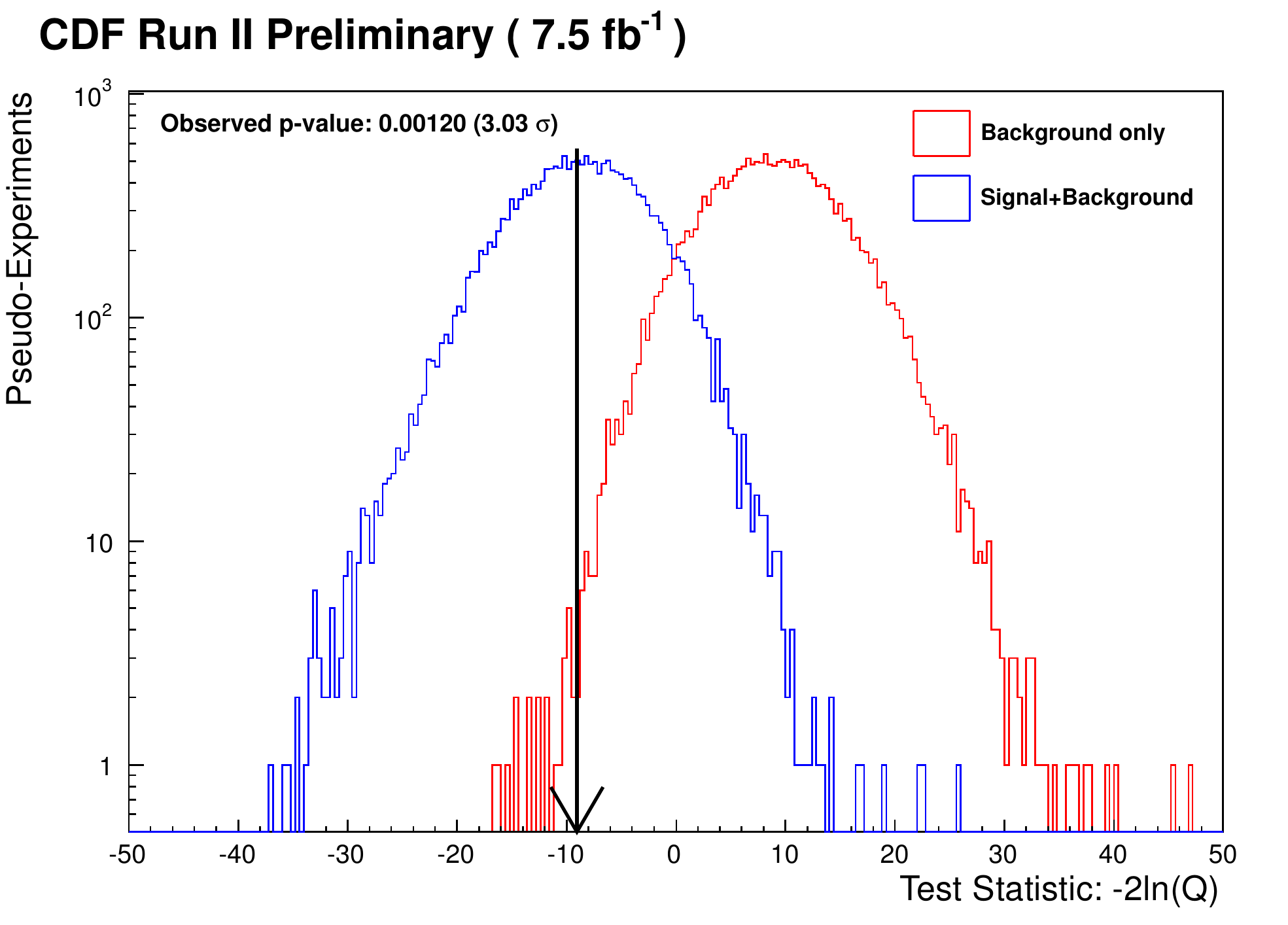}
\caption[Significance Estimate of Diboson Signal with $ - 2 \ln Q$]{Distributions  of $ - 2 \ln Q$  for the {\em test} hypothesis $H_1$, which assumes the estimated backgrounds plus SM diboson production and decay into $HF$ (blue histogram), and for the {\em null} hypothesis, $H_0$, which assumes no diboson  (red histogram). The observed value of $ - 2 \ln Q$  is indicated with a solid, vertical line and the $p$-value is the fraction of the $H_0$ curve integral to the left of the data.}\label{fig:gaussians}
\end{figure}

To perform the hypothesis test we generated two sets of PEs, one assuming $H_0$ and a second one assuming $H_1$ and we evaluated Equation~\ref{eq:l_ratio} for each pseudo-data outcome. The distributions of the  $- 2 \ln Q$ values are shown in Figure~\ref{fig:gaussians}. The integral of the $H_0$ $- 2 \ln Q$ distribution below the $- 2 \ln Q$ value of the real data gives the observed $p-$value of the signal.

We obtained an observed $p-$value of 0.00120, corresponding to a 3.03$\sigma$ excess and producing the {\em evidence} of the diboson signal.  The result is compatible with the expectation as the {\em test} hypothesis $p-$value, obtained from the median of the $H_1$ distribution of $ - 2 \ln Q$, is $0.00126$.

In order to measure the diboson production cross section, a Bayesian marginalization technique is applied as described in Section~\ref{sec:likelihood_meas}. The nuisance parameters are integrated, the maximum of the posterior distribution returns the cross section measurement while the smallest interval containing 68\% of the integral gives the 1-$\sigma$ confidence interval. The resulting cross section measurement is:
\begin{equation}
 \sigma_{Diboson}^{Obs}=\left(1.08^{+0.26}_{-0.40} \right)\times \sigma_{Diboson}^{SM} =  \left(19.9^{+4.8}_{-7.4}\right)\textrm{~pb,}
\end{equation}
where the errors include statistical and systematic uncertainties and $\sigma^{SM}_{Diboson}$ is the SM predicted cross section derived from Table~\ref{tab:mc_dib}

\clearpage

\chapter{Kinematic Distribution of the Signal Regions}

\begin{sidewaysfigure}
  \includegraphics[width=0.49\textwidth]{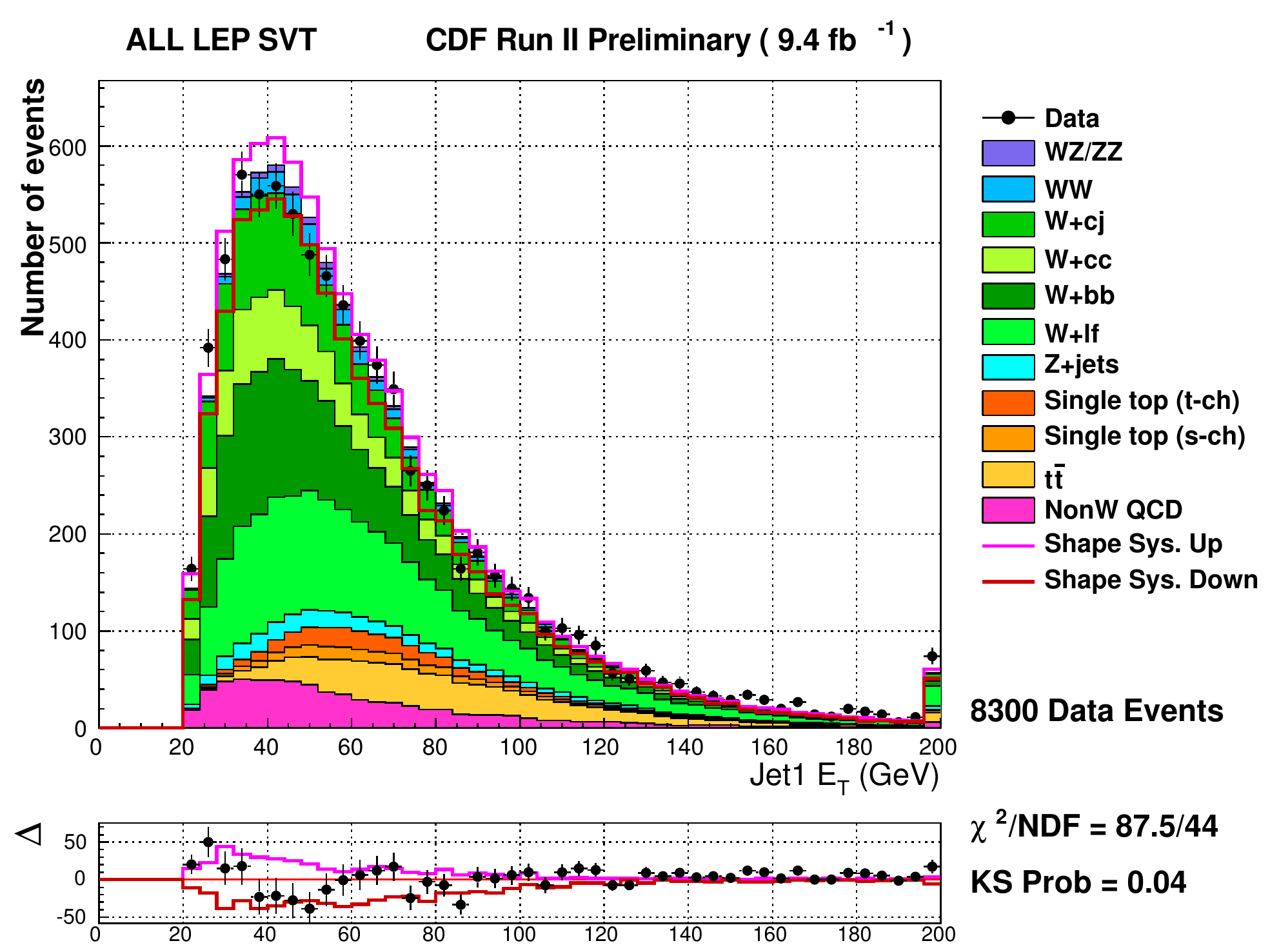}
  \includegraphics[width=0.49\textwidth]{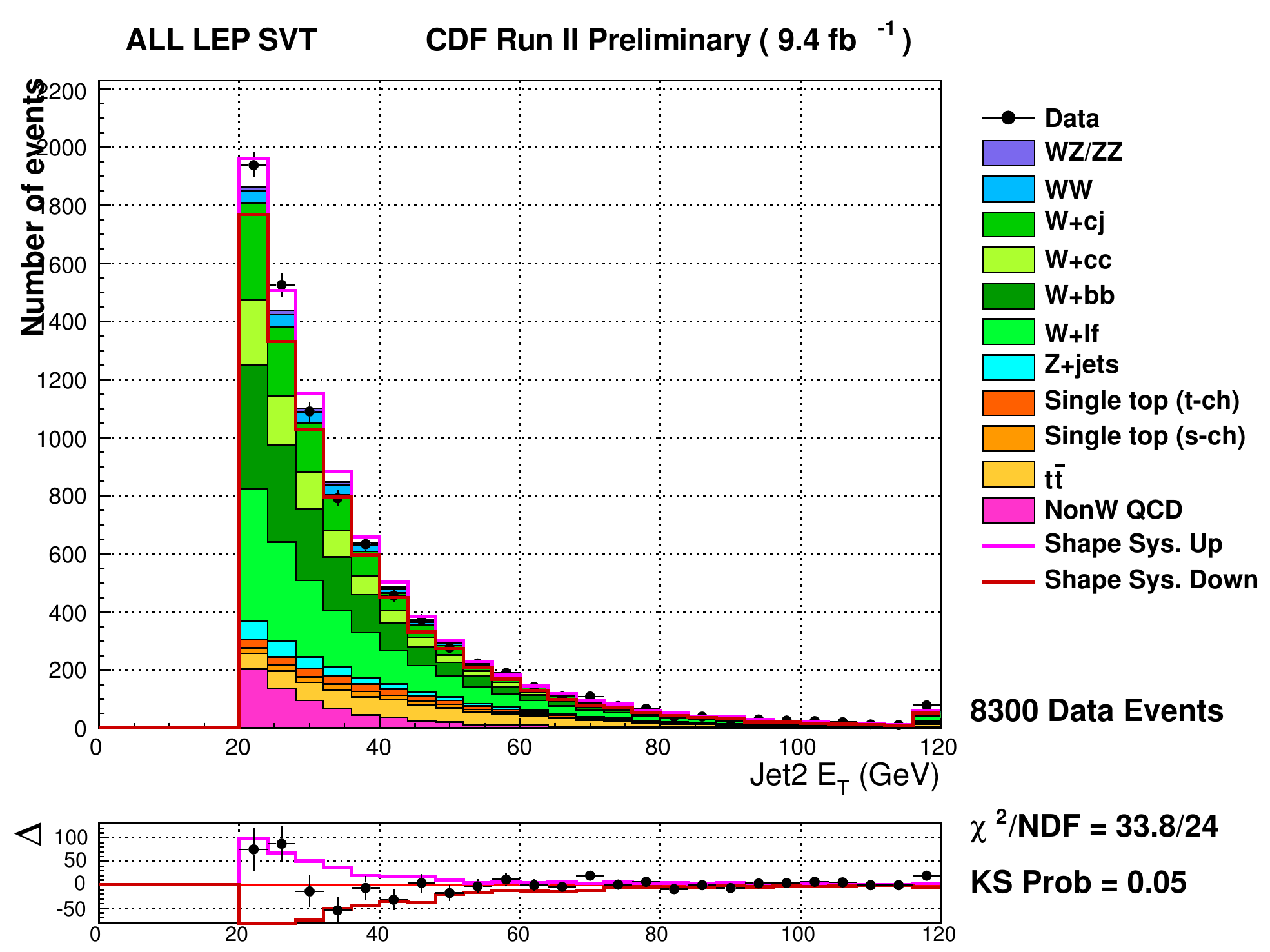}\\
  \includegraphics[width=0.49\textwidth]{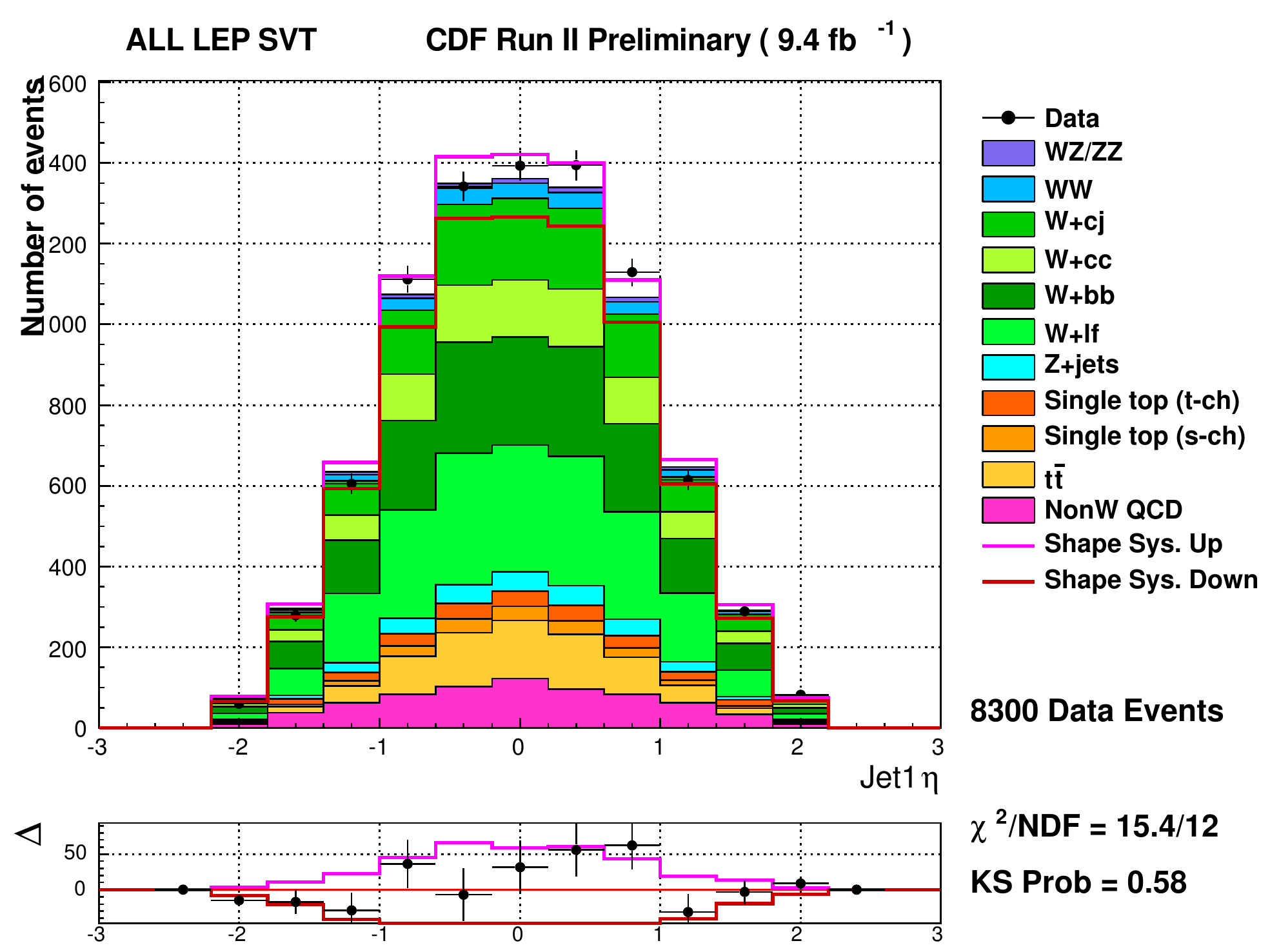}
  \includegraphics[width=0.49\textwidth]{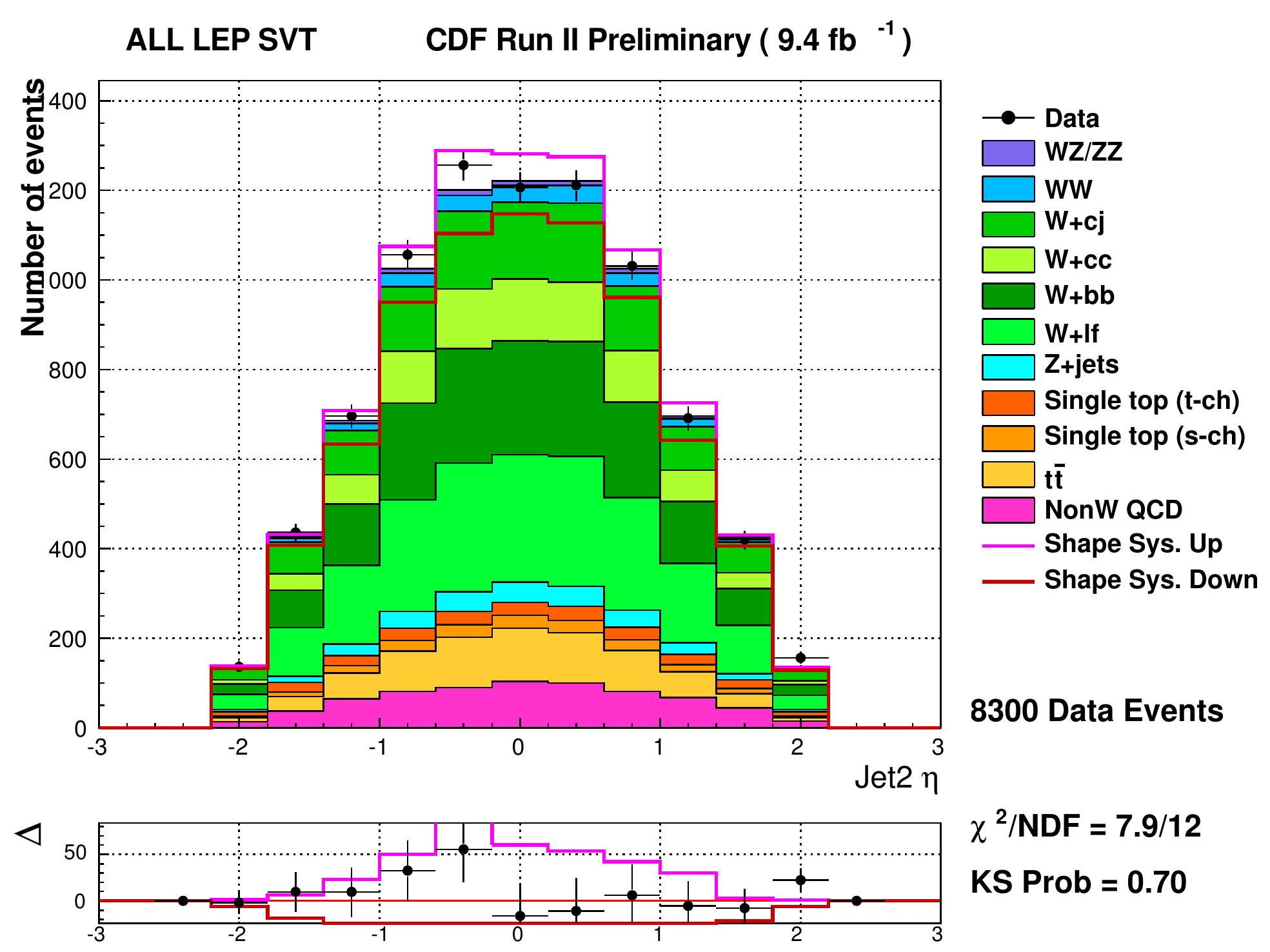}
  \caption[Single Tag Region Jet Kinematic Variables]{Variables relative to the jet kinematic for all the lepton categories combined in the single-tag signal region. Jet 1 $E_T$ (top left), jet 2 $E_T$ (top right), jet 1 $\eta$ (bottom left), jet 2 $\eta$ (bottom left).}\label{fig:1svt_jet}
\end{sidewaysfigure}

\begin{sidewaysfigure}
  \includegraphics[width=0.49\textwidth]{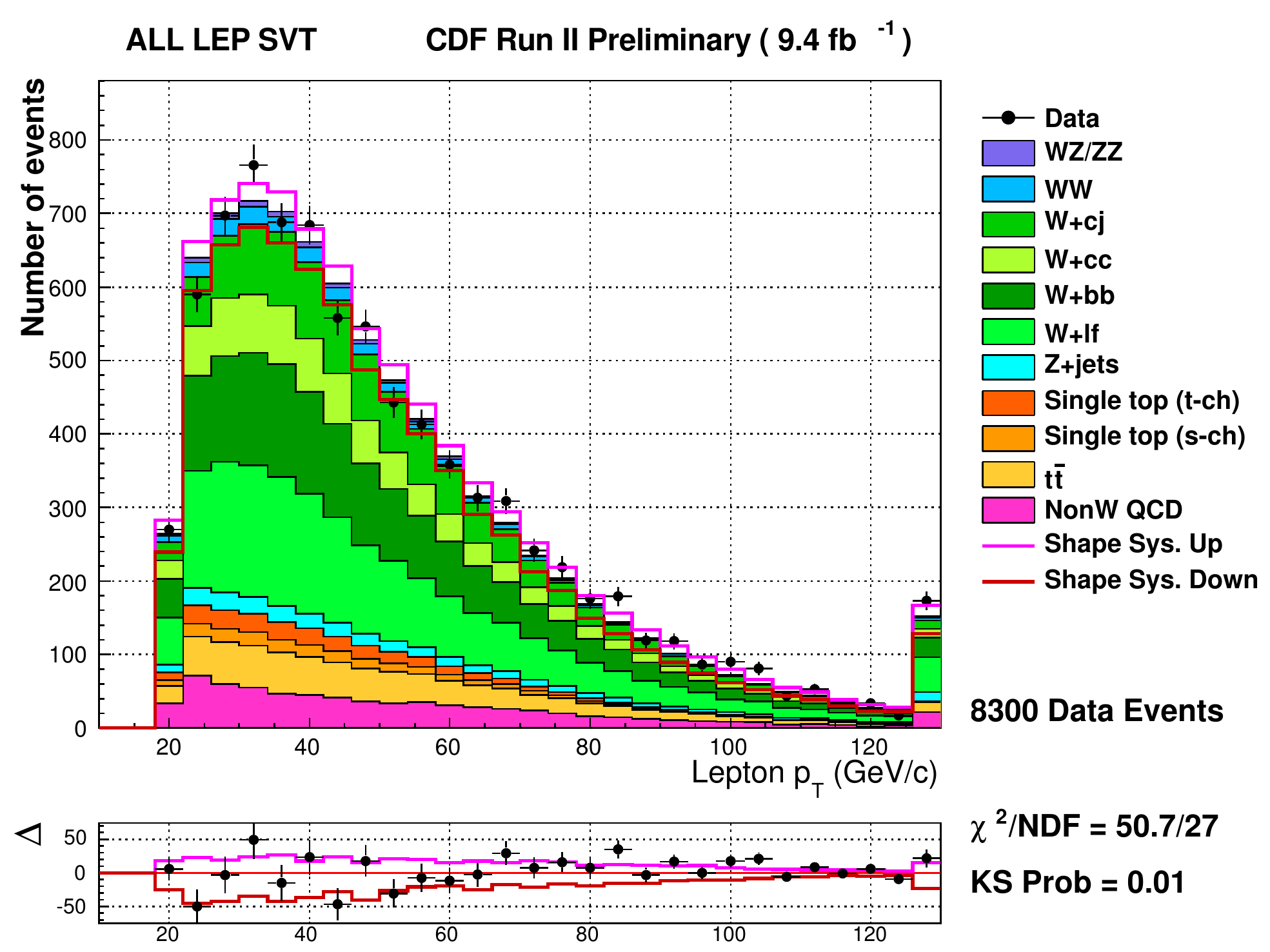}
  \includegraphics[width=0.49\textwidth]{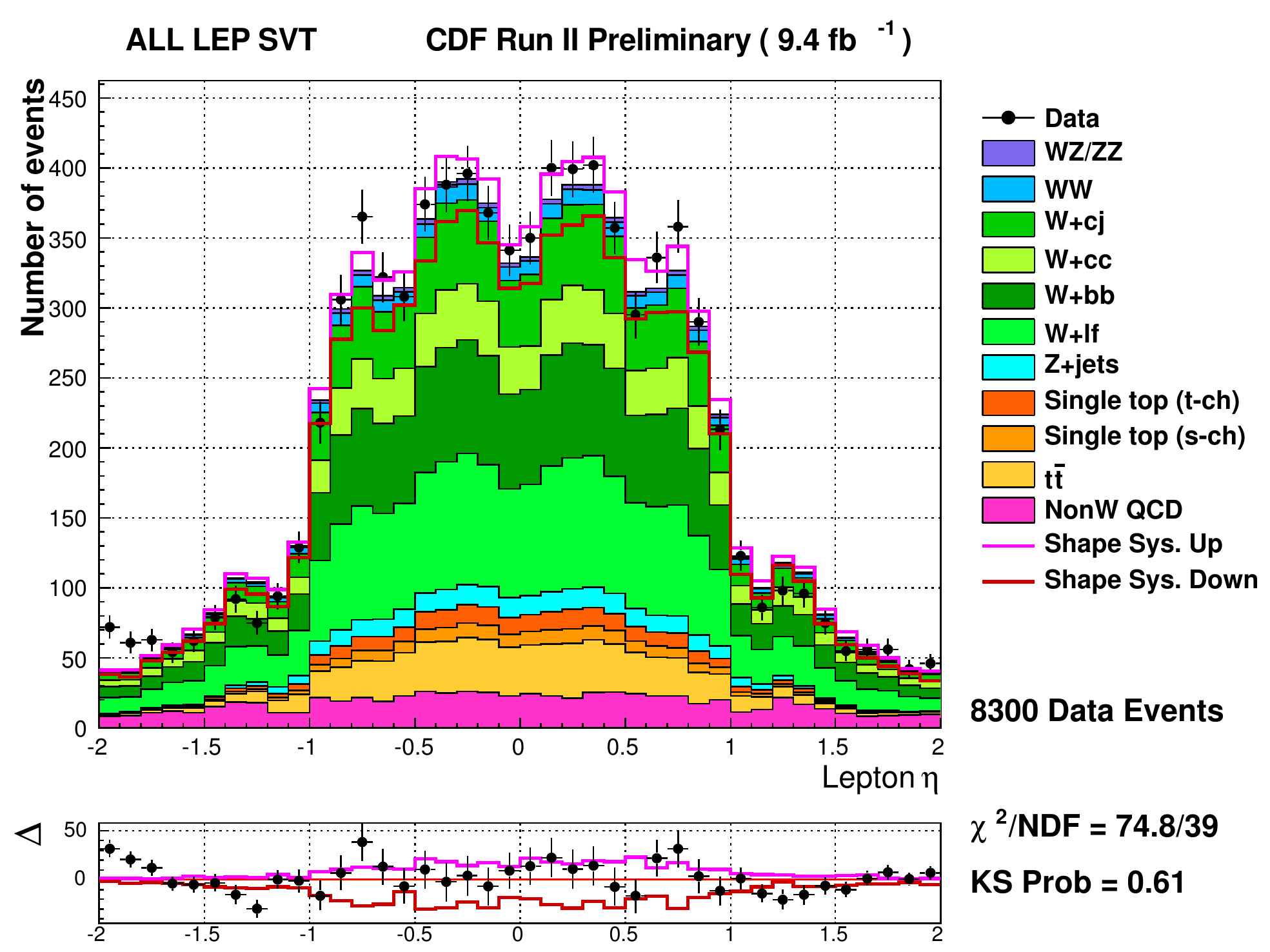}\\
  \includegraphics[width=0.49\textwidth]{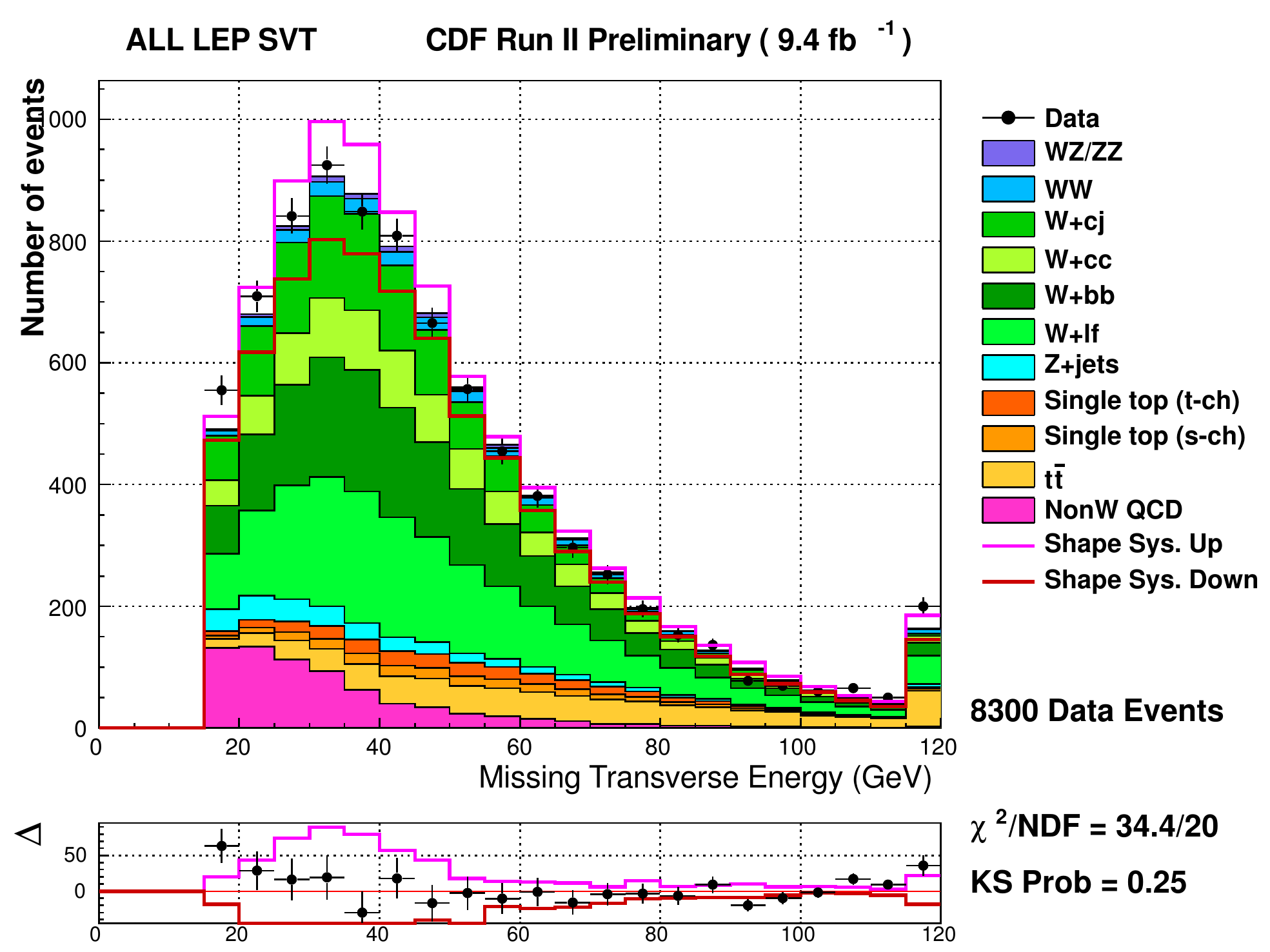}
  \includegraphics[width=0.49\textwidth]{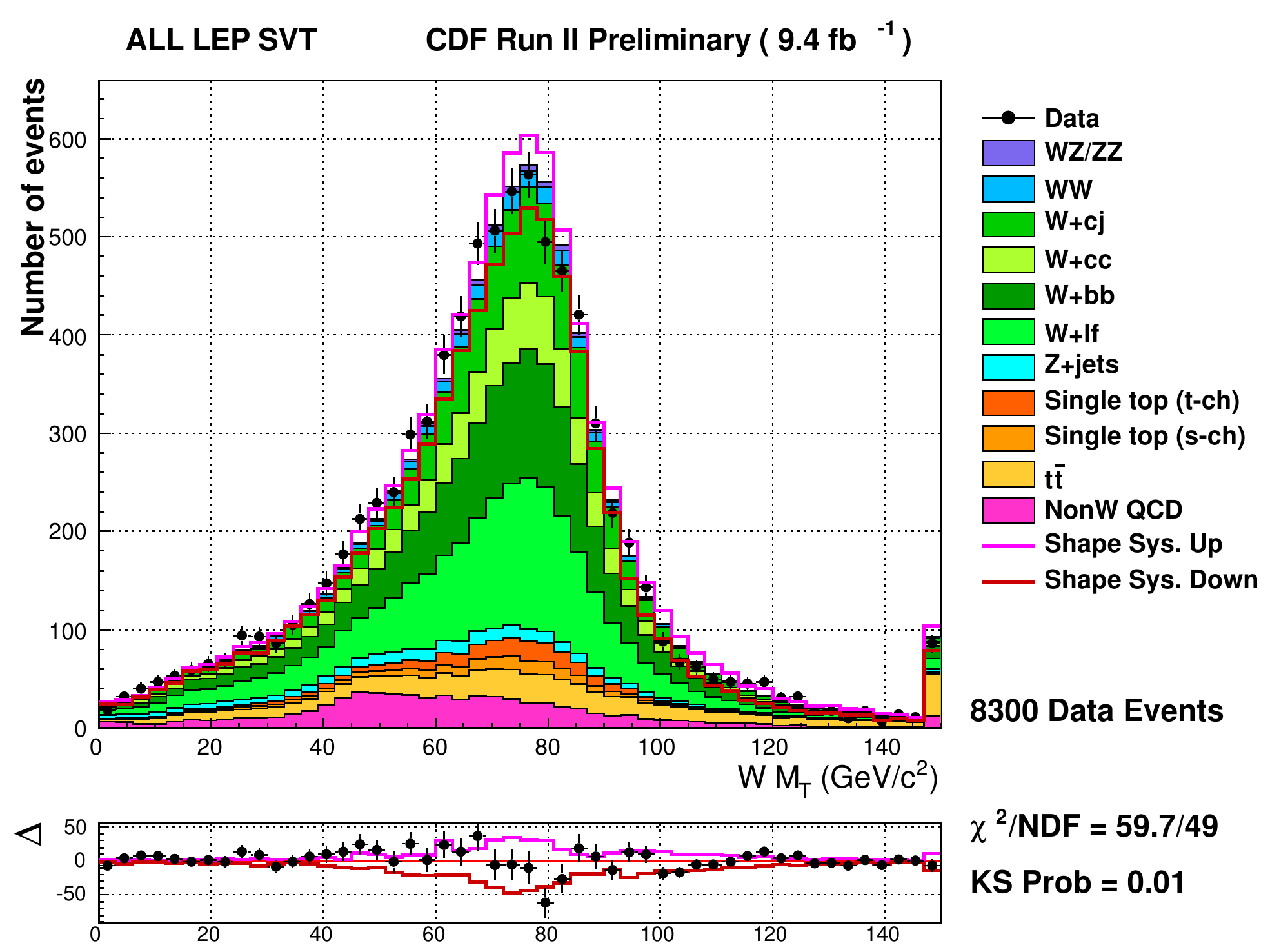}
  \caption[Single Tag Region Lepton Kinematic Variables]{Variables relative to the lepton kinematic for all the lepton categories combined in the single-tag signal region. Lepton $P_T$ (top left), lepton $\eta$ (top right), \met (bottom left), $M_T^W$ (bottom left).}\label{fig:1svt_lep}
\end{sidewaysfigure}

\begin{sidewaysfigure}
\includegraphics[width=0.49\textwidth]{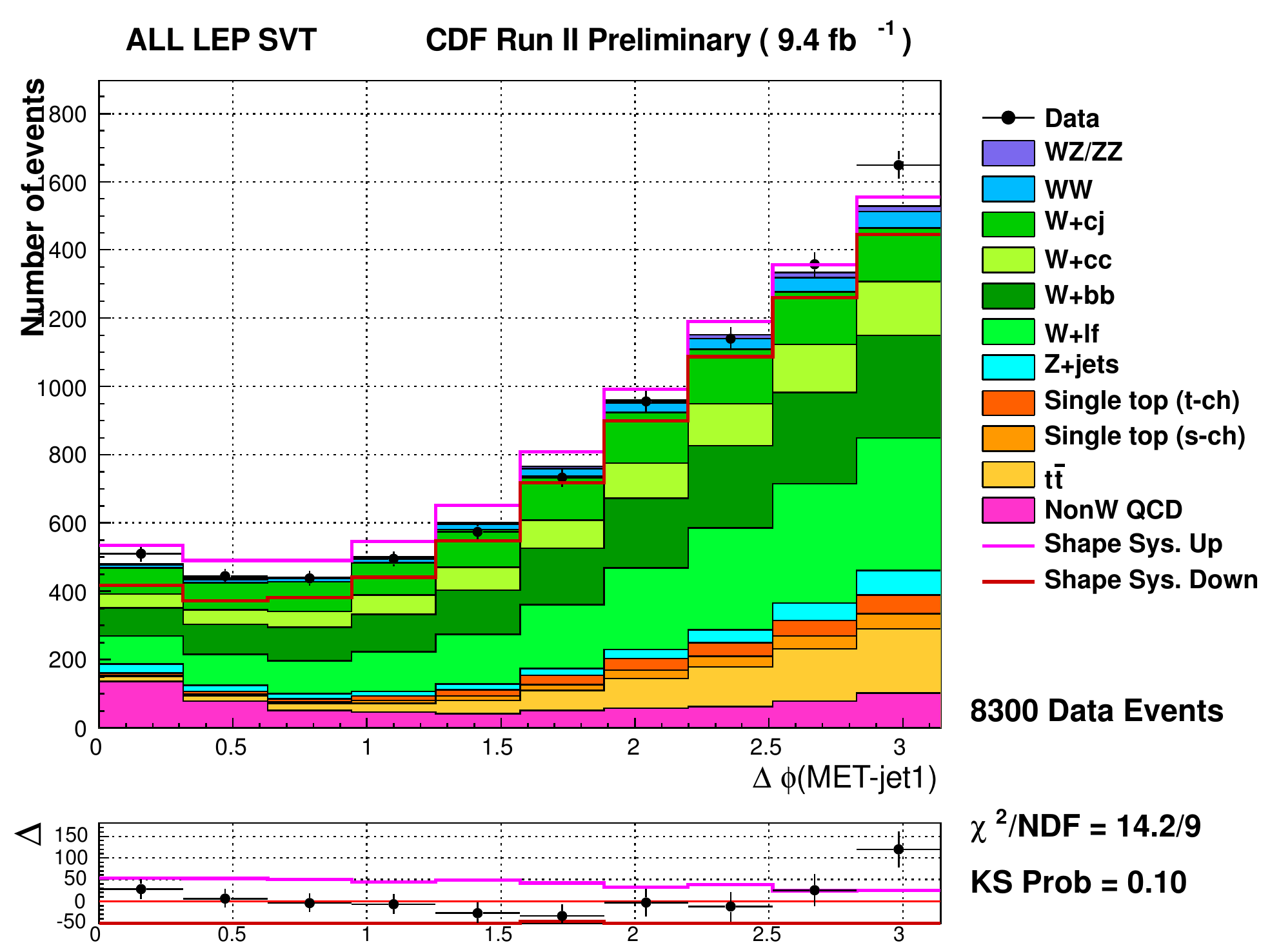}
\includegraphics[width=0.49\textwidth]{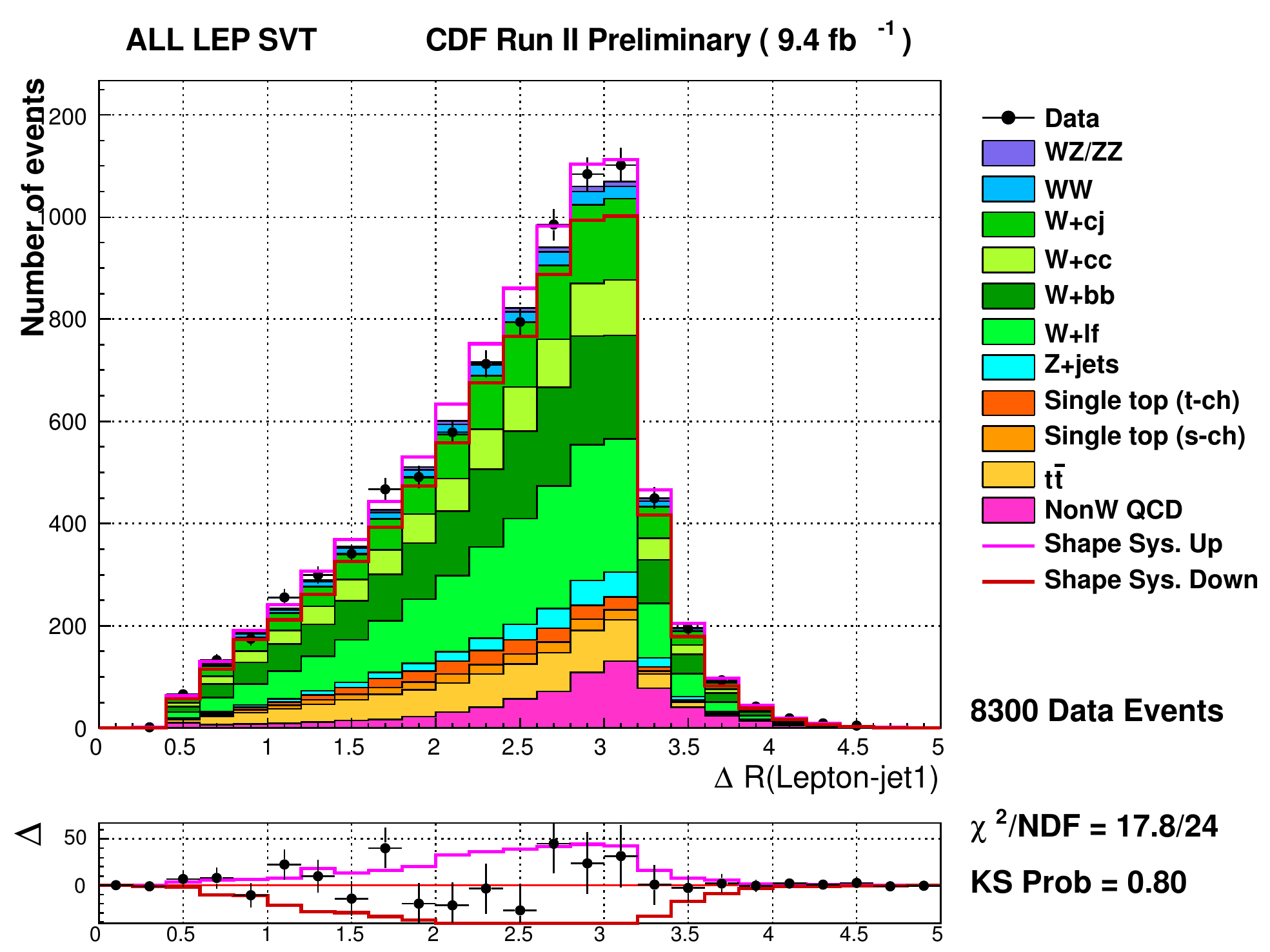}
\includegraphics[width=0.49\textwidth]{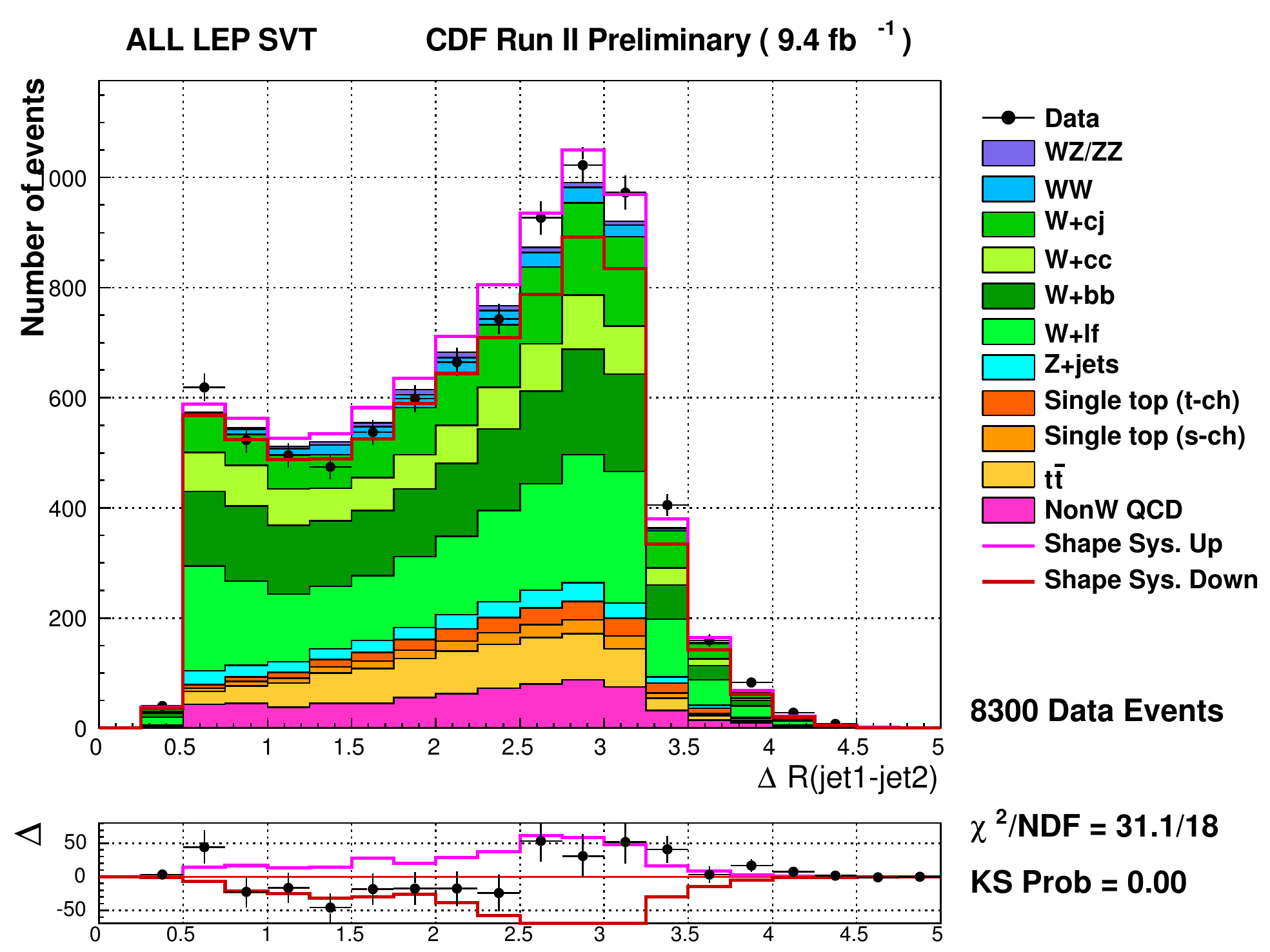}
\includegraphics[width=0.49\textwidth]{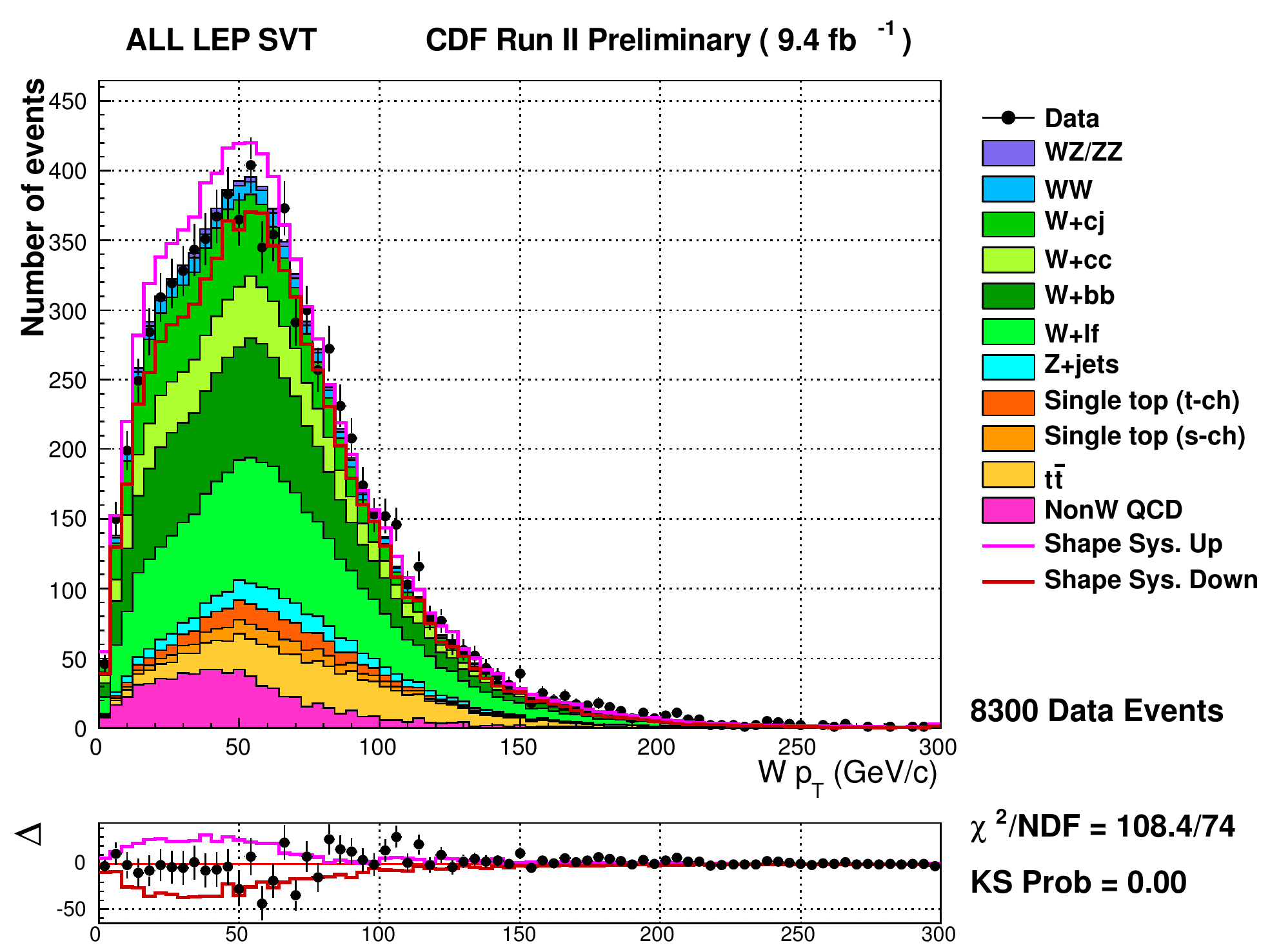}\\
\caption[Single Tag Region Angular Kinematic Variables]{Angular variables for all the lepton categories combined in the single-tag signal region. $\Delta\phi($\met$,jet1)$ (top left), $\Delta R(Lep,jet1)$ (top right), $\Delta R(jet1,jet2)$, (bottom left), $P_T^W$ (bottom left).}\label{fig:1svt_dphi}
\end{sidewaysfigure}

\begin{sidewaysfigure}
  \includegraphics[width=0.49\textwidth]{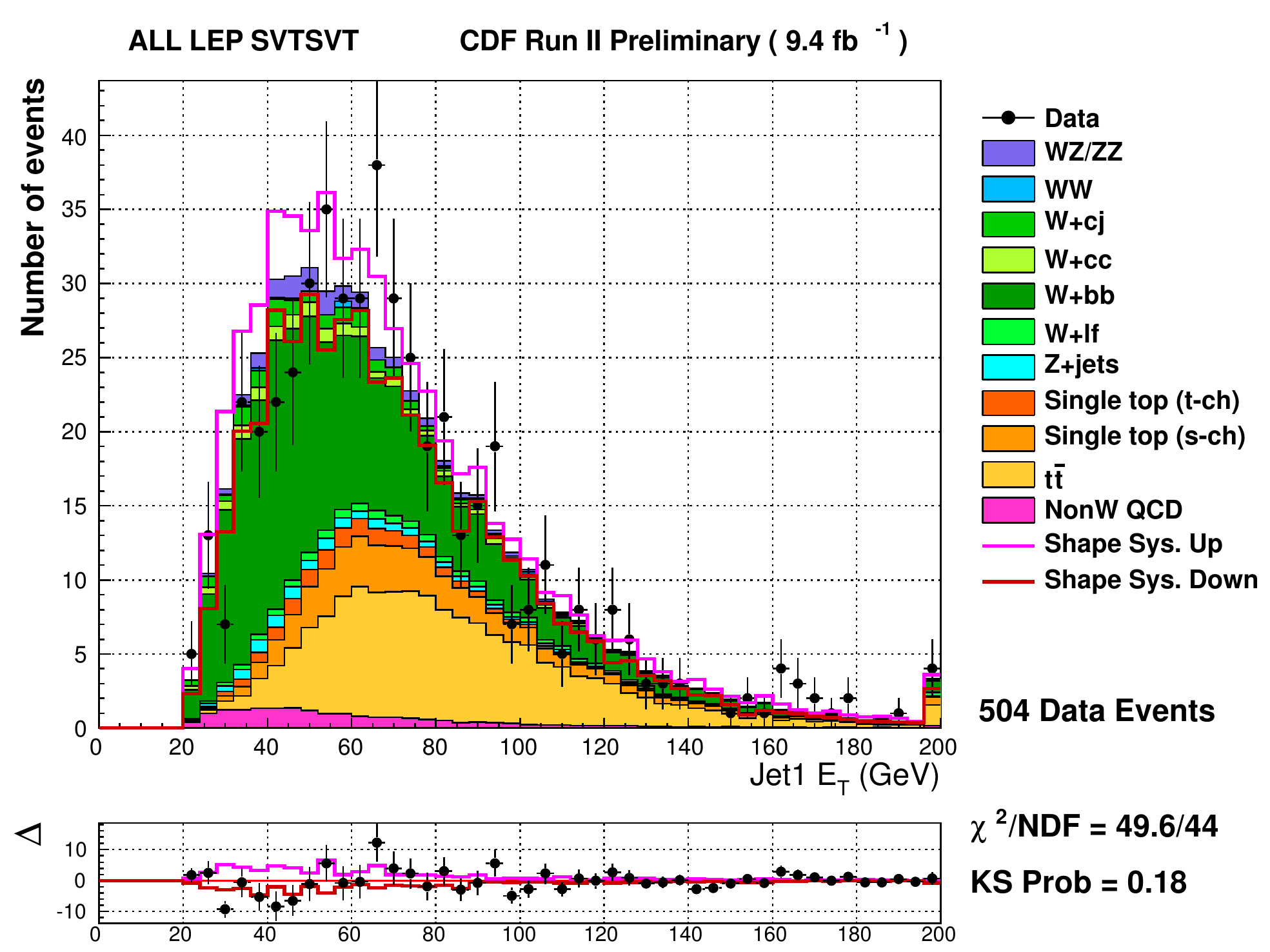}
  \includegraphics[width=0.49\textwidth]{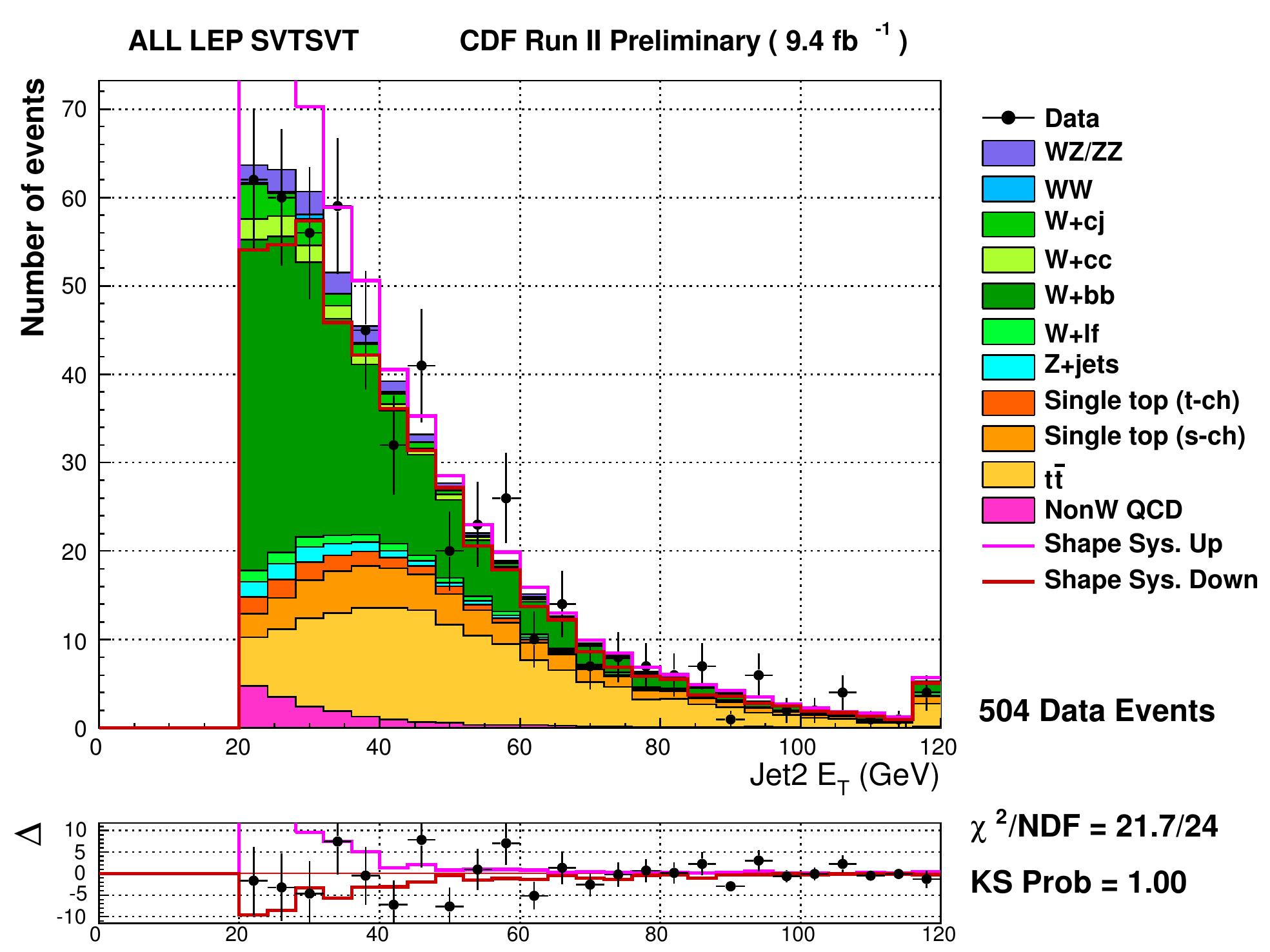}\\
  \includegraphics[width=0.49\textwidth]{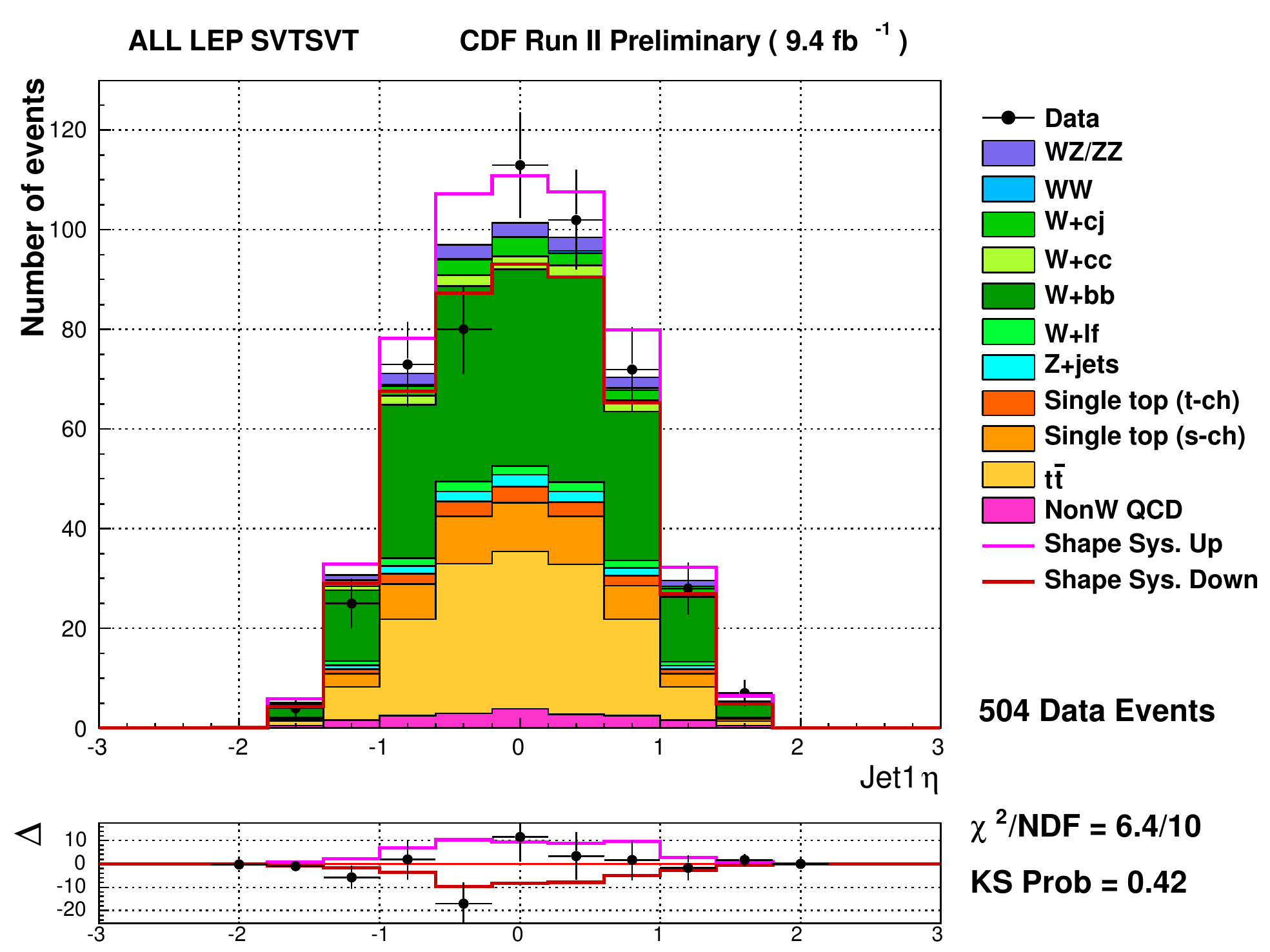}
  \includegraphics[width=0.49\textwidth]{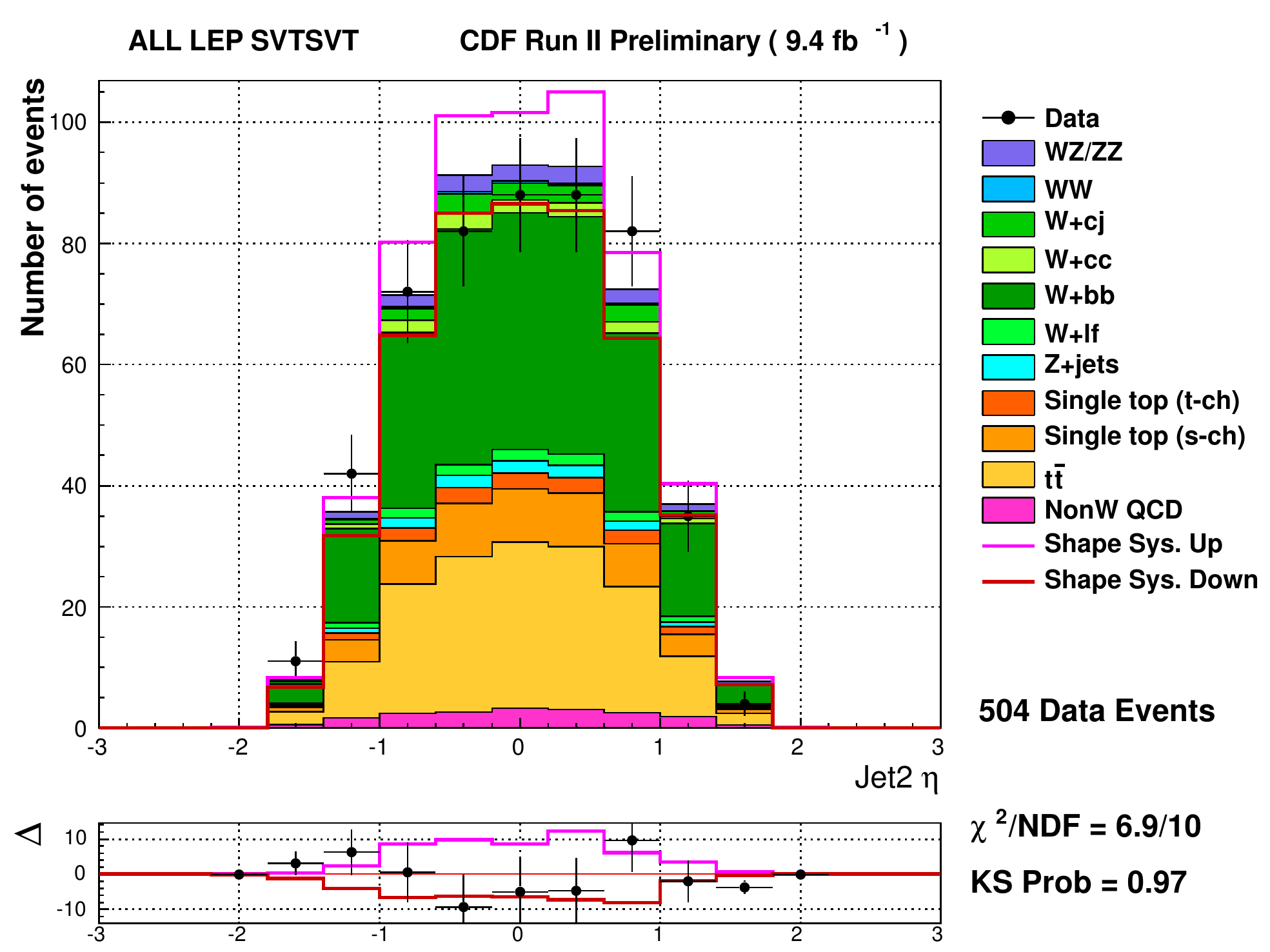}
  \caption[Double Tag Region Jet Kinematic Variables]{Variables relative to the jet kinematic for all the lepton categories combined in the double tag-signal region. Jet 1 $E_T$ (top left), jet 2 $E_T$ (top right), jet 1 $\eta$ (bottom left), jet 2 $\eta$ (bottom left).}\label{fig:2svt_jet}
\end{sidewaysfigure}

\begin{sidewaysfigure}
  \includegraphics[width=0.49\textwidth]{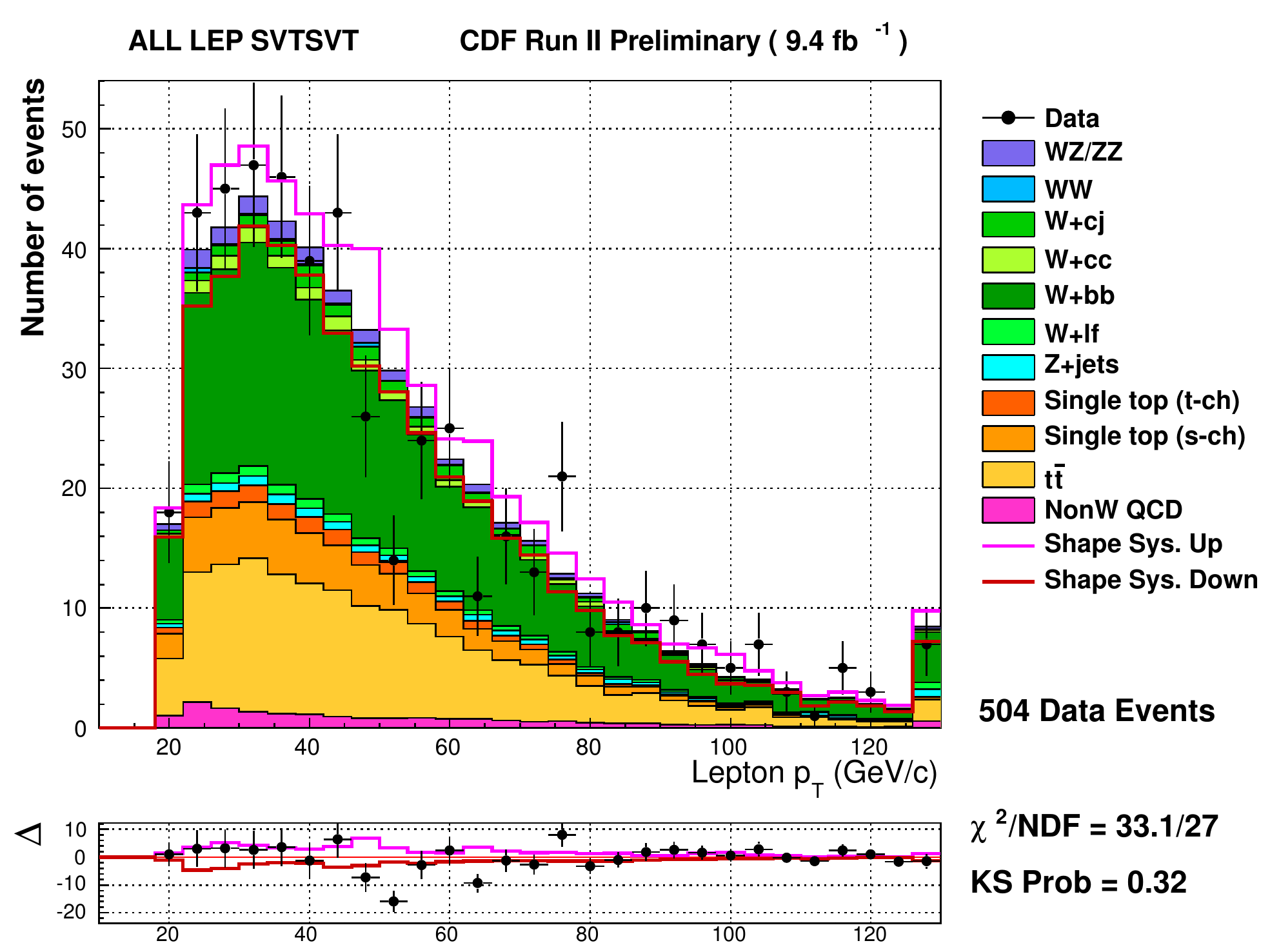}
  \includegraphics[width=0.49\textwidth]{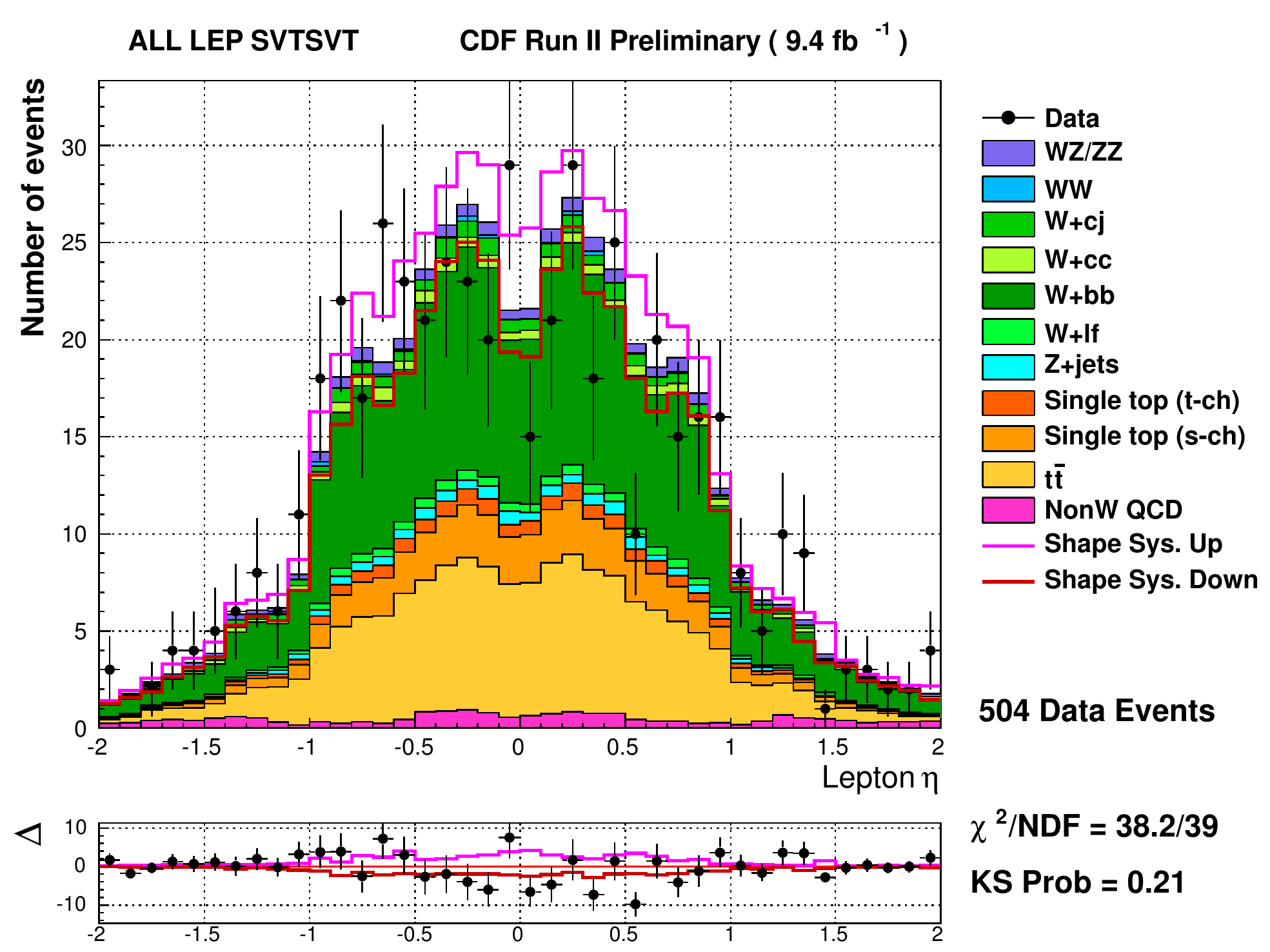}\\
  \includegraphics[width=0.49\textwidth]{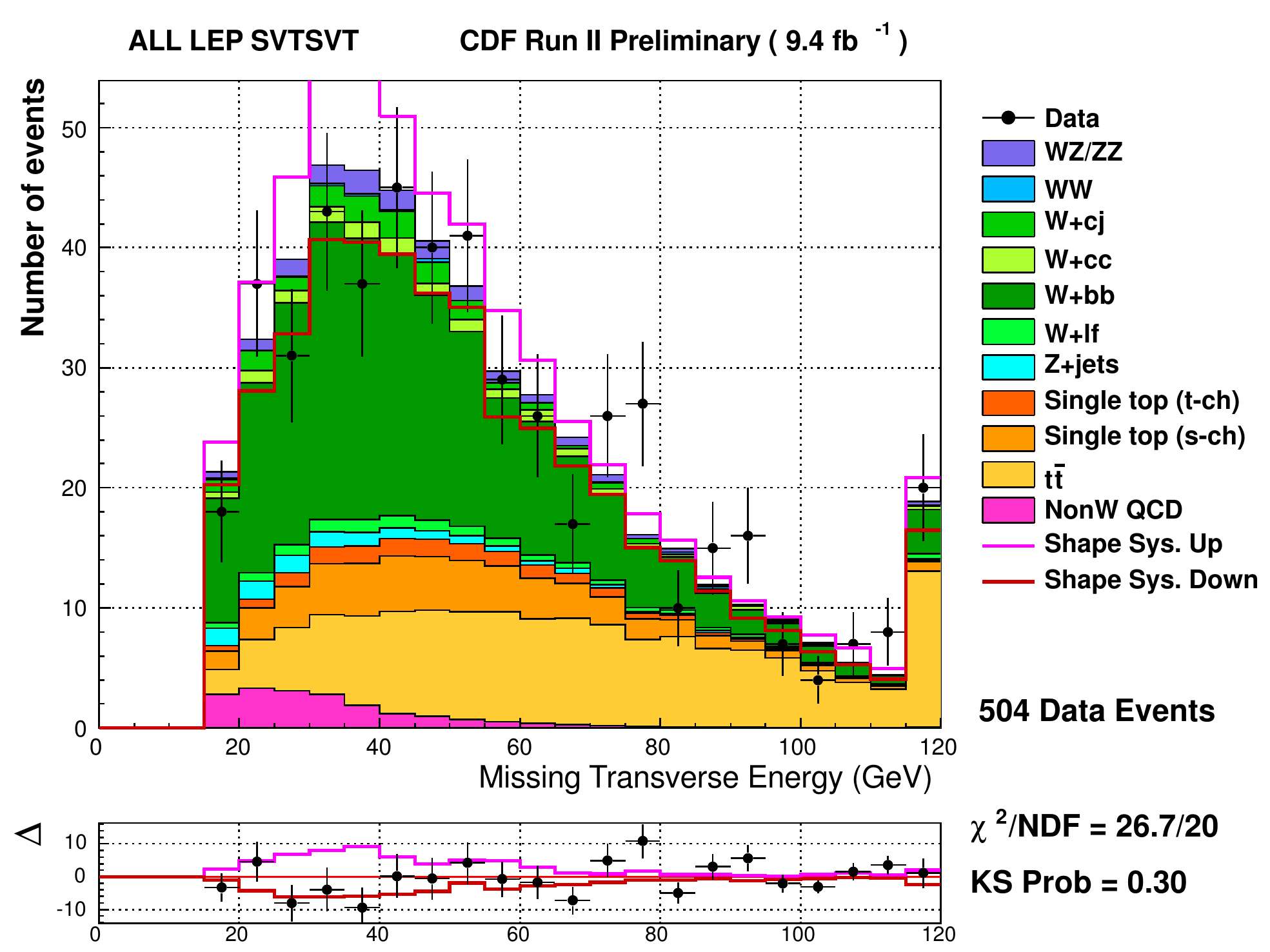}
  \includegraphics[width=0.49\textwidth]{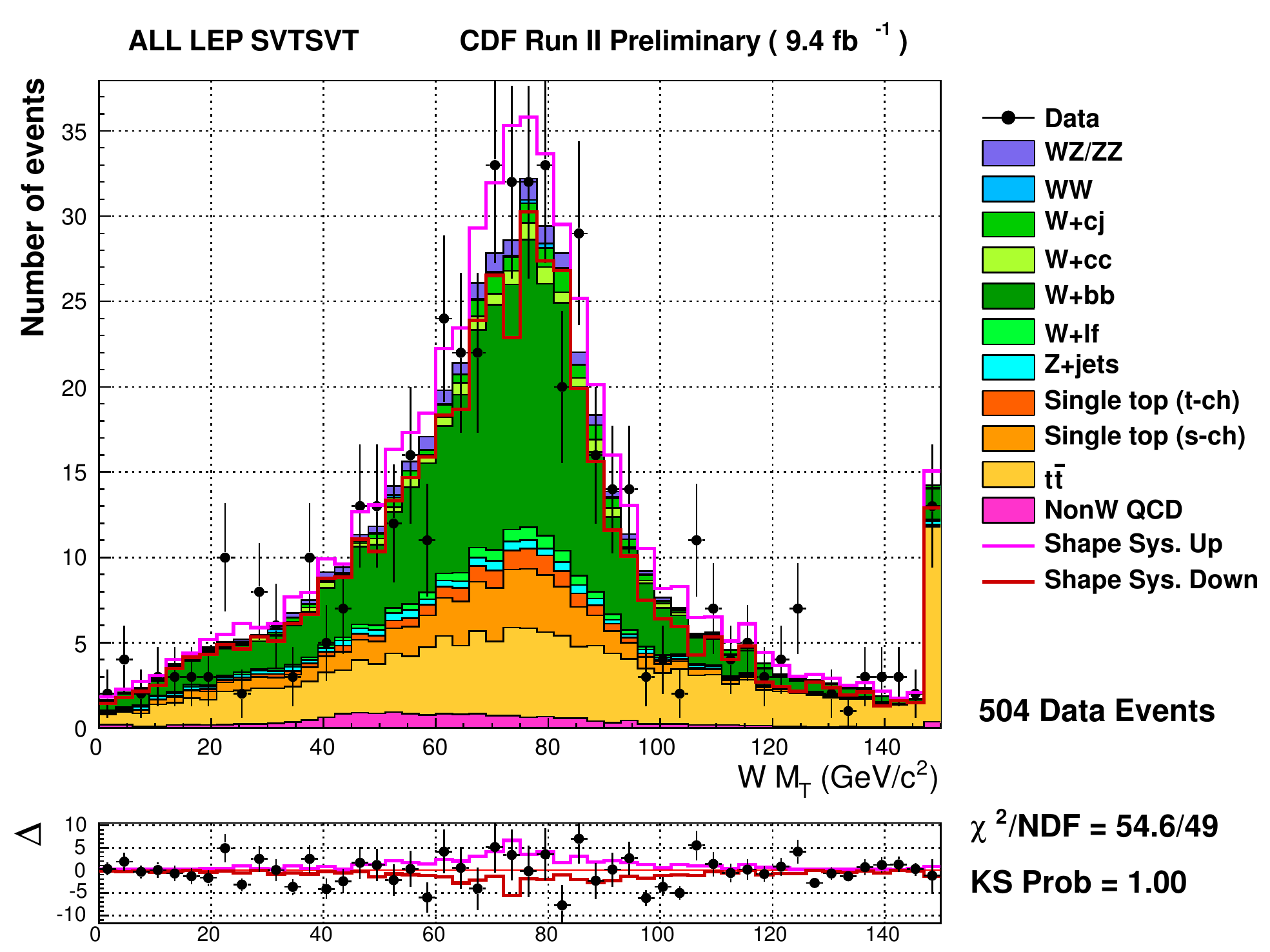}
  \caption[Double Tag Region Lepton Kinematic Variables]{Variables relative to the lepton kinematic for all the lepton categories combined in the double tag-signal region. Lepton $P_T$ (top left), lepton $\eta$ (top right), \met (bottom left), $M_T^W$ (bottom left).}\label{fig:2svt_lep}
\end{sidewaysfigure}

\begin{sidewaysfigure}
\includegraphics[width=0.49\textwidth]{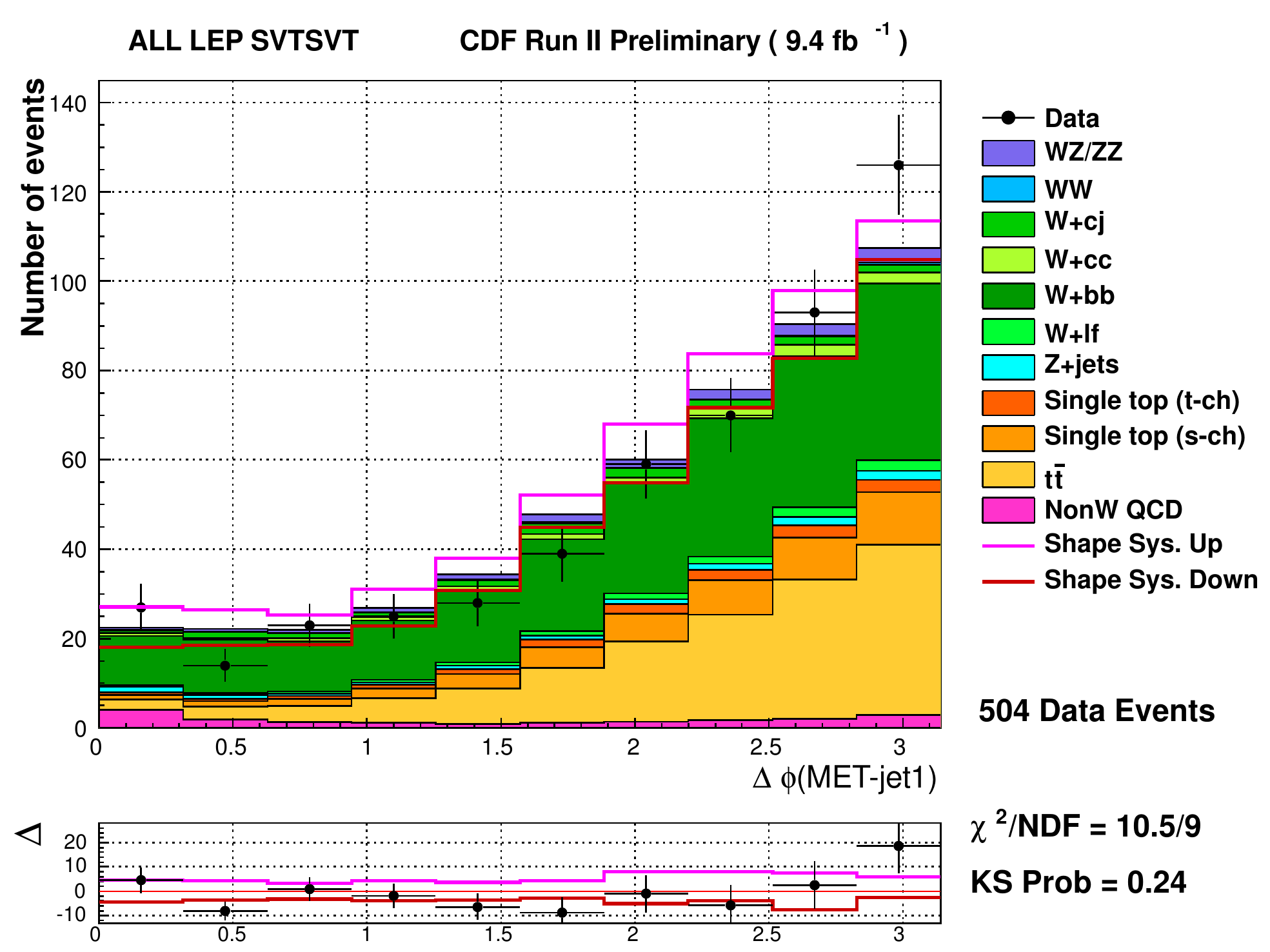}
\includegraphics[width=0.49\textwidth]{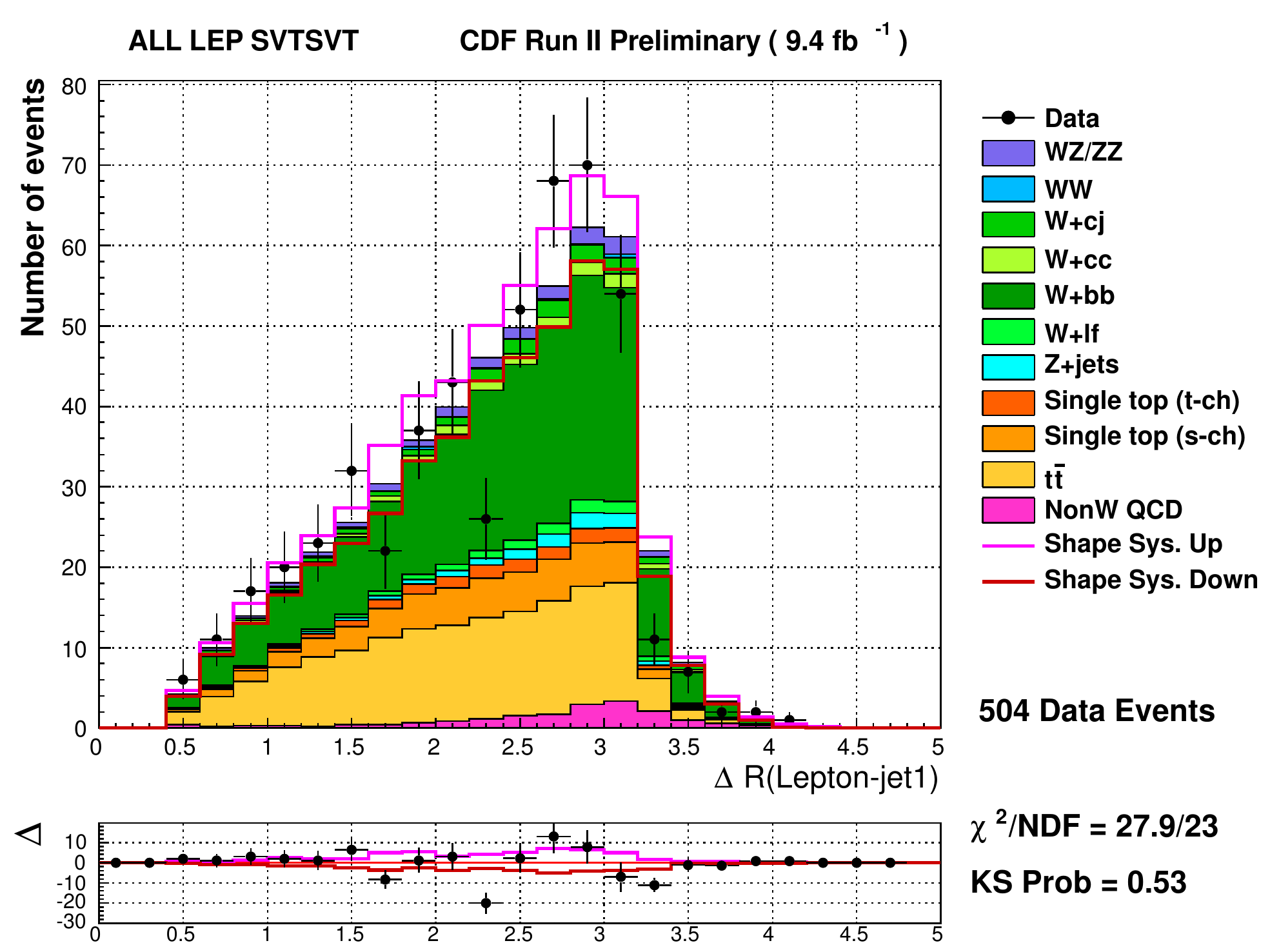}
\includegraphics[width=0.49\textwidth]{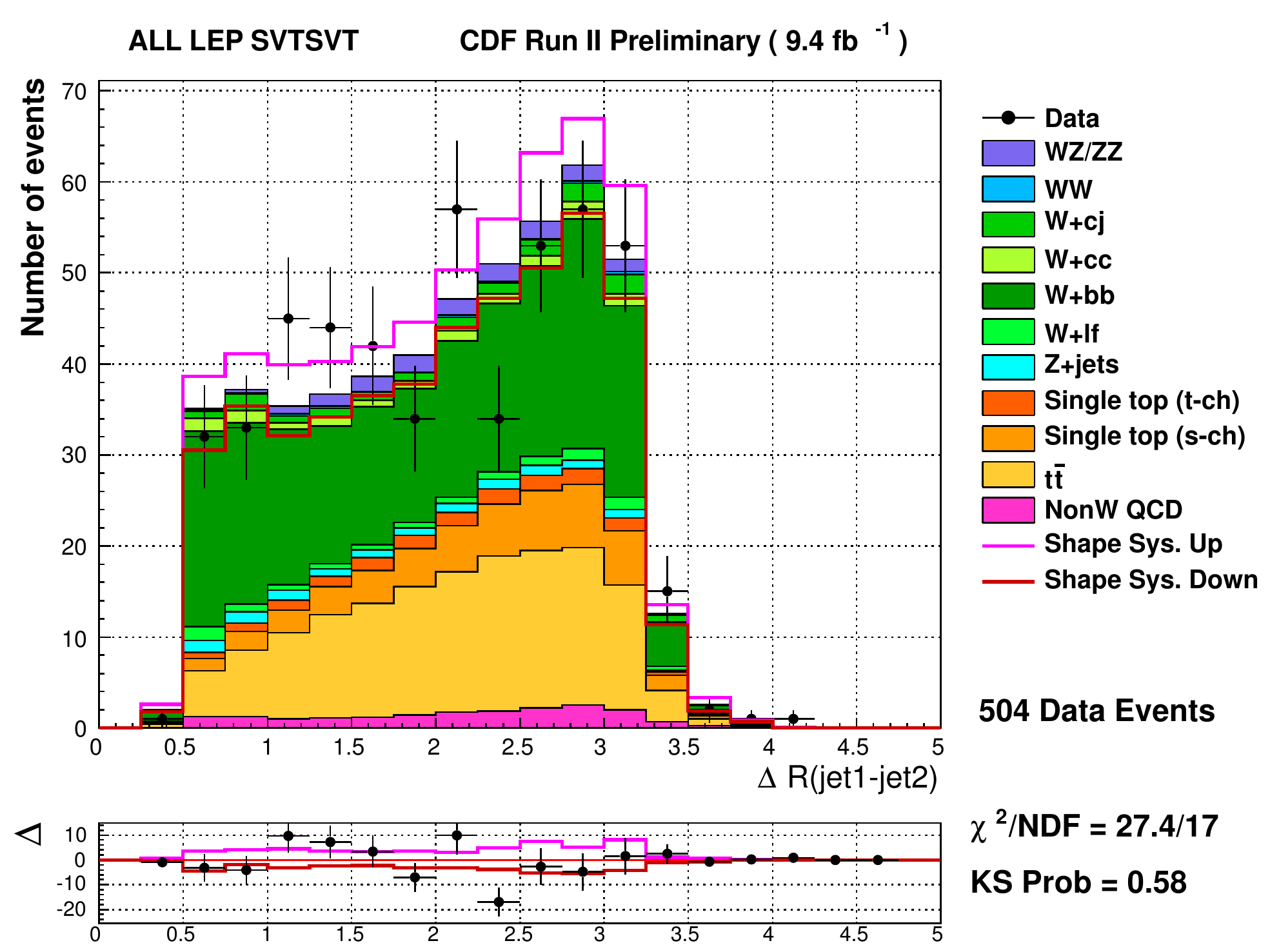}
\includegraphics[width=0.49\textwidth]{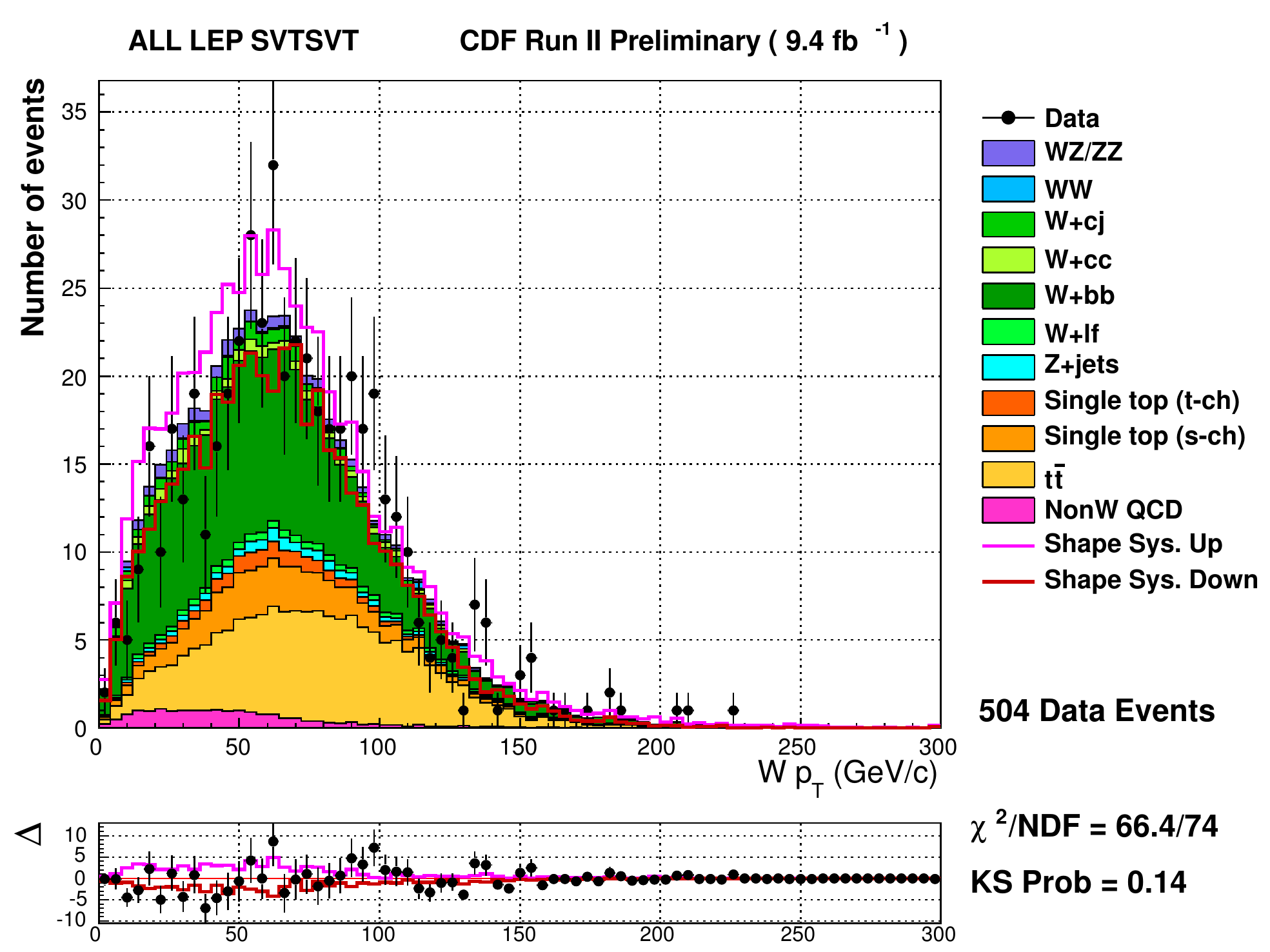}\\
\caption[Double Tag Region Angular Kinematic Variables]{Angular variables for all the lepton categories combined in the double-tag signal region. $\Delta\phi($\met$,jet1)$ (top left), $\Delta R(Lep,jet1)$ (top right), $\Delta R(jet1,jet2)$, (bottom left), $P_T^W$ (bottom left).}\label{fig:2svt_dphi}
\end{sidewaysfigure}

Figures from~\ref{fig:1svt_jet} to~\ref{fig:2svt_dphi} show several kinematic variables for the single-tag and double-tag signal regions after the composition of all the background and all the lepton categories. The normalization of each background has been changed to match the {\em best fit} values returned by statistical analysis of the likelihood described in by Equation~\ref{eq:likelihood_full}, in Chapter~\ref{chap:StatRes}. For each variable we show:
\begin{itemize}
\item the total background prediction overlaid to the selected data.
\item The reduced $\chi^2$ of the prediction against the data distribution. As we use the best-fit normalization is used for the backgrounds, no rate uncertainty is applied in the $\chi^2$ evaluation, however MC statistical uncertainty is accounted.
\item The probability of the Kolmogorov-Smirnov test derived from the predicted and observed shapes.
\item Background subtracted data histogram.
\item JES and $Q^2$ shape variations added in quadrature.
\end{itemize}

\clearpage
\chapter{Rate Systematics Summary}\label{app:rate_sys}

Tables from~\ref{tab:1tag_cem} to~\ref{tab:2tag_emc}: rate systematics variations for each channel entering in the statistical analysis (see Chapter~\ref{chap:StatRes}), same name systematics are fully correlated.

\renewcommand{\arraystretch}{0.97}
\renewcommand{\tabcolsep}{1.5pt}

\begin{table}[!h]
  \begin{scriptsize}
    \begin{tabular}{ c  c c c c c c c c c c c } \toprule
      & nonW & $t\bar{t}$ & s-top $s$ & s-top $t$ & Zjets & W+LF & Wbb & Wcc & Wcj & WW & WZ/ZZ \\\midrule
$\texttt{ XS\_ttbar}$ & $0/0$  & $10/-10$  & $10/-10$  & $10/-10$  & $0/0$  & $0/0$  & $0/0$  & $0/0$  & $0/0$  & $0/0$  & $0/0$  \\
$\texttt{ CDFBTAGSF}$ & $0/0$  & $3/-3$  & $3/-3$  & $5/-5$  & $0/0$  & $0/0$  & $0/0$  & $0/0$  & $0/0$  & $11/-11$  & $5/-5$  \\
$\texttt{ CDFLUMI}$ & $0/0$  & $6/-6$  & $6/-6$  & $6/-6$  & $6/-6$  & $0/0$  & $0/0$  & $0/0$  & $0/0$  & $6/-6$  & $6/-6$  \\
$\texttt{ LEPACC\_CEM}$ & $0/0$  & $2/-2$  & $2/-2$  & $2/-2$  & $2/-2$  & $0/0$  & $0/0$  & $0/0$  & $0/0$  & $2/-2$  & $2/-2$  \\
$\texttt{ QCD\_CEM}$ & $40/-40$  & $0/0$  & $0/0$  & $0/0$  & $0/0$  & $0/0$  & $0/0$  & $0/0$  & $0/0$  & $0/0$  & $0/0$  \\
$\texttt{ CDFVHF}$ & $0/0$  & $0/0$  & $0/0$  & $0/0$  & $0/0$  & $0/0$  & $30/-30$  & $30/-30$  & $0/0$  & $0/0$  & $0/0$  \\
$\texttt{ CDFWC}$ & $0/0$  & $0/0$  & $0/0$  & $0/0$  & $0/0$  & $0/0$  & $0/0$  & $0/0$  & $30/-30$  & $0/0$  & $0/0$  \\
$\texttt{ CDFMISTAG}$ & $0/0$  & $0/0$  & $0/0$  & $0/0$  & $0/0$  & $11/-11$  & $0/0$  & $0/0$  & $0/0$  & $0/0$  & $0/0$  \\
$\texttt{ ZJETS}$ & $0/0$  & $0/0$  & $0/0$  & $0/0$  & $45/-45$  & $0/0$  & $0/0$  & $0/0$  & $0/0$  & $0/0$  & $0/0$  \\
$\texttt{ ISRFSRPDF}$ & $0/0$  & $0/0$  & $0/0$  & $0/0$  & $0/0$  & $0/0$  & $0/0$  & $0/0$  & $0/0$  & $4/-4$  & $4/-4$  \\
$\texttt{ CDFJES}$ & $0/0$  & $-12/12$  & $-3/2$  & $-2/1$  & $1/6$  & $1/-1$  & $-6/6$  & $-3/6$  & $-5/6$  & $-3/-4$  & $2/-2$  \\
$\texttt{ CDFQ2}$ & $0/0$  & $0/0$  & $0/0$  & $0/0$  & $0/0$  & $-2/-1$  & $4/11$  & $5/8$  & $3/9$  & $0/0$  & $0/0$  \\
\hline
$\texttt{TOT}$ & $40/40$  & $17/17$  & $12/12$  & $13/13$  & $45/46$  & $11/11$  & $31/33$  & $31/32$  & $30/32$  & $14/14$  & $9/11$  \\
\hline
    \end{tabular}
  \end{scriptsize}
  \caption[Rate Uncertainties: Single-Tag, CEM Channel]{Per-sample rate uncertainty (\% up/down). Single-tag, CEM  channel.}\label{tab:1tag_cem}
  \end{table}

  \begin{table}[!h]
    \begin{scriptsize}
  \begin{tabular}{ c c c c c c c c c c c c} \toprule
    & nonW & $t\bar{t}$ & s-top $s$ & s-top $t$ & Zjets & W+LF & Wbb & Wcc & Wcj & WW & WZ/ZZ \\\midrule
$\texttt{ XS\_ttbar}$ & $0/0$  & $10/-10$  & $10/-10$  & $10/-10$  & $0/0$  & $0/0$  & $0/0$  & $0/0$  & $0/0$  & $0/0$  & $0/0$  \\
$\texttt{ CDFBTAGSF}$ & $0/0$  & $3/-3$  & $3/-3$  & $5/-5$  & $0/0$  & $0/0$  & $0/0$  & $0/0$  & $0/0$  & $11/-11$  & $5/-5$  \\
$\texttt{ CDFLUMI}$ & $0/0$  & $6/-6$  & $6/-6$  & $6/-6$  & $6/-6$  & $0/0$  & $0/0$  & $0/0$  & $0/0$  & $6/-6$  & $6/-6$  \\
$\texttt{ LEPACC\_PHX}$ & $0/0$  & $2/-2$  & $2/-2$  & $2/-2$  & $2/-2$  & $0/0$  & $0/0$  & $0/0$  & $0/0$  & $2/-2$  & $2/-2$  \\
$\texttt{ QCD\_PHX}$ & $40/-40$  & $0/0$  & $0/0$  & $0/0$  & $0/0$  & $0/0$  & $0/0$  & $0/0$  & $0/0$  & $0/0$  & $0/0$  \\
$\texttt{ CDFVHF}$ & $0/0$  & $0/0$  & $0/0$  & $0/0$  & $0/0$  & $0/0$  & $30/-30$  & $30/-30$  & $0/0$  & $0/0$  & $0/0$  \\
$\texttt{ CDFWC}$ & $0/0$  & $0/0$  & $0/0$  & $0/0$  & $0/0$  & $0/0$  & $0/0$  & $0/0$  & $30/-30$  & $0/0$  & $0/0$  \\
$\texttt{ CDFMISTAG}$ & $0/0$  & $0/0$  & $0/0$  & $0/0$  & $0/0$  & $11/-11$  & $0/0$  & $0/0$  & $0/0$  & $0/0$  & $0/0$  \\
$\texttt{ ZJETS}$ & $0/0$  & $0/0$  & $0/0$  & $0/0$  & $45/-45$  & $0/0$  & $0/0$  & $0/0$  & $0/0$  & $0/0$  & $0/0$  \\
$\texttt{ ISRFSRPDF}$ & $0/0$  & $0/0$  & $0/0$  & $0/0$  & $0/0$  & $0/0$  & $0/0$  & $0/0$  & $0/0$  & $4/-4$  & $4/-4$  \\
$\texttt{ CDFJES}$ & $0/0$  & $-8/12$  & $-1/2$  & $-3/3$  & $2/-6$  & $-1/-1$  & $-4/4$  & $-5/4$  & $-8/6$  & $-7/-5$  & $-10/-3$  \\
$\texttt{ CDFQ2}$ & $0/0$  & $0/0$  & $0/0$  & $0/0$  & $0/0$  & $-1/-2$  & $0/15$  & $0/18$  & $2/10$  & $0/0$  & $0/0$  \\
\hline
$\texttt{TOT}$ & $40/40$  & $14/17$  & $12/12$  & $13/13$  & $45/46$  & $11/11$  & $30/34$  & $30/35$  & $31/32$  & $15/14$  & $14/10$  \\
\hline
\end{tabular}
    \end{scriptsize}
  \caption[Rate Uncertainties: Single-Tag, PHX Channel]{Per-sample rate uncertainty (\% up/down). Single-tag, PHX channel.}\label{tab:1tag_phx}
  \end{table}

  \begin{table}
    \begin{scriptsize}

\begin{tabular}{ c c c c c c c c c c c c } \toprule
    & nonW & $t\bar{t}$ & s-top $s$ & s-top $t$ & Zjets & W+LF & Wbb & Wcc & Wcj & WW & WZ/ZZ \\\midrule
$\texttt{ XS\_ttbar}$ & $0/0$  & $10/-10$  & $10/-10$  & $10/-10$  & $0/0$  & $0/0$  & $0/0$  & $0/0$  & $0/0$  & $0/0$  & $0/0$  \\
$\texttt{ CDFBTAGSF}$ & $0/0$  & $3/-3$  & $3/-3$  & $5/-5$  & $0/0$  & $0/0$  & $0/0$  & $0/0$  & $0/0$  & $11/-11$  & $5/-5$  \\
$\texttt{ CDFLUMI}$ & $0/0$  & $6/-6$  & $6/-6$  & $6/-6$  & $6/-6$  & $0/0$  & $0/0$  & $0/0$  & $0/0$  & $6/-6$  & $6/-6$  \\
$\texttt{ LEPACC\_MU}$ & $0/0$  & $1/-1$  & $1/-1$  & $1/-1$  & $1/-1$  & $0/0$  & $0/0$  & $0/0$  & $0/0$  & $1/-1$  & $1/-1$  \\
$\texttt{ QCD\_MU}$ & $40/-40$  & $0/0$  & $0/0$  & $0/0$  & $0/0$  & $0/0$  & $0/0$  & $0/0$  & $0/0$  & $0/0$  & $0/0$  \\
$\texttt{ CDFVHF}$ & $0/0$  & $0/0$  & $0/0$  & $0/0$  & $0/0$  & $0/0$  & $30/-30$  & $30/-30$  & $0/0$  & $0/0$  & $0/0$  \\
$\texttt{ CDFWC}$ & $0/0$  & $0/0$  & $0/0$  & $0/0$  & $0/0$  & $0/0$  & $0/0$  & $0/0$  & $30/-30$  & $0/0$  & $0/0$  \\
$\texttt{ CDFMISTAG}$ & $0/0$  & $0/0$  & $0/0$  & $0/0$  & $0/0$  & $11/-11$  & $0/0$  & $0/0$  & $0/0$  & $0/0$  & $0/0$  \\
$\texttt{ ZJETS}$ & $0/0$  & $0/0$  & $0/0$  & $0/0$  & $45/-45$  & $0/0$  & $0/0$  & $0/0$  & $0/0$  & $0/0$  & $0/0$  \\
$\texttt{ ISRFSRPDF}$ & $0/0$  & $0/0$  & $0/0$  & $0/0$  & $0/0$  & $0/0$  & $0/0$  & $0/0$  & $0/0$  & $4/-4$  & $4/-4$  \\
$\texttt{ CDFJES}$ & $0/0$  & $-11/12$  & $-4/3$  & $-1/1$  & $9/-6$  & $1/1$  & $-8/6$  & $-5/5$  & $-6/7$  & $-1/-4$  & $1/-2$  \\
$\texttt{ CDFQ2}$ & $0/0$  & $0/0$  & $0/0$  & $0/0$  & $0/0$  & $-3/-4$  & $2/16$  & $6/12$  & $2/15$  & $0/0$  & $0/0$  \\
\hline
$\texttt{TOT}$ & $40/40$  & $17/17$  & $13/12$  & $13/13$  & $46/46$  & $11/12$  & $31/35$  & $31/33$  & $31/34$  & $13/14$  & $9/16$  \\
\hline
\end{tabular}
    \end{scriptsize}
\caption[Rate Uncertainties: Single-Tag, CMUP $+$ CMX Channel]{Per-sample rate uncertainty (\% up/down). Single-tag, CMUP $+$ CMX channel.}\label{tab:1tag_emc}
  \end{table}

  \begin{table}
    \begin{scriptsize}
\begin{tabular}{ c c c c c c c c c c c c } \toprule
    & nonW & $t\bar{t}$ & s-top $s$ & s-top $t$ & Zjets & W+LF & Wbb & Wcc & Wcj & WW & WZ/ZZ \\\midrule
$\texttt{ XS\_ttbar}$ & $0/0$  & $10/-10$  & $10/-10$  & $10/-10$  & $0/0$  & $0/0$  & $0/0$  & $0/0$  & $0/0$  & $0/0$  & $0/0$  \\
$\texttt{ CDFBTAGSF}$ & $0/0$  & $3/-3$  & $3/-3$  & $5/-5$  & $0/0$  & $0/0$  & $0/0$  & $0/0$  & $0/0$  & $11/-11$  & $5/-5$  \\
$\texttt{ CDFLUMI}$ & $0/0$  & $6/-6$  & $6/-6$  & $6/-6$  & $6/-6$  & $0/0$  & $0/0$  & $0/0$  & $0/0$  & $6/-6$  & $6/-6$  \\
$\texttt{ LEPACC\_EMC}$ & $0/0$  & $5/-5$  & $5/-5$  & $5/-5$  & $5/-5$  & $0/0$  & $0/0$  & $0/0$  & $0/0$  & $5/-5$  & $5/-5$  \\
$\texttt{ QCD\_EMC}$ & $40/-40$  & $0/0$  & $0/0$  & $0/0$  & $0/0$  & $0/0$  & $0/0$  & $0/0$  & $0/0$  & $0/0$  & $0/0$  \\
$\texttt{ CDFVHF}$ & $0/0$  & $0/0$  & $0/0$  & $0/0$  & $0/0$  & $0/0$  & $30/-30$  & $30/-30$  & $0/0$  & $0/0$  & $0/0$  \\
$\texttt{ CDFWC}$ & $0/0$  & $0/0$  & $0/0$  & $0/0$  & $0/0$  & $0/0$  & $0/0$  & $0/0$  & $30/-30$  & $0/0$  & $0/0$  \\
$\texttt{ CDFMISTAG}$ & $0/0$  & $0/0$  & $0/0$  & $0/0$  & $0/0$  & $11/-11$  & $0/0$  & $0/0$  & $0/0$  & $0/0$  & $0/0$  \\
$\texttt{ ZJETS}$ & $0/0$  & $0/0$  & $0/0$  & $0/0$  & $45/-45$  & $0/0$  & $0/0$  & $0/0$  & $0/0$  & $0/0$  & $0/0$  \\
$\texttt{ ISRFSRPDF}$ & $0/0$  & $0/0$  & $0/0$  & $0/0$  & $0/0$  & $0/0$  & $0/0$  & $0/0$  & $0/0$  & $4/-4$  & $4/-4$  \\
$\texttt{ CDFJES}$ & $0/0$  & $-10/9$  & $0/-2$  & $0/-3$  & $11/-8$  & $0/-0$  & $-5/6$  & $-8/4$  & $-3/2$  & $9/-6$  & $8/-6$  \\
$\texttt{ CDFQ2}$ & $0/0$  & $0/0$  & $0/0$  & $0/0$  & $0/0$  & $-3/-2$  & $11/13$  & $12/11$  & $11/16$  & $0/0$  & $0/0$  \\
\hline
$\texttt{TOT}$ & $40/40$  & $17/16$  & $13/13$  & $14/14$  & $47/46$  & $11/11$  & $32/33$  & $33/32$  & $32/34$  & $17/15$  & $14/19$  \\
\hline
\end{tabular}
    \end{scriptsize}
\caption[Rate Uncertainties: Single-Tag, EMC Channel]{Per-sample rate uncertainty (\% up/down). Single-tag, EMC channel.}\label{tab:1tag_mu}
  \end{table}

  \begin{table}
    \begin{scriptsize}
\begin{tabular}{ c c c c c c c c c c c c } \toprule
    & nonW & $t\bar{t}$ & s-top $s$ & s-top $t$ & Zjets & W+LF & Wbb & Wcc & Wcj & WW & WZ/ZZ \\\midrule
$\texttt{ XS\_ttbar}$ & $0/0$  & $10/-10$  & $10/-10$  & $10/-10$  & $0/0$  & $0/0$  & $0/0$  & $0/0$  & $0/0$  & $0/0$  & $0/0$  \\
$\texttt{ CDFBTAGSF}$ & $0/0$  & $11/-11$  & $11/-11$  & $12/-12$  & $0/0$  & $0/0$  & $0/0$  & $0/0$  & $0/0$  & $24/-24$  & $11/-11$  \\
$\texttt{ CDFLUMI}$ & $0/0$  & $6/-6$  & $6/-6$  & $6/-6$  & $6/-6$  & $0/0$  & $0/0$  & $0/0$  & $0/0$  & $6/-6$  & $6/-6$  \\
$\texttt{ LEPACC\_CEM}$ & $0/0$  & $2/-2$  & $2/-2$  & $2/-2$  & $2/-2$  & $0/0$  & $0/0$  & $0/0$  & $0/0$  & $2/-2$  & $2/-2$  \\
$\texttt{ QCD\_CEM}$ & $40/-40$  & $0/0$  & $0/0$  & $0/0$  & $0/0$  & $0/0$  & $0/0$  & $0/0$  & $0/0$  & $0/0$  & $0/0$  \\
$\texttt{ CDFVHF}$ & $0/0$  & $0/0$  & $0/0$  & $0/0$  & $0/0$  & $0/0$  & $30/-30$  & $30/-30$  & $0/0$  & $0/0$  & $0/0$  \\
$\texttt{ CDFWC}$ & $0/0$  & $0/0$  & $0/0$  & $0/0$  & $0/0$  & $0/0$  & $0/0$  & $0/0$  & $30/-30$  & $0/0$  & $0/0$  \\
$\texttt{ CDFMISTAG}$ & $0/0$  & $0/0$  & $0/0$  & $0/0$  & $0/0$  & $21/-21$  & $0/0$  & $0/0$  & $0/0$  & $0/0$  & $0/0$  \\
$\texttt{ ZJETS}$ & $0/0$  & $0/0$  & $0/0$  & $0/0$  & $45/-45$  & $0/0$  & $0/0$  & $0/0$  & $0/0$  & $0/0$  & $0/0$  \\
$\texttt{ ISRFSRPDF}$ & $0/0$  & $0/0$  & $0/0$  & $0/0$  & $0/0$  & $0/0$  & $0/0$  & $0/0$  & $0/0$  & $4/-4$  & $4/-4$  \\
$\texttt{ CDFJES}$ & $0/0$  & $-4/10$  & $-5/3$  & $-4/2$  & $-9/-1$  & $3/-3$  & $-7/8$  & $-15/2$  & $-11/9$  & $-18/6$  & $-4/-5$  \\
$\texttt{ CDFQ2}$ & $0/0$  & $0/0$  & $0/0$  & $0/0$  & $0/0$  & $0/1$  & $18/9$  & $6/17$  & $5/-9$  & $0/0$  & $0/0$  \\
\hline
$\texttt{TOT}$ & $40/40$  & $17/19$  & $17/16$  & $17/17$  & $46/45$  & $21/21$  & $35/32$  & $34/35$  & $32/32$  & $31/26$  & $14/14$  \\
\hline
\end{tabular}
    \end{scriptsize}
\caption[Rate Uncertainties: Double-Tag, CEM Channel]{Per-sample rate uncertainty (\% up/down). Double-tag, CEM channel.}\label{tab:2tag_cem}
  \end{table}

  \begin{table}
    \begin{scriptsize}
\begin{tabular}{ c c c c c c c c c c c c } \toprule
    & nonW & $t\bar{t}$ & s-top $s$ & s-top $t$ & Zjets & W+LF & Wbb & Wcc & Wcj & WW & WZ/ZZ \\\midrule
$\texttt{ XS\_ttbar}$ & $0/0$  & $10/-10$  & $10/-10$  & $10/-10$  & $0/0$  & $0/0$  & $0/0$  & $0/0$  & $0/0$  & $0/0$  & $0/0$  \\
$\texttt{ CDFBTAGSF}$ & $0/0$  & $11/-11$  & $11/-11$  & $12/-12$  & $0/0$  & $0/0$  & $0/0$  & $0/0$  & $0/0$  & $24/-24$  & $11/-11$  \\
$\texttt{ CDFLUMI}$ & $0/0$  & $6/-6$  & $6/-6$  & $6/-6$  & $6/-6$  & $0/0$  & $0/0$  & $0/0$  & $0/0$  & $6/-6$  & $6/-6$  \\
$\texttt{ LEPACC\_PHX}$ & $0/0$  & $2/-2$  & $2/-2$  & $2/-2$  & $2/-2$  & $0/0$  & $0/0$  & $0/0$  & $0/0$  & $2/-2$  & $2/-2$  \\
$\texttt{ QCD\_PHX}$ & $40/-40$  & $0/0$  & $0/0$  & $0/0$  & $0/0$  & $0/0$  & $0/0$  & $0/0$  & $0/0$  & $0/0$  & $0/0$  \\
$\texttt{ CDFVHF}$ & $0/0$  & $0/0$  & $0/0$  & $0/0$  & $0/0$  & $0/0$  & $30/-30$  & $30/-30$  & $0/0$  & $0/0$  & $0/0$  \\
$\texttt{ CDFWC}$ & $0/0$  & $0/0$  & $0/0$  & $0/0$  & $0/0$  & $0/0$  & $0/0$  & $0/0$  & $30/-30$  & $0/0$  & $0/0$  \\
$\texttt{ CDFMISTAG}$ & $0/0$  & $0/0$  & $0/0$  & $0/0$  & $0/0$  & $21/-21$  & $0/0$  & $0/0$  & $0/0$  & $0/0$  & $0/0$  \\
$\texttt{ ZJETS}$ & $0/0$  & $0/0$  & $0/0$  & $0/0$  & $45/-45$  & $0/0$  & $0/0$  & $0/0$  & $0/0$  & $0/0$  & $0/0$  \\
$\texttt{ ISRFSRPDF}$ & $0/0$  & $0/0$  & $0/0$  & $0/0$  & $0/0$  & $0/0$  & $0/0$  & $0/0$  & $0/0$  & $4/-4$  & $4/-4$  \\
$\texttt{ CDFJES}$ & $0/0$  & $-7/11$  & $-5/-2$  & $-9/0$  & $1/-5$  & $3/-2$  & $-6/5$  & $-17/8$  & $-35/32$  & $-50/-50$  & $-4/-4$  \\
$\texttt{ CDFQ2}$ & $0/0$  & $0/0$  & $0/0$  & $0/0$  & $0/0$  & $5/0$  & $6/-7$  & $-0/35$  & $-41/9$  & $0/0$  & $0/0$  \\
\hline
$\texttt{TOT}$ & $40/40$  & $17/20$  & $17/16$  & $19/17$  & $45/46$  & $22/21$  & $31/31$  & $34/47$  & $62/45$  & $56/56$  & $14/14$  \\
\hline
\end{tabular}
    \end{scriptsize}
\caption[Rate Uncertainties: Double-Tag, PHX Channel]{Per-sample rate uncertainty (\% up/down). Double-tag, PHX channel.}\label{tab:2tag_phx}
  \end{table}

  \begin{table}
    \begin{scriptsize}
\begin{tabular}{ c c c c c c c c c c c c } \toprule
    & nonW & $t\bar{t}$ & s-top $s$ & s-top $t$ & Zjets & W+LF & Wbb & Wcc & Wcj & WW & WZ/ZZ \\\midrule
$\texttt{ XS\_ttbar}$ & $0/0$  & $10/-10$  & $10/-10$  & $10/-10$  & $0/0$  & $0/0$  & $0/0$  & $0/0$  & $0/0$  & $0/0$  & $0/0$  \\
$\texttt{ CDFBTAGSF}$ & $0/0$  & $11/-11$  & $11/-11$  & $12/-12$  & $0/0$  & $0/0$  & $0/0$  & $0/0$  & $0/0$  & $24/-24$  & $11/-11$  \\
$\texttt{ CDFLUMI}$ & $0/0$  & $6/-6$  & $6/-6$  & $6/-6$  & $6/-6$  & $0/0$  & $0/0$  & $0/0$  & $0/0$  & $6/-6$  & $6/-6$  \\
$\texttt{ LEPACC\_MU}$ & $0/0$  & $1/-1$  & $1/-1$  & $1/-1$  & $1/-1$  & $0/0$  & $0/0$  & $0/0$  & $0/0$  & $1/-1$  & $1/-1$  \\
$\texttt{ QCD\_MU}$ & $40/-40$  & $0/0$  & $0/0$  & $0/0$  & $0/0$  & $0/0$  & $0/0$  & $0/0$  & $0/0$  & $0/0$  & $0/0$  \\
$\texttt{ CDFVHF}$ & $0/0$  & $0/0$  & $0/0$  & $0/0$  & $0/0$  & $0/0$  & $30/-30$  & $30/-30$  & $0/0$  & $0/0$  & $0/0$  \\
$\texttt{ CDFWC}$ & $0/0$  & $0/0$  & $0/0$  & $0/0$  & $0/0$  & $0/0$  & $0/0$  & $0/0$  & $30/-30$  & $0/0$  & $0/0$  \\
$\texttt{ CDFMISTAG}$ & $0/0$  & $0/0$  & $0/0$  & $0/0$  & $0/0$  & $21/-21$  & $0/0$  & $0/0$  & $0/0$  & $0/0$  & $0/0$  \\
$\texttt{ ZJETS}$ & $0/0$  & $0/0$  & $0/0$  & $0/0$  & $45/-45$  & $0/0$  & $0/0$  & $0/0$  & $0/0$  & $0/0$  & $0/0$  \\
$\texttt{ ISRFSRPDF}$ & $0/0$  & $0/0$  & $0/0$  & $0/0$  & $0/0$  & $0/0$  & $0/0$  & $0/0$  & $0/0$  & $4/-4$  & $4/-4$  \\
$\texttt{ CDFJES}$ & $0/0$  & $-7/9$  & $-3/2$  & $-2/0$  & $7/2$  & $-0/0$  & $-10/9$  & $-6/10$  & $-14/7$  & $-1/-6$  & $-2/-3$  \\
$\texttt{ CDFQ2}$ & $0/0$  & $0/0$  & $0/0$  & $0/0$  & $0/0$  & $-2/-5$  & $14/10$  & $-6/-8$  & $-8/-11$  & $0/0$  & $0/0$  \\
\hline
$\texttt{TOT}$ & $40/40$  & $18/19$  & $16/16$  & $17/17$  & $46/45$  & $21/22$  & $35/33$  & $31/33$  & $34/33$  & $25/26$  & $13/13$  \\
\hline
\end{tabular}
    \end{scriptsize}
\caption[Rate Uncertainties: Double-Tag, CMUP $+$ CMX Channel]{Per-sample rate uncertainty (\% up/down). Double-tag, CMUP $+$ CMX channel.}\label{tab:2tag_mu}
  \end{table}

  \begin{table}
    \begin{scriptsize}
\begin{tabular}{ c  c c c c c c c c c c c  } \toprule
    & nonW & $t\bar{t}$ & s-top $s$ & s-top $t$ & Zjets & W+LF & Wbb & Wcc & Wcj & WW & WZ/ZZ \\\midrule
$\texttt{ XS\_ttbar}$ & $0/0$  & $10/-10$  & $10/-10$  & $10/-10$  & $0/0$  & $0/0$  & $0/0$  & $0/0$  & $0/0$  & $0/0$  & $0/0$  \\
$\texttt{ CDFBTAGSF}$ & $0/0$  & $11/-11$  & $11/-11$  & $12/-12$  & $0/0$  & $0/0$  & $0/0$  & $0/0$  & $0/0$  & $24/-24$  & $11/-11$  \\
$\texttt{ CDFLUMI}$ & $0/0$  & $6/-6$  & $6/-6$  & $6/-6$  & $6/-6$  & $0/0$  & $0/0$  & $0/0$  & $0/0$  & $6/-6$  & $6/-6$  \\
$\texttt{ LEPACC\_EMC}$ & $0/0$  & $5/-5$  & $5/-5$  & $5/-5$  & $5/-5$  & $0/0$  & $0/0$  & $0/0$  & $0/0$  & $5/-5$  & $5/-5$  \\
$\texttt{ QCD\_EMC}$ & $40/-40$  & $0/0$  & $0/0$  & $0/0$  & $0/0$  & $0/0$  & $0/0$  & $0/0$  & $0/0$  & $0/0$  & $0/0$  \\
$\texttt{ CDFVHF}$ & $0/0$  & $0/0$  & $0/0$  & $0/0$  & $0/0$  & $0/0$  & $30/-30$  & $30/-30$  & $0/0$  & $0/0$  & $0/0$  \\
$\texttt{ CDFWC}$ & $0/0$  & $0/0$  & $0/0$  & $0/0$  & $0/0$  & $0/0$  & $0/0$  & $0/0$  & $30/-30$  & $0/0$  & $0/0$  \\
$\texttt{ CDFMISTAG}$ & $0/0$  & $0/0$  & $0/0$  & $0/0$  & $0/0$  & $21/-21$  & $0/0$  & $0/0$  & $0/0$  & $0/0$  & $0/0$  \\
$\texttt{ ZJETS}$ & $0/0$  & $0/0$  & $0/0$  & $0/0$  & $45/-45$  & $0/0$  & $0/0$  & $0/0$  & $0/0$  & $0/0$  & $0/0$  \\
$\texttt{ ISRFSRPDF}$ & $0/0$  & $0/0$  & $0/0$  & $0/0$  & $0/0$  & $0/0$  & $0/0$  & $0/0$  & $0/0$  & $4/-4$  & $4/-4$  \\
$\texttt{ CDFJES}$ & $0/0$  & $-7/9$  & $-0/-0$  & $1/-5$  & $5/-3$  & $-1/1$  & $-10/11$  & $-5/19$  & $-7/6$  & $-10/-14$  & $8/-8$  \\
$\texttt{ CDFQ2}$ & $0/0$  & $0/0$  & $0/0$  & $0/0$  & $0/0$  & $-3/-1$  & $17/5$  & $15/0$  & $27/20$  & $0/0$  & $0/0$  \\
\hline
$\texttt{TOT}$ & $40/40$  & $18/19$  & $17/17$  & $18/18$  & $46/46$  & $21/21$  & $36/32$  & $34/36$  & $41/37$  & $27/29$  & $16/16$  \\
\hline
\end{tabular}
    \end{scriptsize}
\caption[Rate Uncertainties: Double-Tag, EMC Channel]{Per-sample rate uncertainty (\% up/down). Double-tag, EMC channel.}\label{tab:2tag_emc}
  \end{table}

\renewcommand{\tabcolsep}{6pt}

\backmatter 

\newpage
\sloppy
{\raggedright
\printbibliography
}
\fussy

\newpage
\chapter{Acknowledgement}\label{Acknowledgement}

It seems incredible but the Ph.D. really comes to the end! 

For sure this has been the most intense period of my life both by the scientific point of view, with {\em my} Di-Bosoni, the Higgs, tons of new things learned and discovered, and also by the personal point of view,
with thousands of miles across the World and several houses and cities where I spent my time. 

The real risk was to get lost along the way but, luckily, I have been continuously surrounded, helped and supported by so so many nice people that it is really hard to thank everybody in one page.

For sure I couldn't had made it without the support of Prof. Giorgio Chiarelli (who now knows me by the far 2007); the work wouldn't had been possible without him but neither without the continuous exchange of ideas with Dr. Sandra Leone and within all the CDF Pisa Group.

Fermilab has been the other indispensable ingredient: the best place in the Chicago suburbs to do Physics and also all the rest. I met there a great research team (all the WHAM'ers), wonderful scientists and researchers from whom I learned so much.

However there have not just been Physics in these years\footnote{Although often friendship, study, work, spare time and Physics all overlaps in a indistinguishable way...}, my family has been always with me also from the other side of the ocean or farther away. No possibility to get anywhere without their help. Then there is Celine, ``la Piccolina'', that I met once, lost, and luckily I met again and now we are together!

And finally there are all the friends from {\em Sancasciani}, {\em Cottonwood}, {\em Siena}, {\em Chicago}, {\em Pisa}, the {\em old} and the {\em new Summer Students}, the {\em ControTV} group and others I am surely missing. I spend with all of you a very good time and first or later I will meet again everybody scattered all around the World.

This travel lasted three years (well almost four), a pretty long period in the lifespan of somebody that is not even thirty but it flowed rapidly like a mountain torrent.

\end{document}